\documentclass[11pt,twoside,a4paper,openright]{book}

\usepackage{floatflt} 
\usepackage{pdfsync} 
\usepackage[bookmarks=true, pdfpagelabels=true, colorlinks=true, linkcolor=black, urlcolor=black, citecolor=blue, anchorcolor=blue]{hyperref}
\usepackage[toc,page]{appendix}

\usepackage[cyr]{aeguill}
\usepackage{booktabs}
\usepackage[utf8]{inputenc}
\usepackage[T1]{fontenc}
\usepackage{lmodern}
\usepackage{tabularx}
\usepackage{slashed}

\usepackage{amsmath,amsthm,amssymb,bm,fixcmex}	  
\usepackage{graphics,xcolor}			  
\usepackage{times}
\usepackage[pdftex]{graphicx}
\usepackage{emptypage}
\usepackage{cite}
\usepackage[export]{adjustbox}
\usepackage{multirow}
\usepackage{array}
\usepackage{pifont}
\newcommand{\cmark}{\text{\ding{51}}}

\allowdisplaybreaks

\usepackage{etoolbox}

\usepackage[font=small,labelfont=bf]{caption}

\usepackage[a4paper,total={12cm,20cm},centering,headsep=.5cm]{geometry}
\usepackage{fancyhdr}
\usepackage{sectsty}
\sectionfont{\fontsize{16}{18}\usefont{OT1}{phv}{m}{n}\selectfont}
\subsectionfont{\fontsize{13}{15}\usefont{OT1}{phv}{m}{n}\selectfont}
\subsubsectionfont{\fontsize{11}{13}\usefont{OT1}{phv}{m}{n}\selectfont}

\usepackage[Lenny]{fncychap}
\ChNameVar{\fontsize{14}{16}\usefont{OT1}{phv}{m}{n}\selectfont}
\ChNumVar{\fontsize{60}{62}\usefont{OT1}{phv}{m}{n}\selectfont}
\ChTitleVar{\fontsize{24}{26}\usefont{OT1}{phv}{m}{n}\selectfont}

\makeatletter
\renewcommand\ps@plain{\let\@mkboth\@gobbletwo
     \let\@oddhead\@empty
     \def\@oddfoot{\reset@font\hfil}
     \let\@evenhead\@empty\let\@evenfoot\@oddfoot}

\patchcmd{\@makechapterhead}{\vspace*{50\p@}}{}{}{}
\patchcmd{\@makeschapterhead}{\vspace*{50\p@}}{}{}{}

\g@addto@macro\normalsize{%
  \setlength\abovedisplayskip{10pt}
  \setlength\belowdisplayskip{10pt}
  \setlength\abovedisplayshortskip{10pt}
  \setlength\belowdisplayshortskip{10pt}
}

\makeatother

\usepackage{makeidx} 
\makeindex

\usepackage{mdframed} 
\usepackage{lipsum} 
\usepackage{xcolor}
\usepackage{xparse} 
\definecolor{graylight}{cmyk}{.30,0,0,.67} 

\newmdenv[ 
  linecolor=graylight,
  topline=false,
  bottomline=false,
  rightline=false,
  skipabove=\topsep,
  skipbelow=\topsep
]{leftrule}

\NewDocumentEnvironment{example}{O{\textbf{Example:}}} 
{\begin{leftrule}\noindent\textcolor{graylight}{#1}\par}
{\end{leftrule}}

\def\Lag{\mathcal{L}}
\def\smefit{{\tt SMEFiT}~}

\def\tev{\mathrm{TeV}}
\def\gev{\mathrm{GeV}}
\newcolumntype{P}[1]{>{\centering\arraybackslash}p{#1}}
\newcolumntype{C}[1]{>{\centering\arraybackslash}p{#1}}
\newcommand{\bra}[1]{\ensuremath{\left\langle#1\right|}}
\newcommand{\ket}[1]{\ensuremath{\left|#1\right\rangle}}

\newcommand{\revised}[1]{\textcolor{black}{{ #1}}}

\newcommand\sss{\scriptscriptstyle}
\newcommand{\MG}{ {\sc Madgraph5\_aMC@NLO} }

\newcommand{\ie} {{\it i.e.}\,}

\newcommand {\beq} {\begin{equation}}
\newcommand {\eeq} {\end{equation}}
\newcommand {\bea} {\begin{eqnarray}}
\newcommand {\eea} {\end{eqnarray}}
\newcommand{\GeV}{{\rm\ GeV}}

\newcommand{\TeV}{{\rm\ TeV}}

\definecolor{darkred}{rgb}{0.7, 0.0, 0.0}

\def\be{\begin{equation}}
\def\ee{\end{equation}}
\def\bsp#1\esp{\begin{split}#1\end{split}}

\newcommand{\OO}{\ensuremath{\mathcal{O}}}
\newcommand{\pdp}{\ensuremath{\phi^\dagger\phi}}
\renewcommand{\phi}{\ensuremath{\varphi}}

\newcommand{\bpm}{\begin{pmatrix}}      
\newcommand{\epm}{\end{pmatrix}}

\newcommand\mw{m_{\sss W}}
\newcommand\mz{m_{\sss Z}}

\newcommand\mt{m_{\sss t}}

\newcommand*{\tmp}[4]{\ensuremath{%
    {#4%
    \ifx\empty#3\empty\ifx\empty#1\empty\else^{#1}\fi\else^{#1(#3)}\fi%
    \ifx\empty#2\empty\else_{#2}\fi}%
}}
\newcommand*{\qq }[4][]{\tmp{#2}{#3}{#4}{#1{\mathcal{O}}}}
\newcommand*{\ccc}[4][]{\tmp{#2}{#3}{#4}{#1{c}}}

\newcommand{\Op}[1]{\OO_{\sss #1}}
\newcommand{\Opp}[2]{\OO_{\sss #1}^{\sss #2}}

\newcommand{\red}[1]{ \textcolor{red}{#1} }
 \def\lra#1{\overset{\text{\scriptsize$\leftrightarrow$}}{#1}}
\newcommand{\gth}{g_{\sss th}}
\newcommand{\gwh}{g_{\sss Wh}}
\newcommand{\gzh}{g_{\sss Zh}}
\newcommand{\gztr}{g^{\sss Z}_{t_{\sss R}}}
\newcommand{\gztl}{g^{\sss Z}_{t_{\sss L}}}
\newcommand{\gzbr}{g^{\sss Z}_{b_{\sss R}}}
\newcommand{\gzbl}{g^{\sss Z}_{b_{\sss L}}}
\newcommand{\gbtw}{g_{\sss btW}}
\newcommand{\gta}{g_{\sss t\gamma}}
\newcommand{\gwa}{g_{\sss W\gamma}}
\newcommand{\gwz}{g_{\sss WZ}}

\newcommand{\lp}{\left(}
\newcommand{\rp}{\right)}
\newcommand{\lc}{\left[}
\newcommand{\rc}{\right]}
\newcommand{\la}{\left\langle}
\newcommand{\ra}{\right\rangle}

\newcommand{\mpmm}{\mu^+\mu^-}

\newcommand{\invab}{{\rm ~ab^{-1}}}

\begin{document}  

%
%
%


\begin{figure}[t!]
\centering
\includegraphics[scale=0.3]{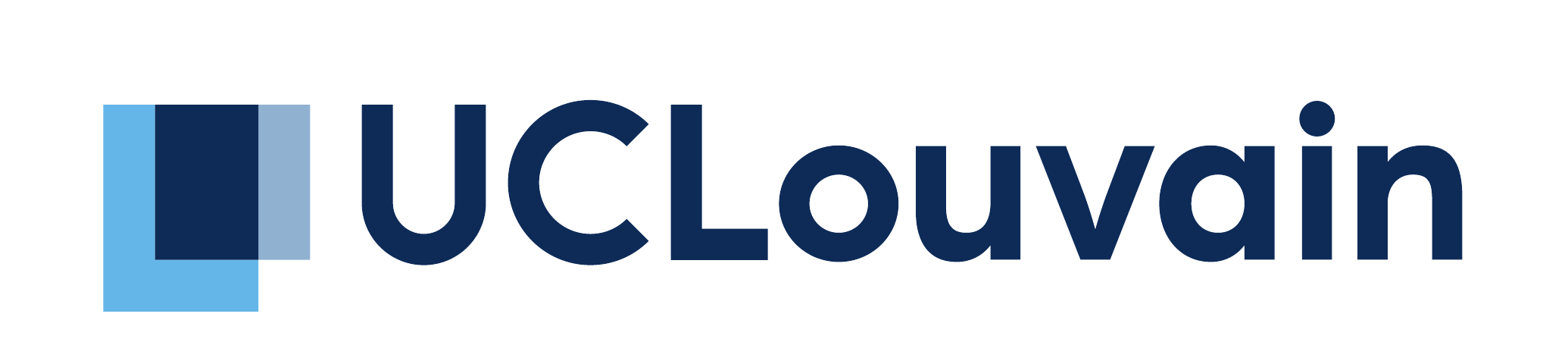}
\end{figure}

\vspace{1.5cm}
\begin{center}
\parbox{0.95\textwidth}{\fontsize{22}{30}\selectfont\centering{Searches for new interactions within the SMEFT framework at present and future colliders}}
\end{center}
\vspace{0.6cm}
\begin{center}
Doctoral dissertation presented by \\
\vspace{2mm}
{\Large Luca Mantani}\\
\vspace{2mm}
in fulfilment of the requirements for the degree of Doctor in Sciences.
\end{center}
\vspace{\fill}
\begin{center}

\begin{tabular}{llc} 
\multicolumn{3}{c}{\large Thesis Jury}                                  \\
                    &               &                           \\
\toprule 
Prof. Vincent Lemaître       & President     & Université Catholique de Louvain  \\
Prof. Fabio Maltoni      & Promotor    & Université Catholique de Louvain  \\
Prof. Céline Degrande      &    & Université Catholique de Louvain  \\
Prof. Alberto Mariotti       &     & Vrije Universiteit Brussel  \\
Dr. Eleni Vryonidou      &     & University of Manchester  \\
Prof. Verónica Sanz      &    & University of Sussex  \\
\bottomrule
\end{tabular}

\vspace{44pt}

\vspace*{0.5cm}
\textsl{August, 2021}\\[1pt]
\end{center}
\thispagestyle{empty}


\hfill
\newenvironment{acknowledgements}%
    {\cleardoublepage\thispagestyle{empty}\null\vfill\begin{center}%
    \bfseries Acknowledgements\end{center}}%
    {\vfill\null}
        \begin{acknowledgements}
        First and foremost, I would like to thank my supervisor Fabio Maltoni for the guidance and support during my years
        at UCLouvain. Not only he provided me with guidance in research, but gave me advice throughout my PhD on numerous
        occasions, helping me to find my way in the complex world of Academia. Without him, this thesis would have not been possible.

        My deepest gratitude goes to Ambresh, Manoj and Ken, for their patience and kindness when I started my PhD. Those
        days would have been much more difficult without their invaluable help.

        I would also like to extend my gratitude to the irreplaceable supervision of Eleni and Chiara. Half of the things I know, I learned
        from them.

        I would also particularly like to thank all members of my thesis committee, Veronica Sanz, Celine Degrande, Vincent Lemaitre, Alberto Mariotti, Eleni Vryonidou as well as Fabio. Their comments and questions have been very insightful and helped in 
        improving this thesis significantly.

        I am grateful for the wonderful research environment I found at CP3, both from a research perspective and a social one.
        In particular, I want to thank Ankit, Andrew, Hesham, Angela, Rafael, Philipp, Claudio, Marina, Pietro and Richard.
        I was expecting to find many colleagues, but not so many friends.

        I would like to express my deepest gratitude to Alessia. You have been by my side all the way, both in good and bad times.
        I found the best friend that one can hope for.

        Last but not least, I will be eternally grateful to my family for the unconditional support during all these years.
        Needless to say, this thesis is for them.

        This work has received funding from the European Union’s Horizon 2020 research and innovation programme as part of the Marie Skłodowska-Curie Innovative Training Network MCnetITN3 (grant agreement no. 722104).
        \end{acknowledgements}


\hfill
\newenvironment{Abstract}%
    {\cleardoublepage\thispagestyle{empty}\null\vfill\begin{center}%
    \bfseries Abstract\end{center}}%
    {\vfill\null}
        \begin{Abstract}
        The existence of 
        Beyond Standard Model (BSM) physics is firmly suggested by both experimental observations (Dark Matter, neutrino masses) and theoretical
        arguments.
        In the hypothesis that the scale of new physics is considerably
        higher than the energies probed at colliders, we can parametrise modified interactions induced by BSM effects among SM particles
        in a model-independent framework, the Standard Model Effective Field Theory (SMEFT). Searches for indirect evidence of
        new physics are conceptually different from the direct ones that have characterised the first part of the LHC program, and
        both experimental and phenomenological studies are needed in order to maximise the chances of uncovering a BSM signal.
        In this thesis, several phenomenological aspects of the SMEFT are discussed, both at present and future colliders.
        A characteristic feature of modified interactions is that they can induce unitarity violating effects which can be exploited to gain sensitivity. In this direction, a thorough study of the top quark electroweak
        sector will be presented, focusing on $2\to2$ scatterings and their embeddings in physical processes at colliders. This analysis allows us to identify
        several final states that have a good potential to explore the SMEFT parameter space and that could be particularly relevant in a global analysis.
        One of the key features of the SMEFT is indeed that deviations from the SM interactions are correlated and global interpretations
        are therefore of fundamental importance. A combined interpretation of the Higgs, top and diboson data from the LHC is here presented and
        the interplay between the various datasets discussed.
        Finally, the physics potential of a futuristic muon collider will be analysed, focusing in particular on the prospects to
        determine the Higgs self-interactions, a task that is arduous even in proposed $100$ TeV proton
        colliders.
        \end{Abstract}

\clearpage

\begin{center}

\vspace{121pt}

{\LARGE Associated Publications:}
\vspace{22pt}
\begin{description}
\item[\cite{Maltoni:2019aot}] 
Fabio Maltoni, Luca Mantani and Ken Mimasu,“Top-quark electroweak interactions at high energy”, JHEP, vol. 10, pp. 004, 2019.
\item[\cite{Ethier:2021bye}]
Jacob J. Ethier, Fabio Maltoni, Luca Mantani, Emanuele R. Nocera, Juan Rojo, Emma Slade, Eleni Vryonidou, and Cen Zhang, “Combined SMEFT interpretation of Higgs, diboson, and top quark data from the LHC”, arXiv:2105.00006 [hep-ph].
\item[\cite{Costantini:2020stv}]
Antonio Costantini, Federico De Lillo, Fabio Maltoni, Luca Mantani, Olivier Mattelaer, Richard Ruiz, and Xiaoran Zhao, “Vector boson fusion at multi-TeV muon colliders”, JHEP, vol. 09, pp. 080, 2020.
\item[\cite{Chiesa:2020awd}]
Mauro Chiesa, Fabio Maltoni, Luca Mantani, Barbara Mele, Fulvio Piccinini, and Xiaoran Zhao, “Measuring the quartic Higgs self-coupling at a multi-TeV muon collider”, JHEP, vol. 09, pp. 098, 2020.
\end{description}

\vspace{22pt}

During my PhD, I have published other works of research not presented in this manuscript:

\begin{description}
\item[\cite{Ambrogi:2018jqj}] Federico Ambrogi, Chiara Arina, Mihailo Backovic, Jan Heisig, Fabio Maltoni, Luca Mantani, Olivier Mattelaer, and Gopolang Mohlabeng, “MadDM v.3.0: a Comprehensive Tool for Dark Matter Studies”, Phys. Dark Univ., vol. 24, pp. 100249, 2019.
\item[\cite{Arina:2020udz}] Chiara Arina, Benjamin Fuks, and Luca Mantani, “A universal frame- work for t-channel dark matter models”, Eur. Phys. J. C, vol. 80, no. 5, pp. 409, 2020.
\item[\cite{Arina:2020tuw}] Chiara Arina, Benjamin Fuks, Luca Mantani, Hanna Mies, Luca Panizzi, and Jakub Salko, “Closing in on t-channel simplified dark matter models”, Phys. Lett. B, vol. 813, pp. 136038, 2021.
\item[\cite{Cermeno:2021rtk}] M. Cermeño, C. Degrande, and L. Mantani, “Circular polarisation of gamma rays as a probe of dark matter interactions with cosmic ray electrons,” arXiv:2103.14658 [hep-ph].
\end{description}

\end{center}

\pagestyle{empty}
\tableofcontents


%
%
%
\phantomsection
\chapter*{Introduction}
\setcounter{page}{1}
\addcontentsline{toc}{chapter}{Introduction}

\hfill
\begin{minipage}{10cm}

{\small\it 
``If I have seen further it is by standing on the shoulders of Giants.''}

\hfill {\small Isaac Newton, \textit{Letter to Robert Hooke}}
\end{minipage}

\vspace{0.5cm}

The human endeavour to find an explanation for natural phenomena has its roots in ancient Greece (650 BC – 480 BC), when the first 
philosophers started to adopt logic, reason and observations in order to interpret and model Reality. Their ideas and hypotheses 
stimulated the rise of an intellectual movement that lead, many centuries later, to the formulation of the scientific method
by Galileo Galilei and the birth of modern physics.

The leitmotif that characterised the development of our understanding of Nature is the need to find an order to what appears
to be an extremely chaotic world. The present interpretation of Reality is that all of the macroscopic and microscopic phenomena we observe
are ultimately described by the interactions among fundamental constituents which we call particles. Their existence has been 
confirmed with a plethora of experiments in the last century and has brought us to the formulation of the Standard Model
of Particle Physics (SM), a theory which provides a framework to describe, interpret and predict most of the current phenomenological
observations.

In order to gain insight on the world of fundamental particles, we need to be able to explore Nature at exceptionally small
distances. The most straightforward way to do this is by building particle accelerators and collide beams of known particles
at very high energies, with the objective of producing new unobserved particles. It is in this line of thought, that the most
powerful collider humankind has ever created, the Large Hadron Collider, has been put into operation in 2009, leading us 
to the discovery of the Higgs boson in 2012. The SM is now complete, partially satisfying our primordial need for an order.

However, our endeavour is not yet over as we aspire to an even more fundamental understanding. Indications of Beyond Standard Model (BSM) physics have been observed (Dark Matter, neutrino masses, etc.)
and puzzles originating from theoretical arguments require more satisfying explanations. Despite its success, the SM is not able to provide answers to any of these enigmas,
motivating the need for a new and more fundamental theory of Nature. Nonetheless, the evidence and information of BSM physics at our disposal
is still moderate and a profusion of SM extensions have been proposed over the last decades unsuccessfully. The lack of 
direct discoveries of new fundamental states at the LHC to-date, and the looming hypothesis that BSM particles are considerably heavier
than the energies that we can currently probe at colliders, suggest that a different perspective might be needed.

At the beginning of the 19th century, it was observed that the orbit of Uranus was exhibiting some irregularities with 
respect to the predictions of Newtonian gravitational laws. Further measurements and increased precisions in the theoretical 
predictions, helped in understanding that an unseen planet was responsible for that perturbation and more than 20 years later,
Neptune was directly observed. The indirect prediction of its existence was only possible because of our ability to perform
precise enough measurements and theoretical calculations, which allowed us to spot a deviation from the Solar
System model of the time.

In the same way, we can think of assessing the presence of new particles indirectly, relying on the precision of experiments to
reveal tensions between data and theory. This is a scenario in which instead of searching for new states, we look for modified 
interactions among the known SM particles. The Standard Model Effective Field Theory (SMEFT) is a powerful framework in which the
possible anomalous interactions are parametrised in a model-independent way, while respecting all the symmetries of the 
SM.

This thesis is devoted to the study of the SMEFT framework and the searches for new interactions at present and future colliders.
We are entering an exciting era for precision physics and this calls for an appropriate effort on the theory side to provide
reliable predictions and indicate sensitive observables to measure. In the following, after having introduced the Standard Model (Chapter~\ref{chap:sm}) and the SMEFT paradigm (Chapter~\ref{chap:smeft}), we will discuss many of these aspects,
emphasising in particular the need to exploit the high energy tails of the distributions in the top quark electroweak sector to uncover anomalies in the data (Chapter~\ref{chap:top}).
Furthermore, the complexity of the SMEFT demands a global approach, \ie the simultaneous combination of different observables is of
crucial importance for the success of the strategy and a landmark step in this direction will be presented (Chapter~\ref{chap:globalfit}). Finally, as the community
is currently debating the next experimental directions to take for the post-LHC scenery, we will discuss the discovery potential
of a futuristic and extremely appealing machine, the muon collider, focusing in particular on the prospects to determine the Higgs potential (Chapter~\ref{chap:muon}).


%
%
%

\chapter{The Standard Model of Particle Physics}
\label{chap:sm}
\pagestyle{fancy}

\hfill
\begin{minipage}{10cm}

{\small\it 
``No matter how many instances of white swans we may have observed, this does not justify the conclusion that all swans are white.''}

\hfill {\small Karl Popper, \textit{The Logic of Scientific Discovery}}
\end{minipage}

\vspace{0.5cm}

Over the last century, a combination of extraordinary experimental and theoretical efforts have lead to a deep understanding
of the fundamental structure of the Universe. The result of this endeavour is the formulation of the so-called Standard Model 
of Particle Physics (SM), a theoretical framework that describes the fundamental constituents of Matter and the interactions among them.
The success of the SM can be appreciated for both its remarkable ability to describe most of the observations to-date and for the 
historical role it had in predicting the existence of particles and interactions that were subsequently confirmed by experiments.
One of the most crucial features of the SM is the existence of a scalar boson, the Higgs boson, which is necessary to give 
masses to the fundamental particles in a theoretically consistent way. The discovery of a Higgs-like boson at the LHC in 2012~\cite{Aad_2012,Chatrchyan_2012}
provided us with the last piece of the puzzle and paved the way for future discovery and precision programs at CERN and other facilities.
In the following, we will briefly summarise the SM, focusing in particular on the Electroweak (EW) sector and how
gauge symmetries ensure the unitarity of the theory.

\section{The fundamental building blocks of Nature}

The SM is a Quantum Field Theory~\cite{Glashow:1961tr,Weinberg:1967tq,Salam:1968rm} describing three of the four fundamental interactions (electromagnetic, weak and strong force)
and all the known elementary matter constituents. The latter are spin-$\frac{1}{2}$ fermions, classified in two broad categories: \textit{quarks} and 
\textit{leptons}.
The justification to this distinction is given by the fact that while quarks are carriers of the strong interaction charge (called colour), leptons are subject to the electroweak force only. 
This simple difference has a substantial phenomenological consequence: quarks are not asymptotic states and are confined in colour-neutral bound states called hadrons\footnote{From the greek hadrós which means large, massive.}, such as protons and neutrons. There are six flavours of quarks: up-like quarks (up (u), 
charm (c) and top (t)) carrying electric charge $\frac{2}{3}$, and down-like quarks (down (d), strange(s) and bottom (b)) having electric
charge $-\frac{1}{3}$. These are further pair-related by means of the EW interaction, defining three different generations.
In a similar fashion, leptons are distinguished in charged leptons (electron (e), muon ($\mu$) and tau ($\tau$)) characterised by 
electric charge $-1$ and the corresponding neutrinos $\nu_e$, $\nu_{\mu}$ and $\nu_{\tau}$ with which they form EW pairs.

Interactions among fermionic particles are built upon the fundamental principle of gauge symmetries. The SM is a non-abelian
gauge theory, invariant under the group $G = SU(3)_c \otimes SU(2)_L \otimes U(1)_Y$. Each gauge group introduces in the theory 
massless fundamental gauge bosons (spin-1 particles) in number equal to the dimension of the adjoint representation of the group itself. 
These are the mediating particles of three of the fundamental forces in Nature (gravity is not present in the SM). In particular, $SU(3)_c$ describes the strong force by 
means of eight gauge bosons called \textit{gluons} while $SU(2)_L \otimes U(1)_Y$ describes the EW interactions, which are mediated 
by the electroweak bosons $W^\pm$, $Z$ and the photon ($\gamma$).

While photons and gluons are actually massless particles mediating long-range interactions, as dictated by gauge invariance,
the weak bosons $W^\pm$ and $Z$ are responsible for a short-range force and are massive. In addition to this, the SM gauge group prohibit us
to explicitly assign masses to the fermions, which are nonetheless observed in experiments. The solution to this conundrum is
given by the last missing particle in the SM realm, the \textit{Higgs boson}. The mechanism through which the Higgs boson is responsible for
the masses of the aforementioned particles will be illustrated in the following section, in which a more theoretically sound construction of
the EW sector of the SM will be presented.

\section{The electroweak SM lagrangian}
\label{sec:EWlag}

The path that led us to formulate a theory of EW interactions started with the observation of neutron decays and the discovery of
the weak force. This force was at first successfully described by the Fermi theory and later improved with the famous V-A theory~\cite{PhysRev.109.193} formulated by
Feynman-Gell-Mann, Marshak-Sudarshan and Sakurai. They suggested a generalisation of the Fermi theory in which weak interactions 
were described by an effective Hamiltonian
\begin{equation}
\mathcal{H} = \frac{G_F}{\sqrt{2}} J^\dagger_\mu J^\mu + h.c. \, ,
\end{equation} 
where $J_\mu$ is the weak current which takes the vector minus axial form
\begin{equation}
J_\mu = \bar{\psi^\prime}\gamma_\mu (1 -\gamma_5)\psi
\end{equation}
and $G_F$ is the Fermi constant with dimensions $[M]^{-2}$. This structure is merely dictated by phenomenology, since experimental 
evidence suggest that the weak force is maximally parity violating~\cite{PhysRev.105.1413}: only left-handed fermions are affected.

Despite a very successful description of observations, the theory was unsatisfying from a purely theoretical perspective: 
it is non-renormalisable and violates the unitarity of the S matrix, \ie the conservation of probability.
The two problems are actually deeply related and stem from the fact that the Hamiltonian term is dimension-6\footnote{A more thorough discussion on this will be later presented when illustrating the concept of Effective Field Theories.}.
In particular, at sufficiently high energies, the theory does not provide us any more with reliable predictions. The weak force is just
a low energy manifestation of a more fundamental interaction: the electroweak interaction. The formulation of the unification theory
was provided by Higgs, Englert and Brout~\cite{PhysRevLett.13.321,PhysRevLett.13.508} in 1964. It is important to stress that
this advancement was not driven by an unexpected or unexplained experimental result but it was purely theoretically motivated\footnote{This is in contrast with other paradigm changes and theoretical advancements of the 20th century such as the formulation of Special Relativity by Albert Einstein and the postulation of the existence of quanta of lights by Max Planck, which were respectively driven by the need to find an explanation to the the lack of evidence for the luminiferous aether and the black-body radiation problem.}.

The EW unification is based on the principle of Spontaneous Symmetry Breaking (SSB). In particular, the EW gauge group
\begin{equation}
G_{EW} = SU(2)_L \otimes U(1)_Y \, \to \, U(1)_{EM}
\end{equation}
is broken to the electromagnetic group at low energies. As we will see in the following, the photon is nothing 
but a linear combination of the hypercharge gauge boson of $U(1)_Y$ and the third generator of $SU(2)_L$.
While the number of gauge bosons and structure of the interactions descends directly from the gauge groups, the matter 
content of the model is purely empirically dictated. A summary of the fermion fields and associated representation under the SM gauge groups is
presented in the following Table
\begin{center}
\begin{tabular}{ |c|c|c|c|c|c|c| } 
\hline
Fields & $l_i = (v_L^i, e_L^i)$ & $e_R$ & $q_i = (u_L^i, d_L^i)$ & $u_R$ & $d_R$\\
 \hline
 $SU(3)_c$ & $\bold{1}$ & $\bold{1}$ & $\bold{3}$ & $\bold{3}$ & $\bold{3}$ \\ 
 $SU(2)_L$ & $\bold{2}$ & $\bold{1}$ & $\bold{2}$ & $\bold{1}$ & $\bold{1}$  \\ 
 $U(1)_Y$ & $1/2$ & $-1$ & $1/6$& $2/3$ & $-1/3$ \\ 
 \hline
\end{tabular}
\end{center}
where the index $i$ runs through the three generations.

In order to break the $G_{EW}$ group in the SM, one needs to introduce a complex $SU(2)_L$ doublet $\varphi$ with hypercharge
$1/2$. The Lagrangian for the EW gauge bosons reads

\begin{equation*}
\Lag_{EW} = -\frac{1}{4} W_{\mu\nu}^a W^{a\mu\nu} -\frac{1}{4} B_{\mu\nu} B^{\mu\nu} + (D_\mu \varphi)^\dagger (D^\mu \varphi) + \mu^2 \varphi^\dagger \varphi - \lambda (\varphi^\dagger \varphi)^2 \, ,
\end{equation*}
where $W$ is the $SU(2)_L$ gauge boson field strength, $B$ the hypercharge one and the covariant derivative $D_\mu$ reads
\begin{equation}
D_\mu = \partial_\mu - i g W_{\mu}^a \tau^a + i g^\prime Y B_\mu \, ,
\end{equation}
where $\tau^a$ are the $SU(2)_L$ generators and $Y$ the hypercharge specific to the field.
The Higgs potential induces a vacuum expectation value (vev) $v=\mu/\sqrt{\lambda}$ different from zero, which allows us to recast the
Higgs doublet in the form
\begin{equation}
\varphi = e^{2i\frac{\pi^a \tau^a}{v}} \begin{pmatrix}
0 \\
\frac{v+h}{\sqrt{2}}
\end{pmatrix} \, ,
\label{eq:higgsunitary}
\end{equation}
where $\pi^a$ are the massless Goldstone bosons which will be responsible for the masses of the weak vector bosons, while $h$ is
the celebrated Higgs boson. Plugging the expression in Eq.~\eqref{eq:higgsunitary} in the EW Lagrangian and defining
\begin{align}
Z_\mu &\equiv \cos\theta_w W^3_\mu - \sin\theta_w B_\mu \\
A_\mu &\equiv \sin\theta_w W^3_\mu + \cos\theta_w B_\mu \\
W^{\pm}_\mu &\equiv \frac{1}{\sqrt{2}} (W^1_\mu \mp i W^2_\mu)
\end{align}
with 
\begin{equation}
\tan\theta_w = \frac{g^\prime}{g} \, ,
\end{equation}
we find that the Higgs doublet generates mass terms in the EW Lagrangian
\begin{equation}
\Lag_{mass} = \frac{1}{2} m_Z^2 Z^\mu Z_\mu + m_W^2 W_\mu^+ W_\mu^- - \frac{1}{2} m_h^2 h^2 \, .
\end{equation}
The masses are related to the EW input parameters and read
\begin{align}
m_W &= \frac{v}{2} g \\
m_Z &= \frac{m_W}{\cos\theta_w} \\
m_h &= \sqrt{2 \lambda v^2}
\end{align}
which already makes an unambiguous prediction on the $W$ boson mass: it has to be smaller than the $Z$ mass. Notice also that the whole mechanism has left the field $A_\mu$ mediating the electromagnetic interaction massless.

The EW gauge sector has four free parameters $\mu, \lambda, g, g^\prime$. However, from an experimental perspective it is 
easier to use another set of input parameters, namely
\begin{align}
m_Z &= 91.19~\gev, \quad m_W = 80.38~\gev, \\
\quad m_h &= 125~\gev, \quad G_F = 1.166\cdot10^{-5}~\gev^{-2} \, ,
\end{align}
where $G_F$ is precisely measured from the muon decay and is in particular directly related\footnote{The precise derivation of this relation will be illustrated in the next section dedicated to Effective Field Theories.} 
to the vev $v = (\sqrt{2} G_F)^{-1/2} = 246~\gev$.

In a similar fashion, the Higgs doublet gives masses to the fermion fields. Explicit Dirac mass terms cannot be written since they violate
the gauge symmetries of the SM. We can however write the Yukawa Lagrangian
\begin{equation}
\Lag_{yuk} = -y_d^{ij} \bar{d}_i \phi^\dagger q_j -y_u^{ij} \bar{u}_i \tilde{\phi}^\dagger q_j -y_e^{ij} \bar{e}_i \phi^\dagger l_j + h.c. \, ,
\end{equation}
where the dual $\tilde{\varphi} = i \sigma_2 \varphi^*$ has hypercharge $-1/2$ and $i,j$ go through the three generations. The matrices $y_{ij}$\footnote{The Yukawa matrices are not diagonal and they define a rotation from the interaction basis to the mass basis, the Cabibbo-Kobayashi-Maskawa matrix $V_{CKM}$~\cite{Cabibbo:1963yz,osti_4474667}. For many practical purposes, the assumption that the CKM matrix is diagonal holds well and will be used unless stated otherwise.} are the Yukawa coefficients and after 
SSB they are responsible for both the generation of masses and the coupling to the Higgs boson of the fermions. In particular, one finds that for a generic fermion $\psi$ of the SM
\begin{equation}
\Lag_{yuk}^\psi = -m_\psi \bar{\psi} \psi - \frac{m_\psi}{v}  \bar{\psi} \psi h \, ,
\end{equation}
with $m_\psi = \frac{y}{\sqrt{2}} v$. The direct consequence of this is that the Higgs boson couples strongly to heavy particles,
while massless particles are not coupled to it. None of the fundamental fermions of the SM is strictly massless, but for practical
purposes, one can effectively often consider only the top quark to be coupled to the Higgs\footnote{This working assumption is called 5-flavour scheme, because five of the six flavours of quarks are considered massless. However it is worth mentioning that while this scheme works well in terms of hard scattering processes, it fails to describe the decay of the Higgs boson, since its most relevant decay mode is the one into a bottom pair.}.

\section{Unitarity and the role of the Higgs boson}
\label{sec:unitarity}

An important feature of Quantum Field Theories is that the predictions preserve probability, \ie the S matrix is unitary.
This simple property has a remarkable consequence: the form that the interactions can have is constrained. Not every theory is unitary
and in particular, it often happens that certain theories break unitarity beyond a certain energy scale and the theory cannot provide 
reliable predictions thereafter.

In order to see this, we recall the \textit{optical theorem}, which relates scattering amplitudes and cross sections. The S matrix is given by
\begin{equation}
S = e^{-iHt} \, ,
\end{equation}
where $H$ is the Hamiltonian of the theory and is Hermitian $H = H^\dagger$. This implies that the S matrix is unitary $S S^\dagger = 1$. We can recast it without loss of generality as
\begin{equation}
S = 1 + i T
\end{equation}
and the unitarity requirement translates into a relation for the so-called transfer matrix $T$
\begin{equation}
i(T^\dagger - T) = T^\dagger T \, .
\end{equation}
Defining a generic initial state $\ket{i}$ and a final state $\bra{f}$, we can sandwich the left-hand side and obtain
\begin{equation}
\bra{f}i(T^\dagger - T)\ket{i} = i (2\pi)^4 \delta^4(p_i-p_f) (\mathcal{M}^*(f \to i) - \mathcal{M}(i \to f)) \, .
\end{equation}
Using the completeness relation, we obtain for the right-hand side
\begin{align*}
&\bra{f}T^\dagger T \ket{i} = \sum_X \int d\Pi_X \bra{f}T^\dagger\ket{i} \bra{X}T\ket{i}\\
&= \sum_X (2\pi)^4 \delta^4(p_f-p_X) (2\pi)^4 \delta^4(p_i-p_X) \int d\Pi_X \mathcal{M}(i \to X) \mathcal{M}^*(f \to X) \, .
\end{align*}
Finally, the \textit{generalised optical theorem} is given by the following:
\begin{align}
\nonumber
&\mathcal{M}(i\to f) - \mathcal{M}^*(f\to i) =\\ 
&i \sum_X \int d\Pi_X (2\pi)^4 \delta^4(p_i-p_X) \mathcal{M}(i \to X) \mathcal{M}^*(f \to X) \, ,
\end{align}
which relates scattering amplitudes on the left-hand side with squared matrix elements, \ie cross sections, on the right-hand side.
This relation must hold at all orders in perturbation theory. A special and interesting case is given for $i=f= A$, with $A$ a 2-particles
state. The theorem reduces to
\begin{equation}
Im \mathcal{M}(A\to A) = 2 E_{CM} |\vec{p}_i| \sum_X \sigma(A \to X) \, ,
\end{equation}
telling us that the imaginary part of the forward scattering amplitude is in direct relation with the total cross section. This special case is often dubbed \textit{optical theorem}.

In particular, if we consider a $2 \to 2$ scattering in the centre of mass frame
\begin{equation}
\sigma(2 \to 2) = \frac{1}{32\pi E_{CM}^2} \int d \cos\theta |\mathcal{M}(\theta)|^2
\end{equation}
we can write without loss of generality the amplitude in partial waves
\begin{equation}
\mathcal{M}(\theta) = 16 \pi \sum_{j=0}^\infty a_j (2j + 1) P_j(\cos\theta) \, ,
\end{equation}
with $P_j$ Legendre polynomials. It can be shown~\cite{Schwartz:2014sze} that the optical theorem translates into conditions for the coefficients $a_j$
\begin{equation}
|a_j| \leq 1, \quad 0 \leq Im(a_j) \leq 1 ,  \quad |Re(a_j)| \leq \frac{1}{2} \, .
\end{equation}
The consequence of these relations is that the $2 \to 2$ scattering amplitude cannot grow with energy. If this happens, unitarity is broken beyond an energy scale defined by the parameters of the theory itself.

\begin{figure}
\centering
\includegraphics[scale=0.27]{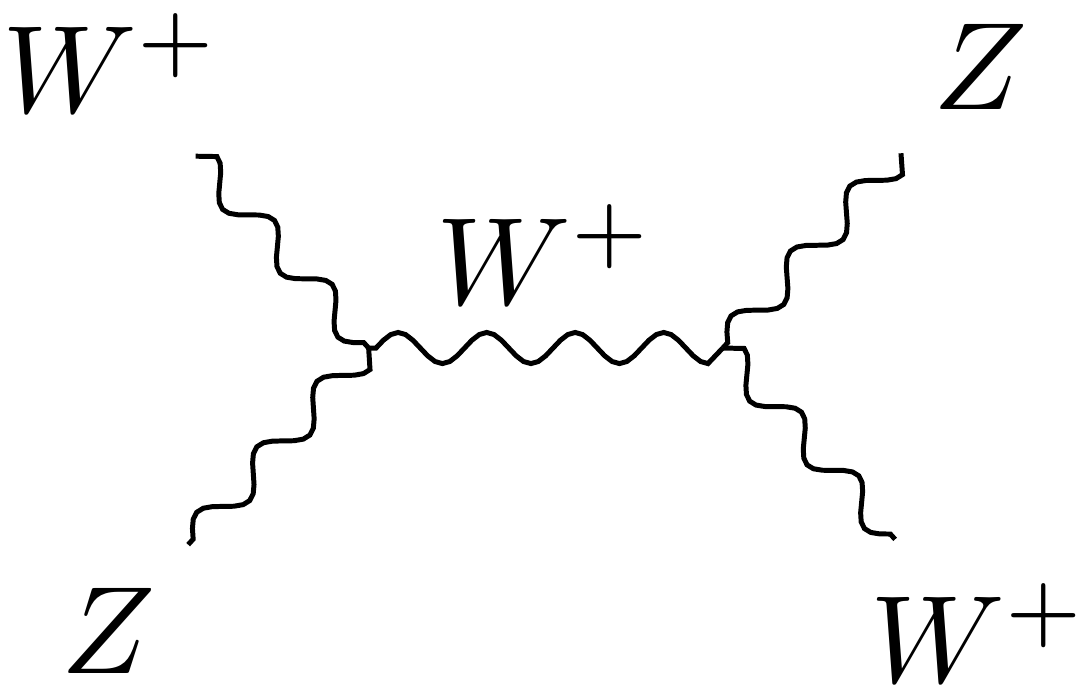} \quad
\includegraphics[scale=0.27]{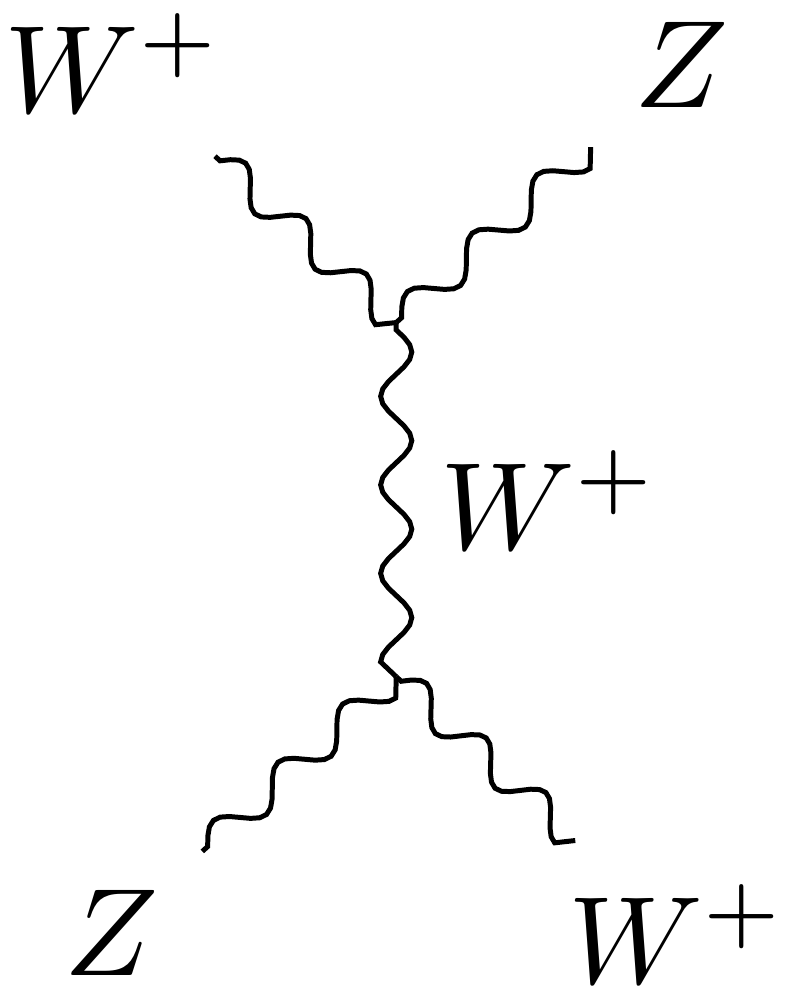} \quad
\includegraphics[scale=0.27]{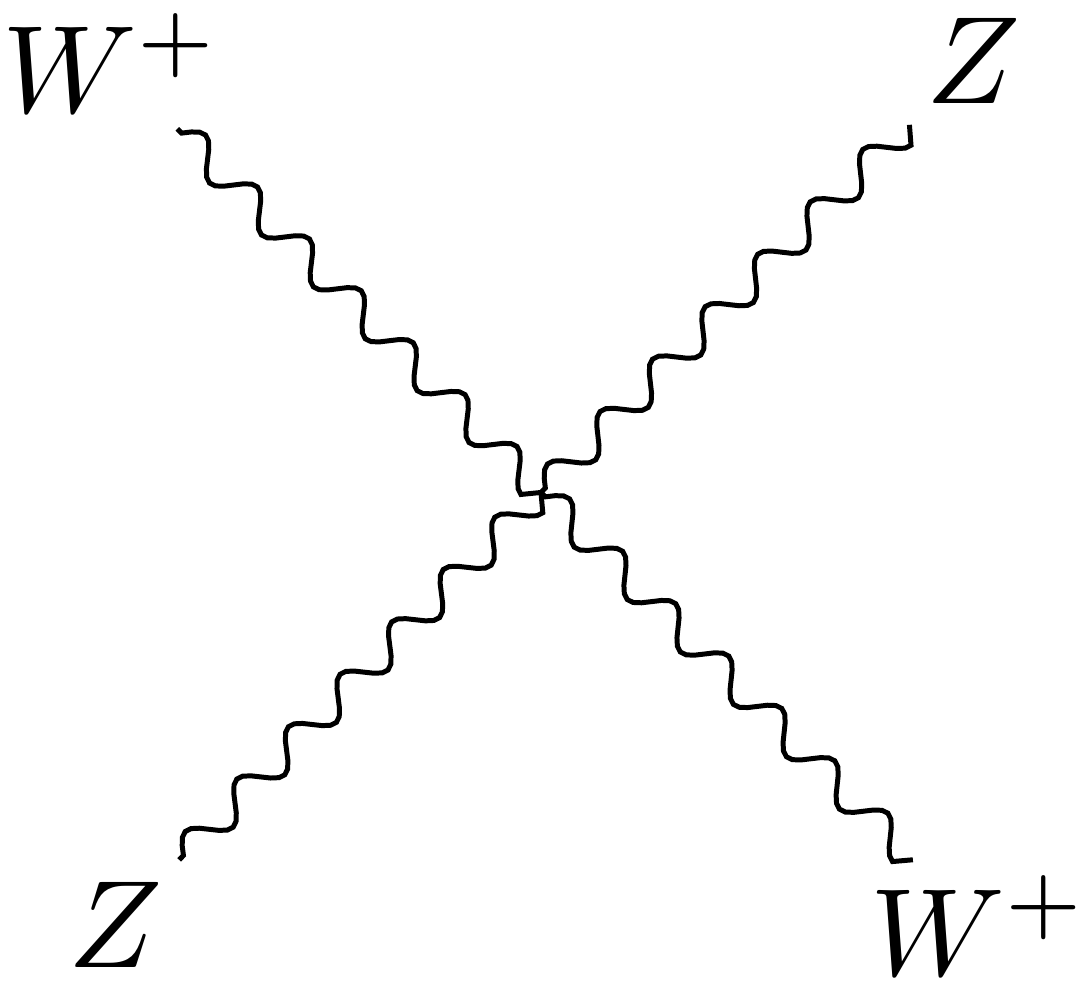} \quad
\includegraphics[scale=0.27]{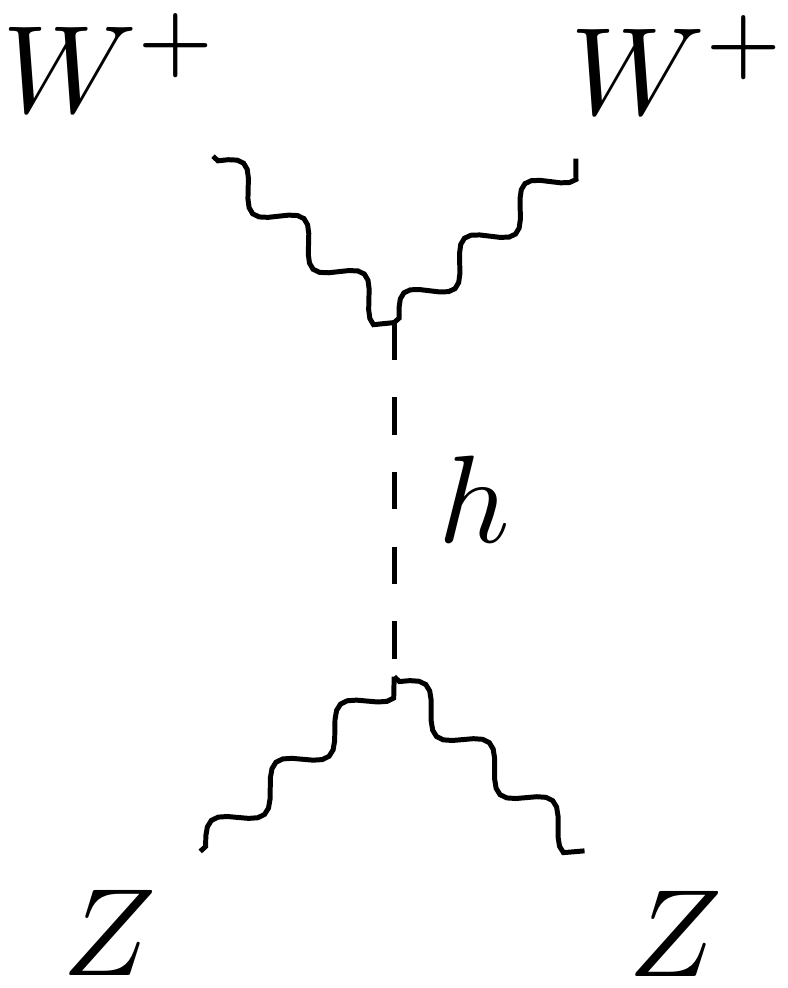} \quad

\caption{The four diagrams for $W^+ Z \to W^+ Z$ in the SM. \label{fig:vbsUnitarity}}
\end{figure}

\revised{Going back to the SM, it is instructive to compute the scattering of longitudinal $W$ and $Z$ bosons in order to understand the importance of the Higgs boson for the theory. Let's consider for instance the physical process $W^+_L Z_L \to W^+_L Z_L$. Diagrams for
this process in the SM are depicted in Fig.~\ref{fig:vbsUnitarity}. If one computes the amplitude neglecting both the quartic gauge coupling and the Higgs boson exchange diagram,
one finds immediately an unacceptable energy dependence $s^2$. This is an effect of explicitly breaking gauge invariance. Once the quartic is included however, unitarity is still broken but with a reduced energy dependence. Indeed, when omitting the diagram with the t-channel Higgs exchange one finds}
\begin{equation}
\mathcal{M}_{noh} = \frac{t}{v^2} + \mathcal{O}(1) \, ,
\end{equation}
which displays a problematic energy growing behaviour. This is a reflection of the fact that gauge theories are broken
if massive gauge bosons are present in the model. The fourth diagram however yields
\begin{equation}
\mathcal{M}_{h} = -\frac{t}{v^2} + \mathcal{O}(1)
\end{equation}
and comes to the rescue by precisely canceling the unitarity violating behaviour, thus restoring unitarity. The Higgs mechanism is the only
theoretically consistent way of assigning masses to gauge bosons while at the same time preserving unitarity. This is a very fascinating 
aspect of the SM: the high energy behaviour of amplitudes is determined by a set of intricate cancellations among various contributions
that would otherwise display unacceptable energy growths. The same holds true for scatterings involving fermions. In particular, the 
scale at which unitarity violation is reached is inversely proportional to the fermion mass~\cite{Appelquist:1987cf,Maltoni:2001dc}. This makes for instance the top quark
potentially the best probe for New Physics effects, as we will discuss in Chapter~\ref{chap:top}. This is the result of the
fact that masses and couplings are all strictly related by the SSB and any spoiling of these relations would yield unitarity violation.

This phenomenon is the reason why we knew that the Higgs boson, or something similarly behaving, should have been there and be found at the
Large Hadron Collider. In particular, one can compute from partial wave unitarity that in order for the theory to be unitary, the Higgs 
boson mass should be $m_h \lesssim 1 \tev$~\cite{PhysRevD.16.1519}.

\revised{When talking of unitarity violating behaviours from an observational perspective, we refer to the fact that
deviations in the data from the theoretical predictions start to follow a pattern which if extrapolated would lead to the breaking of unitarity. However, unitarity violation is just a feature of our theoretical framework and in reality new degrees of freedom will
always come to the rescue and ensure the preservation of probability.}

\section{Motivations for BSM physics}

The SM is arguably the most successful theory in history and is able to explain and describe the vast majority of the phenomena
we observe. However, the job of a scientist is not to validate a theory but to falsify it\footnote{No matter how many positive experimental outcomes one can find, a theory cannot be logically confirmed. The modern approach to science is based on the Falsification Principle first formulated by the philosopher Karl Popper~\cite{Popper:34}.}, in an endless effort to build
approximate models of reality.

In this perspective, there are already several indications that a more general and comprehensive theory is needed. A non-exhaustive
list of unsolved problems is the following:

\begin{itemize}

\item \textbf{Neutrino masses:}
In the SM, neutrinos are massless and right-handed neutrinos are not even present in the model. However, we have clear and conclusive
evidence that neutrinos are light but massive particles. This conclusion stems from the observation of neutrino oscillation 
in solar and atmospheric neutrinos\cite{Bahcall:1964gx,PhysRevLett.12.303,Ahmad:2001an,Ahmad:2002jz,Fukuda:1998mi}. This phenomenon suggests that neutrinos flavour and mass eigenstates are different
(implying the existence of a mass matrix).

\item \textbf{Dark Matter:}
Experimental evidence suggests that only 20\% of the matter content of the Universe is explainable within the SM.
In particular, observations of rotation curves in galaxies\cite{Zwicky:1933gu,1970ApJ...159..379R}, the Cosmological Microwave Background anisotropies\cite{2016} and the Bullet Cluster\cite{Clowe:2003tk} indicate that
a form of unknown, stable and electrically neutral matter is present and we are able to measure its gravitational effects.

\item \textbf{Strong CP problem:}
\revised{The SM symmetries allow a term in the Lagrangian for the gluon strength field of the form $\bar{\theta} G_{\mu \nu} \tilde{G}^{\mu \nu}$. This interaction violates CP in QCD, but no violation of the symmetry has ever been observed in experiments involving strong interactions. 
In particular, the measurement of the electric dipole moment of the neutron poses robust constraints on $\bar{\theta} \lesssim 2\times10^{-10}$ and no explanation
for this can be found within the SM theory\cite{tHooft:1976rip,Peccei:1977ur,Callan:1976je,Belavin:1975fg}.}

\item \textbf{Baryogenesis:}
The Universe is characterised by a matter-antimatter asymmetry, \ie the anti-matter content is almost absent. In order to accommodate this fact, 
one needs C and CP violation in the theory but the SM cannot provide enough of it to explain the observations\cite{Kuzmin:1985mm,Sakharov:1967dj}.

\item \textbf{Gravity:}
The SM describes three of the four fundamental interactions but gravity is not included in the description. A quantum theory of gravity is
needed in order to unify it with the SM, but all efforts so far have not produced a conclusive theory. The SM is expected to break at a scale of $M_{planck} \sim 10^{19}~\gev$, where gravitational effects start to be relevant.

\end{itemize}

Both theoretical and experimental observations seem to lead to the conclusion that the SM is nothing but a low-energy approximation of 
a more fundamental theory. Over the years, many theories have been proposed to address these problems but the lack of experimental
evidence has halted progress in this direction.

%
%
%

\chapter{Standard Model Effective Field Theory}
\pagestyle{fancy}
\label{chap:smeft}

\hfill
\begin{minipage}{10cm}

{\small\it 
``In all fighting, the direct method may be used for joining battle, but indirect methods will be needed in order to secure victory.''}

\hfill {\small Sun Tzu, \textit{The Art of War}}
\end{minipage}

\vspace{0.5cm}

The success of the Standard Model and the lack of clear sign of New Physics (NP) at the LHC pose a dilemma with
respect to future strategies in collider physics. Historically, two approaches have been fruitful to deepen our understanding of Nature.

On the one hand, we have \textit{direct searches}, characterised by the idea that by colliding particles at high energies we can observe the existence of new 
particles by producing them directly. This is for instance the strategy applied for the discovery of the Higgs boson at the LHC. This type
of search is inherently limited by the energy reach of the machines at our disposal, \ie we cannot discover particles that are heavier than
the energy of the collisions. This strategy is however not too dependent on the precision of the SM predictions, since a bump in the data
could in principle be discerned from the background regardless of theory. The absence of 
indications in this perspective, after a decade of data taking at CERN, seem to suggest that NP might be heavier than the reach of the collider.

\begin{figure}
\centering
\includegraphics[width=0.46\textwidth]{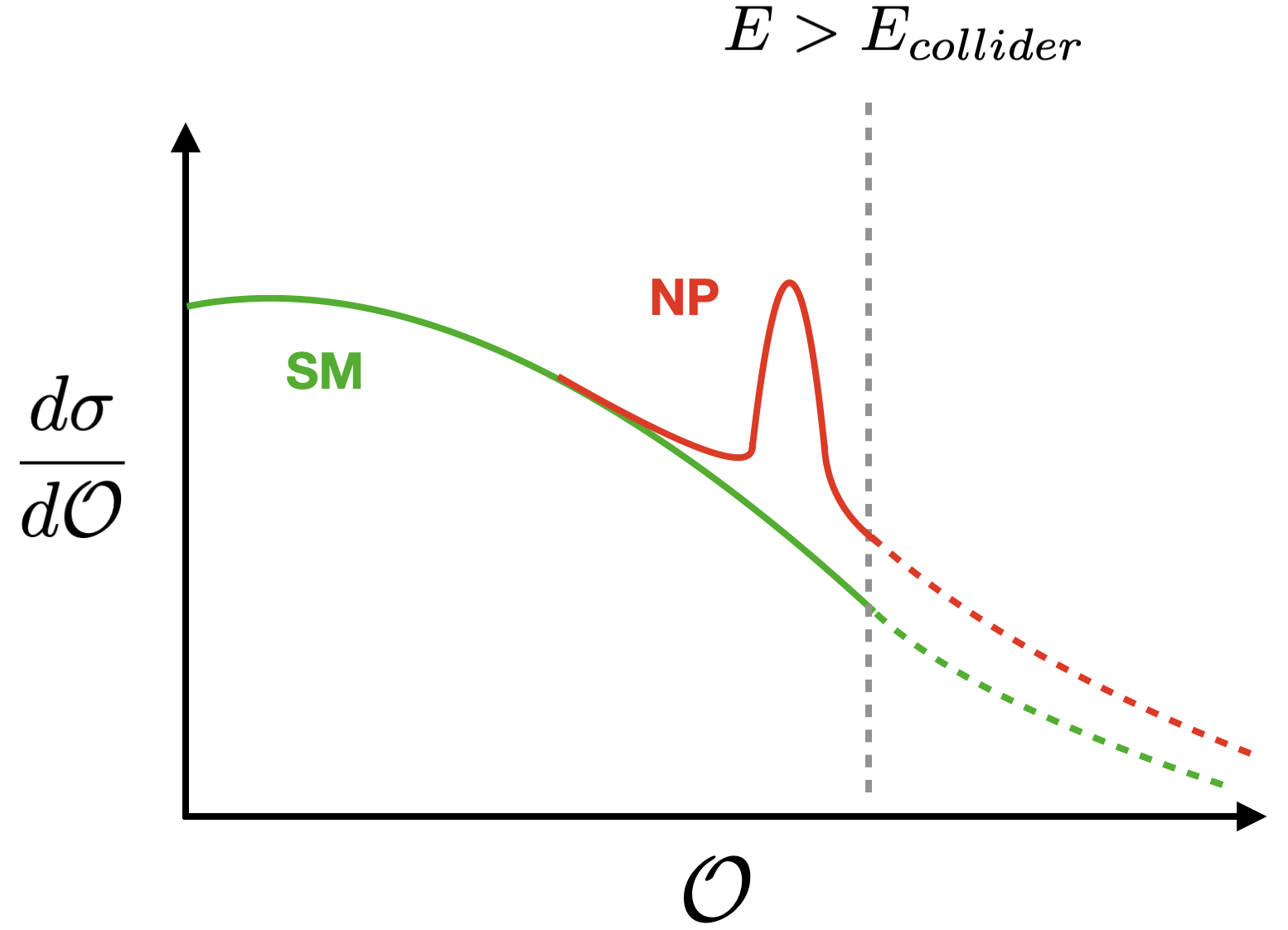} \quad
\includegraphics[width=0.46\textwidth]{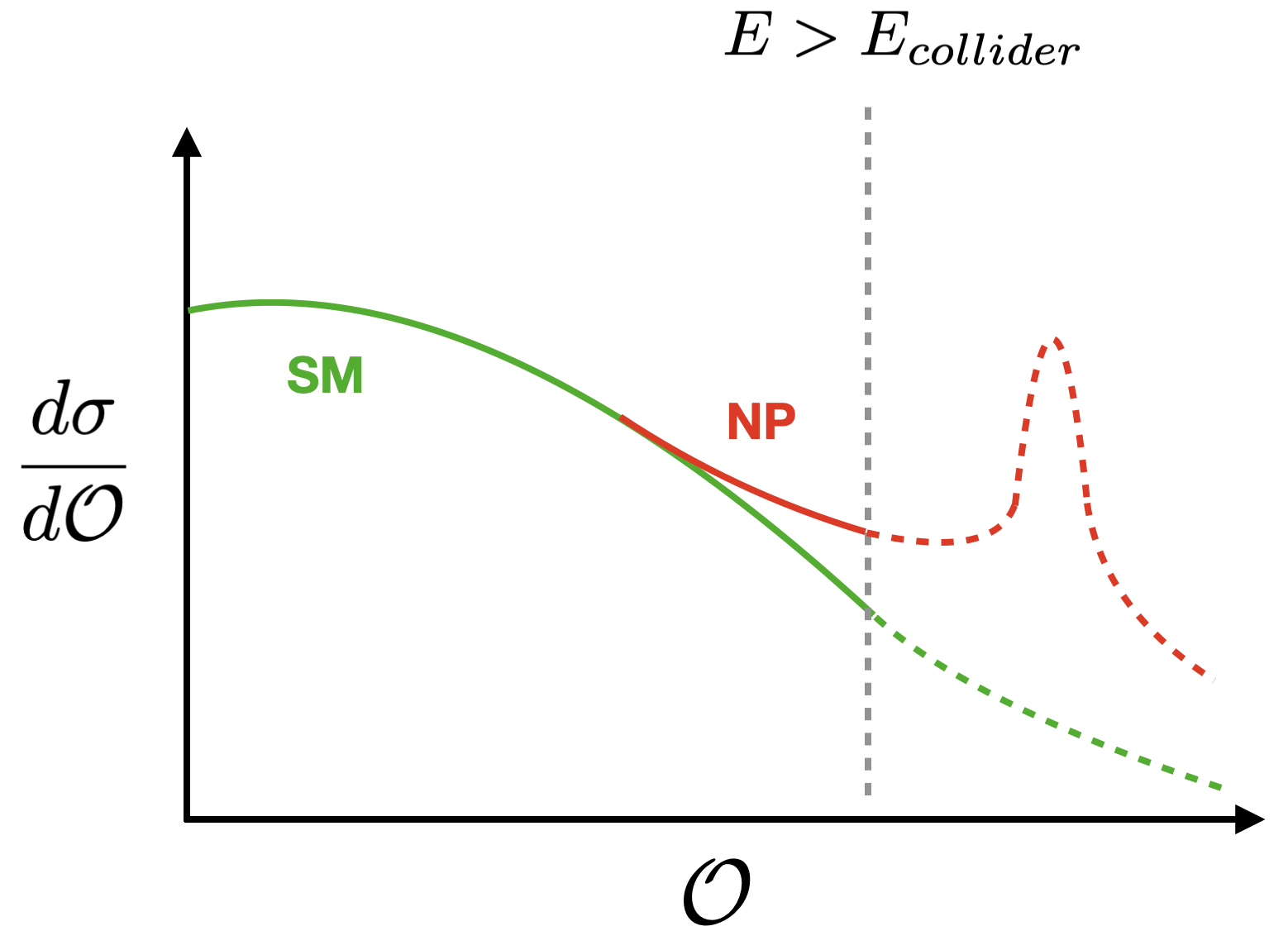}
\caption{\label{fig:directvsindirect} The two main strategies for discovering New Physics at colliders are here shown. 
The observable $\mathcal{O}$ is a generic energy-related quantity. On the left, direct search: the resonance is within the reach
of the machine and can be directly produced. On the right: the new resonance is beyond the energy frontier and the only way to
infer its existence is by scouting the tail of the distribution and find a deviation from the SM.}
\end{figure}

On the other hand, \textit{indirect searches} rely on complementary assumptions. Instead of looking for new particles, the objective is to
look for new interactions by measuring with extreme precision specific observables and comparing them to theoretical predictions.
This approach is not limited by the energy of the collider, but by the precision of both experimental measurements and theoretical calculations. In particular, if NP is heavier than the energy we are probing, we can indirectly assess the existence of it by 
scouting the tails of distributions and observing a clear deviation from the SM (see Fig.~\ref{fig:directvsindirect}).

Since we are entering a new era of precision measurements, with the upgrade at high luminosity of the LHC and possible upcoming lepton colliders,
the second strategy looks particularly appealing. In this perspective, we will now introduce the concept of \textit{Effective Field Theories},
a particularly natural framework to parametrise unknown interactions. This will lead to the formulation of the Standard Model Effective Field Theory (SMEFT),
a model-independent theoretical extension of the SM.

\section{Effective Field Theories}

The concept of Effective Field Theory (EFT) is well addressed in several reviews and lecture notes~\cite{penco2020introduction,skiba2010tasi,Burgess_2007,Brivio_2019,manohar2018introduction,Polchinski:1992ed,McCullough:2018knz}. 
Here we will limit ourselves to briefly report some of the basic concepts that are useful to understand the potential of the framework for our purposes.

The basic idea at the root of Effective Theories is that physics at different energy scales can be described by different theories. For instance,
if one is interested in describing a low energy system such as the movement of billiard balls on a pool table, one can safely neglect 
quantum mechanics effects and only rely on Newtonian mechanics. This simple and intuitive principle allows us to study low energy
phenomena without any knowledge of the full theory. In the realm of Quantum Field Theories (QFTs), the application of this 
vision takes the name of EFT.

In particular, as an effect of energy scale separation, the theory will be characterised by a cut-off scale beyond which it stops
being predictive. There are two approaches to EFT model building:

\begin{itemize}
\item \textbf{Bottom-up:}
one identifies the degrees of freedom and the symmetries characteristic of the system at low energy and
proceeds writing down all the allowed terms in the Lagrangian with no limitation on the dimension of the operators. The higher dimensional operators are a parametrisation of our ignorance on the high-energy theory and the coefficients in front of them need to be determined by experiments.
Examples of this approach are the SMEFT and Chiral Perturbation Theory\footnote{\revised{Despite the fact that the UV theory is known, i.e. QCD, Chiral Perturbation Theory is an effective description for the low-energy degrees of freedom (bound states) consistent with the approximate
chiral symmetry of the UV theory and the coefficients are determined experimentally.}}.

\item \textbf{Top-down:}
the full theory is available, but for practical purposes certain degrees of freedom are irrelevant, \ie particles are heavier than
the energy probed by the experiment. One then proceeds to obtain a low-energy description by integrating out the heavy fields and
generating higher dimensional operators. Examples of this approach are Weak Effective Theory, Heavy Quark Effective Theory and Soft-Collinear
Effective Theory.
\end{itemize}

In order to discuss more formally the above concepts, let us consider a full theory described by the Lagrangian $\Lag(\phi_L, \phi_H)$,
where $\phi_L$ and $\phi_H$ are the light and heavy fields respectively. One can define the effective Lagrangian in the 
path integral formalism
\begin{equation}
\int D\phi_H \, e^{i \int \Lag(\phi_L, \phi_H)} = e^{i \int \Lag_{eff}(\phi_L)} \, ,
\end{equation}
where the heavy degrees of freedom have been integrated out. This formalism helps in understanding the basic idea behind it, but
in practise this can be done as well in perturbation theory using Feynman diagrams. The effective Lagrangian describing the low-energy
regime will now look like
\begin{equation}
\Lag_{eff}(\phi_L) = \Lag_{d \leq 4} + \sum_{i} \frac{\mathcal{O}_i}{\Lambda^{dim(\mathcal{O}_i) - 4}} \, ,
\end{equation}
where $\Lambda$ is the cut-off energy scale and $\mathcal{O}_i$ are higher dimensional operators generated from the removal 
of heavy degrees of freedom. There are an infinite number of generated higher dimensional operators and that could suggest that 
the theory is impractical. However, each of these operators will be suppressed by a power of $\Lambda$ proportional to its dimension 
and this means that their contributions to the calculations will also be tamed by high powers of $\Lambda$. This observation 
allows us to introduce the concept of \textit{power counting}, which is crucial for the development of the theory.

\subsection{Power Counting} 
\label{sec:power_counting}

The existence of an infinite number of operators demands a working criteria to neglect terms before any calculation takes place. This
can be attained with power counting methods. In particular, if we consider natural units\footnote{These are defined by $\hbar = c = 1$, such that energy has the same dimension of masses while lengths and times have dimension $[M]^{-1}$.},
the action $S = \int d^4x \, \Lag$ is dimensionless, implying that the Lagrangian density has to be dimension 4 in units of mass. 
Each term in the Lagrangian density is therefore dimension 4 and from the kinetic terms, one can infer the dimension
of each type of field. Specifically, we have
\begin{equation}
[\psi] = \frac{3}{2} \, , \quad [\phi] = 1 \, , \quad [A_\mu] = 1 \, ,
\label{eq:canonical_dim}
\end{equation}
where $\psi$, $\phi$ and $A_\mu$ are generic fermion, scalar and vector fields respectively. Since derivatives also have mass dimension 1,
we can infer the mass dimension of the composed objects
\begin{equation}
[\bar{\psi} \psi] = 3 \, , \quad [F^{\mu \nu}] = 2\, , \quad [D_\mu] = 1 \, .
\end{equation}
With these basic ingredients, we can assign a mass dimension to any generic term we can write in the Lagrangian. A generic operator
that one can write down will have the structure
\begin{equation}
\mathcal{O} = (\bar{\psi}\psi)^{N_\psi} \, (F^{\mu \nu})^{N_F} \, (D_\mu)^{N_D} \, (\phi)^{N_\phi} \Rightarrow [\mathcal{O}] = 3 N_\psi + 2 N_F + N_D + N_\phi \, .
\end{equation}
It is important to keep in mind that the operator has to be a singlet of all the symmetry groups of the model. In order for the 
terms to be dimension 4, we have to introduce in the model a parameter $\Lambda$, which is the characteristic scale of the system.
If we now consider a process at scale $E$, we can naively expect that each operator will yield a contribution to the action
\begin{equation}
\int d^4 x \frac{\mathcal{O}}{\Lambda^{D - 4}} \sim \left(\frac{E}{\Lambda}\right)^{D-4} \, ,
\end{equation}
where $D$ is the dimension of the operator. We define $\delta = (E/\Lambda)$ the expansion parameter of the EFT.
This allows us to classify the operators in 3 categories

\begin{itemize}
\item \textbf{Relevant:}
operators with dimension $D < 4$. These operators give a contribution which is getting more and more important as the energy of the
process is smaller compared to $\Lambda$.

\item \textbf{Marginal:}
operators with dimension $D = 4$ on the other hand are scale independent in the sense that the importance of the contribution is not dependent on the scale probed by the system.

\item \textbf{Irrelevant:}
operators with dimension $D > 4$ give contributions that are suppressed by positive powers of the expansion parameter. Their
effects are less and less important the bigger the dimension of the operator is.
\end{itemize}

Marginal and Relevant operators are in finite quantity, simply because we can only write a limited amount of terms compatible with the symmetries.
However, irrelevant operators are not and the number of operators actually grows exponentially with the dimension. As long as one
is interested in predictions of order $\delta^n$ in the expansion parameter, the recipe is to truncate the series of operators at 
dimension $n + 4$, neglecting all higher order corrections. It is now manifest that if the energy probed by the experiment approaches
$\Lambda$, the neglected contributions start to be more and more relevant and the power counting simply breaks down, naturally defining a cut-off
scale for the theory.

However, it is not necessarily true that the scale of New Physics $\Lambda$ is the energy at which the new degrees of freedom
appear, \ie the theory becomes UV complete. This ambiguity stems from the fact that in reality masses and scales are two different
things and we lose the distinction because in natural units they have the same dimensions. In order to clarify this concept, we will
present an example.

\begin{example}[\textbf{Example: Masses vs Scales}]

Let us consider a simple model which describes the interaction of a Dirac fermion $\psi$ and a real scalar field $\phi$.
We further assume that the scalar is much heavier than the fermion, \ie $m_\phi \gg m_\psi$.
The Lagrangian is given by
\begin{equation}
\Lag = \bar{\psi} \slashed{\partial} \psi - m_\psi \bar{\psi} \psi + \frac{1}{2} \partial_\mu \phi \partial^\mu \phi - \frac{1}{2} m_\phi^2
\phi^2 + y \, \bar{\psi} \psi \phi \, ,
\end{equation}
where $y$ is the Yukawa coupling which in natural units is dimensionless. However, if we reinstate explicitly the dimensions of $\hbar$ (while keeping c=1),
we find that we can distinguish units of energy $E$ and length $L$. In particular we have
\begin{align}\nonumber
&[\hbar] = E L \, , \qquad [\Lag]=E L^{-3} \, , \qquad [\phi] = E^{1/2} L^{-1/2} \, ,\\ 
&[\psi] = E^{1/2} L^{-1} \, ,
\qquad [y] = E^{-1/2} L^{-1/2} \, , \qquad [M] = E \, . \nonumber
\end{align}
In natural units, $E=L^{-1}$ and the dimensions in Eq.~\eqref{eq:canonical_dim} are recovered.
As we can see, the Yukawa coupling is not dimensionless, but carries the dimensions of $1/\sqrt{\hbar}$.

If we are interested in low energy observables, we can safely integrate out the heavy degree of freedom $\phi$ and describe the 
system with an EFT in which the only degree of freedom is $\psi$, \ie
\begin{equation}
\Lag_{EFT} = \bar{\psi} \slashed{\partial} \psi - m_\psi \bar{\psi} \psi - \frac{1}{\Lambda^2} \bar{\psi} \psi \bar{\psi} \psi \, ,
\end{equation}
where the cut-off scale is given by $\Lambda = m_\phi/y$ and consequently has dimensions
\begin{equation}
[\Lambda] = E^{3/2}L^{1/2} \, .
\end{equation}
Scales do not carry the same units of masses and they are given by the ratio of masses and couplings. Consequently, they carry different meanings. The mass $m_\phi$ is the energy value at which the new degrees of freedom appear, while the 
scale is providing information on when the theory becomes strongly coupled, \ie the perturbative expansion breaks down. For instance,
if $m_\phi = 100$ GeV and $y \sim 0.1$, the cut-off scale is $\Lambda = 1$ TeV. One might therefore erroneously think that the
EFT is UV completed at $1$ TeV, but the new degree of freedom would actually show up at much lower energies.
\end{example}

\subsection{An instructive example: the Fermi theory}

In order to get familiar with the aforementioned concepts, we will now discuss one of the most famous and successful example
of EFT.

When weak interactions were discovered thanks to the observation of beta decays, a theoretical explanation to describe and predict such interactions
was very much needed. In 1933 Enrico Fermi proposed a phenomenological description which was very accurate and explained all the 
observed phenomena~\cite{Fermi:1933jpa}. We can look at the problem both from a bottom-up and a top-down approach.

In the first case, retracing the historical path, let us imagine that we observe a new interaction: muons decay producing an electron
and a pair of neutrinos. The muon mass is $0.1~\gev$ so, in hindsight, we know that we are at much lower energies with respect to the 
weak vector boson masses. In order to describe this phenomenon we can apply the recipe to build an EFT from the bottom-up. We can write down all the possible Lagrangian terms that are compatible with the symmetries and put an unknown coefficient in front of them, which will be determined by experimental measurements. With a completely agnostic attitude on the high-energy theory, but relying on the experimental observations that the involved fermion currents have the $V-A$ structure, we can very simply add to the QED Lagrangian
\begin{equation}
\Lag_{Fermi} = -\frac{c_F}{\Lambda^2} (\bar{\nu}_\mu \gamma^\alpha P_L \mu) (\bar{e} \gamma_\alpha P_L \nu_e) + h.c. \, ,
\label{eq:fermieft}
\end{equation}
which describes a contact interaction between four fermions, and having dimension 6 needs to be suppressed by an unknown energy scale
$\Lambda$. The coefficient $c_F$ is called Wilson coefficient and it is also to be determined. 
We can now experimentally measure the factor\footnote{It is not possible to determine experimentally both $c_F$ and $\Lambda$ but only their ratio.} $c_F/\Lambda^2$ by computing the decay rate
\begin{equation}
\Gamma(\mu \to e \bar{\nu}_e \nu_\mu) = \left(\frac{c_F}{\Lambda^2}\right)^2 \frac{m_\mu^5}{1536\pi^3}
\end{equation}
and find
\begin{equation}
\frac{c_F}{\Lambda^2} = 4 \frac{G_F}{\sqrt{2}} = 3.3 \cdot 10^{-5}~\gev^{-2} \, ,
\end{equation}
where $G_F$ is the Fermi constant and the numerical factors are there for historical reasons. We explicitly see that the decay rate
is suppressed by a power $(m_\mu/\Lambda)^4$ where $m_\mu$ is the characteristic energy of the process, as expected by naive power counting. 
With the bottom-up approach, we end up being able to describe phenomena which are the low-energy effect of some unknown heavier physics,
indirectly assessing its existence. The cut-off scale in particular is there to remind us that as we go to higher energies, the effects of the missing degrees of freedom are
going to be more and more relevant and the theory will start to crumble.

\begin{figure}[t!]
\centering
\includegraphics[scale=0.25, valign=c]{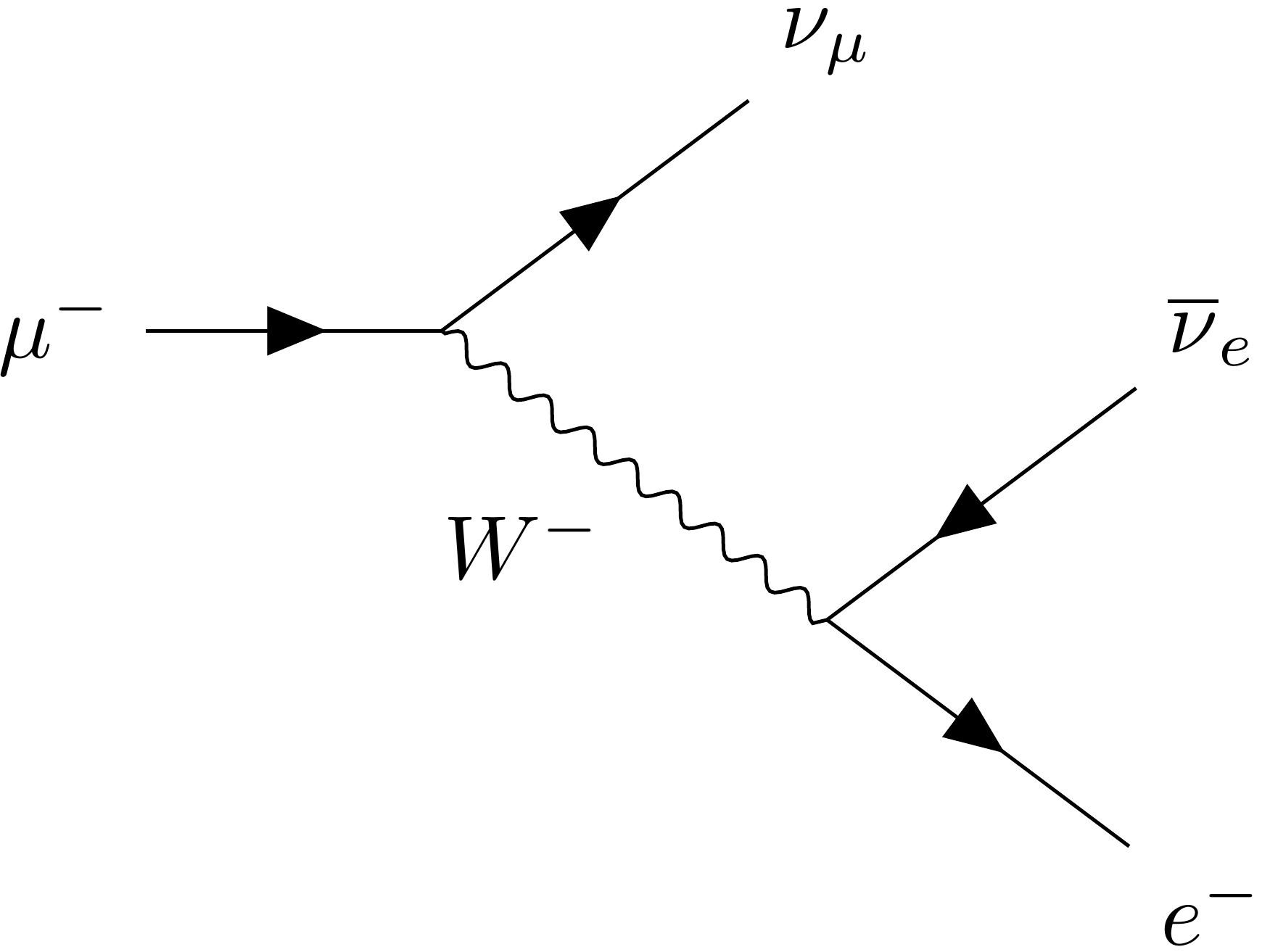} \qquad
\includegraphics[width=0.10\textwidth, valign=c]{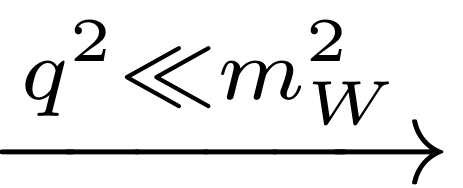} \qquad
\includegraphics[scale=0.25, valign=c]{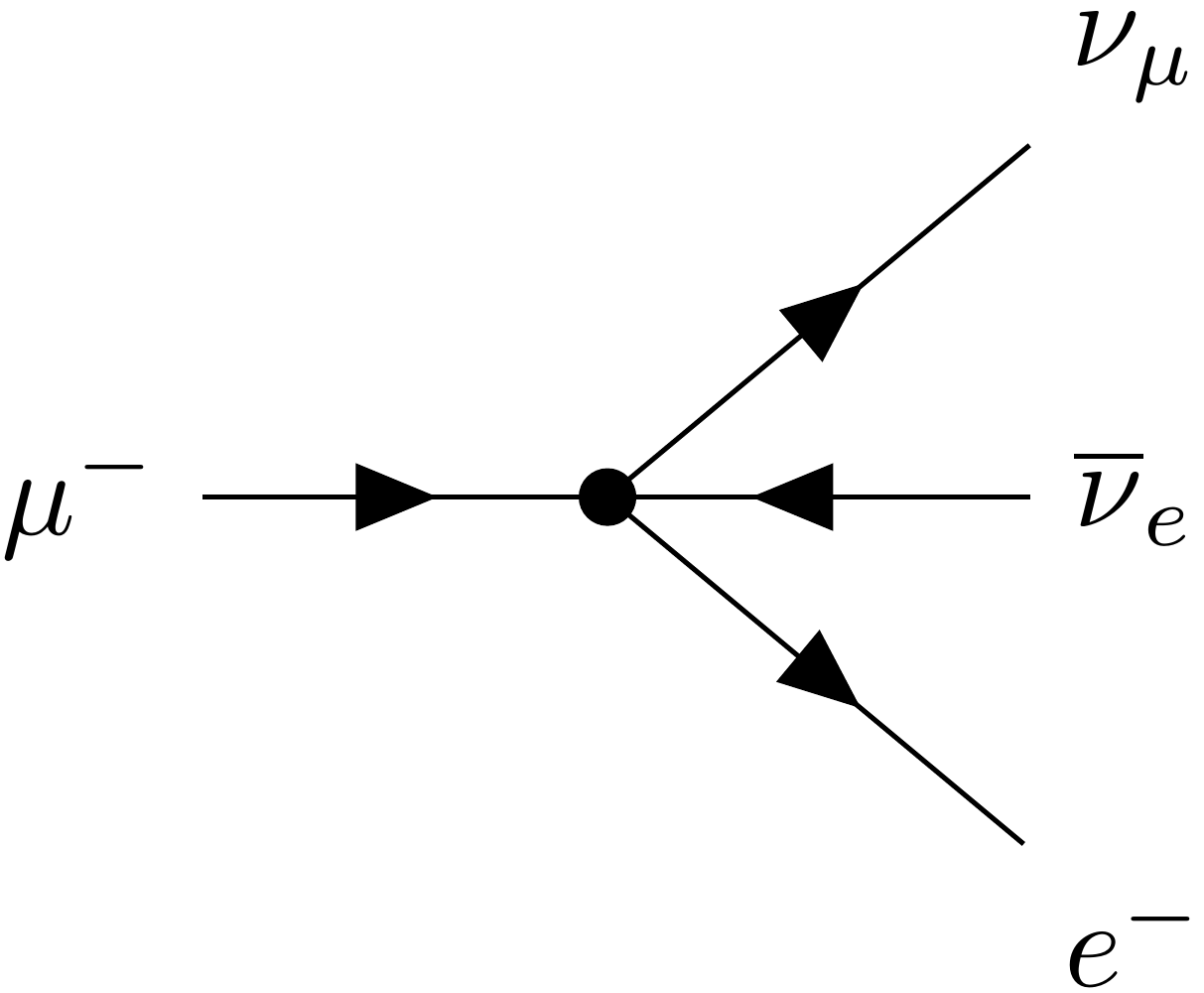}
\caption{A pictorial depiction of the matching procedure to integrate out the $W$ boson in the top-down approach. \label{fig:muonmatching}}
\end{figure}

On the other hand, one can look at the same situation from a top-down perspective. The SM is the full theory, describing the decay
of the muon through the exchange of a $W$ boson. We then proceed to integrate out the heavy degree of freedom with a procedure called 
\textit{matching}, in which we diagrammatically solve for the equation
\begin{equation}
\bra{f} S_{SM} \ket{i}_{E \ll \Lambda} \equiv \bra{f} S_{EFT} \ket{i} \, ,
\label{eq:matching}
\end{equation} 
with $S$ the S-matrix $e^{-i H t}$. Solving this equation will give us the low-energy coefficients as a function of the UV parameters.
The above condition basically is telling us that we expect the two theories to have the same matrix-element when we 
are probing low energies. If we compute now the muon decay amplitude in the SM we have
\begin{equation}
\mathcal{M}_{SM} = \frac{g^2}{2(q^2 - m_W^2)}[\bar{u}(p_e)\gamma^\alpha P_L v(p_{\nu_e})][\bar{u}(p_{\nu_\mu})_\mu\gamma_\alpha P_L u(p_\mu)] \, ,
\end{equation}
where $q$ is the exchanged momentum between the two fermion currents. At low energy $q^2 \ll m_W^2$ and we can approximate the matrix element
\begin{equation}
\mathcal{M}_{SM} \approx -\frac{g^2}{2m_W^2}[\bar{u}(p_e)\gamma^\alpha P_L v(p_{\nu_e})][\bar{u}(p_{\nu_\mu})_\mu\gamma_\alpha P_L u(p_\mu)] \, .
\end{equation}
This has the same exact structure one gets from the EFT calculation in Eq.~\eqref{eq:fermieft} and by solving Eq.~\eqref{eq:matching}, we find
the matching condition
\begin{equation}
\frac{c_F}{\Lambda^2} = \frac{g^2}{2 m_W^2} \, ,
\end{equation}
relating the Wilson coefficient and the cut-off scale with the UV parameters $g$ and $m_W$. 
As we can observe, the distinction between masses and scales discussed in Sec.~\ref{sec:power_counting} is clearly present in the case of the Fermi theory. Specifically, while $G_F^{-1/2} \sim 290$ GeV the energy scale at which the EW gauge boson appear and UV completes
the model is $m_W = 80$ GeV.

This is an example of tree-level matching (see Fig.~\ref{fig:muonmatching} for a pictorial representation), but 
the same can be done at loop level. In general more than one EFT operator might be needed to do a successful matching. For example, if one wants to account also for the correction of order $q^2/m_W^4$, the introduction of another dimension-8 operator would be required.
The top-down approach can in particular be useful if one is interested in describing only a low-energy system, because calculations in the EFT are much simpler than in the complete theory.

\subsection{Equations of motion}
\label{sec:EOM}

When we look at the problem of building an EFT from a bottom-up perspective, one faces the issue of how many operators are needed 
at a given order in $1/\Lambda$. It can be shown that they actually produce a vector space and that there exist a
minimal base of operators we can write. In order to do so and remove redundant operators, equations of motion play a significant role.
The crucial point is that fields redefinitions, while changing the Lagrangian and the correlation functions, leave unchanged
the matrix elements and therefore the cross sections, which are the only observables of scattering experiments\cite{manohar2018introduction}.  
In renormalisable QFTs, these field redefinitions are limited to linear transformations $\phi_i^\prime = c_{ij} \phi_j$, since
we cannot have operators with dimensions higher than 4. This procedure is usually just used to put kinetic terms in canonical forms.
On the other hand, in EFTs we have much more freedom and we can redefine fields as long as we respect the power counting, \ie we use
the cut-off scale $\Lambda$ as mass scale. 
\begin{example}[\textbf{Example: Fields redefinitions}]
Let us now consider an example and define a simple EFT Lagrangian for a real scalar
field
\begin{equation}
\Lag = \frac{1}{2}\partial_\mu \phi \partial^\mu \phi - \frac{1}{2} m^2 \phi^2 -\frac{1}{3!}g \phi^3 -\frac{1}{4!}\lambda \phi^4 + \frac{c_1}{\Lambda}\phi^2\Box \phi + \frac{c_2}{\Lambda}\phi^5 \, .
\end{equation} 
We can now compute the equation of motion for this theory and find
\begin{equation}
\frac{\delta S}{\delta\phi} = -\Box \phi - m^2 \phi - \frac{1}{2} g \phi^2 -\frac{1}{3!}\lambda\phi^3 + \mathcal{O}(\frac{1}{\Lambda}) = 0 \, ,
\end{equation}
where higher order terms in $1/\Lambda$ are being neglected because, as we will see, they simply produce higher order operators.
If we now perform the field redefinition $\phi \to \phi + \frac{c_1}{\Lambda}\phi^2$ in the Lagrangian, we obtain
\begin{align}
\Lag^\prime &= \Lag + \frac{c_1}{\Lambda} \phi^2 \left[-\Box \phi -m^2\phi - \frac{1}{2} g \phi^2 -\frac{1}{3!}\lambda \phi^3 + \mathcal{O}(\frac{1}{\Lambda})\right] \nonumber \\
&= \frac{1}{2}\partial_\mu \phi \partial^\mu \phi - \frac{1}{2} m^2 \phi^2 -\frac{1}{3!}\tilde{g} \phi^3 -\frac{1}{4!}\tilde{\lambda} \phi^4 + \frac{\tilde{c}_2}{\Lambda}\phi^5 \, \,
\end{align}
where we had to redefine 
\begin{equation}
\frac{\tilde{g}}{3!} = \frac{g}{3!} + m^2 \frac{c_1}{\Lambda} \, , \quad
\frac{\tilde{\lambda}}{4!} = \frac{1}{4!} \lambda + \frac{g \,c_1}{2\Lambda} m^2 \, , \quad \frac{\tilde{c_2}}{\Lambda} = \frac{c_2}{\Lambda} - \frac{\lambda \, c_1}{3! \, \Lambda} \, .
\end{equation}
The same could have been obtained by simply using the equation of motion in the original Lagrangian. What we observe here is that
two different Lagrangians are describing the same physics and can therefore be used interchangeably. The conclusion is that the operator
$\phi^2\Box\phi$ was actually redundant and removing it from the Lagrangian is yielding a much simpler theory to deal with.

Specifically, it is interesting to observe that while in the first Lagrangian the operator $\phi^2\Box\phi$ produces a momentum
enhanced vertex, in the second one there is no such vertex and one might wonder how the two theories can have the same
energetic behaviour at the amplitude level. This can be directly inspected by computing the $\phi \phi \to \phi \phi$ scattering
amplitude. In the second theory, the amplitude is simple to compute and given by
\begin{equation}
\includegraphics[scale=0.15, valign=c]{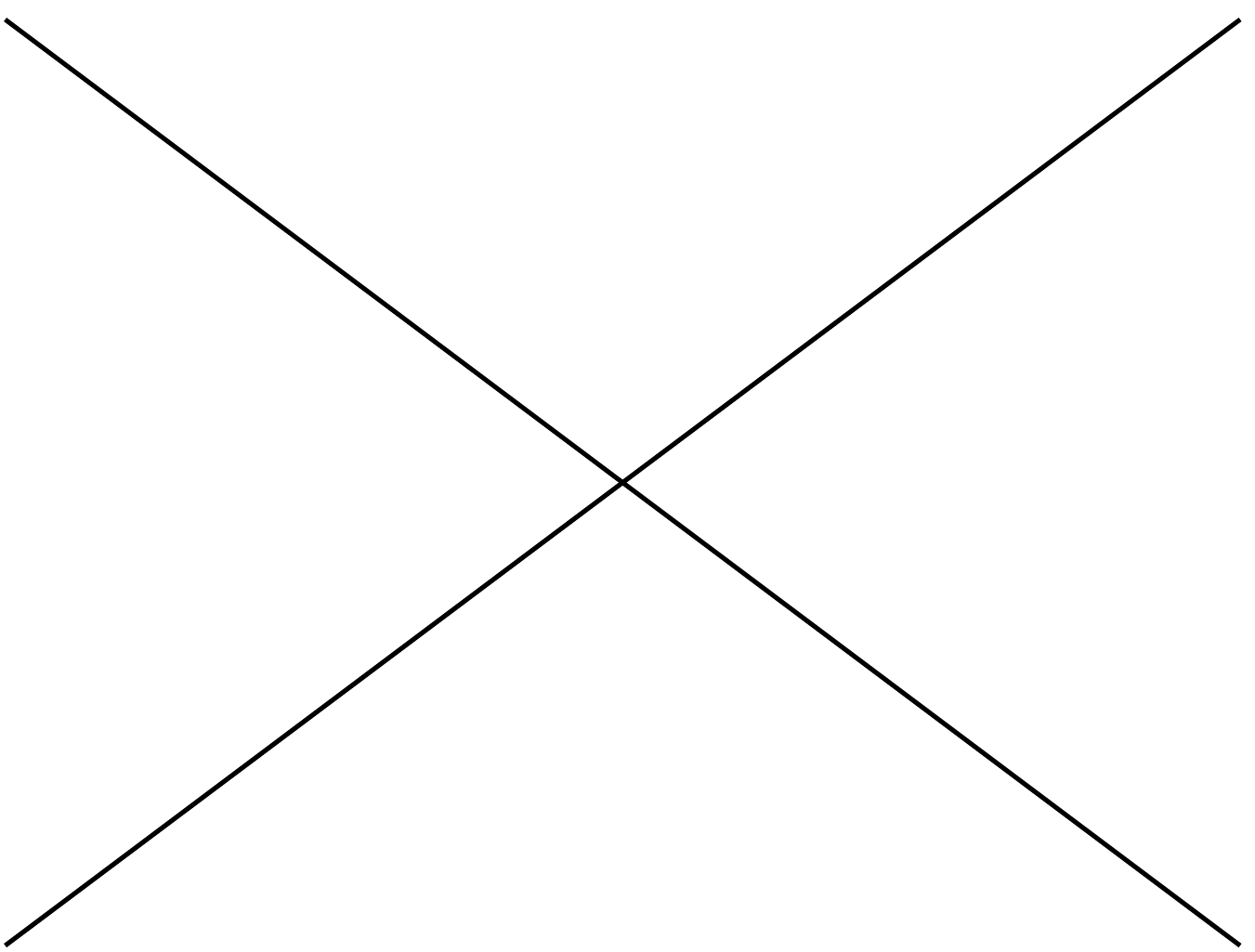} + \includegraphics[scale=0.15, valign=c]{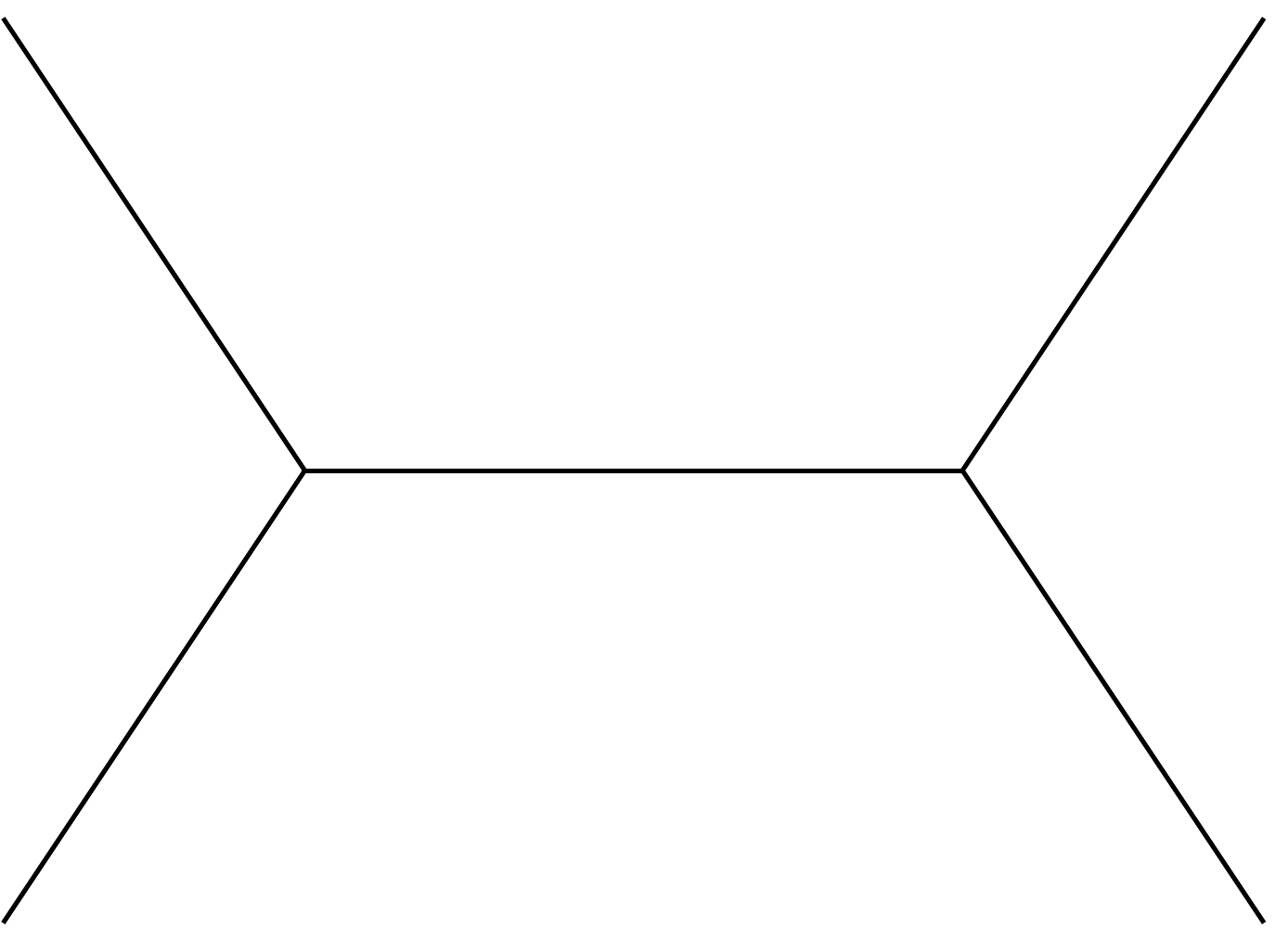} = 
\tilde{\lambda}  - \frac{\tilde{g}^2}{q^2 - m^2} \, .
\end{equation}
On the other hand, for the first Lagrangian we have four diagrams and the amplitude is
\begin{equation}
\renewcommand\arraystretch{3}
\begin{matrix}
\includegraphics[scale=0.15, valign=c]{Diagrams/EFT/diag_1} + \includegraphics[scale=0.15, valign=c]{Diagrams/EFT/diag_2} +\\
\includegraphics[scale=0.15, valign=c]{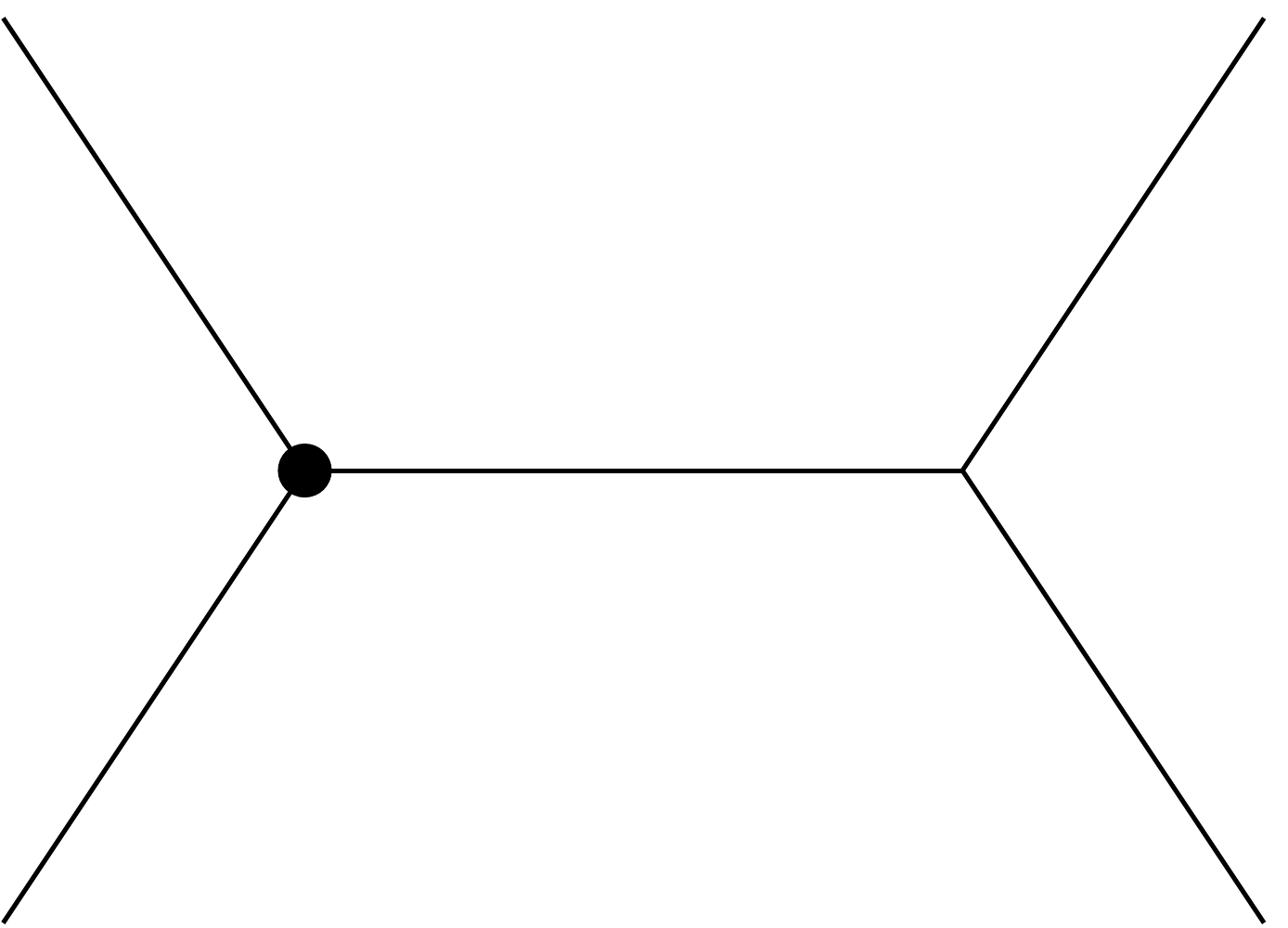} + \includegraphics[scale=0.15, valign=c]{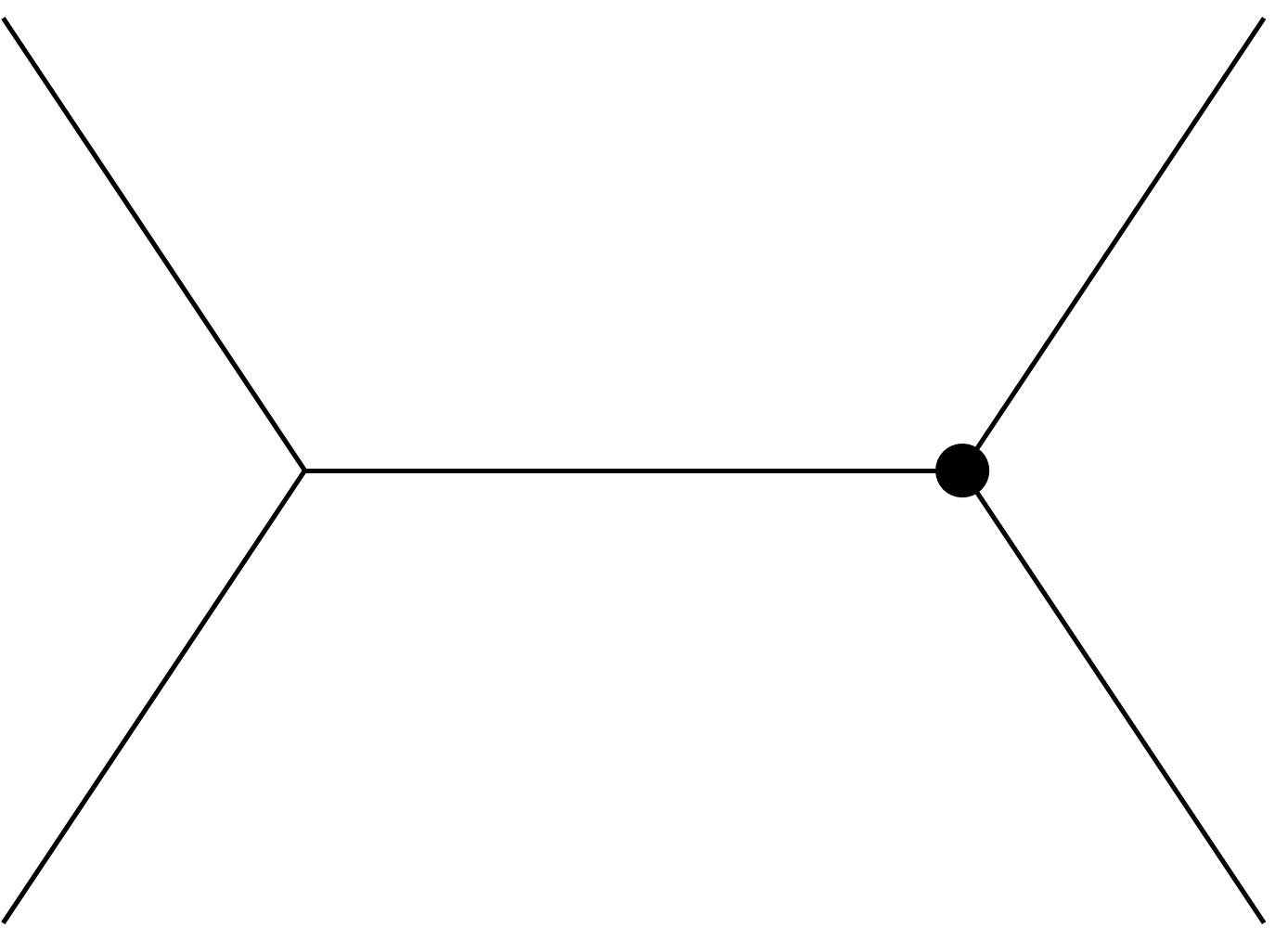}
\end{matrix}
 = 
\lambda - \frac{g^2}{q^2 - m^2} - 12 \frac{c_1 \, g}{\Lambda} \frac{q^2}{q^2 - m^2} \, ,
\end{equation}
where the last two diagrams have the energy enhanced behaviour from the dimension-5 operator $\phi^2\Box\phi$. However, it is simple to see that
with a simple algebraic manipulation in the numerator, adding and subtracting $m^2$, we find
\begin{align}
\nonumber
&\lambda - \frac{g^2}{q^2 - m^2} - 12 \frac{c_1 \, g}{\Lambda} \frac{q^2 -m^2 + m^2}{q^2 - m^2} = \lp \lambda -12 \frac{c_1 \, g}{\Lambda} \rp
-\\ 
&\lp g^2 + 12  \frac{c_1 \, g}{\Lambda} m^2 \rp \frac{1}{q^2 - m^2} = \tilde{\lambda} - \frac{\tilde{g}^2}{q^2 - m^2} + \mathcal{O}(1/\Lambda^2) \, ,
\end{align}
which is the same amplitude we obtain from the previous Lagrangian as expected, up to higher order corrections in $1/\Lambda^2$.
\end{example}
The same procedure can be applied to much more complicated theories and the choice of which operators to keep is completely 
subjective. However, removing operators with derivatives is usually a good option as they make computations easier, especially when dealing with loop calculations.

\subsection{Renormalisation Group Equations}
\label{sec:running}

\revised{In the classical definition of renormalization, one defines as renormalizable a theory in which all the infinities arising at loop level
can be reabsorbed in a finite amount of parameters of the theory. It can be shown, that this criterium leads to the requirement that the
maximum allowed dimension for the operators in the Lagrangian is $4$. However, this vision is now considered outdated, especially from an EFT perspective. In fact, the requirement to have a fully renormalizable theory is dictated by the desire to have a theory which describes physics
at all scales, but in an EFT there is by definition no such expectation. Even if formally, an EFT requires an infinite amount of 
counterterms in order to reabsorb all the infinities, we still have predictive power since we truncate the EFT series at a given order in 
the power expansion over $1/\Lambda$. We can therefore always compute loop contributions in the theory and reabsorb the infinities in the
retained operators while throwing away all the contributions coming from higher order ones.\footnote{In the SMEFT, if one sticks to dimension-6 operators and single-insertion at the amplitude level no higher order contribution needs to be thrown away.} In this sense, the EFT is said to be renormalisable order by order in the $1/\Lambda$ expansion.
In particular, the concepts of unitarity violation and renormalizability are closely related, as higher order operators induce 
energy growing behaviours at the amplitude level which eventually lead to the breaking of perturbative unitarity. So, the same
terms in the Lagrangian that lead to the classical non-renormalizability of the theory, also induce unitarity violation.}

When we computed the matching for the muon decay, we limited ourselves to a tree-level computation. Because of this, it was not manifest that
the matching has to be done at a specific scale, usually the one indicated by the integrated out particle. This was the case since
at tree-level no scale dependence appears in the Wilson coefficients. If the same matching procedure is performed at 1-loop level,
one finds that the Wilson coefficients are scale dependent. This is very important because when we perform the matching, we
fix the Wilson coefficients at the cut-off scale but we might be interested in computing observables 
at much lower scales and the running of the coefficients would be needed.

\revised{The matching at loop level has to be done using the same renormalisation scheme for both the EFT and the UV theory. A standard choice is
the $\overline{MS}$ renormalisation scheme\footnote{In this scheme only the divergent parts of the radiative corrections and a universal constant $e^{\gamma_E}/4\pi$ (with $\gamma_E$ Euler-Mascheroni constant) are absorbed in the counterterms.} and dimensional regularisation.
Other schemes such as the on-shell renormalisation scheme can be used, but being mass dependent, extra care needs to be taken to not spoil the power counting of the theory when performing loop calculations.}

It is interesting to infer a priori that no large logarithms
of the kind $\log{(M/m)}$ (with $M$ the matching scale and $m$ the mass of a light particle) can appear in this procedure, since
the IR behaviour of the two theories is identical and they will therefore cancel in the matching procedure~\cite{skiba2010tasi}. One way to study the
evolution of the Wilson coefficients at low energies is by means of Renormalisation Group Equations (RGEs). These allow us to resum
all the large logs of the theory. As an example~\cite{manohar2018introduction}, let us consider the $\beta$ function of a Wilson coefficient
\begin{equation}
\beta(\mu) = \mu \frac{d c(\mu)}{d\mu} = \gamma_0 \frac{g(\mu)^2}{16 \pi^2} c(\mu) \, ,
\end{equation}
where $g$ is the UV theory coupling which also depends on the renormalisation scale $\mu$. In order to solve the equation, 
one needs to know the $\beta$ function of the coupling $g$. If for the sake of the argument we have
\begin{equation}
\mu \frac{d g(\mu)}{d\mu} = - b_0 \frac{g(\mu)^3}{16 \pi^2} \, ,
\end{equation}
we can solve the RGE for the Wilson coefficient and find
\begin{equation}
\frac{c(\mu)}{c(M)} = \left( \frac{\alpha(\mu)}{\alpha(M)}\right)^{-\frac{\gamma_0}{2 b_0}} \approx 1 + \gamma_0 \frac{\alpha}{4\pi} \log{\left(\frac{\mu}{M}\right)} + \mathcal{O}\left(\alpha^2 \log^2{\left(\frac{\mu}{M}\right)}\right) \, ,
\end{equation}
where $M$ is the matching scale and $\alpha = g^2/4\pi$. It is clear from the above expression that all the leading large-logs 
of the kind $\alpha^n \log^n$ are resummed. In order to have also the next to leading log resummation, one would need to go to 
2-loops calculations.

This was a basic example for a single EFT operator. In a realistic theory however, multiple operators are generated and
the phenomenon of mixing appears. In practice, the RGEs are matrix equations for all the Wilson coefficients
\begin{equation}
\mu \frac{dc_i(\mu)}{d\mu} = \gamma_{ij}(\mu) c(\mu) \, . 
\end{equation}
The significant consequence of this is that even if at the matching scale some operators are not generated by the UV interactions,
they are needed and will enter the picture at low energy when mixing appears. The matrix $\gamma_{ij}$ is called anomalous-dimension
matrix and is a fundamental object of an EFT.

\section{Extending the SM: the SMEFT framework}

At the beginning of the LHC program, there was a firm hope in the particle physics community that new unknown particles would be
produced, opening a novel golden era in the field. This belief was in particular relying on the supersymmetrical 
extension of the SM (SUSY), which is predicting that each known particle has a "supersymmetric partner" with a spin differing by
a half-integer. The fact that it gave natural solutions to theoretical and experimental problems\footnote{For example, it solves the hierarchy problem of the Higgs mass and provides automatically a particle candidate for Dark Matter.},
led many physicists to expect important discoveries in this direction at CERN. 

While SUSY is certainly not a dead paradigm, the absence of clear indications of BSM effects pushes forward the characteristic energy scale
of NP. Above $10~\tev$, SUSY does not offer any more elegant solutions to the aforementioned problems and the excitement 
around it has been slowly declining in the last decade. There exist, however, a plethora of models and UV-complete extensions of the SM
on the market and the need for a model-independent framework is needed more than ever. In this way, we can parametrise NP and interpret experiments 
in a universal language, so that specific models can be mapped into that and constrained accordingly.

In light of these observations, an EFT extension of the SM seems like the natural way to proceed. Extending the SM from a 
bottom-up approach allows us to parametrise NP in a model-independent way, taking advantage of the fact that we expect new particles
to be heavier than the energies that will be probed in the next decades\footnote{The HL-LHC (2027-2040) upgrade will deliver us
increased luminosity, but an unchanged collider energy. This fact motivates furthermore an EFT approach, where the higher statistics
can be exploited to scout tails of distributions in search of deviations.}. The natural next step is to decide which EFT extension should
be adopted, \ie which low-energy limit should the theory have. Two main options are discussed: SMEFT and HEFT.

\subsection{SMEFT vs HEFT}

The difference between SMEFT and HEFT (Higgs Effective Field Theory) is given by the fact that while in the former the IR theory is the SM, in the latter the Higgs doublet
is not adopted and only a Higgs scalar field $h$ is introduced to take into account the $125~\gev$ discovery at CERN~\cite{Feruglio:1992wf,Grinstein:2007iv}. 
In practice, HEFT is a further generalisation in which additional degrees of freedom in the parameter space of the model are 
assumed. This is motivated by the fact that the assumption of the Higgs doublet is yet to be confirmed and, despite being reasonable,
a purely agnostic approach would therefore leave it out of the framework. 

In HEFT, the goldstone bosons are introduced in interaction terms by making use of the field 
\begin{equation}
U = e^{\frac{i\tau^I \pi^I}{v}} \, ,
\end{equation}
where $\pi^I$ are the fields eaten by the $W$ and $Z$ boson and correspond to the longitudinal degrees of freedom. This field transforms
as a bi-doublet under $SU(2)_L \times SU(2)_R$\footnote{This global symmetry is then spontaneously broken into the custodial
symmetry and the $SU(2)_R$ part is explicitly broken by the Yukawa couplings.}. The scale introduced in the denominator is not the scale of NP, but
the EWSB scale, since that is the characteristic scale of the goldstones. The Higgs field is instead introduced as a scalar gauge
singlet. Higgs couplings are generically encoded in functions of the form
\begin{equation}
F(h) = 1 + a\frac{h}{v} + b\frac{h^2}{v^2} + ... \, ,
\end{equation}
where the coefficients are completely arbitrary and not fixed by the EWSB relations. The theory is then given by the
most general Lagrangian compatible with the gauge group $G = SU(3)_c \otimes SU(2)_L \otimes U(1)_Y$
\begin{equation}
\Lag = \Lag_0 + \Delta\Lag + ... \, ,
\end{equation}
where $\Lag_0$ is the leading order Lagrangian describing the EW sector at low energies and $\Delta\Lag$ is giving the first-order
corrections.

\revised{One key difference between HEFT and SMEFT is that being the Higgs doublet not realised in HEFT, the Higgs 
couplings to the EW bosons are not correlated and dictated by the $SU(2)$ gauge structure. The same is true for the Higgs self-couplings (trilinear and quartic interactions), which at dimension-6 in the SMEFT are correlated while no such constraints on the deviations are predicted in HEFT, since arbitrary polynomials of the Higgs singlet are allowed.}

\revised{The distinction between SMEFT and HEFT can also be seen from a different perspective: the analyticity of the interactions~\cite{Falkowski:2019tft}. In this language, HEFT can be equivalently formulated with a linearly realized EW symmetry and a Higgs doublet $\varphi$, but
in addition to the usual terms in the Lagrangian one allows non-analytic contributions. What makes this perspective particularly 
valuable is that it makes manifest the existence of strong unitarity violating effects in $2 \to n$ scatterings of longitudinal gauge bosons in $n > 2$ Higgs bosons
which lead to a violation of perturbative unitarity around $4\pi v \approx 1$ TeV.}

\revised{In the scenario of non-decoupling, \ie when the heavy particles integrated
out are connected to the remaining degrees of freedom by a symmetry, the interactions would necessarily be non-linear 
at the Lagrangian level and HEFT would be the more appropriate low energy theory. An example of this are composite Higgs models.}

Unfortunately, the agnosticism and generality of HEFT comes at a cost: the high dimension of the model parameter space. While in the near future 
distinctive measurements can be performed that could tell apart HEFT and SMEFT indicating us the correct way to proceed, the lack
of evidence of deviations from the Higgs mechanism to-date favours a simplified framework in which the Higgs is realised in the 
theory through the Higgs doublet and with analytic interactions. For this reason, we will set aside HEFT and focus on the SMEFT for the rest of this thesis.

The building blocks of the SMEFT are the fields described in Chapter~\ref{chap:sm}, \ie the SM fields. The Lagrangian is
simply extended with an infinite tower of higher dimensional operators suppressed by a heavy scale
\begin{equation}
\Lag_{SMEFT} = \Lag_{SM} + \sum_{d > 4} \sum_{i=1}^{N_d} \frac{c_i^d}{\Lambda^{d-4}}\mathcal{O}_i^d \, .
\end{equation}
In particular, the first correction at dimension-5 is just given by a single operator, called the Weinberg operator~\cite{Weinberg:1979sa},
which has the form 
\begin{equation}
\Lag^5 = \frac{c}{\Lambda} (\bar{L^c}\tilde{H}^*)(\tilde{H}^\dagger L) + h.c. \, ,
\end{equation}
where $L$ is the lepton doublet. This operator generates Majorana neutrino masses when the Higgs field acquires the vev and it
might therefore be argued that we have already seen its effect\footnote{The nature of the neutrino masses (Dirac or Majorana) has not been confirmed experimentally.}. The Weinberg operator, however, does not affect too much the phenomenology we study at colliders
and is also expected to be highly tamed by the scale of NP. The first relevant correction to the Lagrangian is then given by dimension-6 terms.
It is important to observe that odd dimension operators break explicitly Lepton and Baryon global symmetries, which are 
accidental symmetries in the SM. This could suggest however that a more fundamental principle is in place in a complete theory and these operators are expected to
be marginal and heavily suppressed. The next order correction in the expansion is therefore expected to come from dimension-8 operators, which
have been already subjects of several studies and very recently an operator basis has been found~\cite{Li:2020gnx,Murphy:2020rsh}
\footnote{For a generic dimension-d Lagrangian the minimal number of operators grows exponentially and can be determined by means of the Hilbert series~\cite{Henning:2015daa,Henning:2015alf,Henning:2017fpj}}.
In the following, we will restrain ourselves to the study of dimension-6 operators, since it is reasonable to expect that the leading SM deviations will come from them\footnote{This statement is however model-dependent. Models exist in which dimension-8 operators
are yielding a much bigger effect despite being more suppressed by $\Lambda$.}.

\subsection{Operator basis}

The space of all the dimension-6 operators is a vector space and therefore a minimal set of operators can be identified.
This task has been cumbersome and required many years since the first studies appeared~\cite{Buchmuller:1985jz}, in which
an over-complete set of operators was identified by Buchmuller and Wyler. More than 20 years later, the first non-redundant basis
of operators was established~\cite{Grzadkowski:2010es}. This basis has been dubbed "Warsaw basis"\footnote{Uncreatively from the 
University affiliation of the authors.} and, in the assumption of flavour universality, is characterised by 59 operators.
Other bases of operators have been developed during the last decade, each of them addressing specific problems. 
For instance we have the SILH\footnote{\revised{In the original formulation the SILH basis was not complete, but it has been later extended~\cite{Falkowski:2001958}.}} basis~\cite{Giudice:2007fh,Contino:2013kra,Elias-Miro:2013eta}, particularly suited for a
strongly interacting Higgs sector, and the HISZ
basis~\cite{PhysRevD.48.2182}. In this work, we will make use of the Warsaw basis, which is also becoming the norm in the 
community. It goes without saying that any result in one basis can be translated into another by means of simple transformations.

The guiding principle leading to the formulation of the Warsaw basis is the desire to minimise the number of derivatives in operators.
This is motivated by the fact that computations are easier, especially at loop level. In practice, one systematically removes
derivatives with equations of motion such as
\begin{equation}
D^\mu G_{\mu \nu}^A = g_s(\bar{q}\gamma_\nu T^A q + \bar{d}\gamma_\nu T^A d + \bar{u}\gamma_\nu T^A u) \, ,
\end{equation}
which is an example for the gluon field strength. As explained in Sec.~\ref{sec:EOM}, this procedure does not change the
physics and we can remove derivatives obtaining perfectly equivalent operators.

Fierz identities can be also used to reduce redundant four-fermion operators. These are identities
that allow one to rewrite products of spinor bilinears made of two spinors with products of bilinears of the same spinor.
For example, one has
\begin{equation}
(\bar{\psi_L} \gamma_\mu \chi_L)(\bar{\chi_L} \gamma^\mu\psi_L) = (\bar{\psi_L} \gamma_\mu \psi_L)(\bar{\chi_L} \gamma^\mu\chi_L) \, ,
\end{equation}
which can be used to remove many possible combinations.

The final tool in the shed is the use of integration by parts, exploiting the fact that a total derivative in the Lagrangian does not
affect the equations of motion and can be dropped out. This means that if two operators differ by a total derivative, they are equivalent:
\begin{equation}
\OO_i = \OO_j + \frac{d}{dx}\OO_k \quad \to \quad \OO_i \equiv \OO_j \, .
\end{equation}
The generic operator in the SMEFT dimension-6 Lagrangian can be written as
\begin{equation}
\mathcal{O} = (\bar{\psi}\psi)^{N_\psi} \, (X^{\mu \nu})^{N_F} \, (D_\mu)^{N_D} \, (\varphi)^{N_\varphi} \Rightarrow [\mathcal{O}] = 3 N_\psi + 2 N_F + N_D + N_\varphi = 6 \, ,
\end{equation}
where $\bar{\psi}\psi$ is a generic bilinear of fermions, $X$ is a gauge field strength, $D$ the covariant derivative and $\varphi$ the Higgs doublet. Let us now explore the possible operators that can be built at dimension-6, given the above constraint.

If we consider only bosonic operators ($N_F = 0$), by naive power counting and taking into account gauge and Lorentz structures, we are left with the possible combinations
\begin{equation}
\{X^3, X^2 D^2, X^2 \varphi^2, X D^2 \varphi^2, \varphi^6, D^4 \varphi^2, D^2 \varphi^4\} \, .
\end{equation} 
It can be shown however that, by using EOM, the classes $D^4 \varphi^2$, $X D^2 \varphi^2$ and $X^2 D^2$ can be reduced to
the classes $X^3$, $X^2 \varphi^2$, $\varphi^6$ and $D^2 \varphi^2$. The same can be done for the class $D^4 \varphi^2$, which is
going to generate also 2-fermion operators. Once this is done, the only remaining possibilities are
\begin{equation}
\{X^3, X^2 \varphi^2, \varphi^6, D^2 \varphi^4\} \, .
\end{equation}
Examples of these classes of operators are
\begin{align}
\OO_\varphi &= \left(\varphi^\dagger \varphi - \frac{v^2}{2}\right)^3 \, ,\\ 
\OO_{\varphi W} &= (\varphi^\dagger \varphi) W_{\mu \nu}^I W^{I \mu \nu} \, , \\
\OO_{G} &= f_{ABC} G_{\mu \nu}^A G^{B \nu \rho} G^{C \mu}_{\rho} \, ,
\end{align}
where $\OO_\varphi$ is an operator that modifies the Higgs potential and therefore its self-interaction, $\OO_{\varphi W}$
modifies the interaction of the Higgs boson with the EW bosons while $\OO_G$ changes the cubic and quartic interactions among gluons.
These are examples of CP-even operators, but one can as well build CP-odd ones. For example, the CP-odd equivalent of $\OO_{\varphi W}$
is
\begin{equation}
\OO_{\varphi \tilde{W}} = (\varphi^\dagger \varphi) \tilde{W}_{\mu \nu}^I W^{I \mu \nu} \, ,
\end{equation}
where $\tilde{W}_{\mu \nu}^I = \frac{1}{2} \epsilon_{\mu \nu \alpha \beta} W^{ I \alpha \beta}$. This set of operators is often
neglected in studies, motivated by the need of simplification and the expectation that these operators can give contributions
to CP-odd observables while being less relevant for CP-even ones. In this work we will not comment further on this class of operators.

On the other hand, if $N_F=2$, we have a class of operators called two-fermion operators. The possible combinations are
\begin{equation}
\{\psi^2 X D, \psi^2 X \varphi, \psi^2 D^3, \psi^2 D^2 \varphi, \psi^2 D \varphi^2, \psi^2 \varphi^3\} \, .
\end{equation}
Once again, following the guiding principle of the Warsaw basis and reducing to a minimum the number of derivatives, the only remaining classes are
\begin{equation}
\{\psi^2 X \varphi, \psi^2 D \varphi^2, \psi^2 \varphi^3\} \, ,
\end{equation}
which are respectively called dipole, current and Yukawa operators. Examples of these classes are
\begin{align}
\OO_{u W} &= i (\bar{Q} \sigma^{\mu \nu} \tau_I u) \tilde{\varphi} W_{\mu \nu}^I + h.c. \, , \\
\OO_{\varphi u}^{(1)} &= i (\varphi^\dagger \overleftrightarrow{D}^\mu \varphi) \bar{u} \gamma_\mu u \, , \\
\OO_{u\varphi} &= \left(\varphi^\dagger \varphi -\frac{v^2}{2}\right) \bar{Q} u \tilde{\varphi} + h.c. \, ,
\end{align}
where $\OO_{u W}$ and $\OO_{\varphi u}$ modify the interaction of up-type quarks to EW bosons (with different Lorentz structures),
while $\OO_{u\varphi}$ shifts the Yukawa coupling of quarks, spoiling the relationship with the mass of the particle.

Finally, the last class of operators is given by $N_F=4$ and are called four-fermion operators. Depending on the flavour assumptions,
the number of these operators can quickly levitate due to combinatorics. The basic Lorentz structures are
\begin{align}
&\bar{\psi}_i \psi_j \bar{\psi}_k \psi_l \qquad \bar{\psi}_i \gamma^\mu T^A \psi_j \bar{\psi}_k \gamma_\mu T^A \psi_l  \\
&\bar{\psi}_i \gamma^\mu \psi_j \bar{\psi}_k \gamma_\mu \psi_l \qquad 
\bar{\psi}_i \sigma^{\mu \nu} \psi_j \bar{\psi}_k \sigma_{\mu \nu} \psi_l \, .
\end{align}
The full set of Warsaw basis operators is listed in Appendix~\ref{app:Warsaw} for reference.

An important feature of a complete basis is the fact that it is closed after one-loop renormalisation, \ie the anomalous dimension
matrix does not involve any new operator and relates only operators of the basis. This has been demonstrated for the Warsaw basis~\cite{Jenkins:2013zja,Jenkins:2013wua,Alonso:2013hga,Grojean:2013kd,Alonso:2014zka} using dimensional regularisation and the $\overline{MS}$ 
scheme, while the efforts for other bases are still in place.
As discussed in~\ref{sec:running}, operators mix when running to scales lower than the matching scale and in particular we find
\begin{equation}
\mu \frac{d}{d\mu} c_i(\mu) = \gamma_{ij} c_j(\mu) \, ,
\end{equation}
where $\gamma_{ij}$ is the anomalous dimension matrix and it has been determined for the Warsaw basis in the most general case, 
\ie with no flavour assumptions whatsoever. In this scenario, $\gamma_{ij}$ is a $2499\times2499$ matrix. The fact that after renormalisation
operators mix with each other is a crucial aspect of the SMEFT. In particular, it implies that even if the UV physics does not
generate certain interactions at the matching scale, these are generated once the RGE is taken into account. To put it in other
terms, it is theoretically inconsistent to switch off operators based on UV assumptions, since they will be nonetheless present
at the lower energies probed by the experiments.

\subsection{Input schemes}

The most striking effect induced by higher-dimensional operators is the introduction of vertex corrections to the
interactions among SM particles. Specifically, both SM couplings and new Lorentz structures are introduced, leading to 
unitarity violation as we will discuss in Chapter~\ref{chap:top}. However, there is another more subtle effect introduced by operators:
the modification of the SM input parameters. 
\begin{example}[\textbf{Example: Fermi constant redefinition}]
If the SM Lagrangian is extended by a tower of higher dimensional operators, a subset of them can affect several observables
that we use to extract the parameters of the theory. For example, let us consider the Fermi constant extraction from
the measurement of muon decay. We put ourselves in a simplified scenario in which the Lagrangian is given by
\begin{equation}
\Lag = \Lag_{SM} + \frac{c_{ll}}{\Lambda^2} \Op{ll}^{1221} \, ,
\end{equation}
where the four lepton dimension-6 operator generates a new contact interaction
\begin{equation}
\Op{ll}^{1221} \supset (\bar{\nu}_\mu \gamma^\alpha P_L \mu) (\bar{e} \gamma_\alpha P_L \nu_e) + h.c. \, .
\end{equation}
When computing the decay rate of the muon in electron and two neutrinos, the operator introduces a new vertex which
shifts the Fermi constant in the following way
\begin{equation}
\Gamma(\mu \to e \bar{\nu}_e \nu_\mu) = \left(G_F^0 + \frac{\sqrt{2}}{4}\frac{c_{ll}}{\Lambda^2}\right)^2 \frac{m_\mu^5}{192\pi^3} = 
G_F^2 \frac{m_\mu^5}{192\pi^3} \, ,
\end{equation}
where $G_F$ is the value that we extract from the measurement while $G_F^0$ is the one related to the SM EW parameters by
the relation
\begin{equation}
G_F^0 = \frac{\pi \alpha_{EW}}{\sqrt{2} m_W^2 (1-m_W^2/m_Z^2)} \, .
\end{equation}
If the Wilson coefficient $c_{ll}$ is zero, we recover the usual relationship between the fine structure constant $\alpha_{EW}$
and $G_F$. However, if the higher dimensional operator affects the observable, when measuring the muon decay rate, we inadvertently
absorbed the effect in the definition of $G_F$. As a consequence, the value of $\alpha_{EW}$ that we deduce is a function of 
$c_{ll}$:
\begin{equation}
\alpha_{EW} = \lp G_F - \frac{\sqrt{2}}{4}\frac{c_{ll}}{\Lambda^2} \rp \frac{\sqrt{2} m_W^2 (1-m_W^2/m_Z^2)}{\pi} \, .
\end{equation}
The effects of the operator are therefore propagated to all the EW observables, \ie whenever there is a dependence on $\alpha_{EW}$.
This phenomenon is particularly subtle since, even if one considers a process in which the operator $\Op{ll}$ does not directly
enter at the amplitude level, its effect comes in from "the back door" through
the shift of $\alpha_{EW}$. Additionally, it is worth noting that contrary to vertex corrections, these effects are "artificial"
and depend on our choice of the input parameters of the theory.
\end{example}
There are many operators at dimension-6 in the SMEFT that modify the definition of the input parameters. For instance, the operators
\begin{equation}
\OO_{\varphi d} = \partial_\mu (\varphi^\dagger \varphi) \partial^\mu (\varphi^\dagger \varphi) \qquad 
\OO_{\varphi D} = (\varphi^\dagger D_\mu \varphi)^\dagger (\varphi^\dagger D^\mu \varphi) \, ,
\end{equation}
induce a field redefinition of the Higgs field $h$ in order to put back the kinetic term in canonical form. One finds in particular
\begin{equation}
h \to h\left(1 - \left(2 c_{\varphi d} - \frac{c_{\varphi D}}{2}\right) \frac{v_0^2}{\Lambda^2}\right)^{-\frac{1}{2}} \, ,
\end{equation}
where $v_0$ is the vev parameter from the SM Lagrangian, which is modified as well from NP contributions. This redefinition is
such that all Higgs interactions are modified by a constant factor. This effect is not kinematical and therefore only total rates are
affected.
Regarding the vev, from the operator $\OO_\varphi$ previously defined, one finds that the minimum of the Higgs potential changes
\begin{equation}
v = v_0 \left(1 + \frac{3 c_\varphi v_0^2}{4 \Lambda^2 \lambda}\right)^{-\frac{1}{2}} = v_0 (1 + \delta_v)^{-\frac{1}{2}} \, .
\end{equation}
\revised{Certain operators proportional to $\varphi^\dagger \varphi$ have been defined with an additional term $-v^2/2$, as is the case of 
$\Op{\varphi}$ and the Yukawa operators. This choice is just a convention, but stems from the desire to simplify the effects of these operators. 
If this extra term is not included in the definition, the operators can induce field redefinitions and additional parametric shifts
of the masses. For instance, if the operator $\Op{\varphi}$ is defined as $(\varphi^\dagger \varphi)^3$, it would not only shift the vev, but induce as well a redefinition of the Higgs mass and the Higgs field.}

Another relevant effect is given by the operator
\begin{equation}
\OO_{\varphi W B} = (\varphi^\dagger \tau^I \varphi) B_{\mu \nu} W_I^{\mu \nu} \, ,
\end{equation}
which induces a kinetic mixing for the neutral $SU(2)_L$ and the hypercharge fields.
Far from being an exhaustive list of the affected coefficients, this effect poses the problem of defining the input parameters of our theory. A summary of the EW parameters shifts is here reported
\begin{align}
e &=\frac{g g^{\prime}}{\sqrt{g^{2}+g^{\prime 2}}} \frac{1}{\sqrt{1+2 s_{W, 0} c_{W, 0} \delta_{W B}}} \, ,\\
v &= v_0 (1 + \delta_v)^{-\frac{1}{2}} \, , \quad
m_W = \frac{g v}{2} \, , \\
m_{Z}^{2} &=\frac{\left(g^{2}+g^{\prime 2}\right) v^{2}}{4}\left(1+\delta_{T}\right) \frac{1+2 s_{W, 0} c_{W, 0} \delta_{W B}}{1-\delta_{W B}^{2}} \, ,
\end{align}
where all the NP effects are implicit and $c_{W, 0}$, $s_{W, 0}$ are the sine and cosine of the Weinberg angle. We have in particular
\begin{align}
\delta_T &= \frac{c_{\varphi D}}{2}\frac{v_0^2}{\Lambda^2} \, ,\\
\delta_{WB} &= c_{\varphi WB} \frac{v_0^2}{\Lambda^2} \, .
\end{align}

Whenever we want to make predictions, the first step is to have a finite amount of measurements that allow us
to fix certain parameters of the theory and then use them to predict other observables to be tested by experiments. As mentioned in Section~\ref{sec:EWlag}, the EW sector is characterised by four input parameters. Once
these have been chosen and fixed by the measured values, one needs to propagate all the NP effects onto the dependent parameters of the theory. Two main input schemes are commonly used: the $\alpha_{EW}$ scheme and the $m_W$ scheme.

\subsubsection{The $\alpha_{EW}$ scheme}

In this scheme, the input parameters of the EW sector are chosen to be $\{\alpha_{EW}, \\G_F, m_h, m_Z\}$. In particular, the fine structure constant
is defined from the electric charge of the electron
\begin{equation}
\alpha_{EW} = \frac{e^2}{4\pi} \, ,
\end{equation}
and can be precisely measured through the electron anomalous magnetic moment in atom interferometry. The Fermi constant on the other hand is
measured from the decay rate of the muon, while the Higgs and $Z$ masses from resonant productions. The big disadvantage of this 
scheme is the fact that $m_W$ becomes a function of the Wilson coefficients. In particular, we have
\begin{equation}
m_W = \left(\frac{e v_0}{2 s_{W, 0}}\right) \left(1 + \delta_v + 2 \frac{c_{W, 0}}{s_{W, 0}} \delta_{WB} - \frac{c_{W, 0}^2}{c_{W, 0}^2 - s_{W, 0}^2} \delta_{m_Z^2}\right)^{\frac{1}{2}} \, .
\end{equation}
This means that the effects of the operators enter in 
the denominator of the propagators, changing the pole structure and making power counting at the amplitude level more cumbersome.
This scheme is particularly indicated however for EW precision tests in which $\alpha_{EW}$ plays a central role in the predictions.

\subsubsection{The $m_W$ scheme}

This scheme is more indicated for LHC physics and is given by the choice of input parameters $\{m_W, G_F, m_h, m_Z\}$.
In this case, it is the fundamental electric charge which is a dependent parameter and will change when turning on the operators, but this is easier to track and deal with
since the gauge coupling always appears in the numerator of amplitudes and therefore affects only total rates. The electric charge as a function of the input parameters reads
\begin{align}
e &=e_{0}\left(1-\frac{\delta_{v}}{2}-\frac{c_{W, 0}}{s_{W, 0}} \delta_{W B}-\frac{c_{W, 0}^{2}}{s_{W, 0}^{2}} \frac{\delta_{T}}{2}\right) \, ,\\
e_0 &= \frac{2 m_W s_{W, 0}}{v_0} \, .
\end{align}
In particular,
loop calculations are found to be easier in this scheme, since poles are not affected. For these reasons, this will be the scheme employed for the rest of the work unless stated otherwise.

\subsection{Flavour assumptions}

As previously mentioned, the number of dimension-6 Wilson coefficients changes considerably according to the flavour assumptions
of the model. In the case of flavour universality the model is invariant for flavour rotations under 
$U(3)^5 = U(3)_u \times U(3)_d \times U(3)_q \times U(3)_l \times U(3)_e$, \ie transformations of the kind
\begin{equation}
\psi_j^\prime = U^{ji} \psi_i \qquad U^\dagger = U^{-1} \, ,
\end{equation}
with $\psi$ one of the five fundamental fermion fields.
In this scenario the number of independent degrees of freedom of the
SMEFT dim-6 Lagrangian is $59$ (excluding Hermitian conjugations and flavour structure). On the other hand, in the most extreme case of complete absence of flavour symmetry, the number of Wilson coefficients is $2499$. The choice of the flavour assumptions is therefore of extreme importance for the practicability of the theory.

In the SM, flavour universality is broken by the Yukawa interactions with the Higgs. Because of this it is reasonable to
rely on the principle of Minimal Flavour Violation (MFV), stating that the Yukawa matrices are the only source of flavour violation
of the theory. The idea is to classify the operators according to their properties under $SU(3)^5$ and, if they break the symmetry, they
must do so proportionally to the Yukawa matrices. For example, the Yukawa or the dipole operators are given by
\begin{align}
c_{uW} \, \OO_{u W} &= i \, c_{uW} \, Y^{ij}_u(\bar{Q}^i \sigma^{\mu \nu} \tau_I u^j) \tilde{\varphi} W_{\mu \nu}^I + h.c. \, , \\
c_{u \varphi} \, \OO_{u\varphi} &= c_{u \varphi} \, Y^{ij}_u (\varphi^\dagger \varphi -\frac{v^2}{2}) \bar{Q}^i u^j \tilde{\varphi} + h.c. \, ,
\end{align}
where $i, j$ run through the three generations.

The MFV scheme is widely used and further simplifications are often employed for practical purposes. For instance, in this work,
we will assume for the SMEFT Lagrangian a flavour symmetry $U(2)_u \times U(3)_d \times U(2)_q \times U(3)_l \times U(3)_e$,
such that operators containing top quarks and/or third generation quark doublets can be singled out. In particular, this symmetry only
allows a Yukawa coupling for the top, while all the other quarks and leptons are assumed to be massless. The symmetry also implies 
flavour universality for current operators, apart for the ones involving top quarks. These flavour assumptions are motivated by
the fact that the top is the only fermion with a Yukawa coupling of $\mathcal{O}(1)$ and all the other fermions are orders of 
magnitude lighter and their interaction with the Higgs is practically irrelevant for high-energy experiments\footnote{The only exceptions
are the Yukawas of the bottom quark and the tau, which are necessary to have a realistic description of the Higgs decay.}.

%
%
%

\chapter{Top quark electroweak interactions at high energy}
\pagestyle{fancy}
\label{chap:top}

\hfill
\begin{minipage}{10cm}

{\small\it 
``In science novelty emerges only with difficulty, manifested by resistance, against a background provided by expectation.''}

\hfill {\small Thomas Kuhn, \textit{The Structure of Scientific Revolutions}}
\end{minipage}

\vspace{0.5cm}

As already discussed in Sec.~\ref{sec:unitarity}, a fascinating aspect of the SM is that, being a spontaneously broken gauge theory,
the high-energy behaviour of scattering amplitudes is characterised by a set of intricate and neat cancellations that ensures the 
unitarity of the theory. In particular, in absence of these, the amplitudes exhibit unacceptable energy growing behaviours.
The most well-known example is the scattering of longitudinal EW bosons, which was briefly presented in the aforementioned
section. As we demonstrated, without the Higgs boson in the theory, the scattering amplitude grows with $E^2$ and only once
the Higgs has been accounted for, unitarity is restored~\cite{LlewellynSmith:1973yud,Lee:1977eg,Lee:1977yc}. This is not a feature of only EW bosons interactions, but it is true 
as well for scatterings involving fermions. For instance, it can be shown that for any process involving fermions, the unitarity bounds
scale with the inverse power of the mass of the fermion~\cite{Appelquist:1987cf, Maltoni:2001dc, Maltoni:2000iq}. This suggests that the top quark, the heaviest particle of the SM realm, 
is a valuable probe of NP interactions.

Recently our knowledge on the top quark Yukawa interaction has deepened. The observation of $ttH$ production at the LHC~\cite{CMS:2018rbc,Sirunyan:2018mvw,Sirunyan:2018shy,Aaboud:2017jvq,Aaboud:2017rss} has in particular explicitly established that the top quark
Yukawa is of order 1. This fact has opened the way for a precise determination of the EW top quark interactions, that are 
still imprecisely known to-date. The study of the top quark EW sector is a very promising place to look for signs of NP,
exploiting the energy growing behaviours that are generated in presence of a deviation from the SM couplings. In this perspective,
the SMEFT is the perfect framework to parametrise the NP effects and quantify the sensitivity of specific processes to
modified interactions.

In this Chapter we study several EW scattering amplitudes involving at least a top quark and EW bosons, investigating the
degree of energy growth induced by dimension-6 operators in each helicity configuration. The objective is to identify realistic
processes at present and future colliders that can be exploited to increase our sensitivity to NP~\cite{Degrande:2018fog,Corbett:2014ora,Corbett:2017qgl,Dror:2015nkp}.

\section{Modified interactions in the EW sector}

Being interested in the EW interactions involving top quarks, we employ the SMEFT with imposed flavour symmetry $U(3)_{\ell}\times U(3)_{e}\times U(3)_{d}\times U(2)_{q}\times U(2)_{u}$,
which allows us to diminish considerably the number of dimension-6 operators and focus on the relevant interactions for the study.
In particular, operators involving top quarks or third generation quark doublets are singled out, while flavour universality is
assumed for the others (see~\cite{AguilarSaavedra:2018nen}). The list of dimension-6 operators that affect top-EW interactions is presented in Table~\ref{tab:topOps}.

\begin{table}
{\centering
\renewcommand{\arraystretch}{1.4}
\begin{tabular}{|ll|ll|}
    \hline
     $\Op{W}$&
     $\varepsilon_{\sss IJK}\,W^{\sss I}_{\mu\nu}\,
                             {W^{{\sss J},}}^{\nu\rho}\,
                             {W^{{\sss K},}}^{\mu}_{\rho}$&
     $\Op{t\phi}$&
     $\left(\pdp-\tfrac{v^2}{2}\right)
     \bar{Q}\,t\,\tilde{\phi} + \text{h.c.}$
     \tabularnewline
     $\Op{\phi W}$&
     $\left(\pdp-\tfrac{v^2}{2}\right)W^{\mu\nu}_{\sss I}\,
                                    W_{\mu\nu}^{\sss I}$&
     $\Op{tW}$&
     $i\big(\bar{Q}\sigma^{\mu\nu}\,\tau_{\sss I}\,t\big)\,
     \tilde{\phi}\,W^I_{\mu\nu}
     + \text{h.c.}$
     \tabularnewline
     $\Op{\phi B}$&
     $\left(\pdp-\tfrac{v^2}{2}\right)B^{\mu\nu}\,
                                    B_{\mu\nu}$&
     $\Op{tB}$&
     $i\big(\bar{Q}\sigma^{\mu\nu}\,t\big)
     \,\tilde{\phi}\,B_{\mu\nu}
     + \text{h.c.}$
     \tabularnewline
     \cline{3-4}
     $\Op{\phi WB}$&
     $(\phi^\dagger \tau_{\sss I}\phi)\,B^{\mu\nu}W_{\mu\nu}^{\sss I}\,$&
     $\Op{\phi Q}^{\sss(3)}$&
     $i\big(\phi^\dagger\lra{D}_\mu\,\tau_{\sss I}\phi\big)
     \big(\bar{Q}\,\gamma^\mu\,\tau^{\sss I}Q\big)$
     \tabularnewline

     $\Op{\phi D}$&
     $(\phi^\dagger D^\mu\phi)^\dagger(\phi^\dagger D_\mu\phi)$&
     $\Op{\phi Q}^{\sss(1)}$&
     $i\big(\phi^\dagger\lra{D}_\mu\,\phi\big)
     \big(\bar{Q}\,\gamma^\mu\,Q\big)$
     \tabularnewline
     $\Op{\phi d}$&
     $(\varphi^\dagger\varphi)\square(\varphi^\dagger\varphi)$ &
     $\Op{\phi t}$&
     $i\big(\phi^\dagger\lra{D}_\mu\,\phi\big)
     \big(\bar{t}\,\gamma^\mu\,t\big)$
      \tabularnewline
      &&     
      $\Op{\phi tb}$&
     $i\big(\tilde{\phi}^\dagger\,{D}_\mu\,\phi\big)
     \big(\bar{t}\,\gamma^\mu\,b\big)
     + \text{h.c.}$
     \tabularnewline
      \hline
      %

     %

 
  \end{tabular}

\caption{\label{tab:topOps}
SMEFT operators describing new interactions in the EW top quark 
sector, consistent with a $U(3)^3\times U(2)^2$ flavour symmetry.  
$Q,\,t$ and $b$ denote the third generation components of $q,\,u$ and $d$.
}
}
 \end{table}

The operator $\Op{\varphi t b}$ formally violates the chosen flavour symmetry, but we decide to retain it due to
its interesting helicity structure\footnote{For this reason also the operator does not mix with the others under RGE evolution.}. We also assume the $m_W$ input scheme and absorb all the NP effects in the appropriate 
physical quantities. In order to perform the analysis, we implemented the model in FeynRules~\cite{Alloul:2013bka,Degrande:2011ua} so that we could make use of 
tools such as FeynArts~\cite{HAHN2001418}, FeynCalc~\cite{Shtabovenko:2016sxi} and \MG~\cite{Alwall:2011uj}.
The operators listed in Table~\ref{tab:topOps} modify all the interactions of the top quark with EW bosons, including the Higgs, but leave untouched the
coupling to the photon, which is protected by the $U(1)_{EW}$ gauge symmetry. New Lorentz structures are however introduced
by the dipole operators $\Op{t B}$ and $\Op{t W}$ which induce a new $t\bar{t}\gamma$ vertex.

A general but fundamental feature of the SMEFT is that most operators modify multiple vertices at the same time, correlating
the effects of NP. In particular, operators that modify 3-point interactions, often produce also 4-point vertices, as a consequence of
the underlying gauge symmetry which is preserved in the SMEFT. These contact terms are particularly interesting since they often induce 
maximal energy growth, and a de-correlation from the 3-point vertices is impossible unless higher dimensional operators
are accounted for. This is in contrast with the anomalous coupling framework and we will discuss in the following how these differences can
potentially lead to drastically different interpretations.

It is important to stress that Table~\ref{tab:topOps} is not an exhaustive list of all the operators that affect
top quarks processes at colliders. For instance, the operators
\begin{align}
    \Op{tG} & = i\big(\bar{Q}\sigma^{\mu\nu}\,T_{\sss A}\,t\big)
     \,\tilde{\phi}\,G^{\sss A}_{\mu\nu}
     + \text{h.c.},\\
     \Op{G} & = g_{\sss S}f_{\sss ABC}\,G^{\sss A}_{\mu\nu}\,
                             {G^{{\sss B},}}^{\nu\rho}\,
                             {G^{{\sss C},}}^{\mu}_{\rho},\\
     \Op{\phi G} & = \left(\pdp-\tfrac{v^2}{2}\right)G^{\mu\nu}_{\sss A}\,
                                             G_{\mu\nu}^{\sss A},
\end{align}
are all relevant for gluon-initiated production modes. However, in this study we focus on the operators that directly modify
EW sub-scatterings in order to exploit their energy enhanced sensitivity. Moreover, these operators influence the QCD
component of the processes, which is therefore characterised by completely different kinematical traits. Their study is relevant 
from a global fit perspective and will be taken into account in Chapter~\ref{chap:globalfit}, where a joint fit of
top, Higgs and diboson data will be presented.

\subsection{Anomalous couplings}

Although we parametrise NP effects in the SMEFT framework, we find it useful to compare it to the general anomalous couplings (AC)
Lagrangian, which also helps in understanding the effects of dimension-6 operators. It is in this context that the assessment of unitarity violating behaviour induced by spoiling of cancellations in the SM is evident. However, it is worth noting that
in this framework deviations are not correlated and higher-point interactions, as well as new Lorentz structures, might be missing 
unless explicitly introduced. We employ a general AC Lagrangian given by
\begin{align}
    \label{eq:L_AC}
    \nonumber
    \mathcal{L} \supset &-\gth \, \bar{t} \, t \, h + \gwh \, W^\mu W_\mu \, h 
    + \gzh \, Z^\mu  Z_\mu \, h 
    + \gbtw \,( \bar{t} \, \gamma^\mu \, P_L \, b \, W_\mu + \text{h.c})
    \\\nonumber
    &+ \bar{t} \, \gamma^\mu (\gztr \, P_R + \gztl \, P_L) \, t \, Z_\mu 
    + \bar{b} \, \gamma^\mu (\gzbr \, P_R + \gzbl \, P_L) \, t \, Z_\mu
    \\\nonumber
    &- \gta \, \bar{t} \, \gamma^\mu \, t \, A_\mu 
    + \gwa \, (W^\mu \, W^\nu \, \partial_\mu \, A_\nu + \text{perm.})\\
    &+ \gwz \, (W^\mu \, W^\nu \, \partial_\mu \, Z_\nu + \text{perm.}),
\end{align}
where the SM couplings are
\begin{equation*}
    \label{eq:AC_SM}
    \begin{gathered}
    \gth = \frac{g \, \mt}{2 \, \mw} \, ,\quad
    \gwh = g \, \mw \, , \quad
    \gzh = \frac{g \, \mz}{\cos\theta_W} \, , \\
    \gztr = -\frac{2 \, g \, \sin^2\theta_W}{3 \, \cos\theta_W} \, , \quad
    \gztl = \frac{g}{\cos\theta_W}\left(\frac{1}{2} - \frac{2 \sin^2\theta_W}{3} \right) \, , \\
    \gzbr = \frac{g \, \sin^2\theta_W}{3 \, \cos\theta_W} \, ,\quad
    \gzbl = -\frac{g}{\cos\theta_W}\left(\frac{1}{2} - \frac{\sin^2\theta_W}{3} \right) \, , \\
    \gbtw= \frac{g}{\sqrt{2}} \, ,\quad 
    \gta = \frac{2}{3} \, g \, \sin\theta_W \, , \quad
    \gwa = g \, \sin\theta_W \, , \quad
    \gwz = g \, \cos\theta_W \, .
    \end{gathered}
\end{equation*}

\section{High energy scatterings}

\begin{figure}[t!]
\centering
\includegraphics[width=0.4\textwidth]{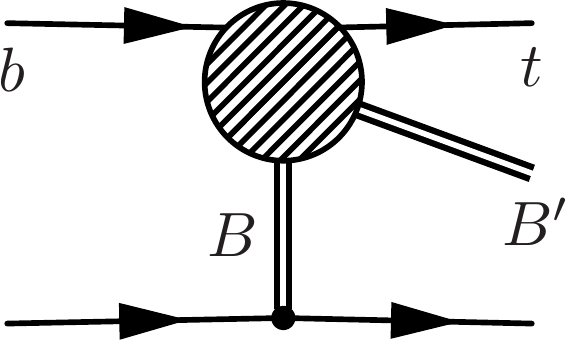}
\caption{\label{fig:topology}
Schematic Feynman diagram for the embedding of an EW top scattering amplitude into a physical, single-top process at a hadron collider. $B$ and $B^\prime$ can be many combinations of $Z,\,\gamma,\,W$ and $h$.
}
\end{figure}

In this section we proceed in studying the $2 \to 2$ generic scattering process involving a top quark. In particular, the interactions
considered are of the kind $f B \to f^\prime B^\prime$, with $f, f^\prime = b, t$ and $B, B^\prime = h, W, Z, \gamma$ and at least
one of the fermions is a top quark. We classify the processes in four categories, characterised by the number of top quarks involved and
the presence of the Higgs boson in the external legs (see Table~\ref{tab:amplitude_organisation}).
\begin{table}[t!]
\centering
\begin{tabular}{|r|p{4.5cm}|p{4.5cm}|}
\hline
&Single-top &Two-top ($t\bar{t}$) \tabularnewline\hline
w/o Higgs  &  $b \, W \to t \, (Z/\gamma)$ \hfill
& $t \, W \to t \, W$
\hfill\newline 
$t \, (Z/\gamma) \to t \, (Z/\gamma)$ \hfill 
\tabularnewline\hline
w/\phantom{o}  Higgs & $b \, W \to t \, h$ 
\hfill & 
$t \, (Z/\gamma) \to t \, h$ \hfill \newline  
$t \, h \to t \, h$ \hfill\tabularnewline\hline
\end{tabular}
\caption{The ten $2\to 2$ scattering amplitudes whose high-energy behaviour we study in this work.
\label{tab:amplitude_organisation}}
\end{table}
The $2\to2$ scattering amplitudes cannot be directly probed at colliders, but they can be embedded in physical processes. For instance,
in the case of the sub-amplitude class $b B \to t B^\prime$, we have that they can be probed with single-top processes (see Fig.~\ref{fig:topology}).
We define the high-energy limit of the amplitudes in terms of Mandelstam variables $s \sim -t \gg v^2$. For each of the 
processes listed in Table~\ref{tab:amplitude_organisation} we compute all the possible helicity amplitudes retaining only the leading
terms with the objective of identifying the sources of energy enhancement. We do so for both the SM and the dimension-6 contributions,
finding as expected a maximum degree of growth of $E^0$ and $E^2$ respectively. A summary of our findings is collected in Appendix~\ref{app:helamp_tables}.

\subsection{Energy growth and interference}

The leading order contribution to the cross section comes from the interference of the EFT amplitude with the SM one. If we limit ourselves
to $\mathcal{O}(1/\Lambda^2)$ corrections at the amplitude level, we indeed have
\begin{equation}
\mathcal{M} = \mathcal{M}_{SM} + \frac{c_i}{\Lambda^2}\mathcal{M}_i^6 \, ,
\end{equation}
where a sum over $i$ is intended for all the dimension-6 operators affecting the process. Being the cross section proportional 
to the squared matrix element, we find
\begin{equation}
\sigma = \sigma_{SM} + \frac{c_i}{\Lambda^2} \sigma_i + \frac{c_i c_j}{\Lambda^4} \sigma_{ij} \, ,
\end{equation}
where $\sigma_i$ is given by the interference between the SM amplitude and the NP contribution. On the other hand the quantity
$\sigma_{ij}$ is produced by squaring the purely NP terms and is the next to leading correction in the EFT expansion, suppressed by a higher power of $\Lambda$. This could suggests that the 
linear correction (the interference) will provide the best sensitivity to NP but it turns out that this is not always the case.
In particular, while square contributions are always expected to grow with energy, this is not the case for the interference.
For example, if we look at the table for $b \, W^+ \to t \, h$ in Appendix~\ref{app:helamp_tables}, we find that the current operator
$\Op{\phi Q}^{\sss (3)}$ has an energy growing contribution proportional to the energy $E$ for the $(-, 0, +)$ helicity configuration (left-handed $b$, right-handed $t$, longitudinal $W$).
However, the SM amplitude is suppressed in energy ($\propto 1/E$) and the interference is therefore not expected to exhibit
an enhanced energy behaviour. On the other hand, if we look at the helicity configuration $(-, 0, -)$ we are in the best possible
case, in which the operator effect is maximally enhanced ($E^2$) and the SM is constant in energy. This suggests that the interference 
will also grow in energy and it is a particularly rare and interesting situation. As there are no external transverse gauge bosons, this finding is consistent with the studies in~\cite{Azatov:2016sqh}.
In particular we confirm that whenever a transverse gauge boson is present in any of the processes under study, the SM compensates the growth from the modified interaction,
yielding an interference that is constant in energy.

Concerning the squared contributions, these are always guaranteed to grow in energy, if the corresponding amplitude does so.
These terms are expected to become increasingly less relevant as the precision of experiments improve, since the Wilson coefficients
will likely be bound to be smaller and smaller. However, the naive power counting can be broken in the scenario in which the interference 
is suppressed and the impact of the squared contributions can overcome the linear ones. For this and other reasons, we find useful to retain these
terms and study their effects.

\subsection{Energy growth and contact terms}

Before getting down to do any calculation, the maximal energy growth induced by an operator can be guessed by the eventual 
contact term generated by it. As a matter of fact, for an operator of dimension $K$, the corresponding coupling
must have dimension $4-K$
\begin{align}
    \mathcal{L}\supset\frac{1}{M^{\sss K-4}}\mathcal{O}_{\sss K},
\end{align}
where $M$ is a generic mass parameter. If we consider the most general case of $2 \to N$ scattering, the amplitude must have dimension
$2-N$. If it is mediated by an insertion of the operator $\mathcal{O}_{\sss K}$, the maximum energy scale has to compensate the mass
dimensions, so
\begin{align}
    \mathcal{M}\propto\frac{1}{M^{K-4}}E^{K-N-2}.
\end{align}
For the specific case of $2 \to 2$ amplitudes, the amplitude can at most scale as $E^{K-4}$, but if massive gauge bosons are
involved, things can get more convoluted. For instance, the longitudinal polarisation can be approximated with the derivative 
of a Goldstone boson $\frac{\partial_\mu G}{M_{\sss V}}$, contributing effectively as a dimension 2 degree of
freedom. This is a consequence of the Goldstone equivalence theorem~\cite{Cornwall:1974km}. Moreover, in the case of the SMEFT, 
many operators involve the Higgs doublet and after SSB, an insertion of the vacuum expectation value can take place. This can consume some of the available energy 
dependence. A naive and generalised formula for the energy dependence of a $2 \to N$ scattering is therefore 
\begin{equation}
    \label{eq:naive_e_general}
    \mathcal{M} \propto \frac{v^m}{\Lambda^{K-4}} \frac{E^{K-N-m-2+n}}{M_{\sss V} ^n} \, ,
\end{equation}
where $n$ is the number of longitudinal gauge bosons in the external legs and $m$ the number of vev insertions to generate
the contact term of interest. For a $2 \to 2$ scattering this reduces to
\begin{equation}
    \label{eq:naive_e}
    \mathcal{M} \propto \frac{v^m}{\Lambda^{2}} \frac{E^{2-m+n}}{M_{\sss V} ^n}.
\end{equation}
This formula gives us the maximal energy growth for a specific helicity configuration. Reduced degrees of growth can be often
understood as the helicity flipped versions of that, where one pays the flip by introducing a proportionality to the mass of the particle.

In order to generate the maximal energy growth, one has to minimise the vev insertions with the objective of generating the 
highest point contact interaction possible for a given operator, reducing as much as possible the number of propagators. It is often instructive to look at the problem in the Feynman
gauge, where the longitudinal polarisations of the EW bosons are explicitly represented by the Goldstone bosons. In this specific
gauge the Higgs doublet can be recasted as
\begin{align}
    \phi = \frac{1}{\sqrt{2}}\begin{pmatrix}
        -iG^+\\
        v+h+i G^{\sss 0}
    \end{pmatrix},\quad
    \tilde{\phi}=i\tau_2\phi^\ast= \frac{1}{\sqrt{2}}\begin{pmatrix}
    v+h-i G^{\sss 0}\\
    iG^-
    \end{pmatrix}.\quad
\end{align}
In current operators for instance, we have that neutral fermion vector currents, $\bar{t}\gamma^\mu t$ are coupled to
\begin{align}
    i\big(\phi^\dagger\lra{D}_\mu\tau_{\sss 3}\phi\big)&\supset
    (v+h)\lra{\partial}_\mu G^{\sss 0} - 
    2\mz h Z_\mu + 
    iG^-\lra{\partial}_\mu G^+,\\
    i\big(\phi^\dagger\lra{D}_\mu \phi \big)&\supset 
    G^{\sss 0}\lra{\partial}_\mu h + 2\mz h Z_\mu +\big(i G^-\partial_\mu G^+ + 2i\mw G^- W^+_\mu  + \text{ h.c.}\big)
\end{align}
and the neutral scalar currents, $\bar{t}_L t_R$ instead to
\begin{align}
    (\pdp) \,\tilde{\phi} \supset \frac{v}{2\sqrt{2}}\left(3h^2 + G^{\sss 0}G^{\sss 0} + 2G^+G^- - 2 i G^{\sss 0} h\right).
\end{align}
On the other hand the right-handed charged current, $\bar{t}\gamma^\mu b$ couples to
\begin{align}
    \nonumber
    i\sqrt{2}(\tilde{\phi}^\dagger\,D_\mu\phi)\supset&
    -(h+i G^{\sss 0})\lra{\partial}_\mu G^+
    -2\mw (h+iG^{\sss 0})\,W_\mu^+
    -2i\mz c^2_{\sss W} G^+Z_\mu\\
    \label{eq:phitDphi_expand}
    &-v(\partial_\mu-ieA_\mu) G^+,
\end{align}
and for the left,
\begin{align}
    \nonumber
    \frac{i}{\sqrt{2}}\big(\phi^\dagger\lra{D}_\mu\tau_{+}\phi\big)\supset &
    (h-i G^{\sss 0})\lra{\partial}_\mu G^+
    +2\mw h\,W^+_\mu
    +2i\mz s_{\sss W}^2 G^+Z_\mu\\
    \label{eq:phiD3phi_expand_charged}
    &+v(\partial_\mu-ieA_\mu) G^+,
\end{align}
where $\tau_{+}=\tau{\sss 1}-i \tau{\sss 2}$. Finally the charged scalar current, $\bar{t}_L b_R$ couples to
\begin{align}
    (\pdp) \,\tilde{\phi} \supset -i v h G^+.
\end{align}
From this we can see that contact terms are generated, leading to $E$ and $E^2$ energy growths as expected.

As an example, looking again at the table for $b W^+ \to t h$ in Appendix~\ref{app:helamp_tables}, we find an $E^2$ energy growth
from the $\Op{\varphi t b}$ operator in the $(+, 0, +)$ helicity configuration. In unitary gauge, the operator generates indeed
a dimension-5 contact term, \ie
\begin{align}
    \Op{\phi tb} = i(\tilde{\varphi} D_\mu \varphi)(\bar{t} \gamma^\mu b) +\text{h.c.}\quad \to \quad v\,h \, W^{\sss +} \, \bar{t}_{\sss R} \gamma^\mu b_{\sss R} +\text{h.c.},
\end{align}
Although this would naively suggest at most a $E$ energy enhancement, the longitudinal degree of freedom of the $W$ allows
us to gain one more power of energy, cancelling the vev insertion we had to pay. On the other hand, in Feynman gauge the 
$E^2$ energy growth can be immediately understood from the dimension-6 contact term generated
\begin{equation}
    \Op{\phi tb} \quad \to \quad h \, \partial_\mu G^{\sss +} \, \bar{t}_{\sss R} \gamma^\mu b_{\sss R} +\text{h.c.}.
\end{equation}
Flipping the top or $W$ helicities, yields a $\sqrt{-t}m_t$ and $\sqrt{-t}m_W$ dependence in the $(+, 0, -)$ and $(+, +, +)$ configurations respectively.

In unitary gauge the sources of unitarity violations might not always be clear. Sometimes however they can be understood in terms of spoiling of 
SM cancellations, \ie the fact that gauge invariance and the Higgs mechanism relate specific parameters of the theory. This is
for instance the case of the Yukawa operator $\Op{t\varphi}$, which affects the relationship between the top mass and its coupling
to the Higgs boson, shifting the $g_{th}$ coupling. However, the same effect can
be understood instead in terms of contact operators in the Feynman gauge, where
\begin{align}
   \Op{t\phi}=\left(\pdp-\tfrac{v^2}{2}\right)
     \bar{Q}\,t\,\tilde{\phi} + \text{h.c.} \quad \to \quad v\,\bar{b}_{\sss L}\,t_{\sss R}\,G^-h +\text{h.c.}\, 
\end{align}
is responsible for the generation of a contact term of effective dimension-5. In a $2 \to 2$ scattering this induces only a linear 
growth, but the operator can be maximally exploited in 5-point interactions~\cite{Henning:2018kys}.

The vast majority of the top quark operators in Table~\ref{tab:topOps} generate a maximal growth already in the $2\to2$ scatterings.
For instance, the bosonic operators $\Op{\phi W}$, $\Op{\phi B}$, $\Op{\phi WB}$, $\Op{\phi W}$, $\Op{\phi D}$ and $\Op{\phi d}$ cannot be 
maximally exploited since they generate contact terms only for fully bosonic amplitudes. However, they contribute to the scattering 
amplitudes and can generate growths by spoiling the SM cancellations.

\subsection{Energy growth and gauge invariance}

While the SMEFT is respecting gauge invariance by construction, in the AC formalism this is not automatically true. Unless
ad hoc parameter dependence is enforced on the various couplings and contact terms added to the Lagrangian, the theory violates
gauge variance leading to energy growths bigger than the ones predicted by our naive formula. As an example, we can have a look at the
dipole operator $\Op{t W}$ which, apart for modifying the vertices for neutral and charged currents, generates contact terms
\begin{equation}
    \label{eq:dipolecontact}
    \Op{tW} =  i\big(\bar{Q}\sigma^{\mu\nu}\,\tau_{\sss I}\,t\big)\,
     \tilde{\phi}\,W^I_{\mu\nu}
     + \text{h.c.}\quad 
     \to \quad 
     \begin{cases}
     gv\,\bar{t}_{\sss L} \sigma^{\mu\nu} t_{\sss R} \, W_\mu^+ W_\nu^-\, \\
     gv\,\bar{b}_{\sss L} \sigma^{\mu\nu} t_{\sss R} \, Z_\mu W_\nu^- \, 
     \end{cases}     
\end{equation}
These affect in particular $b \, W^+ \to t \, Z$ and $t\,W\to t\,W$ scatterings. If we use our naive formula for these contact terms,
we find that the expected energy dependence is $E^3$, which is clearly higher than the naively expected one from dimension-6 operators.
In the SMEFT, this energy dependence is cancelled by the other vertex modifications induced by the same operator, which also produce
a $E^3$ growth. In the AC framework however, the contact term is not included in the description and in the conventional Lagrangian 
for the dipole interaction~\cite{AguilarSaavedra:2008zc} we have
\begin{align}
    \mathcal{L}_{\text{dip.}} \supset -\frac{g}{\sqrt{2}}\bar{b} \, \sigma^{\mu\nu}\left( 
    g_{\sss L} P_{\sss L}+g_{\sss R} P_{\sss R}
\right) t \, \partial_\mu \, W_\nu \, ,
\end{align}
which produces indeed the $E^3$ energy enhancement for the $(-, 0, +, 0)$ helicity configuration. As a matter of fact, not
only the maximum degree of growth is higher, but many helicity configurations exhibit enhanced energy dependence.

In the SMEFT, the correct maximal degree of growth is manifest in the Feynman gauge, in which the field strength is only responsible for the transverse polarisations
of the EW bosons. The contact term for this interaction is indeed given by
\begin{align}
    \label{eq:dipole_contact}
    G\,\bar{f}_{\sss L}\sigma^{\mu\nu}f_{\sss R}\partial_\mu V_
\nu
\end{align}
and as we can see in the table for $t W \to t W$ in the Appendix~\ref{app:helamp_tables}, it is responsible for the maximal 
behaviour for the mixed transverse-longitudinal helicity configuration.

The conclusion of this argument is that we can in principle face different and incompatible predictions when comparing the AC framework and 
the SMEFT. This can potentially lead to overestimation of the NP effects, unless the proper contact terms are accounted for.
For instance, the study of a process such as $tZj$ is affected by the previously mentioned contact term and the two formalisms would disagree.
On the other hand, when considering top quark decay, the contact term does not play a role and the two approaches agree perfectly.
Extra care is therefore needed when dealing with the AC Lagrangian, as gauge violation can give unwarranted effects.

\section{Embedding the amplitudes in collider processes}
\label{sec:EWA}

As already mentioned, the $2 \to 2$ scattering amplitudes under study cannot be directly probed at colliders, but they 
need to be embedded in physical processes. Before doing any full numerical simulation, we can however gain an analytic insight on this
procedure by making use of the Effective W Approximation (EWA)~\cite{Dawson:1984gx,Kunszt:1987tk,Borel:2012by}. Thanks to this,
we can describe the emission of EW bosons off light fermion legs, allowing us to factorise the full process in two distinct parts:
one describing the emission of the $W$ by means of splitting functions and one describing the hard scattering collision. In this
way one can reduce higher multiplicity scatterings in lower ones and access directly the $2 \to 2$ amplitudes we are interested in,
with the objective of understanding how the energy dependence is propagated to the full process.
In particular, this is relevant for processes such as $tXj$, $t\bar{t}W j$ and $t\bar{t}$ through VBF.
We present now the formalism used for the EWA and then move on to a relevant example for our study.
We consider the generic process $q X \to q^\prime Y$, with $q$ and $q^\prime$ massless fermions while $X$ and $Y$ two generic initial and
final states. In particular, the fermion current $q-q^\prime$ emits a t-channel W boson that carries a fraction $x$ of the longitudinal 
momentum of $q$. The four momenta are
\begin{equation}
\begin{aligned}
P_q &= (E,\vec{0},E) \, , \\
P_X &= (E_X, \vec{0}, -E) \, ,\\
P_{q\prime} &=  \left(\sqrt{(1-x)^2 E^2 +p_\perp^2},\, \vec{p}_\perp,\, E(1-x)\right) \,.
\end{aligned}
\end{equation}
The $W$ has momentum 
\begin{equation}
\begin{aligned}
K &= P_q - P_{q\prime} = \left(\sqrt{x^2E^2+p_\perp^2 + m_W^2 - V^2}, - \vec{p}_\perp, xE\right) \, , \\
V^2 &= m_W^2 -K^2 \approx m_W^2 +\frac{p_\perp^2}{1-x}\left[1 + O\left(\frac{p_\perp^2}{E^2}\right)\right] \, ,
\end{aligned}
\end{equation}
with $V$ standing for the virtuality of the $W$. Being the EWA an approximation, it is only valid when
\begin{equation}
    \label{eq:ewa_approx}
E \sim x E \sim (1-x) E, \qquad \frac{m_W}{E} \ll 1, \qquad \frac{p_\perp}{E} \ll 1 \, .
\end{equation}
The full differential cross section in the EWA approximation can be written as
\begin{equation}
\frac{d\sigma_{EWA}}{dxdp_\perp}(q X \to q^\prime Y) = \frac{C^2}{2\pi^2}{\sum_{i=+,-,0} f_i \times d\sigma(W_i X \to Y)} \, ,
    \label{eq:EWA}
\end{equation}
where $C$ is a parameter characteristic of the specific fermion considered and is equal to $C=g/2\sqrt{2}$ for quarks emitting $W$s.
In the sub-amplitude, the equivalent $W$ is now treated as an on-shell particle with momentum
\begin{equation}
p_{\sss W} = \left(\sqrt{x^2 E^2 + m_W^2}, \vec{0}, xE\right) \,,
\end{equation}
and the various splitting functions are given by
\begin{equation}
\label{eq:splitting}
\begin{aligned}
f_+ &= \frac{(1-x)^2}{x}\frac{p_\perp^3}{(m_W^2(1-x)+p_\perp^2)^2} \, , \\
f_- &= \frac{1}{x}\frac{p_\perp^3}{(m_W^2(1-x)+p_\perp^2)^2} \, , \\
f_0 &= \frac{(1-x)^2}{x}\frac{2m_W^2p_\perp}{(m_W^2(1-x)+p_\perp^2)^2} \, .
\end{aligned}
\end{equation}
We can now apply this formalism to study the production of a top quark in association to a Higgs boson (see Fig.~\ref{fig:topology}), which embeds the
$b W^+ \to t h$ scattering amplitude. A characteristic parton level process that gives rise to the signal is $u b \to d t h$,
and since the Higgs cannot be radiated from the light fermion leg, we do not have to worry about extra diagrams that might pollute the
EWA description. This would be the case for instance in $tZj$, where the $Z$ boson can be radiated from the massless quarks. In
this study we also do not account for the PDFs, which would yield a further convolution and is not relevant for the assessment
we are interested in.
We define $\hat{s}$ the invariant mass of the top-Higgs pair, which is a proxy of the energy scale of the $2 \to 2$ sub-amplitude.
The differential cross section in the EWA can be written as
\begin{equation}
d\sigma_{EWA} = \frac{g^2}{16\pi^2}\sum_{i=+,-,0} f_i \frac{|\mathcal{M}^i_{2 \to 2}|^2}{16 p_{\sss W} \cdot p_b}  \frac{\sqrt{E^2_t - m_t^2} \sin\theta d\theta  \, d\phi \, dx \, dp_\perp}{(2\pi)^2 E_h} \, ,
\end{equation}
where $\theta$ and $\varphi$ are the polar and azimuthal angles of the top quark.
\begin{figure}[t]
\centering
\includegraphics[width=0.32\linewidth]{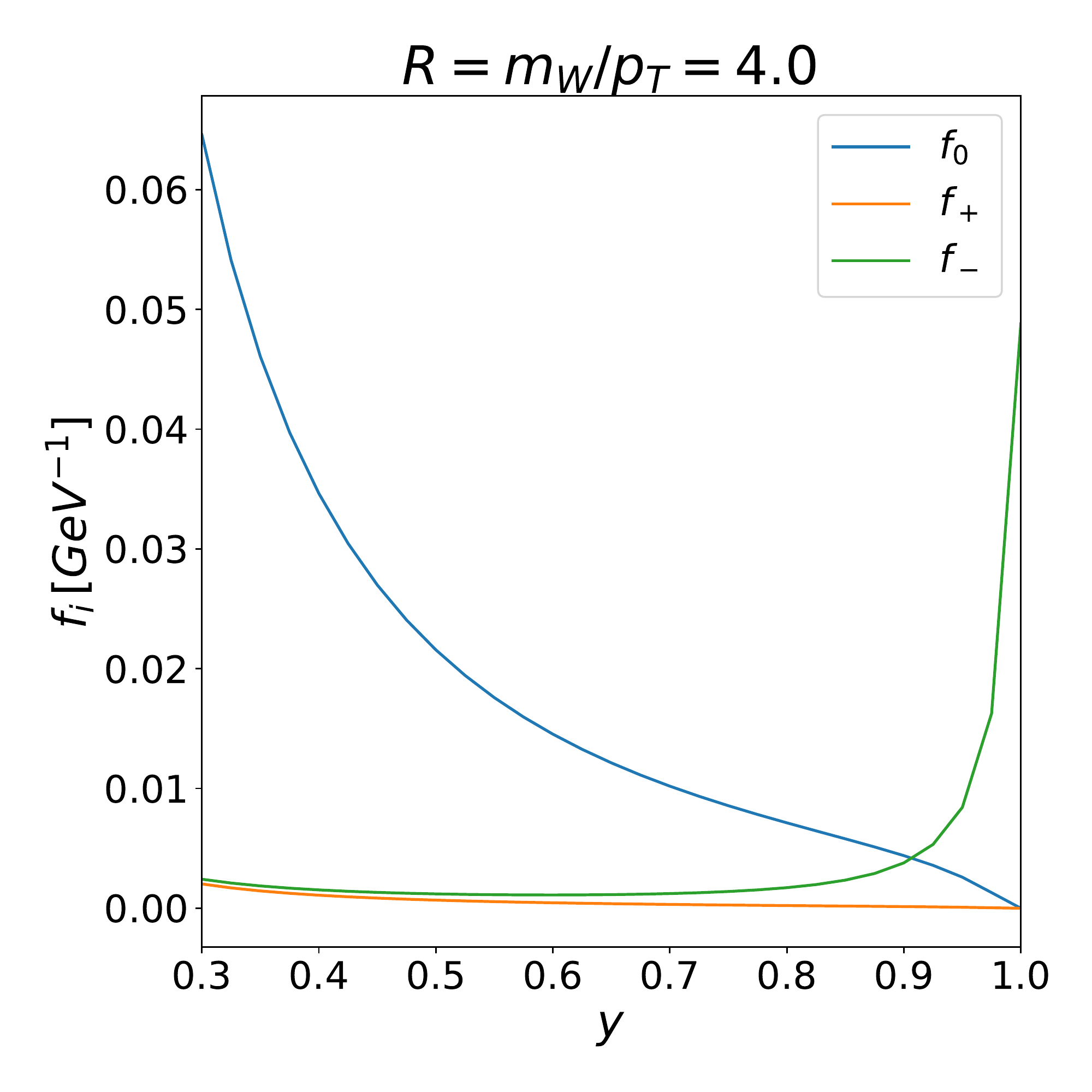}
\includegraphics[width=0.32\linewidth]{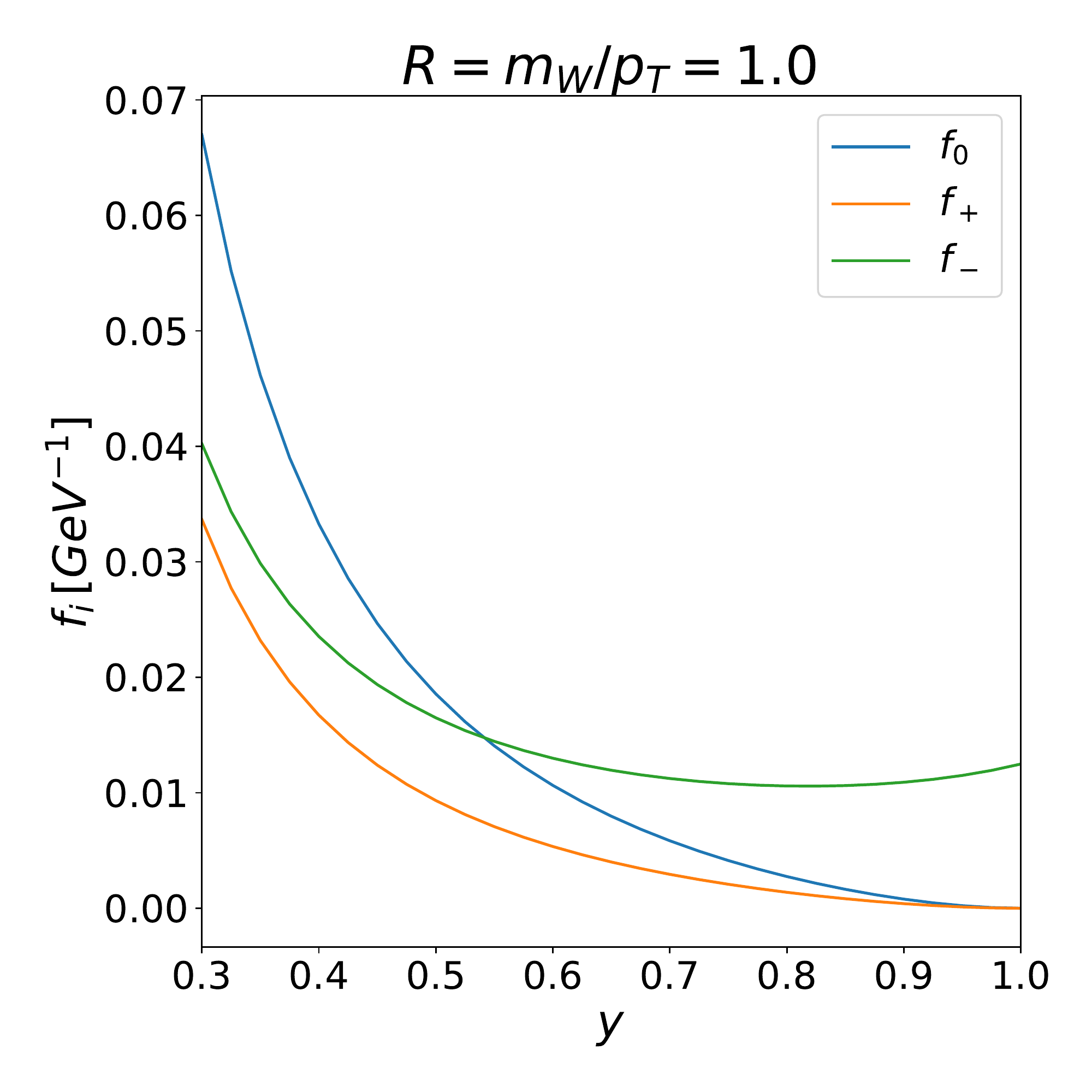}
\includegraphics[width=0.32\linewidth]{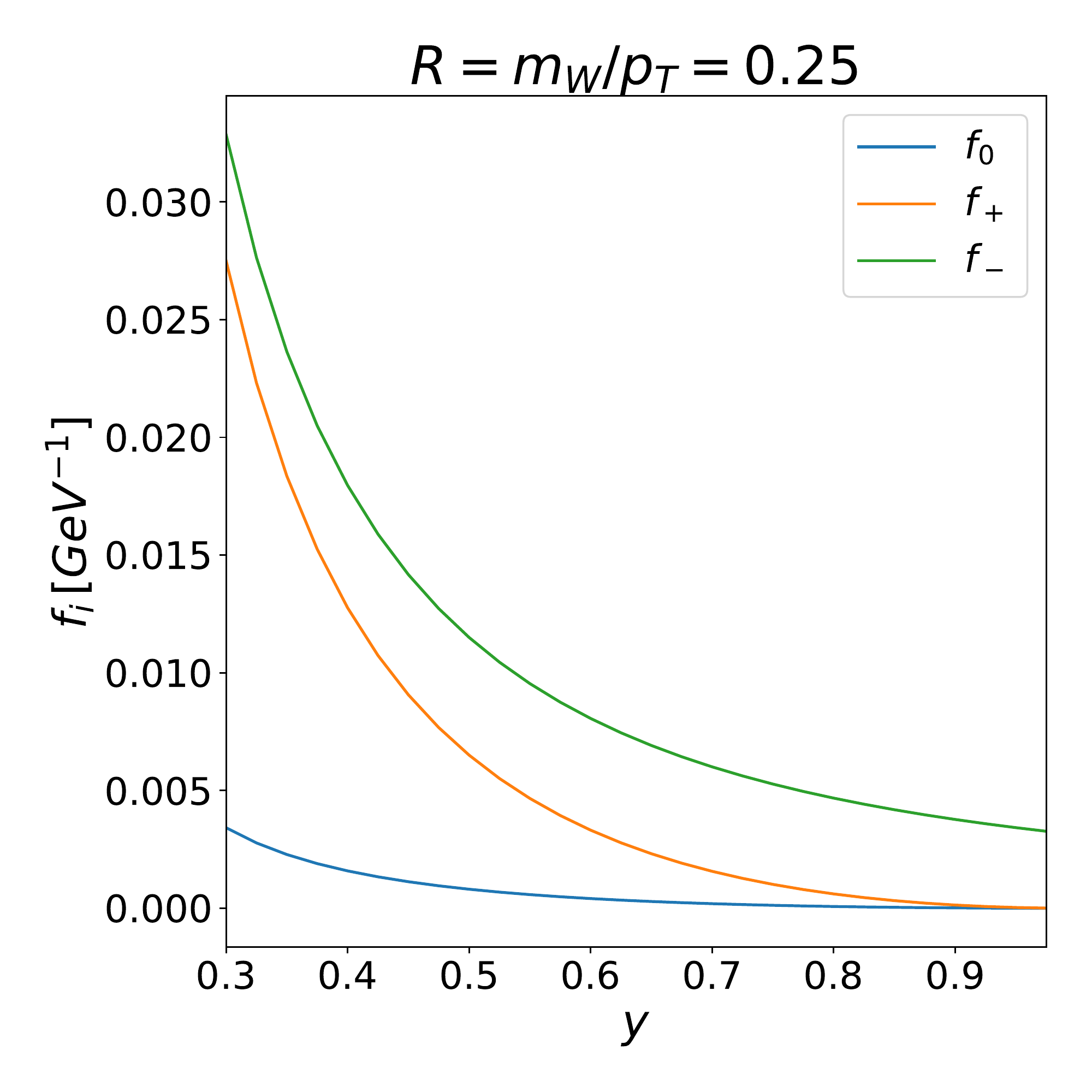}
\caption{Plots of the splitting functions for three different values of $R=\mw/p_\perp$ as a function of $y=\sqrt{\hat{s}}/2E$. \label{fig:splittingFunc}}
\end{figure}
Since the $2 \to 2$ matrix element only depends on $\theta$, we can integrate $\varphi$ and changing variable from $x$ to $\hat{s}$,
we find
\begin{equation}
\frac{d\sigma_{EWA}}{d\theta \, dp_\perp \, d\hat{s}} = \frac{g^2}{512\pi^3}\sum_{i=+,-,0} f_i(\hat{s}, p_\perp) \frac{1}{p_{\sss W} \cdot p_b} |\mathcal{M}^i_{2 \to 2}|^2  \frac{\sqrt{E^2_t - m_t^2} \sin\theta}{E_h} J(\hat{s}) \, ,
\label{eq:ewacross}
\end{equation}
with $E_h$ and $E_t$ the energies of the Higgs and top quark respectively and the Jacobian of the transformation is
\begin{align}
   J(\hat{s})\equiv \frac{1}{(2E)^2}+\frac{\mw^2}{(\hat{s}-\mw^2)^2} \, .
\end{align}
\begin{figure}[t]
\centering
\includegraphics[width=0.45\linewidth]{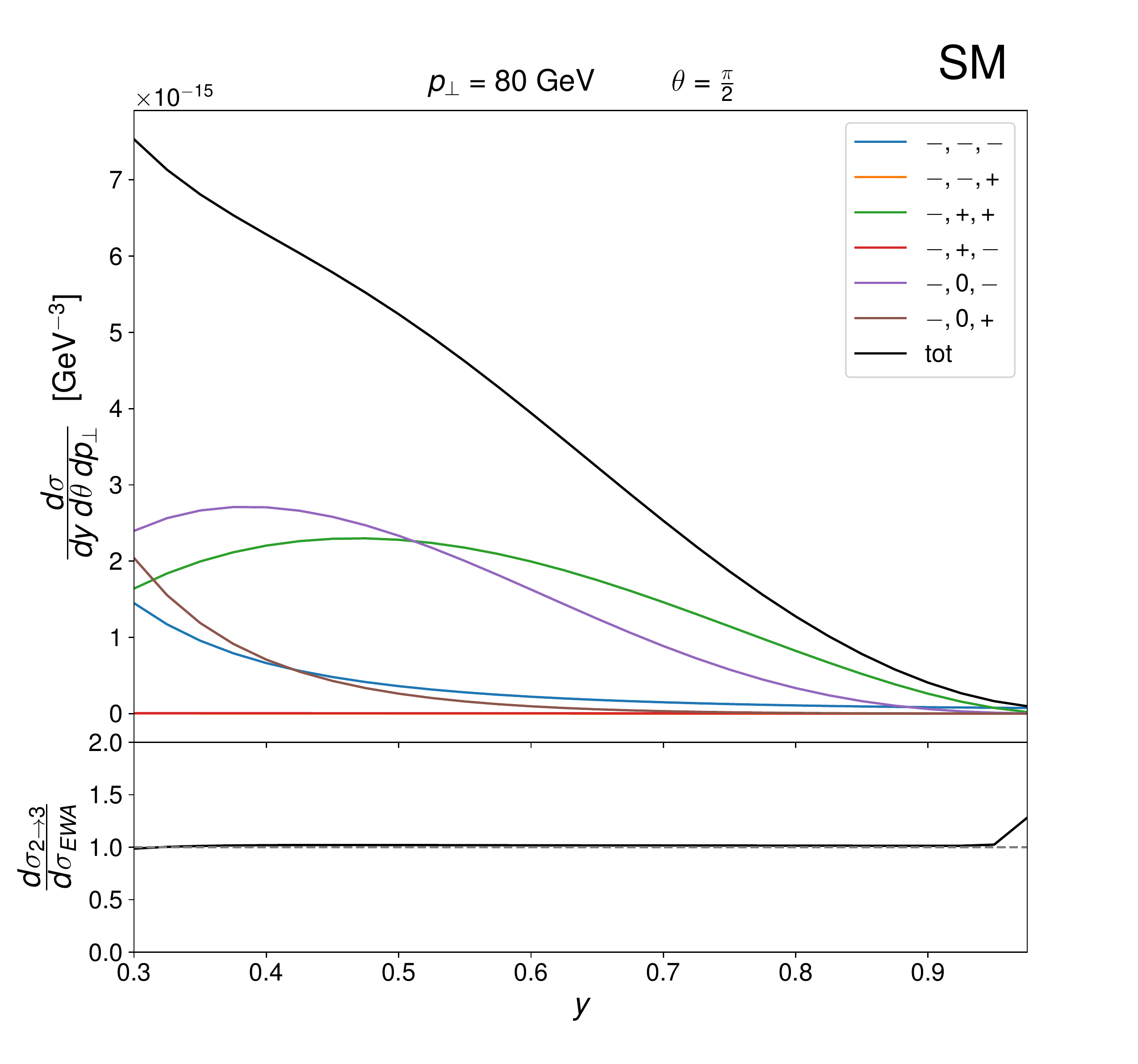}
\caption{The EWA results for the SM contribution in various helicity configurations $(\lambda_{\sss b},\lambda_{\sss W},\lambda_{\sss t})$. We plot the differential cross section as a function of the hard-scattering energy fraction $y=\sqrt{\hat{s}}/2E$ fixing $p_\perp=80~\gev$. In the lower panel, the ratio of the full differential cross section and the EWA expression is displayed, showing the goodness of the approximation. \label{fig:ewa_SM}}
\end{figure}
Instead of integrating in $p_\perp$ and $\theta$, we study the behaviour of the differential cross section by fixing a phase space point,
making sure we are both in the high energy regime $s \sim -t \gg v^2$ and in the validity regime of EWA. In particular, we fix
\begin{align}
E = 2 \text{ TeV} \, ,\quad \theta = \frac{\pi}{2} \, .
\end{align}
In Fig.~\ref{fig:splittingFunc} we show the splitting functions in Eq.~\eqref{eq:splitting} for three different values of $R=m_W/p_\perp$ as a function
of $y=\sqrt{\hat{s}}/2E$. As we can see, as we vary the hard-scattering energy, the probability of emitting a $W$ changes considerably and is
very different for the three polarisations. In particular, the parameter that controls the relative importance of one over the other is
$R$. If we take the ratio of the expressions in Eq.~\eqref{eq:splitting}, we find indeed
\begin{align}
    \frac{f_0}{f_+}=2R^2\,,\quad \frac{f_0}{f_-}=2R^2(1-x)^2.
\end{align}
As it is evident, the smaller $R$ is, the less important the longitudinal polarisation becomes. For $R > 1$ instead, especially at low energies, the longitudinal polarisation is the dominant degree of freedom. At high energies it is always the `$-$' polarisation that
dominates instead, no matter the value of $R$. Unfortunately, when $p_\perp$ becomes too small, Ref.~\cite{Borel:2012by} identifies 
another source of discrepancy between the EWA and the full calculation. Because of this, one cannot hope to gain access to 
the longitudinal degree of freedom by lowering arbitrarily $p_\perp$. The reason why the negative polarisation dominates at high energies
has to be traced back to helicity conservation, which is manifested in the splitting functions by an additional suppression
$(1-x)^2$ for the positive and longitudinal polarisation.

Let us now consider the generic $2 \to 2$ matrix element in the SMEFT.
\begin{figure}[t]
\centering
\includegraphics[width=0.45\linewidth]{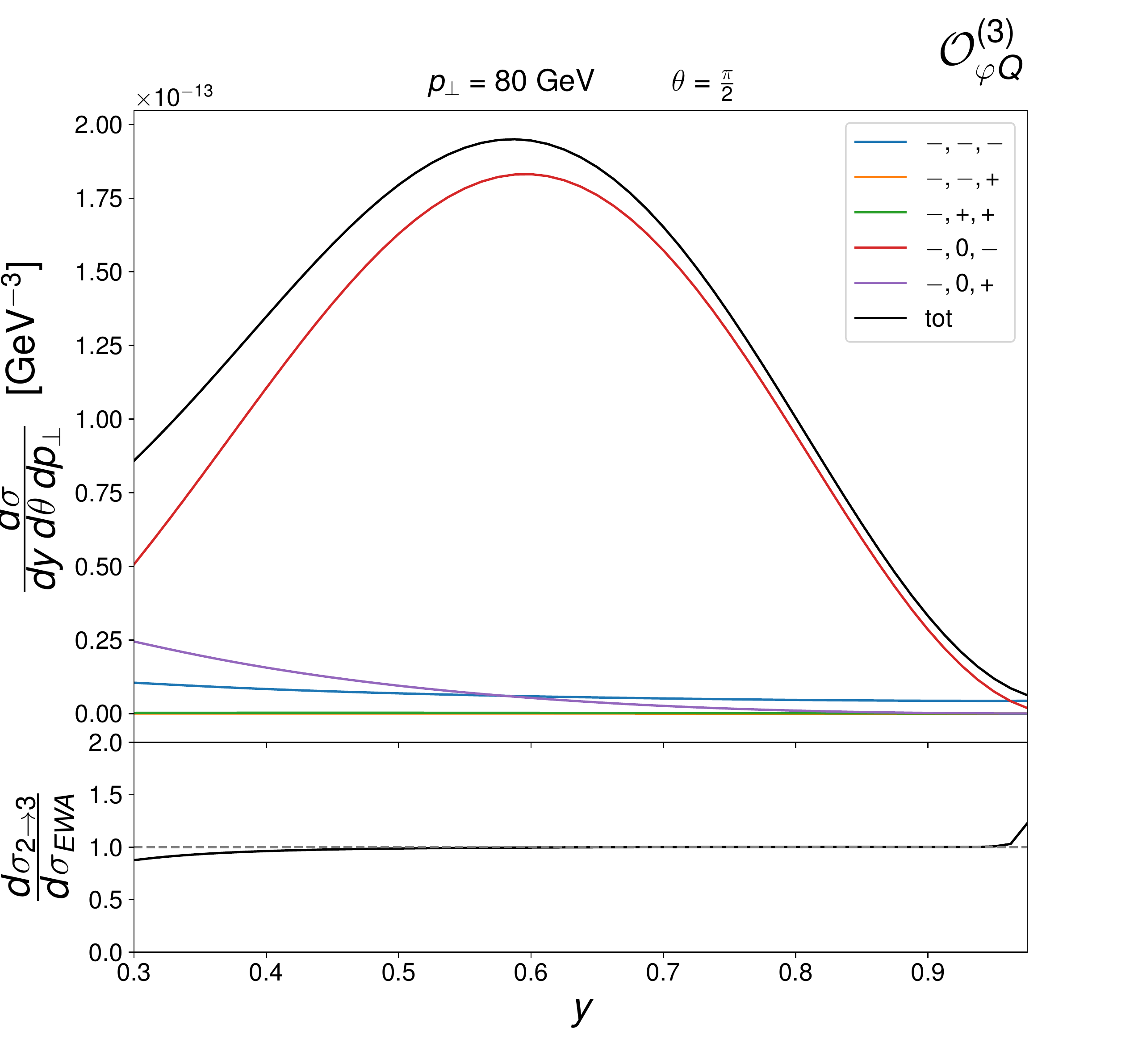}
\includegraphics[width=0.45\linewidth]{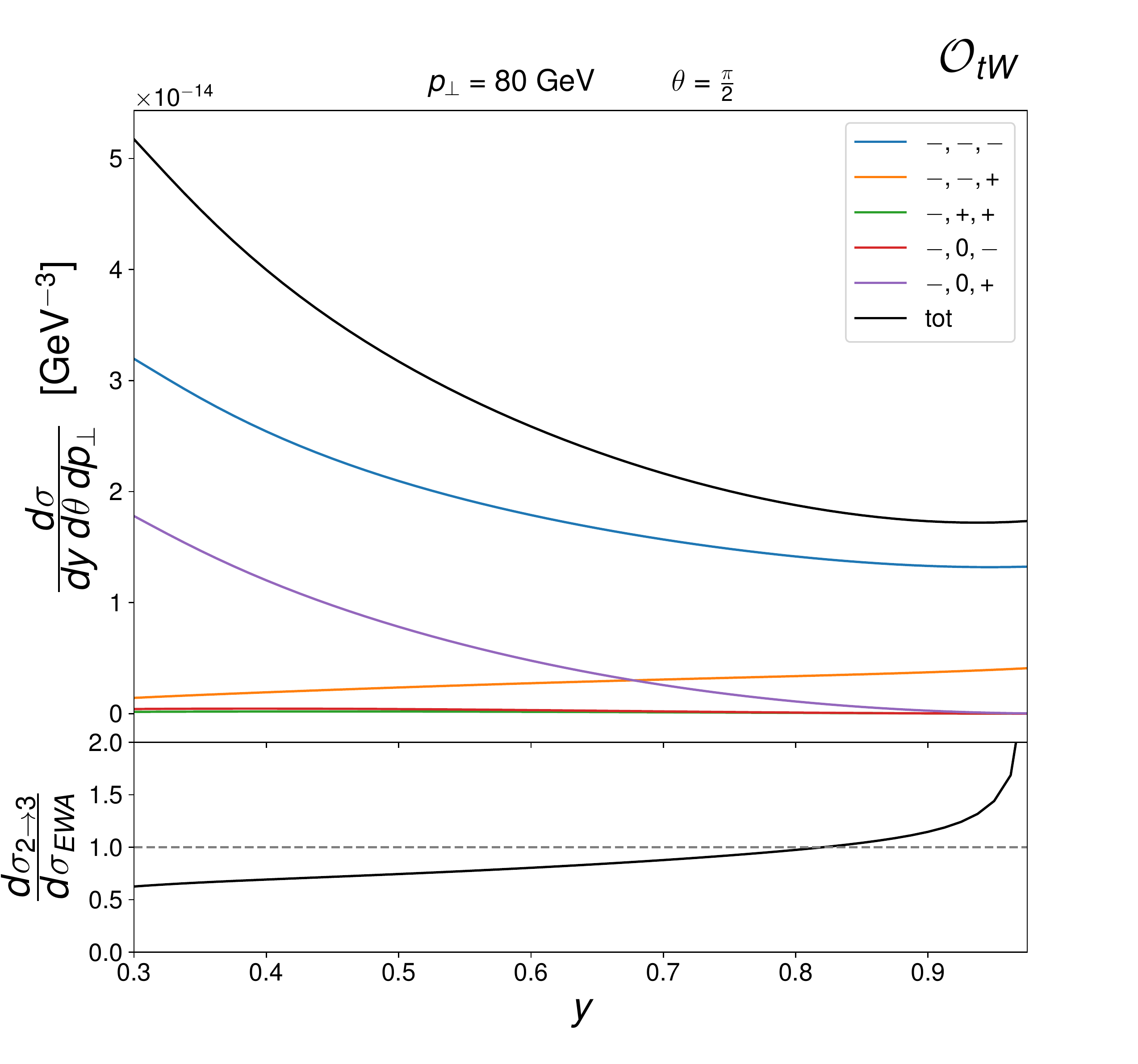} \\
\includegraphics[width=0.45\linewidth]{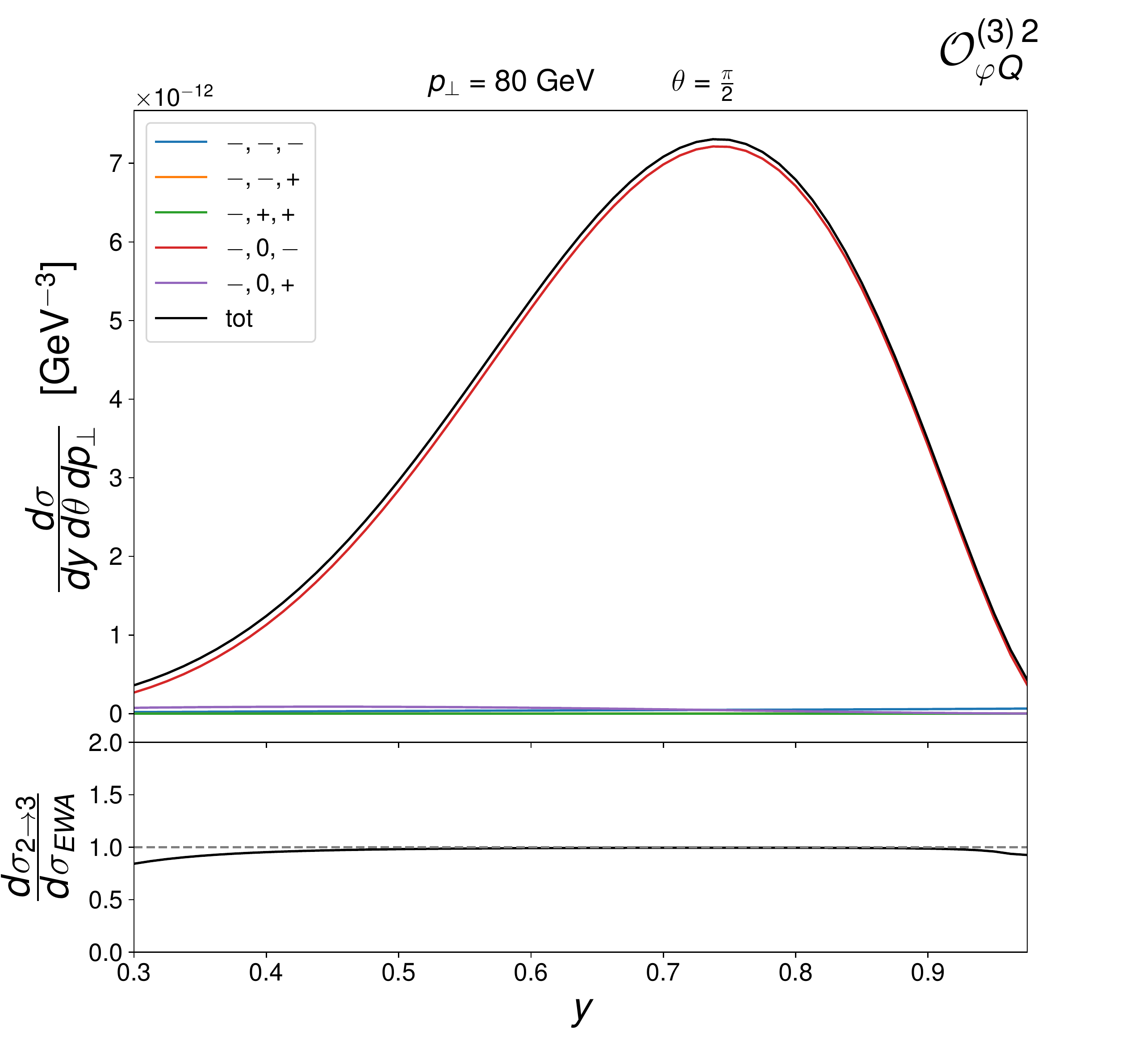}
\includegraphics[width=0.45\linewidth]{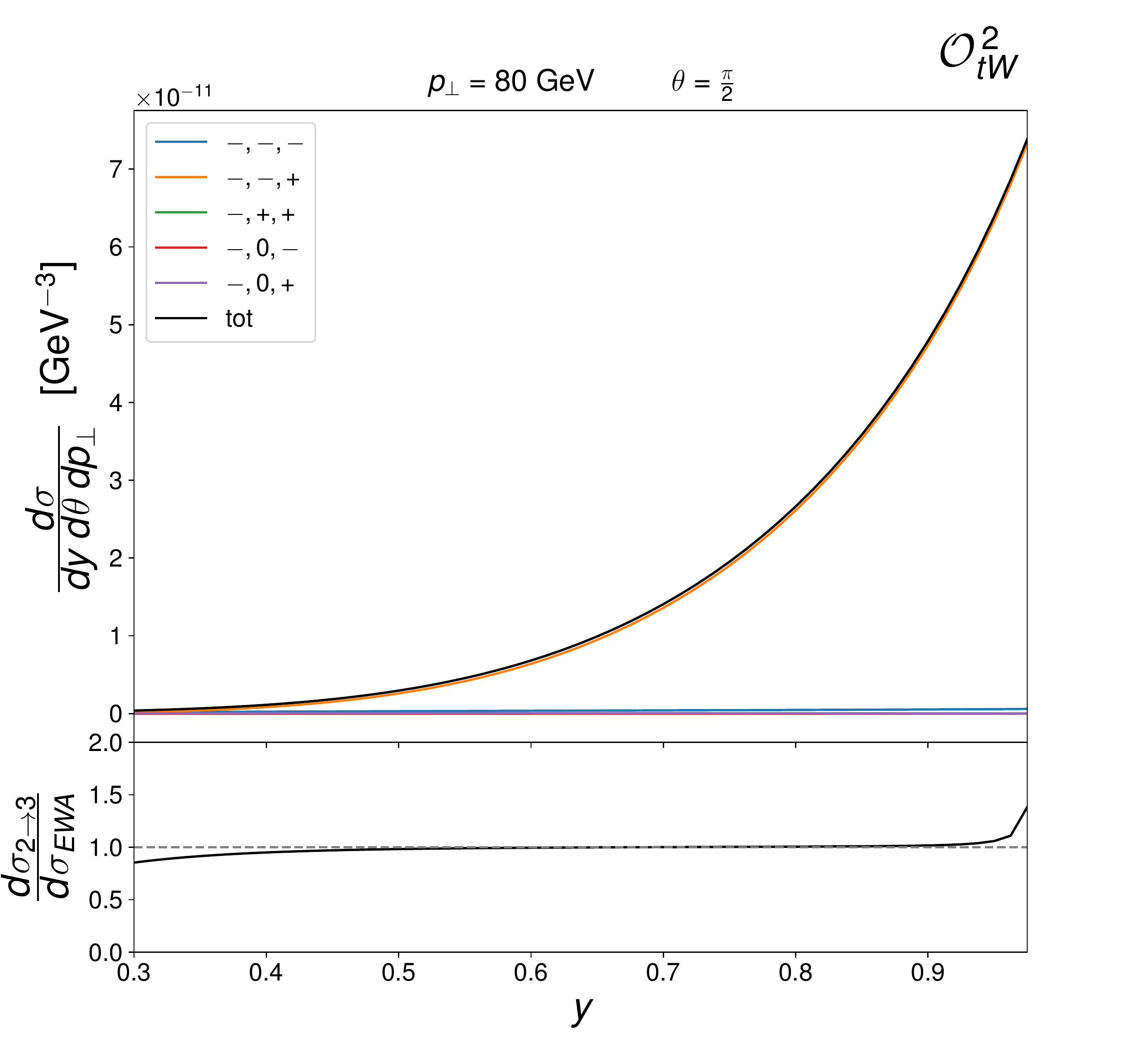}
\caption{Plots of the EWA differential cross section as a function of the hard-scattering energy fraction $y=\sqrt{\hat{s}}/2E$ of the interference and square contribution in various helicity configurations $(\lambda_{\sss b},\lambda_{\sss W},\lambda_{\sss t})$. \label{fig:ewa}}
\end{figure}
The $2 \to 2$ matrix element squared is
\begin{equation}
|\mathcal{M}_{2 \to 2}|^2 = |\mathcal{M}^{SM}_{2 \to 2}+\mathcal{M}^{NP}_{2 \to 2}|^2 = |\mathcal{M}^{SM}_{2 \to 2}|^2+\mathcal{M}^{int}_{2 \to 2}+|\mathcal{M}^{NP}_{2 \to 2}|^2 \, ,
\end{equation}
where $\mathcal{M}^{NP}$ is the contribution coming from dimension-6 operators and whose energy dependence is listed in the tables
in Appendix~\ref{app:helamp_tables}. We can now insert the matrix element in Eq.~\ref{eq:ewacross} and inspect the energy dependence of 
both interference and square contribution for each helicity configuration. We in particular show the results for the operators
$\Op{tW}$ and $\Op{\phi Q}^{\sss (3)}$ fixing $p_\perp = 80~\gev$. The plots for these cases are shown in Fig.~\ref{fig:ewa_SM} for the SM
and in Fig.~\ref{fig:ewa} for the two dimension-6 operators chosen.

As expected, in Fig.~\ref{fig:ewa_SM} we see that the dominant helicity configurations are the ones which are found to be constant in
energy in the helicity table, \ie $(-,0,-)$ and $(-,+,+)$. The case of the SMEFT is quite interesting. One would naively expect that the operator 
$\Op{\phi Q}^{\sss (3)}$ could yield a energy growing contribution in the helicity configuration $(-,0,-)$, since the $\mathcal{M}^{NP}$
has a maximal degree of growth and the SM counterpart is constant in energy. However, the splitting function for a longitudinally polarised
$W$ is taming this effect. While growing considerably in the low-energy regime, it is thereafter suppressed by the factor 
$(1-x)^2$ of the $f_0$ splitting function. On the other hand, in the case of $\Op{t W}$, no energy growing interference is
expected at the interference level, since all the energy growing helicity amplitudes are counterbalanced by a correspondingly
suppressed SM amplitude. 

If we now look at the square contributions, we see a confirmation of this picture, with the current operator not able to access
high energies due to the splitting function suppression. On the other hand, the dipole does not suffer from that since the dominant helicity
configuration is the negative transverse one, which is not suppressed by the $(1-x)^2$ factor, and is actually decisively growing.

In the insets of Fig.~\ref{fig:ewa_SM} and Fig.~\ref{fig:ewa} we can see a validation of the approximation. We plotted the ratio
of the EWA differential cross section and the full $2 \to 3$ process, showing a very good agreement in the intermediate $y$
range, which is where the approximation is expected to hold. This is however not the case for the $\Op{t W}$ interference.
This can be traced back to the fact that when deriving the EWA, one integrates over the azimuthal angle of the jet. This 
procedure however can lead to problematic outcomes when the matrix element is not positive definite, as in the case of the interference.
Cancellations might happen leading to anomalously small terms comparable to the EWA error, \ie proportional to $p_\perp/E$ and
$m/E$. We indeed verified that this cancellation happens for the $\Op{t W}$, while it does not for $\Op{\phi Q}^{\sss (3)}$. 
We conclude that the procedure of splitting the contributions a priori might be flawed and careful checks should be undertaken. 
The complete SMEFT squared amplitude however is always found in good agreement with the full $2\to3$ process.

This study allowed us to have an insight on how things can translate when going to physical processes at collider. Despite a naive
expectation that the longitudinal modes could dominate at high energy, we find that the effects of the splitting functions can be
quite prominent, leading to transverse modes which dominate. Considering this, operators that directly affect transverse helicity amplitudes 
such as the dipole operators are expected to give sizable contributions.

\section{Blueprint for the analysis}
\label{sec:blueprint}

We now move to discuss how we can exploit the $2 \to 2$ unitarity violating behaviours in real collider processes. In particular,
we want to understand whether looking at high energy observables in top quark physics can enhance our sensitivity to NP and
how much the energy growths are retained when going to physical processes.

If we look at Table.~\ref{tab:topOps}, all the operators on the left-side do not explicitly contain a top quark. As a consequence, these operators
are affecting non-top processes as well and strong constraints can actually be derived from electroweak precision observables,
diboson and Higgs measurements. For this reason, in the following, we will restrain ourselves to the study of the set of operators
\begin{align}
\Op{t\phi}, \Op{tB}, \Op{tW}, \Op{\phi Q}^{\sss(1)}, \Op{\phi Q}^{\sss(3)}, \Op{\phi t} \text{ and } \Op{\phi tb}.
\end{align}
In Ref.~\cite{Maltoni:2019aot}, we considered both present and future colliders, discussing the High Luminosity phase of the LHC,
as well as a possible 27 TeV proton collider and a high energy lepton collider operating at 380 GeV, 1.5 TeV and 3 TeV. 
The computations are performed with {\sc MadGraph5\_aMC@NLO}~\cite{Alwall:2011uj,Alwall:2014hca} and using the SMEFT@NLO UFO model~\cite{SMEFTatNLO}.
For each of the considered processes, we define a naive measure for the sensitivity by
\begin{equation}
R(c_i)\equiv \frac{\sigma}{\sigma_{ SM}}= 1+ c_i\frac{\sigma_{ Int}^i}{\sigma_{ SM}} + c_{i}^2 \frac{\sigma_{ Sq}^{i,i}}{\sigma_{ SM}} = 1 + c_i \, r_i + c_i^2 \, r_{i,i} \, ,
\end{equation}
where for each operator $\Op{i}$, we compute the quantities $r_i$ (interference) and $r_{i,i}$ (square) by setting the corresponding
Wilson coefficient to $1$ TeV$^{-2}$ and taking the ratio to the SM cross section. In particular, being interested in the high
energy behaviour, we compute these numbers for both inclusive cross sections and in a restricted high energy region of the 
phase space. We do so by defining a process specific cut on the kinematic variables such that the $2 \to 2$ embedded amplitude 
is in the high energy regime $s \sim -t \gg v^2$. This cut is most of the times defined by simply asking for $p_T > 500$ GeV
for the two final states particles that are produced by the hard scattering sub-amplitude. The relative impact of the inclusive and
exclusive phase space regions are denoted with $r^{tot}$ and $r^{HE}$ respectively. The observation of a higher impact in 
$r^{HE}$ might be indication that the unitarity violating behaviour caused by the operator insertion is effectively accessed.

While this is most of the times clear when dealing with squared contributions, the interference is more subtle and conclusions might
need further inspections. This is due to the fact that the interference is not positive definite and changes of sign in phase space
can lead to cancellations when integrating, resulting in anomalously small contributions which obfuscate the high energy enhancement.
We identify these situations in the survey of the processes but dedicated phase space cuts to remove the cancellations are not explored,
as this is beyond the scope of this analysis.

Whenever a process has a charged particle in the final state, we always include both particle and anti-particle in the numerical
results. Additionally, in case the process has a QCD counterpart that does not probe the $2 \to 2$ EW sub-amplitude we quote the 
corresponding rate.

\revised{When a process has a light jet in the final state, we define them by excluding b-jets. Even if we are technically employing the
5-flavour scheme, the b-quarks can be experimentally distinguished and their inclusion introduces additional dependence on different scattering amplitudes not present in other channels. In addition to this, processes with light jets can be obtained as QCD radiation from some of the other Born level processes considered. These are therefore not infrared finite,
being only a subset of the full NLO computation. When in this situation, we define, in a process dependent manner, a suitable jet $p_T$ cut so that the infrared
pole is avoided but the ratio between EW and QCD components has flattened.}

In the following we will discuss in more details a specific sub-amplitude which we found to be particularly interesting in terms of
sensitivity to NP, presenting ways to embed it in physical processes. The full discussion of the several processes considered (see Tables~\ref{tab:singletop}, \ref{tab:toppairnoh} and \ref{tab:toppairwithh})
can be found in Ref.~\cite{Maltoni:2019aot}. Some of them have been recently measured at the LHC~\cite{Aaboud:2017ylb,Sirunyan:2017nbr,Sirunyan:2018zgs,Sirunyan:2018bsr,Sirunyan:2018lzm} and studied with more in-depth analyses~\cite{Etesami:2017ufk,Bylund:2016phk,Abramowicz:2018rjq,Maltoni:2016yxb,Liu:2015aka}.

\begin{table}[t!]
{\footnotesize
\setlength{\tabcolsep}{2pt}
\begin{center}
\begin{tabular}{|P{2cm}|P{1cm}|P{1cm}|P{1cm}|P{1cm}|P{1cm}|P{1cm}|P{1cm}|}
     \hline
                              & $tWj$  & $tZj$ & $t\gamma j$ & $tWZ$ & $tW\gamma$ & $thj$ & $thW$
     \tabularnewline\hline             
                                       
     $b\,W\to t\,Z$           & \cmark & \cmark &             & \cmark &          &        &        
     \tabularnewline\hline                                                      
                                                                                  
     $b\,W\to t\,\gamma$      & \cmark &        & \cmark      &        &  \cmark  &        &         
     \tabularnewline\hline                                                      
                                                                                 
     $b\,W\to t\,h$           &        &        &             &        &           & \cmark & \cmark
     \tabularnewline\hline
\end{tabular}
\end{center}
}
\caption{Single top processes studied and relevant sub-amplitude accessed. \label{tab:singletop}}
\end{table}

\begin{table}[t!]
{\footnotesize
\setlength{\tabcolsep}{2pt}
\centering
\begin{tabular}{|p{2cm}|P{1.1cm}|P{1.1cm}|P{1.1cm}|P{1.1cm}|P{1.1cm}|P{1.1cm}|P{1.1cm}|P{1.1cm}|}
     \hline
     
                              &$t\bar{t}W(j)$ & $t\bar{t}WW$ & $t\bar{t}Z(j)$ & $t\bar{t}\gamma (j)$ & $t\bar{t}\gamma\gamma$ & $t\bar{t}\gamma Z$ & $t\bar{t}ZZ$ & $VBF$ 
     \tabularnewline\hline             
                                       
     $t\,W\to t\,W$           & \cmark        & \cmark       &                &                      &                        &                    &              & \cmark
     \tabularnewline\hline

     $t\,Z\to t\,Z$           &               &              & \cmark         &                      &                        &                    & \cmark       & \cmark 
     \tabularnewline\hline

     $t\,Z\to t\,\gamma$      &               &              & \cmark         & \cmark               &                        & \cmark             &              & \cmark
     \tabularnewline\hline
 
     $t\,\gamma\to t\,\gamma$ &               &              &                & \cmark               & \cmark                 &                    &              & \cmark
     \tabularnewline\hline
\end{tabular}
}
\caption{Top pair processes without Higgs studied and relevant sub-amplitude accessed. \label{tab:toppairnoh}}
\end{table}

\begin{table}[t!]
{\footnotesize
\setlength{\tabcolsep}{2pt}
\begin{center}
\begin{tabular}{|p{2cm}|P{1.1cm}|P{1.1cm}|P{1.1cm}|P{1.1cm}|}
     \hline
     
                              &$t\bar{t}h(j)$ & $t\bar{t}Zh$ & $t\bar{t}\gamma h$ &  $t\bar{t}hh$
     \tabularnewline\hline             
                                       
     $t\,Z\to t\,h$           & \cmark        & \cmark       &                     &       
     \tabularnewline\hline                                                                

     $t\,\gamma\to t\,h$      & \cmark        &              &  \cmark             &       
     \tabularnewline\hline                                                                                       
                                                                                 
     $t\,h\to t\,h$           &               &              &                     & \cmark       
     \tabularnewline\hline                                                                
                                                                                                                                    
\end{tabular}
\end{center}
}
\caption{Top pair in association with a Higgs processes studied and relevant sub-amplitude accessed. \label{tab:toppairwithh}}
\end{table}

\section{An interesting example: \texorpdfstring{$b \, W^+ \to t Z$}{b W -> t Z} scattering}

An interesting class of processes is the one in which a single-top quark is produced in association with EW bosons. A summary
of the single-top processes relevant for our study is reported in the Table~\ref{tab:singletop},
where we can observe that the separation between the processes and the $2 \to 2$ EW sub-amplitudes probed is almost complete.
In all these processes a $W$ boson is present and therefore $b$ quarks are always left-handed. As a consequence, right-handed
helicity configurations are suppressed and do not lead to interferences with operators that act on that specific helicity amplitude,
\ie $\Op{\varphi t b}$. In particular, we can observe in the following table a summary of the maximum energy growths for the three
single-top sub-amplitudes.
\begin{table}[h!]
{\footnotesize
\setlength{\tabcolsep}{2pt}
\renewcommand{\arraystretch}{1.2}
\begin{center}
  \begin{tabular}{|c|c|c|c|c|c|c|c|c|c|c|c|c|c|}
 \hline
                          & $\Op{\phi D}$ & $\Op{\phi d}$ &  $\Op{\phi B}$ & $\Op{\phi W}$  & $\Op{\phi WB}$  & $\Op{W}$ & $\Op{t \phi}$ & $\Op{tB}$ & $\Op{tW}$ & $\Op{\phi Q}^{\sss (1)}$ & $\Op{\phi Q}^{\sss (3)}$ &  $\Op{\phi t}$ & $\Op{\phi tb}$ 
 \tabularnewline\hline
 
 $b\,W\to t\,Z$           & $E$           & $-$              & $-$            & $-$            & $E$             & $E^2$    & $-$           & $E^2$     & $E^2$     & $E$                      & \red{$E^2$}              & $E$            & $E^2$           
 \tabularnewline\hline

 $b\,W\to t\,\gamma$      & $-$           & $-$              & $-$            & $-$            & $E$             & $E^2$    & $-$           & $E^2$     & $E^2$     & $-$                      & $-$                      & $-$            & $-$           
 \tabularnewline\hline
 
 $b\,W\to t\,h$           & $-$           & $-$              & $-$            & $E$            & $-$             & $-$      & $E$           & $-$       & $E^2$     & $-$                      & \red{$E^2$}              & $-$            & $E^2$            
 \tabularnewline\hline
 
  \end{tabular}
\end{center}
\renewcommand{\arraystretch}{1.}
}
\end{table}
\\
A red entry means that the corresponding interference seems to lead to an energy growing behaviour, being the SM counterpart not suppressed.
Only the operator $\Op{\phi Q}^{(3)}$ leads to such effects in the sub-amplitudes $b\,W\to t\,Z$ and $b\,W\to t\,h$. In particular,
$b\,W\to t\,Z$ seems to be particularly promising since it can benefit from two longitudinal degrees of freedom, \ie $W_L$ and $Z_L$.
\begin{figure}[t!]
    \centering
    \includegraphics[width=0.30\linewidth]{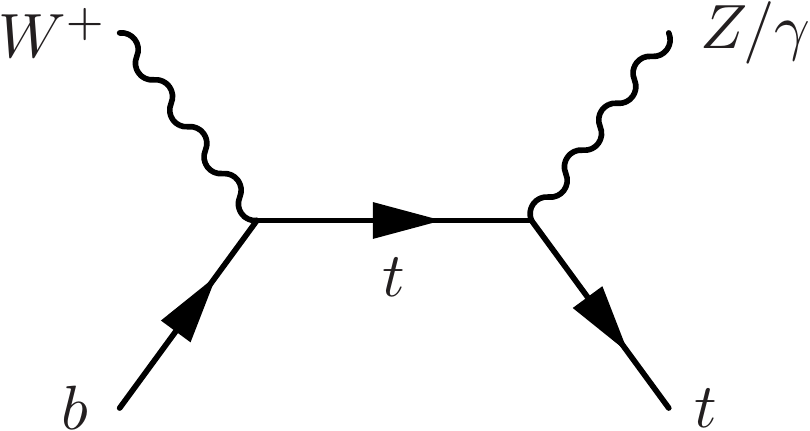} \quad
    \includegraphics[width=0.30\linewidth]{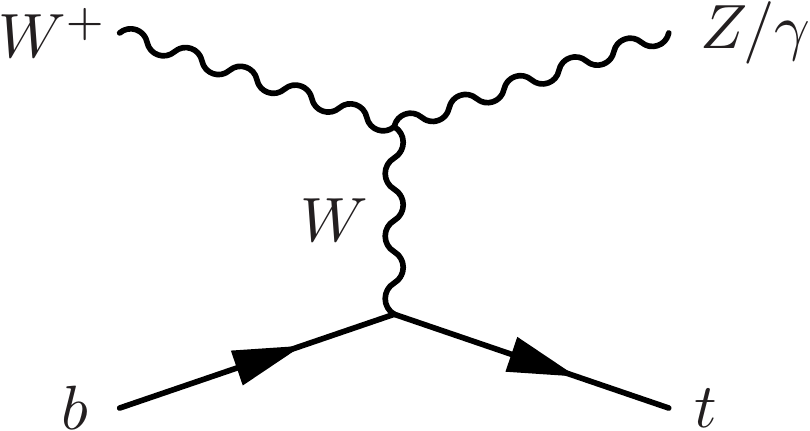} \quad
    \includegraphics[width=0.30\linewidth]{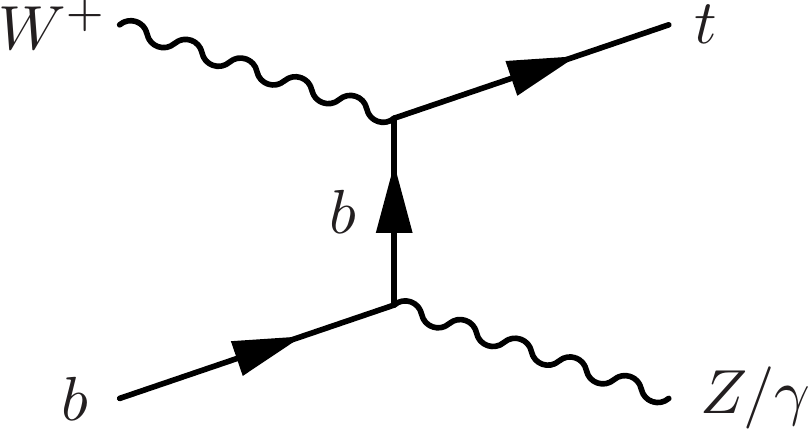}

    \caption{Samples of diagrams for $b W \to t Z$ sub-amplitude.\label{fig:bwtz_sm}}
\end{figure}
Diagrams for the aforementioned process can be seen in Fig.~\ref{fig:bwtz_sm} and one can explicitly notice that it probes the coupling of
the top and bottom with the EW bosons, as well as the triple gauge coupling $WWZ$. Noticeably, the $btW$ vertex is present in every diagram
and therefore a shift in the value of the coupling cannot lead to unitarity violating behaviours. Its effect can only be a 
global rescale of the full amplitude. If we consider the AC framework, we find that for the $(-,0,-,0)$ helicity configuration
any shift in the other vertices lead to a $E^2$ growth
\begin{equation}
        \label{eq:bwtz_max}
        \sqrt{s(s+t)}\,(\gzbl - \gztl + \gwz) \, 
\end{equation}
at the amplitude level, as expected from the previously reported table. While in the SM these three couplings cancel perfectly, NP effects
can lead to non-unitary behaviours. In particular, the aforementioned current operator $\Op{\phi Q}^{(3)}$ shifts both the 
top and bottom coupling to the $Z$ boson, but it does so in a correlated manner. A less relevant degree of growth can be found by
considering non-fully longitudinal configurations, \ie $(-,-,-,0)$ or $(-,0,-,-)$, where the amplitude leading term is given by
\begin{equation}
        \label{eq:bwtz_max_flip}
        \sqrt{-t}\,(\gzbl - \gztl + \gwz) \,.
\end{equation}
It is worth noting that the same cancellation is present, but the helicity flip has to be paid with a lower energy dependence.
A less trivial cancellation is instead seen for the configuration $(-,0,+,0)$
\begin{equation}
        \label{eq:bwtz_subl}
    \sqrt{-t}\,(2m_W^2(\gzbl - \gztr + \gwz) - \gwz m_Z^2) \,,
\end{equation}
where despite both EW bosons being longitudinal, the presence of a helicity flip for the final state top-quark has a suppressing
effect. 
On the other hand, the fully transverse configurations can only be subject of energy growing behaviours if the operator
introduces a new Lorentz structure. This is the case for instance of the dipole operators. 

We now turn to explore ways of embedding
this amplitude at colliders with the objective of maximising the potential for NP.

\begin{figure}[t!]
  \centering 
  \includegraphics[width=0.35\linewidth]{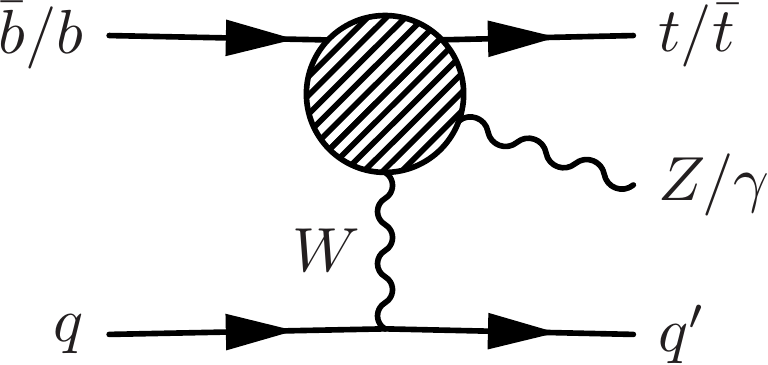}
  \hspace{1cm}
  \includegraphics[width=0.35\linewidth]{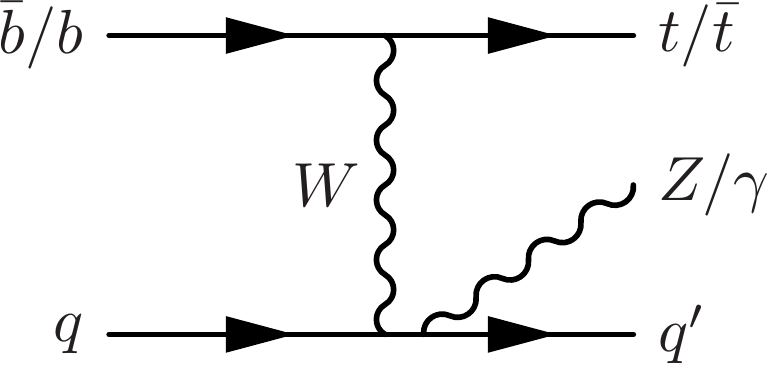}
  \caption{ 
    On the left a schematic picture of the embedding of the $b \, W \to t \, Z$ amplitude in the full $tZj$ production mode, while on the right a sample diagram which is not accessing the aforementioned sub-process.
  \label{fig:diag_tZj}}
\end{figure} 

\subsection{\texorpdfstring{$tZj$}{tZj}}

A simple way to access the $b \, W \to t \, Z$ amplitude is to produce a single top quark in association with a $Z$ boson and a jet.
In a similar fashion to what was described in section~\ref{sec:EWA}, a $W$ boson can be emitted from a light fermion leg and
scatter with an initial state b quark, accessing the desired sub-process. In particular, $tZj$ has been recently observed at the LHC for the first
time~\cite{Aaboud:2017ylb,Sirunyan:2017nbr,Sirunyan:2018zgs}. However, contrary to the example studied in section~\ref{sec:EWA} for
$b \, W \to t \, h$, not every diagram probes the sub-amplitude (see Fig.~\ref{fig:diag_tZj}). The SM cross section for this process
is quite big, $\sim 600$ fb, and prospects of differential measurements can be foreseen. This process has also been extensively studied in 
Ref.~\cite{Degrande:2018fog} and was found to show promising sensitivities to SMEFT interactions. While being affected by four fermion operators 
(that we do not include in our analysis), it has been shown by the authors that these are best constrained by single-top production and therefore can be 
safely neglected in our preliminary analysis.

\begin{figure}[t!]
  \centering \includegraphics[width=\linewidth]{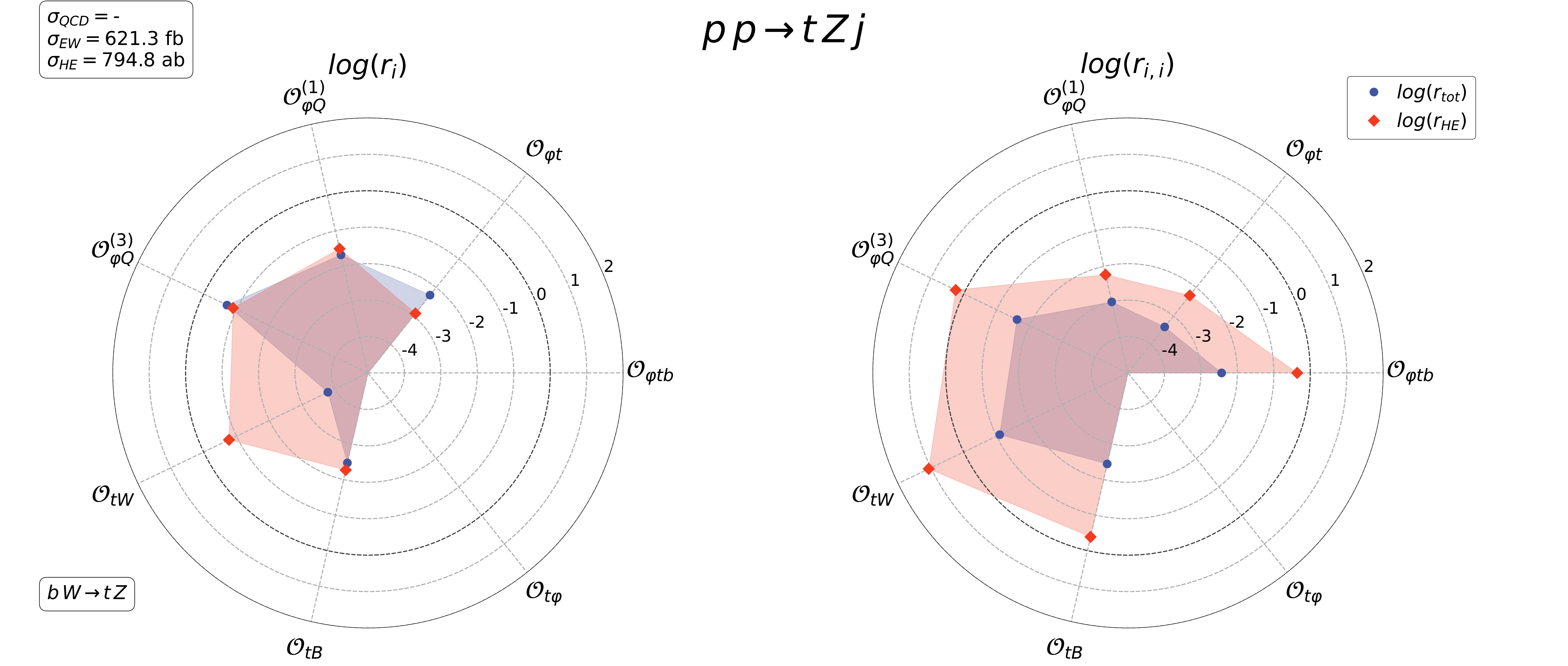}
  \caption{A compact depiction of the sensitivity for NP of the process $p \, p \to t \, Z \, j$  at the 13 TeV LHC.
 \label{fig:radar_tzj_LHC13}}
\end{figure}

The sensitivity results of our study are reported in compact format by making use of radar plots, the one for $tZj$ being shown in Fig.~\ref{fig:radar_tzj_LHC13}.
In the top left corner of the figure we report the inclusive QCD, EW and high energy EW cross sections. In the bottom
left we instead report the $2 \to 2$ sub-amplitude probed by the physical process. The naive sensitivity measures $r_i$ and $r_{i,i}$ 
are plotted in logarithmic scale for each of the operators that affect the process. Concerning the interferences, since they are non-positive definite, we plot the absolute value of the ratio. Blue points correspond to the inclusive values, while red points
are for the high energy phase space region. In particular, for this process the latter is defined by requiring the $p_T$ of both the 
top quark and the $Z$ to be higher than $500$ GeV. The shading is carrying no specific information and just serves as a visual aid.

The process of interest is sensitive to almost all of the operators considered in our study, with the top Yukawa operator being the only exception. Looking at Fig.~\ref{fig:radar_tzj_LHC13}, a particularly surprising outcome is the fact that $\Opp{\phi Q}{(3)}$ is not yielding
the expected energy growing interference. However, as discussed in section~\ref{sec:EWA}, the energy enhancement induced by the operator
affects the fully longitudinal configuration, which is suppressed by the longitudinal splitting function $f_0$ to emit the $W$. After further investigations,
we indeed discovered that the energy growth is present for the longitudinal $Z$ but the full process is dominated by the transverse modes.

On the other hand, the operator $\Op{t W}$ seems to display an enormous energy enhancement. This is however only an artefact of
the non-positivity of the interference term, which at inclusive level exhibits a sensible cancellation.

Overall, the process does not seem to be able to fully exploit the unitarity violating behaviours at linear level, while the
square contributions do so as naively expected. Future dedicated optimisations and the prospect of measuring energy dependent 
differential distributions, can nonetheless be extremely beneficial and help access the high energy behaviour of the EW sub-scattering.  

\subsection{\texorpdfstring{$tWZ$}{tWZ}}

\begin{figure}[t!]
  \centering 
  \includegraphics[width=0.35\linewidth]{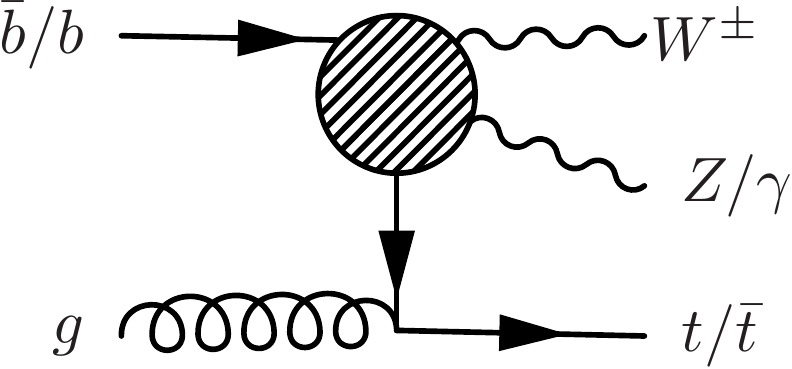}
    \hspace{1cm}
  \includegraphics[width=0.35\linewidth]{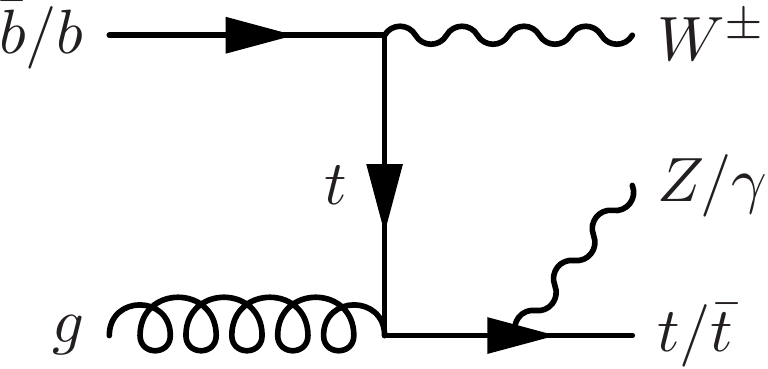}
  \caption{On the left a schematic picture of the embedding of the $b W \to t Z$ amplitude in the full $tZW$ production mode, while on the right a sample diagram which is not accessing the aforementioned sub-process.
\label{fig:diag_tZW_taW}}
\end{figure}

It is clear that if we want to maximally exploit the energy growth of this sub-amplitude and gain sensitivity to NP, we need to be able to access 
the fully longitudinal helicity configuration. However, as we saw in Section.~\ref{sec:EWA}, if the longitudinal EW boson is emitted from a fermion leg, the process suffers a suppression brought in by the splitting function, making it more difficult to
access the unitarity violating enhancement. For this reason, processes such as $tZj$ do not exhibit the desired growth, especially at linear level, as we previously seen. On the other hand, if both EW bosons are in the final state, the problem is not there anymore
and nothing prevents us to access the most promising helicity configuration. This is the case for $tZW$ production and a schematic
sample of the characteristic Feynman diagrams is presented in Fig.~\ref{fig:diag_tZW_taW}. As the process has a rate $6$ times smaller
than $tZj$, it is a bit more complicated to measure and it has not been observed yet, but prospects of measuring it already at the LHC
are not impossible.

An additional experimental caveat, however, is that, even if a pure $tZW$ QCD background is not present, the final state is similar enough to $t\bar{t}Z$ production.
For this reason, it might not be trivial to distinguish the two channels experimentally, since when the top quarks decay, the two processes only differ by a b-jet. The QCD-induced production has a much bigger cross section, but the different kinematics might be enough to
safely distinguish them already at the LHC.

\revised{As it can be seen in Fig.~\ref{fig:diag_tZW_taW}, there are topologies in which the $Z$ boson is not radiated from the EW sub-process, but instead by the top quark.
Since this set of diagrams do not probe the EW $2 \to 2$ amplitude, they are expected to become less relevant when the kinematic cuts
to access the energy enhancement are applied. Another interesting feature of this process is that the polarisation of the $W$ and $Z$
could be measured, as both the EW bosons are in the final state. This can in principle help in singling out the more sensitive configurations and gain discriminating power. In order to do so, the decay products of the EW bosons must be well measured and the polarisation inferred from the angular distributions of the particles detected. The cleanest channel to do so is the fully leptonic one, since background at the LHC in that case is
smaller compared to the hadronic decays and with only one $W$ in the final state the ambiguity given by the neutrino is manageable. However, this also means that the total cross section needs to be multiplied by the branching ratios ($\approx 10\%$ for the $Z$ and $30\%$ for the $W$ boson considering decays into $e$, $\mu$ and $\tau$) lowering even more the rate of expected events.}

\begin{figure}[t!]
  \centering \includegraphics[width=\linewidth]{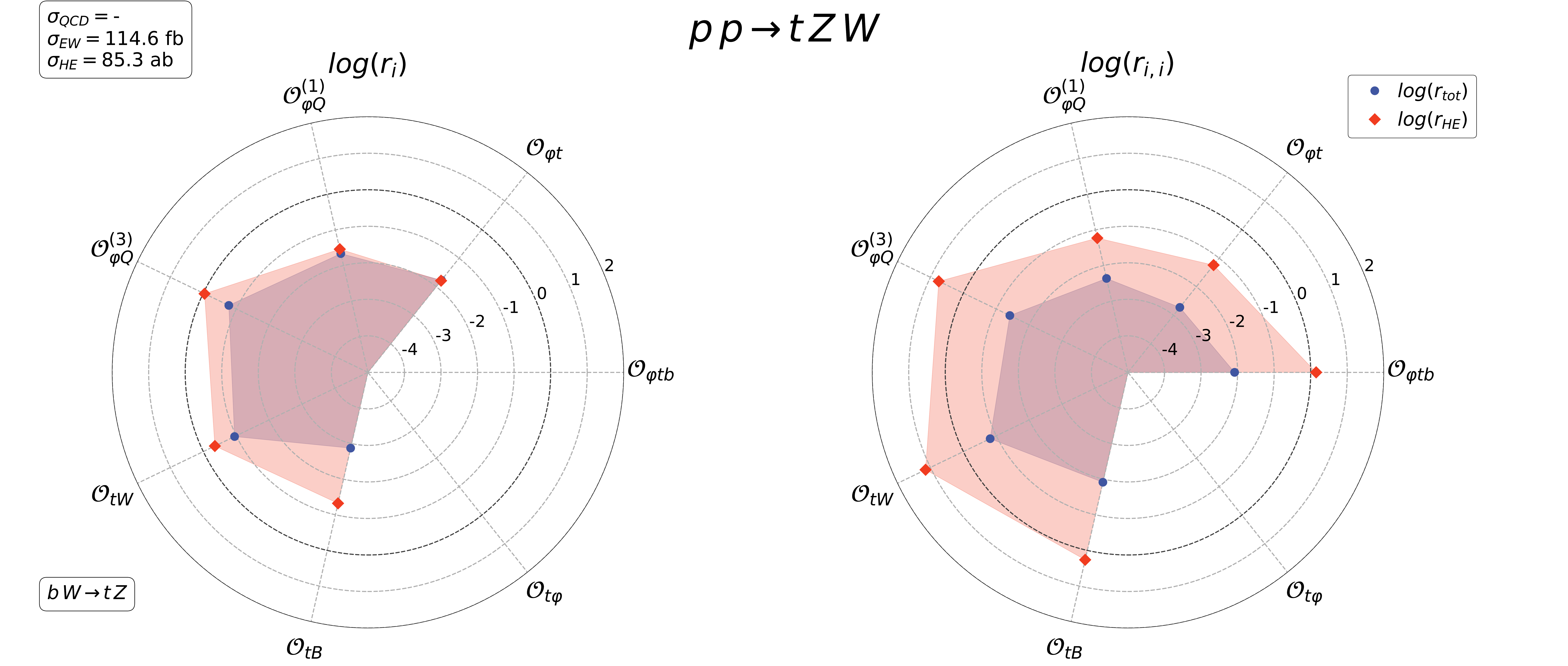}
  \caption{A compact depiction of the sensitivity for NP of the process $p \, p \to t \, Z \, W$  at the 13 TeV LHC.
 \label{fig:radar_tzw_LHC13}}
\end{figure}
In Fig.~\ref{fig:radar_tzw_LHC13} we show the radar chart for $tZW$. The high energy region is defined by requiring both EW bosons to have a $p_T > 500$ GeV. This time, there is a clear improvement when accessing the high energy phase space region, both for the
interference and the squared terms. In particular, the current operator $\Op{\phi Q}^{(3)}$ yields the expected energy growth contrary to $tZj$. 
In addition to this, no clear sign of
cancellations can be seen in the interferences contributions for all the operators except $\Op{t B}$. We see indeed that at the inclusive level the
contribution of the operator is anomalously small, especially compared to the respective squared contribution. As soon as the high energy
cut is implemented, the cancellation seems to disappear and a more natural sensitivity restored.
All the squared sensitivities are extremely enhanced and we find that $tZW$ is therefore an interesting candidate
to gain information on NP. Especially in the context of a global fit, it would be very important to have at our disposal the measurement 
of such process.

\section{Summary}

In this section we briefly discuss a summary of our broad study. The objective of the numerical study was to offer a guide for future directions
in top quark measurements, especially in the context of the SMEFT. No detailed phenomenological analysis was performed, but we have focused
on the assessment of whether the unitarity violating behaviour of the $2 \to 2$ scattering amplitudes is retained when going to colliders.
Often, the presence of a QCD component dilutes the potential to exploit the sensitivity of the EW production, since it does not
embed any of these sub-amplitudes. The QCD component is often roughly two orders of magnitude bigger and this suggests that it might
require some time before we will be able to overcome the sensitivity provided by the overall rate of it. On the other hand, once
the EW component becomes systematics dominated, the large statistics can be used to define optimal differential observables
and measure quantities extremely sensitive to NP effects.

Not every operator however leads to good sensitivities in the high energy phase space regions. Some operators do not seem to lead to explicit energy growths, but rather to overall rescaling of the processes.
The dipole operators on the other hand, by means of a new Lorentz structure, seem to always lead to maximal energy growths, both at the interference and
squared level. 

In particular, while the expected energy growth can be seen almost always at the square level, the same cannot be said for the 
interference. This was partially expected, since the SM contributes with suppressing factors, but is made worse by potential cancellations
that need to be removed with dedicated cuts on a process-dependent basis. In this perspective, we report in Appendix~\ref{app:topbarplots} a summary of the numerical study for each individual operator, including both LHC predictions and a future lepton collider ones.
In particular, in these figures we added additional informations with respect to the radar charts before. Firstly, a second
high energy region is included by increasing the $p_T$ cut from $500$ GeV to $1$ TeV. Secondly, the sign of the interference
is reported, allowing us to better understand the presence of cancellations. In many cases one can indeed see that the apparent
lack of sensitivity after the first cut is completely restored once a even higher energy region is accessed. This is the clear sign that a cancellation is going on and that the energy growth is actually present.

Overall, the numerical study shows a consistency with the $2 \to 2$ analytic analysis and all the expected sources of energy growth are 
observed, especially in the quadratic contributions. When operators do not affect the longitudinal polarisations of the EW
bosons, we observe that, at the sub-amplitude level, no energy growth is expected. However, the dipole operators seem to contradict this
statement, since we observe a relative enhancement with respect to the SM when accessing the high energy phase space regions.
This happens because it is the relative energy behaviour that is truly relevant for the study we have performed. Even if the
interference itself does not grow, the SM component might go down in energy and the relative impact consequently increase.
Additionally, our analytic study with the EWA seems to indicate that emissions of transversely polarised states might be favoured in high energy phase space regions.  

In summary, our study suggests that a combination of $tZj$ and $tZW$ is optimal to access the $b \, W \to t \, Z$ sub-amplitude. Analogously,
$b \, W \to t \, h$ is best probed by $thj$ and $thW$ production. While single top processes seem to be very promising, $t\bar{t}$
production modes look more challenging. In particular all of the $t\bar{t}X$ processes are unable to access the high energy 
violating behaviour because suppressed by the s-channel propagator. The addition of a hard jet in the final state
leads to an increased sensitivity and the class $t\bar{t}X j$ shows an interesting potential. In particular, $t\bar{t} W j$ benefits
from a peculiar enhancement with respect to $t\bar{t}W$ that is not observed in other processes of the same class (see Ref.~\cite{Dror:2015nkp} for details). We investigated as well processes of the kind $t\bar{t} XY$ and VBF, which while showing a promising
sensitivity, we find them to be rate limited. Future machines however might be able to access them and unleash their potential. Finally, 
the $2 \to 2$ scatterings $t \, Z \to t \, h$ and $t \, h \to t \, h$ are the most difficult to access and only at future colliders
we may be able to explore them.

In conclusion, each of the interesting identified processes will need to be further studied and a dedicated phenomenological 
analysis is needed to assess the true sensitivity to NP. Our work offers a broad perspective that can help identify promising
experimental directions to pursue in order to uncover the BSM landscape.

%
%
%

\chapter{Combined EFT interpretation of Higgs, EW and top data}
\pagestyle{fancy}
\label{chap:globalfit}

\hfill
\begin{minipage}{10cm}

{\small\it 
``For however many things have a plurality of parts and are not merely a complete aggregate but instead some kind of a whole beyond its parts.''}

\hfill {\small Aristotle, \textit{Metaphysics}}
\end{minipage}

\vspace{0.5cm}

In the previous chapter we have discussed how NP could be uncovered by taking advantage of the enhanced sensitivity that characterises the top quark sector. However, an attractive feature of the SMEFT is the fact that it predicts correlated deviations from the SM, a property that can be 
exploited to find glimpses of modified interactions in the current data (correlating for example LHCb flavour anomalies~\cite{Pich:2019pzg} and high-$p_T$ tails of LHC distributions~\cite{Greljo:2017vvb}). In order to make the most out of a model-independent framework,
one needs therefore to perform global interpretations taking into account all the measurements at our disposal, with the objective of
both constraining the Wilson coefficients and possibly find evidence of BSM effects in the data.

In particular, the lack of clear and distinctive evidence of NP so far at the LHC suggests that NP could be heavier than the energy
reach of the machine and an indirect approach is therefore justified. This is further encouraged by the upcoming High Luminosity upgrade of the machine, which will
allow us to increase the precision of the measurements and gain discriminating power to discover new particles indirectly.
Unfortunately, the model-independence implies that
the parameter space of the SMEFT can be quite big, unless specific UV assumptions are made. As previously discussed, according to the 
flavour assumptions, we can have between a few tens of operators to several hundreds. 

The quest towards a full-fledged global
interpretation of all the measurements from current and past colliders is of extreme importance for the field and has recently
attracted the attention of many collaborations. To-date, most studies have been focusing on specific sectors, assuming that the SM is a valid theory outside of it. The targets of these studies have been the potentially most sensitive ones, such as the top sector~\cite{Buckley:2016cfg,Buckley:2015lku,Hartland:2019bjb,Brivio:2019ius}, the Higgs and EW sector~\cite{Biekotter:2018rhp,Ellis:2018gqa,Almeida:2018cld}, diboson and vector boson scattering~\cite{Baglio:2020ibv,Gomez-Ambrosio:2018pnl}, jet-production~\cite{Krauss:2016ely,Alte:2017pme}, flavour physics~\cite{Aebischer:2018iyb}, 
neutrinos~\cite{Falkowski:2019xoe} and low-energy observables~\cite{Falkowski:2017pss}. Additionally, in Ref.~\cite{Ellis:2020unq}
a combination of Top, Higgs and diboson data has been used to perform a first combined interpretation at linear level in the EFT expansion.
Recently EFT interpretations have been also undertaken by the experimental collaborations ATLAS and CMS~\cite{ATLAS-CONF-2020-053,CMS-PAS-TOP-19-001}.

While these studies are of indisputable value, the ideal objective is to combine many of these sectors and exploit the model-independence
feature of the framework. As a first step towards this goal, the \smefit collaboration presented an interpretation of many top quark
observables~\cite{Hartland:2019bjb}. The fitting methodology was inspired by the successful techniques used by the NNPDF
collaboration~\cite{Ball:2008by,Ball:2010de,Ball:2014uwa} and allowed to constrain $34$ independent Wilson coefficients affecting
top quark production processes, leading to improvements with respect to previous bounds~\cite{AguilarSaavedra:2018nen} on several operators.

In this chapter, we expand the previous analysis in many ways. First of all, the dataset is substantially extended, combining top, Higgs and 
EW observables from LHC and past colliders. In particular, both signal strengths and differential distributions are considered,
making use of the most recent top-quark and Higgs measurements from ATLAS and CMS. LEP diboson data are also included, as well as electroweak
precision observables (EWPO)~\cite{ALEPH:2005ab} which are taken into account by means of restrictions in the parameter space. In total,
$50$ dimension-6 operators are considered.

In terms of fitting methodology, a novel and independent approach which relies on Nested Sampling (NS) and MultiNest~\cite{Feroz:2013hea} has been implemented. Contrary to the Monte Carlo replica method used in Ref.~\cite{Hartland:2019bjb}, the NS method is a 
Bayesian method which given data and theory predictions, reconstructs a posterior probability distribution for the Wilson coefficients. The
two methods have been cross validated, leading to equivalent results and strengthening the conclusions of our study.

\section{SMEFT description of the Higgs, top and EW sectors}

In this section we discuss the conventions and the definitions used, focusing in particular on the degrees of freedom chosen
for the analysis. Instead of using the Warsaw basis, we fit the degrees of freedom recommended by the LHC Top Working Group~\cite{AguilarSaavedra:2018nen}, 
which are defined to be closely related to experimental results. These are linear combinations of the Warsaw basis Wilson coefficients
which being aligned with specific physical directions in the parameter space, have a more natural interpretation. As a matter of fact, for
certain processes, they are well aligned with the principal components.

\subsection{Flavour assumptions and degrees of freedom}

\subsubsection{Flavour assumptions}

As we already discussed, the number of degrees of freedom at dimension-6 varies substantially according to the flavour assumptions,
from $59$ in the case of flavour universality to $2499$ in the most general case. In this analysis we implement the MFV hypothesis~\cite{DAmbrosio:2002vsn} but we single out the top quark operators, in the same spirit as in Chapter~\ref{chap:top}. The flavour symmetry employed
in the quark sector is therefore $U(2)_q\times U(2)_u \times U(3)_d$ and the only Yukawa coupling different from zero is the
top quark one. These assumptions are also compatible with the UFO model SMEFT@NLO~\cite{Degrande:2020evl}, which is used to
produce automated one-loop predictions for most of the observables, as well as LO order ones for the rest.

Moreover, in order to take into account the decay rates measurements of the Higgs boson, we relax these conditions to include 
the bottom and charm Yukawa operators. These would not be allowed by the flavour symmetry chosen, but the sensitivity of the LHC to 
the coupling of the Higgs to the aforementioned quarks motivates their inclusion. All the other light quark Yukawa are instead set to zero.

With respect to the leptonic sector, we adopt the flavour symmetry $U(1)_l^3 \times
U(1)_e^3$, \ie flavour universality. This is only relaxed to include the $\tau$ Yukawa operator,
for which sensitivity is expected. In practice, the leptonic flavour assumptions do not impact the fit because of the constraints
from $Z$-pole measurements at LEP and SLC.

\subsubsection{Bosonic operators}

\begin{table}[t] 
{\footnotesize
  \begin{center}
    \renewcommand{\arraystretch}{1.90}
        \begin{tabular}{lll}
          \toprule
          Operator $\qquad$ & Coefficient $\qquad\qquad\qquad$ & Definition \\
        \midrule
        $\Op{\varphi G}$ & $c_{\varphi G}$  & $\left(\pdp\right)G^{\mu\nu}_{\sss A}\,
        G_{\mu\nu}^{\sss A}$  \\ \hline
        $\Op{\varphi B}$ & $c_{\varphi B}$ & $\left(\pdp\right)B^{\mu\nu}\,B_{\mu\nu}$\\ \hline
        $\Op{\varphi W}$ &$c_{\varphi W}$ & $\left(\pdp\right)W^{\mu\nu}_{\sss I}\,
        W_{\mu\nu}^{\sss I}$ \\ \hline
        $\Op{\varphi W B}$ &$c_{\varphi W B}$ & $(\varphi^\dagger \tau_{\sss I}\varphi)\,B^{\mu\nu}W_{\mu\nu}^{\sss I}\,$ \\ \hline
        $\Op{\varphi d}$ & $c_{\varphi d}$ & $\partial_\mu(\pdp)\partial^\mu(\pdp)$ \\ \hline
        $\Op{\varphi D}$ & $c_{\varphi D}$ & $(\varphi^\dagger D^\mu\varphi)^\dagger(\varphi^\dagger D_\mu\varphi)$ \\ \hline
         $\mathcal{O}_{W}$&   $c_{WWW}$ & $\epsilon_{IJK}W_{\mu\nu}^I W^{J,\nu\rho} W^{K,\mu}_\rho$ \\
       \bottomrule
        \end{tabular}
        \caption{List of purely bosonic operators at dimension-$6$ affecting Higgs and diboson physics. We indicate
        both the notation for the operators and the Wilson coefficients.
	  %
           \label{tab:oper_bos}}
\end{center}
}
\end{table}


Purely bosonic operators are responsible for modification of the interactions between EW bosons, affecting in particular Higgs
production and decay, as well as diboson production. The list of operators that affect the processes considered in our analysis is given 
in Table~\ref{tab:oper_bos}. The operators $\Op{\varphi WB}$ and $\Op{\varphi D}$ are usually identified with the $S$ and $T$
oblique parameters. This identification is however not strictly true in the Warsaw basis, since it is a basis-dependent correspondence. As
such and together with several two-fermion operators in Table~\ref{tab:oper_ferm_bos} they are severely constrained by $Z$ and $W$ precision measurements
at LEP and SLC. Among these degrees of freedom, two linear combinations are left unconstrained by EWPO and are responsible for
deviations in the triple gauge couplings (TGC) and Higgs EW interactions. These will therefore be constrained by the LEP and LHC
diboson and Higgs measurements. The operator $\Op{W}$ is also responsible for TGC modifications, but affecting only transverse modes
cannot be constrained by Higgs processes and diboson processes are crucial to gain sensitivity on it.

The remaining degrees of freedom affect only Higgs physics. For instance, the operators $\Op{\varphi B}$ and $\Op{\varphi W}$
can be probed by means of Higgs decays ($h\to ZZ^*$ and $h \to W^+W^-$), VBF production or Higgs associated production with a EW boson.
The operator $\Op{\varphi G}$ introduces a direct coupling between the Higgs boson and gluons, modifying therefore Higgs production
via gluon fusion (ggF) and in association with a top pair. Ultimately, the operator $\Op{\varphi d}$ generates a global rescaling of 
the Higgs interactions by means of a different wave-function normalisation.

\subsubsection{Two-fermion operators}

\begin{table}[t!]
{\scriptsize
  \begin{center}
    \renewcommand{\arraystretch}{1.0}
    \begin{tabular}{lll}
      \toprule
      Operator $\qquad$ & Coefficient & Definition \\
                \midrule \midrule
		&3rd generation quarks&\\
                \midrule \midrule
    $\Op{\varphi Q}^{(1)}$ & --~~$c_{\varphi Q}^{(1)}$ & $i\big(\varphi^\dagger\lra{D}_\mu\,\varphi\big)
 \big(\bar{Q}\,\gamma^\mu\,Q\big)$ \\\hline
    $\Op{\varphi Q}^{(3)}$ & $c_{\varphi Q}^{(3)}$  & $i\big(\varphi^\dagger\lra{D}_\mu\,\tau_{\sss I}\varphi\big)
 \big(\bar{Q}\,\gamma^\mu\,\tau^{\sss I}Q\big)$ \\ \hline
    $\Op{\varphi t}$ & $c_{\varphi t}$& $i\big(\varphi^\dagger\,\lra{D}_\mu\,\,\varphi\big)
 \big(\bar{t}\,\gamma^\mu\,t\big)$ \\ \hline
      $\Op{tW}$ & $c_{tW}$ & $i\big(\bar{Q}\tau^{\mu\nu}\,\tau_{\sss I}\,t\big)\,
 \tilde{\varphi}\,W^I_{\mu\nu}
 + \text{h.c.}$ \\  \hline
 $\Op{tB}$ & --~~$c_{tB}$ &
 $i\big(\bar{Q}\tau^{\mu\nu}\,t\big)
 \,\tilde{\varphi}\,B_{\mu\nu}
 + \text{h.c.}$ \\\hline
    $\Op{t G}$ & $c_{tG}$ & $i\,\big(\bar{Q}\tau^{\mu\nu}\,T_{\sss A}\,t\big)\,
 \tilde{\varphi}\,G^A_{\mu\nu}
 + \text{h.c.}$ \\  \hline
    $\Op{t \varphi}$ & $c_{t\varphi}$ & $\left(\pdp\right)
 \bar{Q}\,t\,\tilde{\varphi} + \text{h.c.}$  \\\hline
    $\Op{b \varphi}$ & $c_{b\varphi}$ & $\left(\pdp\right)
 \bar{Q}\,b\,\varphi + \text{h.c.}$ \\
                \midrule \midrule
		&1st, 2nd generation quarks&\\
                \midrule \midrule
    $\Op{\varphi q}^{(1)}$ & --~~$c_{\varphi q}^{(1)}$ & $\sum\limits_{\sss i=1,2} i\big(\varphi^\dagger\lra{D}_\mu\,\varphi\big)
 \big(\bar{q}_i\,\gamma^\mu\,q_i\big)$ \\\hline
    $\Op{\varphi q}^{(3)}$ & $c_{\varphi q}^{(3)}$ & $\sum\limits_{\sss i=1,2} i\big(\varphi^\dagger\lra{D}_\mu\,\tau_{\sss I}\varphi\big)
 \big(\bar{q}_i\,\gamma^\mu\,\tau^{\sss I}q_i\big)$ \\  \hline
    $\Op{\varphi u}$ &$c_{\varphi u}$ & $\sum\limits_{\sss i=1,2} i\big(\varphi^\dagger\,\lra{D}_\mu\,\,\varphi\big)
 \big(\bar{u}_i\,\gamma^\mu\,u_i\big)$ \\ \hline
    $\Op{\varphi d}$ & $c_{\varphi d}$ & $\sum\limits_{\sss i=1,2,3} i\big(\varphi^\dagger\,\lra{D}_\mu\,\,\varphi\big)
 \big(\bar{d}_i\,\gamma^\mu\,d_i\big)$\\ \hline
    $\Op{c \varphi}$ & $c_{c \varphi}$ & $\left(\pdp\right)
 \bar{q}_2\,c\,\tilde\varphi + \text{h.c.}$ \\
                \midrule \midrule
		&two-leptons&\\
                \midrule \midrule
    $\Op{\varphi l_i}^{(1)}$ & $c_{\varphi l_i}^{(1)}$ & $ i\big(\varphi^\dagger\lra{D}_\mu\,\varphi\big)
   \big(\bar{l}_i\,\gamma^\mu\,l_i\big)$ \\\hline 
    $\Op{\varphi l_i}^{(3)}$ & $c_{\varphi l_i}^{(3)}$ & $ i\big(\varphi^\dagger\lra{D}_\mu\,\tau_{\sss I}\varphi\big)
 \big(\bar{l}_i\,\gamma^\mu\,\tau^{\sss I}l_i\big)$ \\  \hline
    $\Op{\varphi e}$ & $c_{\varphi e}$ & $ i\big(\varphi^\dagger\lra{D}_\mu\,\varphi\big)
 \big(\bar{e}\,\gamma^\mu\,e\big)$ \\\hline
    $\Op{\varphi \mu}$ & $c_{\varphi \mu}$ & $ i\big(\varphi^\dagger\lra{D}_\mu\,\varphi\big)
 \big(\bar{\mu}\,\gamma^\mu\,\mu\big)$ \\  \hline
    $\Op{\varphi \tau}$ & $c_{\varphi \tau}$ & $i\big(\varphi^\dagger\lra{D}_\mu\,\varphi\big)
 \big(\bar{\tau}\,\gamma^\mu\,\tau\big)$ \\  \hline
    $\Op{\tau \varphi}$ & $c_{\tau \varphi}$ & $\left(\pdp\right)
 \bar{l_3}\,\tau\,{\varphi} + \text{h.c.}$ \\
                \midrule \midrule
		&four-lepton &\\
                \midrule \midrule
 $\Op{ll}$ & $c_{ll}$ & $\left(\bar l_1\gamma_\mu l_2\right) \left(\bar l_2\gamma^\mu l_1\right)$ \\
 \hline
  \bottomrule
\end{tabular}
\caption{Same as Table~\ref{tab:oper_bos}
  for two fermion operators.
  The coefficients indicated with ``--'' in the second column are operators of the Warsaw basis which are not directly used in the fit. Instead they are replaced by  $c_{\varphi q_i}^{(-)}$, $c_{\varphi Q_i}^{(-)}$, and
  $c_{tZ}$ defined in Table~\ref{tab:oper_ferm_bos2}.
\label{tab:oper_ferm_bos}}
\end{center}
}
\end{table}


In Table~\ref{tab:oper_ferm_bos} we show the operators containing two-fermion fields and a single four-lepton operator.
We separate explicitly the degrees of freedom relative to the third generation quarks, the light quarks and the leptonic ones.
The operators which involve a top quark field, either left-handed ($Q$) or right-handed ($t$), are of crucial importance for
top observables, as we have discussed extensively in Chapter~\ref{chap:top}. It must be noted that all of them involve a 
Higgs field and induce therefore modified interactions with the Higgs boson, causing an interplay between top and Higgs measurements.
For example the chromo-magnetic operator $\Op{tG}$ and the top Yukawa $\Op{t \varphi}$ affect both $t\bar{t} h$ associated
production and ggF Higgs production. Moreover, the EW dipoles, $\Op{t W}$ and $\Op{t B}$, and the current operators, $\Op{\varphi Q}^{(3)}$ and $\Op{\varphi t}$, are responsible for deviations in loop-induced Higgs decays, as well as single-top production.

The light quark operators are instead entering in $Vh$ and VBF Higgs production, while they are less relevant for top physics. Moreover, many of them, 
together with the leptonic operators, affect the Higgs width and branching ratios.
Additionally, $\mathcal{O}_{\varphi l_1}^{(3)}$, $\mathcal{O}_{\varphi l_2}^{(3)}$, and $\mathcal{O}_{ll}$ operators have an
indirect and universal effect on most of the observables considered, since they modify the measurement of the Fermi constant $G_F$ from muon decay.
\begin{table}[t!]
{\footnotesize
  \begin{center}
    \renewcommand{\arraystretch}{1.49}
    \begin{tabular}{ll}
      \toprule
      DoF $\qquad$  & Definition \\
                \midrule
$c_{\varphi Q}^{(-)}$ &  $c_{\varphi Q}^{(1)} - c_{\varphi Q}^{(3)}$ \\
\midrule
$c_{tZ}$ &   $-\sin\theta_W c_{tB} + \cos\theta_W c_{tW} $\\
\midrule
$c_{\varphi q}^{(-)}$ & $ c_{\varphi q}^{(1)} - c_{\varphi q}^{(3)}$ \\
  \bottomrule
\end{tabular}
    \caption{Definitions of the three additional degrees of freedom which substitute the ones indicated with ``--'' in Table~\ref{tab:oper_ferm_bos}.
\label{tab:oper_ferm_bos2}}
\end{center}
}
\end{table}

Three of the degrees of freedom listed in Table~\ref{tab:oper_ferm_bos}, indicated with ``--'', are not the ones directly used in
the fit. Instead, linear combinations of them are used, following the indication of Ref.~\cite{AguilarSaavedra:2018nen}, and are 
listed in Table~\ref{tab:oper_ferm_bos2}. These are defined in such a way that the top quark coupling to the $Z$ boson is explicitly
separated.

Finally, we note that despite many leptonic operators for the various generations are listed as independent degrees of freedom, 
the assumption of flavour universality is maintained and the coefficients of those operators are set to be equal.

\subsubsection{Electroweak precision observables}

A substantial subset of the operators listed before are severely constrained by EWPO~\cite{Han:2004az} from $Z$-pole~\cite{ALEPH:2005ab} and $W$-pole measurements
at LEP and SLC. The accuracy of the LEP measurements in particular still dominates over the limits provided by LHC cross-sections
and one can, in first approximation, neglect them. The operators that are sensitive to the EWPO are
\be
\mathcal{O}_{\varphi WB},
\mathcal{O}_{\varphi D},
\mathcal{O}_{\varphi q}^{\sss(1)},
\mathcal{O}_{\varphi q}^{\sss(3)}, \mathcal{O}_{\varphi u}, \mathcal{O}_{\varphi d},
\mathcal{O}_{\varphi l_i}^{\sss(3)}, \mathcal{O}_{\varphi l_i}^{\sss(1)},
\mathcal{O}_{\varphi e/\mu/\tau}, \mathcal{O}_{ll} \, . \label{eq:LEPconstrainedDoFs}
\ee
Considering the various generations independent, these amount to $16$ degrees of freedom, of which $14$ are constrained by
EWPO~\cite{Falkowski:2014tna}. This leaves $2$ linear combinations that escape the bounds and that are sensitive to other measurements such as diboson
production~\cite{Grojean:2006nn,Alonso:2013hga,Brivio:2017bnu}, Higgs cross sections and decays.
The $14$ constrained directions in the Wilson coefficients parameter space are
\begin{align} \nonumber 
&\frac{1}{4} g_1^2 \left(- 2c_{\varphi l_1}^{(3)}-2 c_{\varphi l_2}^{(3)}+c_{ll}\right)-\frac{c_{\varphi D}g_w^2}{4}-g_1 g_w c_{\varphi W B}, \\ \nonumber 
&c_{\varphi l_i}^{(3)}-f\left(-\frac{1}{2},-1\right)+f\left(\frac{1}{2},0\right)\, \quad  i=1,2,3, \\ \nonumber
&f\left(-\frac{1}{2},-1\right)-\frac{c_{\varphi l_i}^{(3)}}{2}-\frac{c_{\varphi l_i}^{(1)}}{2} \, \quad  i=1,2,3, \nonumber \\
&f(0,-1)-\frac{c_{\varphi e}}{2}, \,\, \quad f(0,-1)-\frac{c_{\varphi \mu}}{2}, \,\, \quad f(0,-1)-\frac{c_{\varphi \tau}}{2}, \label{LEPconstraints}\\
&f\left(\frac{1}{2},\frac{2}{3}\right)-\frac{c_{\varphi q}^{(-)}}{2}, \,\,\quad f\left(-\frac{1}{2},-\frac{1}{3}\right)-\frac{c_{\varphi q}^{(-)}}{2}-c_{\varphi q}^{(3)} \, , \nonumber\\
&f\left(0,\frac{2}{3}\right)-\frac{c_{\varphi u}}{2}, \,\, \quad f\left(0,-\frac{1}{3}\right)-\frac{c_{\varphi d}}{2}, \nonumber
\end{align} 
where the function $f$ is given by:
\begin{align*}
f(T_3,Q)&=\left(-\frac{c_{\varphi l_1}^{(3)}}{2}-\frac{c_{\varphi l_2}^{(3)}}{2}+\frac{c_{ll}}{4}-\frac{c_{\varphi D}}{4}\right) \left(\frac{g_1^2 Q}{g_w^2-g_1^2}+T_3\right)\\
&-c_{\varphi W B}\frac{ Q g_1 g_w}{g_w^2-g_1^2} \, ,
\end{align*}
where $g_1$ and $g_w$ are the corresponding electroweak couplings.
In the following analysis we do not include EWPO in the fit machinery but we want to take them into account. In order to do so
we emulate the impact of these measurements by setting the $14$ linear combinations to zero. The operators in Table~\ref{tab:oper_bos} and~\ref{tab:oper_ferm_bos} are then reduced to $17$ degrees of freedom left to be constrained by LHC and LEP diboson data.
We note however that diboson measurements at the LHC can in the future, in principle, compete with the EWPO~\cite{Zhang:2016zsp,Grojean:2018dqj} and the correct thing to
do would therefore be to include them explicitly in the fit. We leave this for future work.

\subsubsection{Four-fermion operators}

\begin{table}[t!] 
{\footnotesize
  \begin{center}
    \renewcommand{\arraystretch}{1.53}
        \begin{tabular}{ll}
          \toprule
          DoF $\qquad$ &  Definition (in  Warsaw basis notation) \\
          \midrule
      $c_{QQ}^1$    &   $2\ccc{1}{qq}{3333}-\frac{2}{3}\ccc{3}{qq}{3333}$ \\ \hline
    $c_{QQ}^8$       &         $8\ccc{3}{qq}{3333}$\\  \hline
 $c_{Qt}^1$         &         $\ccc{1}{qu}{3333}$\\   \hline
 $c_{Qt}^8$         &         $\ccc{8}{qu}{3333}$\\   \hline
  $c_{tt}^1$         &     $\ccc{}{uu}{3333}$  \\    \hline
            \midrule      
  $c_{Qq}^{1,8}$       &  	 $\ccc{1}{qq}{i33i}+3\ccc{3}{qq}{i33i}$     \\   \hline
  $c_{Qq}^{1,1}$         &   $\ccc{1}{qq}{ii33}+\frac{1}{6}\ccc{1}{qq}{i33i}+\frac{1}{2}\ccc{3}{qq}{i33i} $   \\    \hline
   $c_{Qq}^{3,8}$         &   $\ccc{1}{qq}{i33i}-\ccc{3}{qq}{i33i} $   \\   \hline
  $c_{Qq}^{3,1}$          & 	$\ccc{3}{qq}{ii33}+\frac{1}{6}(\ccc{1}{qq}{i33i}-\ccc{3}{qq}{i33i}) $   \\     \hline
   $c_{tq}^{8}$         &  $ \ccc{8}{qu}{ii33}   $ \\    \hline
   $c_{tq}^{1}$       &   $  \ccc{1}{qu}{ii33} $\\    \hline
   $c_{tu}^{8}$      &   $2\ccc{}{uu}{i33i}$  \\     \hline
    $c_{tu}^{1}$        &   $ \ccc{}{uu}{ii33} +\frac{1}{3} \ccc{}{uu}{i33i} $ \\   \hline
    $c_{Qu}^{8}$         &  $  \ccc{8}{qu}{33ii}$\\     \hline
    $c_{Qu}^{1}$     &  $  \ccc{1}{qu}{33ii}$  \\     \hline
    $c_{td}^{8}$        &   $\ccc{8}{ud}{33jj}$ \\    \hline
    $c_{td}^{1}$          &  $ \ccc{1}{ud}{33jj}$ \\     \hline
    $c_{Qd}^{8}$        &   $ \ccc{8}{qd}{33jj}$ \\     \hline
    $c_{Qd}^{1}$         &   $ \ccc{1}{qd}{33jj}$\\
         \bottomrule
  \end{tabular}
  \caption{\small List of the four-fermion operators used in the fit as functions of Warsaw basis operators of Eq.~(\ref{eq:FourQuarkOp}).
\label{tab:summaryOperatorsTop}}
  \end{center}
  }
\end{table}


Ultimately, we discuss the four-fermion operators included in the fit. The relevant ones are those that involve a top-quark
field explicitly and which affect several top production modes at hadron colliders. These operators can be further classified in 
two classes depending on whether they are made of $4$ heavy fields (third generation) or two-light-two-heavy ones. The degrees of 
freedom used in this work are obtained from linear combinations of the Warsaw basis operators
\begin{align}
    \qq{1}{qq}{ijkl}
    &= (\bar q_i \gamma^\mu q_j)(\bar q_k\gamma_\mu q_l)
     \nonumber
    ,\\
    \qq{3}{qq}{ijkl}
    &= (\bar q_i \gamma^\mu \tau^I q_j)(\bar q_k\gamma_\mu \tau^I q_l)
 \nonumber
    ,\\
    \qq{1}{qu}{ijkl}
    &= (\bar q_i \gamma^\mu q_j)(\bar u_k\gamma_\mu u_l)
         \nonumber
    ,\\
    \qq{8}{qu}{ijkl}
    &= (\bar q_i \gamma^\mu T^A q_j)(\bar u_k\gamma_\mu T^A u_l)
         \nonumber
    ,\\
    \qq{1}{qd}{ijkl}
    &= (\bar q_i \gamma^\mu q_j)(\bar d_k\gamma_\mu d_l)
         \nonumber
    ,\\
    \qq{8}{qd}{ijkl}
    &= (\bar q_i \gamma^\mu T^A q_j)(\bar d_k\gamma_\mu T^A d_l)
        \label{eq:FourQuarkOp} 
    ,\\
    \qq{}{uu}{ijkl}
    &=(\bar u_i\gamma^\mu u_j)(\bar u_k\gamma_\mu u_l)
         \nonumber
    ,\\
    \qq{1}{ud}{ijkl}
    &=(\bar u_i\gamma^\mu u_j)(\bar d_k\gamma_\mu d_l)
         \nonumber
    ,\\
    \qq{8}{ud}{ijkl}
    &=(\bar u_i\gamma^\mu T^A u_j)(\bar d_k\gamma_\mu T^A d_l)
         \nonumber \, ,
\end{align}
and are listed in Table~\ref{tab:summaryOperatorsTop}. In total, because of our flavour assumptions, we find ourselves with
$5$ four-heavy operators and $14$ two-light-two-heavy ones.

\subsubsection{Summary of the degrees of freedom}

\begin{table}[htbp] 
{\scriptsize
  \begin{center}
    \renewcommand{\arraystretch}{1.70}
        \begin{tabular}{cccc}
          \toprule
  Class  &  $N_{\rm dof}$ &  Independent DOFs  & DoF in EWPOs\\
          \midrule
          \multirow{4}{*}{four-quark}    &  \multirow{5}{*}{14}
          & $c_{Qq}^{1,8}$, $c_{Qq}^{1,1}$, $c_{Qq}^{3,8}$   \\
          \multirow{4}{*}{(two-light-two-heavy)}   &
          &  $c_{Qq}^{3,1}$,  $c_{tq}^{8}$,  $c_{tq}^{1}$,  \\
             &    & $c_{tu}^{8}$, $c_{tu}^{1}$, $c_{Qu}^{8}$,\\
            &    & $c_{Qu}^{1}$, $c_{td}^{8}$, $c_{td}^{1}$,   \\
          &    &  $c_{Qd}^{8}$, $c_{Qd}^{1}$   \\
          \midrule
                    \multirow{1}{*}{four-quark}      &  \multirow{2}{*}{5}
                    & $c_{QQ}^1$, $c_{QQ}^8$, $c_{Qt}^1$ &   \\
          \multirow{1}{*}{(four-heavy)}      &   & $c_{Qt}^8$, $c_{tt}^1$ &  \\
\midrule
                    \multirow{1}{*}{four-lepton}      &  \multirow{1}{*}{1}
                    &   &  $c_{ll}$ \\
	  \midrule
      \multirow{4}{*}{two-fermion}     &  \multirow{5}{*}{23} &  $c_{t\varphi}$, $c_{tG}$,   $c_{b\varphi}$,    &  
                     $c_{\varphi l_1}^{(1)}$, $c_{\varphi l_1}^{(3)}$, $c_{\varphi l_2}^{(1)}$ \\
       \multirow{4}{*}{(+ bosonic fields)}      &    & $c_{c\varphi}$, $c_{\tau\varphi}$,   $c_{tW}$,   &
       $c_{\varphi l_2}^{(3)}$, $c_{\varphi l_3}^{(1)}$, $c_{\varphi l_3}^{(3)}$,\\
            &    &  $c_{tZ}$,  $c_{\varphi Q}^{(3)}$, $c_{\varphi Q}^{(-)}$,     &
            $c_{\varphi e}$, $c_{\varphi \mu}$, $c_{\varphi \tau}$,  \\
       &    &  $c_{\varphi t}$     & $c_{\varphi q}^{(3)}$, $c_{\varphi q}^{(-)}$, \\
                       &    &       &  $c_{\varphi u}$, $c_{\varphi d}$ \\
       \midrule
      \multirow{2}{*}{Purely bosonic}     &  \multirow{2}{*}{7} &
      $c_{\varphi G}$, $c_{\varphi B}$, $c_{\varphi W}$,   &  $c_{\varphi W B}$, $c_{\varphi D}$   \\
               &   &  $c_{\varphi d}$,  $c_{WWW}$   &  \\
            \midrule
          Total  & 50 (36 independent)   & 34   & 16 (2 independent)   \\
         \bottomrule
  \end{tabular}
  \caption{\small \label{tab:operatorbasis} List of all the operators considered in this analysis divided in different categories.
 }
  \end{center}
  }
\end{table}


In Table~\ref{tab:operatorbasis} we provide an overview of all the relevant degrees of freedom for our analysis. Not all of them
are directly corresponding to Warsaw basis operators but they are at most linear combinations of them. In total, we end up with $50$
degrees of freedom, $14$ of which fixed by EWPO, leaving $36$ independent. Regarding the two directions left unconstrained by LEP,
in the following we will select the operators $\Op{\varphi W B}$ and $\Op{\varphi D}$ as representatives, but it is worth noting
that this is an arbitrary choice.

\revised{It is also interesting to observe that the two directions left unconstrained by EWPO correspond precisely to two operators
in the HISZ basis:
\begin{align}
\mathcal{O}_{H W}&=\left(D^{\mu} \varphi\right)^{\dagger} \tau_{I}\left(D^{\nu} \varphi\right) W_{\mu \nu}^{I} \, , \\ 
\mathcal{O}_{H B}&=\left(D^{\mu} \varphi\right)^{\dagger}\left(D^{\nu} \varphi\right) B_{\mu \nu} \, .
\end{align}
These two degrees of freedom can then be constrained with diboson and Higgs observables, since they modify both the Higgs couplings to the EW bosons
and the triple gauge couplings, but they do not enter in $Z$ pole measurements.}

\subsection{Cross section positivity}

Cross sections are defined to be semi-positive and as such we have to ensure that SMEFT predictions respect this requirement.
In particular, at the linear level $\mathcal{O}(\Lambda^{-2})$, this is not automatically ensured since NP terms can be negative and 
grow linearly with the Wilson coefficients. Requiring cross sections to be positive results in one-sided bounds on them and this 
can be easily implemented in the fit machinery. However, we find that the sensitivity of the experiments is already high enough that
this is effectively not needed.

On the other hand, in the case of expansion up to quadratic terms $\mathcal{O}(\Lambda^{-4})$, the cross section is theoretically
fulfilling the requirement automatically and no bounds on the Wilson coefficients is required. However, taking into account that predictions
are computed with a certain Monte Carlo uncertainty, it might happen that semi-positivity is not respected. For this reason
it is worth to verify the positivity as a sanity check. Considering the SMEFT Lagrangian
\begin{equation}
\mathcal{L} =   \mathcal{L}_{\rm SM}  + \sum^{ n_{\rm op}}_{i=1} \frac{c_i}{\Lambda^2}\mathcal{O}_i \,,
\end{equation}
we can compute observables at the quadratic level and they can be written as a generic quadratic form
\begin{eqnarray}
\Sigma&=&c_0^2 \Sigma_{00} \nonumber \\
 &+&c_0 c_1 \Sigma_{01} + c_1 c_0 \Sigma_{10}  +  c_0 c_2 \Sigma_{02} +\dots \nonumber \\
 &+&c_1^2 \Sigma_{11} + c_1 c_2 \Sigma_{12} +  c_1 c_3 \Sigma_{13} + \ldots \nonumber \\
&=& {\mathbf c^T} \cdot  \mathbf{\Sigma} \cdot {\mathbf c}. 
\end{eqnarray}
In this expression $c_0$ is an auxiliary coefficient that can be set to $1$. The $\mathbf{\Sigma}$ matrix is by construction
symmetric\footnote{It is worth noting that in comparison to
the convention where $\sigma= \sigma_{\rm SM} + \sum^{n_{\rm op}}_{i=1} c_i \sigma_i  + \sum^{n_{\rm op}}_{i<j} c_i c_j \sigma_{ij}$, one has to account for factors of 2, {\it e.g.}, $\sigma_i=2 \Sigma_{i0}$ and $\sigma_{ij}=2 \Sigma_{ij}$ for $i\neq  j$.}.
In order for the cross section to be semi-positive definite, the Sylvester criterion tells us that the matrix $\mathbf{\Sigma}$
has to be semi-positive definite, \ie all principal minors are greater or equal to zero. For instance, for the $2\times2$ minors we have
\begin{equation}
  \label{eq:sylvester}
   \left(\Sigma_{ii}\Sigma_{jj} - \Sigma_{ij}^2\right)  \ge 0 \, , \qquad i,j=0,\ldots\, n_{\rm op}\,.
\end{equation}
We verified that our calculations respect the conditions and the theoretical inputs provide semi-positive cross sections for
all the values of the Wilson coefficients.

\section{Experimental data and theoretical calculations}
\label{sec:dataset}

In this section we discuss the experimental data considered and the corresponding theoretical calculations.

\subsection{Top-quark production data}

Initially we present the top quark datasets considered for the present analysis. We will in particular focus on the ones
that have been added with respect to the work in Ref.~\cite{Hartland:2019bjb}. We refer the reader to that publication for discussions
on the datasets that were already included at the time.

The top quark observables are further categorised into four classes: inclusive top pair production, associated pair production
with EW bosons or heavy quarks, inclusive single top production and associated single top production with vector bosons.
We arbitrarily decide to classify $t\bar{t}h$ as a Higgs production process.

In regards to the theoretical calculations, the SM cross sections are computed at NLO with {\tt MCFM}~\cite{Boughezal:2016wmq} and supplemented with
NNLO $K$-factors~\cite{Czakon:2016dgf,Czakon:2016olj}. In order not to have EFT contamination in the PDFs fit, we use the NNPDF set
NNPDF3.1NNLO no-top~\cite{Ball:2017nwa}, \revised{which also do not include data from diboson observables}. The EFT contributions are computed with \MG~\cite{Alwall:2014hca}
and SMEFT@NLO~\cite{Degrande:2020evl}. The same EFT machinery is used to compute all the observables, unless 
stated otherwise. Whenever possible, calculations have been performed taking into account NLO-QCD corrections. 

\subsubsection{Inclusive top-quark pair production}

\begin{table}[t]
{\fontsize{7pt}{8pt}\selectfont
  \centering
   \renewcommand{\arraystretch}{0.50}
   \setlength{\tabcolsep}{1.0pt}
  \begin{tabular}{c|c|c|c|c|c}
 Dataset   &  $\sqrt{s}$, $\mathcal{L}$  & Info  &  Observables  & $n_{\rm dat}$ & Ref   \\
    \toprule
      {\tt ATLAS\_tt\_8TeV\_ljets}
      & { \bf 8 TeV, 20.3~fb$^{-1}$}
      & lepton+jets
      & $d\sigma/dm_{t\bar{t}}$
      & 7
      & \cite{Aad:2015mbv} \\
    \midrule
      {\tt CMS\_tt\_8TeV\_ljets}
      & {\bf 8 TeV, 20.3~fb$^{-1}$}
      & lepton+jets
      & $1/\sigma d\sigma/dy_{t\bar{t}}$
      & 10
      & \cite{Khachatryan:2015oqa} \\
    \midrule
      {\tt CMS\_tt2D\_8TeV\_dilep}
      & {\bf 8 TeV, 20.3~fb$^{-1}$}
      & dileptons
      & $1/\sigma d^2\sigma/dy_{t\bar{t}}dm_{t\bar{t}}$
      & 16
      & \cite{Sirunyan:2017azo} \\
    \midrule
      {\tt ATLAS\_tt\_8TeV\_dilep} {\bf(*)}
      & {\bf 8 TeV, 20.3~fb$^{-1}$}
      & dileptons
      & $d\sigma/dm_{t\bar{t}}$
      & 6
      & \cite{Aaboud:2016iot}  \\
    \midrule
    \midrule
      {\tt CMS\_tt\_13TeV\_ljets\_2015 }
      & {\bf 13 TeV, 2.3~fb$^{-1}$}
      & lepton+jets
      & $d\sigma/dm_{t\bar{t}}$
      & 8
      & \cite{Khachatryan:2016mnb}  \\
    \midrule
      {\tt CMS\_tt\_13TeV\_dilep\_2015 }
      & {\bf 13 TeV, 2.1~fb$^{-1}$}
      & dileptons
      & $d\sigma/dm_{t\bar{t}}$
      & 6
      & \cite{Sirunyan:2017mzl}  \\
    \midrule
      {\tt CMS\_tt\_13TeV\_ljets\_2016 }
      & {\bf 13 TeV, 35.8~fb$^{-1}$}
      & lepton+jets
      & $d\sigma/dm_{t\bar{t}}$
      & 10
      & \cite{Sirunyan:2018wem}  \\
    \midrule
      {\tt CMS\_tt\_13TeV\_dilep\_2016 } {\bf(*)}
      & {\bf 13 TeV, 35.8~fb$^{-1}$}
      & dileptons
      & $d\sigma/dm_{t\bar{t}}$
      & 7
      & \cite{Sirunyan:2018ucr}  \\
    \midrule
      {\tt ATLAS\_tt\_13TeV\_ljets\_2016 } {\bf (*)}
      & {\bf 13 TeV, 35.8~fb$^{-1}$}
      & lepton+jets
      & $d\sigma/dm_{t\bar{t}}$
      & 9
      & \cite{Aad:2019ntk}  \\
   \midrule
\midrule
 {\tt ATLAS\_WhelF\_8TeV}  & {\bf 8 TeV, 20.3~fb$^{-1}$}  & $W$ hel. fract &
$F_0, F_L, F_R$  &  3  &  \cite{Aaboud:2016hsq}  \\
\midrule
{\tt CMS\_WhelF\_8TeV}  & {\bf 8 TeV, 20.3~fb$^{-1}$}  & $W$ hel. fract &
$F_0, F_L, F_R$  &  3  &  \cite{Khachatryan:2016fky}  \\
\midrule
\midrule
    {\tt ATLAS\_CMS\_tt\_AC\_8TeV}  {\bf(*)} & {\bf 8 TeV, 20.3~fb$^{-1}$} & charge asymmetry &
    $ A_C$ & 6 & \cite{Sirunyan:2017lvd}\\
\midrule   
    {\tt ATLAS\_tt\_AC\_13TeV
}  {\bf(*)} & {\bf 8 TeV, 20.3~fb$^{-1}$} & charge asymmetry &
    $ A_C$ & 5 & \cite{ATLAS:2019czt}\\
\bottomrule
  \end{tabular}
  \caption{\small The list of the inclusive top quark pair production datasets included in this analysis.
    The datasets indicated with {\bf (*)} are new
    as compared to the analysis of~\cite{Hartland:2019bjb}.
\label{tab:input_datasets}
}
}
\end{table}


The list of the top quark pair production measurements included in the present analysis is shown in Table~\ref{tab:input_datasets}.
For each measurement we report the type of process, informations on the measurement such as the centre of mass energy, the
observables and the final states, and the number of data points $n_{dat}$. In particular, given that many distributions within the same 
datasets are correlated, one needs to be careful in choosing which data points to consider to avoid double counting.

With respect to Ref.~\cite{Hartland:2019bjb}, we have augmented the datasets by including several LHC measurements from 
both CMS and ATLAS. For instance, we included the $8$ TeV ATLAS distributions with dilepton final state~\cite{Aaboud:2016iot}, while for
$13$ TeV collisions we added the CMS dilepton distributions~\cite{Sirunyan:2018ucr} as well as the ATLAS lepton $+$ jet distributions
from the same data taking period~\cite{Aad:2019ntk}. We also include top-quark pair charge asymmetry measurements.
While these datasets do not add to the kinematic coverage in the EFT parameter space, they provide additional weight for the
global fit. Everything considered, the total number of data points for this class of processes is around $90$.

As discussed in Chapter~\ref{chap:top}, top quark observables could be an ideal place to look for SM deviations in the high energy 
phase space regions. Because of this, greater SMEFT sensitivity could be achieved by integrating the datasets with $m_{t\bar{t}}$ or transverse momentum $p_T^t$ distributions. Unfortunately, at the moment analyses with luminosities higher than $\mathcal{L}\simeq 36$~fb$^{-1}$ for differential distributions of this kind are not available. ATLAS published an analysis at $\mathcal{L}=139$~fb$^{-1}$~\cite{Aad:2020tmz}, but only inclusive 
cross sections are presented.

\subsubsection{Associated top-quark pair production}


\begin{table}[t]
  \centering
  \scriptsize
   \renewcommand{\arraystretch}{0.5}
   \setlength{\tabcolsep}{1.0pt}
  \begin{tabular}{c|c|c|c|c|c}
 Dataset   &  $\sqrt{s}, \mathcal{L}$ & Info  &  Observables  & $N_{\rm dat}$ & Ref   \\
    \toprule
 {\tt CMS\_ttbb\_13TeV}  & {\bf 13 TeV, 2.3~fb$^{-1}$}  & total xsec & $\sigma_{\rm tot}(t\bar{t}b\bar{b})$  &  1  &  \cite{Sirunyan:2017snr}  \\
\midrule
{\tt CMS\_ttbb\_13TeV\_2016}  {\bf (*)}  & {\bf 13 TeV, 35.9~fb$^{-1}$}  & total xsec & $\sigma_{\rm tot}(t\bar{t}b\bar{b})$  &  1  &  \cite{Sirunyan:2019jud}  \\
\midrule
{\tt ATLAS\_ttbb\_13TeV\_2016}  {\bf (*)}  & {\bf 13 TeV, 35.9~fb$^{-1}$}  & total xsec & $\sigma_{\rm tot}(t\bar{t}b\bar{b})$  &  1  &  \cite{Aaboud:2018eki}  \\
\midrule
 {\tt CMS\_tttt\_13TeV}  & {\bf 13 TeV, 35.9~fb$^{-1}$}  & total xsec & $\sigma_{\rm tot}(t\bar{t}t\bar{t})$  &  1  &  \cite{Sirunyan:2017roi}  \\
\midrule
 {\tt CMS\_tttt\_13TeV\_run2} {\bf (*)} & {\bf 13 TeV, 137~fb$^{-1}$}  & total xsec & $\sigma_{\rm tot}(t\bar{t}t\bar{t})$  &  1  &  \cite{Sirunyan:2019wxt}  \\
 \midrule
 {\tt ATLAS\_tttt\_13TeV\_run2} {\bf (*)} & {\bf 13 TeV, 137~fb$^{-1}$}  & total xsec & $\sigma_{\rm tot}(t\bar{t}t\bar{t})$  &  1  &  \cite{Aad:2020klt}  \\
\midrule
\midrule
  {\tt CMS\_ttZ\_8TeV}  & {\bf 8 TeV, 19.5~fb$^{-1}$}  & total xsec & $\sigma_{\rm tot}(t\bar{t}Z)$  &  1  &  \cite{Khachatryan:2015sha}  \\ \midrule
   {\tt CMS\_ttZ\_13TeV}  & {\bf 13 TeV, 35.9~fb$^{-1}$ }  & total xsec & $\sigma_{\rm tot}(t\bar{t}Z)$  &  1  &  \cite{Sirunyan:2017uzs}  \\
   \midrule
    {\tt CMS\_ttZ\_ptZ\_13TeV}  {\bf (*)} & {\bf 13 TeV, 77.5~fb$^{-1}$ }  & total xsec & $\sigma_{\rm tot}(t\bar{t}Z)$, $d\sigma(t\bar{t}Z)/dp_T^Z $  &  1, 4  &  \cite{CMS:2019too}  \\
   \midrule
  {\tt ATLAS\_ttZ\_8TeV}  & {\bf 8 TeV, 20.3~fb$^{-1}$}  & total xsec & $\sigma_{\rm tot}(t\bar{t}Z)$  &  1  &  \cite{Aad:2015eua}  \\
    \midrule
        {\tt ATLAS\_ttZ\_13TeV}  & {\bf 13 TeV, 3.2~fb$^{-1}$}  & total xsec & $\sigma_{\rm tot}(t\bar{t}Z)$  &  1  &  \cite{Aaboud:2016xve}  \\
     \midrule
   {\tt ATLAS\_ttZ\_13TeV\_2016} {\bf (*)} & {\bf 13 TeV, 36~fb$^{-1}$}  & total xsec & $\sigma_{\rm tot}(t\bar{t}Z)$  &  1  &  \cite{Aaboud:2019njj}  \\
  \midrule
  \midrule
    {\tt CMS\_ttW\_8\_TeV}  & {\bf 8 TeV, 19.5~fb$^{-1}$}  & total xsec & $\sigma_{\rm tot}(t\bar{t}W)$  &  1  &  \cite{Khachatryan:2015sha}  \\ \midrule
     {\tt CMS\_ttW\_13TeV}  & {\bf 13 TeV, 35.9~fb$^{-1}$}  & total xsec & $\sigma_{\rm tot}(t\bar{t}W)$  &  1  &  \cite{Sirunyan:2017uzs}  \\
   \midrule
   {\tt ATLAS\_ttW\_8TeV}  & {\bf 8 TeV, 20.3~fb$^{-1}$}  & total xsec & $\sigma_{\rm tot}(t\bar{t}W)$  &  1  &  \cite{Aad:2015eua}  \\
    \midrule
   {\tt ATLAS\_ttW\_13TeV}  & {\bf 13 TeV, 3.2~fb$^{-1}$}  & total xsec & $\sigma_{\rm tot}(t\bar{t}W)$  &  1  &  \cite{Aaboud:2016xve}  \\
     \midrule
   {\tt ATLAS\_ttW\_13TeV\_2016} {\bf (*) } & {\bf 13 TeV, 36~fb$^{-1}$} & total xsec & $\sigma_{\rm tot}(t\bar{t}W)$  &  1  &  \cite{Aaboud:2019njj}  \\
\bottomrule
  \end{tabular}
  \caption{Same as Table~\ref{tab:input_datasets} for associated top-quark pair production.
     \label{tab:input_datasets2}
  }
\end{table}


In Table~\ref{tab:input_datasets2} we report the measurements considered for top-quark pair production in association with
EW bosons or heavy quarks. With respect to Ref.~\cite{Hartland:2019bjb}, several new datasets have been implemented. For instance,
the most updated four tops and $t\bar{t} b \bar{b}$ measurements at $13$ TeV from ATLAS and CMS are now included. These are performed 
at higher luminosities than the previous ones and the augmented statistics leads to an improved statistical uncertainties which
is then reflected in a higher sensitivity to NP effects.

Concerning associated production with an EW boson, we include here total cross section measurements of $t\bar{t}Z$ and $t\bar{t}W$
at $\mathcal{L}=36$ fb$^{-1}$~\cite{Aaboud:2019njj}, as well as $p_T^Z$ differential distribution $\mathcal{L}=78$ fb$^{-1}$~\cite{CMS:2019too}. At higher luminosities, ATLAS has recently performed a preliminary analysis of $t\bar{t}Z$ at 139 fb$^{-1}$~\cite{ATLAS-CONF-2020-028} but this has not been included in the current dataset and will be in future works. 

Overall, $20$ data points are considered for this class of processes.

\subsubsection{Inclusive single top production}


\begin{table}[t]
  \centering
  \scriptsize
    \renewcommand{\arraystretch}{1.0}
   \setlength{\tabcolsep}{1.0pt}
  \begin{tabular}{c|c|c|c|c|c}
 Dataset   &  $\sqrt{s}, \mathcal{L}$ & Info  &  Observables  & $N_{\rm dat}$ & Ref   \\
\toprule
    {\tt CMS\_t\_tch\_8TeV\_inc}
    & {\bf 8 TeV, 19.7~{\rm \bf fb}$^{-1}$ }
    & $t$-channel
    & $\sigma_{\rm tot}(t),\sigma_{\rm tot}(\bar{t})$  & 2
    & \cite{Khachatryan:2014iya}  \\
\midrule
    {\tt ATLAS\_t\_tch\_8TeV}
    & {\bf 8 TeV, 20.2~{\rm \bf fb}$^{-1}$}
    & $t$-channel
    & $d\sigma(tq+\bar{t}q)/dy_t,d\sigma(\bar{t}q)/dy_t$
    & 4
    & \cite{Aaboud:2017pdi}  \\
\midrule
    {\tt CMS\_t\_tch\_8TeV\_dif}
    & {\bf 8 TeV, 19.7~{\rm \bf fb}$^{-1}$}
    & $t$-channel
    & $d\sigma/d|y^{(t+\bar{t})}|$
    & 6
    & \cite{CMS-PAS-TOP-14-004}  \\
\midrule
    {\tt CMS\_t\_sch\_8TeV}
    & {\bf 8 TeV, 19.7~{\rm \bf fb}$^{-1}$}
    & $s$-channel
    & $\sigma_{\rm tot}(t+\bar{t})$
    & 1
    & \cite{Khachatryan:2016ewo}  \\
\midrule
    {\tt ATLAS\_t\_sch\_8TeV}
    & {\bf 8 TeV, 20.3~{\rm \bf fb}$^{-1}$ }
    & $s$-channel
    & $\sigma_{\rm tot}(t+\bar{t})$
    & 1
    & \cite{Aad:2015upn}  \\
\midrule
\midrule
    {\tt ATLAS\_t\_tch\_13TeV}
    & {\bf 13 TeV, 3.2~{\rm \bf fb}$^{-1}$}
    & $t$-channel
    & $\sigma_{\rm tot}(t),\sigma_{\rm tot}(\bar{t})$
    & 2
    & \cite{Aaboud:2016ymp}  \\
    \midrule
    {\tt CMS\_t\_tch\_13TeV\_inc}
    & {\bf 13 TeV, 2.2~{\rm \bf fb}$^{-1}$}
    & $t$-channel
    & $\sigma_{\rm tot}(t+\bar{t})$
    & 1
    & \cite{Sirunyan:2016cdg}  \\
    \midrule
    {\tt CMS\_t\_tch\_13TeV\_dif}
    & {\bf 13 TeV, 2.3~{\rm \bf fb}$^{-1}$}
    & $t$-channel
    & $d\sigma/d|y^{(t+\bar{t})}|$
    & 4
    & \cite{CMS:2016xnv}  \\
    \midrule
    {\tt CMS\_t\_tch\_13TeV\_2016} {\bf (*)}
    & {\bf 13 TeV, 35.9~{\rm \bf fb}$^{-1}$}
    & $t$-channel
    & $d\sigma/d|y^{(t)}|$
    & 5
    & \cite{Sirunyan:2019hqb}  \\
\bottomrule
  \end{tabular}
  \caption{Same as Table~\ref{tab:input_datasets} for inclusive single top production.
     \label{tab:input_datasets3}
  }
\end{table}


We now discuss the single top production processes included in the fit, both in the t-channel and s-channel. The full list
of datasets included is displayed in Table~\ref{tab:input_datasets3}.

With respect to the previous work, the dataset has been extended with only a new measurement, the CMS differential cross section
at $13$ TeV for t-channel single top production at $\mathcal{L}=35.9$ fb$^{-1}$~\cite{Sirunyan:2019hqb}. Both normalised
and absolute distributions are presented at particle and parton level. Since theoretical predictions are more easily produced 
at parton level, we focus on those, implementing in particular the rapidity and $p_T$ distributions of the top quark.
To-date, no Run II analyses have been published on single-top quark production by ATLAS. In total, we find ourselves with $27$
data points in this category.

Regarding the theoretical calculations, it is important to state that we work in the 5-flavour scheme and therefore bottom quarks
are considered massless, see~\cite{Nocera:2019wyk}
for more details. The K-factors at NNLO-QCD are obtained from Ref.~\cite{Berger:2016oht}.

\subsubsection{Associated single top production}


\begin{table}[t]
  \centering
  \scriptsize
   \renewcommand{\arraystretch}{1.0}
   \setlength{\tabcolsep}{1.0pt}
  \begin{tabular}{c|c|c|c|c|c}
 Dataset   &  $\sqrt{s}, \mathcal{L}$ & Info  &  Observables  & $N_{\rm dat}$ & Ref   \\
\toprule
\multirow{2}{*}{ {\tt ATLAS\_tW\_8TeV\_inc}}      &
\multirow{2}{*}{{\bf 8 TeV, 20.2}~{\rm \bf fb}$^{-1}$}   & \multirow{1}{*}{inclusive}   &
\multirow{2}{*}{$\sigma_{\rm tot}(tW)$}  &  1  &
\multirow{2}{*}{\cite{Aad:2015eto}}  \\
    &
  & \multirow{1}{*}{(dilepton)}   &
  &   &\\
\toprule
\multirow{2}{*}{ {\tt ATLAS\_tW\_inc\_slep\_8TeV} {\bf (*)}}      &
\multirow{2}{*}{{\bf 8 TeV, 20.2}~{\rm \bf fb}$^{-1}$}   & \multirow{1}{*}{inclusive}   &
\multirow{2}{*}{$\sigma_{\rm tot}(tW)$}  &  1  &
\multirow{2}{*}{\cite{Aad:2020zhd}}  \\
    &
  & \multirow{1}{*}{(single lepton)}   &
  &   &\\
\midrule
      \multirow{1}{*}{ {\tt CMS\_tW\_8TeV\_inc}}      &
 \multirow{1}{*}{{\bf 8 TeV, 19.7}~{\rm \bf fb}$^{-1}$}   & \multirow{1}{*}{inclusive}   &
\multirow{1}{*}{$\sigma_{\rm tot}(tW)$}  &  1  &
\multirow{1}{*}{\cite{Chatrchyan:2014tua}}  \\
\midrule
        \multirow{1}{*}{ {\tt ATLAS\_tW\_inc\_13TeV}}      &
 \multirow{1}{*}{{\bf 13 TeV, 3.2}~{\rm \bf fb}$^{-1}$}   & \multirow{1}{*}{inclusive}   &
\multirow{1}{*}{$\sigma_{\rm tot}(tW)$}  &  1  &
\multirow{1}{*}{\cite{Aaboud:2016lpj}}  \\
\midrule
     \multirow{1}{*}{ {\tt CMS\_tW\_13TeV\_inc}}      &
 \multirow{1}{*}{{\bf 13 TeV, 35.9}~{\rm \bf fb}$^{-1}$}   & \multirow{1}{*}{inclusive}   &
\multirow{1}{*}{$\sigma_{\rm tot}(tW)$}  &  1  &
\multirow{1}{*}{\cite{Sirunyan:2018lcp}}  \\
\midrule
\midrule
      \multirow{1}{*}{ {\tt ATLAS\_tZ\_13TeV\_inc}}      &
 \multirow{1}{*}{{\bf 13 TeV, 36.1}~{\rm \bf fb}$^{-1}$}    & \multirow{1}{*}{inclusive}   &
\multirow{1}{*}{$\sigma_{\rm tot}(tZq)$}  &  1  &
\multirow{1}{*}{\cite{Aaboud:2017ylb}}  \\
\midrule
 \multirow{1}{*}{ {\tt ATLAS\_tZ\_13TeV\_run2\_inc} {\bf (*)}}      &
 \multirow{1}{*}{{\bf 13 TeV, 139.1}~{\rm \bf fb}$^{-1}$}    & \multirow{1}{*}{inclusive}   &
\multirow{1}{*}{$\sigma_{\rm fid}(tl^+l^-q)$}  &  1  &
\multirow{1}{*}{\cite{Aad:2020wog}}  \\
\midrule
       \multirow{1}{*}{ {\tt CMS\_tZ\_13TeV\_inc}}      &
 \multirow{1}{*}{{\bf 13 TeV, 35.9}~{\rm \bf fb}$^{-1}$}   & \multirow{1}{*}{inclusive}   &
\multirow{1}{*}{$\sigma_{\rm fid}(Wbl^+l^-q)$}  &  1  &
\multirow{1}{*}{\cite{Sirunyan:2017nbr}}  \\
\midrule
       \multirow{1}{*}{ {\tt CMS\_tZ\_13TeV\_2016\_inc}  {\bf (*)}}      &
 \multirow{1}{*}{{\bf 13 TeV, 77.4}~{\rm \bf fb}$^{-1}$ }   & \multirow{1}{*}{inclusive}   &
\multirow{1}{*}{$\sigma_{\rm fid}(tl^+l^-q)$}  &  1  &
\multirow{1}{*}{\cite{Sirunyan:2018zgs}}  \\
\bottomrule
  \end{tabular}
  \caption{Same as Table~\ref{tab:input_datasets} for associated single top production.
     \label{tab:input_datasets4}
  }
\end{table}


The last class of top quark processes is single top production in association with EW bosons. The list of the experimental measurements considered
in this analysis is given in Table~\ref{tab:input_datasets4}. With respect to the work in Ref.~\cite{Hartland:2019bjb}, three
new measurements have been added. Regarding $tZ$ production, we included both a Run II analysis from ATLAS at the full luminosity
139 fb$^{-1}$ in the trilepton channel~\cite{Aad:2020wog} and the corresponding one from CMS~\cite{Sirunyan:2018zgs} but at a lower luminosity. On the other hand, for $tW$ production, we included an inclusive cross section measurement from ATLAS at $8$ TeV~\cite{Aad:2020zhd}.
No differential measurement for these processes has been published so far and the total number of data points for this category 
is $9$.

Combining all the top quark measurements, we end up with a grand total of $153$ data points as compared to the $103$ from Ref.\cite{Hartland:2019bjb}.

\subsection{Higgs production and decay}

We now discuss the Higgs measurements introduced in the present work. In particular, we classify them in two categories:
inclusive cross sections and differential distributions. For the case of inclusive cross sections we present them in the form
of signal strengths, where the measured cross section is normalised by the SM prediction.

\subsubsection{Signal strengths}


\begin{table}[t]
  \centering
  \scriptsize
   \renewcommand{\arraystretch}{1.0}
   \setlength{\tabcolsep}{1.0pt}
  \begin{tabular}{c|c|c|c|c|c}
 Dataset   &  $\sqrt{s},~\mathcal{L}$ & Info  &  Observables  & $n_{\rm dat}$ & Ref.   \\
    \toprule
    \multirow{2}{*}{ {\tt ATLAS\_CMS\_SSinc\_RunI} {\bf (*)}}  &\multirow{2}{*}{ {\bf 7+8 TeV, 20~fb$^{-1}$}}  &
    \multirow{2}{*}{Incl. $\mu_i^f$} &  $gg$F, VBF, $Vh$, $t\bar{t}h$
    &  \multirow{2}{*}{20}    &  \multirow{2}{*}{\cite{Khachatryan:2016vau} } \\
    &   &     & $h\to \gamma\gamma, VV, \tau\tau, b\bar{b}$   &  &    \\ \midrule
    \multirow{1}{*}{ {\tt ATLAS\_SSinc\_RunI} {\bf (*)}}  &\multirow{1}{*}{ {\bf 8 TeV, 20~fb$^{-1}$}}  &
    \multirow{1}{*}{Incl. $\mu^f_i$} &  $h\to Z\gamma, \mu\mu$
    &  \multirow{1}{*}{2}    &  \multirow{1}{*}{\cite{Aad:2015gba} } \\ \midrule
    \midrule
 \multirow{2}{*}{ {\tt ATLAS\_SSinc\_RunII} {\bf (*)}}  &\multirow{2}{*}{ {\bf 13 TeV, 80~fb$^{-1}$}}  &
 \multirow{2}{*}{Incl. $\mu_i^f$} &  $gg$F, VBF, $Vh$, $t\bar{t}h$
 &  \multirow{2}{*}{16}    &  \multirow{2}{*}{\cite{Aad:2019mbh} } \\
 &   &     & $h\to \gamma\gamma, WW, ZZ, \tau\tau, b\bar{b}$   &  &    \\ \midrule
 \multirow{2}{*}{ {\tt CMS\_SSinc\_RunII} {\bf (*)}}  &\multirow{2}{*}{ {\bf 13 TeV, 36.9~fb$^{-1}$}}  &
 \multirow{2}{*}{Incl. $\mu_i^f$} &  $gg$F, VBF, $Wh$, $Zh$ $t\bar{t}h$
 &  \multirow{2}{*}{24}    &  \multirow{2}{*}{\cite{Sirunyan:2018koj} } \\
 &   &     & $h\to \gamma\gamma, WW, ZZ, \tau\tau, b\bar{b}$   &  &    \\ 
    \bottomrule
    \end{tabular}
  \caption{\small Same as Table~\ref{tab:input_datasets} for Higgs signal strengths.
     \label{tab:input_datasets_higgsSS}
  }
\end{table}


Regarding the Higgs signal strengths we considered both Run I and Run II measurements from CMS and ATLAS. In terms of cross sections
$\sigma_i$ and branching ratios $B_i$, these are defined as
\be
\mu_i^f \equiv \frac{\sigma_i \cdot B_f}{\lp \sigma_i\rp_{\rm SM} \cdot
  \lp B_f \rp_{\rm SM}} = \mu_i \cdot \mu^f = \lp \frac{\sigma_i }{\lp \sigma_i\rp_{\rm SM} } \rp
\lp  \frac{  B_f}{ \lp B_f \rp_{\rm SM}}\rp \, ,
\ee
separating explicitly the production cross section and the decay channels. More explicitly as a function of the decay widths, this can be expressed as
\be
\mu_i^f = \lp \frac{\sigma_i }{\lp \sigma_i\rp_{\rm SM} } \rp
\lp  \frac{  \Gamma(h \to f) }{  \Gamma(h \to f)\big|_{\rm SM} }\rp
\lp  \frac{  \Gamma(h \to {\rm all}) }{  \Gamma(h \to {\rm all})\big|_{\rm SM} }\rp^{-1} \, .
\ee
The list of inclusive measurements included in the analysis is given in Table~\ref{tab:input_datasets_higgsSS}.

Regarding the Run I measurements, we take the inclusive measurements from the ATLAS and CMS combination at $7$ and $8$ TeV~\cite{Khachatryan:2016vau}. Multiple production channels ($gg{\rm F}$, VBF, $Wh$, $Zh$, and $t\bar{t}h$) are considered, as well as many decay channels ($\gamma\gamma$, $ZZ$, $WW$, $\tau\tau$, and $b\bar{b}$). Correlations between measurements are taken into account, as provided in the same reference. Additionally, we consider the signal strengths on the $Z\gamma$ and $\mu\mu$ decays from~\cite{Aad:2015gba}.

With respect to the Run II data, we include the ATLAS combination at $\sqrt{s}=13$ TeV~\cite{Aad:2019mbh} and the CMS corresponding ones
at a lower luminosity~\cite{Sirunyan:2018koj}.

Overall, a total of $62$ data points are considered for inclusive Higgs signal strengths. The theoretical predictions for the SM are
taken from the associated experimental publications, which are the ones provided by the LHC Higgs Cross-Section Working Group~\cite{deFlorian:2016spz,Dittmaier:2012vm}.

\subsubsection{Differential distributions and STXS}


\begin{table}[t]
  \centering
  \scriptsize
     \renewcommand{\arraystretch}{1.0}
   \setlength{\tabcolsep}{0.2pt}
  \begin{tabular}{c|c|c|c|c|c}
 Dataset   &  $\sqrt{s}, \mathcal{L}$ & Info  &  Observables  & $N_{\rm dat}$ & Ref   \\
 \toprule
  \multirow{2}{*}{ {\tt CMS\_H\_13TeV\_2015} {\bf (*)}}  &\multirow{2}{*}{ {\bf 13 TeV, 35.9~fb$^{-1}$}}  &
  \multirow{1}{*}{$gg$F, VBF, $Vh$, $t\bar{t}h$} &  $d\sigma/d|y_h|$    &   \multirow{1}{*}{6}    &  \multirow{2}{*}{ \cite{Sirunyan:2018sgc} } \\
     &  & $h\to ZZ,\gamma\gamma,b\bar{b}$
  &  $d\sigma/dp_T^h$    &  9   &    \\
  \midrule
   \multirow{2}{*}{ {\tt ATLAS\_ggF\_13TeV\_2015} {\bf (*)}}  &\multirow{2}{*}{ {\bf 13 TeV, 36.1~fb$^{-1}$}}  &
  \multirow{1}{*}{$gg$F, VBF, $Vh$, $t\bar{t}h$} &  $d\sigma/d|y_h|$    &   \multirow{1}{*}{6}    &  \multirow{2}{*}{ \cite{Aaboud:2018ezd} } \\
     &  & $h\to ZZ(\to 4l)$
  &  $d\sigma/dp_T^h$    &  9   &    \\
  \midrule
   \midrule
    \multirow{2}{*}{ {\tt ATLAS\_Vh\_hbb\_13TeV} {\bf (*)}}  &\multirow{2}{*}{ {\bf 13 TeV, 79.8~fb$^{-1}$}}  &
    \multirow{2}{*}{$Wh,~Zh$} &  $d\sigma^{\rm (fid)}/dp_T^W$   &  2    &  \multirow{2}{*}{\cite{Aaboud:2019nan} } \\
    &   &     & $d\sigma^{\rm (fid)}/dp_T^Z$   & 3  &    \\ \midrule
     \multirow{1}{*}{ {\tt ATLAS\_ggF\_ZZ\_13TeV} {\bf (*)}}  &\multirow{1}{*}{ {\bf 13 TeV, 79.8~fb$^{-1}$}}  &
     \multirow{1}{*}{$gg$F, $h\to ZZ$} &  $\sigma_{\rm ggF}(p_T^h,N_{\rm jets})$    &  6    &  \multirow{1}{*}{\cite{Aad:2019mbh} } \\
     \midrule
     \multirow{1}{*}{ {\tt CMS\_ggF\_aa\_13TeV} {\bf (*)}}  &\multirow{1}{*}{ {\bf 13 TeV, 77.4~fb$^{-1}$}}  &
    \multirow{1}{*}{$gg$F, $h\to \gamma\gamma$} &   $\sigma_{\rm ggF}(p_T^h,N_{\rm jets})$    &  6    &  \multirow{1}{*}{\cite{CMS:1900lgv} } \\
    \midrule
     \multirow{1}{*}{ {\tt CMS\_ggF\_tautau\_13TeV} {\bf (*)}}  &\multirow{1}{*}{ {\bf 13 TeV, 77.4~fb$^{-1}$}}  &
    \multirow{1}{*}{$gg$F, $h\to \tau\tau$} &   $\sigma_{\rm ggF}(p_T^h,N_{\rm jets})$    &  5    &  \multirow{1}{*}{\cite{CMS:2019pyn} } \\
    \midrule
    \midrule
     \multirow{1}{*}{ {\tt ATLAS\_h\_ZZ\_13TeV\_RunII} {\bf (*)}}  &\multirow{1}{*}{ {\bf 13 TeV, 139~fb$^{-1}$}}  &
     \multirow{1}{*}{$h\to ZZ\to 4l$} &  \multirow{1}{*}{$d\sigma^{\rm (fid)}/dp_T^{4l}$}   &  10    &  \multirow{1}{*}{\cite{ATLAS:2020wny} } \\
    \bottomrule
    \end{tabular}
  \caption{Same as Table~\ref{tab:input_datasets} for
   differential distributions and STXS for Higgs production and decay.
     \label{tab:input_datasets_higgs}
  }
\end{table}


In Table~\ref{tab:input_datasets_higgs} we summarise the differential and Simplified Template Cross Sections (STXS) included in
the analysis. It is worth pointing out that whenever there is a possible double counting between signal strengths and differential
measurements, we always choose to use the latter, as they provide a higher sensitivity to NP effects. 

Since SMEFT effects are particularly prominent in the high energy tails of the distributions, here we consider $p_T^h$ and $|y_h|$
Higgs distributions from ATLAS and CMS~\cite{Aaboud:2018ezd,Sirunyan:2018sgc}. These are performed at $13$ TeV and a luminosity
$\mathcal{L}=36$ fb$^{-1}$ in the Higgs decay channels $h\to \gamma\gamma$, $h\to ZZ$, and (in the CMS case)
$h \to b\bar{b}$.

In Ref.~\cite{Aaboud:2019nan} measurements of associated production of a Higgs and EW boson are provided within the 
kinematical regions defined by the STXS. In particular, bins are defined as a function of the $p_T$ of the EW bosons, which are reconstructed by means of their leptonic decays.

Higgs boson production STXS from Run II are instead taken from Ref.~\cite{Aad:2019mbh, CMS:1900lgv,CMS:2019pyn}, where bins are defined in $p_T^h$ and the number of jets. These are dominated by gluon fusion production, but other channels are relevant especially in the 
high $p_T^h$ regions.

Finally, we include two ATLAS measurements with the full Run II luminosity $\mathcal{L}=139$ fb$^{-1}$, where the Higgs boson is
reconstructed from the four lepton $h\to ZZ^* \to 4l$ final state~\cite{ATLAS:2019ssu,ATLAS:2020wny}.

Other measurements have been published recently by both ATLAS~\cite{Aad:2020jym} and CMS~\cite{Sirunyan:2020tzo}, but we do not expect them to be too impactful on the fit and
their inclusion is left for future work. 

In total, the Higgs dataset put together for the study has more than $100$ data points, $50$ of which from differential measurements.

\subsection{Diboson production}


\begin{table}[t]
  \centering
  \scriptsize
   \renewcommand{\arraystretch}{1.0}
   \setlength{\tabcolsep}{0.2pt}
  \begin{tabular}{c|c|c|c|c|c}
 Dataset   &  $\sqrt{s}, ~\mathcal{L}$ & Info  &  Observables  & $N_{\rm dat}$ & Ref   \\
    \toprule
        {\tt LEP2\_WW\_diff} {\bf (*)}   & $\lc 182,296\rc$ GeV   & LEP-2 comb   & $d^2\sigma(WW)/dE_{\rm cm}d\cos\theta_{W}$  & 40  &  \cite{Schael:2013ita} \\ \toprule
    {\tt ATLAS\_WZ\_13TeV\_2016} {\bf (*)}  & {\bf 13 TeV, 36.1~fb$^{-1}$}  &
    fully leptonic &  $d\sigma^{\rm (fid)}/dm_T^{WZ}$  &  6    & \cite{Aaboud:2019gxl}  \\
    \midrule
    {\tt ATLAS\_WW\_13TeV\_2016} {\bf (*)}  &{\bf 13 TeV, 36.1~fb$^{-1}$}  &
    fully leptonic &  $d\sigma^{\rm (fid)}/dm_{e\mu}$  &  13    &  \cite{Aaboud:2019nkz} \\
     \midrule
    {\tt CMS\_WZ\_13TeV\_2016} {\bf (*)}  & {\bf 13 TeV, 35.9~fb$^{-1}$}  &
    fully leptonic &  $d\sigma^{\rm (fid)}/dp_T^{Z}$  &  11    &  \cite{Sirunyan:2019bez} \\
    \bottomrule
    \end{tabular}
  \caption{Same as Table~\ref{tab:input_datasets} for
    diboson pair
    production from LEP-2 and the LHC.
     \label{tab:input_datasets_diboson}
  }
\end{table}


In Table~\ref{tab:input_datasets_diboson} we report the diboson inclusive and differential cross-section measurements
included in the present work. As previously discussed, the EWPO leave two unconstrained directions in the $16$ dimensional
Wilson coefficient space sensitive to LEP-1 measurements. These can however be bounded with Higgs and diboson observables. 

To begin with, we consider the LEP-2 measurement of $WW$ production~\cite{Schael:2013ita}, including the differential distributions in
$\cos\theta_W$ in different centre of mass energies, from $\sqrt{s}=182$ GeV up to $\sqrt{s}=206$ GeV. The SM theoretical predictions
used are precise at NLO-EW and are taken from the same reference.

From the LHC, we consider the $WZ$ CMS~\cite{Sirunyan:2019bez} and ATLAS~\cite{ATLAS-CONF-2018-034} differential distributions at $\mathcal{L}=36.1$ fb$^{-1}$, where both EW bosons are reconstructed from fully leptonic final states. Several differential cross-sections
are provided ($p_T^W$, $p_T^Z$ and $m_T^{WZ}$), but only one of them is ultimately included in order to avoid double counting and maximise sensitivity to NP at the same time.

Production of $W$ pairs from ATLAS~\cite{Aaboud:2019nkz} is also included, and a similar reasoning is applied to the differential 
distributions ($m_{e\mu}$, $p_T^{e\mu}$ and $|y_{e\mu}|$).

The SM theoretical predictions are computed with {\tt MATRIX}~\cite{Grazzini:2017mhc}. Regarding the EFT contributions,
in order to take into account constraints from Eq.~\eqref{eq:LEPconstrainedDoFs}, we computed them in terms of only three degrees
of freedom ($c_{WWW}$, $c_{\varphi D}$ and $c_{\varphi W B}$).

\subsection{EFT sensitivity}

Before discussing the fit methodology and the results, it is instructive to take a step back and have a comprehensive overview
of the dataset considered in order to assess a priori the EFT sensitivity. In Appendix~\ref{app:dataoverview}, we report a summary of
the datasets and theory calculations taken into account in this analysis (Table~\ref{tab:table_dataset_overview} and~\ref{tab:table-processes-theory} respectively). A total of $317$
cross sections is considered and the vast majority of the calculations in the EFT are performed at NLO-QCD accuracy. Moreover, in Table~\ref{tab:operatorprocess},
we display which operators affect each class of processes, providing a first measure of the kind of sensitivity we can expect.
However, this information does not allow to ascertain quantitatively the sensitivity on the various degrees of freedom brought in by the
experimental data. 

In order to achieve this, we introduce the concept of information geometry~\cite{Brehmer:2017lrt} defining the Fisher information
matrix
\be
\label{eq:FisherDef}
I_{ij}\lp {\boldsymbol c} \rp = -{\rm E}\lc \frac{\partial^2 \ln f \lp {\boldsymbol \sigma}_{\rm exp}|
{\boldsymbol c} \rp}{\partial c_i \partial c_j} \rc \, , \qquad i,j=1,\ldots,n_{\rm op} \, ,
\ee
where $ f \lp {\boldsymbol \sigma}_{\rm exp}|{\boldsymbol c} \rp$ represents the dependence of the experimental measurements
from the true values of the EFT coefficients $ {\boldsymbol c}$. The covariance matrix in the Wilson coefficients space satisfies
\be
C_{ij} \ge \lp I^{-1}\rp_{ij} \, ,
\ee
known as the Cramer-Rao bound. In particular, the diagonal entries give us information on the smallest uncertainty achievable with a set of 
input data on the individual coefficients $\delta_{c_i}^{\rm (best)}=\sqrt{ (I^{-1})_{ii}}$. Because of this, the Fisher information 
provides a measure of which dataset is the most sensitive for a given degree of freedom. The larger the value of the entry, the 
smaller the projected uncertainty on the Wilson coefficient is. In particular, if one has $n_{dat}$ experimental measurements
$\sigma_m^{(exp)}$ whose theoretical predictions depend on Wilson coefficients $ {\boldsymbol c}$, we have
\be
\label{eq:fisherf}
f \lp {\boldsymbol \sigma}_{\rm exp})|
{\boldsymbol c} \rp = \prod_{m=1}^{n_{\rm dat}}\frac{1}{\sqrt{2\pi \delta_{{\rm exp},m}^2}}
\exp \lp -\frac{ \lp \sigma_m^{\rm (exp)}-\sigma_m^{\rm (th)}({\boldsymbol c})\rp^2  }{ 2\delta_{{\rm exp},m}^2}\rp \, ,
\ee
where we have assumed that the measurements are distributed following a Gaussian and $\delta_{{\rm exp},m}$ is the total uncertainty
on each of them.
Since the SMEFT theory predictions are given by
\be
\label{eq:quadraticTHform}
\sigma_m^{\rm (th)}({\boldsymbol c})= \sigma_m^{\rm (sm)} + \sum_{i=1}^{n_{\rm op}}c_i\sigma^{(\rm eft)}_{m,i} +
\sum_{i<j}^{n_{\rm op}}c_ic_j \sigma^{(\rm eft)}_{m,ij} \, ,
\ee
it can be shown that the Fisher information is 
\begin{align}
\label{eq:fisherinformation}
I_{ij} = {\rm E}\Bigg[ \sum_{m=1}^{n_{\rm dat}} \frac{1}{\delta_{{\rm exp},m}^2}  &\Bigg( \sigma_{m,ij}\lp
  \sigma_m^{\rm (th)}-\sigma_m^{\rm (exp)} \rp  
  +\nonumber\\ 
  &\lp  \sigma^{\rm (eft)}_{m,i} + \sum_{l=1}^{n_{\rm op}}
  c_l \sigma^{\rm (eft)}_{m,il} \rp
  \lp \sigma^{\rm (eft)}_{m,j}+ \sum_{l'=1}^{n_{\rm op}} 
 c_{l'}  \sigma_{m,jl'} \rp \Bigg)\Bigg] \, .
\end{align}
In the linear EFT approximation~\cite{Ellis:2018gqa}, the Fisher information matrix becomes especially simple and takes the form
\be
\label{eq:fisherinformation2}
I_{ij} = \sum_{m=1}^{n_{\rm dat}} \frac{\sigma^{\rm (eft)}_{m,i}\sigma^{\rm (eft)}_{m,j}}{\delta_{{\rm exp},m}^2} \, ,
\ee
which is notably independent of the Wilson coefficients and therefore of the fit results.

\begin{figure}[t!]
  \begin{center}
    \includegraphics[width=0.96\linewidth]{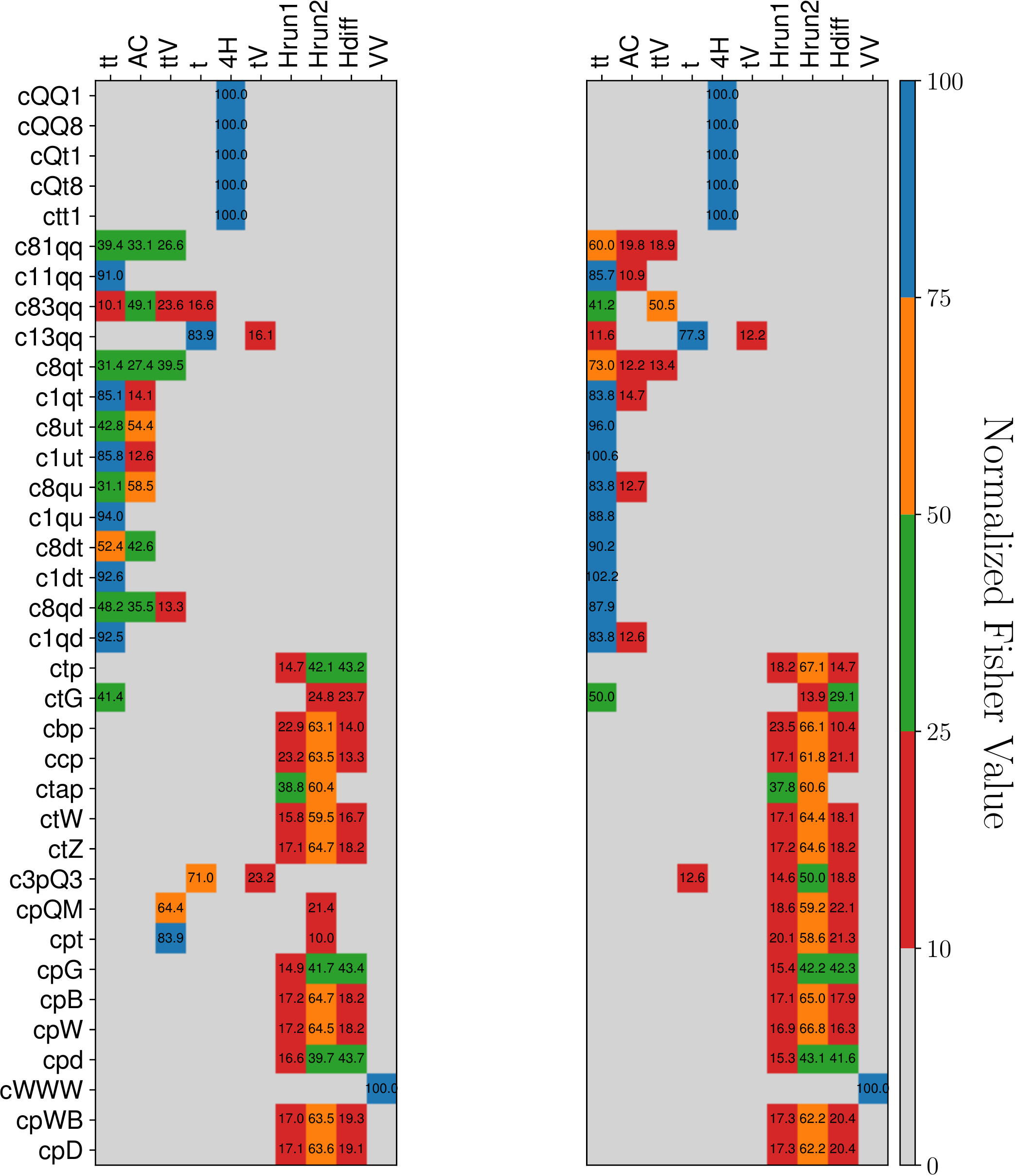}
    \caption{Fisher information matrix for the normalised diagonal entries of Eq.~\eqref{eq:fisherinformation}
    both at linear (left) and quadratic (right) level in the EFT expansion.
     \label{fig:FisherMatrix} }
  \end{center}
\end{figure}

While the absolute size of the Fisher entries does not yield information, the relative values among each of them for
a fixed operator have a physical meaning. In particular, for each diagonal $I_{ii}$, we can normalise so that the sum over the 
experimental contributions is $100$ and extract a measure of the sensitivity of each of them to the individual degree of freedom.
In Fig.~\ref{fig:FisherMatrix} we report the values of the normalised diagonal entries of the Fisher information both at linear and quadratic level,
separating the contributions of the various datasets and indicating in blue datasets that provide more than $75\%$ of the constraining
power.
%
%
As we can observe in the plot, two-light-two-heavy operators are mostly constrained by inclusive top quark measurements, except
for $c_{Qq}^{3,1}$ which is more sensitive to single top.
Two-fermion operators are mostly bounded by Higgs observables, specifically the ones from Run II.
One of the few exception is $c_{tG}$ which receives a relevant bound contribution from $t\bar{t}$ production.

Further, it can be observed that the quadratic corrections do not dramatically change the Fisher information matrix. Two
examples of operators that have a markedly different balance are $c_{\varphi t}$ and $c_{\varphi Q}^{(3)}$, 
which become dominated by Higgs data only when order $\mathcal{O}\lp \Lambda^{-4}\rp$ corrections are included.

\section{Fitting methodology}

In this section we discuss the technical details of the methodology used to fit the dataset described in section~\ref{sec:dataset}.
With respect to Ref.~\cite{Hartland:2019bjb}, in this work we implemented an additional method based on Nested Sampling (NS)~\cite{Feroz:2013hea,Feroz:2007kg} which complements the MCfit algorithm.
First, we will present the log-likelihood function and discuss the treatment of the uncertainties, discussing in particular
the $\chi^2$ profiles and the presence of degenerate minima. Then we briefly discuss the NS algorithm and carry out a principal
component analysis (PCA) which is giving us information on the directions in parameter space that have the most variability 
given the dataset at our disposal.

\subsection{Log-likelihood}

In order to have a measure of the quality of the fit, we define the log-likelihood, or $\chi^2$, as
\begin{equation}
  \chi^2\lp {\boldsymbol c} \rp \equiv \frac{1}{n_{\rm dat}}\sum_{i,j=1}^{n_{\rm dat}}\lp 
  \sigma^{(\rm th)}_i\lp {\boldsymbol c} \rp
  -\sigma^{(\rm exp)}_i\rp ({\rm cov}^{-1})_{ij}
\lp 
  \sigma^{(\rm th)}_j\lp {\boldsymbol c}\rp
  -\sigma^{(\rm exp)}_j\rp
 \label{eq:chi2definition2}
    \; ,
\end{equation}
where $\sigma_i^{\rm (exp)}$ and $\sigma^{(\rm th)}_i\lp {\boldsymbol c}\rp$ are the experimental and theoretical cross sections
as a function of the Wilson coefficients respectively. The total covariance matrix ${\rm cov}_{ij}$ carries both experimental and theoretical
uncertainties. In particular, assuming that they are normally distributed and uncorrelated, we define the matrix as
\be
\label{eq:covmatsplitting}
{\rm cov}_{ij} = {\rm cov}^{(\rm exp)}_{ij} + {\rm cov}^{(\rm th)}_{ij} \, ,
\ee
summing explicitly the two contributions~\cite{AbdulKhalek:2019ihb,AbdulKhalek:2019bux}. 
While the experimental uncertainties are taken directly from the experimental publications, taking into account both statistical and
systematic uncertainties, the theoretical ones differ process by process.
In the case of top quark and diboson production, we compute the SM with the best possible theoretical precision and we associate
PDFs uncertainties to the predictions. Regarding the Higgs measurements on the other hand, SM predictions are taken from the 
experimental papers and are collected from the HXSWG, where both PDF and missing higher order corrections are accounted for in the 
uncertainty. 

\revised{On the EFT side, Monte Carlo uncertainties of the predictions are not currently taken into account, but their implementation in the
machinery is underway and will be included in the next release. The same is true for scale uncertainties, which have not been included and can provide an assessment on the missing higher order effects. However, we do not expect a priori that the inclusion of the theory uncertainties for the SMEFT predictions will
have a huge impact on the fit currently, since the covariance matrix is still mostly dominated by statistical uncertainties. In the future, when the precision of the measurements will be improved and will start to be dominated by systematics, taking into account the theory uncertainties
for the EFT will prove to be fundamental for a consistent interpretation.}

While the multi-dimensional problem is complex and requires sophisticated methods to solve, minimisation of the $\chi^2$
in the case of fitting one single operator is straightforward. Even if switching on only one Wilson coefficient and setting the others
to zero is not physically meaningful, it provides a baseline for the complete fit, since the constraining power on individual
fits is maximal. In particular, if we only have one operator $\Op{j}$, the theoretical cross section for the measurement $m$ is
\be
\label{eq:quadraticTHform}
\sigma_m^{\rm (th)}(c_j)= \sigma_m^{\rm (sm)} + c_j\sigma^{(\rm eft)}_{m,j} +
c_j^2 \sigma^{(\rm eft)}_{m,jj} \, ,
\ee
while the $\chi^2$ will be a quartic function
\be
\label{eq:quartic-chi2}
\chi^2(c_j) = \sum_{k=0}^4 a_k \lp c_j\rp^k \, .
\ee
If we want to determine the coefficients $a_k$, we just need to scan the $\chi^2$ and subsequently fit the functional form.
The $95\%$ CL is then obtained by requiring
\be
\label{eq:deltachi2}
\chi^2(c_j)-\chi^2(c_{j,0}) \equiv \Delta \chi^2 \le 5.91 \, .
\ee
%
\begin{figure}[h!]
  \begin{center}
\includegraphics[width=0.297\linewidth]{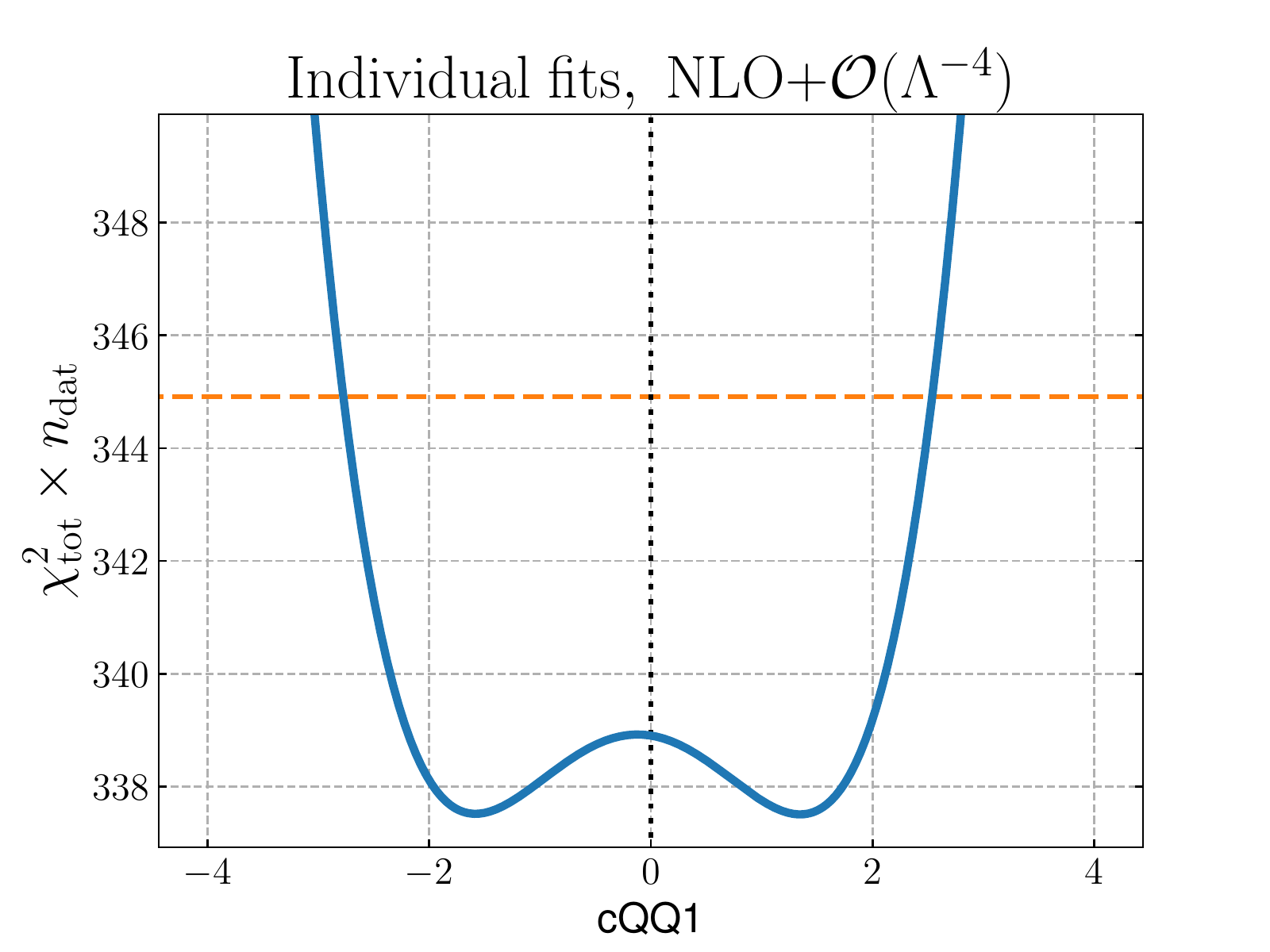}
\includegraphics[width=0.297\linewidth]{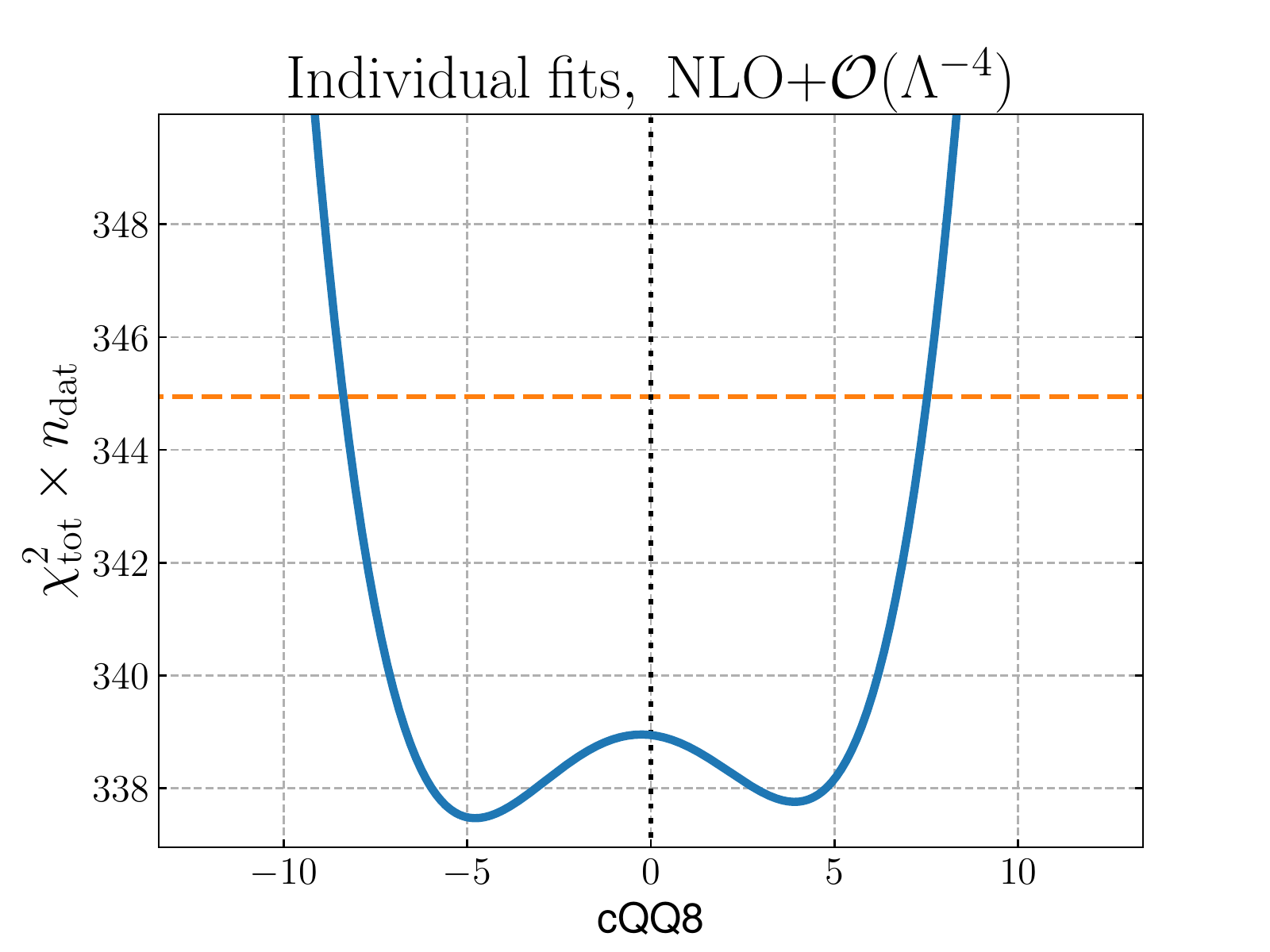}
\includegraphics[width=0.297\linewidth]{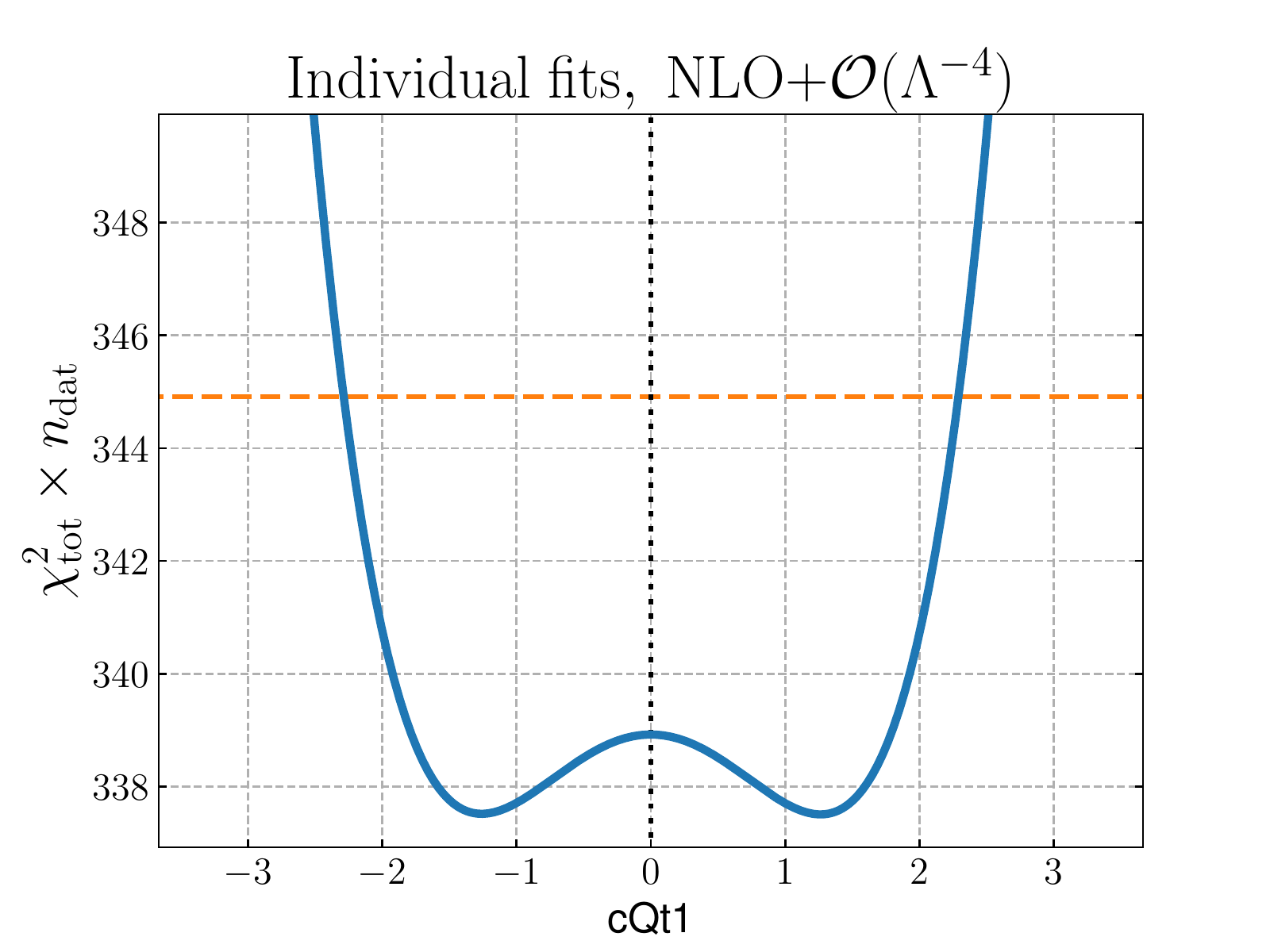}
\includegraphics[width=0.297\linewidth]{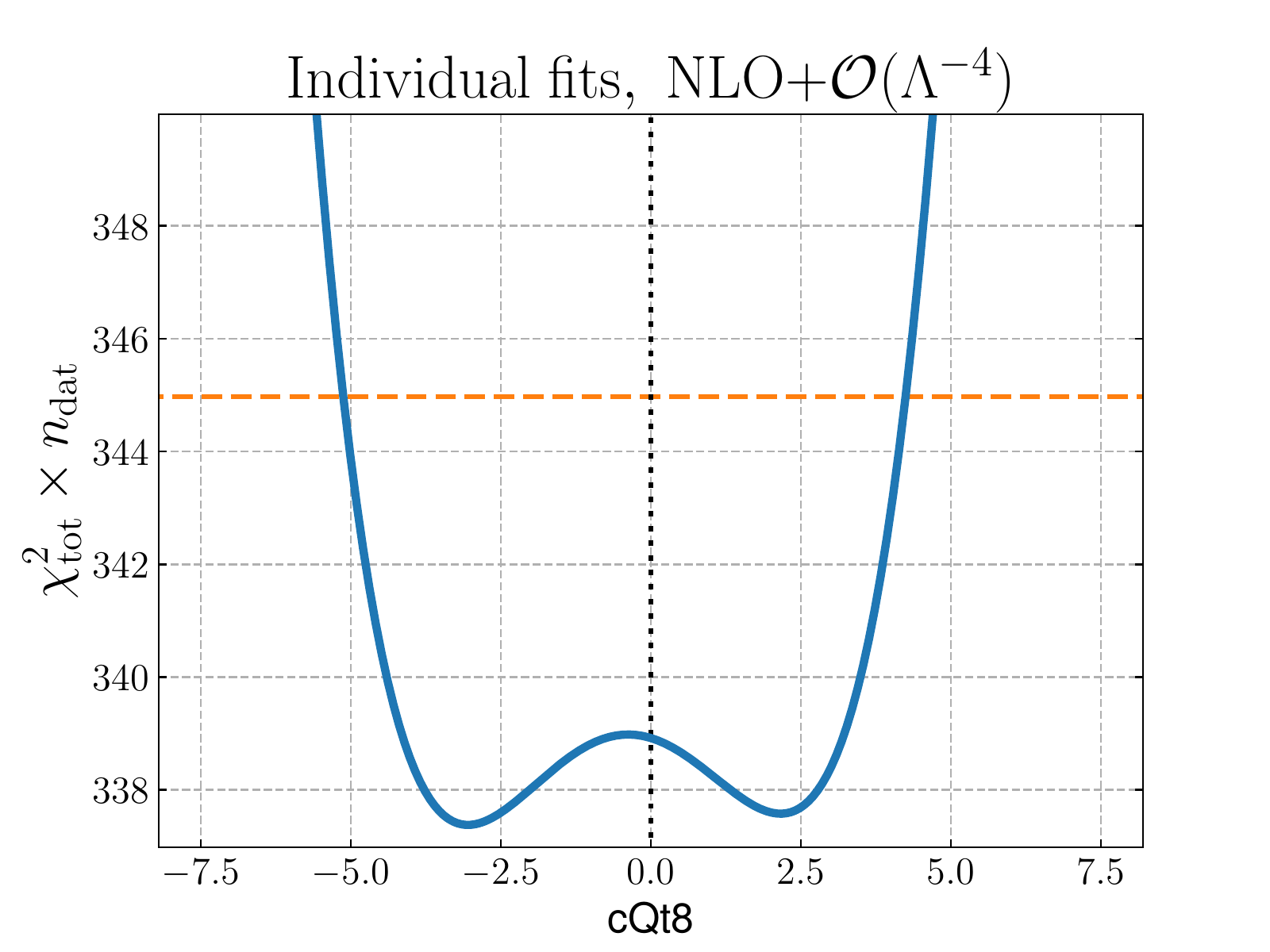}
\includegraphics[width=0.30\linewidth]{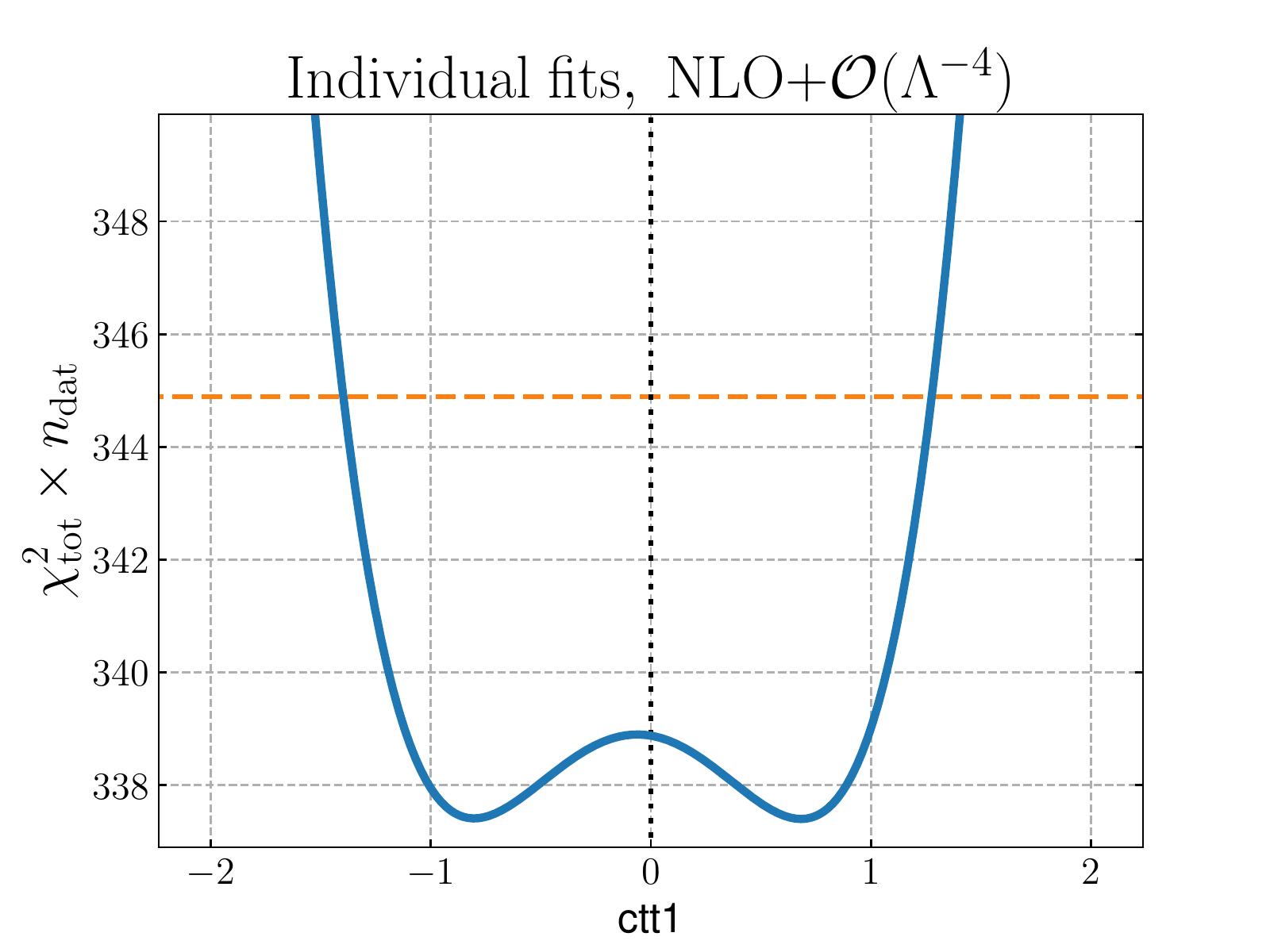}
\includegraphics[width=0.297\linewidth]{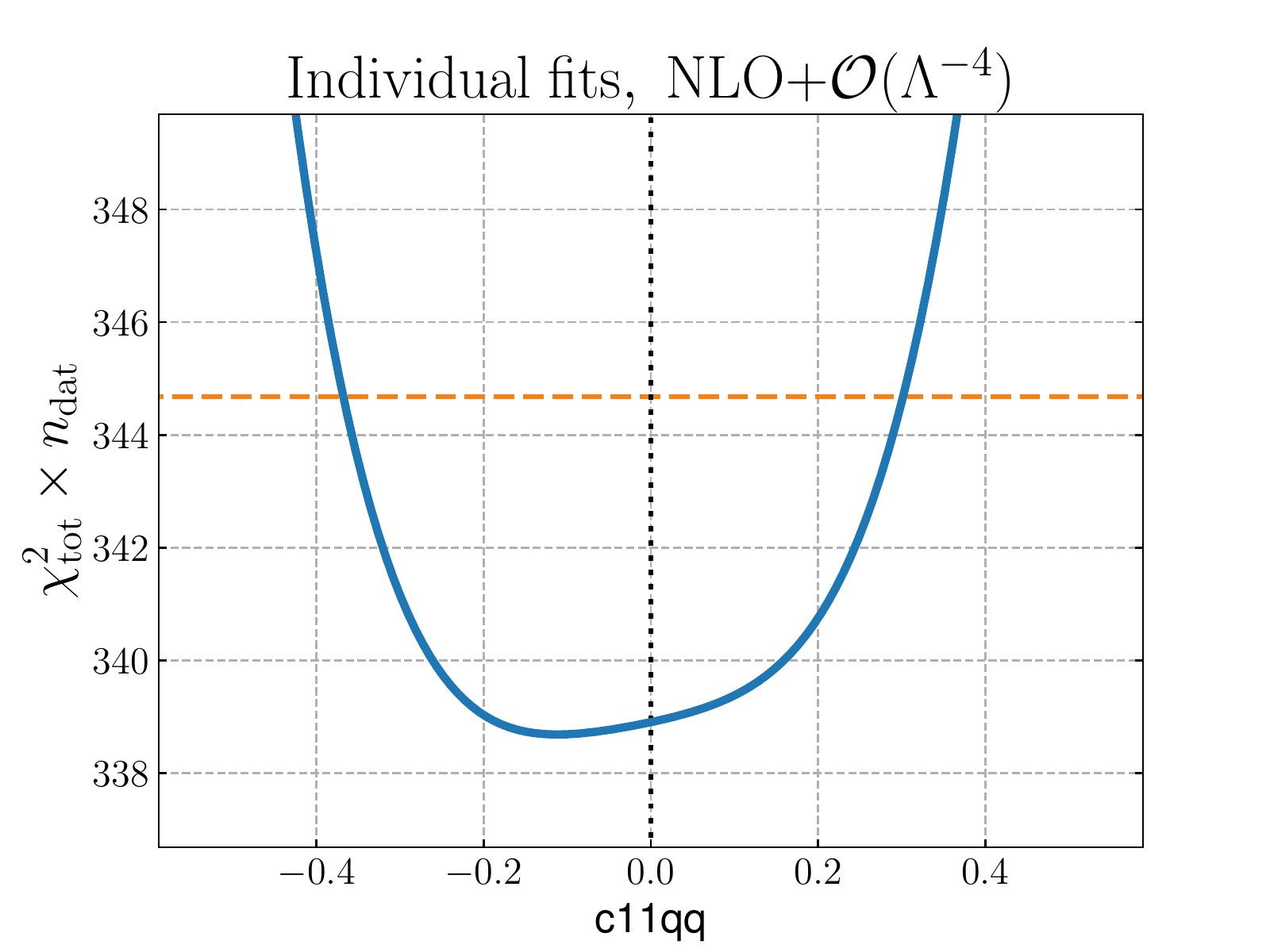}
\includegraphics[width=0.297\linewidth]{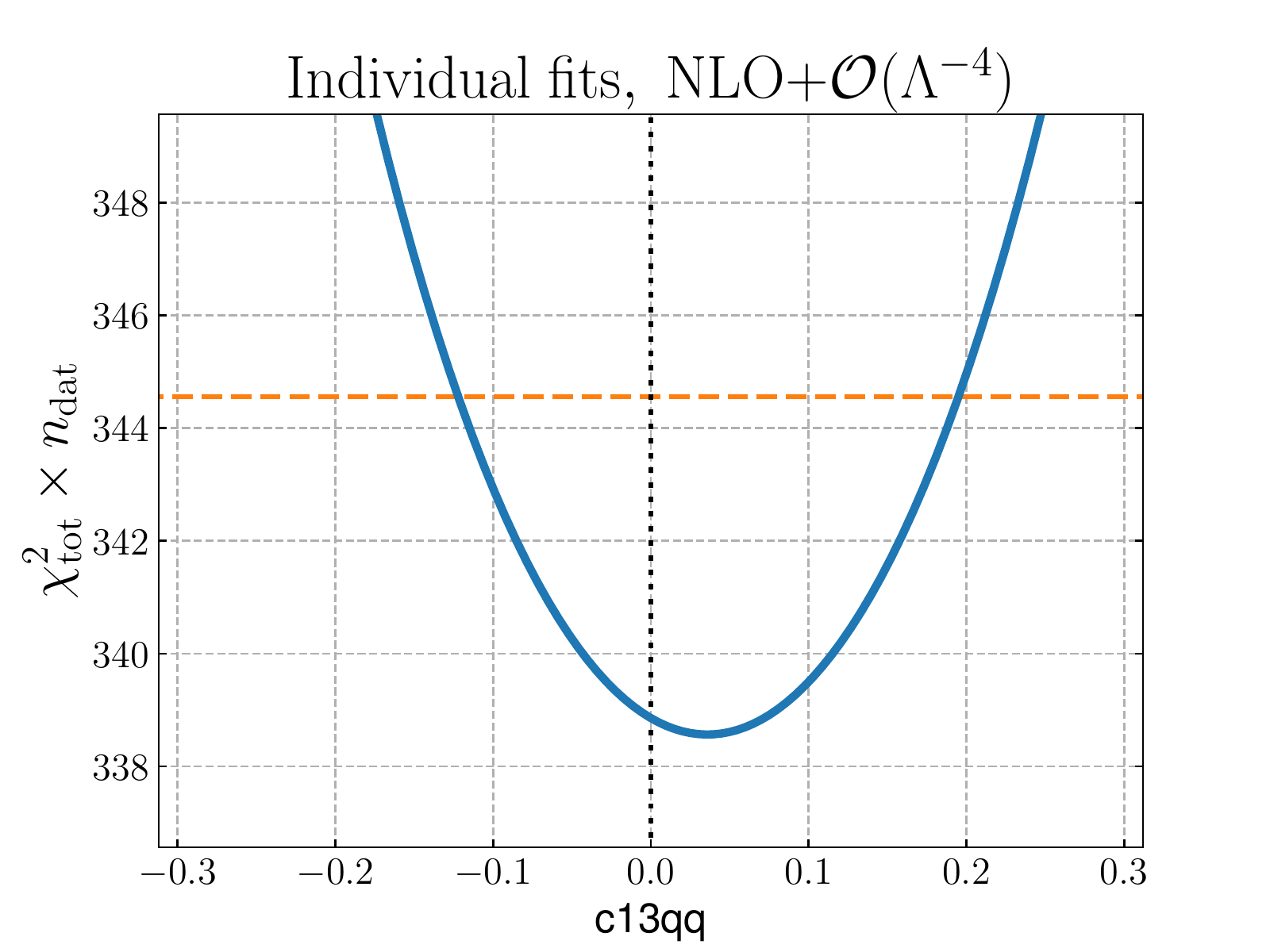}
\includegraphics[width=0.297\linewidth]{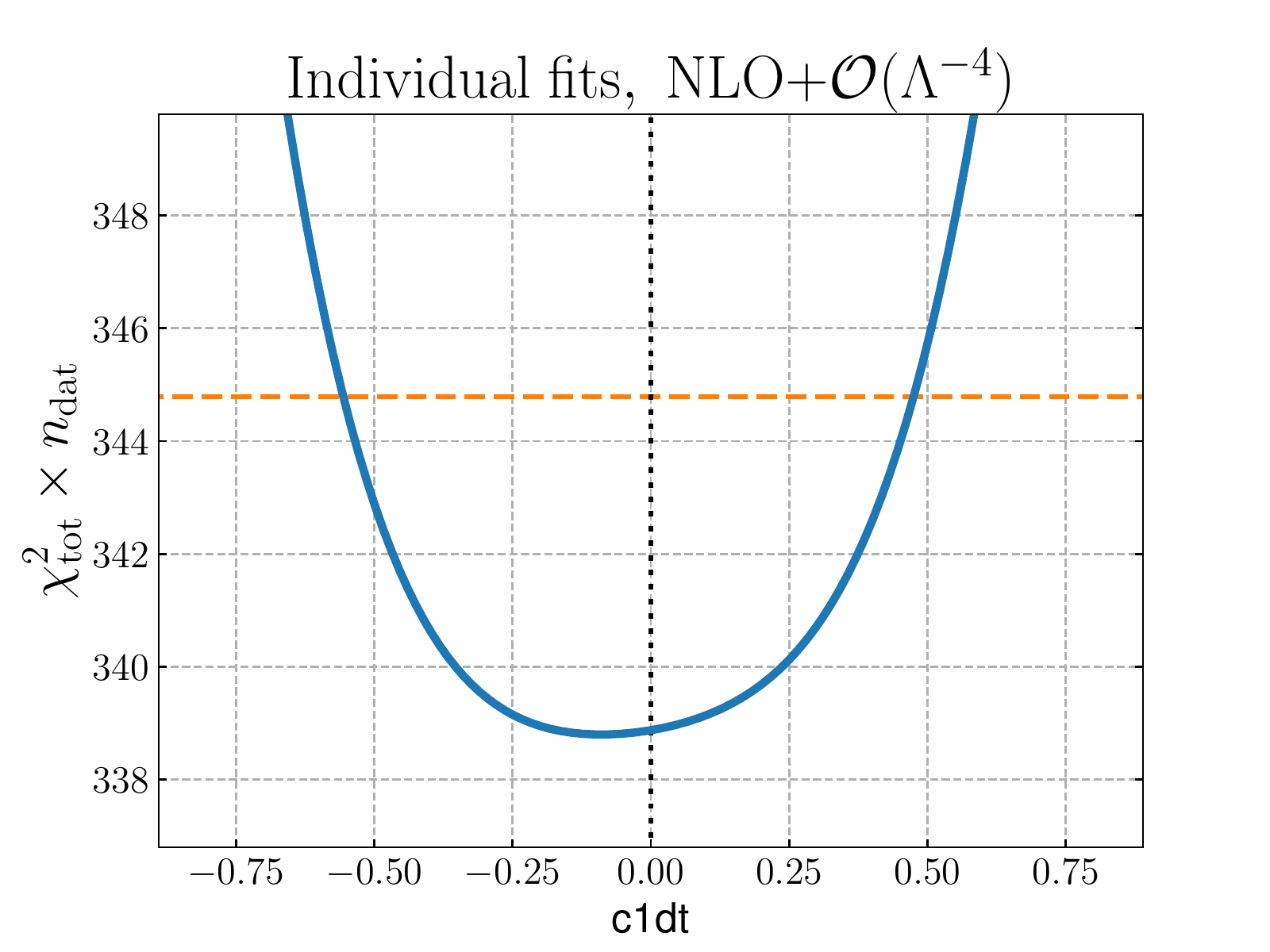}
\includegraphics[width=0.297\linewidth]{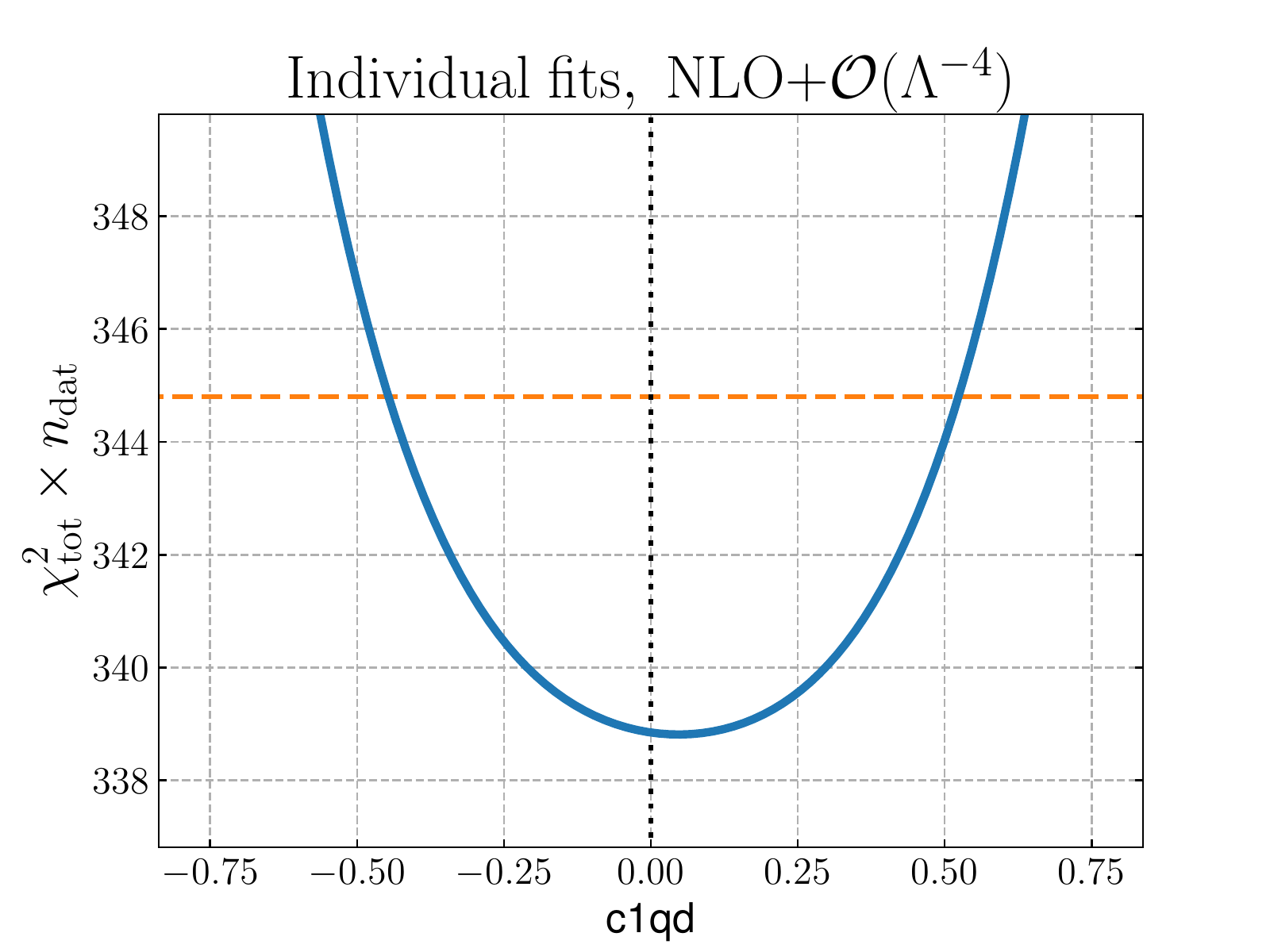}
\includegraphics[width=0.297\linewidth]{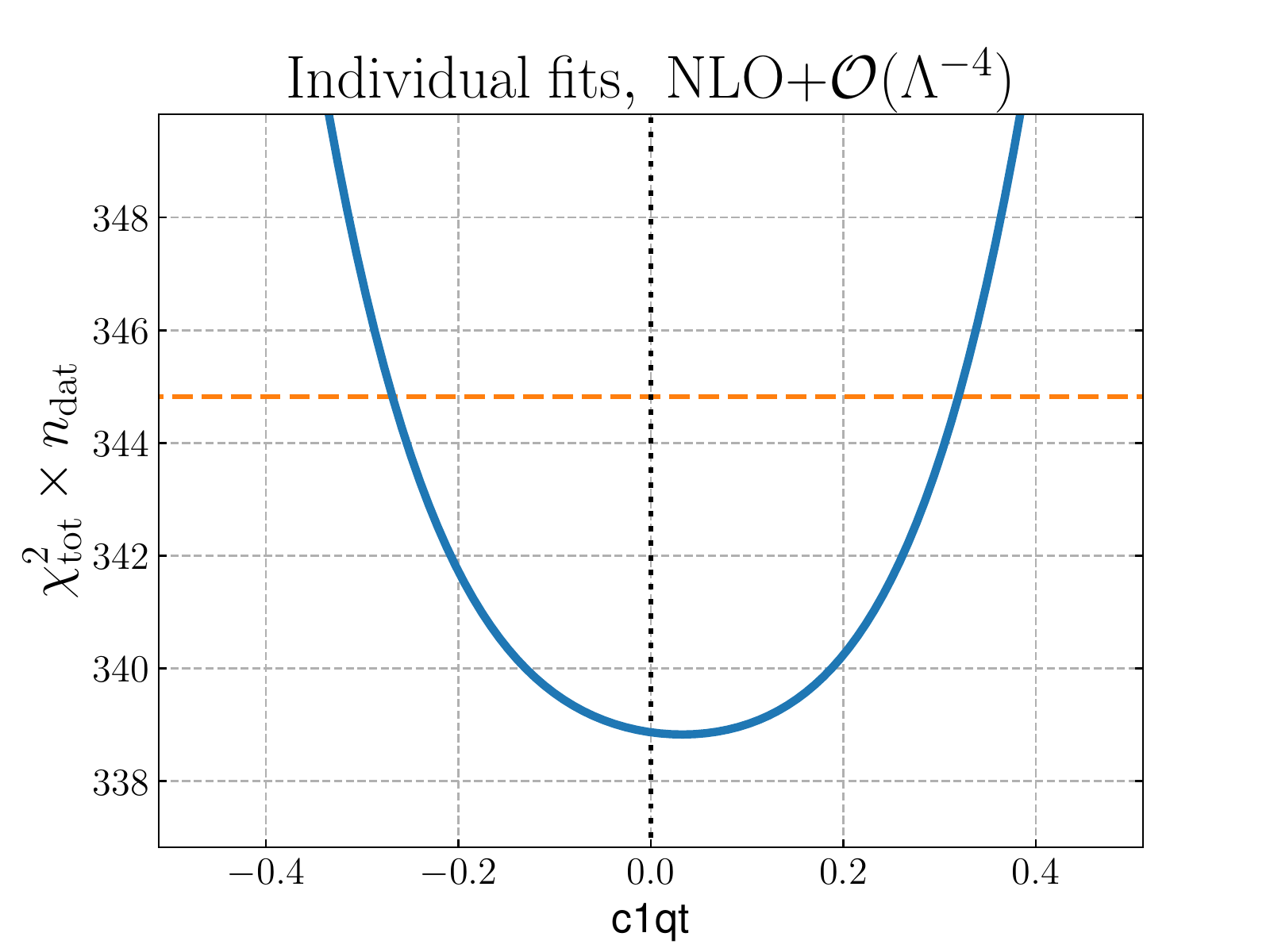}
\includegraphics[width=0.297\linewidth]{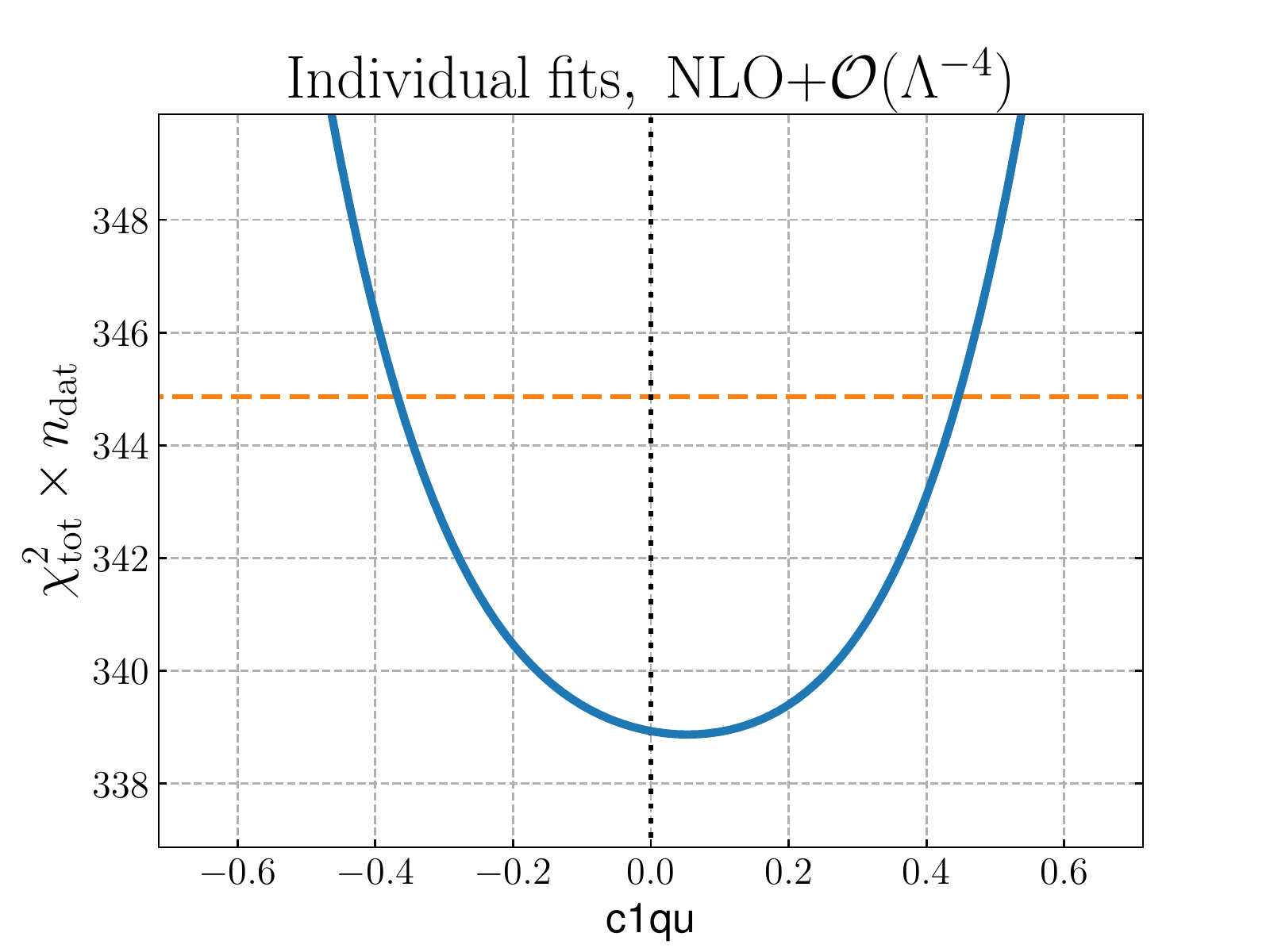}
\includegraphics[width=0.297\linewidth]{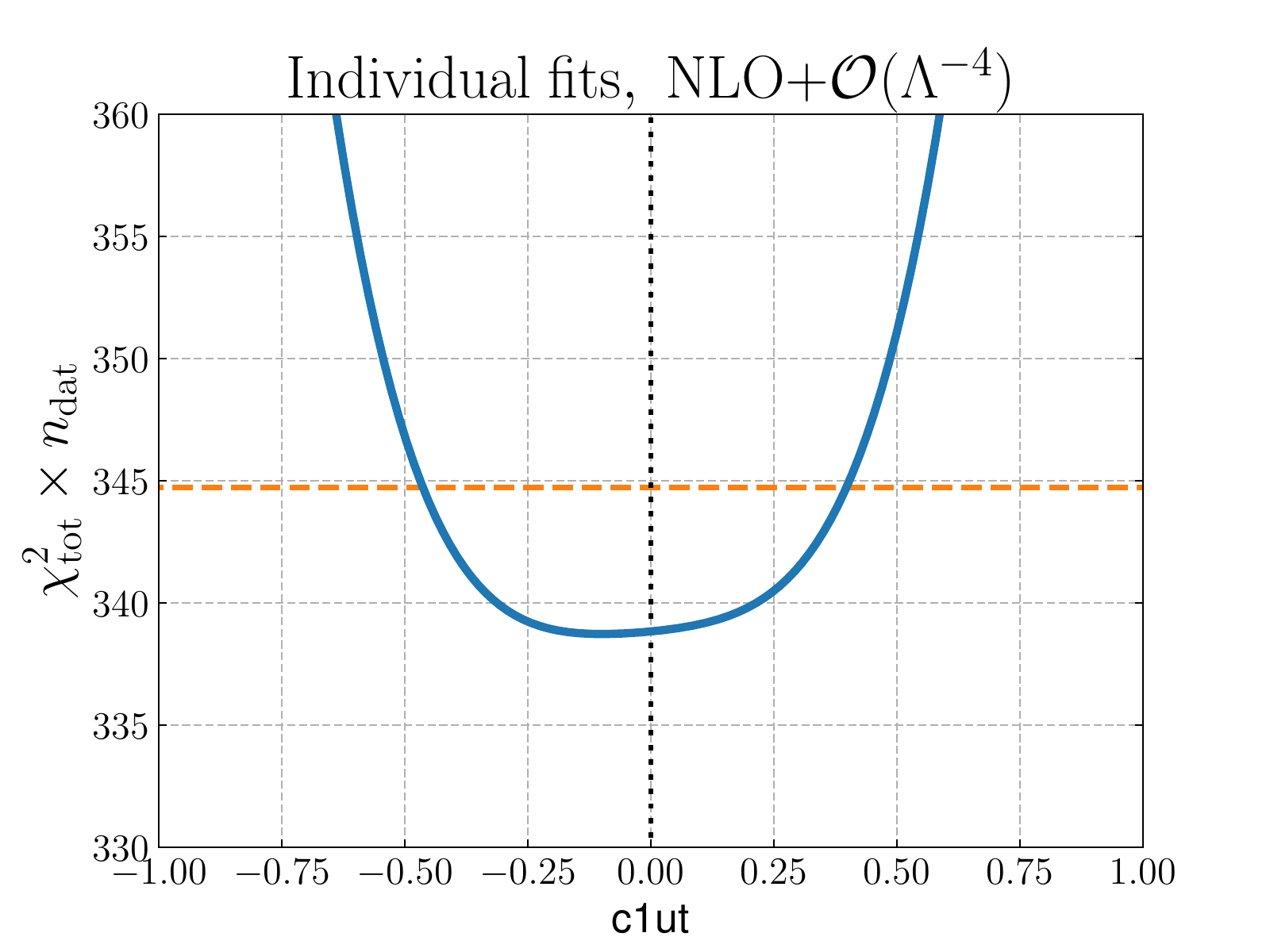}
\includegraphics[width=0.297\linewidth]{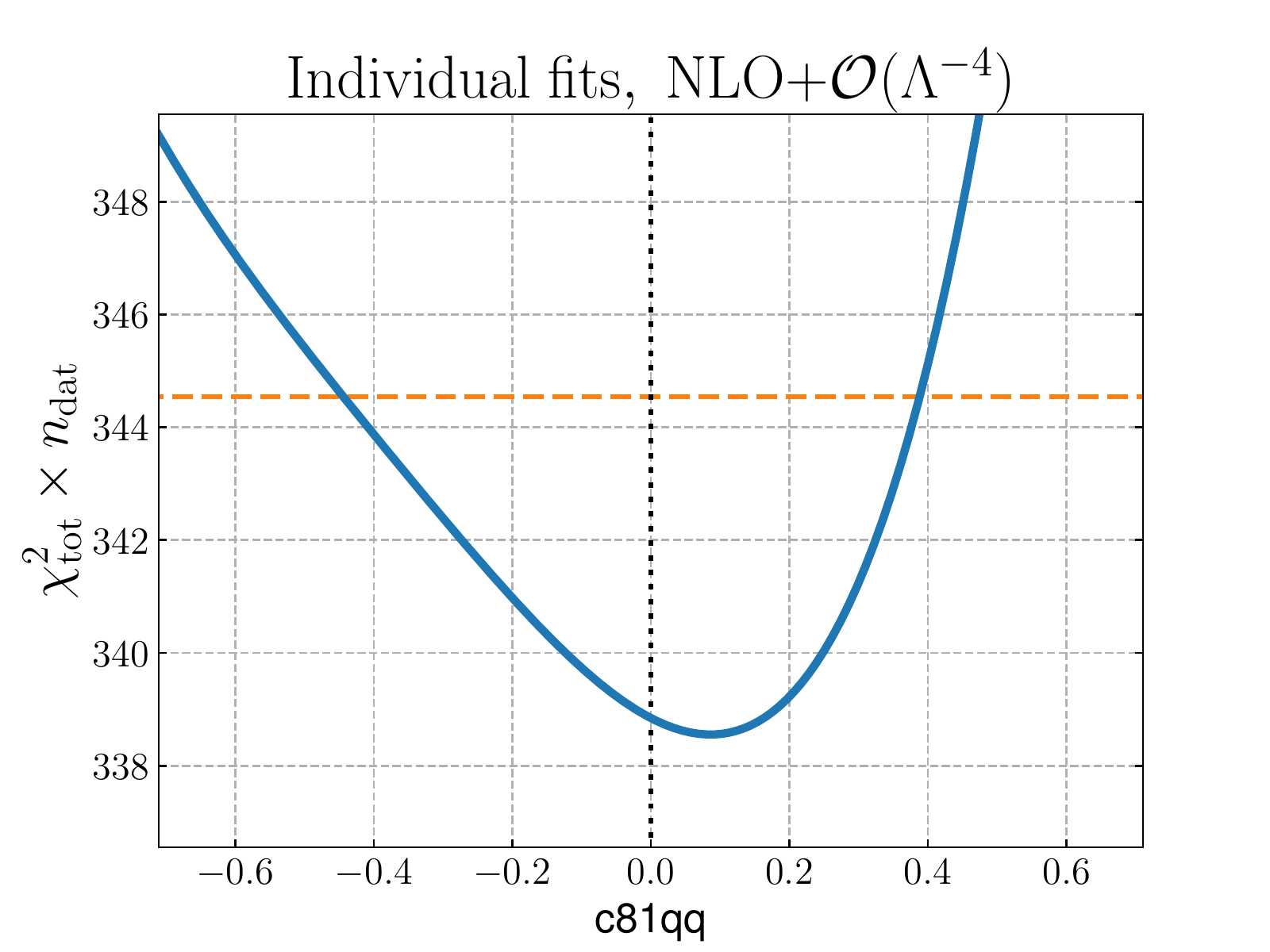}
\includegraphics[width=0.297\linewidth]{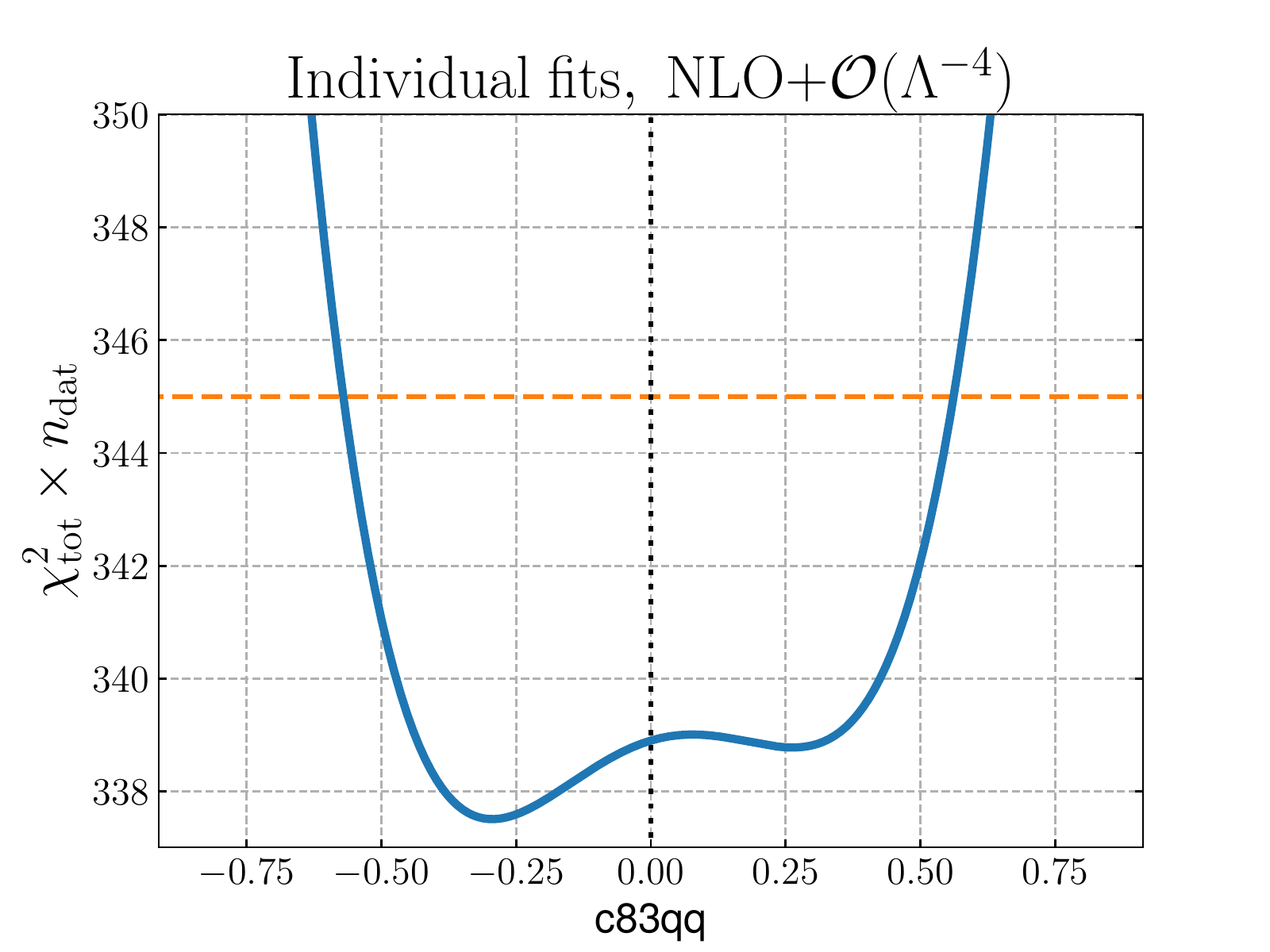}
\includegraphics[width=0.297\linewidth]{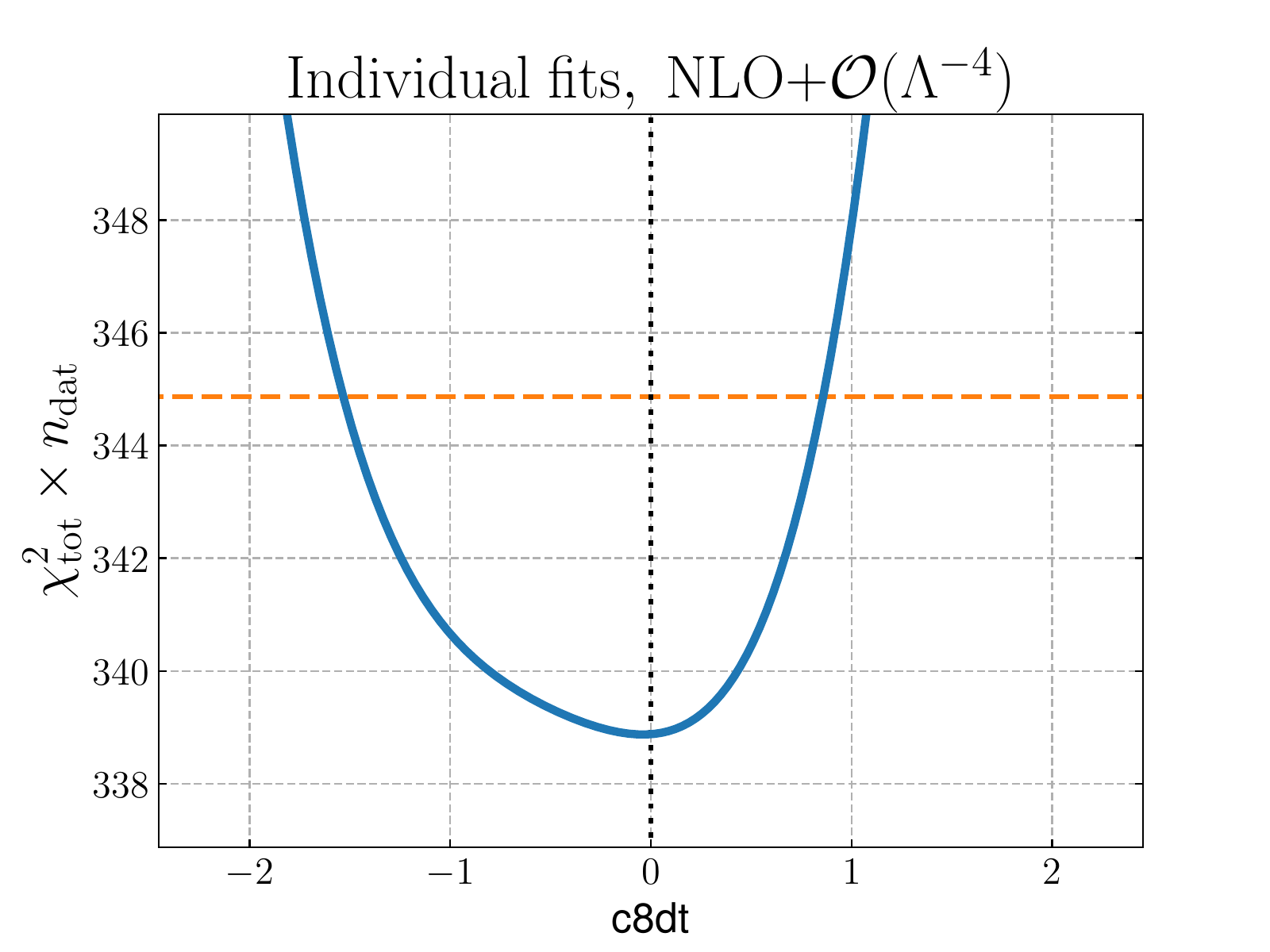}
\includegraphics[width=0.297\linewidth]{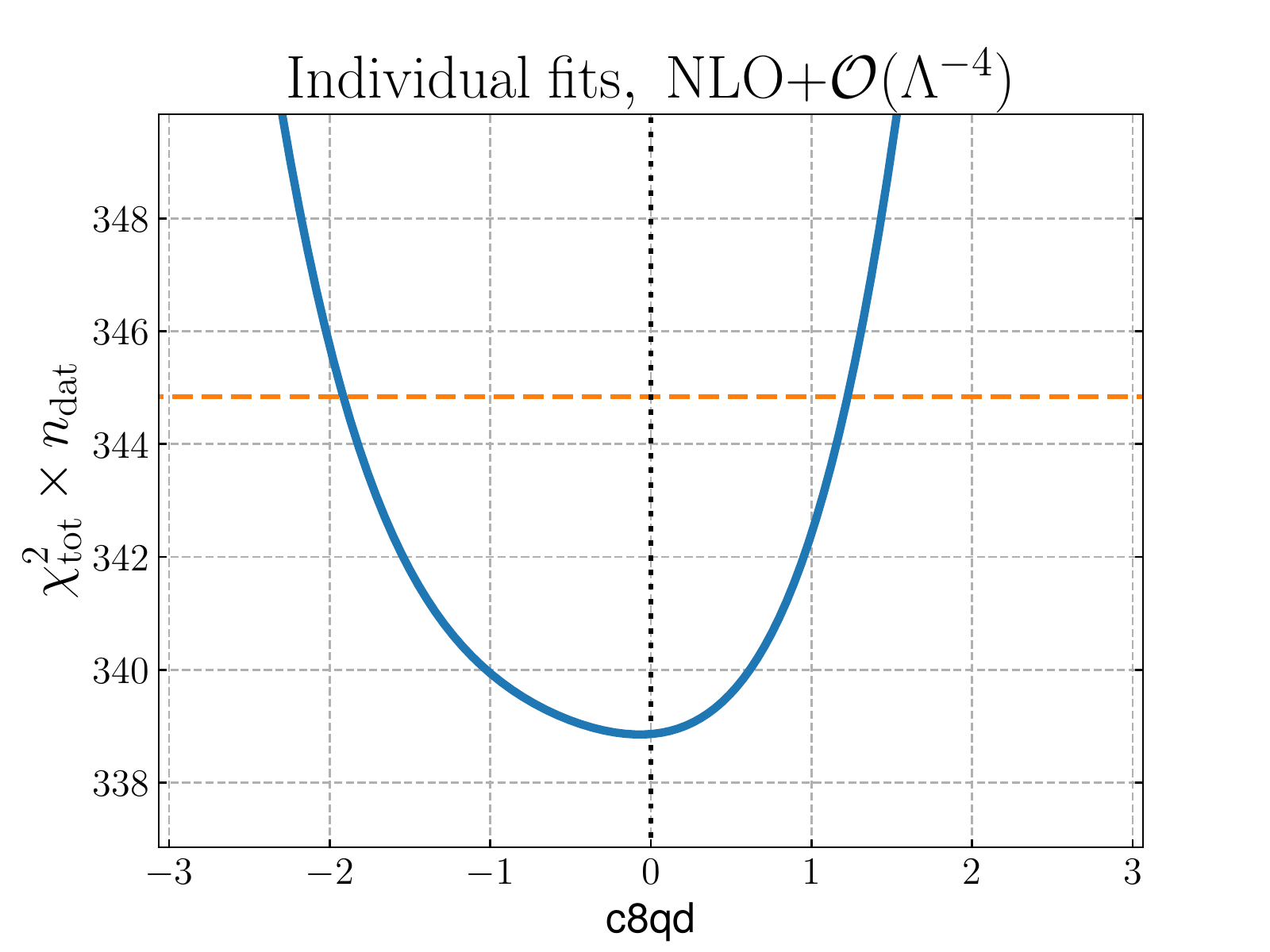}
\includegraphics[width=0.297\linewidth]{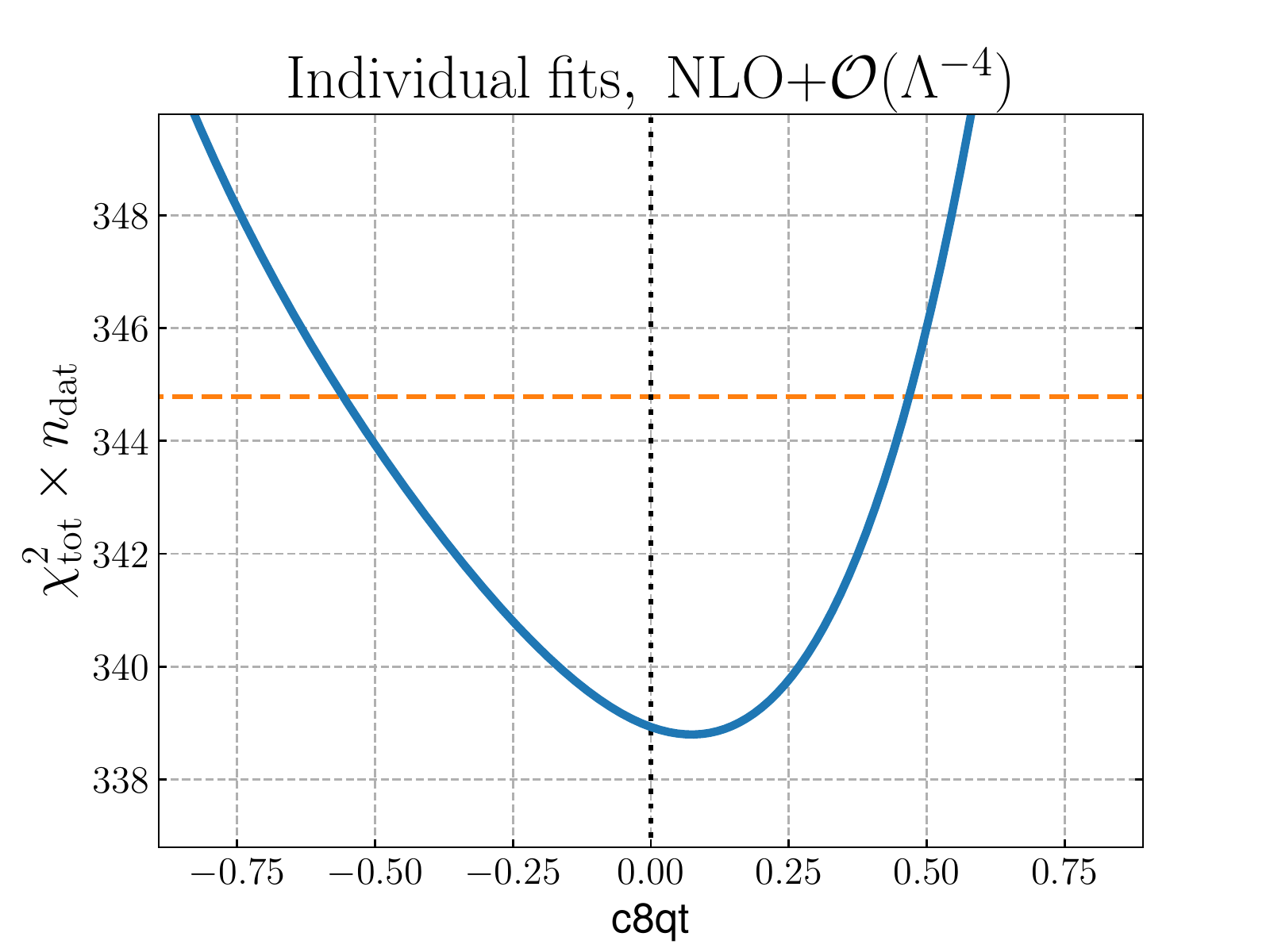}
\includegraphics[width=0.297\linewidth]{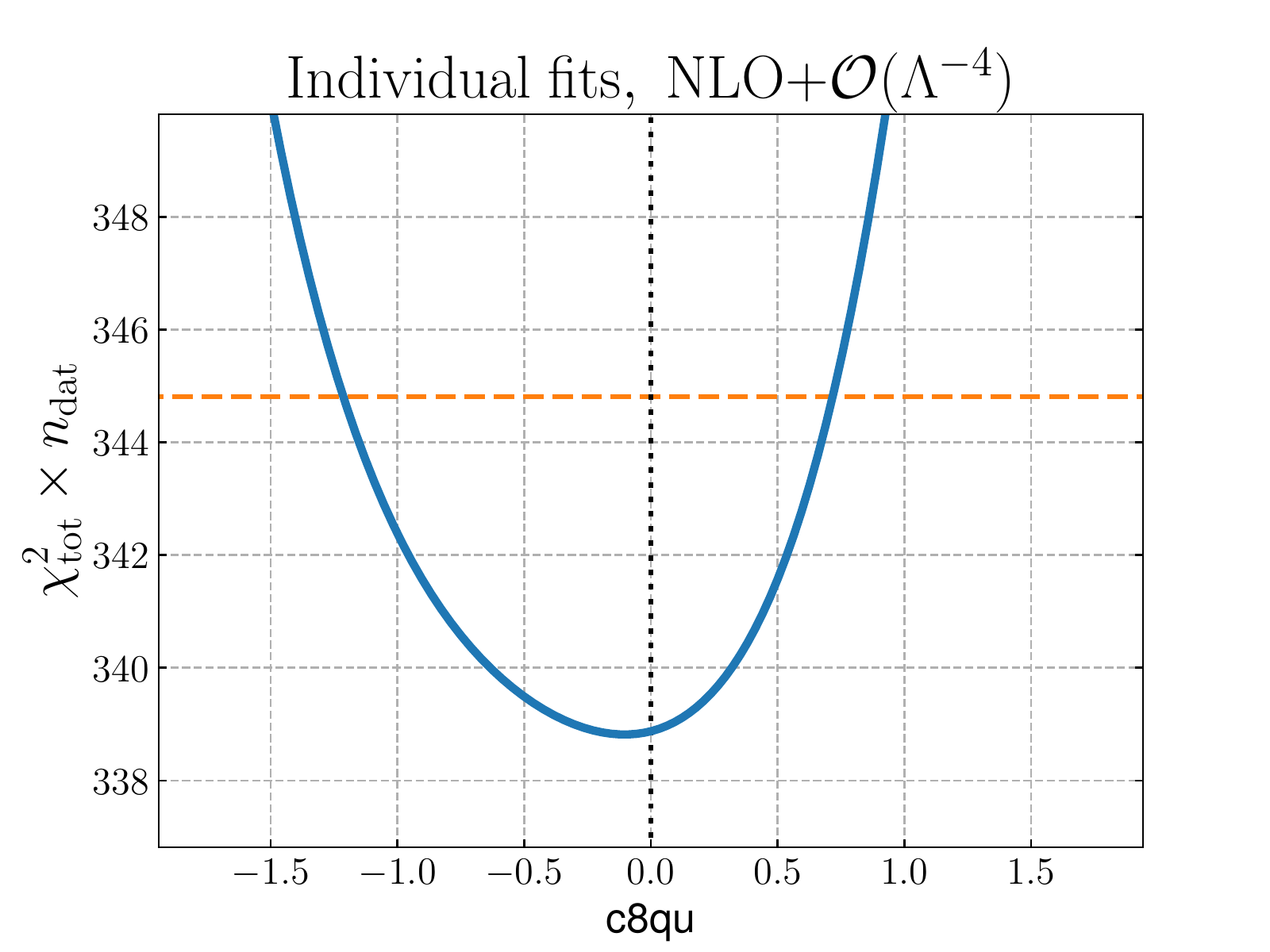}
\caption{The log-likelihood profiles for the individual fits, taking
into account the global dataset. The orange dashed line represents the $95\%$ CL interval.
     \label{fig:quartic-individual-fits} }
  \end{center}
\end{figure}

\begin{figure}[h!]
  \begin{center}
\includegraphics[width=0.297\linewidth]{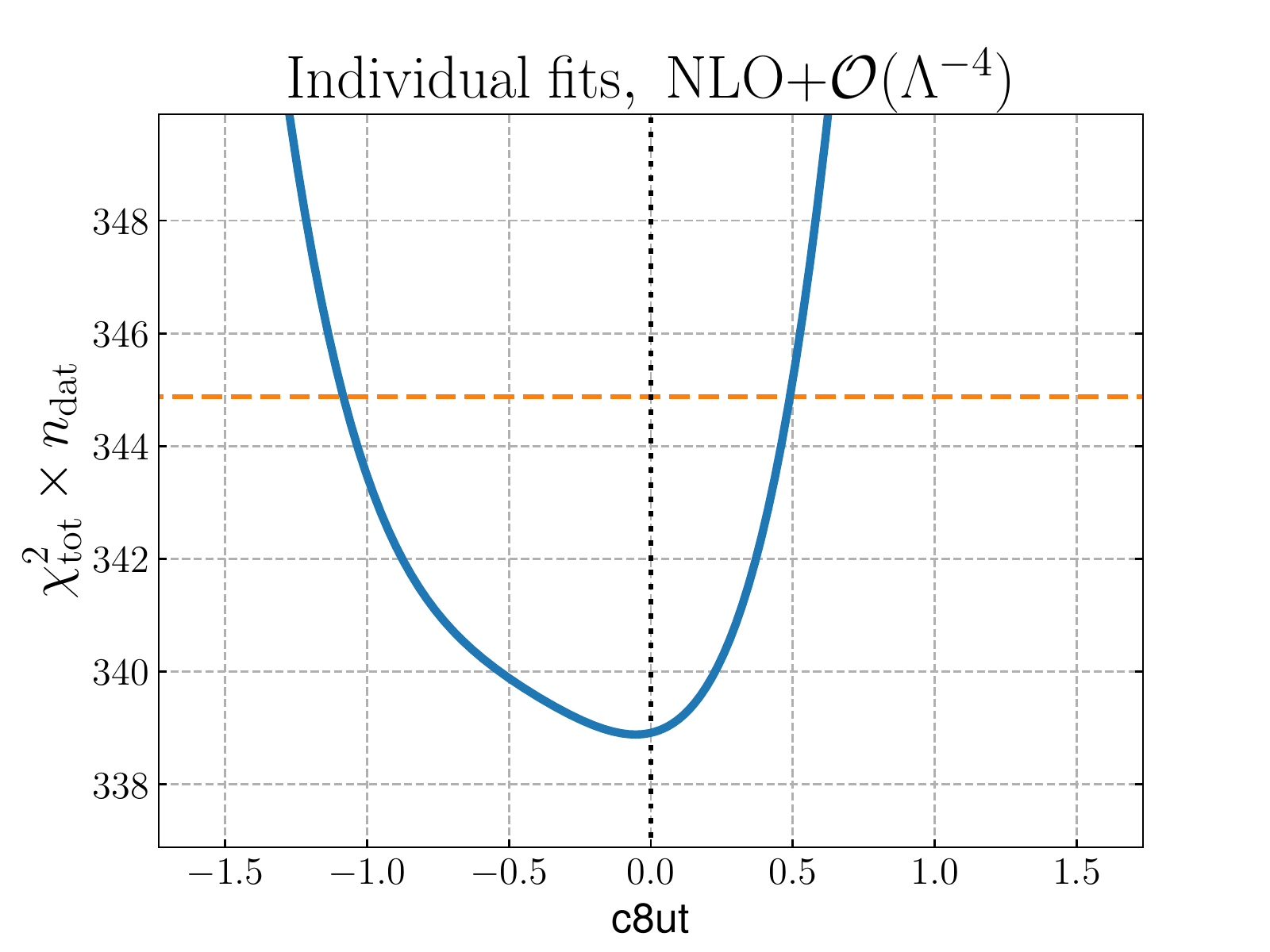}
\includegraphics[width=0.30\linewidth]{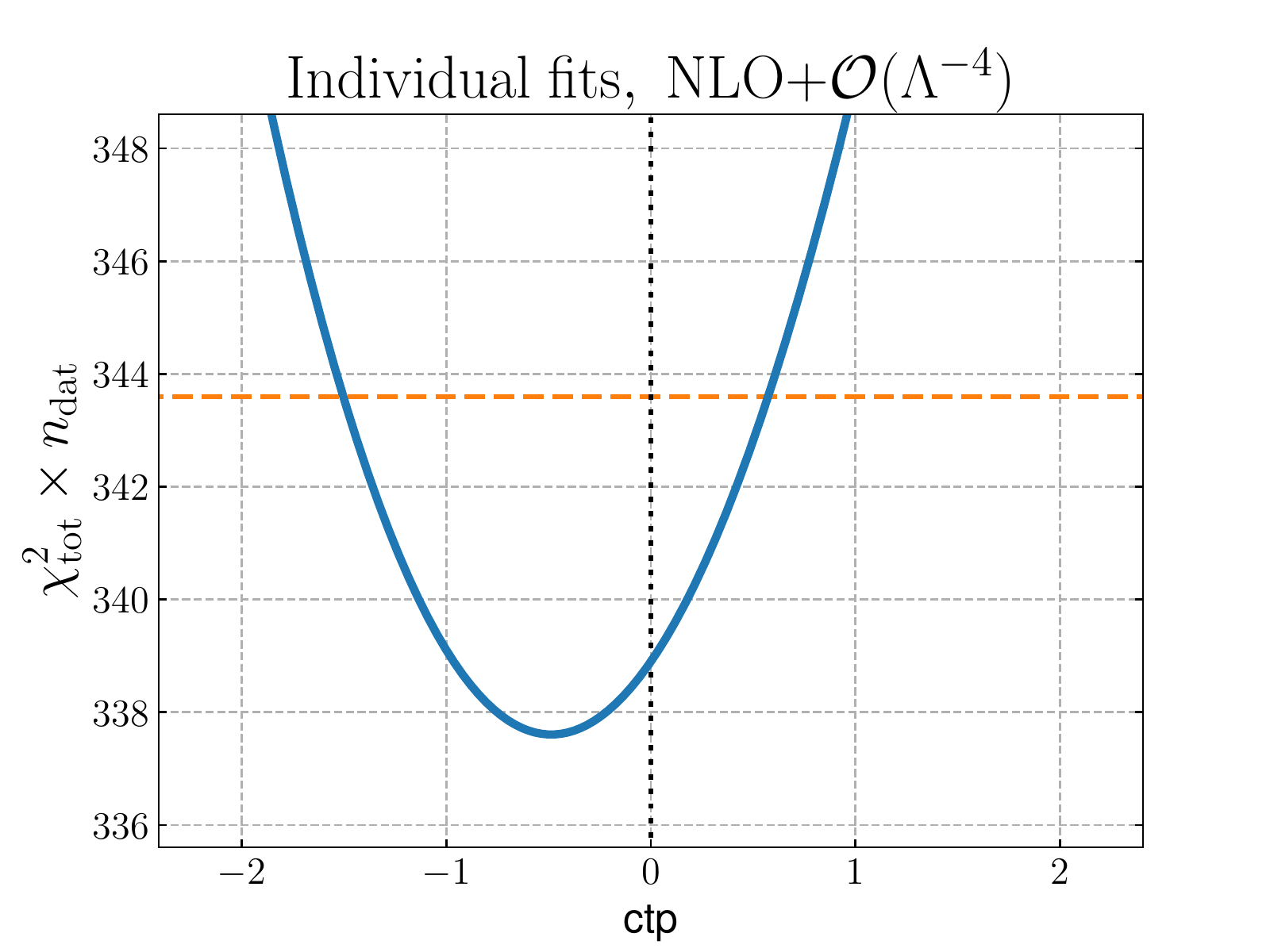}
\includegraphics[width=0.30\linewidth]{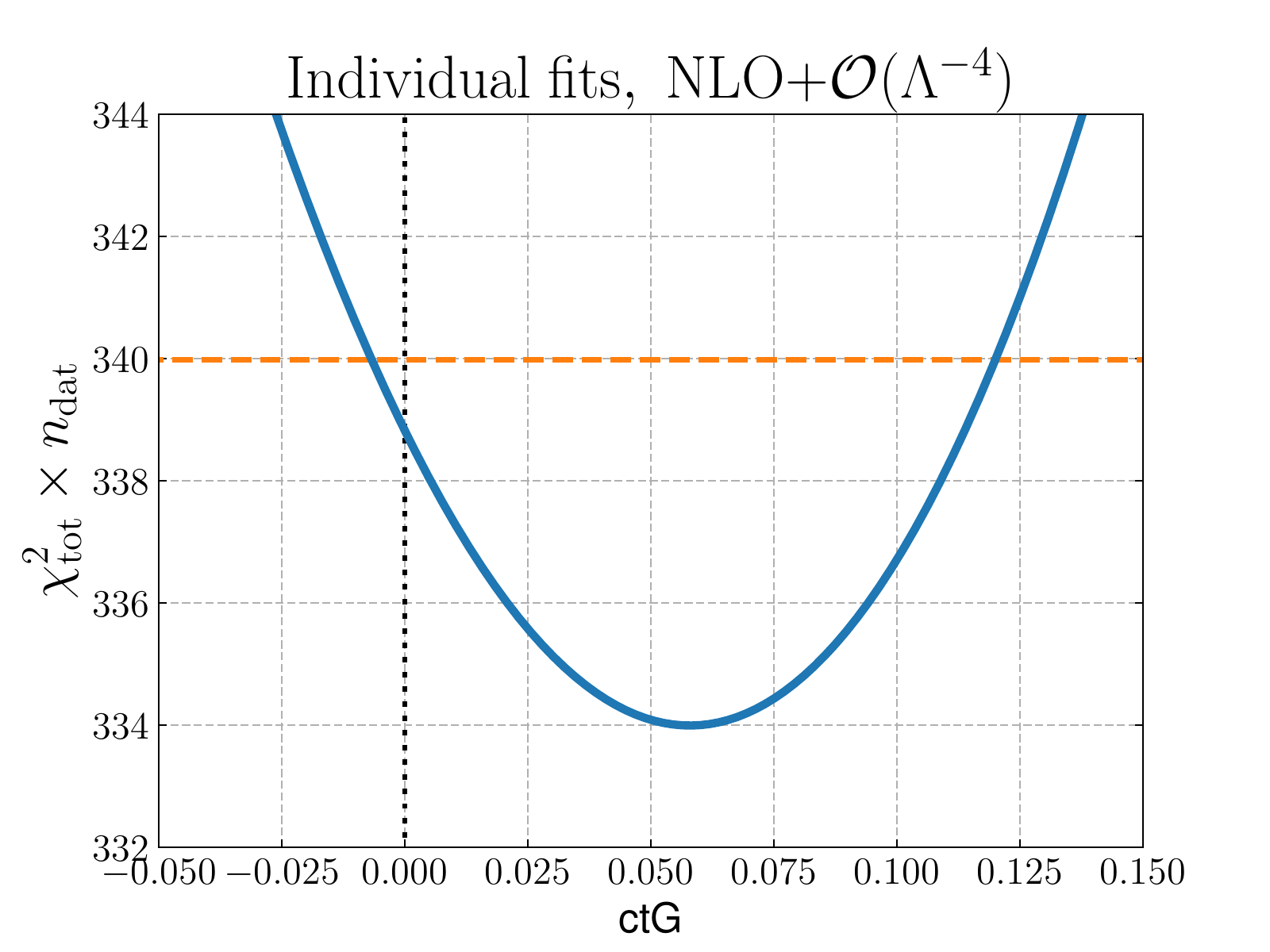}
\includegraphics[width=0.30\linewidth]{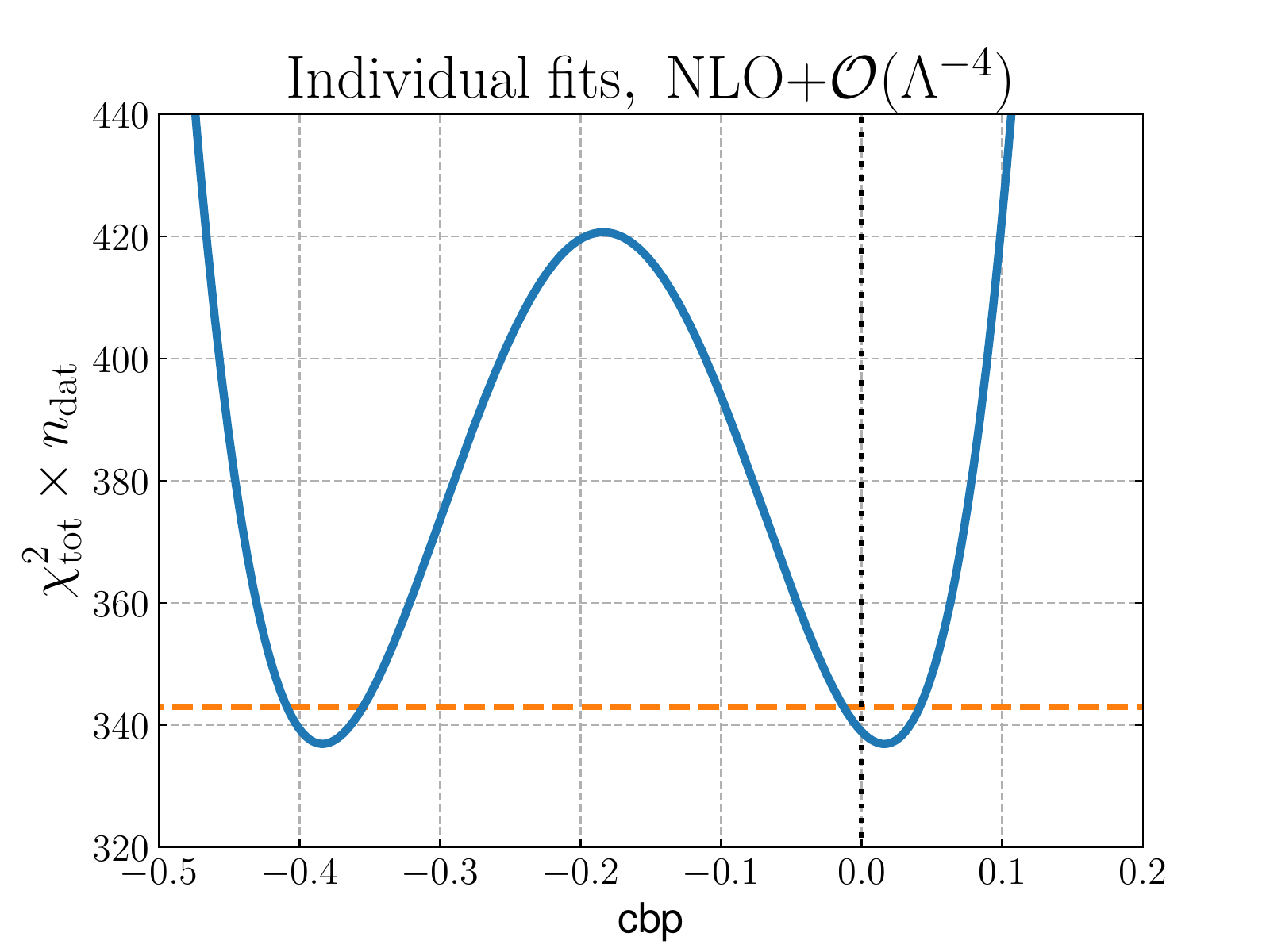}
\includegraphics[width=0.30\linewidth]{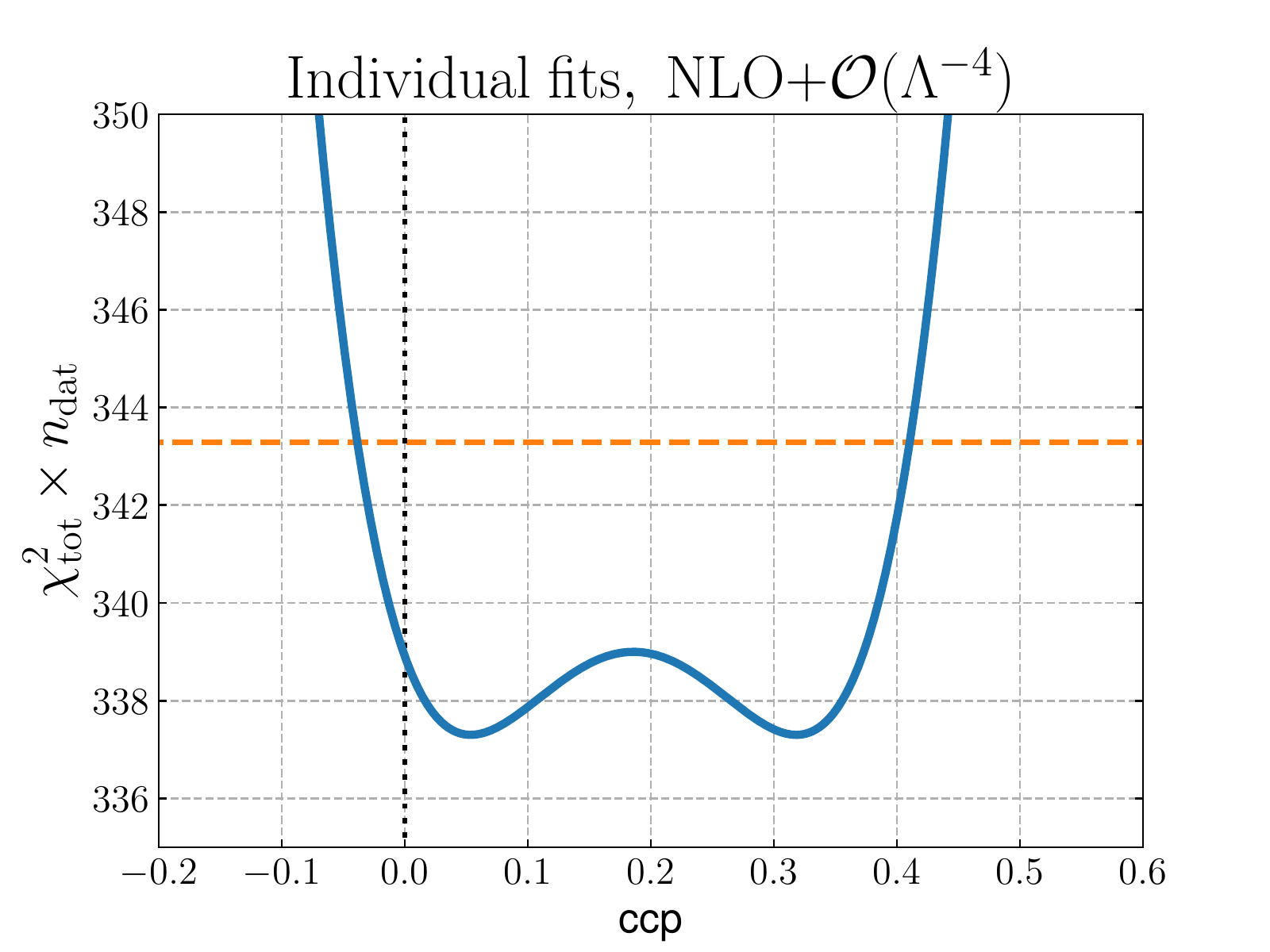}
\includegraphics[width=0.30\linewidth]{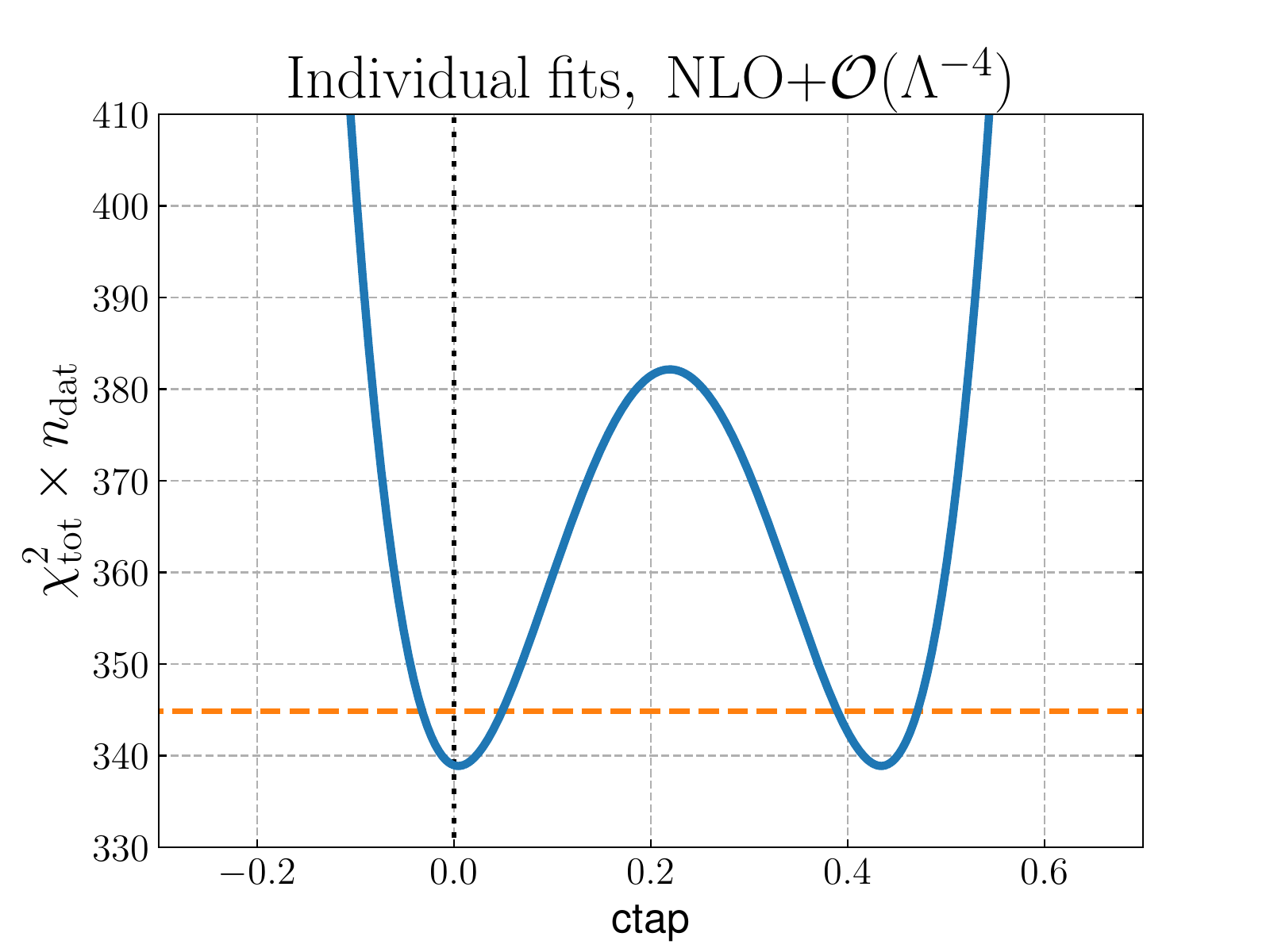}
\includegraphics[width=0.30\linewidth]{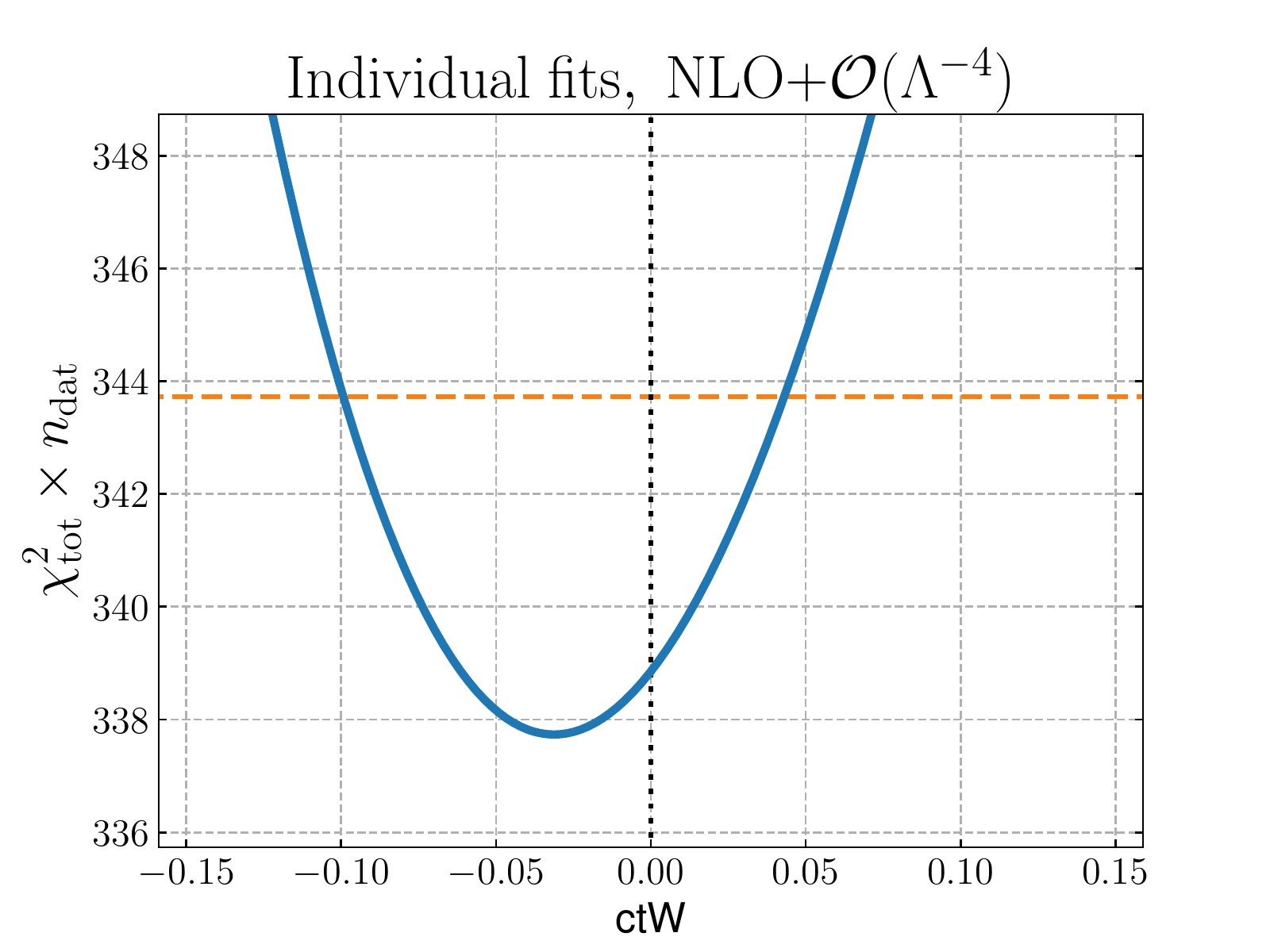}
\includegraphics[width=0.30\linewidth]{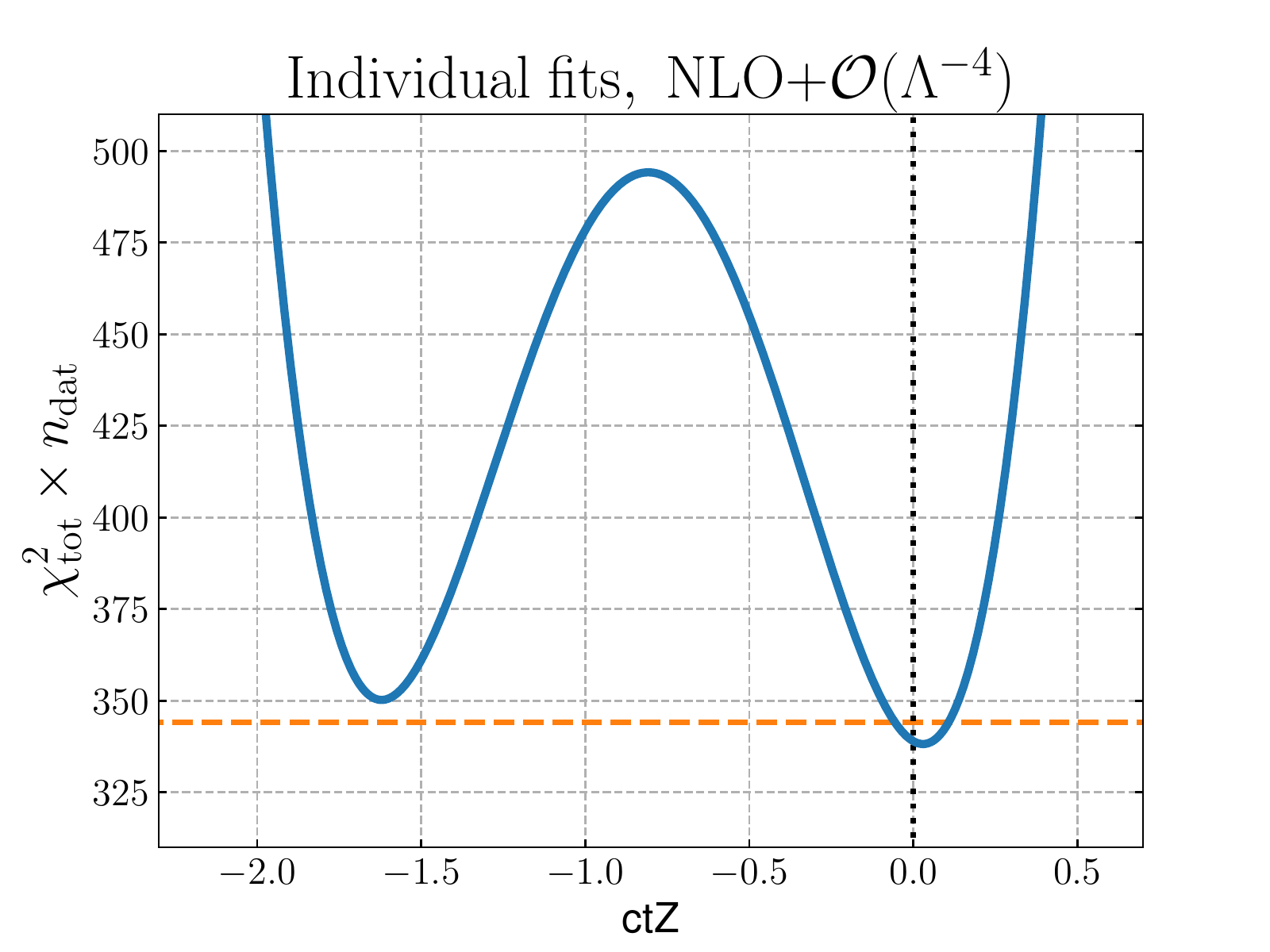}
\includegraphics[width=0.30\linewidth]{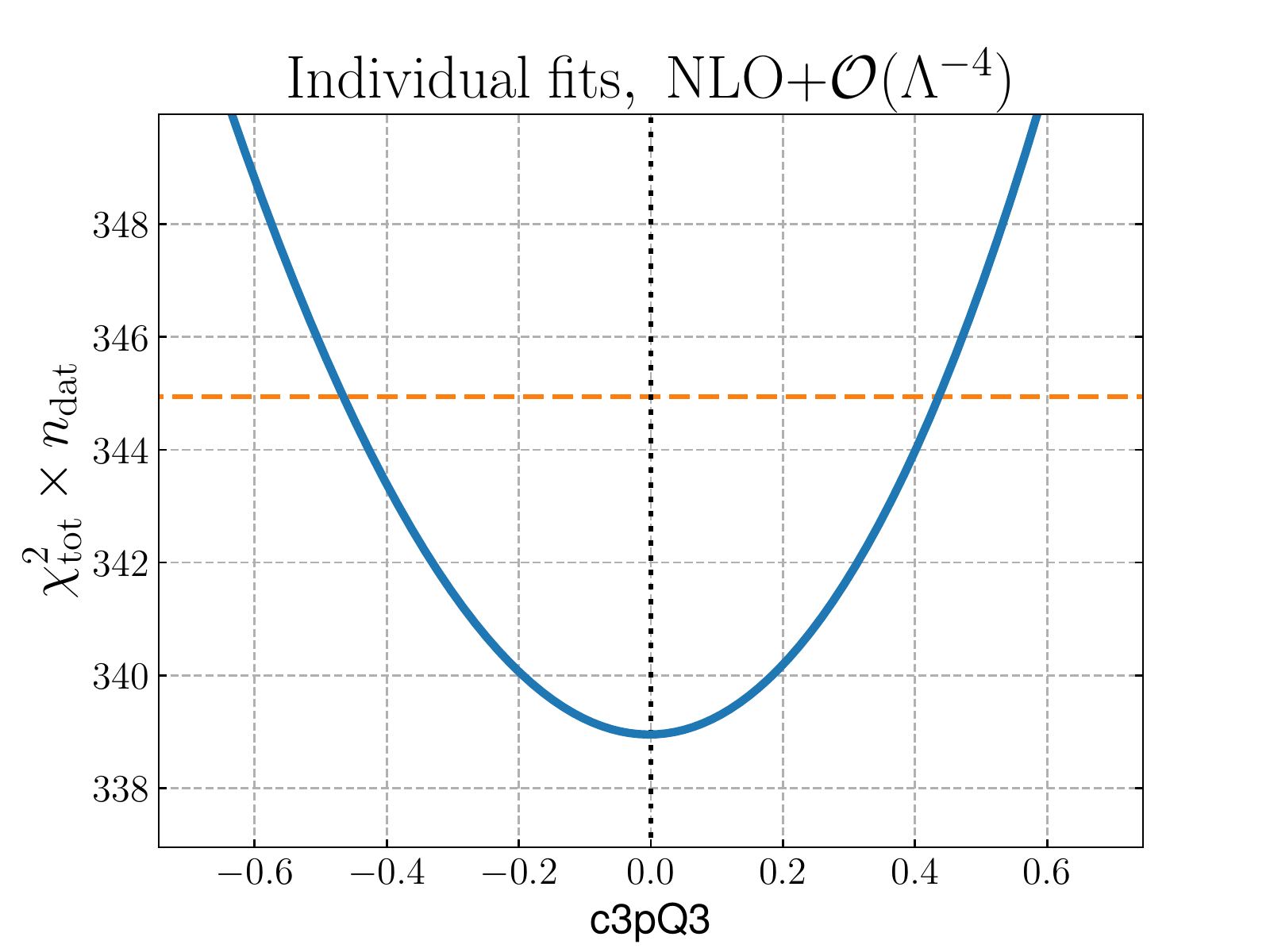}
\includegraphics[width=0.30\linewidth]{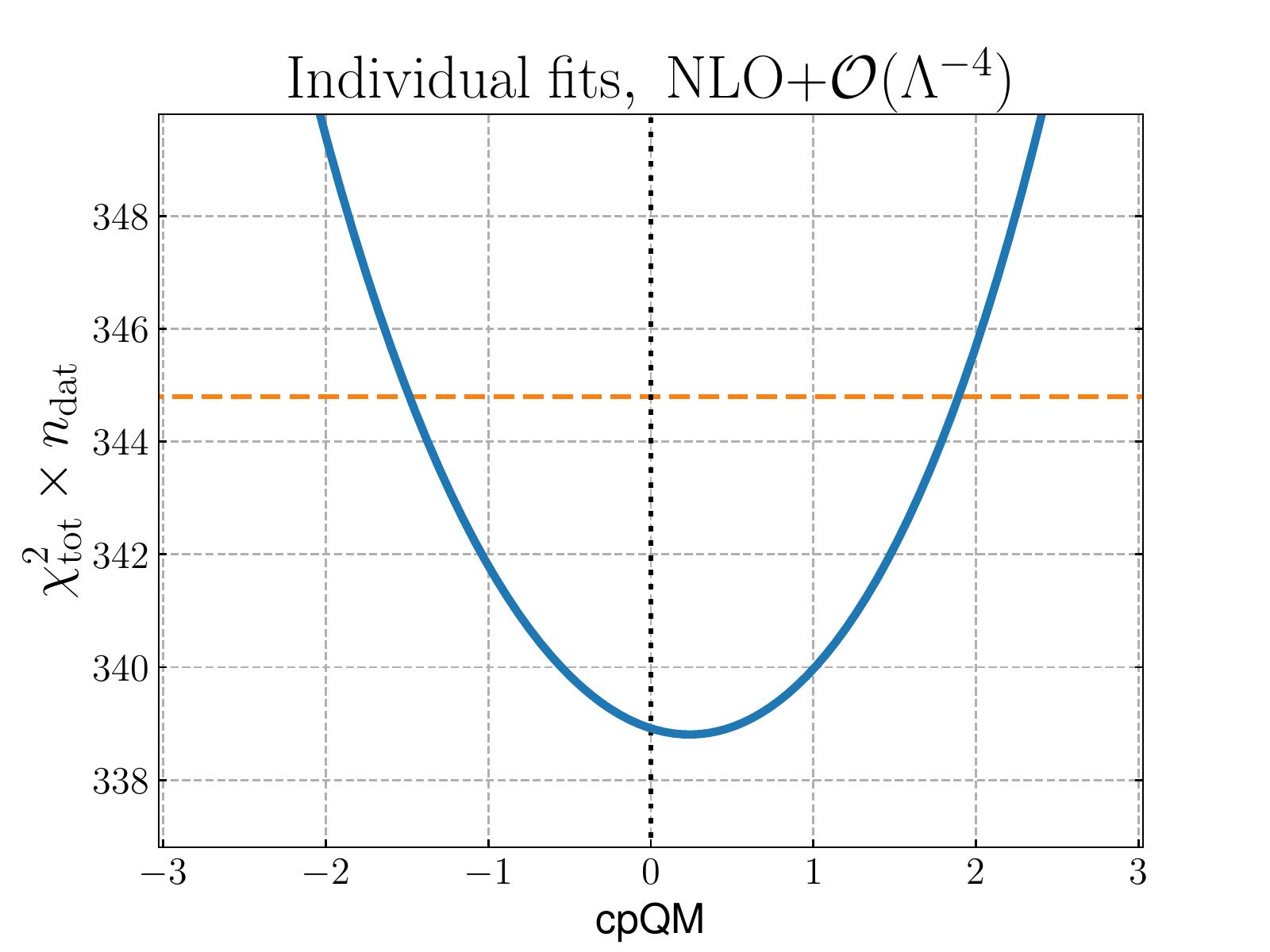}
\includegraphics[width=0.30\linewidth]{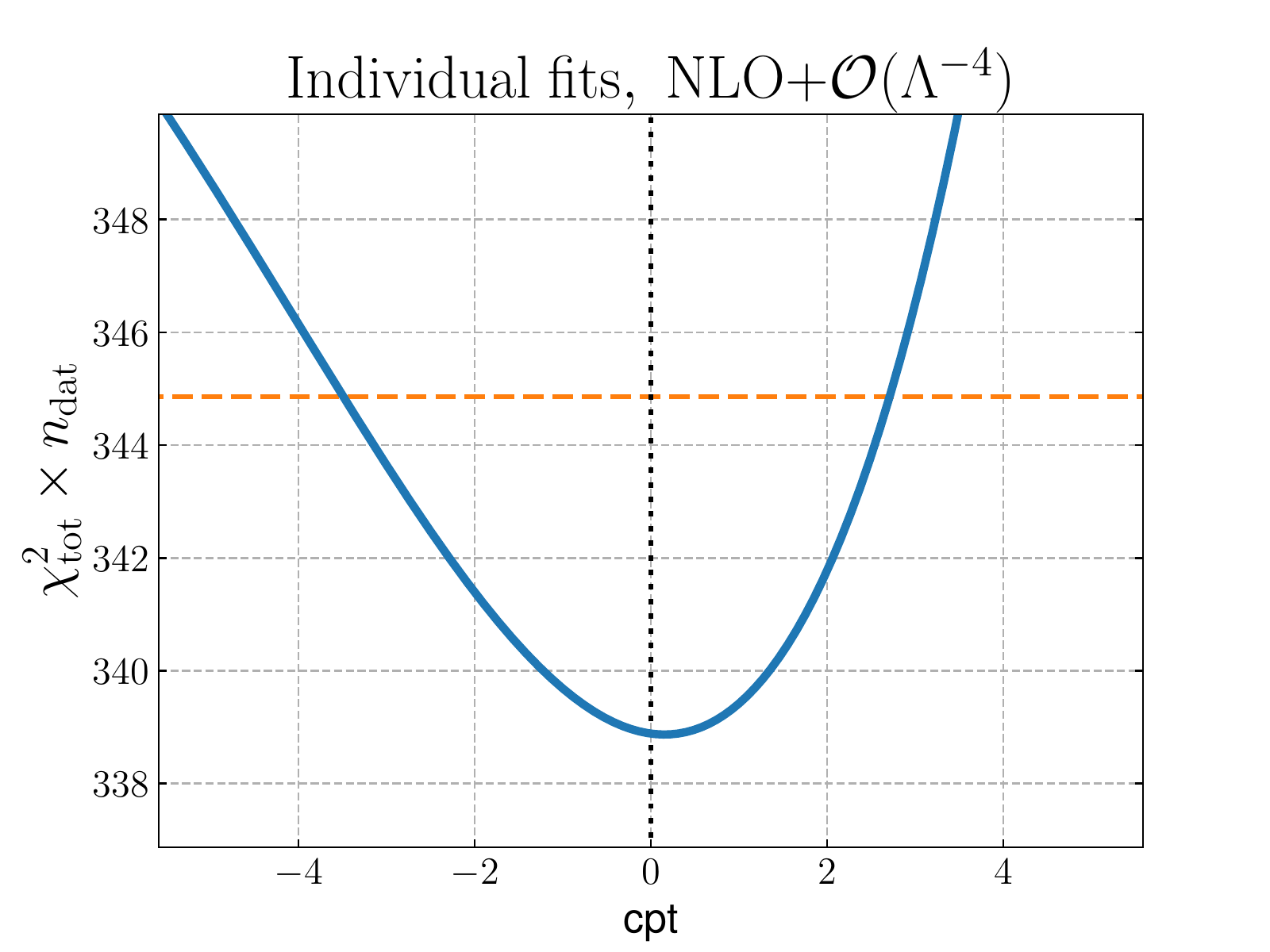}
\includegraphics[width=0.30\linewidth]{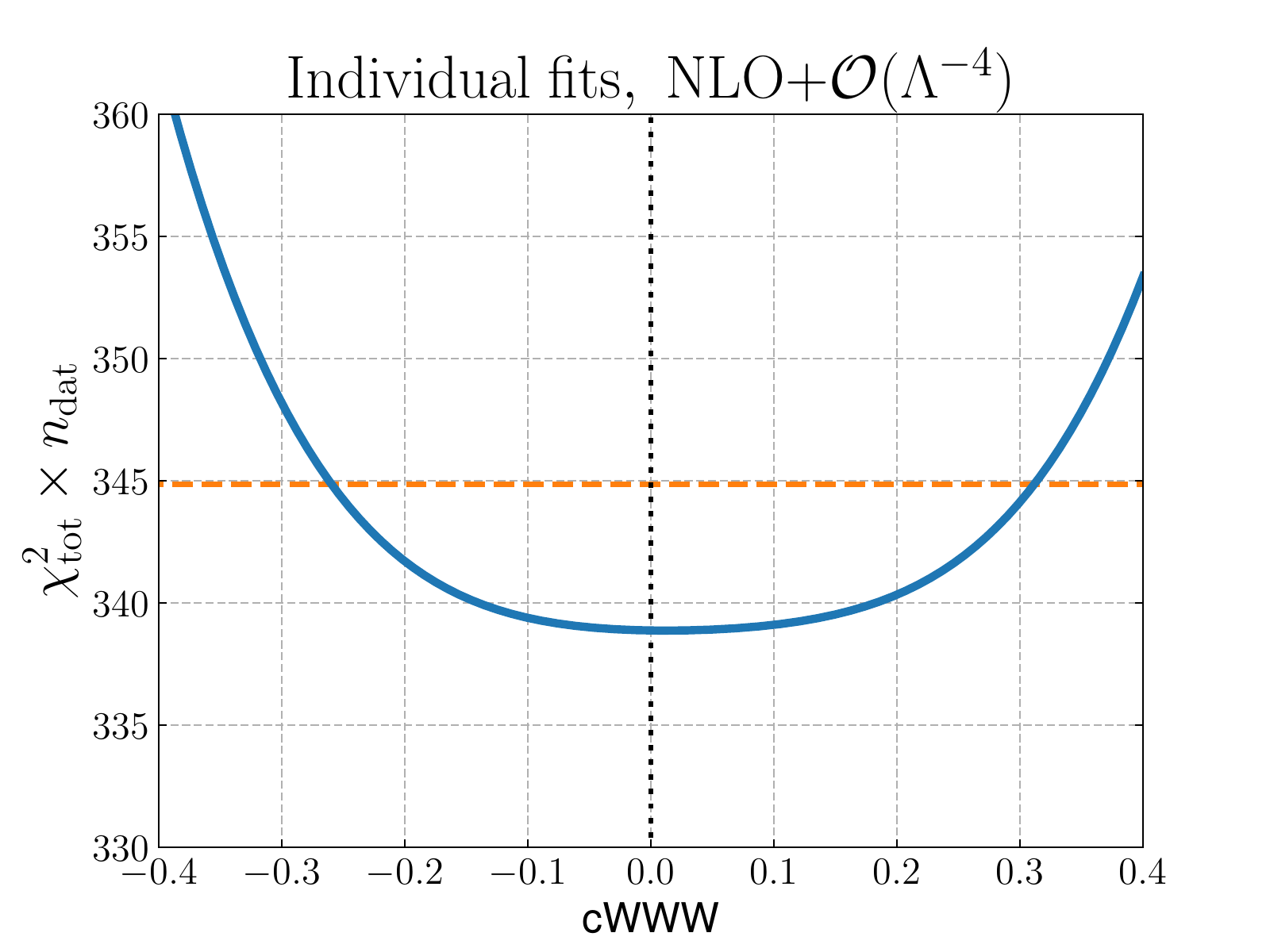}
\includegraphics[width=0.30\linewidth]{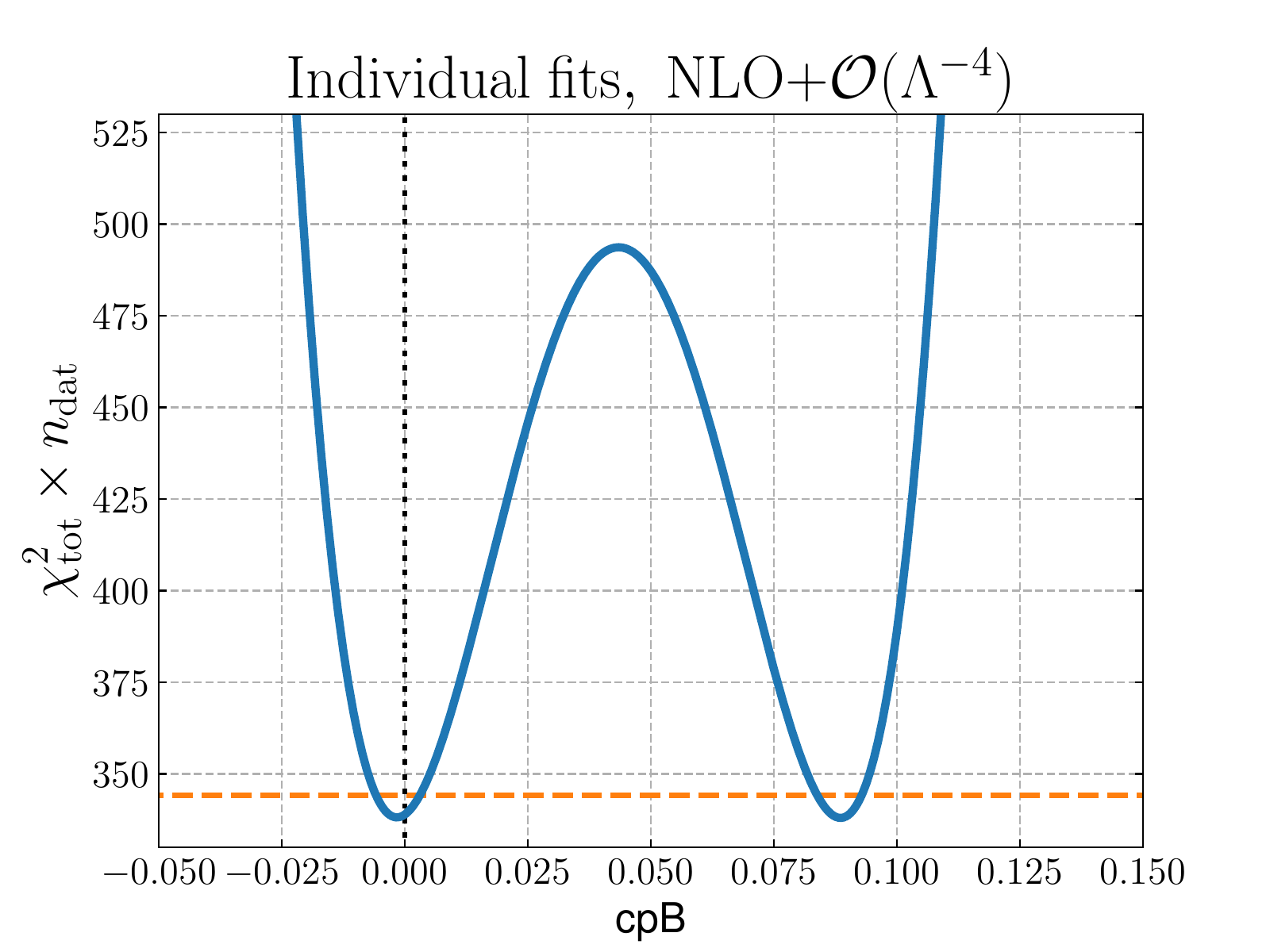}
\includegraphics[width=0.30\linewidth]{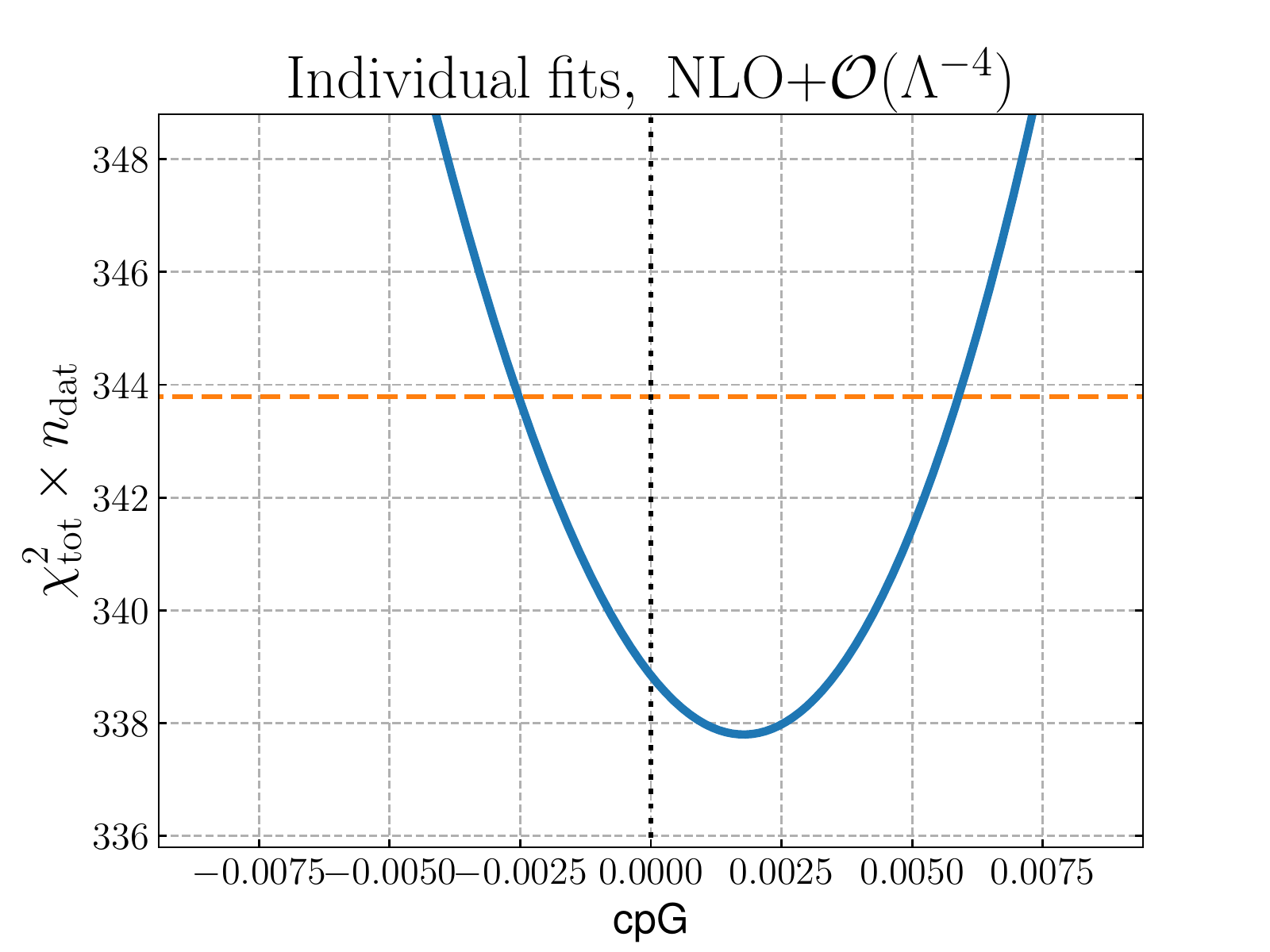}
\includegraphics[width=0.30\linewidth]{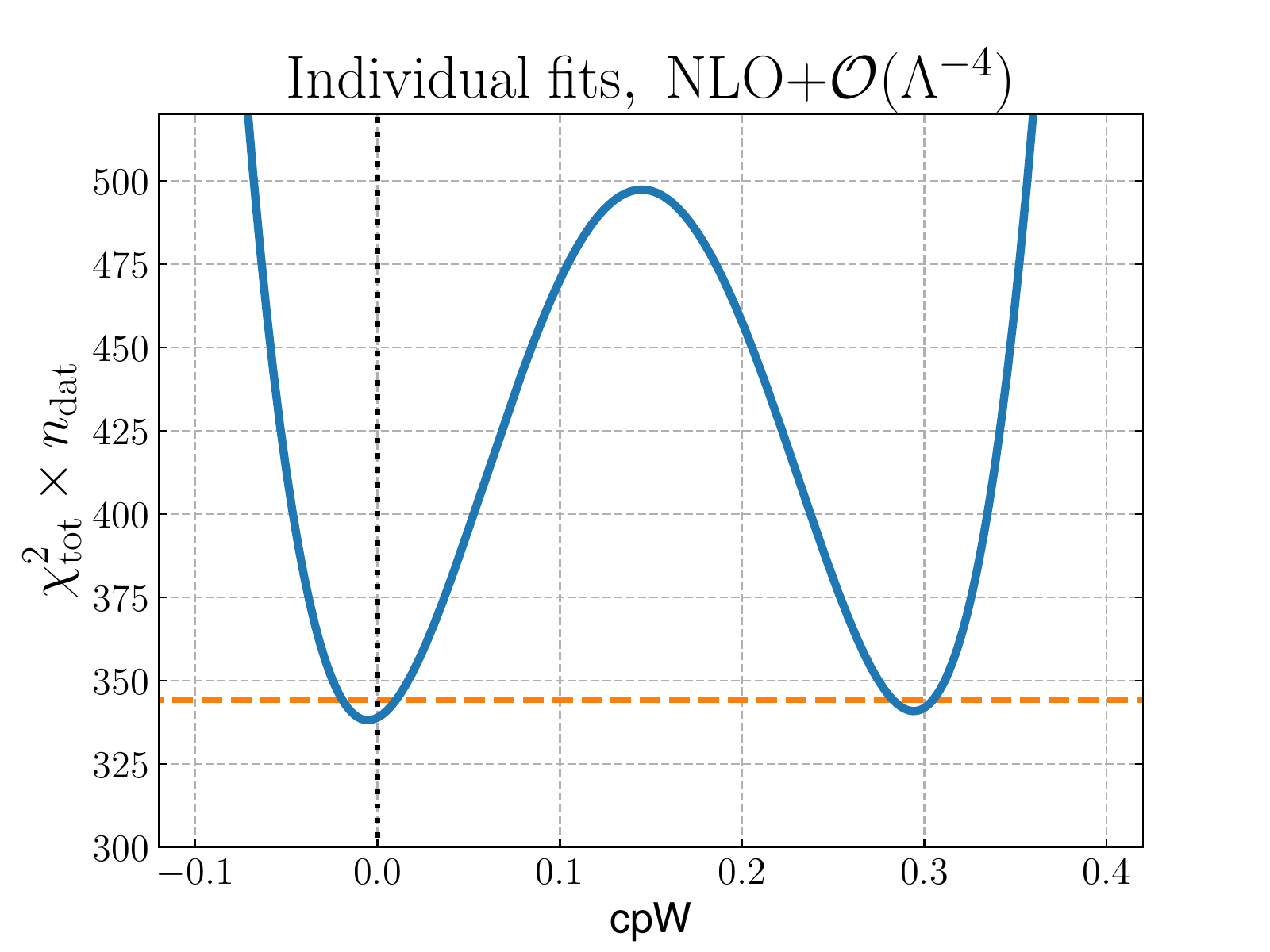}
\includegraphics[width=0.30\linewidth]{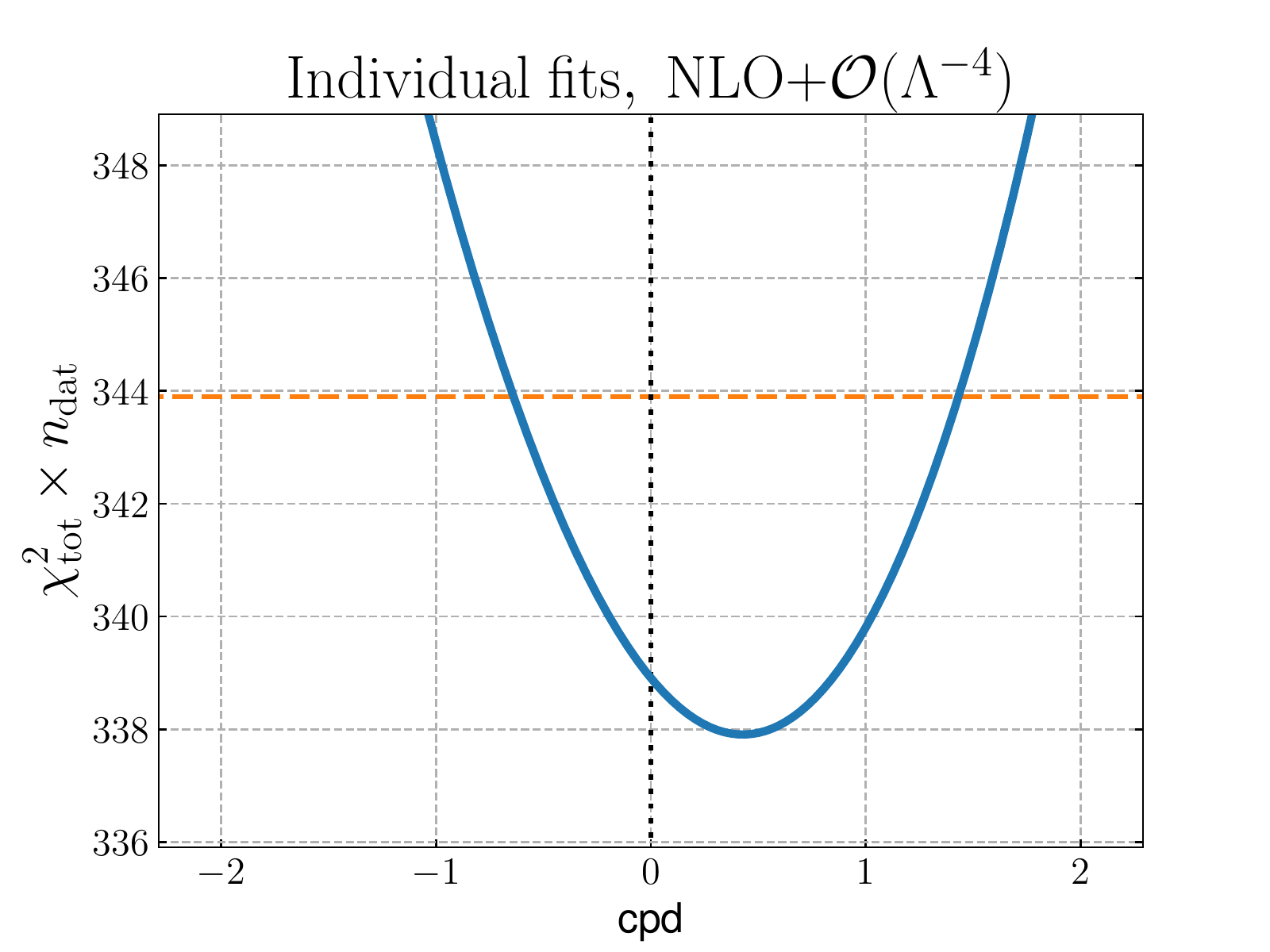}
\caption{\small Fig.~\ref{fig:quartic-individual-fits} continued.
     \label{fig:quartic-individual-fits-2} }
  \end{center}
\end{figure}
It is worth pointing out that if the quadratic corrections are included and sizable, there could be more than one minimum
and the CL interval might be given by two disjoint intervals. This can be observed in Fig.~~\ref{fig:quartic-individual-fits} and~\ref{fig:quartic-individual-fits-2},
where the $\chi^2$ profiles for the individual operators are presented. These profiles are obtained assuming that the SM is the underlying theory.
While many operators exhibit a parabolic behaviour, which is symptomatic of irrelevant quadratic contributions, some are
characterised by the presence of two solutions, one of which SM-like. Three main categories can be observed.
First, we see that for the four-heavy operators, the two degenerate minima are very close and the $\chi^2$ is relatively flat
for a large range of $c_j$. This category of degenerate minima does not affect the evaluation of the CL intervals.
Secondly, there are degrees of freedom that present a second minima far from the SM one, but at a higher value of $\chi^2$. These 
minima do not affect the CL as well.
Lastly, some coefficients have degenerate minima that lead to disjoint intervals. This is the case for example of the Yukawa
operators $c_{\varphi b}$, $c_{\varphi c}$, and $c_{\varphi \tau}$. This situation has to be handled with care and cross-checked, but the 
degeneracy can be in principle broken once the full fit is performed.

These individual fits are also useful to define suitable sampling ranges for the NS algorithm and for this reason they are performed automatically 
before running the global procedure.

\subsection{Nested Sampling}

The NS method is based on the idea of sampling the log-likelihood $\chi^2$ in order to determine its dependence from the 
Wilson coefficients and minimise it. Given a set of experiments $\mathcal{D}$ and a model $\mathcal{M}(\mathbf{c})$, the NS algorithm
exploits Bayes theorem
\be
\label{eq:bayestheorem}
P\lp\mathbf{c}| \mathcal{D},\mathcal{M} \rp = \frac{P\lp\mathcal{D}|\mathcal{M},\mathbf{c}
  \rp P\lp \mathbf{c}|\mathcal{M}  \rp
}{P(\mathcal{D}|\mathcal{M})} \, ,
\ee
which allows us to define the posterior probability $P\lp\mathbf{c}| \mathcal{D},\mathcal{M} \rp$ given the likelihood
function $P\lp\mathcal{D}|\mathcal{M},\mathbf{c}\rp = \mathcal{L}\lp\mathbf{c} \rp$ and the prior distribution $P\lp \mathbf{c}|\mathcal{M}  \rp = \pi \lp  \mathbf{c} \rp$.
The denominator $P(\mathcal{D}|\mathcal{M}) = \mathcal{Z}$ is called Bayesian evidence and ensures the normalisation of the 
posterior probability
\be
\mathcal{Z} = \int \mathcal{L}\lp  \mathbf{c} \rp
\pi \lp  \mathbf{c} \rp d \mathbf{c} \, .
\ee
NS objective is to map the multi-dimensional integral over the prior density into a one-dimensional quantity
\be
\label{eq:NS1}
X(\lambda) = \int_{\{ \mathbf{c} : \mathcal{L}\lp\mathbf{c} \rp > \lambda \}}
\pi(\mathbf{c} ) d\mathbf{c} \,. 
\ee
In terms of the prior mass $X(\lambda)$, the Bayesian evidence can be recasted as
\be
\label{eq:bayesianevidence}
\mathcal{Z} = \int_0^1  \mathcal{L}\lp X\rp dX \, .
\ee
The likelihood is then evaluated at a series of values of $X_i$ and the integral is performed.
While the end result is the computation of the evidence $\mathcal{Z}$, we can extract a sampling of the posterior probability
for the Wilson coefficients and therefore evaluate associated expectation values and variances.
For more details on the algorithm we refer to Ref.~\cite{Ethier:2021bye}.

Regarding the choice of the prior distributions, we use flat ones defined in a suitable way to speed up the computations.
While too narrow ranges for the coefficients might lead to biases, a wide range can make the whole algorithm inefficient.
In order to define proper ranges automatically, we scan the $\chi^2$ profile for individual operators and choose the range of coefficients
such that $\chi^2/{\rm ndat}=4$, ending up with a set of pairs $\lp c_i^{\rm (min)},c_i^{\rm (max)} \rp$ with $i=1,\ldots, N_{\rm op}$.

The advantages of this method are mostly the fact that it does not require any particular tuning of hyper-parameters and it is
not affected by the possibility to end up in a local minimum. However, a limitation is that, being a sampling method, it 
does not scale well with the dimension of the parameter space and performs optimally for moderate dimensions. This is the case for
the problem at hand, but for higher dimensional studies the algorithm might become impractical.

\subsection{Principal component analysis}

Principal component analysis (PCA) is a valuable tool to understand which directions in parameter space are characterised by
high and low variability, leading to dimensional reductions and the identification of flat directions. In particular, the latter
is of crucial importance for us, since the presence of flat directions would translate into a poor constraining power, suggesting that the 
dataset could be augmented to break degeneracies.

In particular, we applied PCA with Singular Value Decomposition (SVD) to the EFT calculations including corrections of order $\mathcal{O}\lp \Lambda^{-2}\rp$
\be
\label{eq:linearTHform}
\sigma_m^{\rm (th)}({\boldsymbol c})= \sigma_m^{\rm (sm)} + \sum_{i=1}^{n_{\rm op}}c_i\sigma^{(\rm eft)}_{m,i}\, ,
\qquad m =1\,\ldots, n_{\rm dat} \, .
\ee
Defining then a matrix $K_{mi}=\sigma^{(\rm eft)}_{m,i}/\delta_{\rm exp,m}$, we can decompose it in principal components 
by defining two unitary matrices $U$ and $V$ such that
\be
\label{eq:SVD}
K = U W V^\dagger \, ,
\ee
and $W$ is a diagonal matrix. The entries of $W$ are semi-positive and are called singular values. The larger the singular value,
the larger the variability of the associated principal component is. These in particular correspond to a linear combination of the initial
Wilson coefficients
\be
\label{eq:PCdef}
{\rm PC}_k = \sum_{i=1}^{n_{\rm op}} a_{ki}c_i \, , \quad k=1,\ldots,n_{\rm op} \, \qquad \lp~ \sum_{i=1}^{n_{\rm op}} a_{ki}^2=1\,~\forall k \rp \, .
\ee
%
\begin{figure}[htbp]
  \begin{center}
  \includegraphics[width=0.95\linewidth]{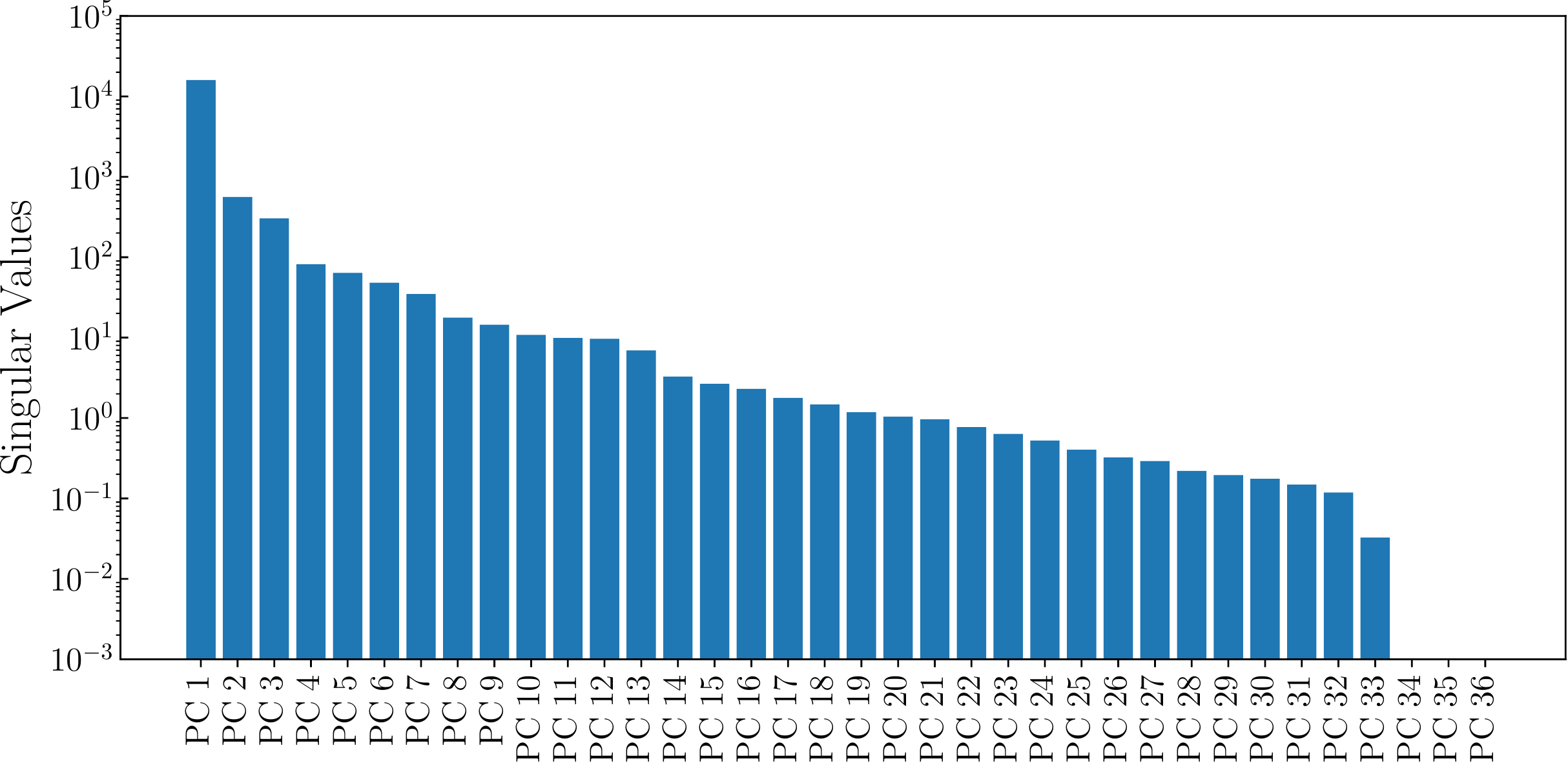}
  \includegraphics[width=0.80\linewidth]{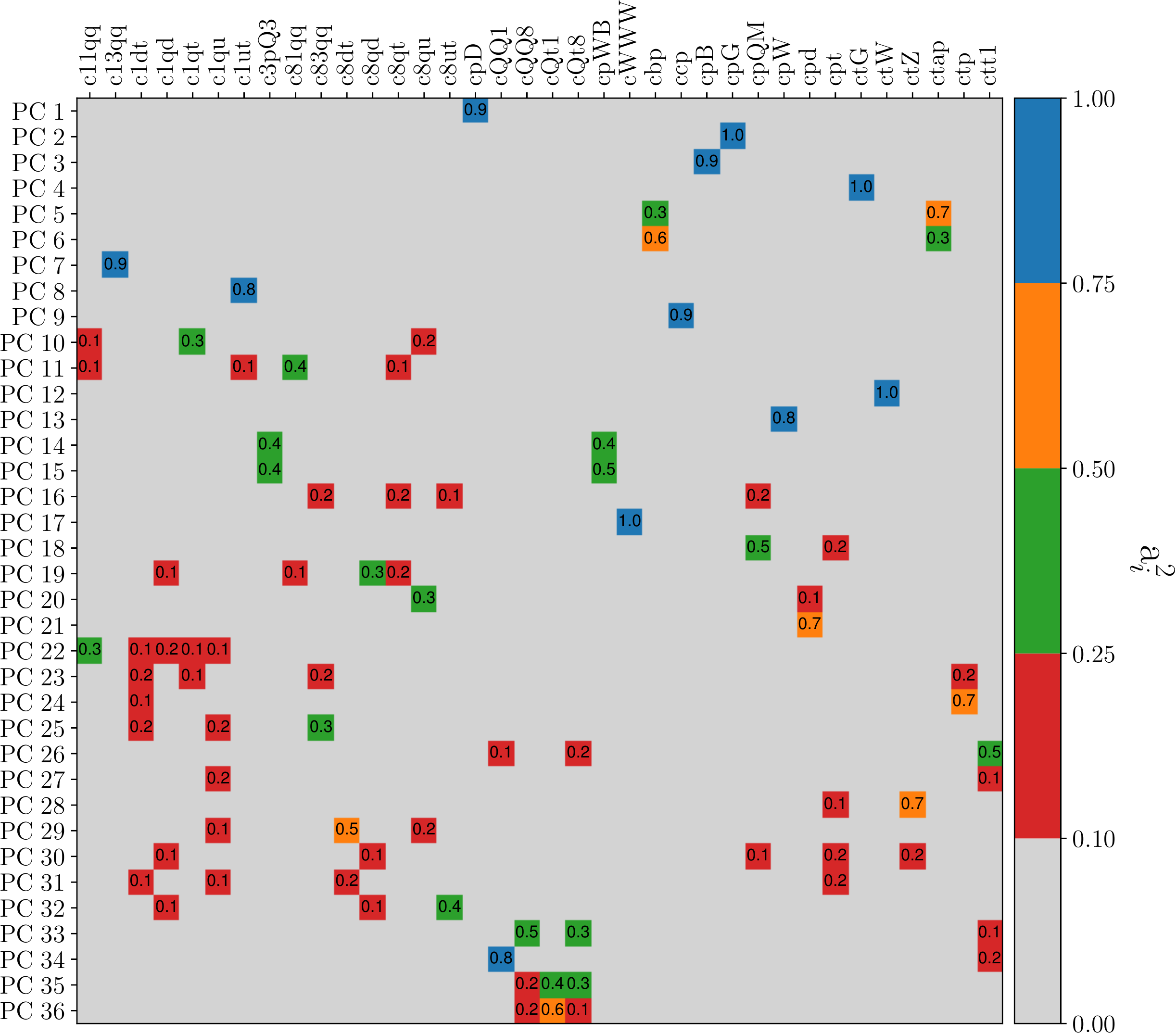}
\caption{In the upper part we report the singular values of the $n_{\rm op}=36$
principal components of the dataset and EFT coefficients. In the lower panel,
a heat-map of the values of the coefficients $a_{i}^2$ is displayed, providing information
on which Wilson coefficient is associated with each principal component.
  \label{fig:PCA_SVs}
  }
  \end{center}
\end{figure}

In Fig.~\ref{fig:PCA_SVs} we report the singular values and a heat-map of the PCA decomposition in the Wilson coefficients. We observe that 
there are three principal components with vanishing singular values, \ie flat directions. These correspond to four-heavy 
quark operators. However, except for these, no large hierarchies are present for the rest, indicating that the physical dimensionality
is similar to the one of the adopted fitting basis.

The highest variability is associated to the bosonic operator $c_{\varphi D}$, which affects Higgs-gauge couplings and input parameters.
This means that we can expect high constraining power on it. From the heat-map we can observe that while many directions
are given by linear combinations of degrees of freedom, some are dominated by specific Wilson coefficients. For instance, $c_{\varphi D}$ ($k=1$), $c_{\varphi G}$ ($k=2$), $c_{tG}$ ($k=4$), $c_{tu}^1$ ($k=8$), $c_{c \varphi }$ ($k=9$), and $c_{tW}$ ($k=12$) are very close to the principal components.

Regarding the flat directions, these are given by
\bea
   {\rm PC}_{34} &=& 0.91 c_{QQ}^{1} - 0.42 c_{tt}^1 \, , \nonumber \\
    {\rm PC}_{35} &=& 0.62 c_{Qt}^{1} - 0.56 c_{Qt}^{8} + 0.49 c_{QQ}^8 \, , \\
    {\rm PC}_{36} &=& 0.78 c_{Qt}^1 -0.50 c_{QQ}^{8}+0.38 c_{tQ}^8 \, .\nonumber
\eea
While these directions cannot be constrained in the linear fit, once quadratic terms are accounted for, pure flat directions
are not possible and the problem disappears. However, when performing the fit at order $\mathcal{O}\lp \Lambda^{-2}\rp$
we will remove the four-heavy operators.

The lack of large hierarchies in the singular values suggests that there is no clear advantage in performing the fit in the rotated basis.
We will therefore limit the use of PCA to a diagnostic tool, postponing an eventual application for the fit to future works.

\section{Results}

After having discussed the fitting methodology, we finally present the results of the global interpretation of top, Higgs and 
EW diboson data. In particular we report the best fit values for the $50$ Wilson coefficients of the analysis, as well as the 
confidence intervals. We first discuss the quality of the fit and then move on to present the bounds, studying in particular the dependence
of the results from the choice of input datasets and the theory settings. For a more complete analysis of the interpretation we refer
to Ref.~\cite{Ethier:2021bye}, while here we will focus on specific aspects relevant for this thesis.

\subsection{Fit quality}

In order to discuss the quality of the fit, we define a slightly different $\chi^2$ with respect to Eq.~\eqref{eq:chi2definition2}.
In particular we have
\begin{equation}
  \chi^2 \equiv \frac{1}{n_{\rm dat}}\sum_{i,j=1}^{n_{\rm dat}}\lp 
  \la \sigma^{(\rm th)}_i\lp {\boldsymbol c} \rp \ra
  -\sigma^{(\rm exp)}_i\rp ({\rm cov}^{-1})_{ij}
\lp 
 \la \sigma^{(\rm th)}_j\lp {\boldsymbol c} \rp \ra
  -\sigma^{(\rm exp)}_j\rp
 \label{eq:chi2definition3}
    \; ,
\end{equation}
where the average on the theory cross sections is performed over the NS samples. It is worth mentioning that since the 
cross sections are in general quadratic in the EFT parameters, these do not correspond to the theoretical cross section
evaluated for the average value of the Wilson coefficients, \ie 
\be
\la \sigma^{(\rm th)}_i\lp {\boldsymbol c} \rp \ra \ne
\sigma^{(\rm th)}_i\lp \la  {\boldsymbol c}  \ra \rp \, .
\ee
%
\begin{table}[t]
  \centering
  \footnotesize
   \renewcommand{\arraystretch}{1.0}
   \setlength{\tabcolsep}{1.0pt}
   \begin{tabular}{l|C{1.0cm}|C{1.4cm}|C{2.2cm}|C{2.2cm}}
        \multirow{2}{*}{ Dataset}   & \multirow{2}{*}{$ n_{\rm dat}$} & \multirow{2}{*}{ $\chi^2_{\rm SM}$} &  $\chi^2_{\rm EFT}$   & $\chi^2_{\rm EFT}$     \\
      &   &   & $\mathcal{O}\lp \Lambda^{-2}\rp$ &  $\mathcal{O}\lp \Lambda^{-4}\rp$  \\
        \toprule
 $t\bar{t}$ inclusive  & 83 &  1.46    &  1.32    &  1.42       \\
 $t\bar{t}$ charge asymmetry  & 11 &  0.60    &  0.39    &  0.59       \\
 $t\bar{t}+V$  & 14  &  0.65     &  0.48     &  0.65       \\
 single-top inclusive  &  27  &   0.43    &  0.44     &  0.41       \\
 single-top $+V$ & 9  &  0.71     &  0.55     &  0.75       \\
 $t\bar{t}b\bar{b}$ \& $t\bar{t}t\bar{t}$   & 6   &  1.68     &  1.09     &  2.12       \\
 Higgs signal strenghts (Run I)  &  22  &   0.86     &  0.85     &  0.90       \\
 Higgs signal strenghts (Run II)  &  40 &   0.67    &  0.64     &  0.63       \\
 Higgs differential \& STXS  &  35  & 0.88     &  0.85     &   0.83      \\
 Diboson (LEP+LHC)  & 70  &  1.31     &  1.31     &  1.30       \\
 \midrule
 {\bf Total}  & {\bf 317}  &  {\bf 1.05 }   & {\bf 0.98 }  & {\bf 1.04 } \\
\bottomrule
\end{tabular}
\caption{Overview of the fit quality in terms of the figure of merit defined in Eq.~\eqref{eq:chi2definition3} for the various
classes of processes.
\label{tab:chi2-baseline-grouped}
}
\end{table}

%
In Table~\ref{tab:chi2-baseline-grouped} we collect the values of the figure of merit defined in Eq.~\eqref{eq:chi2definition3},
showing both the value for the SM predictions and the EFT ones at linear and quadratic accuracy obtained from the global fit. The various 
datasets are grouped in classes in order to have an overall picture.
The global $\chi^2$ slightly improves from $1.05$ in the SM case to $0.96$ and $1.01$ for the linear and quadratic SMEFT fit respectively. 
In particular, we observe relevant improvements in the top-quark sector, especially for top pair production, while in the case of the Higgs and diboson datasets the figure of
merit is very similar.

\subsection{Constraints on the SMEFT parameter space}

After discussing the fit quality, we now present the constraints on the $50$ Wilson coefficients in Table~\ref{tab:operatorbasis} that can be derived from
the dataset under study. We remind the reader that not all of the EFT parameters are independent as $16$ operators are related by
means of the EWPO constraints. We present results for both $\mathcal{O}\lp \Lambda^{-2}\rp$ and $\mathcal{O}\lp \Lambda^{-4}\rp$
predictions in the EFT expansion.
\begin{figure}[t!]
  \begin{center}
    \includegraphics[width=0.32\linewidth]{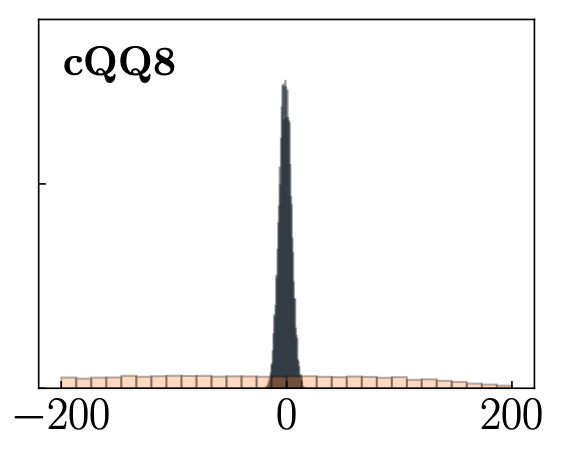}
    \includegraphics[width=0.32\linewidth]{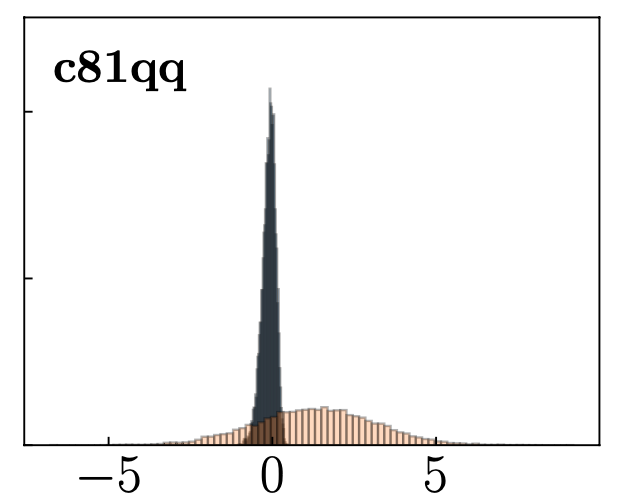}
    \includegraphics[width=0.32\linewidth]{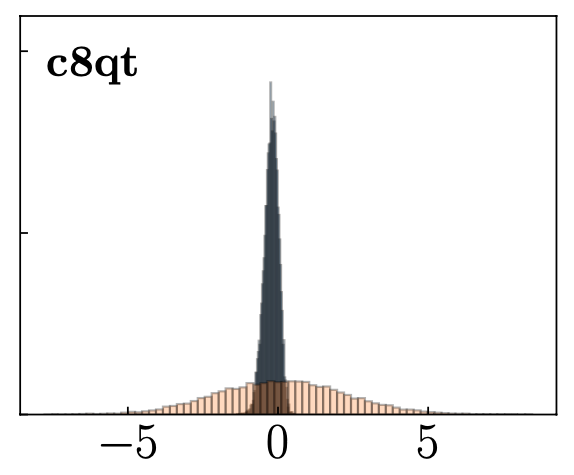}\\
    \includegraphics[width=0.32\linewidth]{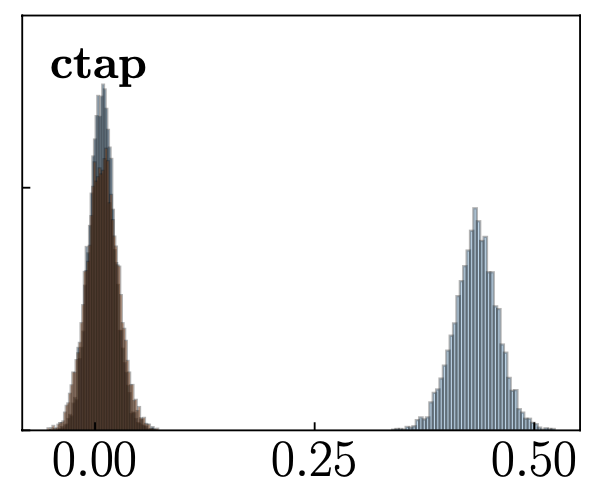}
    \includegraphics[width=0.32\linewidth]{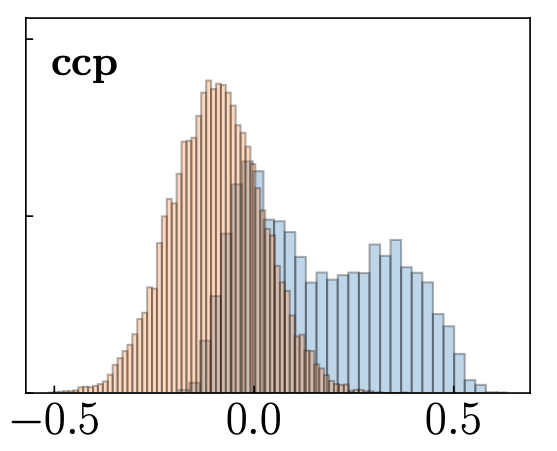}
    \includegraphics[width=0.32\linewidth]{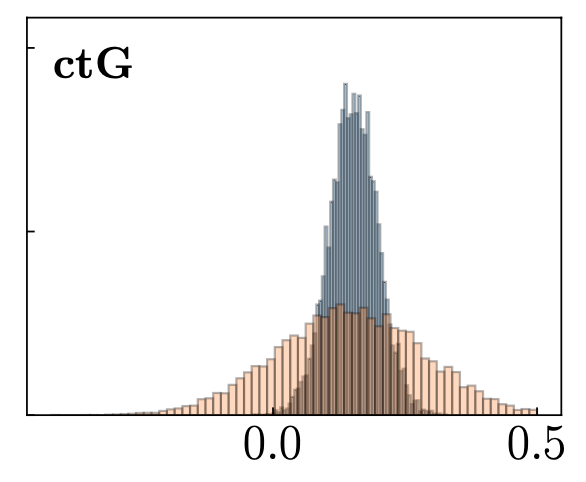}\\
    \includegraphics[width=0.32\linewidth]{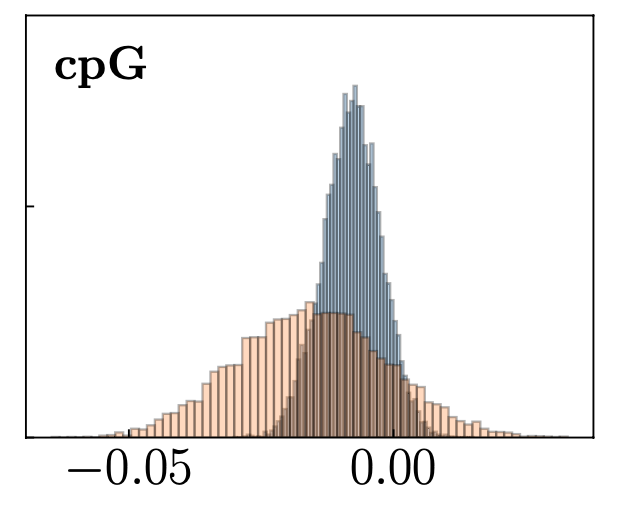}
    \includegraphics[width=0.32\linewidth]{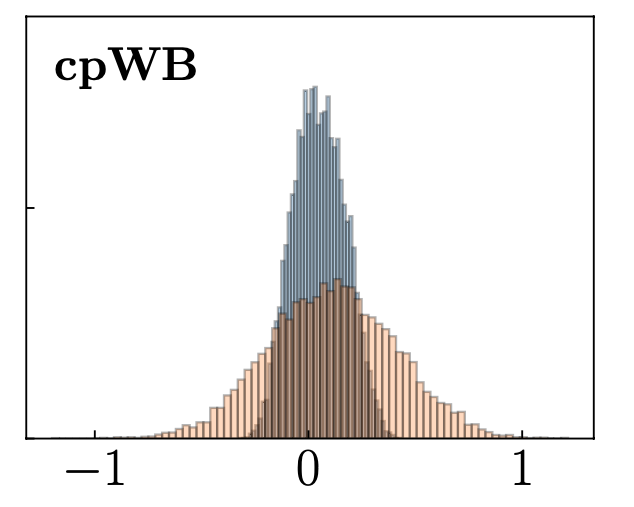}
    \includegraphics[width=0.32\linewidth]{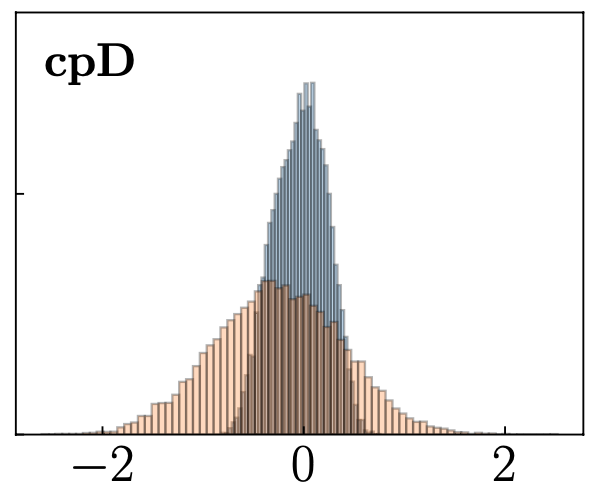}
    \caption{Normalised posterior probability distributions of a sample of Wilson coefficients. We show in orange the results from the linear fit
    while in blue the results for the quadratic one.
     \label{fig:posterior_coeffs} }
  \end{center}
\end{figure}

In Fig.~\ref{fig:posterior_coeffs} we show the posterior probability distributions of a representative subset of operators.
One thing that can be immediately observed in general is that the quadratic effects lead to narrower constraints, as can be seen
in particular for four-fermion operators. Other coefficients, such as the ones affected by EWPO, do not show a marked improvement from
the inclusion of $\mathcal{O}\lp \Lambda^{-4}\rp$ corrections. 
On the other hand, if we look at the charm and $\tau$ Yukawa, we notice that the degenerate
minima that were reported in Fig.~\ref{fig:quartic-individual-fits-2} are propagated to the fit and result in a multi-modal
distribution.
\begin{figure}[t!]
  \begin{center}
    \includegraphics[width=0.90\linewidth]{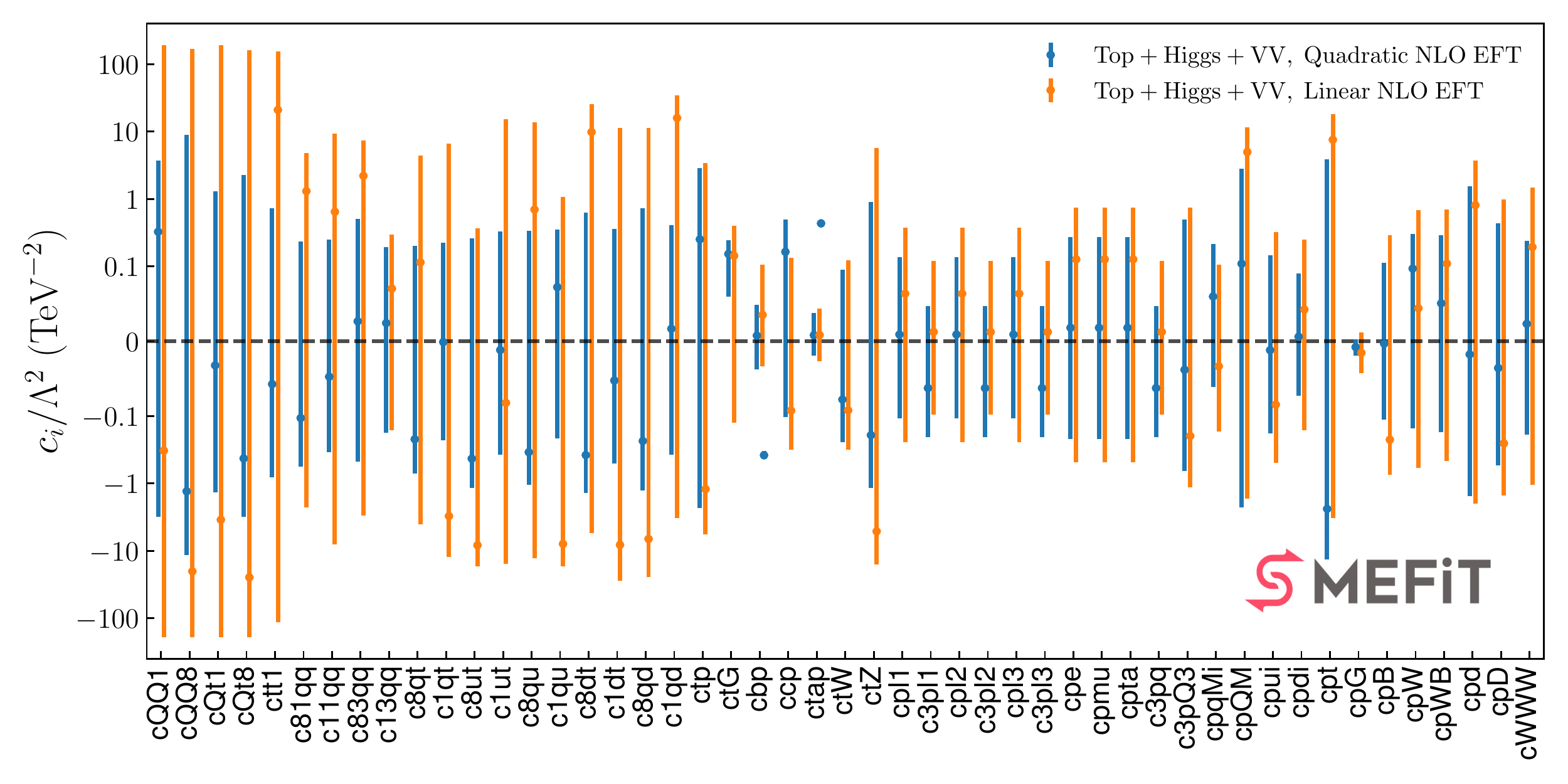}
    \includegraphics[width=0.90\linewidth]{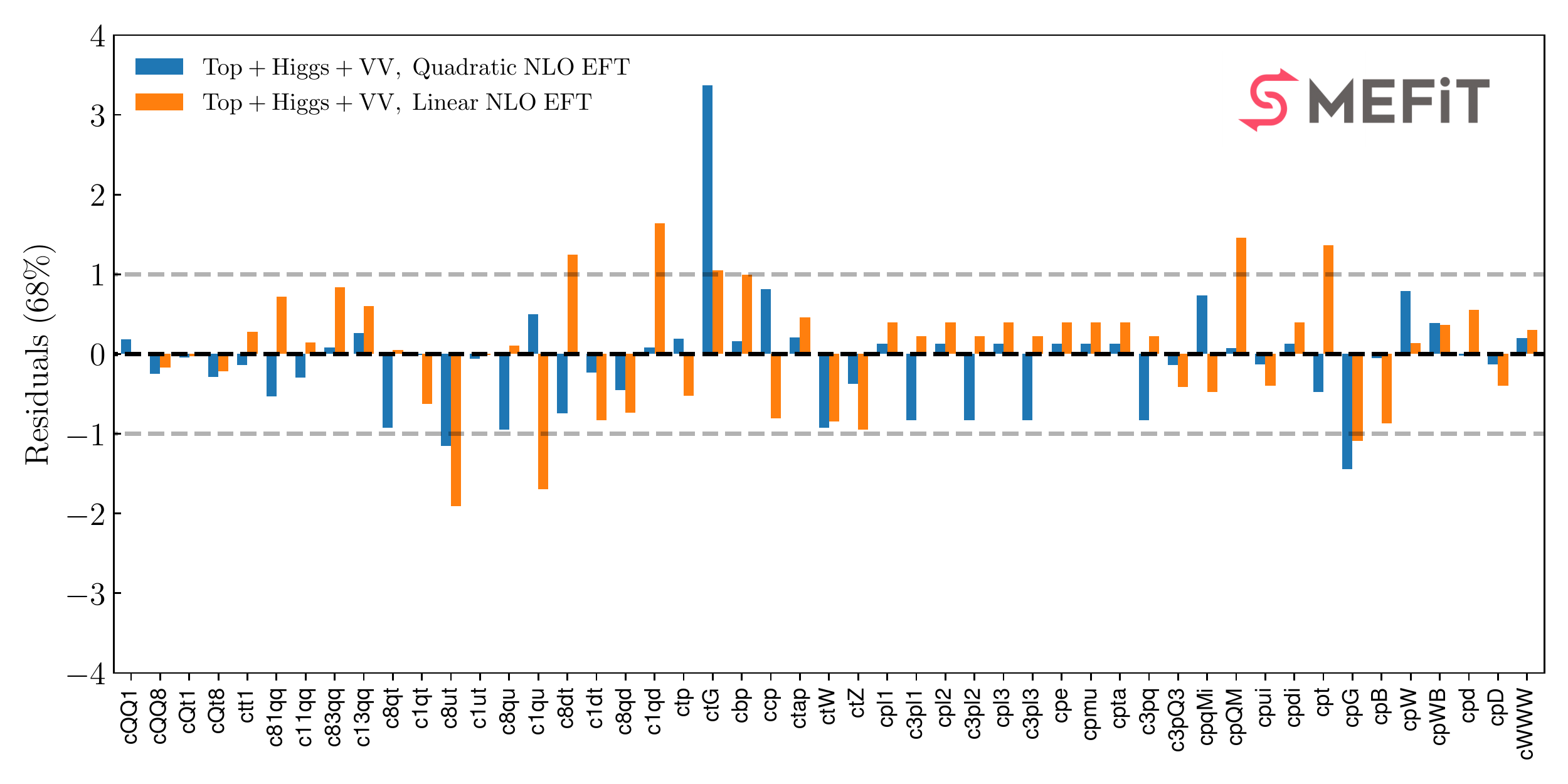}
    \caption{In the upper panel, best fit and $95\%$ CL intervals for the $50$ Wilson coefficients. In the lower panel
    the associated residuals in units of fit uncertainty.
     \label{fig:globalfit-baseline-coeffsabs-lin-vs-quad} }
  \end{center}
\end{figure}

In the upper panel of Fig.~\ref{fig:globalfit-baseline-coeffsabs-lin-vs-quad} we display the full marginalised CL intervals at $95\%$ and the best fit values for the $50$ coefficients $c_i/\Lambda^2$ that can be derived from the posterior
probability distributions (the numerical values are reported in Appendix~\ref{app:coeffs_bounds}). The impact of the quadratic corrections to the bounds is quite relevant for most of the operators,
suggesting that the precision of the experimental results is not yet good enough to safely neglect them. While in the linear fit we 
find that all of the coefficients are in agreement with the SM, this is not the case for the gluon dipole operator $c_{tG}$ when quadratic corrections are included. This deviation is consistent with what has been observed in other similar works~\cite{Ellis:2020unq,Brivio:2019ius}

In the lower panel of Fig.~\ref{fig:globalfit-baseline-coeffsabs-lin-vs-quad} we instead report the $68\%$ residuals in units of fit uncertainty, defined as
\be
\label{eq:fit_residual}
R(c_i)\equiv \frac{\lp c_i|_{\rm EFT} -c_i|_{\rm SM} \rp }{\delta c_i} \, ,\qquad
i=1,\ldots,n_{\rm op} \, ,
\ee
with $c_i|_{\rm EFT}$ the median of the posterior distribution, $c_i|_{\rm SM}=0$, and $\delta c_i$ the total fit uncertainty
for this parameter.
As expected, for most of the coefficients, the residuals are within $1$ sigma, with the exception of the aforementioned dipole
operator. This is however not surprising, as for a large number of EFT coefficients one expects a fraction of these to be above unity 
even if the SM is the underlying theory.

\subsection{Dataset dependence}

\begin{figure}[t!]
  \begin{center}
    \includegraphics[width=0.90\linewidth]{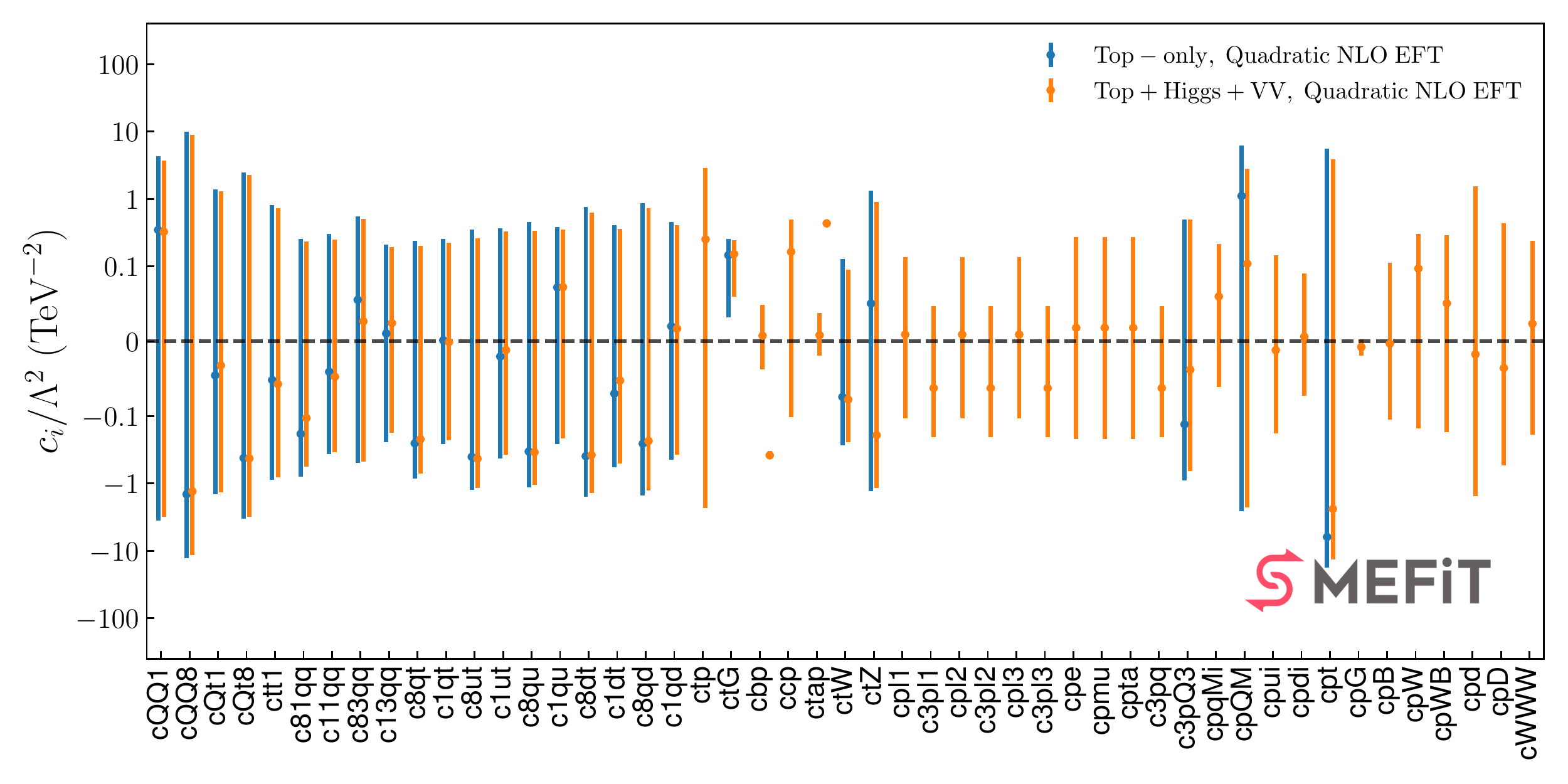}
    \includegraphics[width=0.90\linewidth]{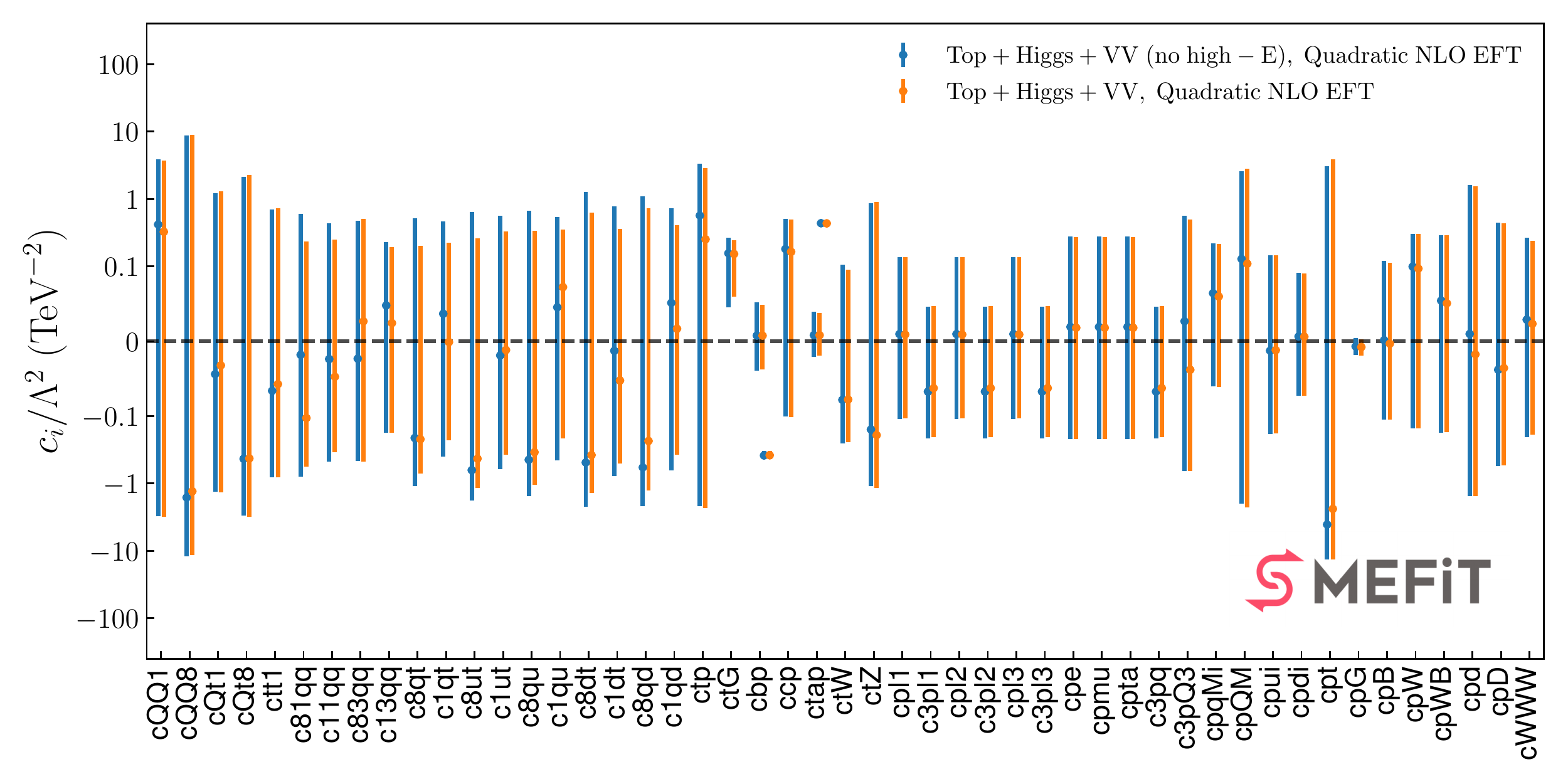}
    \caption{Best fit and $95\%$ CL intervals for the $50$ Wilson coefficients. In the upper panel we show a comparison between the global fit
   and the top-quark only fit. In the lower panel we show a comparison between the global dataset and a restricted one in which high-energy
   bins $E\gtrsim E_{\rm max}$, with $E_{\rm max}\simeq 1$ TeV are removed from the dataset. \revised{However, inclusive cross sections which might probe high energy scales (e.g. four tops) are retained in the fit.}
     \label{fig:globalfit-dataset-dependence} }
  \end{center}
\end{figure}

In this section, we discuss the input dataset dependence of the SMEFT fit. In particular, we compare the results in two scenarios:
the global fit and the top-quark only dataset. This way we can assess the importance of the added Higgs and EW diboson
production datasets with respect to the previous study in Ref.~\cite{Hartland:2019bjb}. Furthermore, we study the importance of 
high energy bins in the experimental measurements by removing them from the fit. In particular, we choose to neglect the
bins in which the energy transfer is $E\gtrsim E_{\rm max}$, with $E_{\rm max}\simeq 1$ TeV. For this comparison, the EFT effects
are computed at order $\mathcal{O}\lp \Lambda^{-4}\rp$.
The results of the fit with modified datasets can be seen in Fig.~\ref{fig:globalfit-dataset-dependence}.
\begin{figure}[t!]
  \begin{center}
    \includegraphics[width=0.45\linewidth]{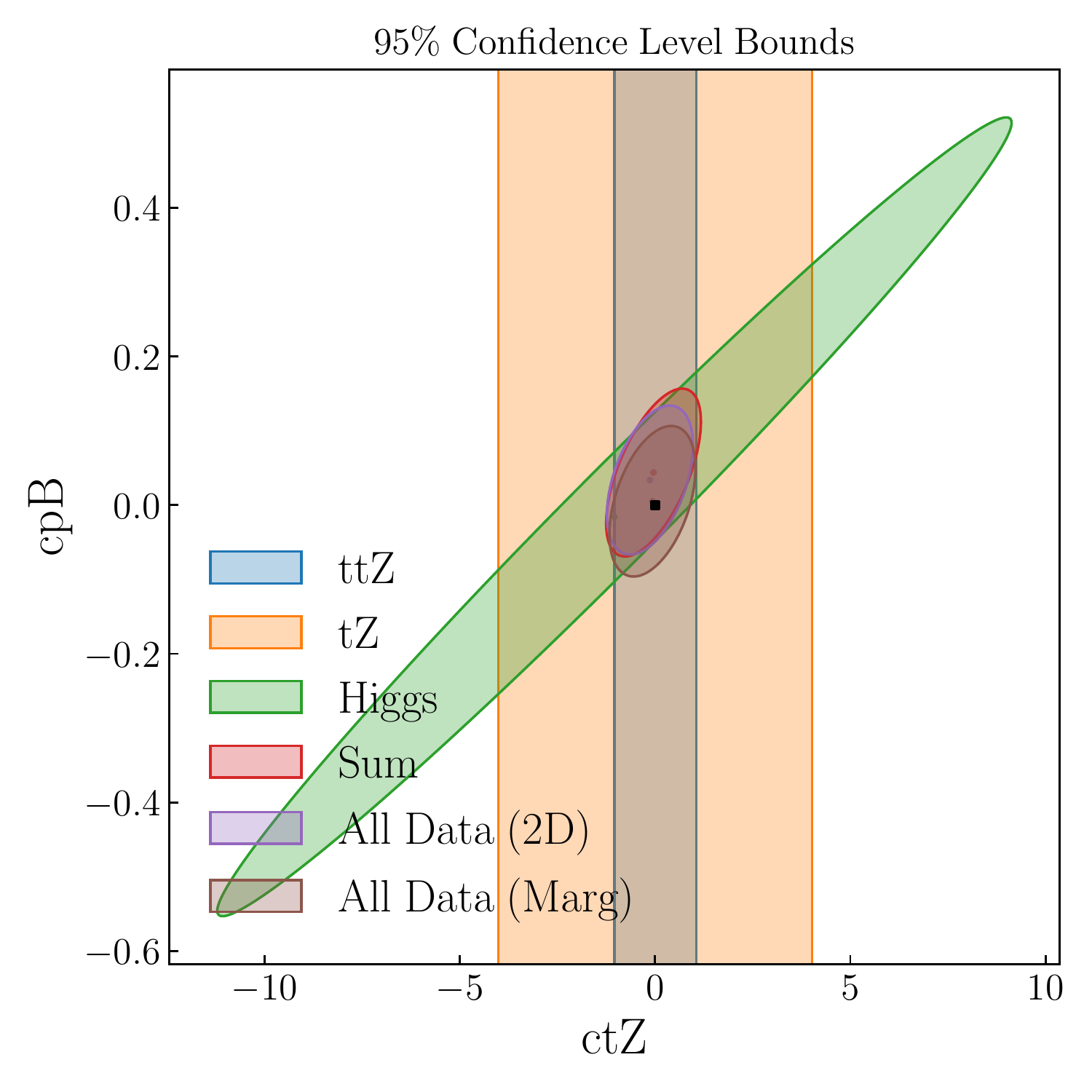}
    \includegraphics[width=0.45\linewidth]{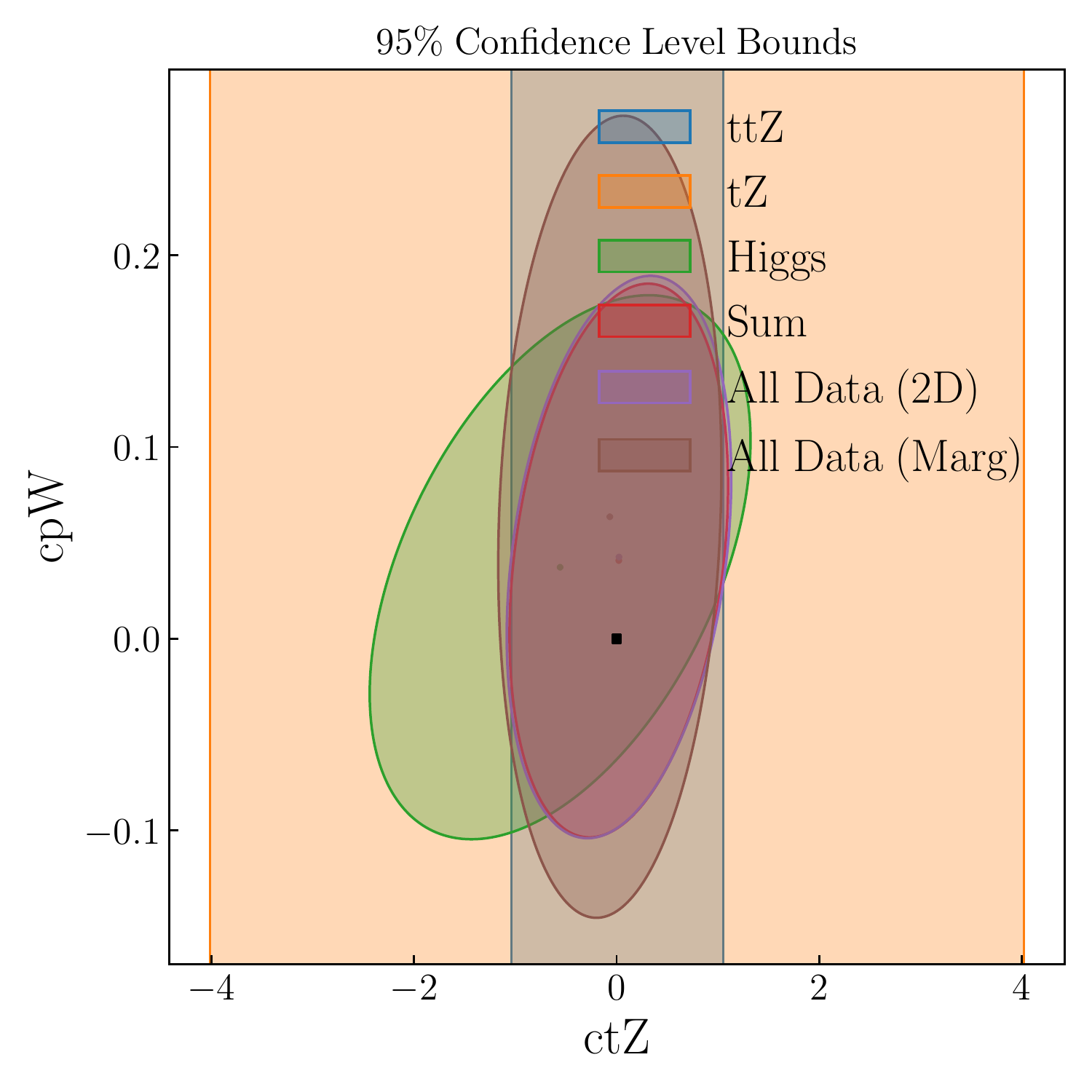}
    \caption{\small
      Two-parameter fits
      of $(c_{tZ},c_{\varphi B})$ (left) and $(c_{tZ},c_{\varphi W})$ (right) at quadratic level in the EFT expansion. All the other Wilson coefficients are set to zero. In brown the 2D-projection of the marginalised fit is displayed. 
      \label{fig:2d_fits_ctZ}}
  \end{center}
\end{figure}
In the upper panel, a comparison between the global fit and the top-only fit is displayed. First of all, the Higgs and diboson
measurements provide sensitivity to several operators that would otherwise be completely unconstrained. In addition to this, the 
global fit bounds are more stringent, sign of the fact that the interplay between the datasets is helpful to gain sensitivity. It is also interesting to observe that, while according to the Fisher information heat-map in Fig~\ref{fig:FisherMatrix} the Higgs datasets is expected to provide the constraints on $c_{t Z}$ and $c_{t W}$, this does not seem to be reflected in the global interpretation.
The bounds seem to come mainly from the top dataset, while the interplay with the Higgs is just leading to a mild improvement.
The Fisher matrix offers additional information and helps in getting a feeling a priori on the sensitivity, but fails to 
account for correlations among operators, since it only considers the diagonal contributions. In particular, we observed 
that while 1-D fits are in agreement with the picture displayed in the Fisher heat-map, in the global parameter space the constraints
come mostly from $ttZ$ production, since the Higgs sector is affected by major correlations among operators which degrade the constraining power (see the 2-D fits displayed in Fig.~\ref{fig:2d_fits_ctZ}).

With respect to the fit without the high-energy bins, in the lower panel of Fig.~\ref{fig:globalfit-dataset-dependence}, we see that the results are very similar to the full global fit. Overall, we see that the fit is
not dominated by high energy bins, a region in which the validity of the EFT might be questioned.

\subsection{Impact of NLO-QCD corrections}

\begin{figure}[t!]
  \begin{center}
    \includegraphics[width=0.90\linewidth]{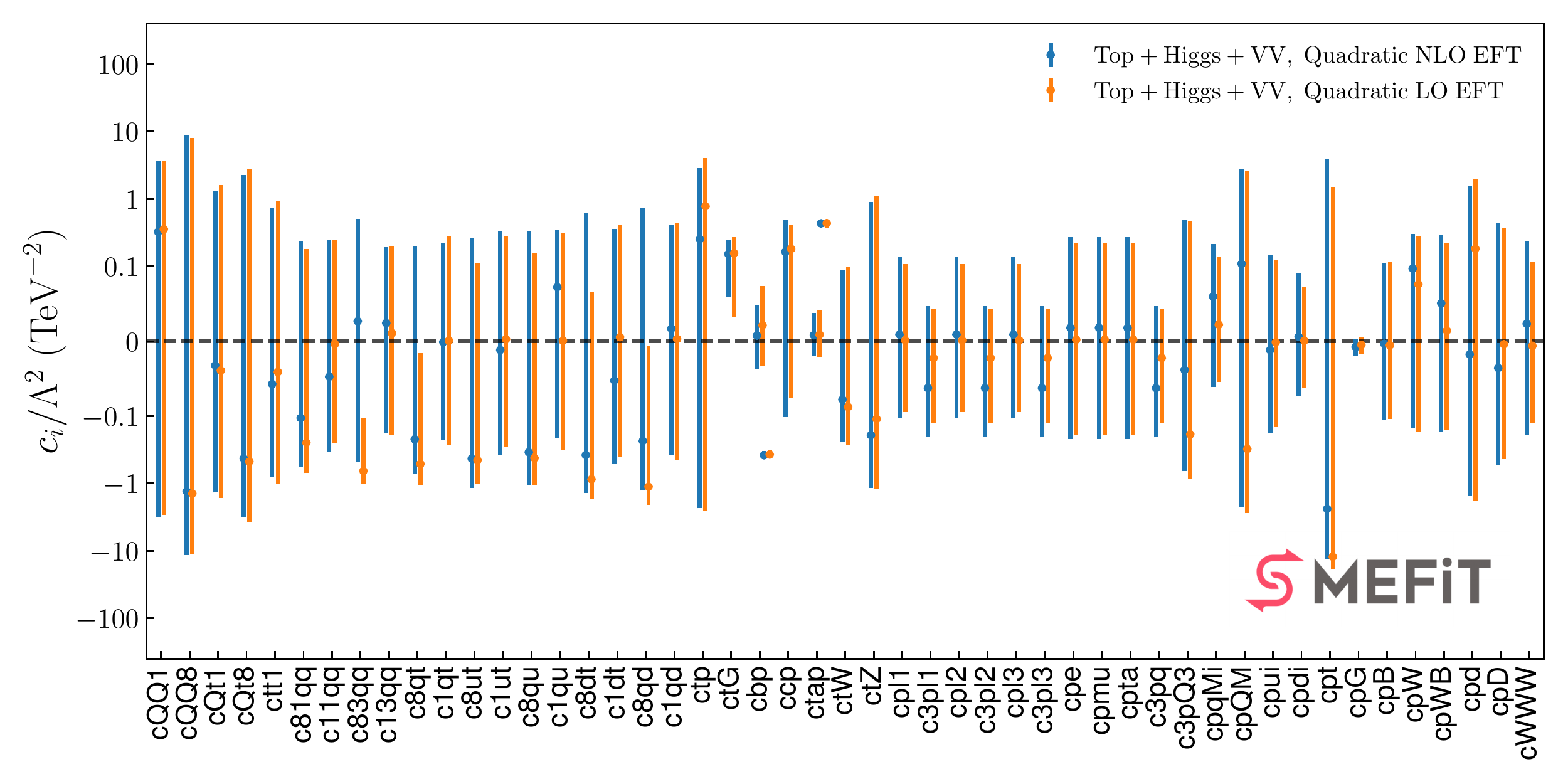}\\
    \includegraphics[width=0.32\linewidth]{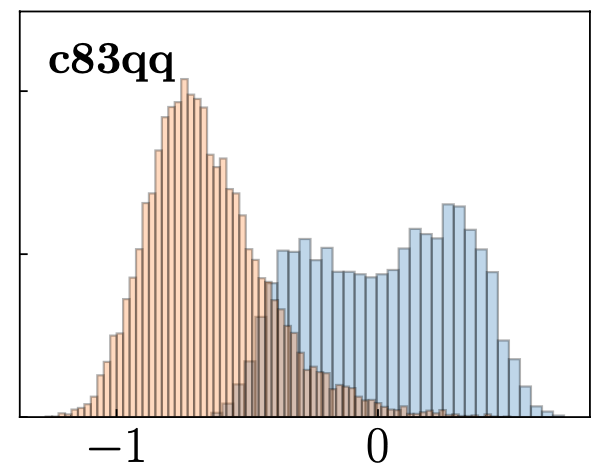}
    \includegraphics[width=0.32\linewidth]{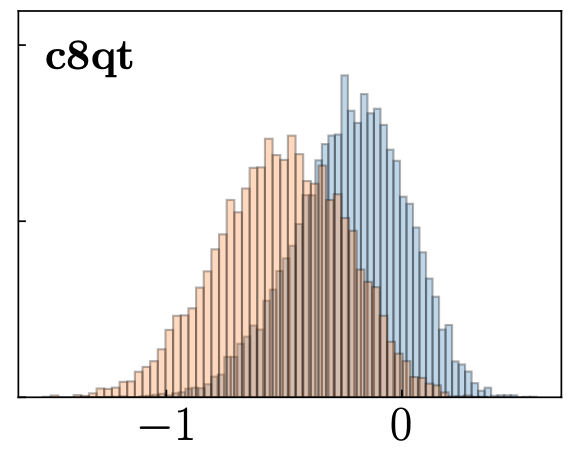}
    \includegraphics[width=0.32\linewidth]{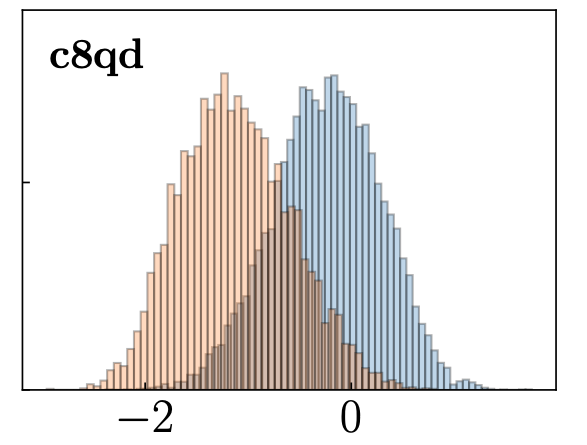}
    \caption{In the upper panel we show the effects of including the NLO-QCD corrections to the theoretical cross sections in the EFT. In the lower panel we report the posterior probability distributions for three four fermion operators to highlight the effects of including the higher order corrections.
     \label{fig:globalfit-theory-dependence} }
  \end{center}
\end{figure}

Finally, we discuss the impact of theoretical settings of the EFT calculations. Specifically, we want to gauge the importance of the
NLO-QCD corrections to the SMEFT predictions. In order to do so, while keeping the SM predictions at the same precision, we perform the fit without
accounting for the NLO corrections and compare the results to the complete global fit. For both cases we include
quadratic EFT corrections.

In Fig.~\ref{fig:globalfit-theory-dependence} the comparison of the fit results is displayed. In particular, in the upper panel,
we show the best fit values and $95\%$ CL intervals in the two theoretical settings. Overall, as we would expect, the effect of NLO corrections is not
dramatic, especially for the Higgs and diboson operators. However, the missing higher order terms seem to have a conspicuous
impact on the four fermion operators bounds, changing both the best fit value and the bounds ranges. In the lower panel of Fig.~\ref{fig:globalfit-theory-dependence} we display the
posterior probability distribution of three of the aforementioned operators, in order to make more visible the effects. These effects
are often the result of an additional sensitivity to the operators introduced at NLO. For instance, colour singlet operators do not interfere at LO with the SM amplitudes, but this suppression is partly lifted once QCD corrections are included.
For certain operators, the effects on the best fit value can be quite relevant. Understanding these effects is however non-trivial,
since the complex interplay and correlations among the whole set of coefficients can produce convoluted effects at the global fit level.
For example, the solution for $c^{3,8}_{Qq}$ in the LO fit is far from the SM but it is back to be consistent with the SM
in the NLO case.

All in all, we find that the NLO corrections have non-trivial effects and their inclusion is of absolute importance for future
accurate EFT interpretations.

\section{Summary and outlook}

The objective of this analysis was to perform an in-depth interpretation of up-to-date measurements in the SMEFT framework.
We did so by combining top, Higgs and diboson data, as well as state of the art theoretical calculations. 

A preliminary study with information geometry and principal component analysis allowed us to quantify the importance of the combined
dataset, showing their complementarity and motivating the need for a global interpretation. The presence of flat directions was
also discussed, showing in particular that they are essentially absent.

The fit results have then been presented and the dependence from the input dataset and the precision of the theoretical calculations discussed.
In particular, we showed that the addition of the Higgs and diboson datasets is of crucial importance to gain sensitivity on several operators and 
the benefits of the interplay with the top-quark dataset have been highlighted.

This study is a stepping stone towards a complete global interpretation of the LHC measurements. As such, there are several directions 
in which it can be extended.
First of all, the direct inclusion of EWPO will be mandatory in the future when LHC results will become competitive in terms
of precision. Moreover, in order to fully exploit the potential to uncover NP, the addition of high energy observables could be 
of extreme importance. An example of such is the inclusion of vector boson scattering cross sections, which are particularly sensitive to
unitarity violating effects~\cite{Ethier:2021ydt}. 

Other measurements worth considering are high-mass Drell-Yan production and 
flavour data from LHCb and B-factories, as well as lower energy observables such as electric dipole moments. From the theoretical
perspective, this will require taking into account RGEs for the Wilson coefficients (see Sec.\ref{sec:running}) in order to relate physics at different energy scales.

Finally, the inclusion of higher order terms in the EFT expansion, such as double insertion or the effects from subsets of dimension-8 operators,
could be considered and their relative importance assessed.

%
%
%

\chapter{Looking for New Physics at a future muon collider}
\pagestyle{fancy}
\label{chap:muon}

\hfill
\begin{minipage}{10cm}

{\small\it 
``Thoughts without content are empty, intuitions without concepts are blind. The understanding can intuit nothing, the senses can think nothing. Only through their unison can knowledge arise.''}

\hfill {\small Immanuel Kant, \textit{Critique of Pure Reason}}
\end{minipage}

\vspace{0.5cm}

While providing us with several discoveries and measurements, the Large Hadron Collider has not produced yet evidence of
New Physics. In particular, solutions to the stabilisation of the EW scale, Dark Matter and neutrino masses problems still evade our
reach. Despite the fact that the LHC program is not over, the high energy physics community started to discuss the next steps to
take to find answers to these problems. This was for instance the main objective of the European Strategy Update for Particle Physics~\cite{Strategy:2019vxc,EuropeanStrategyGroup:2020pow}.

In this context, two main directions are discussed. The first one is to explore the energy frontier by building a new hadron collider
at centre of mass energy of $\sqrt{s}=100$ TeV. The second is to instead focus on precision measurements with up to few TeVs
$e^+e^-$ colliders. Both have different and complementary features that make them an ideal combination for the future
to come. 

However, since construction of such colliders requires time and they would not be operative for the next 20-25 years from now,
the community started to explore also other options. Among these, one of the most exciting is the realisation of a multi-TeV
muon collider, which could in principle combine the best of both hadron and lepton colliders~\cite{Palmer:1996gs,Ankenbrandt:1999cta}.
In fact, being muons leptonic fundamental particles, one can take advantage of the rather clean environment characterising lepton colliders
while at the same time reach high energy collisions due to the reduced synchrotron radiation with respect to electrons. Moreover,
such a machine might not need a dedicated tunnel and laboratories and could be hosted in pre-existing facilities.

The main challenges for the realisation of this project all stem from the basic fact that muons are unstable particles. Research and 
development are therefore needed in order to be able to produce muon beams with a luminosity high enough to undertake a successful physics
program. Various proposals have been made during the years. The Muon Accelerator Program (MAP)~\cite{Palmer:2014nza}
in particular explores the possibility to produce muon beams from the decay of pions which are produced by collision of protons over
a heavy-material target. Recently, another scheme has been proposed by the Low Emission Muon Accelerator (LEMMA) program~\cite{Antonelli:2013mmk,Antonelli:2015nla},
where muons are produced at threshold from $e^+ e^-$ annihilations.

Given the technological efforts, it is important to study the physics potential of such machines. In particular,
dedicated studies have been presented where processes produced from muon annihilation were discussed~\cite{Barger:1995hr,Palmer:1996gs,Ankenbrandt:1999cta,Han:2012rb,Chakrabarty:2014pja,Greco:2016izi,Buttazzo:2018qqp,Delahaye:2019omf,Ruhdorfer:2019utl}.
However, a striking feature of high-energy muon colliders is the potential to scatter directly EW bosons by means of vector boson
fusion (VBF)\cite{Costantini:2020stv}. In the following we will show that beyond $5-10$ TeV, this becomes the dominant production mechanism for many
SM processes. This was already established for heavy Higgs production~\cite{Dawson:1984ta,Hikasa:1985ee,Altarelli:1987ue,Kilian:1995tr,Gunion:1998jc}, but it holds true also for multi-boson and top quark production.

The scattering of EW boson is a natural portal to indirectly assess the presence of NP at higher scales and in this perspective, muon colliders
are the ideal machines to take advantage of that, by combining high energy reach and precision. In particular, we will present a
discussion of the potential to determine the Higgs potential by measuring its self-interaction couplings~\cite{Chiesa:2020awd}.

\section{Comparing proton and muon colliders}
\label{sec:p_vs_mu}

To start with, it is instructive to discuss and compare the physics potential for discovery of muon and proton colliders. In order to do so,
a suitable measure of reach has to be established, as the two classes of colliders have clearly distinct features. For instance, while
muons are fundamental particles and the whole centre of mass energy is available in the collision, in proton colliders we are effectively 
colliding partons, which carry only a fraction of the total beam energy. In the following we will discuss three different classes
of processes, trying to maintain a model-independent perspective.

\subsection{\texorpdfstring{$2 \to 1$}{2 -> 1} annihilations}

In order to define a measure to compare the different colliders, we make use of parton luminosities~\cite{Quigg:2009gg}.
The generic cross section for a proton collision in this framework is given by
\begin{equation}
    \sigma(p p \to X + \text{anything}) = \int_{\tau_0}^1 d\tau \sum_{ij} \Phi_{ij}(\tau, \mu_f) \, \hat{\sigma}(i j \to X) \, ,
\end{equation}
where $X$ is a generic final state with invariant mass $m_X=\sqrt{\hat{s}}=\sqrt{\tau s}$, $\hat{\sigma}(i j \to X)$ is the partonic cross section and $\tau_0$ the minimum energy required to produce $X$. The parton luminosity $\Phi_{ij}(\tau, \mu_f)$ is defined as a function of the hard scattering energy fraction $\sqrt{\tau}$ 
and the factorisation scale $\mu_f$ as
\begin{equation}\label{eq:parton_lumi_def}
    \Phi_{ij}(\tau, \mu_f) \equiv \frac{1}{1 + \delta_{ij}} \int_{\tau}^1 \frac{d\xi}{\xi} \left[f_{i/p}(\xi, \mu_f) \, f_{j/p}\left(\frac{\tau}{\xi}, \mu_f\right)
    + (i \leftrightarrow j)
    \right] \, .
\end{equation}
The function $f_{i/p}(\xi,\mu_f)$ is the PDF for a parton $i$ carrying a longitudinal momentum fraction $\xi$ of the proton beam.
In the following, we set the factorisation scale to $\mu_f = \sqrt{\hat{s}}/2$.

With this formalism in mind, we define the equivalent proton energy as the energy $\sqrt{s_p}$ needed for a proton collider
to have the same cross section of a $\sqrt{s_\mu}$ muon collider when producing a final state $X$. In the case of a one particle final 
state $X$ with mass $M=\sqrt{\hat{s}}$, we have
\begin{align}
    \sigma_p(s_p)       &= \int_{\tau_0}^1 d\tau ~ \sum_{ij} \Phi_{ij}(\tau, \mu_f) \, ~ [\hat{\sigma}_{ij}]_p \, \delta\left(\tau - \frac{M^2}{s_p}\right) \, , \\
    \sigma_\mu(s_\mu)   &= [\hat{\sigma}]_\mu \, ,
\end{align}
where with $[\hat{\sigma}]$ we denote the partonic cross sections. In the case of the proton collider cross section, we make explicit the 
1-body phase space Dirac delta, while in the case of the muon collider we assume that this has been absorbed by means of the narrow width
approximation. It is worth stressing that in the case of muons, the production can happen only at threshold and therefore $s_\mu = \hat{s}=M^2$.
Equating the cross section expressions we find
\begin{eqnarray}
[\hat{\sigma}]_\mu  &=& \sigma_\mu(s_\mu)  = \sigma_p(s_p)  \\
                &=&   \sum_{ij} \Phi_{ij}\left(\frac{s_\mu}{s_p}, \mu_f\right) \times [\hat{\sigma}_{ij}]_p \approx  
                [\hat{\sigma}]_p \times \sum_{ij} \Phi_{ij}\left(\frac{s_\mu}{s_p}, \frac{\sqrt{s_\mu}}{2}\right) ,
\label{eq:single_prod}
\end{eqnarray}
where in the last step we made the simplifying assumption that each specific partonic cross section can be substituted with a
universal $[\hat{\sigma}]_p$. In terms of the partonic cross section ratio $\beta$, Eq.~\eqref{eq:single_prod} can be expressed as
\begin{equation}
\sum_{ij} \Phi_{ij}\left(\frac{s_\mu}{s_p}, \frac{\sqrt{s_\mu}}{2}\right)    = \cfrac{[\hat{\sigma}]_\mu}{[\hat{\sigma}]_p } \equiv \frac{1}{\beta} \, .
\label{eq:single_prod_beta}
\end{equation}
\begin{figure}[t]
\begin{center}
\includegraphics[width=.75\textwidth]{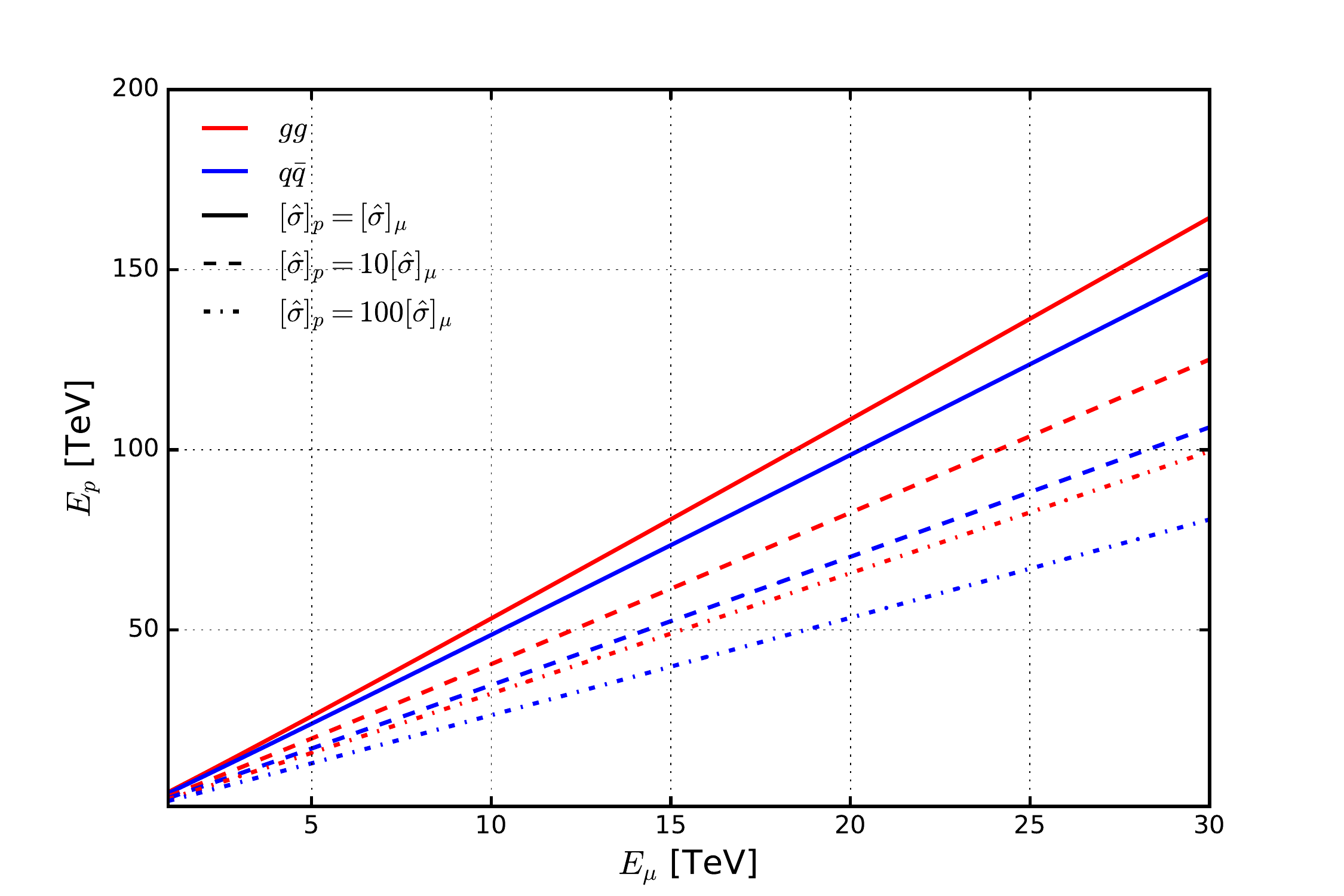}
\end{center}
\caption{\label{fig:p_vs_muon_1body}
Solutions to Eq.~\eqref{eq:single_prod_beta} for the equivalent proton energy as a function of the muon collider energy.
Three different values of $\beta$ are considered and solutions are divided in two partonic classes.
}
\end{figure}
The numerical solutions for Eq.~\eqref{eq:single_prod_beta} are reported in Fig.~\ref{fig:p_vs_muon_1body} for different 
values of the ratio $\beta$ and different classes of partons ($q\bar{q}$ annihilation with $q\in\{u,c,d,s\}$ and gluon fusion).
The different values of $\beta$ are motivated by the need to assess different scenarios in which NP is coupled to quarks and leptons.
For instance, when $\beta=1$ both processes are governed by the same physics, while when $\beta=10$, $ij\to X$ is governed by QCD while $\mpmm\to X$ by EW interactions.

As evident from the figure, the expectation that a higher proton collider energy is needed to match the reach of a muon collider is reproduced.
In particular, the scaling is linear for a fixed $\beta$ and varies from $\sqrt{s_p}\sim 5\times \sqrt{s_\mu}$ to $\sqrt{s_p}\sim 3.3\times \sqrt{s_\mu}$ 
when $\beta=1$ and $\beta=100$ respectively. For example, in the case of equivalent partonic interactions, a muon collider
at $3$ TeV is characterised by a similar single resonance reach of the $14$ TeV LHC.

\subsection{\texorpdfstring{$2 \to 2$}{2 -> 2} annihilations}

Another possibility to compare muon and proton colliders is to consider the pair production of heavy particles. In order to do so,
we assume that the muon collider energy is just beyond threshold and therefore in an optimal configuration. Contrary to the 
previous case, for the proton collider the collisions do not occur at a fixed $\hat{s}$, but instead are suppressed by 
$[\hat{\sigma}_{ij} ]_p\sim1/\hat{s}$ once above threshold. The assumption that we make is therefore that the quantities $[\hat{\sigma}_{ij} \hat{s}]_p$
are energy independent and we can write the cross sections as
\begin{align}
    \sigma_p(s_p) &= \frac{1}{s_p}\int_{\tau_0}^1 d\tau \frac{1}{\tau} ~ \sum_{ij} \Phi_{ij}(\tau, \mu_f) ~ \, [\hat{\sigma}_{ij} \hat{s}]_p \, , \\
    \sigma_\mu(s_\mu) &= \frac{1}{s_\mu}[\hat{\sigma} \hat{s}]_\mu\,.
\end{align}
Assuming once again that the partonic cross sections are flavour independent and equating the two expressions we find
\begin{equation}\label{eq:pair_prod}
    \frac{s_\mu}{s_p} ~ \int_{\frac{s_\mu}{s_p}}^1 d\tau ~ \frac{1}{\tau} ~ \sum_{ij} \Phi_{ij}\left(\tau, \frac{\sqrt{s_\mu}}{2}\right) ~=~ \frac{[\hat{\sigma} \hat{s}]_\mu}{[\hat{\sigma} \hat{s}]_p} ~\equiv~  \frac{1}{\beta} \, .
\end{equation}
\begin{figure}[t]
\begin{center}
\includegraphics[width=.75\textwidth]{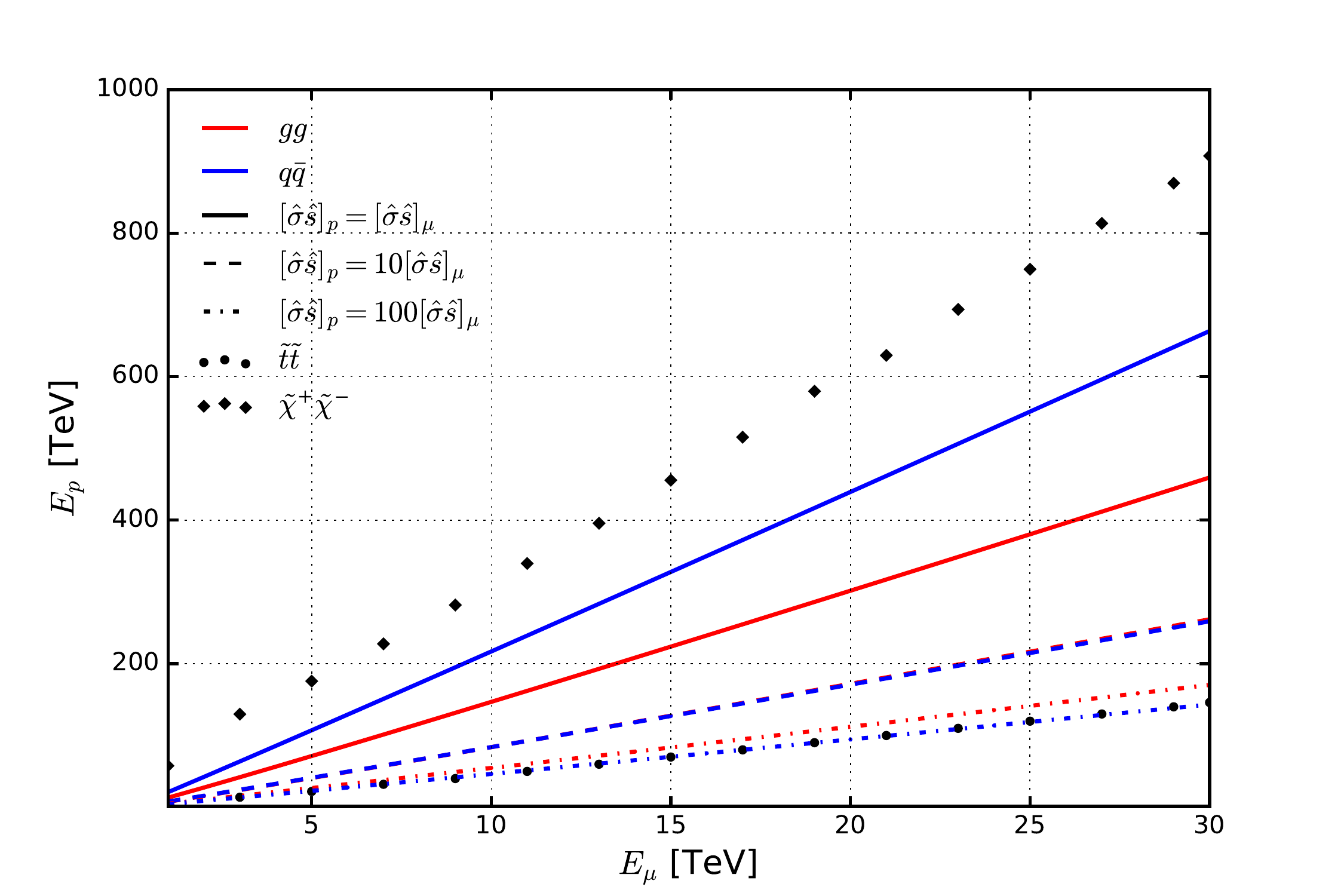}
\end{center}
\caption{\label{fig:p_vs_muon_2body}
Solutions to Eq.~\eqref{eq:pair_prod} for the equivalent proton energy as a function of the muon collider energy.
Three different values of $\beta$ are considered and solutions are divided in two partonic classes. Solutions for
stop and chargino production are also displayed.
}
\end{figure}
This equation can be solved numerically for the equivalent proton collider energy. In Fig.~\ref{fig:p_vs_muon_2body} we display solutions of the
equation for three different values of $\beta$. In addition to the model-independent solutions, we report the cases of LO
stop and chargino pair production\footnote{For these cases we fixed the BSM masses at $90\%$ of the muon collider energy, \ie $M=0.9\times\sqrt{s_\mu}/2$.}.
As in the previous scenario, we observe that the proton collider energy required to match the muon collider cross section is
sensibly higher. The scaling is even more drastic and goes from $\sqrt{s_p}\sim 22\times \sqrt{s_\mu}$ to $\sqrt{s_p}\sim 5.5\times \sqrt{s_\mu}$ when $\beta=1$ and $\beta=100$ respectively. This suggests that in the case of comparable interaction strengths ($\beta=1$),
a muon collider at $10$ TeV has the same reach of a $220$ proton machine.

Regarding the production of specific BSM candidates, we observe that $\tilde{t}\overline{\tilde{t}}$ is in very good agreement
with the model-independent solution for $\beta=100$. Specifically, being the production in a proton collider driven by QCD quarks
annihilation, we see that it is well described by the $ij=q\overline{q}$ solution. This result is not totally unexpected, as $\beta=100$
is precisely the ratio between QCD and EW interactions. On the other hand, the case of chargino production is characterised by 
EW annihilation in both proton and muon collider. Therefore we would expect the $ij=q\overline{q}$ with $\beta=1$ to be a good
description, but we instead find a poor agreement. The reason for this has been traced back to the approximation of flavour 
independence for the partonic cross sections in the derivation of Eq.~\eqref{eq:pair_prod}. While it was holding for the stop,
this approximation is less valid for charginos. Nevertheless, we find that the solutions in Fig.~\ref{fig:p_vs_muon_2body}
provide reliable if not conservative estimates for the equivalent proton energy.

\subsection{Weak boson fusion}

After discussing the potential to produce new resonances, we now analyse the VBF scattering potential. As we will see later,
a high energy muon collider is effectively an EW boson collider and it is therefore interesting to compare it with a proton collider
in this perspective.
In order to do so, we employ the formalism of EWA, already presented in Section~\ref{sec:EWA}, so that we can continue to use the language
of parton luminosities by describing the effective emission of EW bosons.
For a generic vector boson $V$ with helicity $\lambda$ and emitted from a fermion $a$, the splitting functions are given by
\begin{align}
f_{V_\lambda/a}(\xi, \mu_f, \lambda) &= \frac{C}{16\pi^2} \frac{(g_V^a \mp g_A^a)^2 +(g_V^a \pm g_A^a)^2(1-\xi)^2}{\xi}\log{\left(\frac{\mu_f^2}{M_V^2}\right)}, 
\label{eq:ewa_VT}
\\
f_{V_0/a}(\xi, \mu_f) &= \frac{C}{4\pi^2} (g_V^{a~2} + g_A^{a~2})\left( \frac{1-\xi}{\xi}\right)\,,
\label{eq:ewa_V0}
\end{align}
where 
\begin{eqnarray}
\text{for}~V=W      :&  C=\frac{g^2}{8}, \qquad             & g_V^a=-g_A^a=1 \, ,
\\
\text{for}~V=Z      :& C=\frac{g^2}{\cos^2{\theta_W}}, \quad & g_V^a=\frac{1}{2}\left(T^3_L\right)^a- Q^a\sin^2{\theta_W},\\ &g_A^a = -\frac{1}{2} \left(T^3_L\right)^a \nonumber .  
\end{eqnarray}
We further define the spin averaged transverse splitting function by
\begin{equation}
f_{V_T/a}(\xi, \mu_f) \equiv \cfrac{f_{V_{+1}/a}(\xi, \mu_f^2, \lambda=+1) + f_{V_{-1}/a}(\xi, \mu_f^2, \lambda=-1)}{2} \, .
\end{equation}
For the muon collider, the parton luminosity for the scattering of a vector boson $V$ and $V^\prime$ are defined as
\begin{equation}
 \Phi_{V_{\lambda_1}  V^\prime_{\lambda_2} }(\tau, \mu_f) = \int_{\tau}^1 \frac{d\xi}{\xi} ~ f_{V_{\lambda_1} /\mu}\left(\xi, \mu_f\right) \, f_{V^\prime_{\lambda_2} /\mu}\left(\frac{\tau}{\xi}, \mu_f\right)  \, .
\end{equation}
In the case of a proton collider, we have to first extract a quark field from the proton and then a EW boson from the quark.
The expression for the parton luminosity is then given by a double convolution
\begin{eqnarray}
 &\Phi_{V_{\lambda_1} V'_{\lambda_2}}(\tau, \mu_f) =\frac{1}{1+\delta_{V_{\lambda_1} V'_{\lambda_2}}} \int_\tau^1 \frac{d\xi}{\xi}\int_{\tau/\xi}^1 \frac{dz_1}{z_1}\int_{\tau/(\xi z_1)}^1 \frac{dz_2}{z_2} \sum_{q, q'} 
\\
&
f_{V_{\lambda_1}/q}(z_2)f_{V'_{\lambda_2}/q'}(z_1)
\left[
f_{q/p}(\xi)f_{q'/p}\left(\frac{\tau}{\xi z_1 z_2}\right) 
+ 
f_{q/p}\left(\frac{\tau}{\xi z_1 z_2}\right)f_{q'/p}(\xi)\right] \, .
\nonumber
\end{eqnarray}
\begin{figure}[t]
\begin{center}
\includegraphics[width=.48\textwidth]{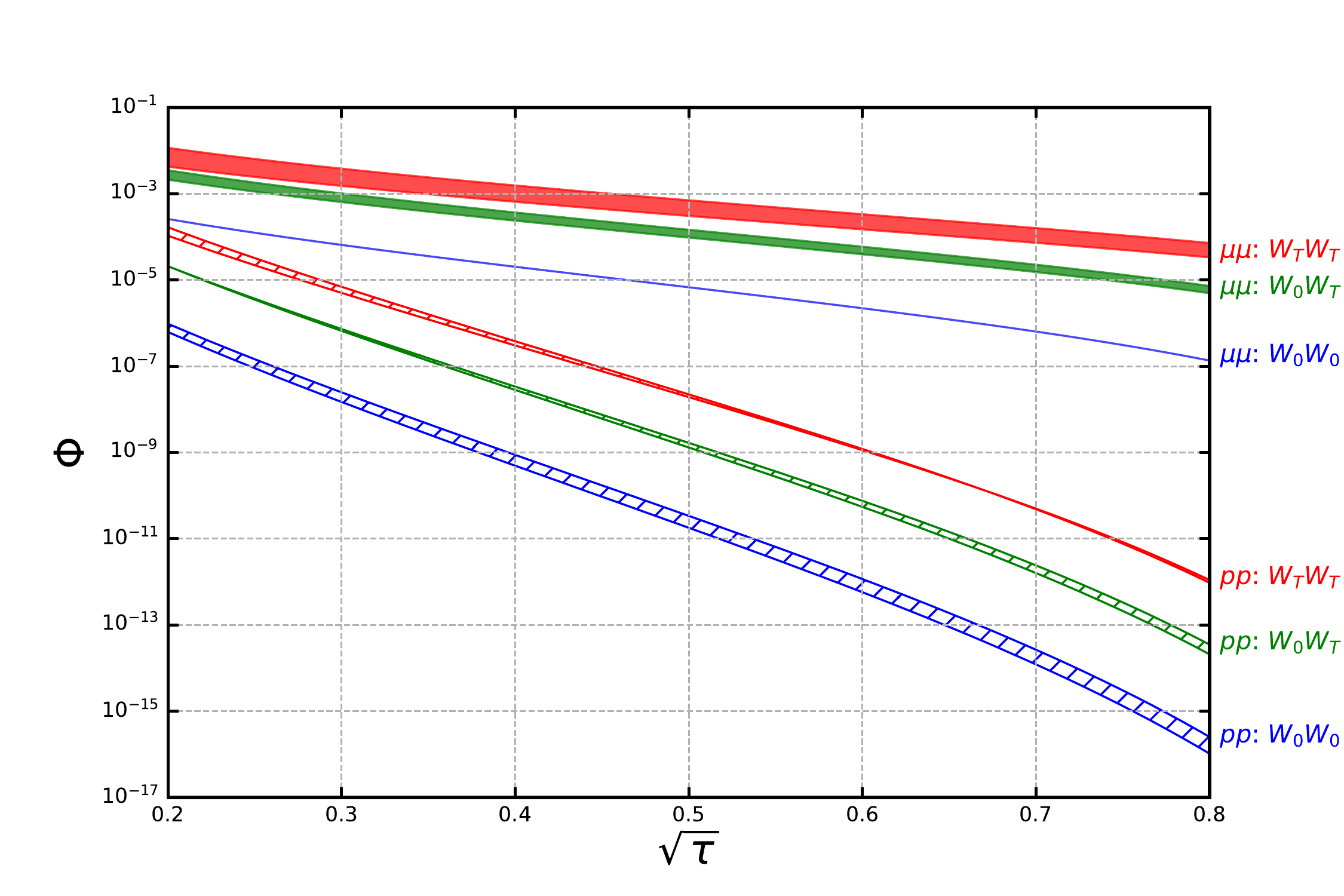}
\includegraphics[width=.48\textwidth]{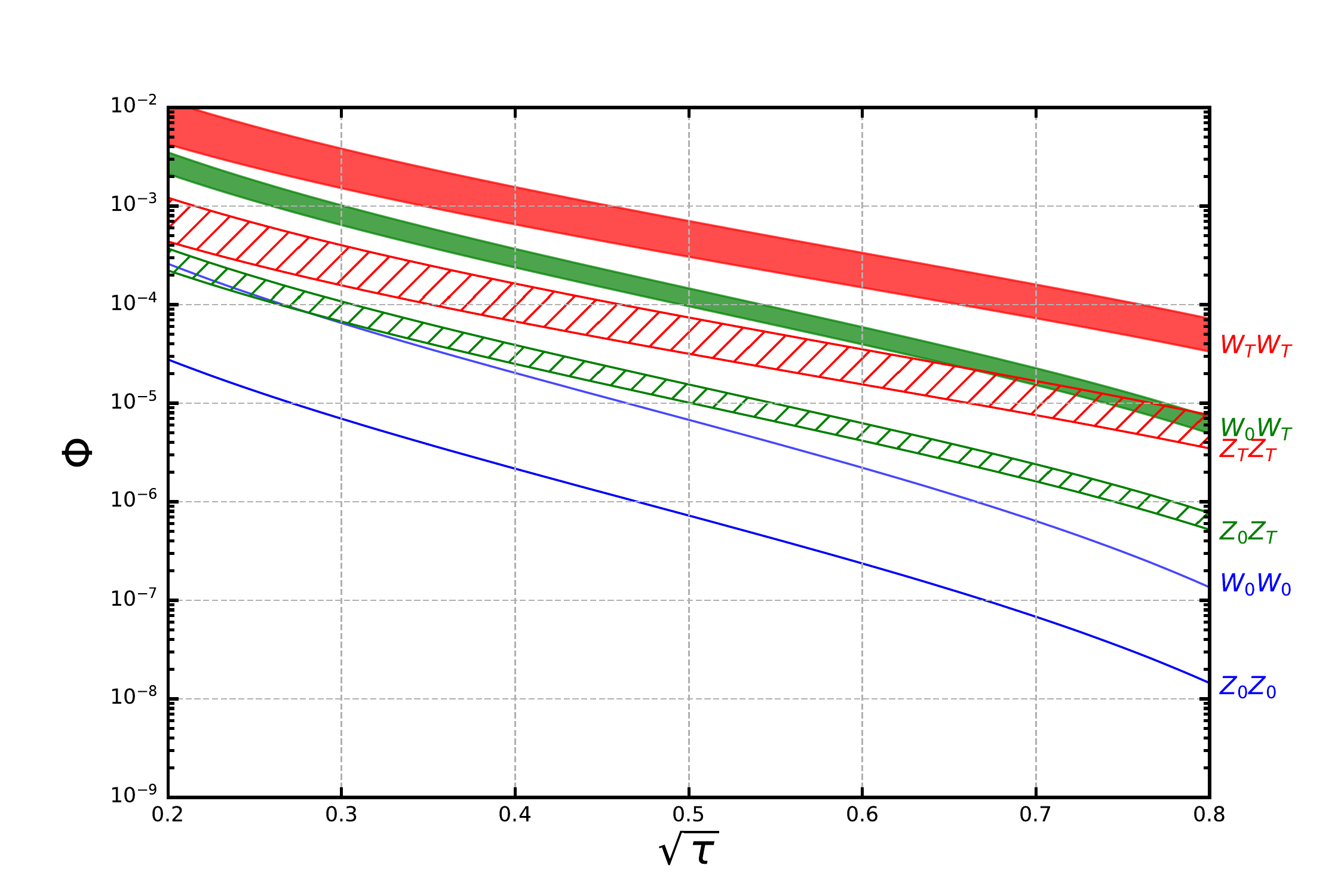}
\end{center}
\caption{
On the left figure, a plot of the parton luminosities of $WW$ bosons for the muon and proton collider.
We separate the contributions in terms of the helicity of the $W$. On the right, we display a comparison between
the $WW$ and $ZZ$ luminosities at a muon collider.
}
\label{fig:p_vs_muon_VBF}
\end{figure}
In Fig.~\ref{fig:p_vs_muon_VBF} we report the parton luminosities as a function of the scattering scale $\sqrt{\tau}=M_{VV'}/\sqrt{s}$ for muon and proton colliders. We separate the various contributions in terms of the helicity states. In addition to
that, because of the choice of the factorisation scale $\mu_f = \sqrt{\tau s}$, there is a dependence on the total collider energy
of the machine. To account for that, we plot envelopes varying the muon collider energy between $3-30$ TeV and the proton collider
energy between $14-200$ TeV. The range of $\sqrt{\tau}$ that we display is chosen to make sure that we are in the appropriate regime
for the EWA to be valid.

In the left panel of Fig.~\ref{fig:p_vs_muon_VBF}, a comparison between the $WW$ parton luminosities of a muon collider and a proton collider are presented.
As expected, the muon ones are not only higher in general, but the suppression with energy is less accentuated, \ie it is easier to
access high energy collisions. This feature is due to the fact that for proton colliders the emission of EW bosons is a
sub-leading phenomenon, while for muons they arise at the lowest order.

On the other hand, in the right panel of Fig.~\ref{fig:p_vs_muon_VBF} we show a comparison between the muon collider $WW$ and $ZZ$ luminosities.
We notice that the behaviour in $\sqrt{\tau}$ is substantially similar and that the difference is given by a constant factor of order
$10$.
This is simply the result of the fact that because of the quantum numbers of muons, the coupling to the $Z$ is suppressed and 
in particular we have
\begin{equation}
\cfrac{\Phi_{W_0W_0}(\tau)}{\Phi_{Z_0Z_0}(\tau)} = 
\left[\cfrac{\cos^2\theta_W}{\left(T^{3~\mu}_L- 2Q^\mu\sin^2{\theta_W}\right)^2 + \left(T^{3~\mu}_L\right)^2 }\right]^2
\approx
\cfrac{\cos^4\theta_W}{\left(T^{3~\mu}_L\right)^4 }
\approx 9.
\end{equation}
The mixed $WZ$ luminosities are not displayed, but they would have similar shapes and be located at the geometric mean of the $WW$
and $ZZ$ curves.

\section{SM processes at muon colliders}

In this section we discuss the production of various final states in the SM at a muon collider. Specifically, we compute cross sections for
generic final states of the form $X= n\, t \bar{t} + m\, V + k\, h $, where $n$, $m$ and $k$ are integers and correspond to the number of
top pairs, EW bosons and Higgs bosons respectively. We compute predictions for both s-channel and VBF production, with the goal
of determining the centre of mass energy at which VBF production becomes dominant. We do so for three different classes of
processes:
\begin{center}
\begin{tabular}{ll}
$\mu^+\mu^-\to X\, {\nu}_{\mu}\overline{\nu}_{\mu}$ &($WW$\, \textrm{fusion}),\nonumber\\
$\mu^+\mu^-\to X\, \mu^+\mu^-$&($ZZ/Z\gamma/\gamma\gamma$\, \rm{fusion}),\nonumber\\
$\mu^+\mu^-\to X\, \mu^\pm \overset{(-)}{\nu_\mu}$& ($WZ$\, \rm{fusion}).\nonumber
\label{procs}
\end{tabular}
\end{center}
\begin{figure}[t]
\centering
\includegraphics[width=0.30\textwidth]{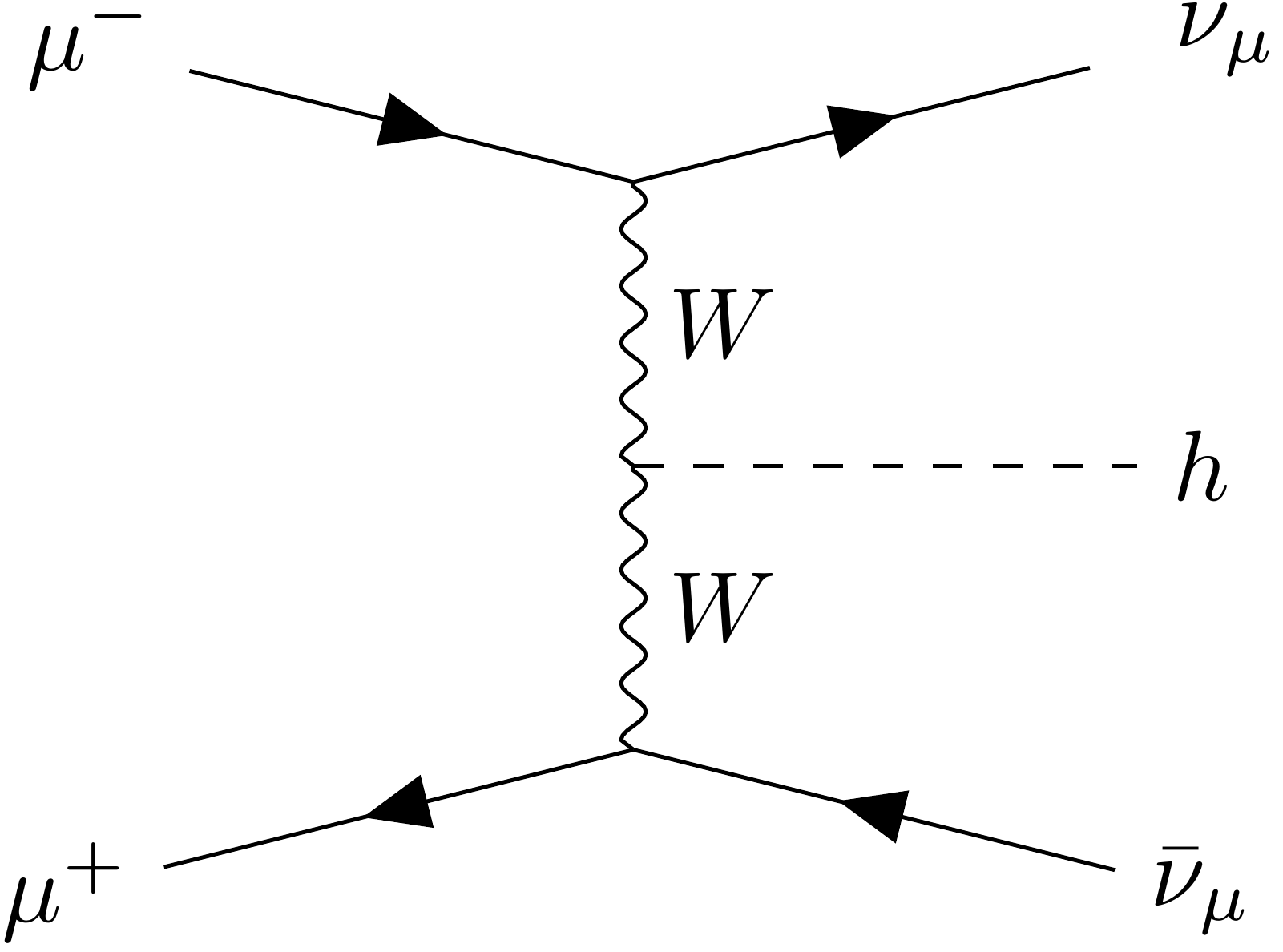} \qquad
\includegraphics[width=0.30\textwidth]{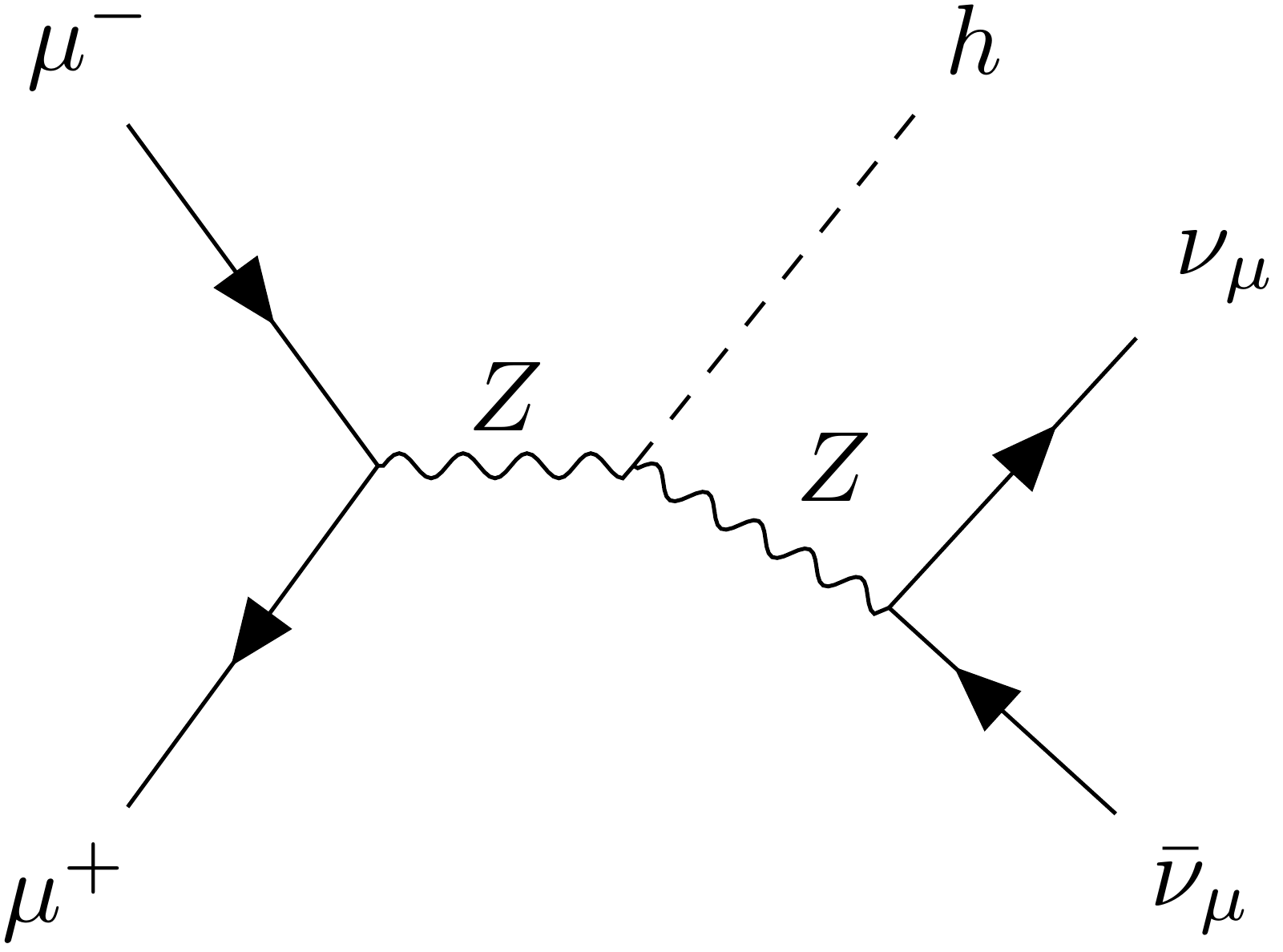}
\caption{\label{fig:singleH} Diagrams for single $h$ production in the $\mu^+\mu^-\to h\, {\nu}_{\mu}\overline{\nu}_{\mu}$ VBF channel. While
the diagram on the left is a pure VBF contribution from the $WW$ channel, the one on the right has a completely different topology
and is resonant when the $Z$ is on-shell.}
\end{figure}

Before presenting the results, a few comments are needed regarding technical aspects of computing cross sections at high energy.
Firstly, the aforementioned VBF topologies are "polluted" by s-channel diagrams with V-strahlung emission that then split in 
lepton pairs (see Fig.~\ref{fig:singleH} for an example). These contributions interfere with the VBF topologies, but are relevant in
completely different regions of phase space. While the former are enhanced at high energies, the latter are resonant when the $V$
vector boson is on-shell and therefore can be experimentally distinguished. In order to avoid double-counting from pure s-channel processes
and to make computations more efficient, we neglect the diagrams with $V$ boson decays by simulating $\mu^- e^+$ collisions,
such that the net muon and electron flavour of the initial state is non-zero and those diagrams are forbidden. Another option,
which we verified yielding the same results, is to apply minimum invariant mass cuts on the final state lepton pairs.

Second, the treatment of unstable particles requires special care in terms of the inclusion of the decay widths $\Gamma$. In the fixed 
width scheme, these are responsible for the breaking of gauge invariance at high energy and lead to incorrect results. One option is
to compute observables in the complex mass scheme~\cite{Denner:1999gp,Denner:2005fg}, an option that is available in MG5\_aMC.
However, in this scheme all unstable particles cannot be external legs and this would considerably complicate our calculations.
We choose instead to set the widths of $W,Z,H,t$ to zero, since no resonant s-channel propagators are expected in VBF production.

\subsection{\texorpdfstring{$W^+ W^-$}{WW} fusion}

\begin{figure}[!t]
\centering
\includegraphics[width=0.45\textwidth]{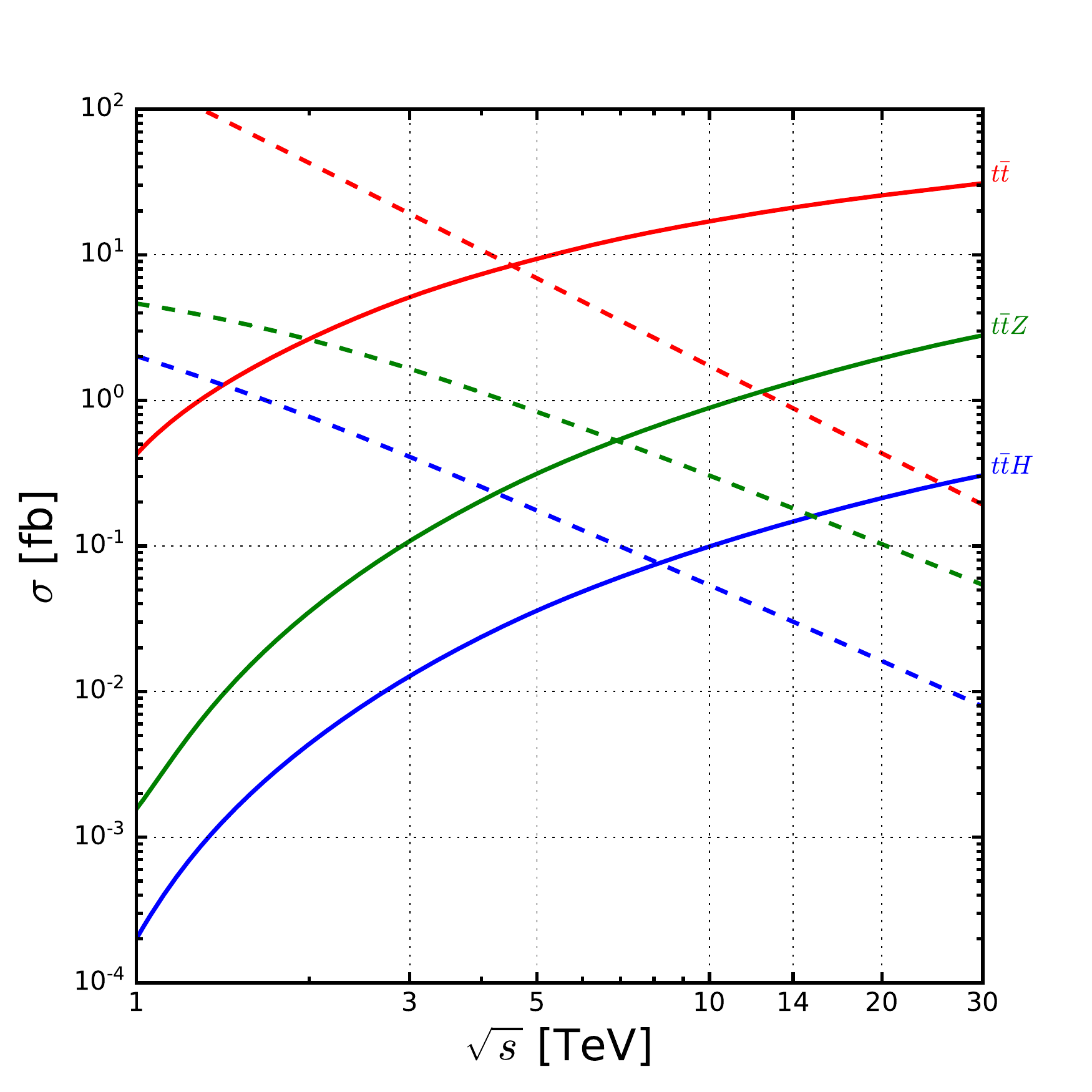}
\includegraphics[width=0.45\textwidth]{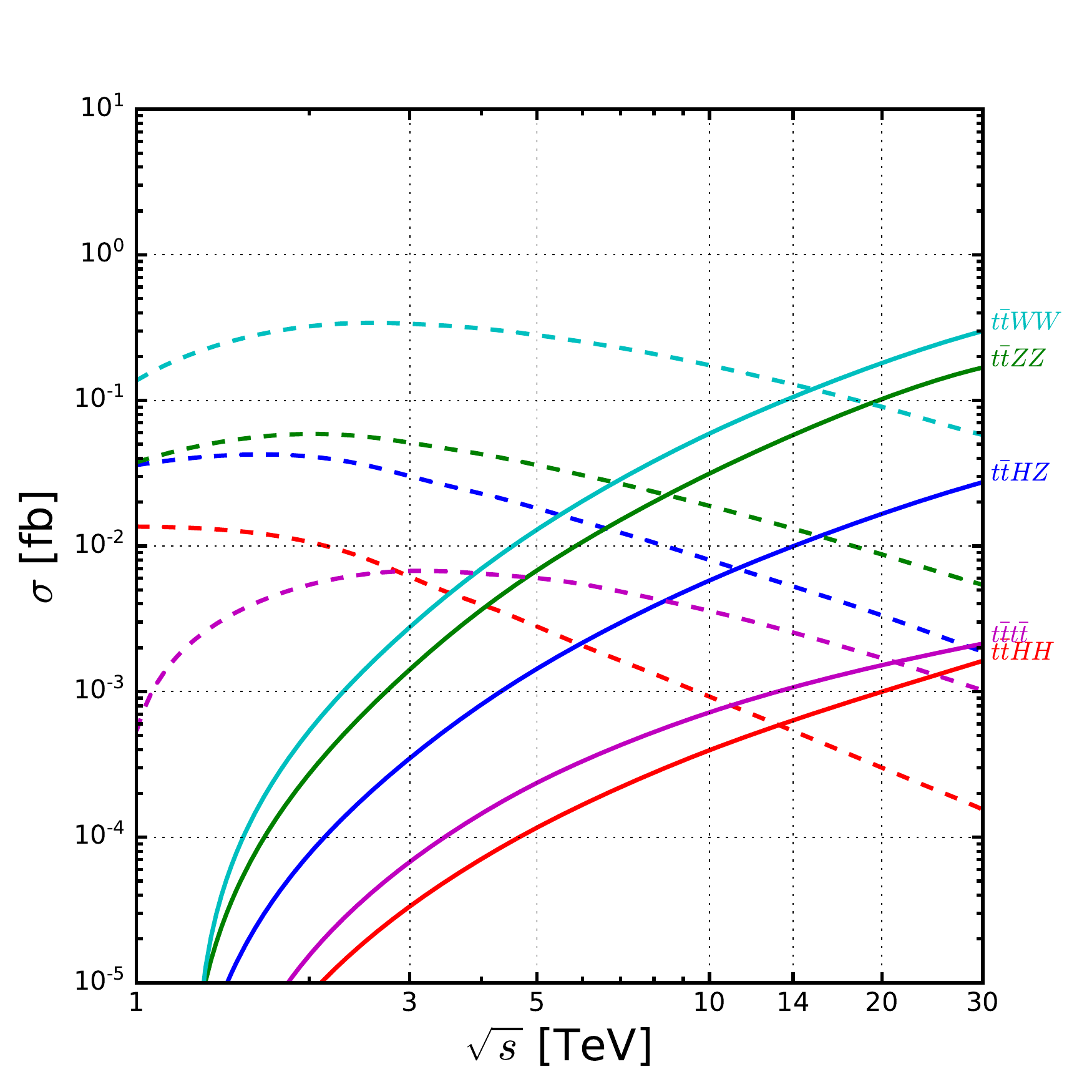}\\
\includegraphics[width=0.45\textwidth]{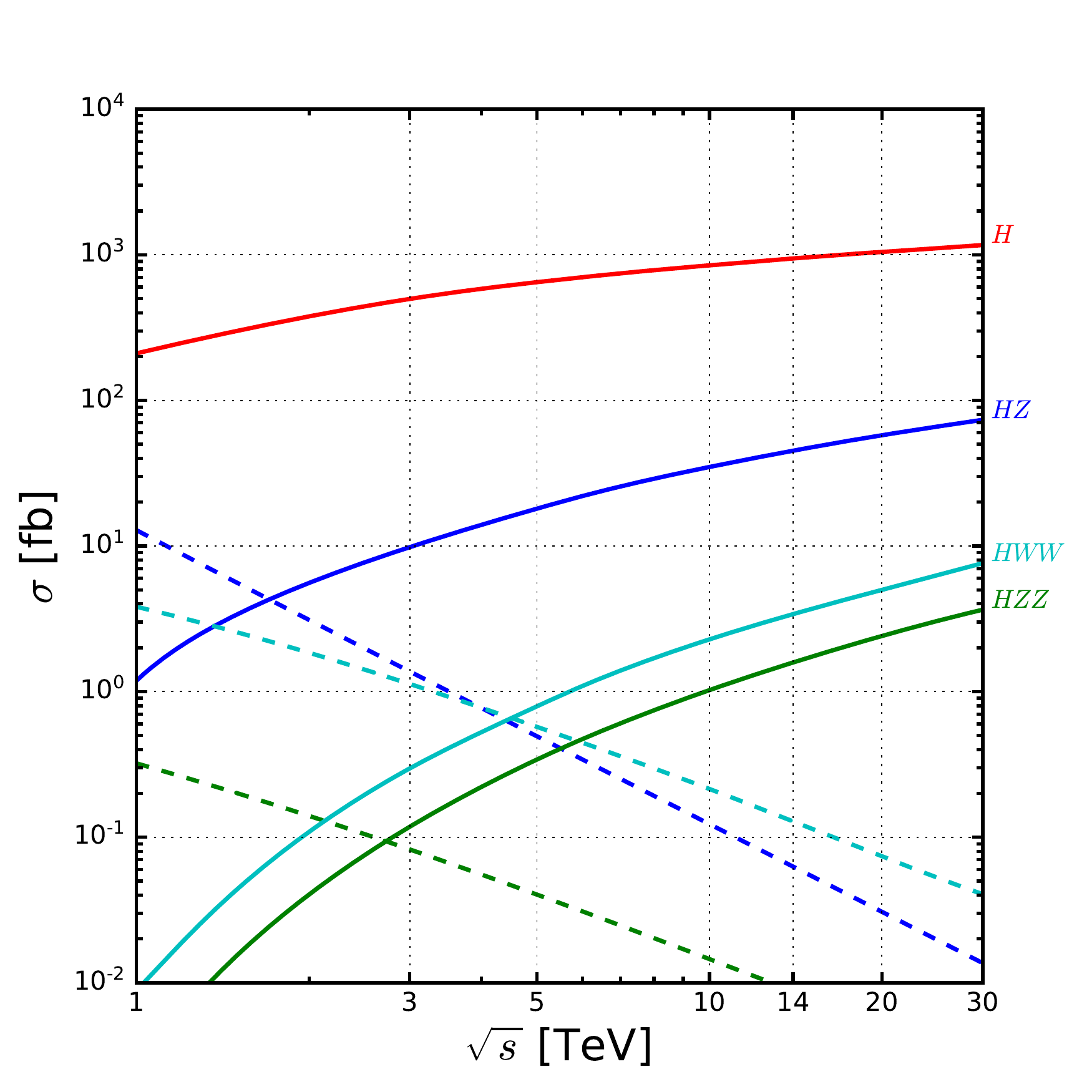}
\includegraphics[width=0.45\textwidth]{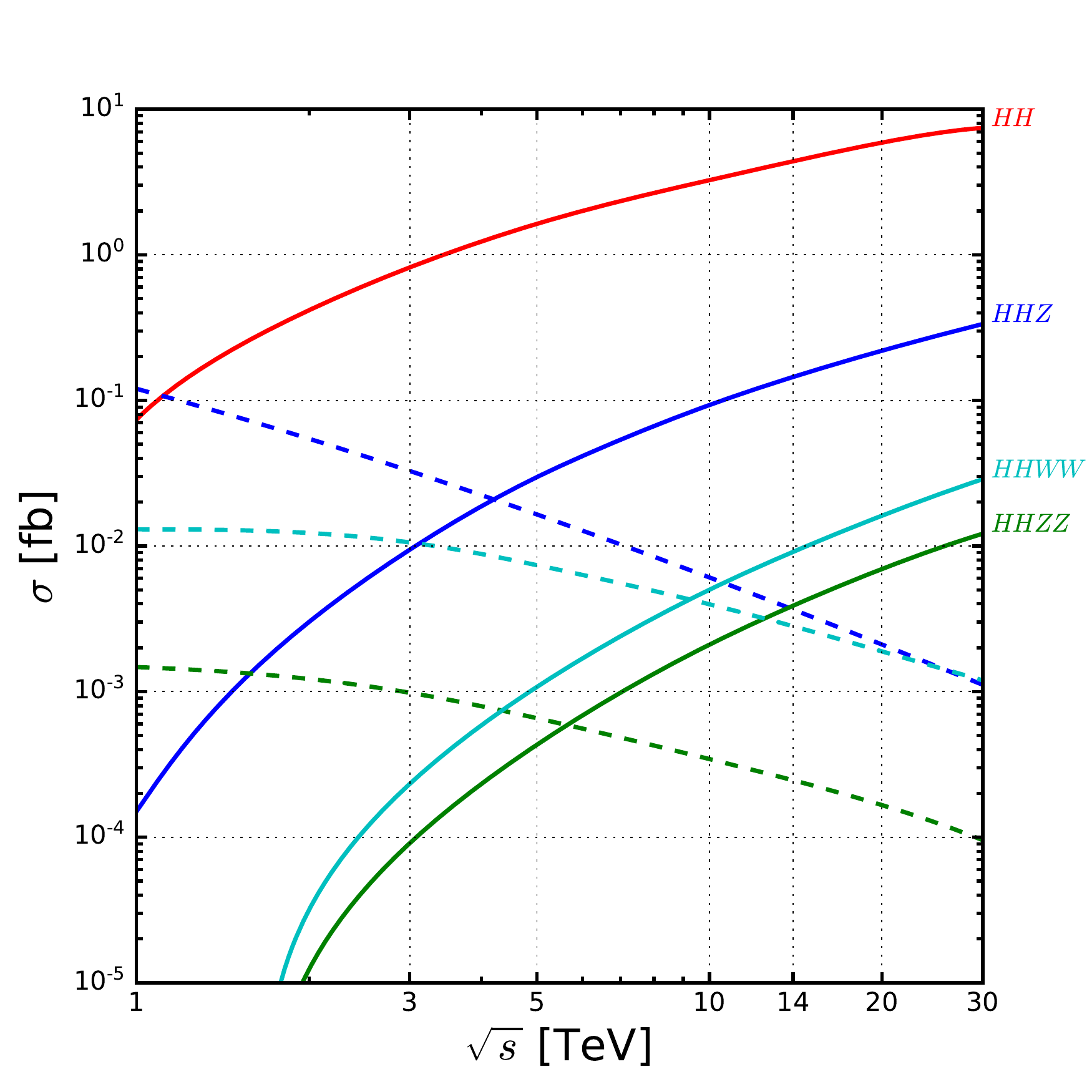}\\
\includegraphics[width=0.45\textwidth]{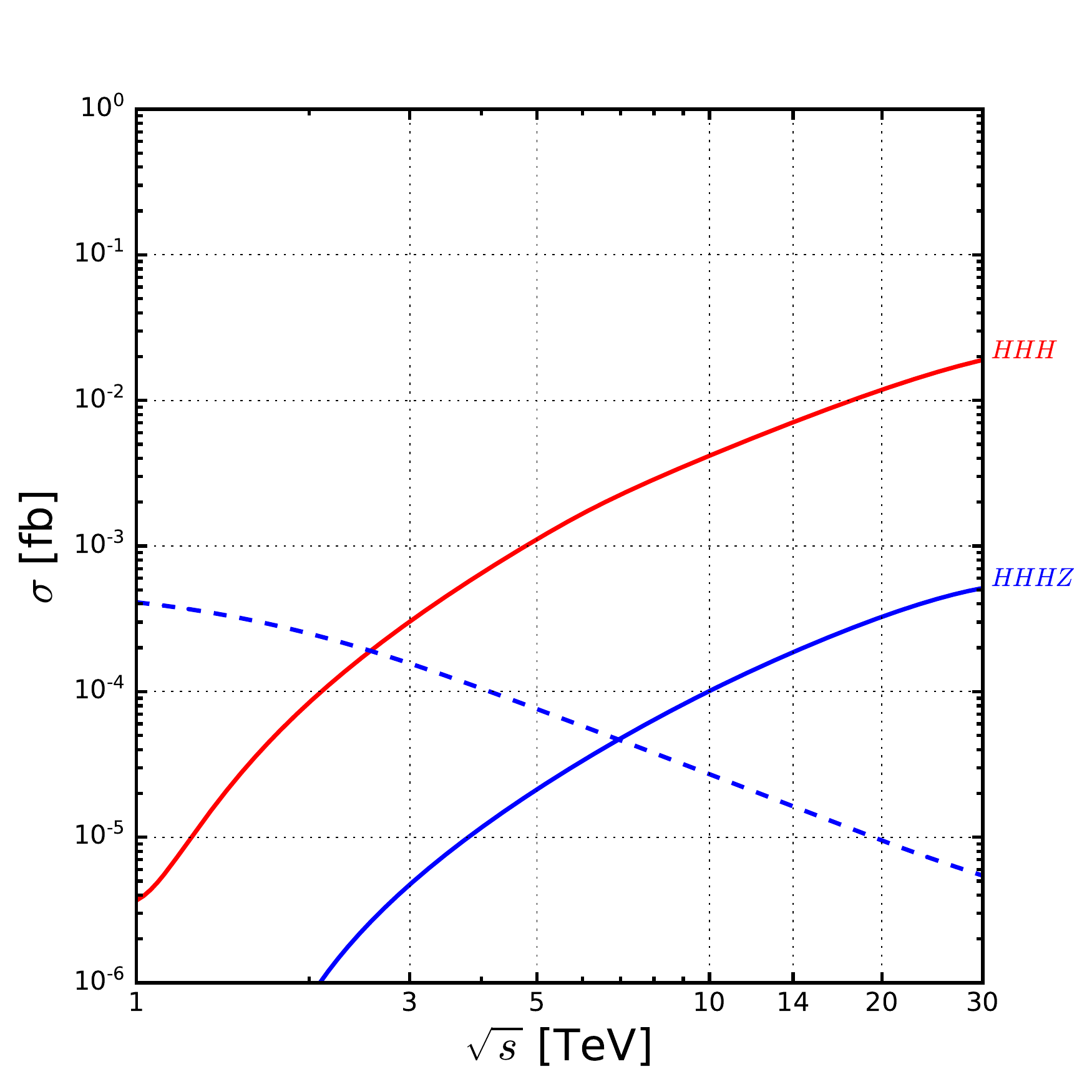}
\includegraphics[width=0.45\textwidth]{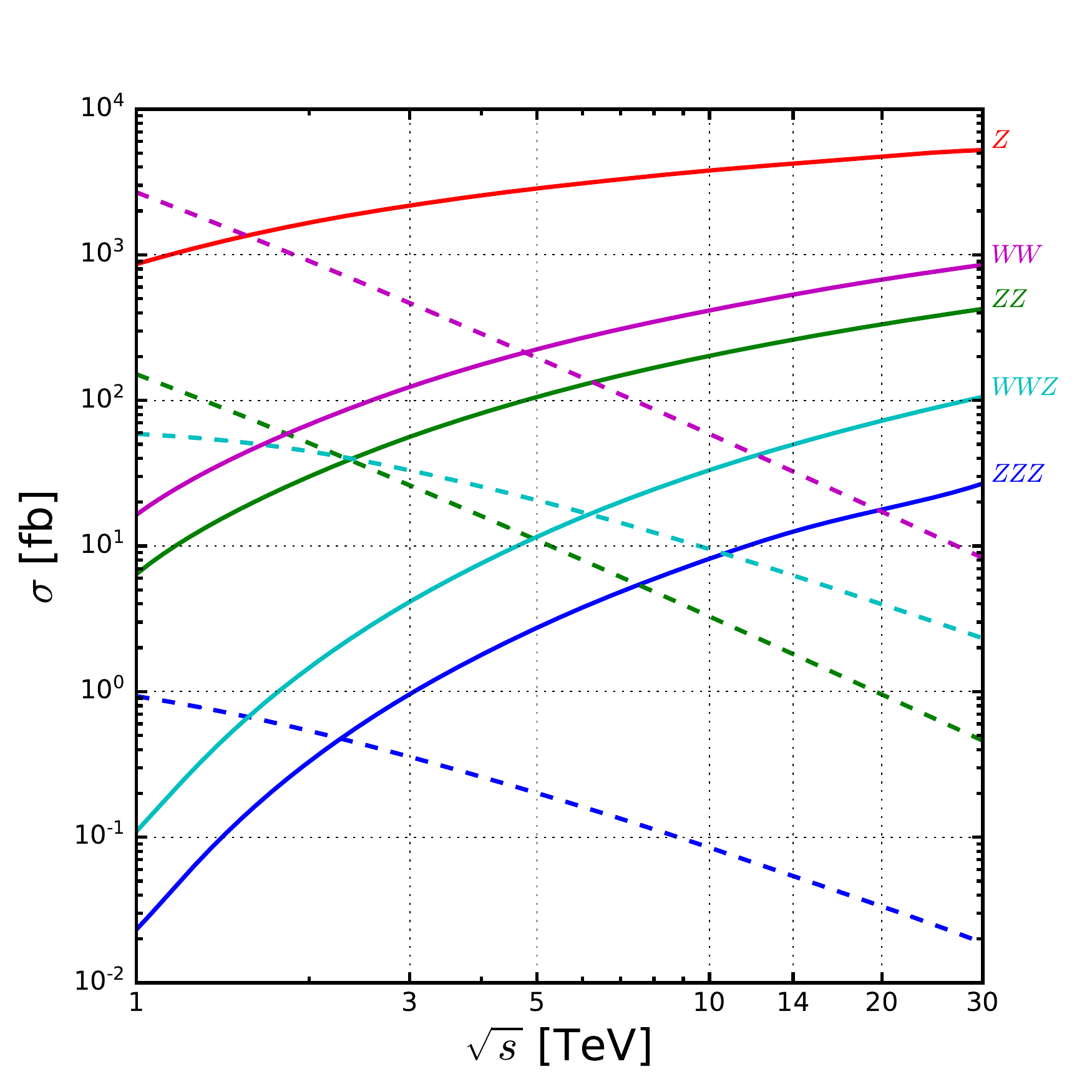}
\caption{SM cross sections for $WW$ initiated processes as a function of the muon collider energy. The corresponding s-channel 
cross sections are displayed with dashed curves.
}
\label{fig:SM_WW}
\end{figure}

\begin{table}[!t]
\begin{center}
\resizebox{\textwidth}{!}{
\begin{tabular}{l||cc||cc||cc||cc}
\toprule
\toprule
\multirow{2}{*}{$\sigma$ [fb]} &        \multicolumn{2}{c}{$\sqrt s=$ 1 TeV} &    \multicolumn{2}{c}{$\sqrt s=$ 3 TeV} &    \multicolumn{2}{c}{$\sqrt s=$ 14 TeV} &      \multicolumn{2}{c}{$\sqrt s=$ 30 TeV}\\
        &VBF&s-ch.&VBF&s-ch.&VBF&s-ch.&VBF&s-ch\\
\midrule
\midrule
$t \bar{t}$             &  4.3$\cdot 10^{-1}$ &1.7$\cdot 10^{2}$ &  5.1$\cdot 10^{0}$ &1.9$\cdot 10^{1}$ &  2.1$\cdot 10^{1}$ &8.8$\cdot 10^{-1}$ &  3.1$\cdot 10^{1}$ &1.9$\cdot 10^{-1}$ \\
\hline
$t \bar{t} Z$           &  1.6$\cdot 10^{-3}$ &4.6$\cdot 10^{0}$ &  1.1$\cdot 10^{-1}$ &1.6$\cdot 10^{0}$ &  1.3$\cdot 10^{0}$ &1.8$\cdot 10^{-1}$ &  2.8$\cdot 10^{0}$ &5.4$\cdot 10^{-2}$ \\
$t \bar{t} H$           &  2.0$\cdot 10^{-4}$ &2.0$\cdot 10^{0}$ &  1.3$\cdot 10^{-2}$ &4.1$\cdot 10^{-1}$ &  1.5$\cdot 10^{-1}$ &3.0$\cdot 10^{-2}$ &  3.1$\cdot 10^{-1}$ &7.9$\cdot 10^{-3}$ \\
\hline
$t \bar{t} W W$         &  4.8$\cdot 10^{-6}$ &1.4$\cdot 10^{-1}$ &  2.8$\cdot 10^{-3}$ &3.4$\cdot 10^{-1}$ &  1.1$\cdot 10^{-1}$ &1.3$\cdot 10^{-1}$ &  3.0$\cdot 10^{-1}$ &5.8$\cdot 10^{-2}$ \\
$t \bar{t} Z Z$         &  2.3$\cdot 10^{-6}$ &3.8$\cdot 10^{-2}$ &  1.4$\cdot 10^{-3}$ &5.1$\cdot 10^{-2}$ &  5.8$\cdot 10^{-2}$ &1.3$\cdot 10^{-2}$ &  1.7$\cdot 10^{-1}$ &5.4$\cdot 10^{-3}$ \\
$t \bar{t} H Z$         &  7.1$\cdot 10^{-7}$ &3.6$\cdot 10^{-2}$ &  3.5$\cdot 10^{-4}$ &3.0$\cdot 10^{-2}$ &  1.0$\cdot 10^{-2}$ &5.3$\cdot 10^{-3}$ &  2.7$\cdot 10^{-2}$ &1.9$\cdot 10^{-3}$ \\
$t \bar{t} H H$         &  7.2$\cdot 10^{-8}$ &1.4$\cdot 10^{-2}$ &  3.4$\cdot 10^{-5}$ &6.1$\cdot 10^{-3}$ &  6.4$\cdot 10^{-4}$ &5.4$\cdot 10^{-4}$ &  1.6$\cdot 10^{-3}$ &1.5$\cdot 10^{-4}$ \\
\hline
$t \bar{t} t \bar{t}\;\;(i)$    &  5.1$\cdot 10^{-8}$ &5.4$\cdot 10^{-4}$ &  6.8$\cdot 10^{-5}$ &6.7$\cdot 10^{-3}$ &  1.1$\cdot 10^{-3}$ &2.5$\cdot 10^{-3}$ &  2.1$\cdot 10^{-3}$ &1.0$\cdot 10^{-3}$ \\
$t \bar{t} t \bar{t}\;\;(ii)$   &  6.2$\cdot 10^{-9}$ &7.9$\cdot10^{-4}$ & 3.7$\cdot 10^{-5}$ &6.9$\cdot10^{-3}$&  1.7$\cdot 10^{-3}$ &2.3$\cdot10^{-3}$ &  4.7$\cdot 10^{-3}$ &9.0$\cdot10^{-4}$ \\
\midrule
\midrule
$H$                     &        2.1$\cdot 10^{2}$ &- &        5.0$\cdot 10^{2}$ &- &        9.4$\cdot 10^{2}$ &- &        1.2$\cdot 10^{3}$ &- \\
$H H$                   &        7.4$\cdot 10^{-2}$ &- &        8.2$\cdot 10^{-1}$ &- &        4.4$\cdot 10^{0}$ &- &        7.4$\cdot 10^{0}$ &- \\
$H H H$                 &        3.7$\cdot 10^{-6}$ &- &        3.0$\cdot 10^{-4}$ &- &        7.1$\cdot 10^{-3}$ &- &        1.9$\cdot 10^{-2}$ &- \\
\hline
$H Z$                   &  1.2$\cdot 10^{0}$ &1.3$\cdot 10^{1}$ &  9.8$\cdot 10^{0}$ &1.4$\cdot 10^{0}$ &  4.5$\cdot 10^{1}$ &6.3$\cdot 10^{-2}$ &  7.4$\cdot 10^{1}$ &1.4$\cdot 10^{-2}$ \\
$H H Z$                 &  1.5$\cdot 10^{-4}$ &1.2$\cdot 10^{-1}$ &  9.4$\cdot 10^{-3}$ &3.3$\cdot 10^{-2}$ &  1.4$\cdot 10^{-1}$ &3.7$\cdot 10^{-3}$ &  3.3$\cdot 10^{-1}$ &1.1$\cdot 10^{-3}$ \\
$H H H Z$               &  1.5$\cdot 10^{-8}$ &4.1$\cdot 10^{-4}$ &  4.7$\cdot 10^{-6}$ &1.6$\cdot 10^{-4}$ &  1.9$\cdot 10^{-4}$ &1.6$\cdot 10^{-5}$ &  5.1$\cdot 10^{-4}$ &5.4$\cdot 10^{-6}$ \\
\hline
$H W W$                 &  8.9$\cdot 10^{-3}$ &3.8$\cdot 10^{0}$ &  3.0$\cdot 10^{-1}$ &1.1$\cdot 10^{0}$ &  3.4$\cdot 10^{0}$ &1.3$\cdot 10^{-1}$ &  7.6$\cdot 10^{0}$ &4.1$\cdot 10^{-2}$ \\
$H H W W$               &  7.2$\cdot 10^{-7}$ &1.3$\cdot 10^{-2}$ &  2.3$\cdot 10^{-4}$ &1.1$\cdot 10^{-2}$ &  9.1$\cdot 10^{-3}$ &2.8$\cdot 10^{-3}$ &  2.9$\cdot 10^{-2}$ &1.2$\cdot 10^{-3}$ \\
\hline
$H Z Z$                 &  2.7$\cdot 10^{-3}$ &3.2$\cdot 10^{-1}$ &  1.2$\cdot 10^{-1}$ &8.2$\cdot 10^{-2}$ &  1.6$\cdot 10^{0}$ &8.8$\cdot 10^{-3}$ &  3.7$\cdot 10^{0}$ &2.5$\cdot 10^{-3}$ \\
$H H Z Z$               &  2.4$\cdot 10^{-7}$ &1.5$\cdot 10^{-3}$ &  9.1$\cdot 10^{-5}$ &9.8$\cdot 10^{-4}$ &  3.9$\cdot 10^{-3}$ &2.5$\cdot 10^{-4}$ &  1.2$\cdot 10^{-2}$ &9.5$\cdot 10^{-5}$ \\
\midrule
\midrule
$W W$       &1.6$\cdot 10^{1}$&2.7$\cdot 10^{3}$&1.2$\cdot 10^{2}$&4.7$\cdot 10^{2}$&5.3$\cdot 10^{2}$&3.2$\cdot 10^{1}$&8.5$\cdot 10^{2}$&8.3$\cdot 10^{0}$\\
$Z Z$       &6.4$\cdot 10^{0}$&1.5$\cdot 10^{2}$&5.6$\cdot 10^{1}$&2.6$\cdot 10^{1}$&2.6$\cdot 10^{2}$&1.8$\cdot 10^{0}$&4.2$\cdot 10^{2}$&4.6$\cdot 10^{-1}$\\
\hline
$W W Z$     &1.1$\cdot 10^{-1}$&5.9$\cdot 10^{1}$&4.1$\cdot 10^{0}$&3.3$\cdot 10^{1}$&5.0$\cdot 10^{1}$&6.3$\cdot 10^{0}$&1.0$\cdot 10^{2}$&2.3$\cdot 10^{0}$\\
$Z Z Z$     &2.3$\cdot 10^{-2}$&9.3$\cdot 10^{-1}$&9.6$\cdot 10^{-1}$&3.5$\cdot 10^{-1}$&1.2$\cdot 10^{1}$&5.4$\cdot 10^{-2}$&2.7$\cdot 10^{1}$&1.9$\cdot 10^{-2}$\\
\bottomrule
\bottomrule
\end{tabular}
}
\caption{Summary of the $WW$ initiated cross sections for specific benchmark collider energies.}
\label{tab:neutralSM}
\end{center}
\end{table}

The most important channel in terms of magnitude in VBF production is $WW$ scattering, as we saw in Fig.~\ref{fig:p_vs_muon_VBF}.
We survey different final states with up to four final state particles, plotting the cross sections $\sigma$ as a function of
the muon collider energy $\sqrt{s}$ (see Fig.~\ref{fig:SM_WW}) for both VBF and s-channel production. Our findings are
also summarised for benchmark collider energies in Table~\ref{tab:neutralSM}.

In summary, we observe as expected that the s-channel production mode is decreasing sharply in energy and is eventually 
overtaken by the VBF production which instead grows logarithmically with powers of $\log(s/M_V^2)$. In general, the larger the multiplicity
of the final state, the later the two curves cross. As a matter of fact, the energy growth is steeper for larger particle
multiplicities, since the collinear enhancement is related to the number of t-channel propagators, but the higher invariant mass threshold
requires more collider energy to open up the phase space and start taking advantage of it. However, at high energies the 
rise in cross section is in general proportional to $\sigma \sim \log^m (s/M_V^2)$, with $m$ number of final state particles.

Lastly, it is worth noting that for top quark pair processes, the s-channel cross sections at $\mathcal{O}(1)$ TeV are higher
than the corresponding ones at several TeV in VBF production. The natural question that arises is whether it makes sense to
go to high $\sqrt{s}$ when in terms of pure statistical precision we are better off at low energy. However, as discussed at length in
Chapter~\ref{chap:top}, the sensitivity to NP effects increases at high energies, especially so for VBF processes. Additionally, VBF is able to access more interactions. For example, in the case of $t\bar{t}$ production, while $\mpmm\to \gamma^*/Z^* \to t\overline{t}$ is sensitive only to $ttZ/\gamma^*$ and $\mu\mu t t$ anomalous couplings, in $W^+W^- \to  t\overline{t}$ we are sensitive to
modified interactions coming from the $WWh$, $tth$, $WWtt$ and $tWb$ vertices.

\subsection{\texorpdfstring{$ZZ$, $Z\gamma$, and $\gamma\gamma$}{ZZ, Zγ and γγ} fusion}

\begin{table}[!t]
\begin{center}
\renewcommand*{\arraystretch}{0.95}
\begin{tabular}{l||cccc}
\toprule
\toprule
$\sigma$ [fb] &  $\sqrt s=$ 1 TeV &  $\sqrt s=$ 3 TeV &  $\sqrt s=$ 14 TeV &  $\sqrt s=$ 30 TeV\\
\midrule
\midrule
$t \bar{t}$     &  1.0 $\cdot 10^{-1}$ &   1.1 $\cdot 10^{0}$ &   4.3 $\cdot 10^{0}$ &   6.2 $\cdot 10^{0}$ \\
\hline
$t \bar{t} Z$   &  1.2 $\cdot 10^{-4}$ &  6.7 $\cdot 10^{-3}$ &  5.2 $\cdot 10^{-2}$ &  8.5 $\cdot 10^{-2}$ \\
$t \bar{t} H$   &  5.3 $\cdot 10^{-5}$ &  2.8 $\cdot 10^{-3}$ &  2.7 $\cdot 10^{-2}$ &  5.0 $\cdot 10^{-2}$ \\
\hline
\hline
$H$             &   1.5 $\cdot 10^{1}$ &   3.8 $\cdot 10^{1}$ &   7.6 $\cdot 10^{1}$ &   9.6 $\cdot 10^{1}$ \\
$H H$           &  5.0 $\cdot 10^{-3}$ &  7.3 $\cdot 10^{-2}$ &  4.3 $\cdot 10^{-1}$ &  7.5 $\cdot 10^{-1}$ \\
$H H H$         &  3.6 $\cdot 10^{-7}$ &  3.1 $\cdot 10^{-5}$ &  8.4 $\cdot 10^{-4}$ &  2.3 $\cdot 10^{-3}$ \\
\hline
$H W W$         &  3.5 $\cdot 10^{-3}$ &  1.4 $\cdot 10^{-1}$ &   1.7 $\cdot 10^{0}$ &   3.8 $\cdot 10^{0}$ \\
$H Z Z$         &  2.5 $\cdot 10^{-5}$ &  4.9 $\cdot 10^{-4}$ &  3.6 $\cdot 10^{-3}$ &  5.9 $\cdot 10^{-3}$ \\
\hline
\hline
$W W$           &   2.2 $\cdot 10^{1}$ &   1.4 $\cdot 10^{2}$ &   5.2 $\cdot 10^{2}$ &   8.1 $\cdot 10^{2}$ \\
$Z Z$           &  1.2 $\cdot 10^{-1}$ &  4.0 $\cdot 10^{-1}$ &  7.4 $\cdot 10^{-1}$ &  8.0 $\cdot 10^{-1}$ \\
\bottomrule
\bottomrule
\end{tabular}
\caption{
As in Table~\ref{tab:neutralSM} for $ZZ$, $Z\gamma$ and $\gamma \gamma$ fusion.
}
\label{tab:ZZSM}
\end{center}
\end{table}

In Table~\ref{tab:ZZSM} we report the cross sections for a selection of neutrally initiated processes.
\revised{In order to regulate infrared divergences related to virtual photons in the t-channel}, we impose a minimum $p_T$ cut of $30$ GeV for the charged final state leptons.
As expected from the plots in Fig.~\ref{fig:p_vs_muon_VBF} for the VBF luminosities, the $ZZ$ initiated processes are roughly
an order of magnitude smaller than the corresponding $WW$ ones. However, these rates are not small enough to be neglected. 
Furthermore, the presence of detectable charged particles in the final state opens a new window to fully characterise the
reconstruction of the event. For instance, in the case of invisible production such as $WW/ZZ \to h \to 4\nu$, we would have
a completely invisible final state in the case of $WW$ initiated production, while $ZZ$ gives ways to tag the process.

\subsection{\texorpdfstring{$WZ$ and $W\gamma$}{WZ and Wγ} fusion}

\begin{table}[t!]
\begin{center}
\renewcommand*{\arraystretch}{0.95}
\begin{tabular}{l||cccccccc}
\toprule
\toprule
\multirow{1}{*}{$\sigma$ [fb]} &        \multicolumn{1}{c}{$\sqrt s=$ 1 TeV} &    \multicolumn{1}{c}{$\sqrt s=$ 3 TeV} &    \multicolumn{1}{c}{$\sqrt s=$ 14 TeV} &      \multicolumn{1}{c}{$\sqrt s=$ 30 TeV}\\
\midrule 
\midrule 
$W$&$9.9\cdot 10^2$& $2.4\cdot 10^3$& $4.6\cdot 10^3$& $5.7\cdot 10^3$\\
\hline
$WZ$&$5.8\cdot 10^0$& $5.0\cdot 10^1$& $2.3\cdot 10^2$& $3.7\cdot 10^2$\\
$WH$&$8.4\cdot 10^{-1}$& $7.2\cdot 10^0$& $3.3\cdot 10^1$& $5.5\cdot 10^1$\\
\hline
$WWW$&$1.4\cdot 10^{-1}$& $4.2\cdot 10^0$& $4.4\cdot 10^1$& $1.0\cdot 10^2$\\
$WZZ$&$1.8\cdot 10^{-2}$& $8.0\cdot 10^{-1}$& $1.0\cdot 10^1$& $2.3\cdot 10^1$\\
$WZH$&$1.7\cdot 10^{-3}$& $8.0\cdot 10^{-2}$& $1.1\cdot 10^0$& $2.5\cdot 10^0$\\
$WHH$&$9.5\cdot 10^{-5}$& $6.2\cdot 10^{-3}$& $9.7\cdot 10^{-2}$& $2.3\cdot 10^{-1}$\\
\hline\hline
$t\bar b$&$4.4\cdot 10^{-1}$& $2.9\cdot 10^0$& $9.5\cdot 10^0$& $1.3\cdot 10^1$\\
\hline
$t\bar b Z$&$1.3\cdot 10^{-3}$& $4.4\cdot 10^{-2}$& $4.1\cdot 10^{-1}$& $8.0\cdot 10^{-1}$\\
$t\bar b H$&$1.5\cdot 10^{-4}$& $6.6\cdot 10^{-3}$& $6.6\cdot 10^{-2}$& $1.3\cdot 10^{-1}$\\
$t\bar tW$&$1.0\cdot 10^{-3}$& $7.6\cdot 10^{-2}$& $9.0\cdot 10^{-1}$& $1.9\cdot 10^0$\\
\bottomrule
\bottomrule
\end{tabular}
\caption{Same as table~\ref{tab:neutralSM} but for $WZ/W\gamma$ fusion.}
\label{tab:chargedSM}
\end{center}
\end{table}

Finally, we discuss the production of charged final states through $WZ$ and $W\gamma$ fusion. Our findings are summarised in 
Table~\ref{tab:chargedSM} for benchmark muon collider energies. A $p_T$ cut of $30$ GeV is applied on charged leptons in order
to regulate the collinear divergences from t-channel photon emission.

As expected, the cross sections for this class of processes is somewhere between the $WW$ and $ZZ$ ones. Since these processes give
rise to charged signals, we can expect qualitatively different final states that cannot be replicated with s-channel production.
For example, processes like single $W$ production, single top or triple gauge boson $WWW$ have all appreciable cross sections and lead
to unique signatures.

\section{Higgs self-couplings}

After discussing the direct discovery potential in Section~\ref{sec:p_vs_mu}, we now turn to explore the indirect detection prospects
at a muon collider. In particular, one of the most important quest that the community has to undertake in the future is the
precise determination of the Higgs boson properties~\cite{Strategy:2019vxc,EuropeanStrategyGroup:2020pow}. Specifically, while the couplings to the third generation fermions and EW bosons are 
in agreement with the SM predictions, interactions with light fermions still evade our scrutiny. Moreover, a determination of the
Higgs potential 
\begin{equation}
V(h)=\frac{1}{2}m_h^2h^2 + \lambda_3 v h^3 + \frac{1}{4} \lambda_4 h^4 \, ,
\end{equation}
where in the SM,  $\lambda_3=\lambda_4=m_H^2/2v^2\equiv\lambda_{SM}$, is of crucial importance to gain understanding
on the EW symmetry breaking mechanism~\cite{Chung:2012vg} and its role in the thermal history of the universe.

Unless sizeable deviations are foreseen in the Higgs self-interactions, measurements at the LHC appear to be arduous~\cite{Sirunyan:2017guj, CMS:2018rig, CMS:2018ccd, ATLAS:2018otd, Aaboud:2018zhh, Aad:2019uzh, Aad:2019yxi, Sirunyan:2018iwt, CMS:2018dvu, ATLAS:2019pbo, ATL-PHYS-PUB-2014-019, ATL-PHYS-PUB-2017-001, Kim:2018cxf}.
This motivates the need to build a future lepton collider, either at low or high energy, with the objective of gaining the precision
needed to determine these crucial Higgs properties. 

Higgs sensitivity studies at CLIC~\cite{Roloff:2018dqu, Vasquez:2019muw, Roloff:2019crr, Liu:2018peg, Maltoni:2018ttu, deBlas:2019rxi}
confirm that a multi-TeV lepton collider would increase dramatically our potential to determine the Higgs self-couplings.
As we saw already in Fig.~\ref{fig:SM_WW}, the production of single, double and even triple Higgs is greatly enhanced at high energies by means of
VBF production.
It is indeed expected that in the SM a $\sqrt{s}=10$ TeV collider with an integrated luminosity of $\mathcal{L}=10 \invab$
would yield $8\cdot10^6$ Higgs bosons. Under these conditions, given that the background is expected to be mostly under control,
a muon collider would effectively act as a Higgs boson factory.

In this section we focus on the prospects of determining the Higgs boson self-interactions at a muon collider in the context of the SMEFT at dimension-6, keeping a model-independent perspective.

\subsection{Higgs potential in the SMEFT}

In the SMEFT framework, at dimension-6, there are three operators that directly affect the Higgs potential:
\begin{equation}
\mathcal{O}_{\phi}, \quad \mathcal{O}_{\phi d}, \quad\text{and}\quad \mathcal{O}_{\phi D}.
\label{eq:eft_higgs_dim6ops}
\end{equation} 
All of the operators affect the Higgs self-couplings and therefore VBF Higgs production through 
the following Lagrangian terms:
\begin{align}
&\mathcal{O}_{\varphi}= \left(\varphi^\dagger \varphi - \frac{v^2}{2}\right)^3 \supset v^3 h^3 + \frac{3}{2}v^2 h^4, \label{eq:op_expansion_varphi}\\
&\mathcal{O}_{\varphi d}= \left(\varphi^\dagger\varphi\right)\Box \left(\varphi^\dagger\varphi\right) \supset 2v \left(h\Box h^2 + h^2\Box h\right) + h^2 \Box h^2, \\
& \mathcal{O}_{\varphi D}= \left(\varphi^\dagger D _\mu \varphi\right)^\dagger\left(\varphi^\dagger D^\mu\varphi\right) \supset \frac{v}{2}h \partial_\mu h \partial^\mu h +\frac{h^2}{4} \partial_\mu h \partial^\mu h \, .
\label{eq:op_expansion}
\end{align}
Notably, $\Op{\varphi}$ produces SM-like vertices, while $\mathcal{O}_{\phi d}$ and $\mathcal{O}_{\phi D}$ introduce derivative
couplings. 

\revised{It is of particular interest to briefly discuss the case of the operator $\mathcal{O}_{\varphi d}$. In Chapter~\ref{chap:globalfit} we
said that this operator only generates a kinetic term redefinition, while here we are saying
that it introduces new derivative vertices. Both statements are true and depend on the conventional choice of how to redefine the Higgs
field. The full expansion of the operator is given by
\begin{align}
\mathcal{O}_{\varphi d} &= \frac{c_{\varphi d}}{\Lambda^2} \left(\varphi^\dagger\varphi\right)\Box \left(\varphi^\dagger\varphi\right) \equiv  -\frac{c_{\varphi d}}{\Lambda^2} \partial_\mu \left(\varphi^\dagger\varphi\right) \partial^\mu \left(\varphi^\dagger\varphi\right) =\\
& -\frac{c_{\varphi d}}{\Lambda^2} \left(h^2 \partial_\mu h \partial^\mu h + 2 v \, h \, \partial_\mu h \partial^\mu h + v^2 \, \partial_\mu h \partial^\mu h \right) \, .
\end{align}
If we decide to retain the first two terms in the Lagrangian while reabsorb the last one in the kinetic term, we have
\begin{equation}
\Lag_{kin} = \left(\frac{1}{2} -\frac{c_{\varphi d} \, v^2}{\Lambda^2} \right)\partial_\mu h \partial^\mu h \, ,
\end{equation}
and if we redefine the Higgs field
\begin{equation}
h = \frac{h^\prime}{\left(1 -\frac{2 \, c_{\varphi d} \, v^2}{\Lambda^2} \right)^{\frac{1}{2}}} \approx h^\prime \left(1 +\frac{c_{\varphi d} \, v^2}{\Lambda^2} \right) \, ,
\end{equation}
the kinetic term is put back in canonical form. This redefinition has to be done in the whole Lagrangian and therefore wherever there is an Higgs field, the operator effect is to shift the coupling. In summary, with this choice we are left with both field redefinitions which 
shift SM Higgs couplings to EW bosons (and fermions) and the introduction of new vertices for Higgs self-couplings.
On the other hand, one could have chosen to reabsorb completely the operator in the kinetic term:
\begin{align}
\Lag_{kin} &= \left(\frac{1}{2} -\frac{c_{\varphi d}}{\Lambda^2} (v^2 + 2 \, v \, h + h^2) \right)\partial_\mu h \partial^\mu h \, ,\\
h &= \frac{h^\prime}{\left(1 -\frac{2 \, c_{\varphi d}}{\Lambda^2} (v^2 + 2 \, v \, h + h^2) \right)^{\frac{1}{2}}} \approx h^\prime \left(1 +\frac{c_{\varphi d}}{\Lambda^2} (v^2 + 2 \, v \, h^\prime + h^{\prime 2}) \right) \, .
\end{align}
This kind of non-linear field redefinition is allowed because we are in an EFT. In this case, we do not have anymore new vertices for the Higgs self-couplings, but the field redefinition is more complex and will introduce for instance a $VVhhh$ vertex not present in the SM.
The two descriptions are equivalent and lead eventually to the same physics and an energy growing behaviour at the amplitude level. In one case this is explicit from the introduction of derivative couplings, while in the other it is an effect of the more complex spoiling of the
cancellations among diagrams involving $VVh$, $VVhh$ and $hhh$ vertices.}

In the following, for conciseness, we will focus only on $\Op{\varphi}$ and $\Op{\varphi d}$ and their effect on the Higgs self-interactions.
We decide to neglect $\Op{\varphi D}$ given that it modifies also gauge boson couplings and is therefore highly constrained by EWPO,
as discussed at length in Chapter~\ref{chap:globalfit}. For the purpose of this analysis, we consider three benchmark scenarios 
for the muon collider in which the centre of mass energy is $\sqrt{s}=3, 14$, and $30$ TeV and the luminosities $\mathcal{L}=6, 20$, and $100$ $\invab$ respectively.
\begin{figure}[t]
\centering
\includegraphics[width=0.30\textwidth]{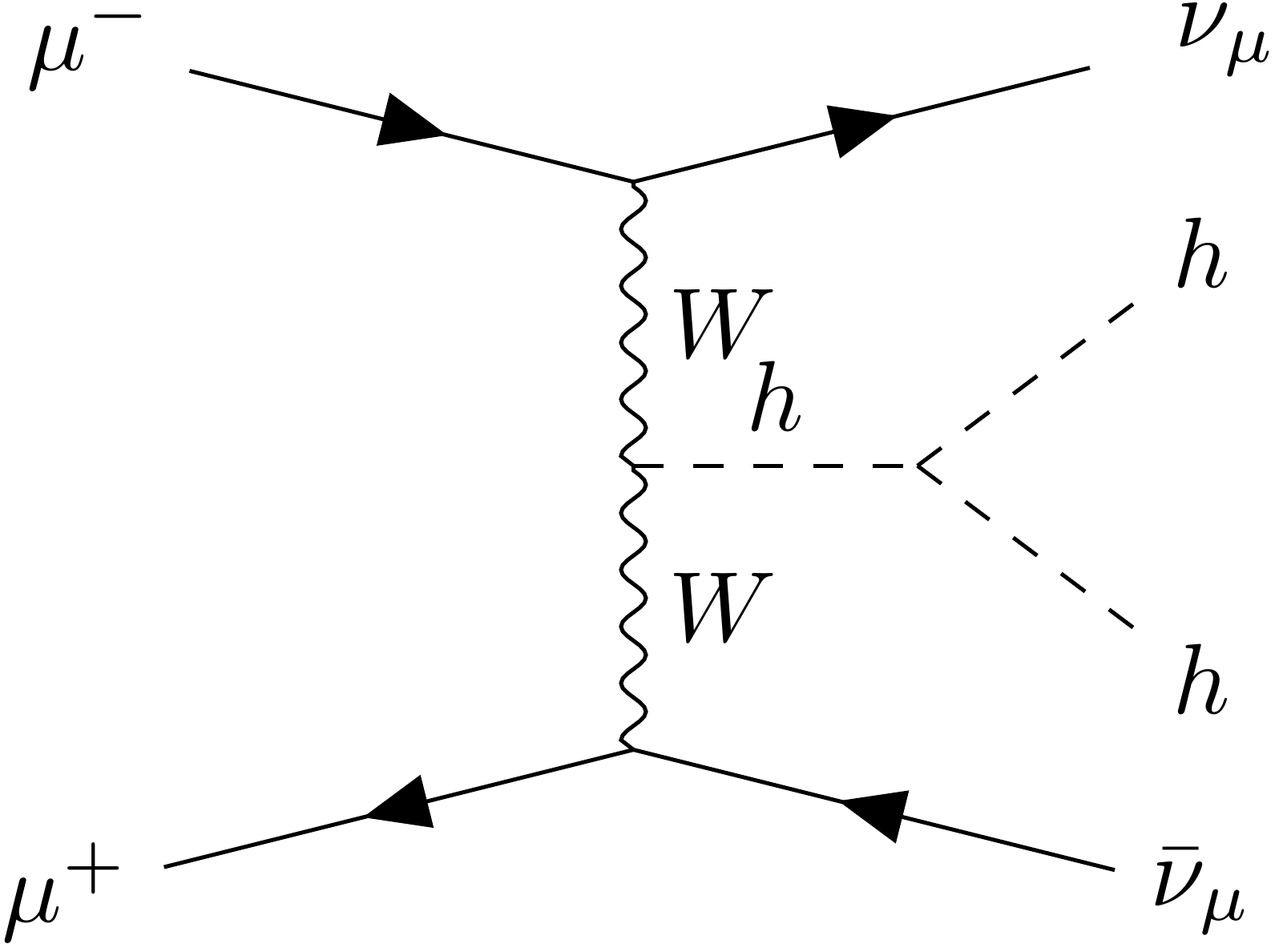} \qquad
\includegraphics[width=0.30\textwidth]{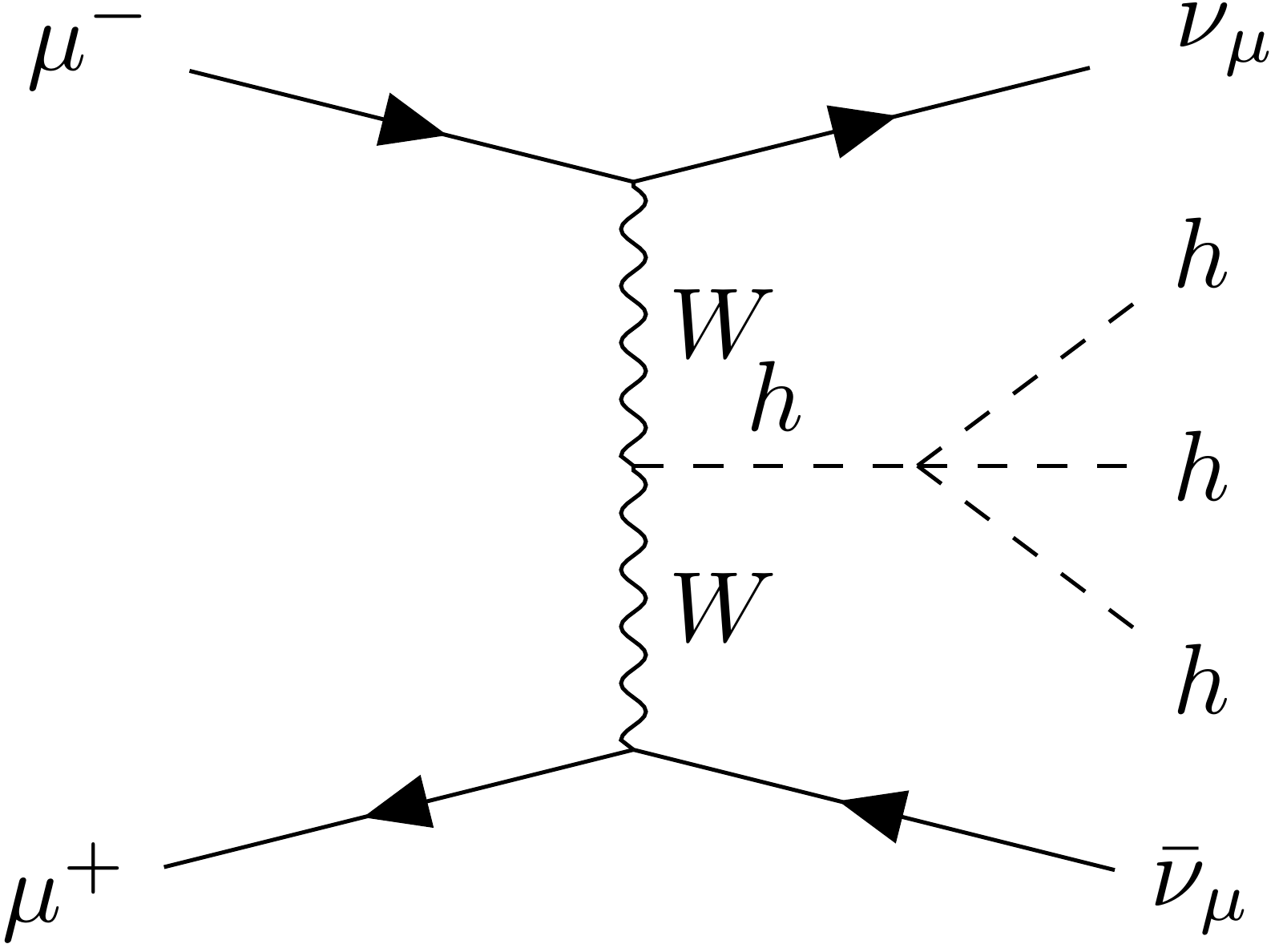}
\caption{\label{fig:higgs_diagrams} Example of diagrams accessing trilinear and quartic coupling in double Higgs production (left) and
triple Higgs production (right).}
\end{figure}

We begin by investigating the effects of the operators individually, fixing the other Wilson coefficients to zero. The total cross section of a
process in the SMEFT is given by
\begin{equation}
\sigma = \sigma_{ SM} + \sum_i c_i \sigma_{ Int}^i + \sum_{i,j} c_{i,j} \sigma_{ Sq}^{i,j} \, ,
\label{eq:eft_xsec_smeft}
\end{equation}
where $\sigma_{ Int}^i$ are the interference terms and $\sigma_{ Sq}^{i,j}$ the purely NP interactions. For concreteness we
define a naive measure of sensitivity by defining the ratio
\begin{equation}
R(c_i)\equiv \frac{\sigma}{\sigma_{ SM}}= 1+ c_i\frac{\sigma_{ Int}^i}{\sigma_{ SM}} + c_{i,i}^2 \frac{\sigma_{ Sq}^{i,i}}{\sigma_{ SM}} = 1 + r_i + r_{i,i} \, ,
\label{Eq:sens_ratio}
\end{equation}
as we previously did in Section~\ref{sec:blueprint}. 
\begin{figure}[t]
\centering
\includegraphics[width=.45\textwidth]{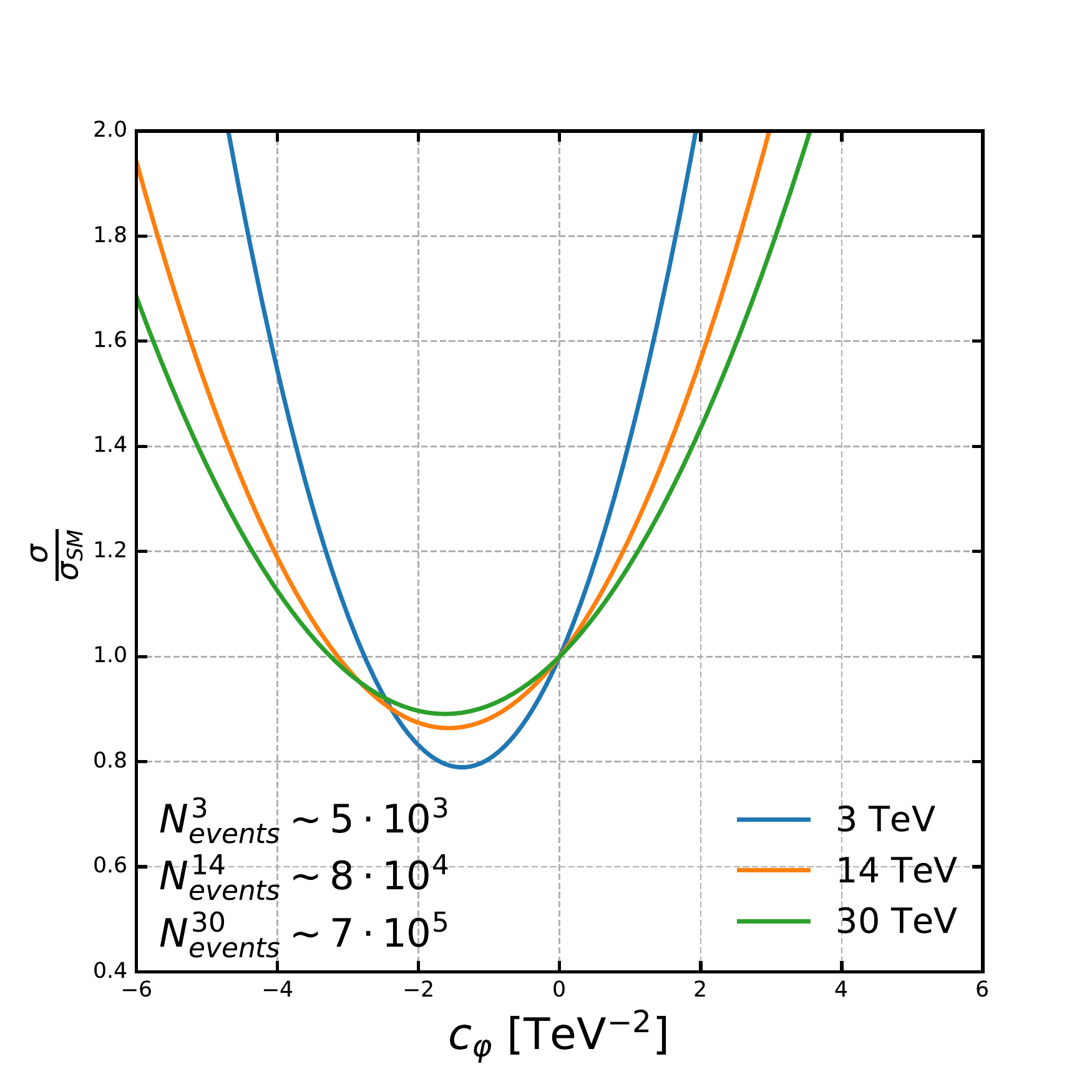}
\includegraphics[width=.45\textwidth]{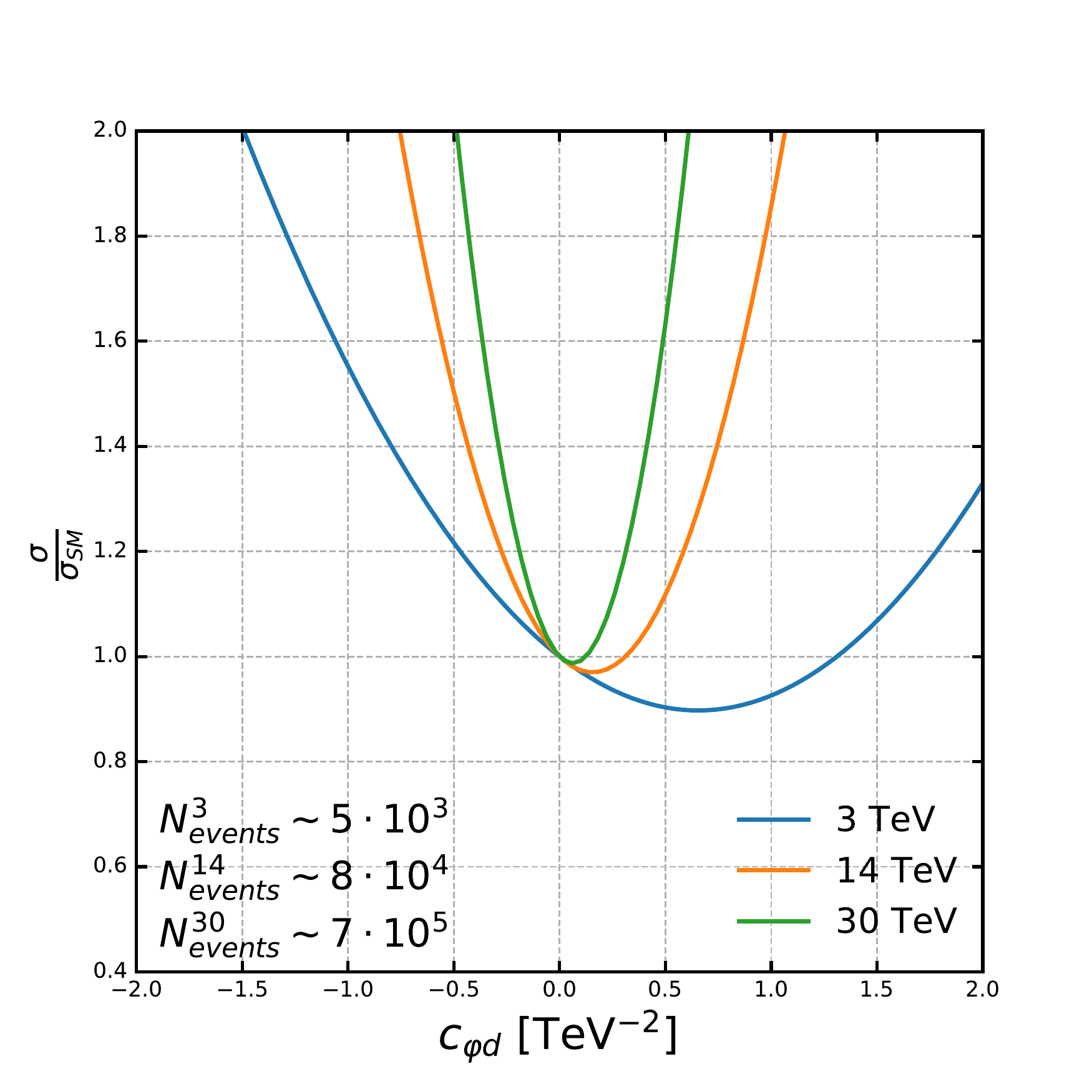}
\caption{\label{fig:WW_HH} Sensitivity ratio for the operators $\Op{\varphi}$ (left) and $\Op{\varphi d}$ (right) to double Higgs production. The expected number of events for the SM is also reported in the figures. }

\end{figure}

\begin{figure}[t]
\centering
\includegraphics[width=.45\textwidth]{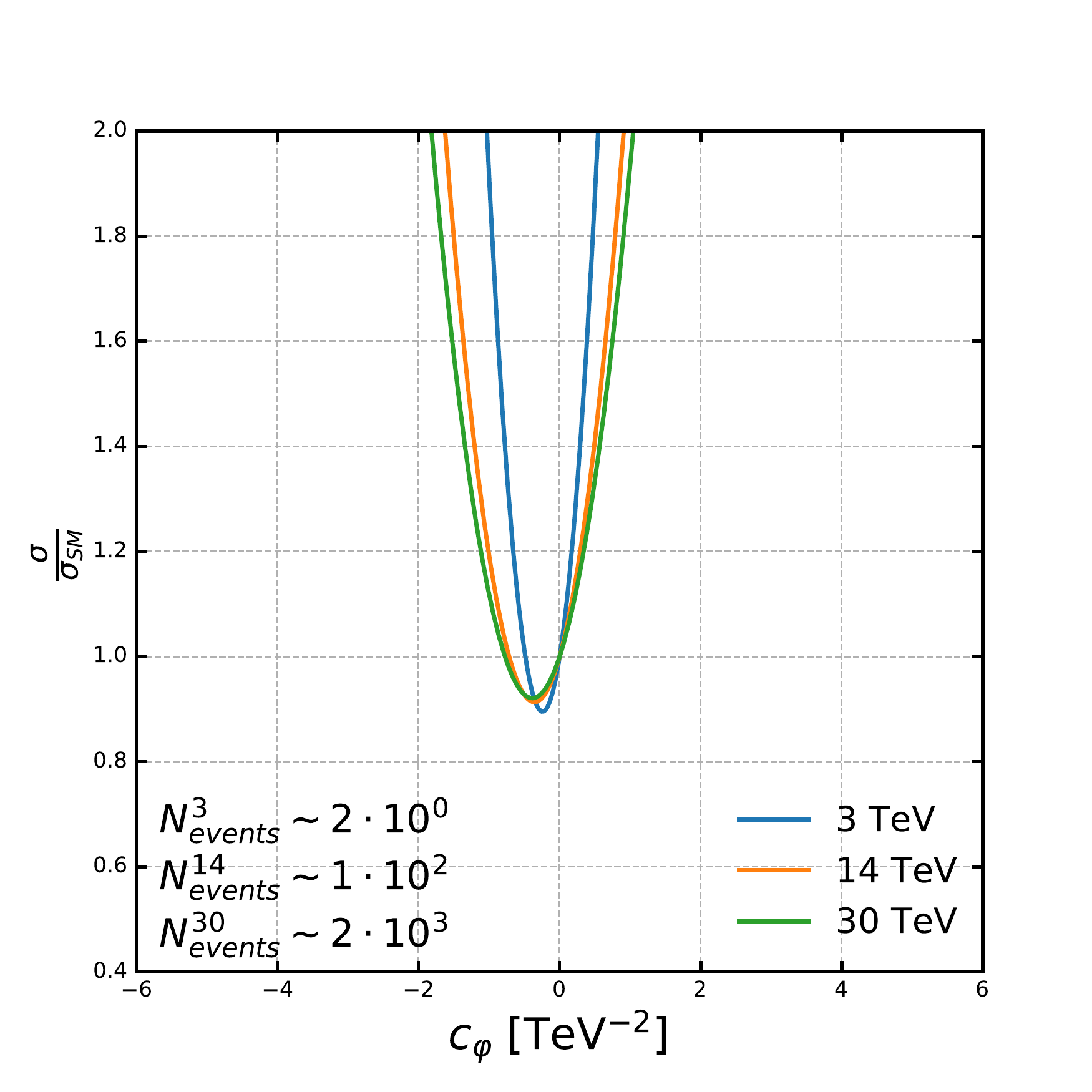}
\includegraphics[width=.45\textwidth]{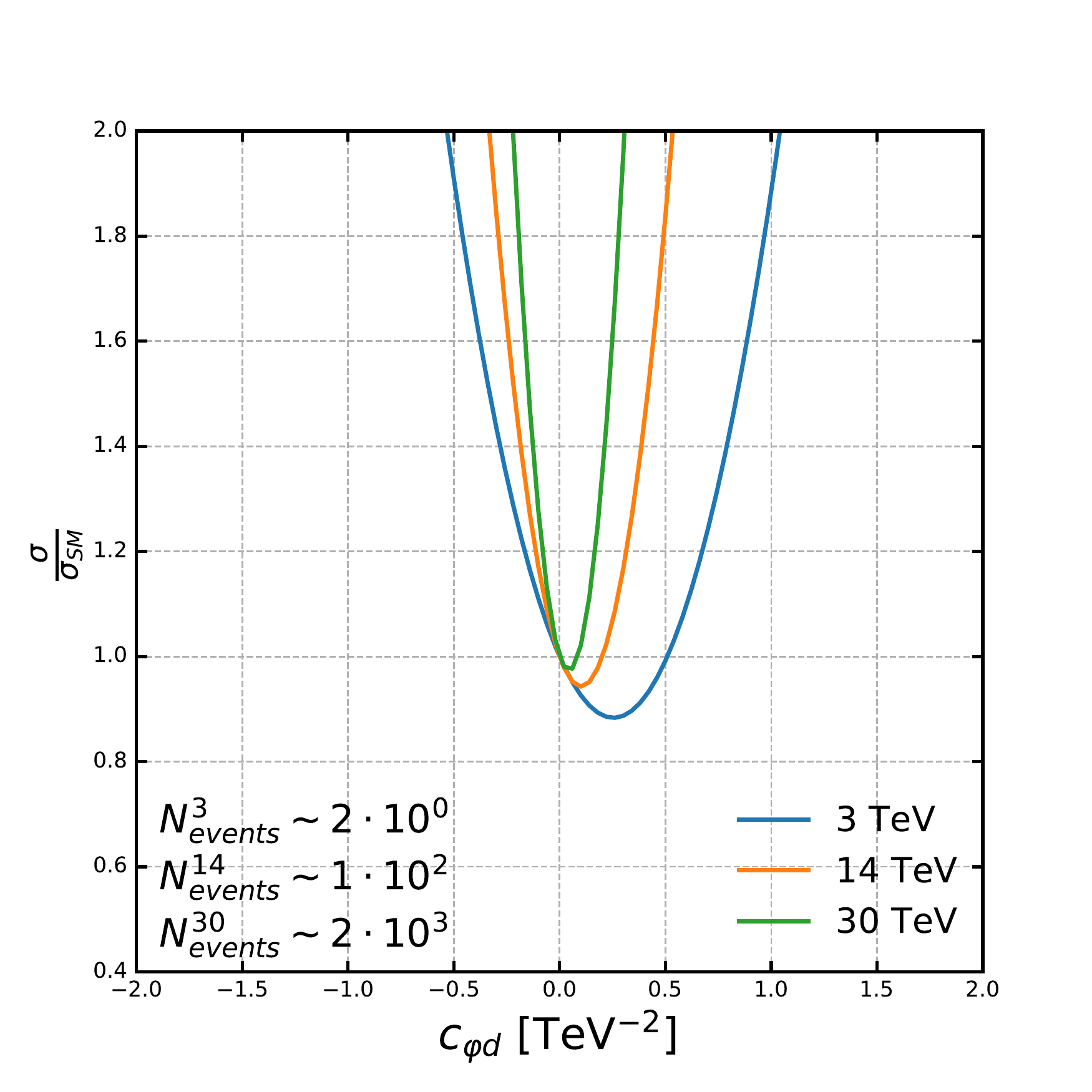}
\caption{\label{fig:WW_HHH} Same as figure~\ref{fig:WW_HH} but for triple Higgs production.}
\end{figure}

The simplest way to access the trilinear and the quartic Higgs couplings is by double and triple Higgs production via VBF (see Fig.~\ref{fig:higgs_diagrams}).
In Fig.~\ref{fig:WW_HH} and~\ref{fig:WW_HHH} we plot the sensitivity ratio for $hh$ and $hhh$ production respectively. Here we consider only the 
$WW$ initiated channel since it is the dominant one, keeping in mind that the $ZZ$ accounts for $\sim 10\%$ of the total cross section.
It is clear by looking at the figures that the two operators under discussion affect the cross sections in two qualitatively
different ways. In particular, $\Op{\varphi}$ does not generate energy dependent vertices and therefore only unitarity violation
effects due to the spoiling of the SM cancellations are expected. However, the highest sensitivity to the trilinear and quartic
couplings come from the near threshold phase space region, both for $hh$ and $hhh$. As a result, an increase in the energy of the collider 
leads to a loss in sensitivity. However, the sensitivity ratio is only a part of the picture and with increased $\sqrt{s}$
a significant higher number of events are expected. The higher statistics can open up the possibility to perform 
differential measurements which can greatly enhance the prospects of detecting NP indirectly by identifying highly sensitive regions
of phase space. It is also worth noting that for triple Higgs production, the number of events that we can expect at $3$ TeV
is completely negligible and no measurement can be undertaken.

On the other hand, the operator $\Op{\varphi d}$, introduces derivative couplings in the Lagrangian which translate to a $p^2$
dependence in the interaction vertices. As a consequence, the effects of the aforementioned operator benefit from higher
collider energies and sensitivity gains are observed. In particular, the square terms become relevant for small values of $c_{\varphi d}$
and quickly start to dominate despite being suppressed by a higher power of the NP scale $\Lambda$. We conclude that the two
operators are probed by complementary parts of the phase space.

\subsection{Projected limits on the parameter space}

In order to have a more realistic picture, we now study the interplay between the operators. As we discussed in Chapter~\ref{chap:globalfit}
in particular, when multiple operators are considered, correlations and cancellations between the various degrees of freedom can
complicate the interpretation. If we assume that the measured cross sections for the processes at hand are consistent with the SM,
we can estimate the potential of the muon collider to constrain the Wilson coefficients. Specifically, we define the $95\%$
CL intervals with the following equation
\begin{equation}
\frac{S}{\sqrt{B}} = \frac{|\mathcal{L} \cdot (\sigma - \sigma_{SM})|}{\sqrt{\mathcal{L} \cdot \sigma_{SM}}} \le 2 \, ,
\label{eq:sig_back}
\end{equation}
where $\sigma$ is the SMEFT cross section defined in Eq.~\eqref{eq:eft_xsec_smeft}, the number of background events $B$ is given
by the expected SM events for a given luminosity $\mathcal{L}$ and the signal $S$ by the excess or lack of events caused by
the modified interactions.
\begin{figure}[t]
\centering
\includegraphics[width=.45\textwidth]{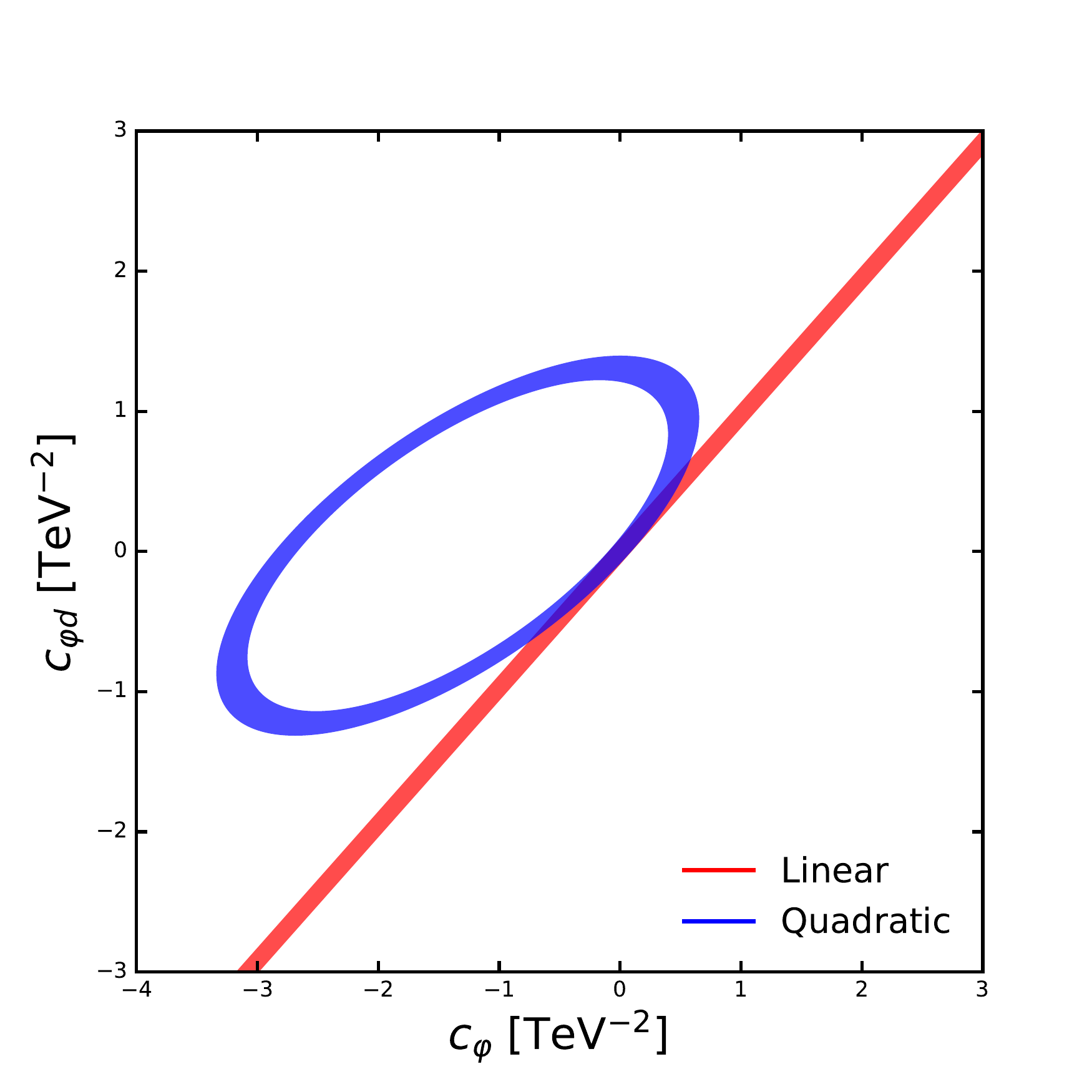}
\includegraphics[width=.45\textwidth]{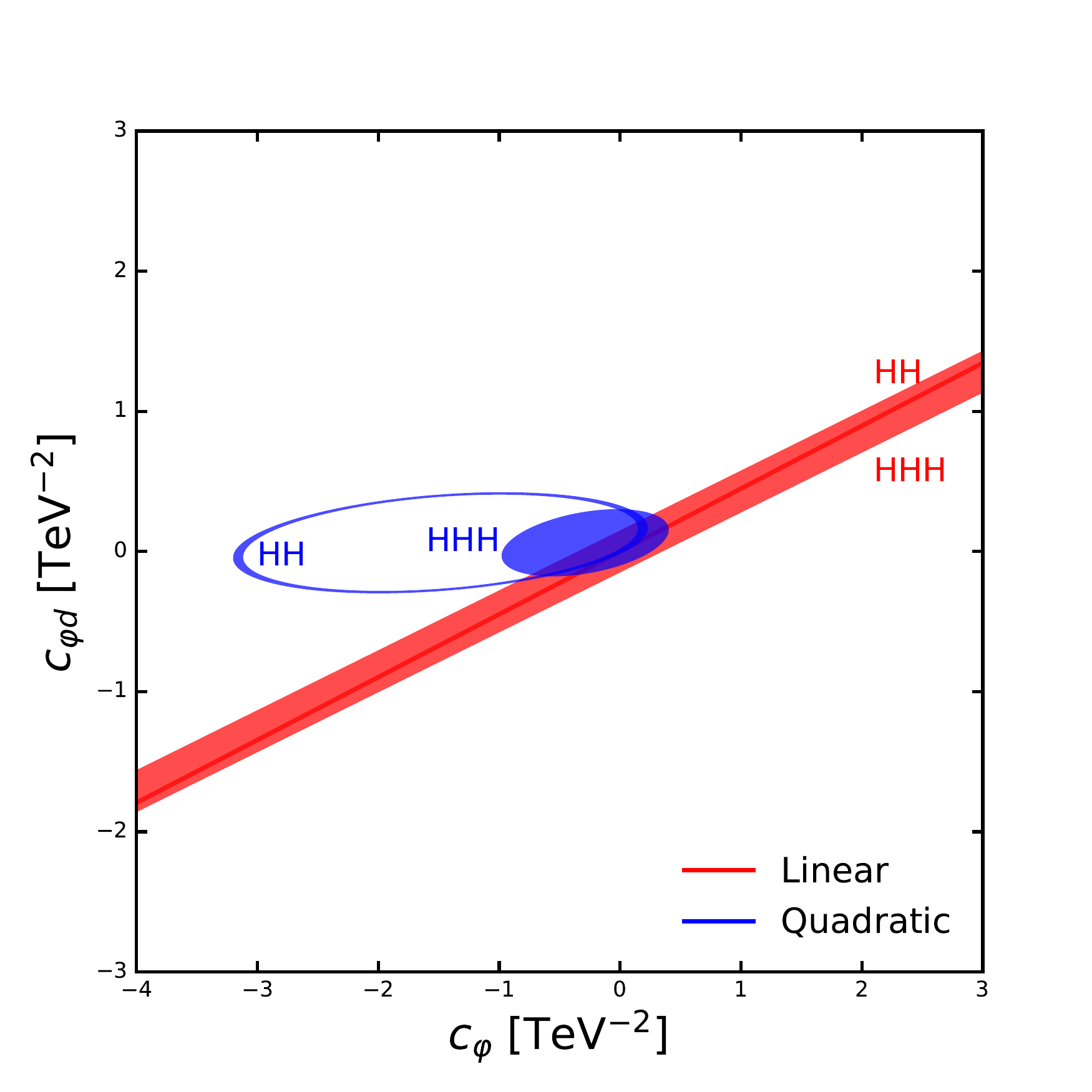}
\caption{\label{fig:2d_lim}
Regions of parameter space allowed from Eq.~\eqref{eq:sig_back} for a $3$ TeV (left) and a $14$ TeV collider (right). In the case of 
the $3$ TeV centre of mass energy, only double Higgs production is considered. Solutions
are displayed for both the case in which only linear corrections are retained and the case in which quadratic terms are included.
}
\end{figure}
In the 2D plane of the parameter space spanned by $c_{\varphi}$ and $c_{\varphi d}$, the solutions of Eq.~\eqref{eq:sig_back}
identify a band and a disk in the case of order $\mathcal{O}(1/\Lambda^2)$ and $\mathcal{O}(1/\Lambda^4)$ corrections respectively.
In Fig.~\ref{fig:2d_lim} we plot the allowed coefficients in the scenario of a $3$ TeV and a $14$ TeV collider. In the lower energy scenario, we notice that a larger portion of the parameter space is left unconstrained with
respect to the $14$ TeV collider. In particular, we observe that the measurement of triple Higgs production at $14$ TeV is enabling us to break the degeneracy in the parameter space. This is a demonstration of the relevance of measuring multiple processes
and further motivates a multi-TeV machine. Measuring both double and triple Higgs production is of critical importance to characterise
the Higgs potential. The marginalised limits obtained from Fig.~\ref{fig:2d_lim} are reported in Table~\ref{tab:2d_lim}.
\begin{table}[!t]
\begin{center}
\begin{tabular}{|c|c|c|}
\hline
 & 3 TeV  & 14 TeV \\
$c_{\varphi}$ & [-3.33, 0.65] & [-0.66, 0.23]\\
$c_{\varphi d}$ & [-1.31, 1.39] & [-0.17, 0.30] \\
\hline
\end{tabular}
\end{center}
\caption{Marginalized projected limits at 95\% confidence level on the Wilson coefficients in TeV$^{-2}$.}
\label{tab:2d_lim}
\end{table}

An important caveat of the SMEFT is that realistic assessments can be obtained only with global analyses. In this perspective,
while $\Op{\varphi}$ effectively only enters in double and triple Higgs production, $\Op{\varphi d}$ shifts universally all the Higgs couplings
and therefore can be constrained from other processes. In particular, single Higgs production can be used to put bounds on the operator.
Even if this is just a total rate effect, the high number of expected events for single Higgs production is enough to put strong limits.
Specifically, we find that at $95\%$ CL they are projected to be $[-0.01, 0.01] \TeV^{-2}$ at 3 TeV and  $[-0.004, 0.004] \TeV^{-2}$ at 14 TeV.
Despite this, we found instructive to include the operator in the study given its enhanced unitarity violating effects.
However, having noticed that we obtain stronger bounds from $h$ production, we conclude that only the operator $\Op{\varphi}$
is relevant for $hh$ and $hhh$ production.

Finally, it is worth providing a comparison with other proposed future colliders. In Ref.~\cite{deBlas:2019rxi}
a combined projection from FCC-ee$_{240}$, FCC-ee$_{365}$, FCC-eh and FCC-hh colliders is reported. 
The first two are $e^+e^-$ colliders with $\mathcal{L}= 5$, $1.5\invab$ at $\sqrt{s}=240$, $365\GeV$ respectively.
The third is an $e^\pm p$ collider with $\mathcal{L}= 2\invab$ at $\sqrt{s}=3.5\TeV$, 
while the last is a $pp$ collider with $\mathcal{L}= 30\invab$ at $\sqrt{s}=100\TeV$.
The projected individual limits at $68\%$ CL are
\begin{equation}
c_{\varphi} \sim [-0.79, 0.79] \textrm{ TeV$^{-2}$} \,  \quad\text{and}\quad c_{\varphi d} \sim [-0.03, 0.03] \textrm{ TeV$^{-2}$} \, .
\end{equation}
On the other hand, for a $14$ TeV muon collider, the projected individual bounds from the measurement of single, double and triple Higgs production are
\begin{equation}
c_{\varphi} \sim [-0.02, 0.02] \textrm{ TeV$^{-2}$} \, \quad\text{and}\quad c_{\varphi d} \sim [-0.002, 0.002] \textrm{ TeV$^{-2}$} \, .
\end{equation}
The improvement is a factor $40$ for $c_{\varphi}$ and a factor $15$ for $c_{\varphi d}$. It is important to stress that these
naive estimates only take into account total cross section rates and are therefore most probably conservative projections.
Nonetheless, this comparison shows the high potential of a multi-TeV muon collider to precisely determine the Higgs self-couplings.

\section{Prospects on the Higgs quartic coupling}

Projected measurements for the trilinear coupling $\lambda_3$ have been vastly considered by the community, both at present
and future colliders, from the measurement of single (through radiative corrections) and double Higgs production~\cite{deBlas:2019rxi,DiMicco:2019ngk}.
In particular, a ${\cal O}(5\%)$ accuracy is expected at FCC~\cite{Abada:2019lih,Abada:2019zxq,Benedikt:2018csr}.

On the other hand, the quartic coupling prospects look rather grim. While in the SMEFT at dimension-6 deviations from these two couplings are still
correlated, an unbiased determination requires a direct measurement. The most simple way to access the quartic interaction is
by means of triple Higgs production. At FCC, the expected constraints from the measurement of the aforementioned process are
quite poor and stand at $\lambda_4/\lambda_4^{\rm SM} \in $ [-2,+13] $95\%$ CL interval~\cite{Papaefstathiou:2015paa,Contino:2016spe,Fuks:2017zkg}.
Another way to access the quartic coupling is by loop corrections to double Higgs production. In this case, projections are
slightly better but still unsatisfactory, with a range of [-2.3,+4.3] at $68\%$ CL~\cite{DiMicco:2019ngk,Maltoni:2018ttu,Bizon:2018syu,Borowka:2018pxx}.

In the following, we will provide a more in-depth phenomenological study of $hhh$ production via VBF at a muon collider, \ie
\begin{equation}
\mu^+ \mu^- \to W^\ast W^\ast \nu_\mu \overline{\nu}_\mu\to h h h \nu_\mu \overline{\nu}_\mu .
\end{equation}
We will work under the assumption that the issues related to machine and detector elements can be resolved after detailed studies. 
At the same time, we will assume that Higgs decays can be efficiently reconstructed and proceed to estimate the potential to
detect deviations from the trilinear and quartic couplings in the anomalous coupling framework.

\subsection{Triple Higgs production in the SM}
\begin{figure}[t]
\begin{center}
  \includegraphics[width=0.30\textwidth]{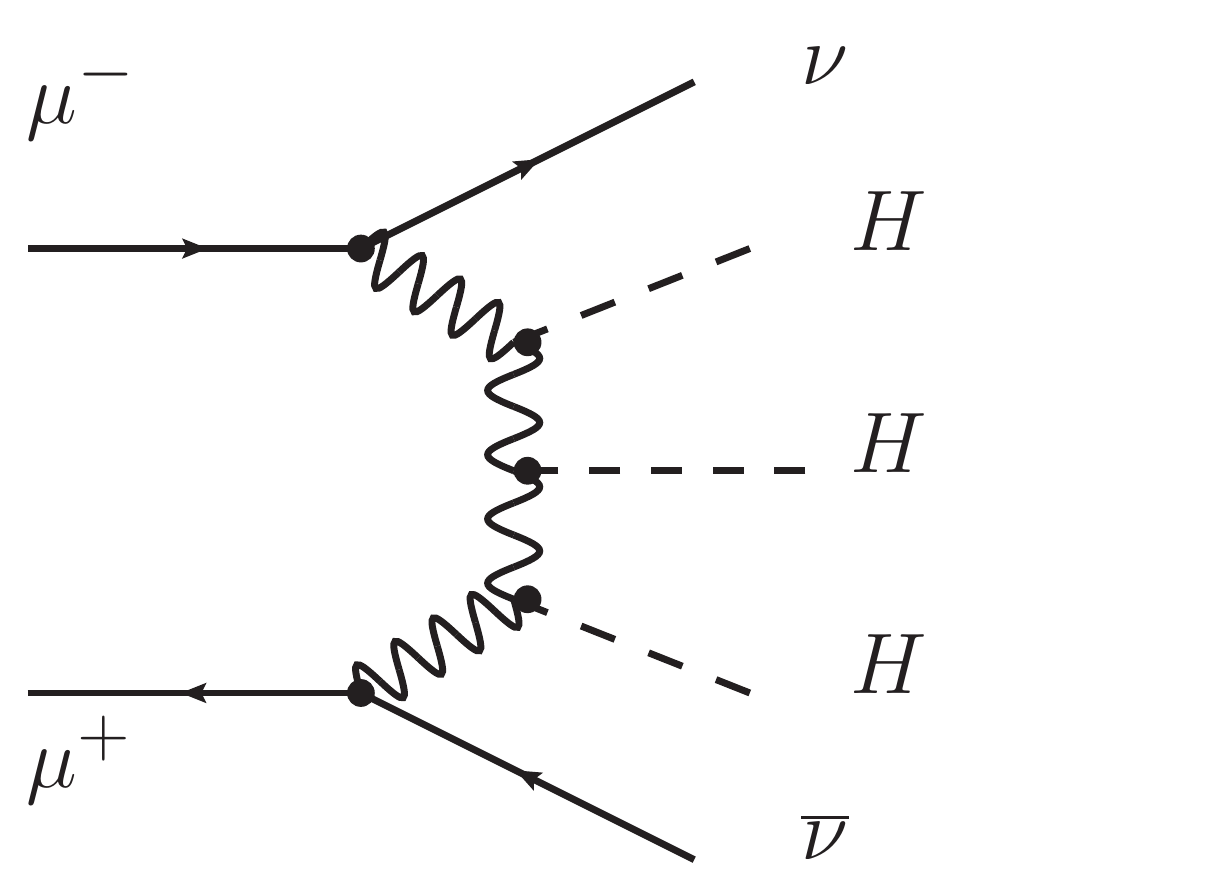}   \includegraphics[width=0.30\textwidth]{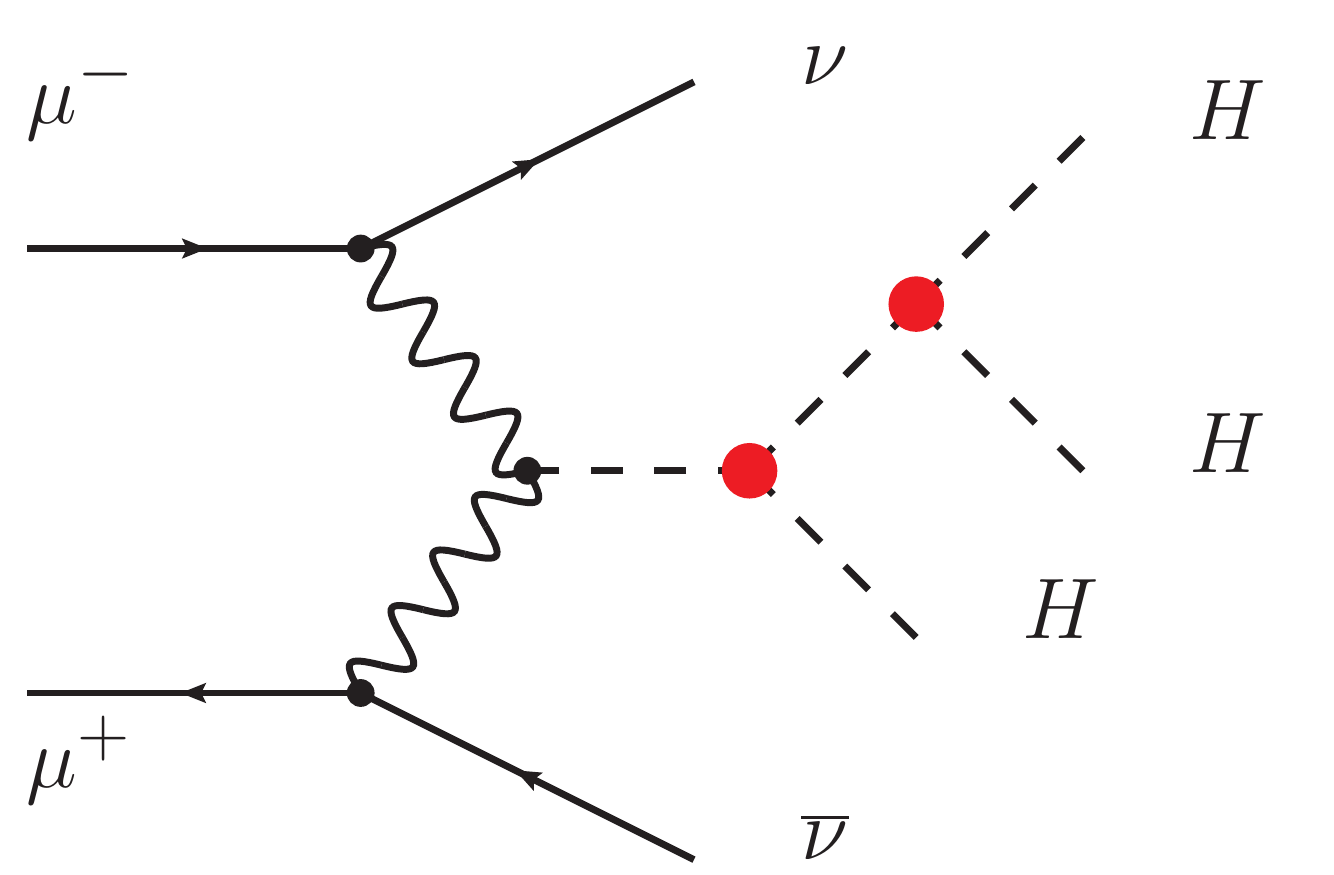}    
  \includegraphics[width=0.30\textwidth]{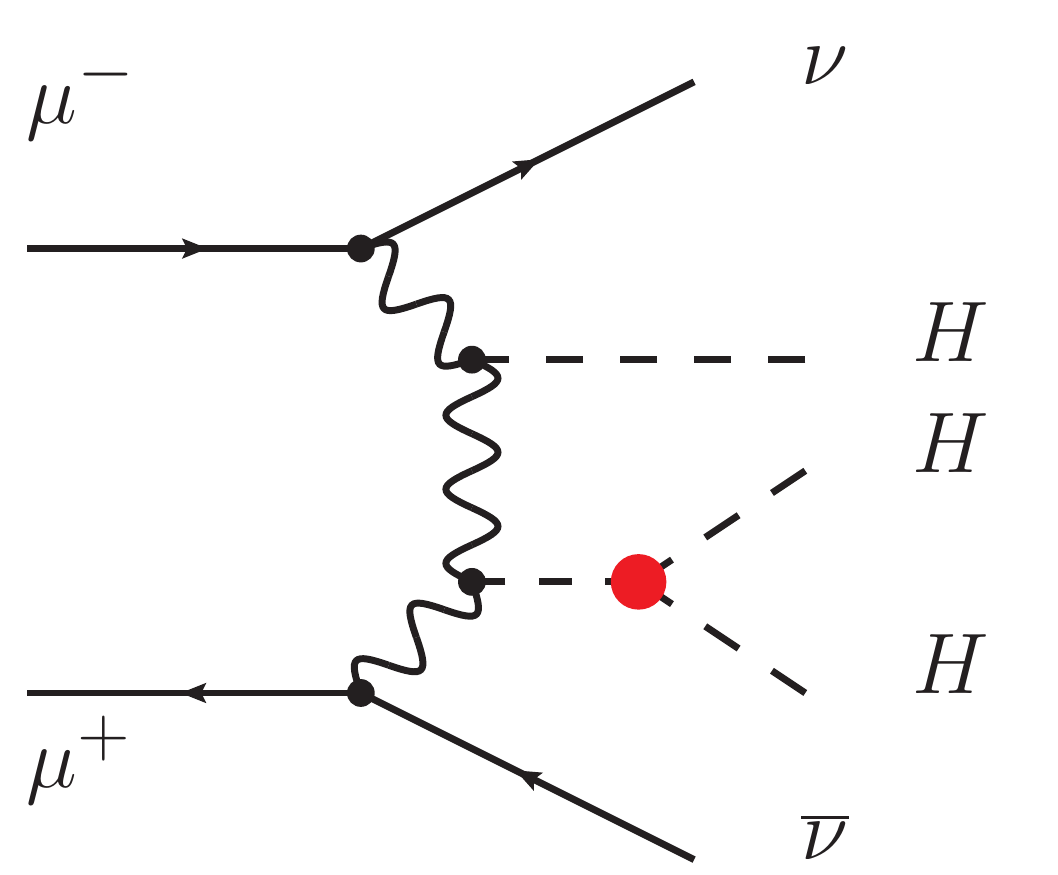}\\   
  \includegraphics[width=0.30\textwidth]{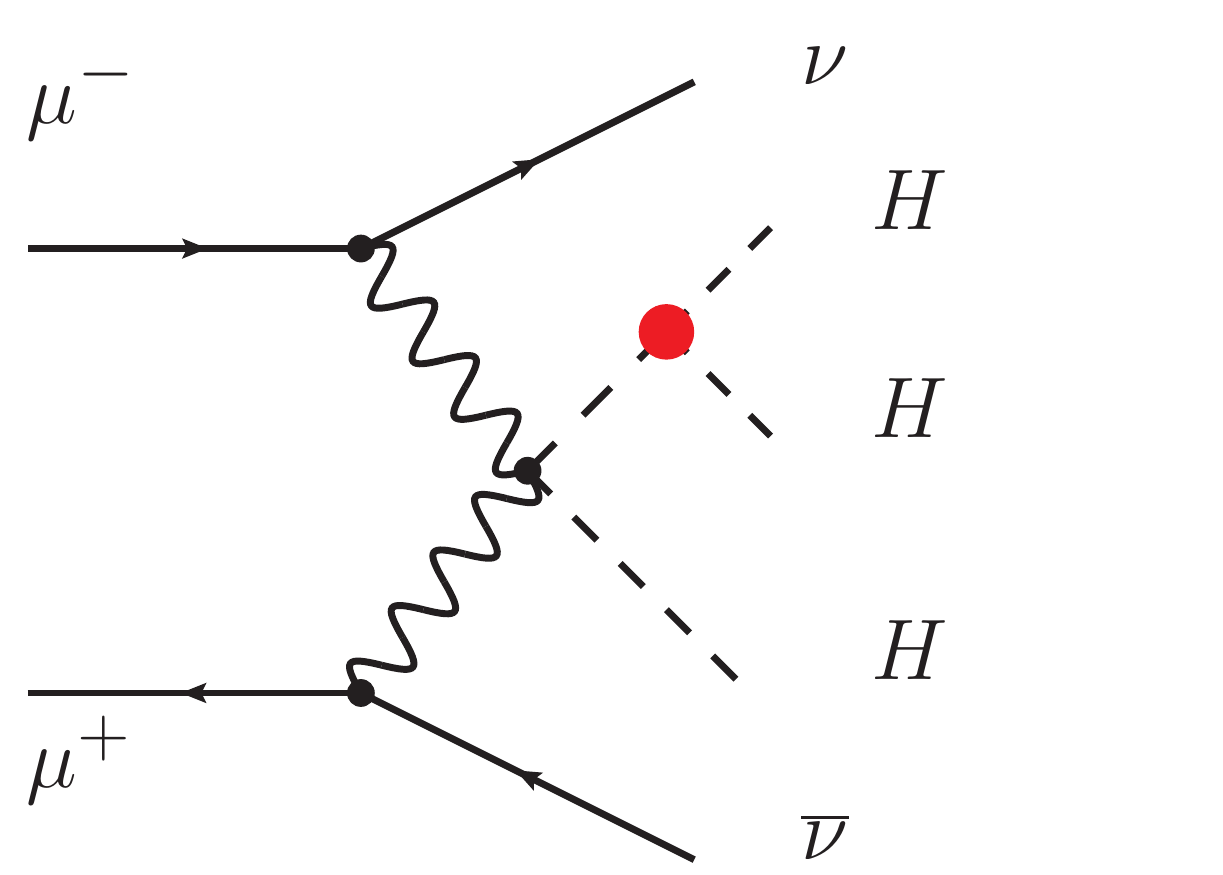}
  \includegraphics[width=0.30\textwidth]{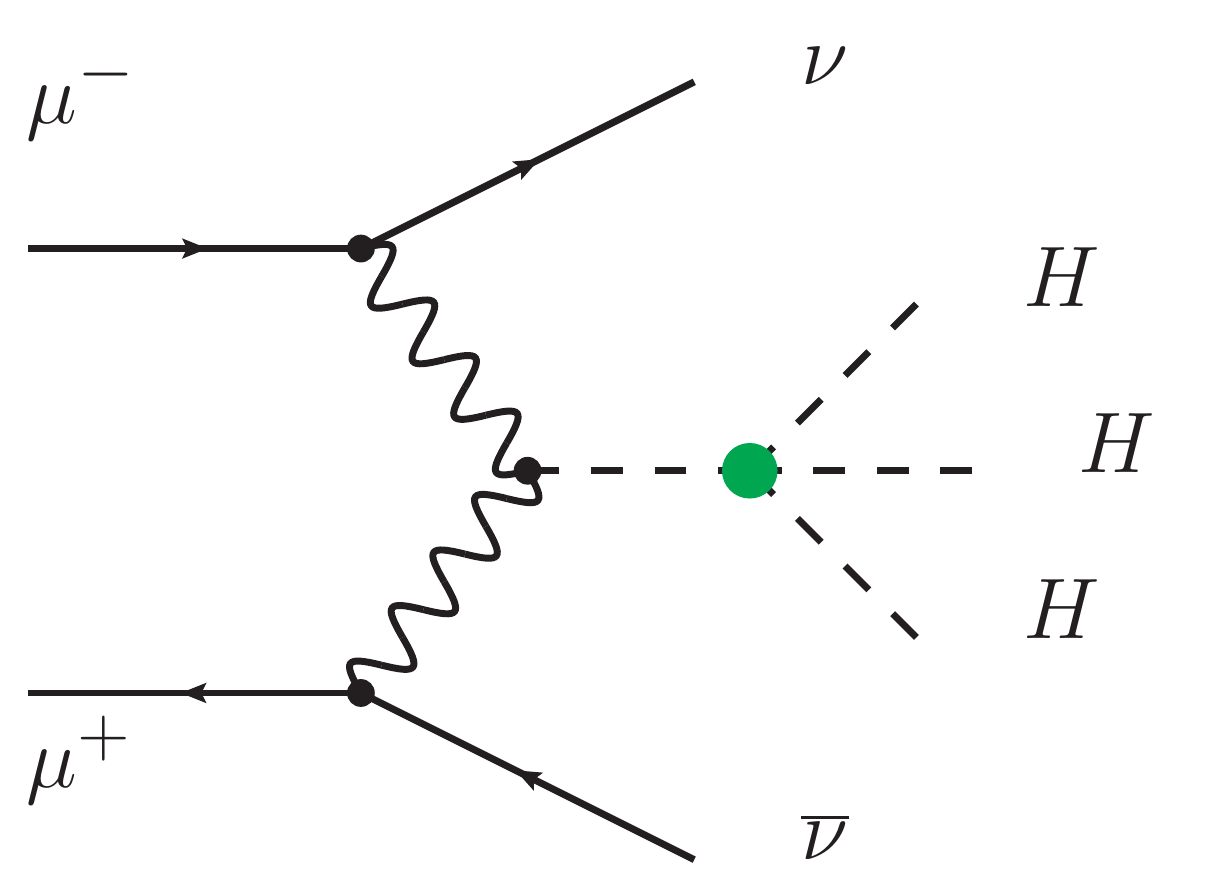}    \includegraphics[width=0.30\textwidth]{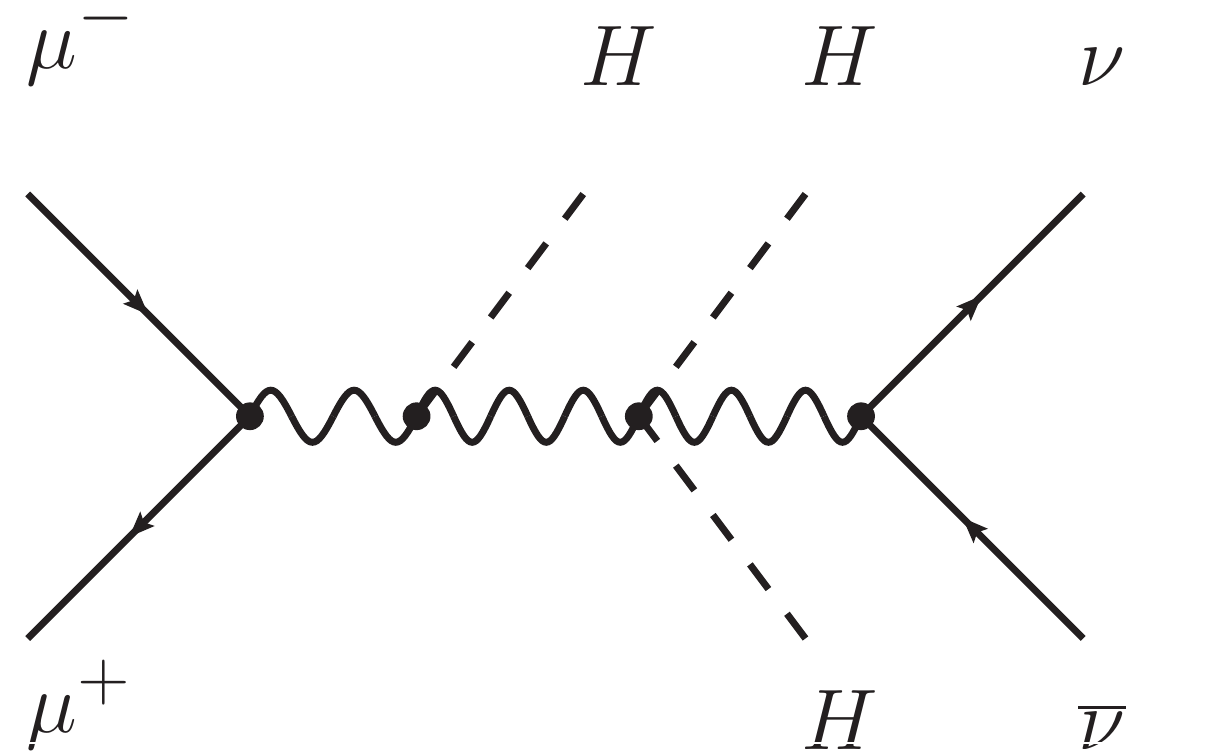}
  \caption{\label{fig:Feyndiags} Selection of diagrams which contribute to triple Higgs production at a muon collider. In red we indicate
  the presence of the trilinear coupling, while in green the one of the quartic. }
    \end{center}
\end{figure}

Before delving into the assessment of anomalous deviations, we will present SM predictions for the production of three Higgses
via VBF. We report in Fig.~\ref{fig:Feyndiags} a few representative diagrams for the considered process. We observe that 
the dependence on $\lambda_3$ at the amplitude level is quadratic and that diagrams exist in which neither the trilinear nor
the quartic coupling are present. In particular, it is of paramount importance that a precise knowledge of the $WWh$ and $WWhh$
is available in order to extract information on the self-couplings.

As previously discussed, the production of multi-Higgs final states happens both from pure VBF scatterings and from s-channel
V-strahlung, \ie 
\begin{equation}
\mu^+ \mu^- \to h h h \;Z^\ast \to h h h \,\nu_{e,\mu,\tau} \overline{\nu}_{e,\mu,\tau}\; .
\end{equation}
This component is however relevant in a different region of phase space and therefore can be efficiently removed with a simple
cut. Specifically, in the further analysis, we always apply a minimum invariant mass cut $M_{\nu \overline{\nu}}>150$~GeV,
so that we force the s-channel $Z$ to be off-shell.

The simulations are performed with two event generators: {\sc Whizard} \cite{Moretti:2001zz,Kilian:2007gr} and {\sc MadGraph5\_aMC@NLO} \cite{Alwall:2014hca}.
In order to avoid unitarity violating behaviours from the fixed width scheme, we set the widths of the Higgs and gauge bosons to zero,
as well as the muon Yukawa.
\begin{table}[t]
 \begin{center}
\resizebox{\textwidth}{!}{
  \begin{tabular}{| c || c || c || c || c || c || c || c || c || c |}
      \hline
    $\sqrt{s}$ (TeV)\;/\;L  (ab$^{-1}\!)$         & 1.5 / 1.2  &   3  / 4.4  & 6 / 12     & 10  / 20    & 14  / 33 & 30  / 100    \\
  \hline\hline
    \multicolumn{7}{|c|}{$\sigma_{SM}$         (ab) \;\;[$N_{\rm ev}$]}\\
 \hline\hline
   $\sigma^{\rm tot}$                        & 0.03 [0] & 0.31 [1] & 1.65 [20] & 4.18 [84] & 7.02 [232] & 18.51 [1851] \\
  $\sigma(M_{HHH}\!<3$\, TeV)          & 0.03 [0] & 0.31 [1] & 1.47 [18] & 2.89 [58] & 3.98 [131] &  6.69 [669]  \\
  $\sigma(M_{HHH}\!<1$\, TeV)                           & 0.02 [0] & 0.12 [1] & 0.26 [3]  & 0.37 [7]  & 0.45 [15]  &  0.64 [64]   \\
 \hline
\end{tabular}
}
\caption{\label{tab:HHHnocut} Cross sections and event numbers in the SM for triple Higgs production at different energies in a muon collider. The cases in which a maximum invariant mass cut is applied on the Higgses is also displayed.}
\end{center}
\end{table}
The total cross sections and expected event numbers for $hhh$ production for a set of benchmark scenarios are reported in Table~\ref{tab:HHHnocut}.
We observe that collider energies smaller than $6$ TeV, for the realistic integrated luminosities, 
do not produce enough events to measure the process and will therefore be neglected in the analysis. In addition to the inclusive cross sections, we report in the
table cross section numbers for regions of phase space closer to threshold, where the sensitivity to the quartic coupling is expected to be the 
highest.
\begin{figure}[t]
  \centering
  \includegraphics[width=0.4\textwidth]{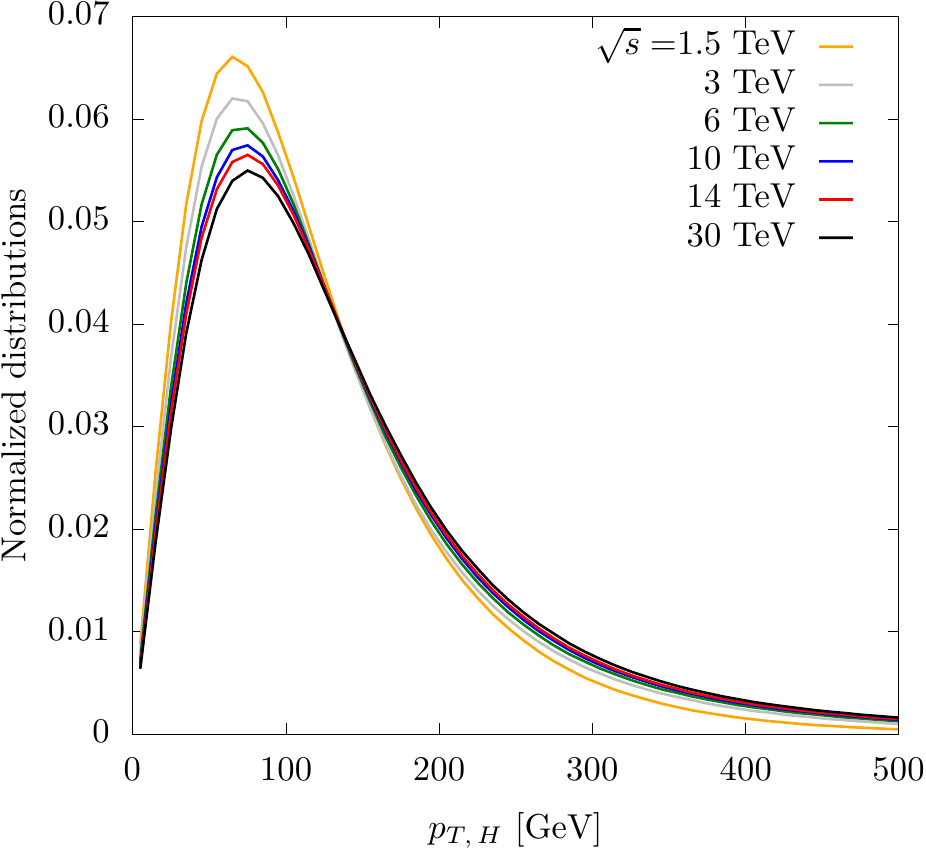}
  \includegraphics[width=0.4\textwidth]{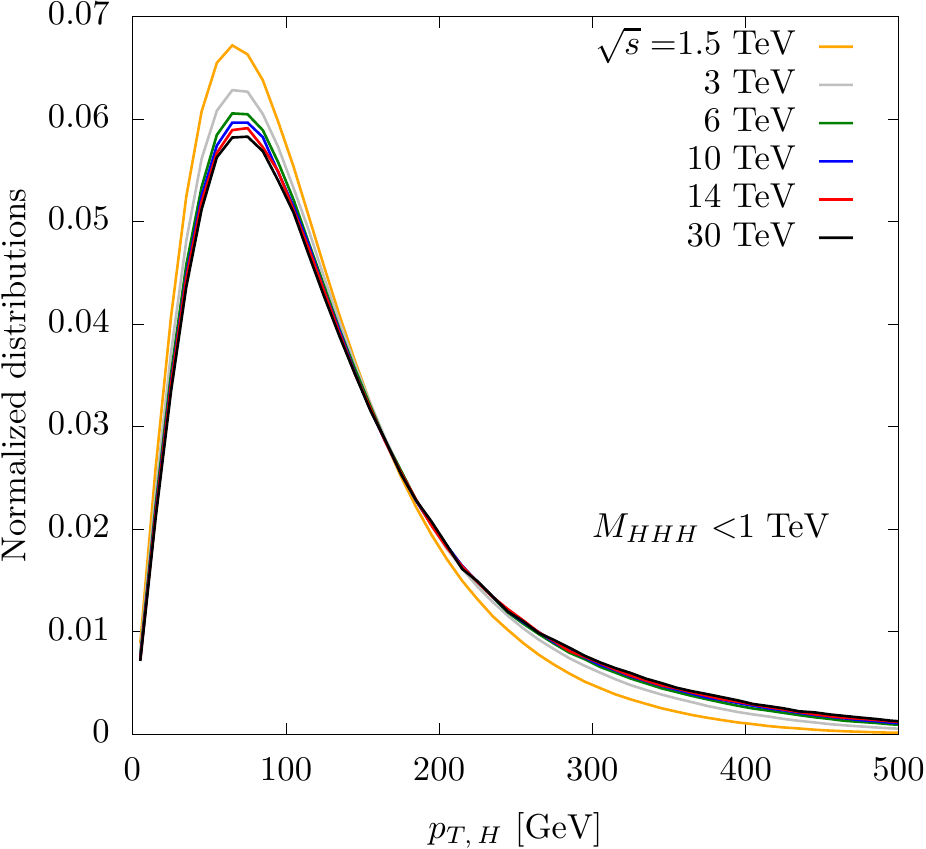}
  
  \caption{Normalised distributions for the $p_T$ of the Higgs bosons in the inclusive scenario (left) and the near threshold phase space region (right).}
  \label{fig:pth}
\end{figure}
\begin{figure}[t]
  \centering
  \includegraphics[width=0.4\textwidth]{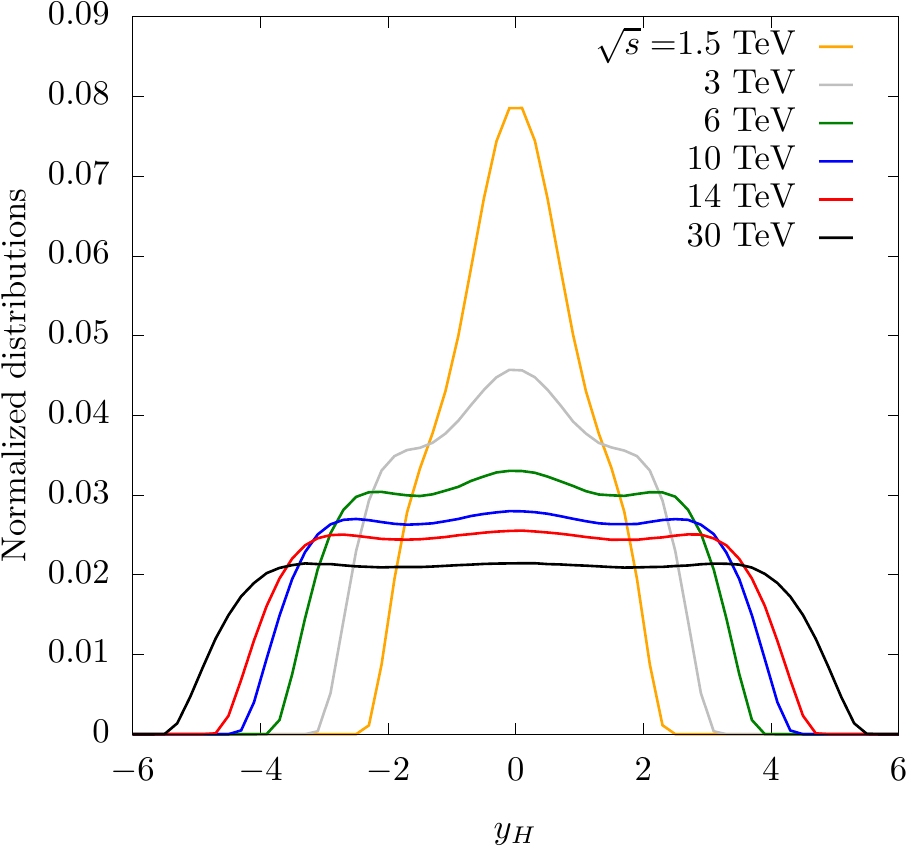}
  \includegraphics[width=0.4\textwidth]{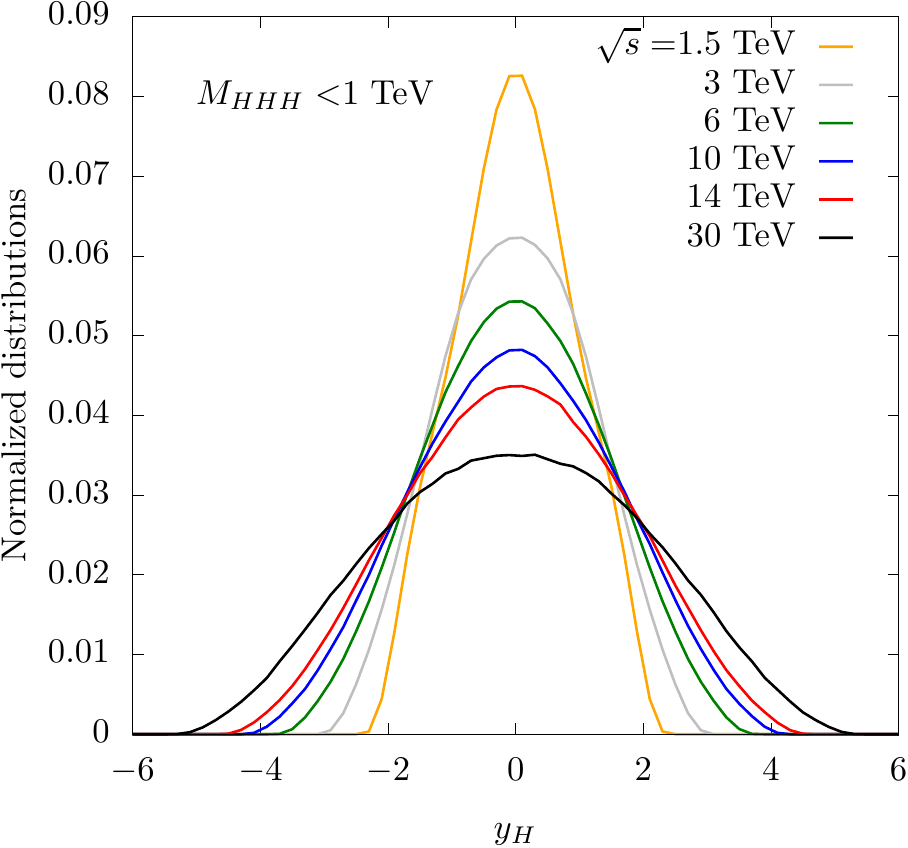}
  \caption{Normalised distributions for the rapidity $y_H$ of the Higgs bosons in the inclusive scenario (left) and the near threshold phase space region (right). }
  \label{fig:yh}
\end{figure}

In Fig.~\ref{fig:pth} and~\ref{fig:yh} we display the $p_T$ and rapidity normalised distribution of the Higgs bosons in the inclusive and 
near threshold scenarios, for the different benchmark collider energies considered.

Remarkably, the peak of the $p_T$ distribution does not seem to be dependent on the energy of the collider, both in the
inclusive and in the near threshold region. However, as the energy increases, the tails of the distributions start to be more and more
populated as expected.

On the other hand, the rapidity distribution exhibits a rather striking dependence on the centre of mass energy. Specifically,
the higher the energy, the more forward the produced Higgs bosons are. This suggests that, in order to detect them, a wide rapidity
coverage is needed for the detector. Nevertheless, in preliminary studies on a $1.5$ TeV muon collider~\cite{Foster:1995ru,Johnstone:1996hp,Mokhov:2011zzd,Alexahin:2011zz,Mokhov:2014hza}, it was proposed to put nozzles with openings of $5-10$ degrees at the entrance of the beam in the interaction region, 
in order to shield the beam-induced radiation. It is however worth observing that in the event of a more energetic beam, this radiation
would be more forward and in principle the angular coverage of the nozzles could be reduced. Having established the crucial
role of the angular acceptance of the detector to measure triple Higgs production, we argue that this should be taken into account in future 
detector designs. This will be further discussed in Section~\ref{sec:real_detector}.

\begin{table}[t]
{\footnotesize
\begin{center}
  \begin{tabular}{| c | c | c | c | c | c | c | c | c | c |}
    \hline
      \hline
    \multicolumn{6}{|c|}{$   \sigma= c_1
     +       c_2 \delta_3 
     +       c_3 \delta_4
     +       c_4 \delta_3  \delta_4
     +       c_5 \delta_3^2
     +       c_6 \delta_4^2
     +       c_7 \delta_3^3
     +       c_8 \delta_3^2 \delta_4
     +       c_9 \delta_3^4$}\\
            \multicolumn{6}{|c|}{ }\\
\hline\hline
    $\sqrt{s}$\;(TeV) & 3 & 6 & 10 & 14 & 30 \\
   \hline \hline
  \multicolumn{6}{|c|}{$c_i\;\;$   (ab) }\\
    \hline
    $c_1$      &    0.3127 &   1.6477 &   4.1820 &   7.0200 &  18.5124  \\
\hline
    $c_2$      &   -0.1533 &  -1.7261 &  -4.4566 &  -7.1000 & -15.9445  \\
    $c_3$      &   -0.0753 &  -0.1159 &  -0.1166 &  -0.1147 &  -0.1117  \\
 \hline
    $c_4$      &   -2.0566 &  -6.3052 & -11.4981 & -15.9807 & -29.2794  \\
    $c_5$      &    4.7950 &  14.9060 &  27.1081 &  37.4658 &  67.7539  \\
    $c_6$      &    0.2772 &   0.8637 &   1.5992 &   2.2455 &   4.2038  \\
\hline
    $c_7$      &   -1.8353 &  -4.3210 &  -6.6091 &  -8.3962 & -13.0964  \\
    $c_8$      &    0.5032 &   1.1861 &   1.8173 &   2.2967 &   3.5217  \\
\hline
    $c_9$      &    0.2943 &   0.5954 &   0.8946 &   1.1611 &   1.9349  \\
   \hline   \hline
  \multicolumn{6}{|c|}{$\bar c_i\equiv c_i(M_{HHH}<1$~TeV)\;\;   (ab) }\\
    \hline
    $\bar c_1$     &  0.1165 &   0.2567 &   0.3743 &   0.4541 &   0.6404 \\
\hline
    $\bar c_2$     &  0.1667 &   0.3003 &   0.4046 &   0.3545 &   0.6972 \\
    $\bar c_3$     & -0.0768 &  -0.1510 &  -0.2105 &  -0.2285 &  -0.3519 \\
 \hline
    $\bar c_4$     & -1.3604 &  -2.8996 &  -4.1522 &  -5.0582 &  -6.9538 \\
    $\bar c_5$     &  3.1017 &   6.6033 &   9.4721 &  11.4547 &  15.9505 \\
    $\bar c_6$     &  0.1842 &   0.3954 &   0.5679 &   0.6931 &   0.9543 \\
\hline
    $\bar c_7$     & -1.5210 &  -3.0591 &  -4.3186 &  -4.8598 &  -7.3196 \\
    $\bar c_8$     &  0.4222 &   0.8550 &   1.2103 &   1.3906 &   2.0398 \\
 \hline
    $\bar c_9$     &  0.2691 &   0.5482 &   0.7720 &   0.9702 &   1.2482 \\
    \hline
  \end{tabular}
  \caption{\label{tab:HHHcoeff2} 
  Coefficients $c_i$ for triple $h$ production via VBF in the anomalous coupling framework.  
  }
  \label{tab:ci}
\end{center}
}
\end{table}

\subsection{Higgs production with anomalous self-couplings}

We now move on to explore the presence of anomalous interactions in the Higgs potential. In the following, we parametrise
the self-couplings as
\begin{eqnarray}
&&\lambda_3 = \lambda_{SM} ( 1 + \delta_3 ) = \kappa_3  \lambda_{SM} \,, \\
&&\lambda_4 = \lambda_{SM} ( 1 + \delta_4 ) = \kappa_4  \lambda_{SM}\,,
\end{eqnarray}
such that we recover the SM when $\delta_3=\delta_4=0$ or $\kappa_3=\kappa_4=1$. In the case of the dimension-6 operator
$\Op{\varphi}$, the deviations from the SM potential are correlated and in particular we have
\begin{equation}
\delta_4= 6 \,\delta_3  , \, \qquad {( \rm SMEFT\;at\dim=6).}
\end{equation}
However, this relation can be broken by adding effects from the dimension-8 operator $c_8 (\varphi^\dagger \varphi)^4/\Lambda^4$.

In this framework, the cross section can be parametrised as
\begin{equation}
     \sigma= c_1
     +       c_2 \delta_3 
     +       c_3 \delta_4
     +       c_4 \delta_3  \delta_4
     +       c_5 \delta_3^2
     +       c_6 \delta_4^2
     +       c_7 \delta_3^3
     +       c_8 \delta_3^2 \delta_4
     +       c_9 \delta_3^4 \,.
\label{eq:sigmak3k4}
\end{equation}
The coefficients $c_i$ are reported in Table~\ref{tab:HHHcoeff2} for different collider energies, in inclusive and near threshold phase space
region.
We will explore the sensitivity to the self-couplings in different scenarios. Mostly, we focus on two cases:
\begin{itemize}
\item[A)] $\delta_3=0$, $\delta_4 \neq 0$, i.e., deviations only in the quartic Higgs coupling; 
\item[B)] $\delta_4=6 \,\delta_3$, i.e., the pattern of deviations as expected from the SMEFT at dim=6. 
\end{itemize}
In particular, in scenario A we assume that no deviation from the SM in the trilinear coupling is observed in other experiments or
processes, while assessing the potential to pinpoint the quartic coupling. On the other hand, in the second scenario, we discuss
the situation in the SMEFT at dimension-6, exploring also the possibility of being sensitive to deviations to this
pattern.
\begin{figure}[t]
  \centering
  \includegraphics[width=0.4\textwidth]{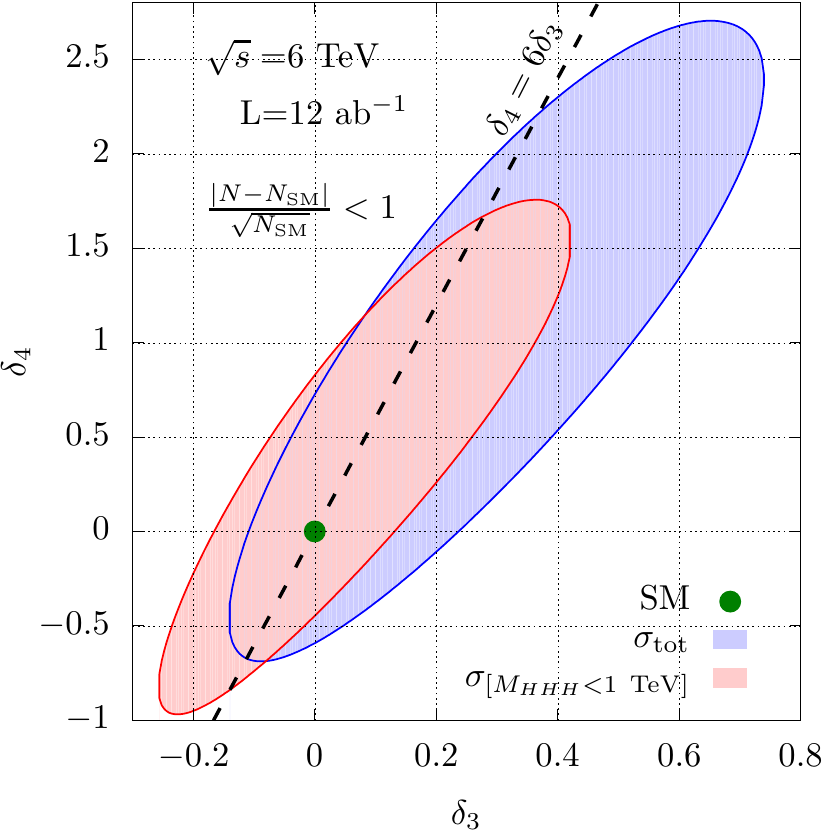} \quad
  \includegraphics[width=0.39\textwidth]{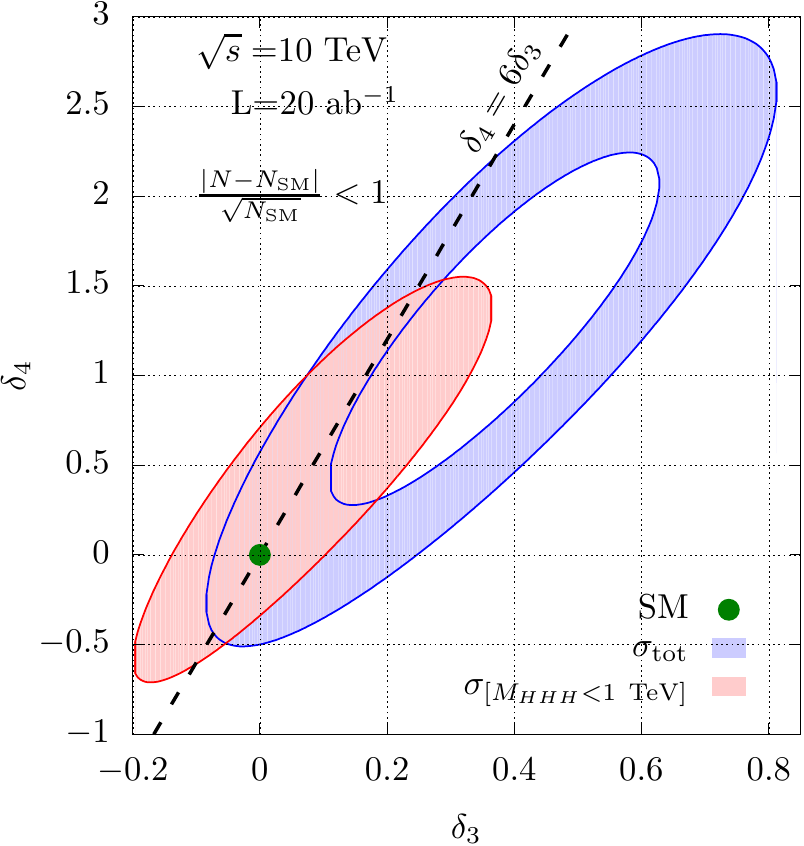}\\
  \includegraphics[width=0.4\textwidth]{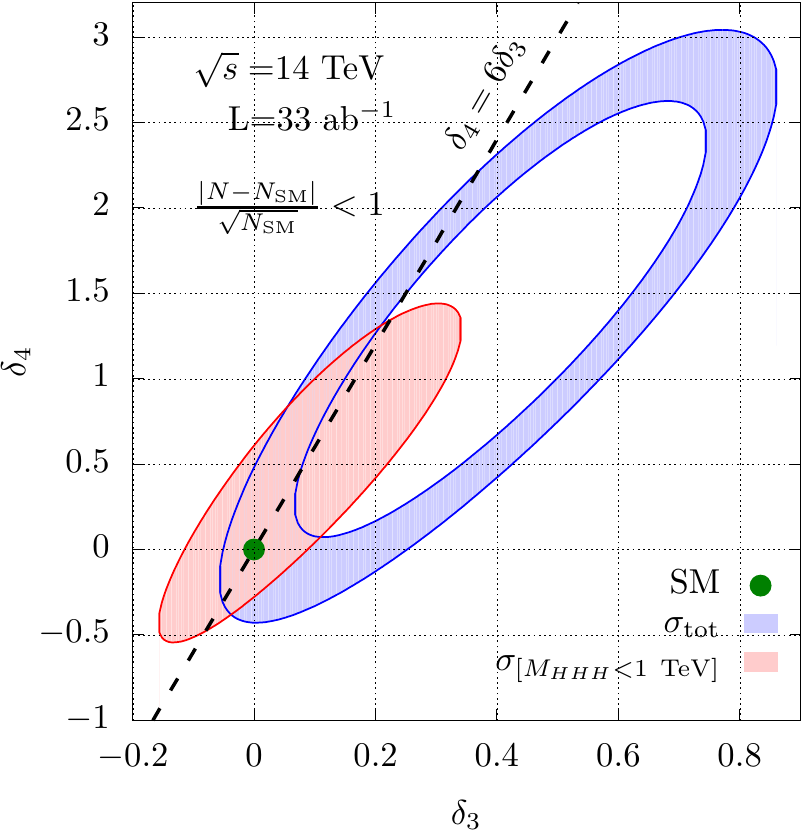}\quad
  \includegraphics[width=0.4\textwidth]{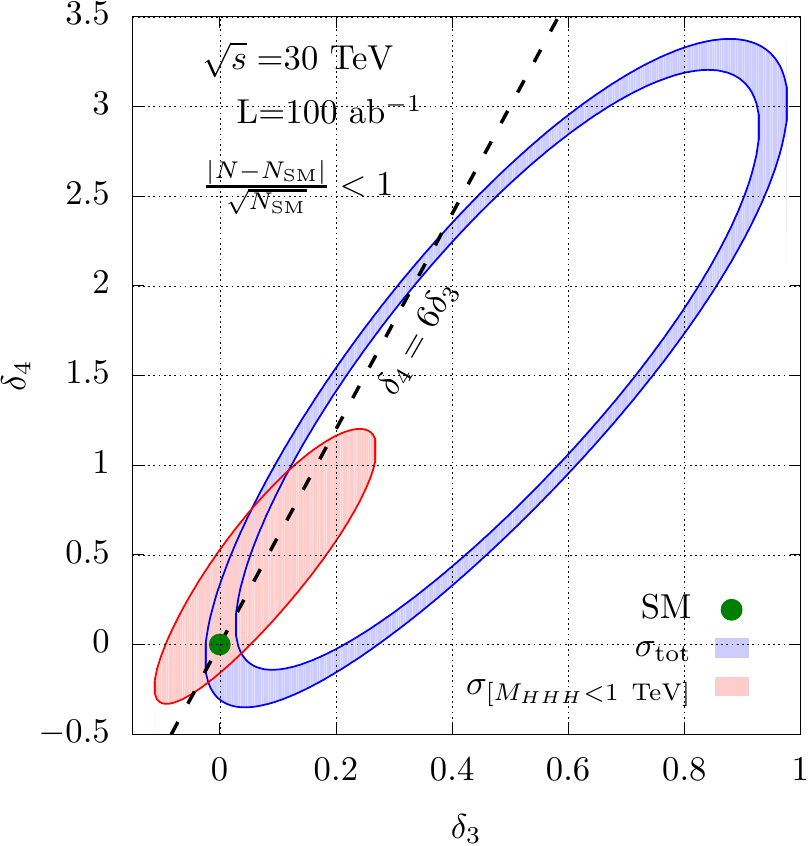}
  \caption{Projections at 1-sigma from triple Higgs production assuming that both $\delta_3$ and $\delta_4$ are independent for a series of benchmark
  muon collider centre of mass energies.
 }
 \label{fig:lim1}
\end{figure}

Defining the sensitivity with the formula in Eq.~\eqref{eq:sig_back}
\begin{equation}
  \frac{|N-N_{\rm SM}|}{\sqrt{N_{\rm SM}}},
  \label{eq:significance}
\end{equation}
where $N$ is the number of expected events in the anomalous interaction scenario and $N_{SM}$ the expected events in the SM, in Fig.~\ref{fig:lim1} we report the contour plots at 1-sigma for four 
benchmark collider energies, combining both projections from inclusive cross section and the region of phase space with
$M_{HHH} < 1$ TeV, assuming that both $\delta_3$ and $\delta_4$ are independent. As expected, as the energy increases, the blue
areas (inclusive) benefit from a substantial enhancement from the higher statistics, forming ring-shaped regions in the parameter
space. This is the consequence of being sensitive to quadratic terms and that bounds are obtained from both excess and lack of events.
The red areas (threshold) on the other hand are only slightly improving with the higher number of events, but have a
higher constraining power as expected. It is interesting to see that a combination of the two observables is able to severely improve the projected 
bounds.

If we now consider scenario A, in Table~\ref{tab:summary} we report the projected constraints on the couplings by considering only the total
cross section and by combining the below and above threshold regions.
\begin{table}[t]
{\footnotesize
\setlength{\tabcolsep}{4.pt}
\begin{center}
  \begin{tabular}{|cc|ccc|}
    \hline
 & & \multicolumn{3}{c|}{Constraints  on $\delta_4$ (with $\delta_3=0$)}    \\
$\sqrt{s}$ (TeV)    &    Lumi (ab$^{-1}$)      &  x-sec only &  x-sec only &  threshold + $M_{HHH}>1$~TeV   \\
  & &   1 $\sigma$ & 2 $\sigma$ &  1 $\sigma$ \\
\hline
6                     & 12                                 & $[-0.60, 0.75]$& $[-0.90, 1.00]$ & $[-0.55, 0.85]$\\
10                    & 20                                 & $[-0.50, 0.55]$& $[-0.70, 0.80]$ & $[-0.45, 0.70]$\\
14                    & 33                                 & $[-0.45, 0.50]$& $[-0.60, 0.65]$ & $[-0.35, 0.55]$\\
30                    & 100                                & $[-0.30, 0.35]$& $[-0.45, 0.45]$ & $[-0.20, 0.40]$\\
\hline
\end{tabular}
\caption{Projected constraints on $\delta_4$ in scenario A from Eq.~\ref{eq:significance} when taking into account only triple Higgs inclusive cross section
and when having two bins $M_{HHH}<1$~TeV and $M_{HHH}>1$~TeV.}
\label{tab:summary}
\end{center}
}
\end{table}
In the $\delta_3=0$ case, the threshold region does not yield anymore a substantial improvement and the limits are mostly dominated by
the total cross section. With respect to the FCC projections, we observe an order $\sim 10$ improvement, depending on the collider
energy.
\begin{figure}[t]
  \centering
  \includegraphics[width=0.4\textwidth]{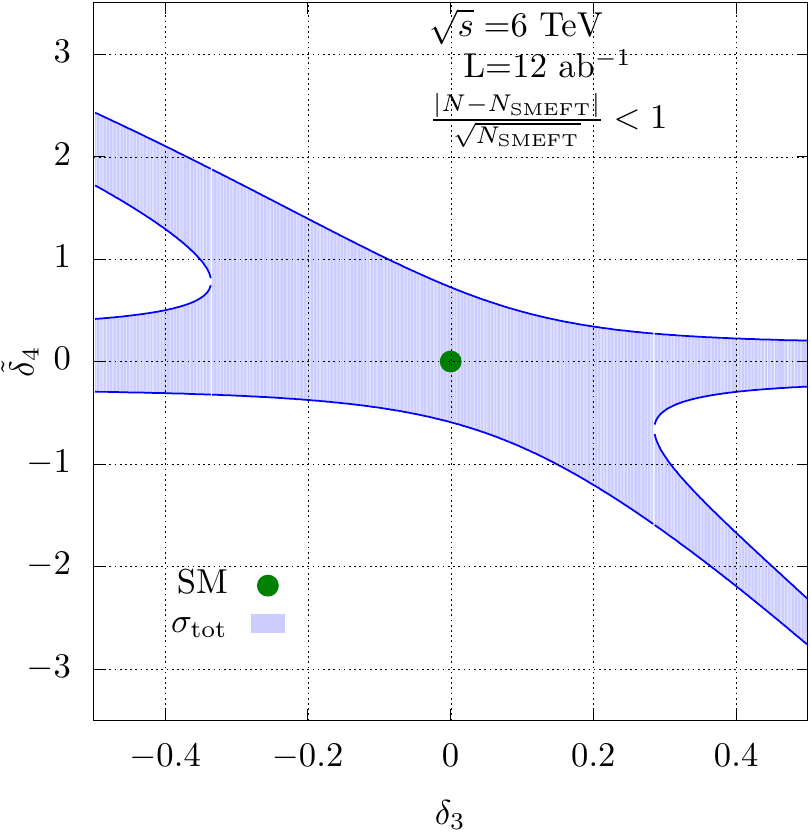}\quad
  \includegraphics[width=0.4\textwidth]{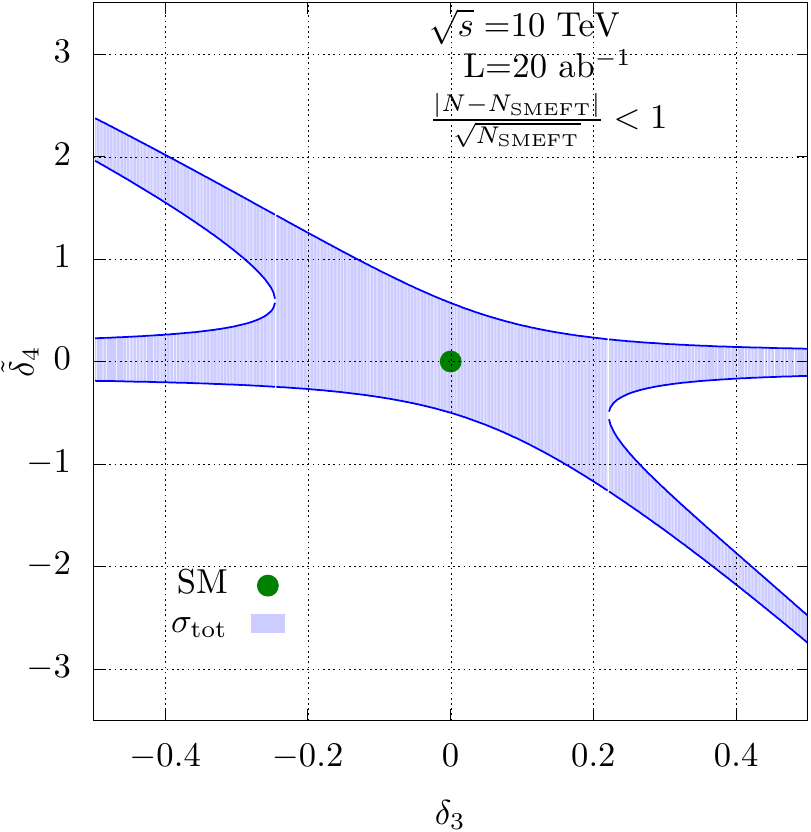}\\
  \includegraphics[width=0.4\textwidth]{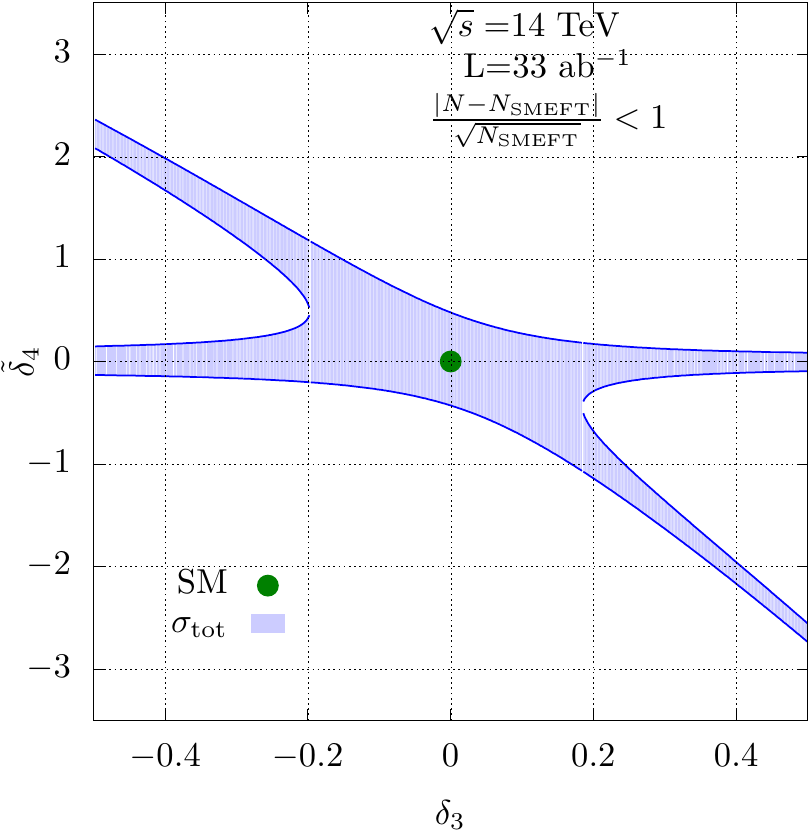}\quad
  \includegraphics[width=0.4\textwidth]{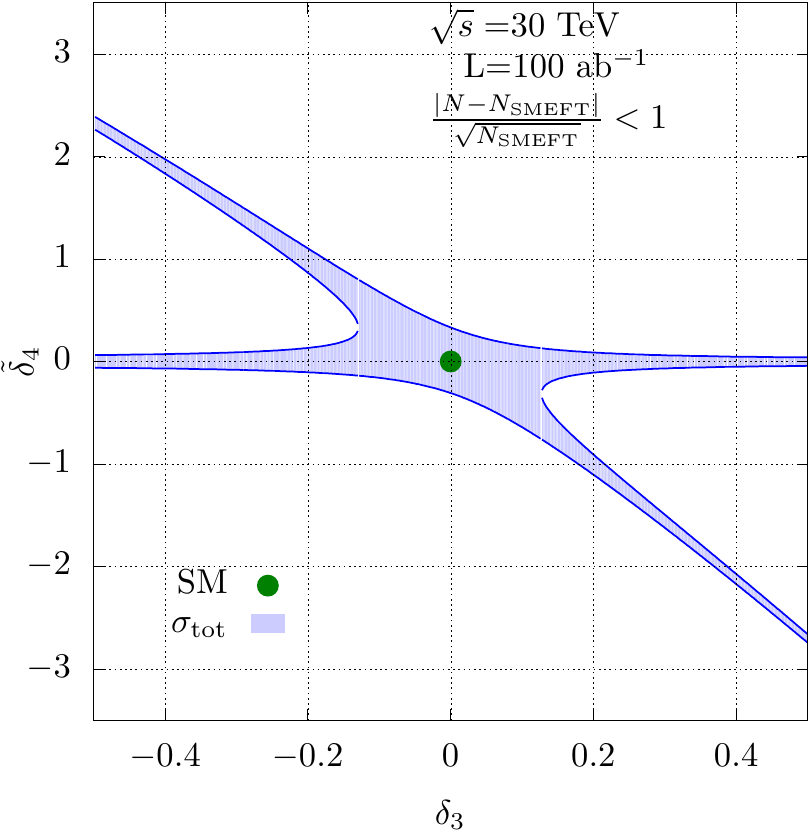}
  \caption{Projections at 1-sigma from triple Higgs production in the ($\delta_3$, $\tilde{\delta}_4$) plane for a series of benchmark
  muon collider centre of mass energies.}
\label{fig:d4tildeCT}
\end{figure}

Regarding scenario B, in the event of a precise measurement of a trilinear coupling deviation, it would be interesting to
assess the potential for the muon collider to establish whether $\delta_4$ follows the dimension-6 SMEFT relation or whether
potential dimension-8 operators come into play. In order to quantify the study, we define $\tilde \delta_4=\delta_4-6 \delta_3$
as the deviation from the dimension-6 pattern and plot in Fig.~\ref{fig:d4tildeCT} the exclusion bounds in the ($\delta_3$, $\tilde{\delta}_4$)
plane. To each $\delta_3$ values, we can associate an interval around $\tilde{\delta}_4 = 0$. In particular, in the scenario
of a large deviation in $\delta_3$, we observe that the interval in $\tilde{\delta}_4$ becomes disjoint in two very narrow intervals. 
\begin{figure}[t!]
  \centering
  \includegraphics[width=0.4\textwidth]{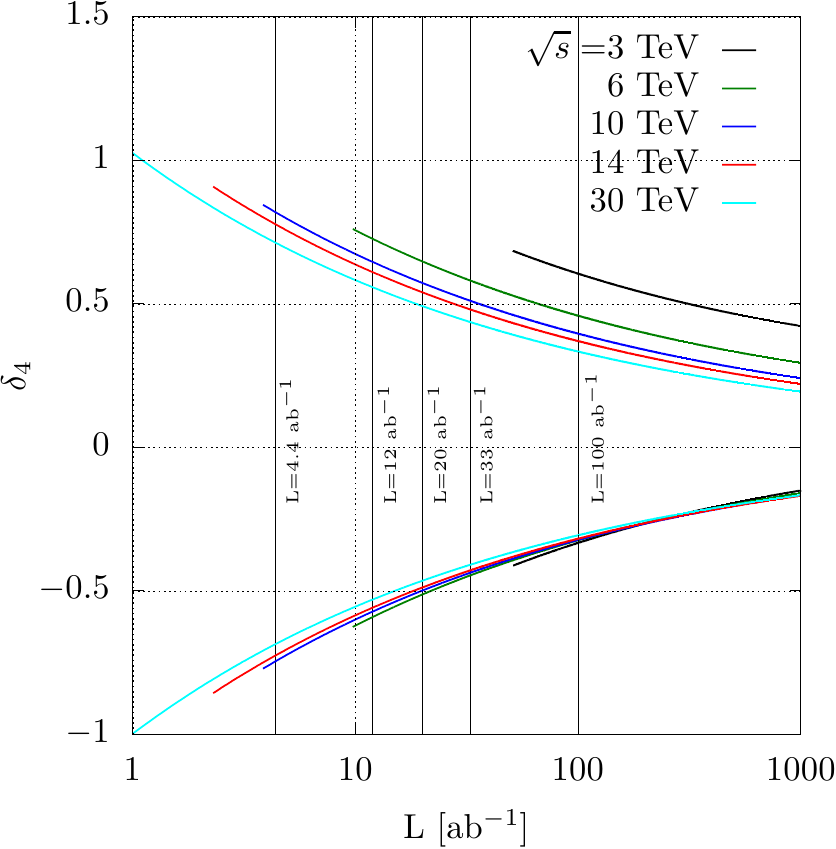}\quad
  \includegraphics[width=0.4\textwidth]{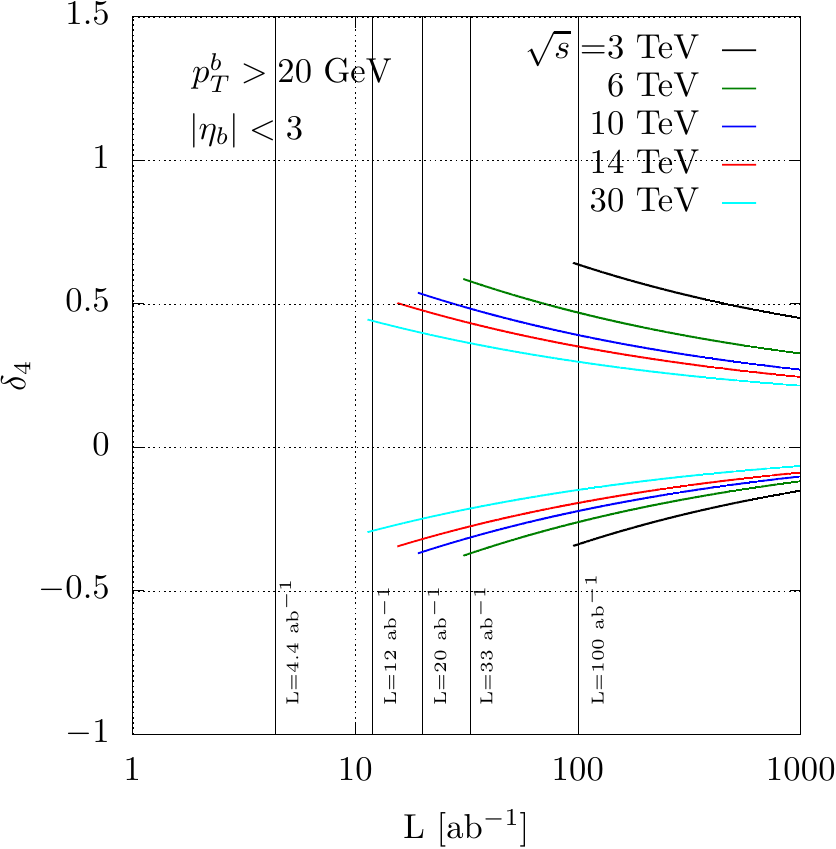}
   \caption{   \label{fig:d4vsL} Constraints at 1-sigma on $\delta_4$ from triple Higgs production in the hypothesis that $\delta_3=0$ as a function of the
   integrated luminosity. In the left panel, the full phase space region is considered while in the right one geometrical acceptance cuts
   on the Higgs decay products are imposed.}
\end{figure}

\subsection{Projected limits in a realistic detector scenario}
\label{sec:real_detector}
Finally, we turn to analyse the impact of a more realistic detector scenario.
As previously discussed, in order to avoid the interaction region from being polluted by the beam-induced radiation, the presence of
nozzles that act like shields at the entrance of the detector has been suggested in the literature. These however, can severely limit
the angular coverage. While technological advancements and novel ideas might find solutions to this problem, enabling us to
exploit a wider range in rapidity, it is interesting to assess the degree to which this is a limitation in the measurement of
the Higgs potential.

In Fig.~\ref{fig:d4vsL}, we show the projected exclusion limits at 1-sigma CL on $\delta_4$ in scenario A as a function of the total
integrated luminosity for the different centre of mass energy scenarios. An additional requirement of at least $20$ events
is imposed. Specifically, on the left we show the limits when considering the full phase space. As expected, the improvement 
with the luminosity is present, but not dramatic. On the other hand, in the right panel, we show the same bounds but requiring the Higgs
decay products to be in the geometrical acceptance region of the detector. In practise, we considered the decay in a bottom quark pair
with the requirement that $p_T^b > 20$ GeV and $|\eta_b| < 3$.

Two main messages can be understood from these plots. First of all, the acceptance cuts have a relevant impact on the statistics at hand, 
since in order to obtain at least $20$ events, considerably higher luminosities have to be considered. However, the sensitivity to
$\delta_4$ improves, \ie for the same number of events the projected limits are better in the more realistic scenario. This counter-intuitive observation stems from the fact that while SM production is mostly forward, the NP effects
are maximal in the central region. As a consequence, by applying the geometrical cuts, we mostly get rid of SM-like events
and find ourselves in a region of phase space that is ideal to detect a deviation.

\section{Summary}

In this chapter, a discussion on the potential of a muon collider to explore new interactions has been presented.
In this phase, in which detailed studies at the accelerator and detector level are not available, it is of stark importance
that phenomenological investigations establish whether this machine could potentially be a dream venue to combine the energy and 
precision frontier.

The main feature of a high-energy lepton collider is that it effectively behaves like an EW boson collider, \ie VBF
production dominates over the s-channel mode. This has been observed to be true for many SM processes and the energy scale
to which this becomes true is dependent on the multiplicity of the final state, ranging in general from $3$ to $5$ TeV.
Knowing that the scattering of EW bosons has a peculiar sensitivity to NP effects, this observation indicates that a muon
collider could be an ideal machine to test the EW sector.

Specifically, it is of paramount importance to explore the EWSB mechanism by characterising the Higgs potential. A preliminary study
has shown that projected bounds on the trilinear and quartic couplings at a muon collider are substantially better than the ones
foreseen at a $100$ TeV proton collider. We have in particular focused on the quartic coupling and verified that even in the
case of geometrical acceptance cuts in line with current detector designs, the prospects to pinpoint the four-point self-interaction
are satisfactory. This result supports the idea that when a more detailed and realistic analysis is performed, taking into account 
background estimation as well as systematic uncertainties, the worsening in statistics caused by the acceptance cuts can be tamed by observable optimisations
and detector improvements.


%
%
%
\phantomsection
\chapter*{Conclusion}
\addcontentsline{toc}{chapter}{Conclusion}
\pagestyle{fancy}

Despite amounting experimental confirmations of the Standard Model (SM) picture, the existence of New Physics (NP) is required in order 
to find an explanation to several observed phenomena, from neutrino masses to the presence of Dark Matter. In the last decades, the
experimental and theoretical efforts to find distinctive signs of Beyond Standard Model (BSM) physics have intensified, but the lack of direct discoveries at 
the LHC is now suggesting that a different approach might be needed. In the event that the energies we probe are not sufficient
to produce new states, we might nonetheless be able to find conclusive evidence of NP indirectly, \ie by scouting the tails of the
distributions in order to find deviations from the SM predictions.

In this scenario, it is of paramount importance to parametrise the modified interactions between SM particles without relying on
specific UV complete models. The Standard Model Effective Field Theory (SMEFT) is a powerful model-independent framework 
to parametrise the effects of heavy states in low-energy observables by extending the SM Lagrangian with a tower of higher dimensional operators. In this context, an in-depth analysis has been presented in 
this thesis, focusing in particular on the possibility to uncover evidence of new interactions at colliders.

After providing a review of the SMEFT framework in Chapter~\ref{chap:smeft}, in Chapter~\ref{chap:top} we discussed how
the top quark electroweak sector could be a window to NP effects. Specifically, we showed how the unitarity violating behaviour at high energy of 
$2 \to 2$ scattering amplitudes involving top quarks can be exploited in physical processes at colliders to gain sensitivity
on the SMEFT operators. Having discussed how the embedding of the amplitudes in full processes results from a complex interplay of
effects, several final states in present and future colliders that exhibit a good potential to
explore the SMEFT parameter space were identified.

In order to fully exploit the various correlations induced by the SMEFT operators, in Chapter~\ref{chap:globalfit} a discussion on a combined interpretation of top quark, Higgs and EW diboson production measurements at the LHC is
presented. In particular, state-of-the-art theoretical calculations in the SMEFT are used to constrain simultaneously $50$ 
Wilson coefficients, among which $36$ are independent degrees of freedom. The interplay and complementarity of the datasets 
is discussed by means of a set of statistical estimators and CL intervals for the various operators are provided.
This study constitutes a pivotal step in the ongoing efforts towards a fully global fit of the SMEFT, laying the ground for the inclusion
of low-energy observables and other processes from the LHC program.

Finally, in Chapter~\ref{chap:muon}, the physics case of a future muon collider is discussed, analysing both its discovery and
precision potential. In particular, we show that at sufficiently high energies, vector boson fusion becomes the dominating 
mode of production for most SM final states and the machine behaves effectively as a high-luminosity EW boson collider.
This feature opens a new window on the possibility to pinpoint the Higgs potential, a task which despite being of the foremost
importance, is challenging even in proposed future $100$ TeV proton colliders. The prospects of measuring the Higgs quartic coupling have been 
extensively analysed assuming modified interactions for different collider centre of mass energies and luminosities. Our findings show that there is a clear indication
that a muon collider is an ideal machine for precision physics and could lead to the determination of the Higgs self-couplings
with unparalleled potential.

The field of particle physics is now entering a precision era. For this reason, the thorough study of modified interactions in the context of the SMEFT
is of crucial relevance if we want to maximise the possibility to observe SM deviations. Indirect searches are indeed not limited
by the energy of the collider, but only by the accuracy of both theoretical calculations and experimental measurements. 
In this context, phenomenological studies as the ones presented in this thesis provide decisive contributions to the ultimate goal of
improving our understanding of the laws of Nature.

\begin{appendices}

\chapter{Warsaw Basis}
\label{app:Warsaw}
\begin{table}[h!]
\begin{center}
\scriptsize
\begin{minipage}[t]{3.4cm}
\renewcommand{\arraystretch}{1.5}
\begin{tabular}[t]{c|c}
\multicolumn{2}{c}{$1:X^3$} \\
\hline
$\mathcal{O}_G$                & $f^{ABC} G_\mu^{A\nu} G_\nu^{B\rho} G_\rho^{C\mu} $ \\
$\mathcal{O}_{\widetilde G}$          & $f^{ABC} \widetilde G_\mu^{A\nu} G_\nu^{B\rho} G_\rho^{C\mu} $ \\
$\mathcal{O}_W$                & $\epsilon^{IJK} W_\mu^{I\nu} W_\nu^{J\rho} W_\rho^{K\mu}$ \\ 
$\mathcal{O}_{\widetilde W}$          & $\epsilon^{IJK} \widetilde W_\mu^{I\nu} W_\nu^{J\rho} W_\rho^{K\mu}$ \\
\end{tabular}
\end{minipage}
\qquad
\begin{minipage}[t]{1.9cm}
\renewcommand{\arraystretch}{1.5}
\begin{tabular}[t]{c|c}
\multicolumn{2}{c}{$2:\varphi^6$} \\
\hline
$\mathcal{O}_\varphi$       & $(\varphi^\dag \varphi)^3$ 
\end{tabular}
\end{minipage}
\begin{minipage}[t]{4.1cm}
\renewcommand{\arraystretch}{1.5}
\begin{tabular}[t]{c|c}
\multicolumn{2}{c}{$3:\varphi^4 D^2$} \\
\hline
$\mathcal{O}_{\varphi\Box}$ & $(\varphi^\dag \varphi)\Box(\varphi^\dag \varphi)$ \\
$\mathcal{O}_{\varphi D}$   & $\ \left(\varphi^\dag D_\mu \varphi\right)^* \left(\varphi^\dag D_\mu \varphi\right)$ 
\end{tabular}
\end{minipage}
\qquad
\begin{minipage}[t]{2.6cm}
\renewcommand{\arraystretch}{1.5}
\begin{tabular}[t]{c|c}
\multicolumn{2}{c}{$4:X^2\varphi^2$} \\
\hline
$\mathcal{O}_{\varphi G}$     & $\varphi^\dag \varphi\, G^A_{\mu\nu} G^{A\mu\nu}$ \\
$\mathcal{O}_{\varphi\widetilde G}$         & $\varphi^\dag \varphi\, \widetilde G^A_{\mu\nu} G^{A\mu\nu}$ \\
$\mathcal{O}_{\varphi W}$     & $\varphi^\dag \varphi\, W^I_{\mu\nu} W^{I\mu\nu}$ \\
$\mathcal{O}_{\varphi\widetilde W}$         & $\varphi^\dag \varphi\, \widetilde W^I_{\mu\nu} W^{I\mu\nu}$ \\
$\mathcal{O}_{\varphi B}$     & $ \varphi^\dag \varphi\, B_{\mu\nu} B^{\mu\nu}$ \\
$\mathcal{O}_{\varphi\widetilde B}$         & $\varphi^\dag \varphi\, \widetilde B_{\mu\nu} B^{\mu\nu}$ \\
$\mathcal{O}_{\varphi WB}$     & $ \varphi^\dag \tau^I \varphi\, W^I_{\mu\nu} B^{\mu\nu}$ \\
$\mathcal{O}_{\varphi\widetilde W B}$         & $\varphi^\dag \tau^I \varphi\, \widetilde W^I_{\mu\nu} B^{\mu\nu}$ 
\end{tabular}
\end{minipage}
\end{center}
\caption{The bosonic operators of the Warsaw basis presented in Ref.~\cite{Grzadkowski:2010es}.}
\end{table}

\begin{table}[h!]
\begin{center}
\scriptsize
\begin{minipage}[t]{2.4cm}

\renewcommand{\arraystretch}{1.5}
\begin{tabular}[t]{c|c}
\multicolumn{2}{c}{$5: \psi^2\varphi^3 + \hbox{h.c.}$} \\
\hline
$\mathcal{O}_{e\varphi}$           & $(\varphi^\dag \varphi)(\bar l_p e_r \varphi)$ \\
$\mathcal{O}_{u\varphi}$          & $(\varphi^\dag \varphi)(\bar q_p u_r \widetilde \varphi )$ \\
$\mathcal{O}_{d\varphi}$           & $(\varphi^\dag \varphi)(\bar q_p d_r \varphi)$\\
\end{tabular}
\end{minipage}
\qquad
\begin{minipage}[t]{3.8cm}
\renewcommand{\arraystretch}{1.5}
\begin{tabular}[t]{c|c}
\multicolumn{2}{c}{$6:\psi^2 X\varphi+\hbox{h.c.}$} \\
\hline
$\mathcal{O}_{eW}$      & $(\bar l_p \sigma^{\mu\nu} e_r) \tau^I \varphi W_{\mu\nu}^I$ \\
$\mathcal{O}_{eB}$        & $(\bar l_p \sigma^{\mu\nu} e_r) \varphi B_{\mu\nu}$ \\
$\mathcal{O}_{uG}$        & $(\bar q_p \sigma^{\mu\nu} T^A u_r) \widetilde \varphi \, G_{\mu\nu}^A$ \\
$\mathcal{O}_{uW}$        & $(\bar q_p \sigma^{\mu\nu} u_r) \tau^I \widetilde \varphi \, W_{\mu\nu}^I$ \\
$\mathcal{O}_{uB}$        & $(\bar q_p \sigma^{\mu\nu} u_r) \widetilde \varphi \, B_{\mu\nu}$ \\
$\mathcal{O}_{dG}$        & $(\bar q_p \sigma^{\mu\nu} T^A d_r) \varphi\, G_{\mu\nu}^A$ \\
$\mathcal{O}_{dW}$         & $(\bar q_p \sigma^{\mu\nu} d_r) \tau^I \varphi\, W_{\mu\nu}^I$ \\
$\mathcal{O}_{dB}$        & $(\bar q_p \sigma^{\mu\nu} d_r) \varphi\, B_{\mu\nu}$ 
\end{tabular}
\end{minipage}
\begin{minipage}[t]{4.2cm}
\renewcommand{\arraystretch}{1.5}
\begin{tabular}[t]{c|c}
\multicolumn{2}{c}{$7:\psi^2\varphi^2 D$} \\
\hline
$\mathcal{O}_{\varphi l}^{(1)}$      & $(\varphi^\dag i\overleftrightarrow{D}_\mu \varphi)(\bar l_p \gamma^\mu l_r)$\\
$\mathcal{O}_{\varphi l}^{(3)}$      & $(\varphi^\dag i\overleftrightarrow{D}^I_\mu \varphi)(\bar l_p \tau^I \gamma^\mu l_r)$\\
$\mathcal{O}_{\varphi e}$            & $(\varphi^\dag i\overleftrightarrow{D}_\mu \varphi)(\bar e_p \gamma^\mu e_r)$\\
$\mathcal{O}_{\varphi q}^{(1)}$      & $(\varphi^\dag i\overleftrightarrow{D}_\mu \varphi)(\bar q_p \gamma^\mu q_r)$\\
$\mathcal{O}_{\varphi q}^{(3)}$      & $(\varphi^\dag i\overleftrightarrow{D}^I_\mu \varphi)(\bar q_p \tau^I \gamma^\mu q_r)$\\
$\mathcal{O}_{\varphi u}$            & $(\varphi^\dag i\overleftrightarrow{D}_\mu \varphi)(\bar u_p \gamma^\mu u_r)$\\
$\mathcal{O}_{\varphi d}$            & $(\varphi^\dag i\overleftrightarrow{D}_\mu \varphi)(\bar d_p \gamma^\mu d_r)$\\
$\mathcal{O}_{\varphi u d}$   & $i(\widetilde \varphi ^\dag D_\mu \varphi)(\bar u_p \gamma^\mu d_r)$\\
\end{tabular}
\end{minipage}
\end{center}
\caption{The current operators of the Warsaw basis presented in Ref.~\cite{Grzadkowski:2010es}.}
\end{table}

\begin{table}
\begin{center}
\scriptsize

\begin{minipage}[t]{3.75cm}
\renewcommand{\arraystretch}{1.5}
\begin{tabular}[t]{c|c}
\multicolumn{2}{c}{$8:(\bar LL)(\bar LL)$} \\
\hline
$\mathcal{O}_{\ell \ell}$        & $(\bar l_p \gamma_\mu l_r)(\bar l_s \gamma^\mu l_t)$ \\
$\mathcal{O}_{qq}^{(1)}$  & $(\bar q_p \gamma_\mu q_r)(\bar q_s \gamma^\mu q_t)$ \\
$\mathcal{O}_{qq}^{(3)}$  & $(\bar q_p \gamma_\mu \tau^I q_r)(\bar q_s \gamma^\mu \tau^I q_t)$ \\
$\mathcal{O}_{\ell q}^{(1)}$                & $(\bar l_p \gamma_\mu l_r)(\bar q_s \gamma^\mu q_t)$ \\
$\mathcal{O}_{\ell q}^{(3)}$                & $(\bar l_p \gamma_\mu \tau^I l_r)(\bar q_s \gamma^\mu \tau^I q_t)$ 
\end{tabular}
\end{minipage}
\qquad
\begin{minipage}[t]{4.25cm}
\renewcommand{\arraystretch}{1.5}
\begin{tabular}[t]{c|c}
\multicolumn{2}{c}{$8:(\bar RR)(\bar RR)$} \\
\hline
$\mathcal{O}_{ee}$               & $(\bar e_p \gamma_\mu e_r)(\bar e_s \gamma^\mu e_t)$ \\
$\mathcal{O}_{uu}$        & $(\bar u_p \gamma_\mu u_r)(\bar u_s \gamma^\mu u_t)$ \\
$\mathcal{O}_{dd}$        & $(\bar d_p \gamma_\mu d_r)(\bar d_s \gamma^\mu d_t)$ \\
$\mathcal{O}_{eu}$                      & $(\bar e_p \gamma_\mu e_r)(\bar u_s \gamma^\mu u_t)$ \\
$\mathcal{O}_{ed}$                      & $(\bar e_p \gamma_\mu e_r)(\bar d_s\gamma^\mu d_t)$ \\
$\mathcal{O}_{ud}^{(1)}$                & $(\bar u_p \gamma_\mu u_r)(\bar d_s \gamma^\mu d_t)$ \\
$\mathcal{O}_{ud}^{(8)}$                & $(\bar u_p \gamma_\mu T^A u_r)(\bar d_s \gamma^\mu T^A d_t)$ \\
\end{tabular}
\end{minipage}
\begin{minipage}[t]{4.75cm}
\renewcommand{\arraystretch}{1.5}
\begin{tabular}[t]{c|c}
\multicolumn{2}{c}{$8:(\bar LL)(\bar RR)$} \\
\hline
$\mathcal{O}_{le}$               & $(\bar l_p \gamma_\mu l_r)(\bar e_s \gamma^\mu e_t)$ \\
$\mathcal{O}_{lu}$               & $(\bar l_p \gamma_\mu l_r)(\bar u_s \gamma^\mu u_t)$ \\
$\mathcal{O}_{ld}$               & $(\bar l_p \gamma_\mu l_r)(\bar d_s \gamma^\mu d_t)$ \\
$\mathcal{O}_{qe}$               & $(\bar q_p \gamma_\mu q_r)(\bar e_s \gamma^\mu e_t)$ \\
$\mathcal{O}_{qu}^{(1)}$         & $(\bar q_p \gamma_\mu q_r)(\bar u_s \gamma^\mu u_t)$ \\ 
$\mathcal{O}_{qu}^{(8)}$         & $(\bar q_p \gamma_\mu T^A q_r)(\bar u_s \gamma^\mu T^A u_t)$ \\ 
$\mathcal{O}_{qd}^{(1)}$ & $(\bar q_p \gamma_\mu q_r)(\bar d_s \gamma^\mu d_t)$ \\
$\mathcal{O}_{qd}^{(8)}$ & $(\bar q_p \gamma_\mu T^A q_r)(\bar d_s \gamma^\mu T^A d_t)$\\
\end{tabular}
\end{minipage}
\begin{minipage}[t]{3.5cm}
\renewcommand{\arraystretch}{1.5}
\begin{tabular}[t]{c|c}
\multicolumn{2}{c}{$8:(\bar LR)(\bar L R)+\hbox{h.c.}$} \\
\hline
$\mathcal{O}_{quqd}^{(1)}$ & $(\bar q_p^j u_r) \epsilon_{jk} (\bar q_s^k d_t)$ \\
$\mathcal{O}_{quqd}^{(8)}$ & $(\bar q_p^j T^A u_r) \epsilon_{jk} (\bar q_s^k T^A d_t)$ \\
$\mathcal{O}_{lequ}^{(1)}$ & $(\bar l_p^j e_r) \epsilon_{jk} (\bar q_s^k u_t)$ \\
$\mathcal{O}_{lequ}^{(3)}$ & $(\bar l_p^j \sigma_{\mu\nu} e_r) \epsilon_{jk} (\bar q_s^k \sigma^{\mu\nu} u_t)$\\
\\
\multicolumn{2}{c}{$8:(\bar LR)(\bar RL)+\hbox{h.c.}$} \\
\hline
$\mathcal{O}_{ledq}$ & $(\bar l_p^j e_r)(\bar d_s q_{tj})$ 
\end{tabular}
\end{minipage}
\end{center}
\caption{\label{tab:4fbasis}
Four fermion operators of the Warsaw basis given in 
Ref.~\cite{Grzadkowski:2010es}.}
\end{table}

\chapter{Helicity amplitudes for top-EW scatterings\label{app:helamp_tables}}

We report here the helicity amplitude computations for the generic $f \, B \to f^\prime \, B^\prime$ processes.
The explicit representations for the spinors and polarisations used are
    \begin{align}
        u_+(p)&=\sqrt{E+m}\begin{pmatrix}
        \cos{\frac{\theta}{2}} ,& \sin{\frac{\theta}{2}} ,& \frac{|\textbf{p}|\cos{\frac{\theta}{2}}}{E+m} ,& \frac{|\textbf{p}|\sin{\frac{\theta}{2}}}{E+m}
        \end{pmatrix} \, ,
        \\
        u_-(p)&=\sqrt{E+m}\begin{pmatrix}
        -\sin{\frac{\theta}{2}} ,& \cos{\frac{\theta}{2}} ,& \frac{|\textbf{p}|\sin{\frac{\theta}{2}}}{E+m} ,& -\frac{|\textbf{p}|\cos{\frac{\theta}{2}}}{E+m}
        \end{pmatrix} \, ,
        \\
        v_+(p)&=\sqrt{E+m}\begin{pmatrix}
    \frac{|\textbf{p}|\sin{\frac{\theta}{2}}}{E+m} ,& -\frac{|\textbf{p}|\cos{\frac{\theta}{2}}}{E+m} ,& -\sin{\frac{\theta}{2}} ,& \cos{\frac{\theta}{2}}
        \end{pmatrix} \, ,
        \\
        v_-(p)&=\sqrt{E+m}\begin{pmatrix}
        \frac{|\textbf{p}|\cos{\frac{\theta}{2}}}{E+m}  ,& \frac{|\textbf{p}|\sin{\frac{\theta}{2}}}{E+m} ,& \cos{\frac{\theta}{2}} ,& \sin{\frac{\theta}{2}}
        \end{pmatrix} \, ,
        \\
        \varepsilon_\pm(p)&=\frac{1}{\sqrt{2}}\begin{pmatrix}
        0 ,& \cos{\theta} ,& \pm i ,& -\sin{\theta}
        \end{pmatrix} \, ,
        \\
        \varepsilon_0(p)&=\frac{1}{M}\begin{pmatrix}
        |\textbf{p}| ,& E \sin{\theta} ,& 0 ,& E \cos{\theta}
        \end{pmatrix} \, .  
    \end{align}
where m, $E$ and $\textbf{p}$ denote the mass, energy and 3-momentum of a particle and $\theta$ is the polar angle. The calculations are done in the high nergy limit $s~\sim -t\gg v$. The results of the calculations are presented in tables containing the SM prediction and the contribution of each operator from Table~\ref{tab:topOps}. For the SM predictions we do not distinguish between $s$ and $t$ and only retained down to $1/s$ dependence.
For the SMEFT operators we report energy dependence down to constant energy behaviour ($s^0$), not including numerical factors.
\clearpage
    \renewcommand{\arraystretch}{1.2}
\begin{table}[h!]
{\tiny
\begin{center}
\begin{tabular}{c|cccccc}
$\lambda_{\sss b}$, $\lambda_{\sss W}$, $\lambda_{\sss t}$, $\lambda_\gamma$&SM&$\Op{\phi WB}$&$\Op{W}$&$\Op{tB}$&$\Op{tW}$\tabularnewline
\hline
$-, 0, -, -$&$\frac{1}{\sqrt{s}}$&$\frac{s m_W}{\sqrt{-t}}$&$\frac{m_W^2 (s+ t)}{\sqrt{-t} v}$&$ \sqrt{-t} m_t$&$\sqrt{-t} m_t$\tabularnewline
$-, 0, -, +$&$\frac{1}{\sqrt{s}}$&$\frac{m_W (s+t)}{\sqrt{-t}}$&$\frac{m_W^2 (s+t)}{\sqrt{-t} v}$&$-$&$-$\tabularnewline
$-, 0, +, -$&$s^0$&$s^0$&$s^0$&$-$&$s^0$\tabularnewline
$-, 0, +, +$&$\frac{1}{s}$&$-$&$-$&$ \sqrt{s (s+t)}$&$ \sqrt{s (s+t)}$\tabularnewline
$+, 0, \,\ast\,, \,\ast\,$&$-$&$-$&$-$&$-$&$-$\tabularnewline
\hline
$-, -, -, -$&$s^0$&$s^0$&$s^0$&$s^0$&$s^0$\tabularnewline
$-,\pm, -, \mp$&$\frac{1}{s}$&$s^0$&$\frac{ m_W \sqrt{s (s+t)}}{v}$&$-$&$-$\tabularnewline
$-, \pm, +, \pm$&$\frac{1}{\sqrt{s}}$&$-$&$-$&$-$&$\frac{m_W ( s+ t)}{ \sqrt{-t}}$\tabularnewline
$-, -, +, +$&$-$&$\sqrt{-t} m_t$&$\frac{ \sqrt{-t} m_t m_W}{v}$&$\sqrt{-t} m_W$&$\sqrt{-t} m_W$\tabularnewline
$-, +, -, +$&$s^0$&$s^0$&$-$&$-$&$-$\tabularnewline
$-, +, +, -$&$\frac{1}{\sqrt{s}}$&$\sqrt{-t} m_t$&$\frac{\sqrt{-t} m_t m_W}{v}$&$-$&$-$\tabularnewline
$+, \,\ast\,, \,\ast\,, \,\ast\,$&$-$&$-$&$-$&$-$&$-$\tabularnewline
\end{tabular}
\end{center}
}
\caption{\label{tab:bwta}
Helicity amplitudes in the high-energy limit ($s\sim-t\gg v$) for $b W^+\to t \gamma$ scattering. Overall multiplicative factors are dropped and only the schematic energy
dependence is retained for the SM contribution. Helicity entries marked by `$\ast$' indicate any combination of $\pm,\mp$.
}
\end{table}
\renewcommand{\arraystretch}{1.}
\renewcommand{\arraystretch}{1.2}
\begin{table}[h!]
{\tiny
\setlength{\tabcolsep}{0.1pt}
\begin{center}
\begin{tabular}{c|cccccccccc}
$\lambda_{\sss b}$, $\lambda_{\sss W}$, $\lambda_{\sss t}$, $\lambda_{\sss Z}$&SM&$\Op{\phi D}$&$\Op{\phi WB}$&$\Op{W}$&$\Op{tB}$&$\Op{tW}$&$\Op{\phi Q}^{\sss (1)}$&$\Op{\phi Q}^{\sss (3)}$&$\Op{\phi t}$&$\Op{\phi tb}$\tabularnewline
\hline
$-, 0, -, 0$&$s^0$&$s^0$&$s^0$&$s^0$&$-$&$s^0$&$-$&$\sqrt{s (s+t)}$&$-$&$-$\tabularnewline
$-, 0, +, 0$&$\frac{1}{\sqrt{s}}$&$\sqrt{-t} m_t$&$-$&$-$&$\sqrt{-t} m_W$&$\frac{m_W (s+t)}{\sqrt{-t}}$&$\sqrt{-t} m_t$&$ \sqrt{-t} m_t$&$\sqrt{-t} m_t$&$-$\tabularnewline
$+, 0, -, 0$&$-$&$-$&$-$&$-$&$-$&$-$&$-$&$-$&$-$&$-$\tabularnewline
$+, 0, +, 0$&$-$&$-$&$-$&$-$&$-$&$-$&$-$&$-$&$-$&$\sqrt{s (s+t)}$\tabularnewline
\hline
$-, 0, -, -$&$\frac{1}{\sqrt{s}}$&$-$&$\frac{s m_W}{\sqrt{-t}}$&$\frac{m_W^2 (s+t)}{\sqrt{-t} v}$&$\sqrt{-t} m_t$&$\sqrt{-t} m_t$&$-$&$\sqrt{-t} m_W$&$-$&$-$\tabularnewline
$-, -, -, 0$&$\frac{1}{\sqrt{s}}$&$-$&$-$&$\frac{m_W^2 (s+t)}{\sqrt{-t} v}$&$-$&$-$&$-$&$\sqrt{-t}m_W$&$-$&$-$\tabularnewline
$-, 0, +, -$&$s^0$&$s^0$&$s^0$&$s^0$&$-$&$s^0$&$s^0$&$s^0$&$-$&$-$\tabularnewline
$-, -, +, 0$&$\frac{1}{s}$&$s^0$&$s^0$&$s^0$&$s^0$&$\sqrt{s (s+t)}$&$s^0$&$s^0$&$s^0$&$-$\tabularnewline
$-, 0, -, +$&$\frac{1}{\sqrt{s}}$&$-$&$\frac{m_W (s+t)}{\sqrt{-t}}$&$\frac{m_W^2 (s+t)}{\sqrt{-t} v}$&$-$&$-$&$-$&$-$&$-$&$-$\tabularnewline
$-, +, -, 0$&$\frac{1}{\sqrt{s}}$&$-$&$-$&$\frac{m_W^2 (s+t)}{\sqrt{-t} v}$&$-$&$-$&$-$&$-$&$-$&$-$\tabularnewline
$-, 0, +, +$&$\frac{1}{s}$&$s^0$&$-$&$-$&$\sqrt{s (s+t)}$&$\sqrt{s (s+t)}$&$s^0$&$s^0$&$s^0$&$-$\tabularnewline
$-, +, +, 0$&$s^0$&$s^0$&$s^0$&$-$&$-$&$s^0$&$-$&$s^0$&$-$&$-$\tabularnewline
$+, 0, -, \pm$&$-$&$-$&$-$&$-$&$-$&$-$&$-$&$-$&$-$&$s^0$\tabularnewline
$+, -, -, 0$&$-$&$-$&$-$&$-$&$-$&$-$&$-$&$-$&$-$&$s^0$\tabularnewline
$+, 0, +, -$&$-$&$-$&$-$&$-$&$-$&$-$&$-$&$-$&$-$&$-$\tabularnewline
$+, \pm, \mp, 0$&$-$&$-$&$-$&$-$&$-$&$-$&$-$&$-$&$-$&$-$\tabularnewline
$+, 0, +, +$&$-$&$-$&$-$&$-$&$-$&$-$&$-$&$-$&$-$&$\sqrt{-t}m_W$\tabularnewline
$+, +, +, 0$&$-$&$-$&$-$&$-$&$-$&$-$&$-$&$-$&$-$&$\sqrt{-t}m_W$\tabularnewline
\hline
$-, -, -, -$&$s^0$&$s^0$&$s^0$&$s^0$&$s^0$&$s^0$&$s^0$&$s^0$&$-$&$-$\tabularnewline
$-, -, -, +$&$\frac{1}{s}$&$-$&$s^0$&$\frac{m_W \sqrt{s (s+t)}}{v}$&$-$&$-$&$-$&$-$&$-$&$-$\tabularnewline
$-, -, +, -$&$\frac{1}{\sqrt{s}}$&$-$&$-$&$-$&$-$&$\frac{m_W (s+t)}{\sqrt{-t}}$&$-$&$-$&$-$&$-$\tabularnewline
$-, -, +, +$&$-$&$-$&$\sqrt{-t} m_t$&$\frac{\sqrt{-t} m_t m_W}{v}$&$\sqrt{-t} m_W$&$\sqrt{-t} m_W$&$-$&$-$&$-$&$-$\tabularnewline
$-, +, -, -$&$\frac{1}{s}$&$-$&$s^0$&$\frac{m_W \sqrt{s (s+t)}}{v}$&$-$&$-$&$-$&$-$&$-$&$-$\tabularnewline
$-, +, -, +$&$s^0$&$s^0$&$s^0$&$-$&$-$&$-$&$s^0$&$s^0$&$-$&$-$\tabularnewline
$-, +, +, -$&$\frac{1}{\sqrt{s}}$&$-$&$\sqrt{-t} m_t$&$\frac{\sqrt{-t} m_t m_W}{v}$&$-$&$-$&$-$&$-$&$-$&$-$\tabularnewline
$-, +, +, +$&$\frac{1}{\sqrt{s}}$&$-$&$-$&$-$&$-$&$\frac{m_W (s+t)}{\sqrt{-t}}$&$-$&$-$&$-$&$-$\tabularnewline
$+, \,\ast\,, -, \,\ast\,$&$-$&$-$&$-$&$-$&$-$&$-$&$-$&$-$&$-$&$-$\tabularnewline
$+, \pm, +, \pm$&$-$&$-$&$-$&$-$&$-$&$-$&$-$&$-$&$-$&$s^0$\tabularnewline
$+, \pm, +, \mp$&$-$&$-$&$-$&$-$&$-$&$-$&$-$&$-$&$-$&$-$\tabularnewline
\end{tabular}
\end{center}

}
\caption{\label{tab:bwtz}
Helicity amplitudes in the high-energy limit ($s\sim-t\gg v$) for $b\,W^+\to t\,Z$ scattering. Overall multiplicative factors are dropped and only the schematic energy
dependence is retained for the SM contribution. Helicity entries marked by `$\ast$' indicate any combination of $\pm,\mp$.
}
\end{table}
\renewcommand{\arraystretch}{1.}
    \renewcommand{\arraystretch}{1.2}
\begin{table}[h!]
\begingroup
\fontsize{4.8pt}{8pt}\selectfont
\setlength{\tabcolsep}{0.pt}
\begin{center}
\begin{tabular}{c|cccccccccccc}
$\lambda_{\sss t}$, $\lambda_{\sss W}$, $\lambda_{\sss t}$, $\lambda_{\sss W}$&SM&$\Op{\phi D}$&$\Op{\phi d}$&$\Op{\phi W}$&$\Op{\phi WB}$&$\Op{W}$&$\Op{t \phi}$&$\Op{tB}$&$\Op{tW}$&$\Op{\phi Q}^{\sss (1)}$&$\Op{\phi Q}^{\sss (3)}$&$\Op{\phi t}$\tabularnewline
\hline
$-, 0, -, 0$&$s^0$&$s^0$&$s^0$&$-$&$s^0$&$s^0$&$s^0$&$s^0$&$s^0$&$ \sqrt{s (s+t)}$&$ \sqrt{s (s+t)}$&$-$\tabularnewline
$\pm, 0, \mp, 0$&$\frac{1}{\sqrt{s}}$&$\sqrt{-t} m_t$&$ \sqrt{-t} m_t$&$-$&$-$&$-$&$\sqrt{-t} v$&$\frac{m_W (s+t)}{\sqrt{-t}}$&$\frac{m_W (s+t)}{\sqrt{-t}}$&$\sqrt{-t} m_t$&$\sqrt{-t} m_t$&$\sqrt{-t} m_t$\tabularnewline
$+, 0, +, 0$&$s^0$&$s^0$&$s^0$&$-$&$s^0$&$-$&$s^0$&$s^0$&$-$&$-$&$s^0$&$\sqrt{s (s+t)}$\tabularnewline
\hline
$-, 0, -, -$&$\frac{1}{\sqrt{s}}$&$-$&$-$&$-$&$\frac{s m_W}{ \sqrt{-t}}$&$\frac{m_W^2 (s+ t)}{\sqrt{-t} v}$&$-$&$-$&$-$&$\sqrt{-t} m_W$&$ \sqrt{-t} m_W$&$-$\tabularnewline
$-, -, -, 0$&$\frac{1}{\sqrt{s}}$&$-$&$-$&$-$&$\frac{s m_W}{ \sqrt{-t}}$&$\frac{m_W^2 (s+ t)}{\sqrt{-t} v}$&$-$&$-$&$-$&$ \sqrt{-t} m_W$&$ \sqrt{-t} m_W$&$-$\tabularnewline
$-, 0, +, -$&$s^0$&$s^0$&$s^0$&$s^0$&$s^0$&$s^0$&$s^0$&$s^0$&$s^0$&$s^0$&$s^0$&$s^0$\tabularnewline
$-, -, +, 0$&$-$&$s^0$&$s^0$&$s^0$&$s^0$&$s^0$&$s^0$&$s^0$&$s^0$&$s^0$&$s^0$&$s^0$\tabularnewline
$-, 0, -, +$&$\frac{1}{\sqrt{s}}$&$-$&$-$&$-$&$\frac{ m_W (s+t)}{ \sqrt{-t}}$&$\frac{ m_W^2 (s+t)}{\sqrt{-t} v}$&$-$&$-$&$-$&$-$&$-$&$-$\tabularnewline
$-, +, -, 0$&$\frac{1}{\sqrt{s}}$&$-$&$-$&$-$&$\frac{ m_W (s+t)}{\sqrt{-t}}$&$\frac{ m_W^2 (s+t)}{\sqrt{-t} v}$&$-$&$-$&$-$&$-$&$-$&$-$\tabularnewline
$-, 0, +, +$&$-$&$s^0$&$s^0$&$s^0$&$s^0$&$-$&$s^0$&$s^0$&$\sqrt{s (s+t)}$&$s^0$&$s^0$&$s^0$\tabularnewline
$-, +, +, 0$&$-$&$s^0$&$s^0$&$s^0$&$s^0$&$-$&$s^0$&$s^0$&$s^0$&$s^0$&$s^0$&$s^0$\tabularnewline
$+, 0, -, -$&$-$&$s^0$&$s^0$&$s^0$&$s^0$&$s^0$&$s^0$&$s^0$&$s^0$&$s^0$&$s^0$&$s^0$\tabularnewline
$+, -, -, 0$&$s^0$&$s^0$&$s^0$&$s^0$&$s^0$&$s^0$&$s^0$&$s^0$&$s^0$&$s^0$&$s^0$&$s^0$\tabularnewline
$+, 0, +, -$&$\frac{1}{\sqrt{s}}$&$-$&$-$&$-$&$\frac{ m_W (s+t)}{\sqrt{-t}}$&$-$&$-$&$-$&$\sqrt{-t} m_t$&$-$&$-$&$-$\tabularnewline
$+, -, +, 0$&$\frac{1}{\sqrt{s}}$&$-$&$-$&$-$&$\frac{ m_W (s+t)}{ \sqrt{-t}}$&$-$&$-$&$-$&$ \sqrt{-t} m_t$&$-$&$-$&$-$\tabularnewline
$+, 0, -, +$&$-$&$s^0$&$s^0$&$s^0$&$s^0$&$-$&$s^0$&$s^0$&$s^0$&$s^0$&$s^0$&$s^0$\tabularnewline
$+, +, -, 0$&$-$&$s^0$&$s^0$&$s^0$&$s^0$&$-$&$s^0$&$s^0$&$ \sqrt{s (s+t)}$&$s^0$&$s^0$&$s^0$\tabularnewline
$+, 0, +, +$&$\frac{1}{\sqrt{s}}$&$-$&$-$&$-$&$\frac{s m_W}{ \sqrt{-t}}$&$-$&$-$&$-$&$\sqrt{-t} m_t$&$-$&$-$&$ \sqrt{-t} m_W$\tabularnewline
$+, +, +, 0$&$\frac{1}{\sqrt{s}}$&$-$&$-$&$-$&$\frac{ s m_W}{\sqrt{-t}}$&$-$&$-$&$-$&$\sqrt{-t} m_t$&$-$&$-$&$ \sqrt{-t} m_W$\tabularnewline
\hline
$-, -, -, -$&$s^0$&$-$&$-$&$-$&$s^0$&$s^0$&$-$&$-$&$s^0$&$-$&$s^0$&$-$\tabularnewline
$-, \pm, -, \mp$&$\frac{1}{s}$&$-$&$-$&$s^0$&$s^0$&$\frac{m_W \sqrt{s (s+t)}}{v}$&$-$&$-$&$-$&$-$&$-$&$-$\tabularnewline
$-, +, -, +$&$s^0$&$-$&$-$&$-$&$s^0$&$-$&$-$&$-$&$-$&$s^0$&$-$&$-$\tabularnewline
$-, \pm, \pm, \mp$&$-$&$-$&$-$&$\sqrt{-t} m_t$&$-$&$\frac{\sqrt{-t} m_t m_W}{v}$&$-$&$-$&$-$&$-$&$-$&$-$\tabularnewline
$\pm, -, \mp, -$&$\frac{1}{\sqrt{s}}$&$-$&$-$&$-$&$-$&$-$&$-$&$-$&$\frac{s m_W}{\sqrt{-t}}$&$-$&$-$&$-$\tabularnewline
$\pm, +, \mp, +$&$\frac{1}{\sqrt{s}}$&$-$&$-$&$-$&$-$&$-$&$-$&$-$&$\frac{ m_W (s+t)}{\sqrt{-t}}$&$-$&$-$&$-$\tabularnewline
$\pm, \pm, \mp, \mp$&$-$&$-$&$-$&$\sqrt{-t} m_t$&$-$&$\frac{\sqrt{-t} m_t m_W}{v}$&$-$&$-$&$ \sqrt{-t} m_W$&$-$&$-$&$-$\tabularnewline
$+, -, +, -$&$\frac{1}{s}$&$-$&$-$&$-$&$s^0$&$-$&$-$&$-$&$s^0$&$-$&$-$&$s^0$\tabularnewline
$+, \pm, +, \mp$&$-$&$-$&$-$&$s^0$&$s^0$&$s^0$&$-$&$-$&$-$&$-$&$-$&$-$\tabularnewline
$+, +, +, +$&$\frac{1}{s}$&$-$&$-$&$-$&$s^0$&$-$&$-$&$-$&$s^0$&$-$&$-$&$-$\tabularnewline
\end{tabular}
\end{center}
\endgroup
\caption{\label{tab:twtw}
Helicity amplitudes in the high-energy limit ($s\sim-t\gg v$) for $t W^+\to t W^+$ scattering. Overall multiplicative factors are dropped and only the schematic energy
dependence is retained for the SM contribution.  Helicity entries marked by `$\ast$' indicate any combination of $\pm,\mp$.
}
\end{table}
\renewcommand{\arraystretch}{1.}

    \renewcommand{\arraystretch}{1.2}
\begin{table}[h!]
{\tiny
\begin{center}
\begin{tabular}{c|cccccc}
$\lambda_{\sss b}$, $\lambda_{\sss W}$, $\lambda_{\sss t}$&SM&$\Op{\phi W}$&$\Op{t \phi}$&$\Op{tW}$&$\Op{\phi Q}^{\sss (3)}$&$\Op{\phi tb}$\tabularnewline
\hline
$-, 0, -$&$s^0$&$s^0$&$s^0$&$s^0$&$\sqrt{s (s+t)}$&$-$\tabularnewline
$-, 0, +$&$\frac{1}{\sqrt{s}}$&$-$&$\sqrt{-t} v$&$\frac{s m_W}{\sqrt{-t}}$&$\sqrt{-t} m_t$&$-$\tabularnewline
$+, 0, -$&$-$&$-$&$-$&$-$&$-$&$\sqrt{-t} m_t$\tabularnewline
$+, 0, +$&$-$&$-$&$-$&$-$&$-$&$ \sqrt{s (s+t)}$\tabularnewline
\hline
$-, -, -$&$\frac{1}{\sqrt{s}}$&$\frac{s m_W}{\sqrt{-t}}$&$-$&$\sqrt{-t} m_t$&$\sqrt{-t} m_W$&$-$\tabularnewline
$-, -, +$&$\frac{1}{s}$&$-$&$s^0$&$ \sqrt{s (s+t)}$&$s^0$&$-$\tabularnewline
$-, +, -$&$\frac{1}{\sqrt{s}}$&$\frac{m_W (s+t)}{\sqrt{-t}}$&$-$&$-$&$-$&$-$\tabularnewline
$-, +, +$&$s^0$&$s^0$&$-$&$s^0$&$s^0$&$-$\tabularnewline
$+, \pm, -$&$-$&$-$&$-$&$-$&$-$&$s^0$\tabularnewline
$+, -, +$&$-$&$-$&$-$&$-$&$-$&$-$\tabularnewline
$+, +, +$&$-$&$-$&$-$&$-$&$-$&$\sqrt{-t} m_W$\tabularnewline
\end{tabular}
\end{center}
}
\caption{\label{tab:bwth}
Helicity amplitudes in the high-energy limit ($s\sim-t\gg v$) for $b \, W^+ \to t \, h$ scattering.
Overall multiplicative factors are dropped and only the schematic energy
dependence is retained for the SM contribution. Helicity entries marked by `$\ast$' indicate any combination of $\pm,\mp$.
}
\end{table}
\renewcommand{\arraystretch}{1.}


    \renewcommand{\arraystretch}{1.2}
\begin{table}[h!]
{\tiny
\setlength{\tabcolsep}{0.2pt}
\begin{center}
\begin{tabular}{c|cccccccccccc}
$\lambda_{\sss t}$, $\lambda_{\sss Z}$, $\lambda_{\sss t}$, $\lambda_{\sss Z}$&SM&$\Op{\phi D}$&$\Op{\phi d}$&$\Op{\phi B}$&$\Op{\phi W}$&$\Op{\phi WB}$&$\Op{t \phi}$&$\Op{tB}$&$\Op{tW}$&$\Op{\phi Q}^{\sss (1)}$&$\Op{\phi Q}^{\sss (3)}$&$\Op{\phi t}$\tabularnewline
\hline
$-, 0, -, 0$&$s^0$&$s^0$&$s^0$&$-$&$-$&$-$&$s^0$&$s^0$&$s^0$&$s^0$&$s^0$&$s^0$\tabularnewline
$\pm, 0, \mp, 0$&$\frac{1}{\sqrt{s}}$&$\sqrt{-t} m_t$&$\sqrt{-t} m_t$&$-$&$-$&$-$&$\sqrt{-t} v$&$-$&$-$&$\sqrt{-t} m_t$&$\sqrt{-t} m_t$&$ \sqrt{-t} m_t$\tabularnewline
$+, 0, +, 0$&$s^0$&$s^0$&$s^0$&$-$&$-$&$-$&$s^0$&$s^0$&$s^0$&$s^0$&$s^0$&$s^0$\tabularnewline
\hline
$\pm, 0, \mp, \mp$&$\frac{1}{s}$&$s^0$&$s^0$&$s^0$&$s^0$&$s^0$&$s^0$&$\sqrt{s (s+t)}$&$\sqrt{s (s+t)}$&$s^0$&$s^0$&$s^0$\tabularnewline
$\pm, \pm, \mp, 0$&$\frac{1}{s}$&$s^0$&$s^0$&$s^0$&$s^0$&$s^0$&$s^0$&$\sqrt{s (s+t)}$&$\sqrt{s (s+t)}$&$s^0$&$s^0$&$s^0$\tabularnewline
$\pm, 0, \pm, \mp$&$\frac{1}{\sqrt{s}}$&$-$&$-$&$-$&$-$&$-$&$-$&$\sqrt{-t} m_t$&$\sqrt{-t} m_t$&$-$&$-$&$-$\tabularnewline
$\pm, \mp, \pm, 0$&$\frac{1}{\sqrt{s}}$&$-$&$-$&$-$&$-$&$-$&$-$&$\sqrt{-t} m_t$&$\sqrt{-t} m_t$&$-$&$-$&$-$\tabularnewline
$\pm, 0, \mp, \pm$&$s^0$&$s^0$&$s^0$&$s^0$&$s^0$&$s^0$&$s^0$&$s^0$&$s^0$&$s^0$&$s^0$&$s^0$\tabularnewline
$\pm, \mp, \mp, 0$&$s^0$&$s^0$&$s^0$&$s^0$&$s^0$&$s^0$&$s^0$&$s^0$&$s^0$&$s^0$&$s^0$&$s^0$\tabularnewline
$\pm, 0, \pm, \pm$&$\frac{1}{\sqrt{s}}$&$-$&$-$&$-$&$-$&$-$&$-$&$-$&$-$&$-$&$-$&$-$\tabularnewline
$\pm, \pm, \pm, 0$&$\frac{1}{\sqrt{s}}$&$-$&$-$&$-$&$-$&$-$&$-$&$-$&$-$&$-$&$-$&$-$\tabularnewline
\hline
$-, \pm, -, \pm$&$s^0$&$s^0$&$-$&$-$&$-$&$s^0$&$-$&$s^0$&$s^0$&$s^0$&$s^0$&$-$\tabularnewline
$+, \pm, +, \pm$&$s^0$&$s^0$&$-$&$-$&$-$&$s^0$&$-$&$s^0$&$s^0$&$-$&$-$&$s^0$\tabularnewline
$-, \pm, -, \mp$&$\frac{1}{s}$&$-$&$-$&$s^0$&$s^0$&$s^0$&$-$&$s^0$&$s^0$&$-$&$-$&$-$\tabularnewline
$+, \pm, +, \mp$&$\frac{1}{s}$&$-$&$-$&$s^0$&$s^0$&$s^0$&$-$&$s^0$&$s^0$&$-$&$-$&$-$\tabularnewline
$\pm, \pm, \mp, \mp$&$-$&$-$&$-$&$\sqrt{-t} m_t$&$\sqrt{-t} m_t$&$\sqrt{-t} m_t$&$-$&$\sqrt{-t} m_W$&$\sqrt{-t} m_W$&$-$&$-$&$-$\tabularnewline
$\pm, \mp, \pm, \mp,$&$\frac{1}{\sqrt{s}}$&$-$&$-$&$\sqrt{-t} m_t$&$\sqrt{-t} m_t$&$\sqrt{-t} m_t$&$-$&$-$&$-$&$-$&$-$&$-$\tabularnewline
$\pm, -, \mp, -$&$\frac{1}{\sqrt{s}}$&$-$&$-$&$-$&$-$&$-$&$-$&$\sqrt{-t} m_W$&$ \sqrt{-t} m_W$&$-$&$-$&$-$\tabularnewline
$\pm, +, \mp, +$&$\frac{1}{\sqrt{s}}$&$-$&$-$&$-$&$-$&$-$&$-$&$\sqrt{-t} m_W$&$\sqrt{-t} m_W$&$-$&$-$&$-$\tabularnewline


\end{tabular}
\end{center}
}
\caption{\label{tab:tztz}
Helicity amplitudes in the high-energy limit ($s\sim-t\gg v$) for $t Z\to t Z$ scattering.  Overall multiplicative factors are dropped and only the schematic energy
dependence is retained for the SM contribution.  Helicity entries marked by `$\ast$' indicate any combination of $\pm,\mp$.
}
\end{table}
\renewcommand{\arraystretch}{1.}

    \renewcommand{\arraystretch}{1.2}
\begin{table}[h!]
{\tiny
\begin{center}
\begin{tabular}{c|cccccc}
$\lambda_{\sss t}$, $\lambda_{\sss Z}$, $\lambda_{\sss t}$, $\lambda_{\sss \gamma}$&SM&$\Op{\phi B}$&$\Op{\phi W}$&$\Op{\phi WB}$&$\Op{tB}$&$\Op{tW}$\tabularnewline
\hline
$\pm, 0, \mp, \mp$&$\frac{1}{s}$&$s^0$&$s^0$&$s^0$&$\sqrt{s(s+t)}$&$\sqrt{s (s+t)}$\tabularnewline
$\pm, 0, \pm, \mp$&$\frac{1}{\sqrt{s}}$&$-$&$-$&$-$&$\sqrt{-t} m_t$&$\sqrt{-t} m_t$\tabularnewline
$\pm, 0, \pm, \pm$&$\frac{1}{\sqrt{s}}$&$-$&$-$&$-$&$-$&$-$\tabularnewline
$\pm, 0, \mp, \pm$&$s^0$&$s^0$&$s^0$&$s^0$&$s^0$&$s^0$\tabularnewline
\hline
$\pm, \pm, \pm, \pm$&$s^0$&$-$&$-$&$s^0$&$s^0$&$s^0$\tabularnewline
$\pm, \mp, \pm, \mp$&$s^0$&$-$&$-$&$s^0$&$s^0$&$s^0$\tabularnewline
$\pm, \pm, \pm, \mp$&$\frac{1}{s}$&$s^0$&$s^0$&$s^0$&$s^0$&$s^0$\tabularnewline
$\pm, \mp, \pm, \pm$&$\frac{1}{s}$&$s^0$&$s^0$&$s^0$&$s^0$&$s^0$\tabularnewline
$\pm, \pm, \mp, \pm$&$\frac{1}{\sqrt{s}}$&$-$&$-$&$-$&$\sqrt{-t} m_W$&$\sqrt{-t} m_W$\tabularnewline
$\mp, \pm, \pm, \pm$&$\frac{1}{\sqrt{s}}$&$-$&$-$&$-$&$\sqrt{-t} m_W$&$\sqrt{-t} m_W$\tabularnewline
$\pm, \pm, \mp, \mp$&$-$&$\sqrt{-t} m_t$&$\sqrt{-t} m_t$&$\sqrt{-t} m_t$&$\sqrt{-t} m_W$&$\sqrt{-t} m_W$\tabularnewline
$\pm, \mp, \mp, \pm$&$\frac{1}{\sqrt{s}}$&$\sqrt{-t} m_t$&$\sqrt{-t} m_t$&$\sqrt{-t} m_t$&$-$&$-$\tabularnewline
\end{tabular}
\end{center}
}
\caption{\label{tab:tzta}
Helicity amplitudes in the high-energy limit ($s\sim-t\gg v$) for $t Z\to t \gamma$ scattering.  Overall multiplicative factors are dropped and only the schematic energy
dependence is retained for the SM contribution.  Helicity entries marked by `$\ast$' indicate any combination of $\pm,\mp$.
}
\end{table}
\renewcommand{\arraystretch}{1.}

    \renewcommand{\arraystretch}{1.2}
\begin{table}[h!]
{\tiny
\begin{center}
\begin{tabular}{c|cccccc}
$\lambda_{\sss t}$, $\lambda_{\sss \gamma}$, $\lambda_{\sss t}$, $\lambda_{\sss \gamma}$&SM&$\Op{\phi B}$&$\Op{\phi W}$&$\Op{\phi WB}$&$\Op{tB}$&$\Op{tW}$\tabularnewline
\hline
$\pm, \pm, \pm, \pm$&$s^0$&$-$&$-$&$s^0$&$s^0$&$s^0$\tabularnewline
$\pm, \pm, \pm, \mp$&$\frac{1}{s}$&$s^0$&$s^0$&$s^0$&$s^0$&$s^0$\tabularnewline
$\pm, \mp, \pm, \pm$&$\frac{1}{s}$&$s^0$&$s^0$&$s^0$&$s^0$&$s^0$\tabularnewline
$\pm, \pm, \mp, \pm$&$\frac{1}{\sqrt{s}}$&$-$&$-$&$-$&$\sqrt{-t} m_W$&$\sqrt{-t} m_W$\tabularnewline
$\mp, \pm, \pm, \pm$&$\frac{1}{\sqrt{s}}$&$-$&$-$&$-$&$\sqrt{-t} m_W$&$\sqrt{-t} m_W$\tabularnewline
$\pm, \pm, \mp, \mp$&$-$&$\sqrt{-t} m_t$&$\sqrt{-t} m_t$&$\sqrt{-t} m_t$&$\sqrt{-t} m_W$&$\sqrt{-t} m_W$\tabularnewline
$\pm, \mp, \pm, \mp$&$s^0$&$-$&$-$&$s^0$&$-$&$-$\tabularnewline
$\pm, \mp, \mp, \pm$&$\frac{1}{\sqrt{s}}$&$\sqrt{-t} m_t$&$\sqrt{-t} m_t$&$\sqrt{-t} m_t$&$-$&$-$\tabularnewline
\end{tabular}
\end{center}
}
\caption{\label{tab:tata}
Helicity amplitudes in the high-energy limit ($s\sim-t\gg v$) for $t \gamma\to t \gamma$ scattering. Overall multiplicative factors are dropped and only the schematic energy
dependence is retained for the SM contribution.  Helicity entries marked by `$\ast$' indicate any combination of $\pm,\mp$.
}
\end{table}
\renewcommand{\arraystretch}{1.}

    \renewcommand{\arraystretch}{1.2}
\begin{table}[h!]
{\tiny
\setlength{\tabcolsep}{0.pt}
\begin{center}
\begin{tabular}{c|ccccccccccc}
$\lambda_{\sss t}$, $\lambda_{\sss Z}$, $\lambda_{\sss t}$&SM&$\Op{\phi D}$&$\Op{\phi B}$&$\Op{\phi W}$&$\Op{\phi WB}$&$\Op{t \phi}$&$\Op{tB}$&$\Op{tW}$&$\Op{\phi Q}^{\sss (1)}$&$\Op{\phi Q}^{\sss (3)}$&$\Op{\phi t}$\tabularnewline
\hline
$-, 0, -$&$s^0$&$s^0$&$s^0$&$s^0$&$s^0$&$s^0$&$s^0$&$s^0$&$\sqrt{s (s+t)}$&$\sqrt{s (s+t)}$&$s^0$\tabularnewline
$-, 0, +$&$\frac{1}{\sqrt{s}}$&$\sqrt{-t} m_t$&$-$&$-$&$-$&$\sqrt{-t} v$&$\frac{s m_W}{\sqrt{-t}}$&$\frac{s m_W}{\sqrt{-t}}$&$\sqrt{-t} m_t$&$\sqrt{-t} m_t$&$-$\tabularnewline
$+, 0, -$&$\frac{1}{\sqrt{s}}$&$\sqrt{-t} m_t$&$-$&$-$&$-$&$\sqrt{-t} v$&$\frac{s m_W}{\sqrt{-t}}$&$\frac{ s m_W}{\sqrt{-t}}$&$-$&$-$&$\sqrt{-t} m_t$\tabularnewline
$+, 0, +$&$s^0$&$s^0$&$s^0$&$-$&$s^0$&$s^0$&$s^0$&$s^0$&$s^0$&$s^0$&$\sqrt{s (s+t)}$\tabularnewline
\hline
$-, -, -$&$\frac{1}{\sqrt{s}}$&$-$&$\frac{s m_W}{\sqrt{-t}}$&$\frac{s m_W}{\sqrt{-t}}$&$\frac{ s m_W}{\sqrt{-t}}$&$-$&$ \sqrt{-t} m_t$&$\sqrt{-t} m_t$&$\sqrt{-t} m_W$&$\sqrt{-t} m_W$&$-$\tabularnewline
$-, -, +$&$-$&$s^0$&$s^0$&$-$&$s^0$&$s^0$&$\sqrt{s (s+t)}$&$ \sqrt{s (s+t)}$&$s^0$&$s^0$&$s^0$\tabularnewline
$-, +, -$&$\frac{1}{\sqrt{s}}$&$-$&$\frac{m_W (s+t)}{\sqrt{-t}}$&$\frac{m_W (s+t)}{\sqrt{-t}}$&$\frac{m_W (s+t)}{\sqrt{-t}}$&$-$&$\sqrt{-t} m_t$&$\sqrt{-t} m_t$&$-$&$-$&$-$\tabularnewline
$-, +, +$&$s^0$&$s^0$&$s^0$&$s^0$&$s^0$&$s^0$&$s^0$&$s^0$&$s^0$&$s^0$&$s^0$\tabularnewline
$+, -, -$&$s^0$&$s^0$&$s^0$&$-$&$s^0$&$s^0$&$s^0$&$s^0$&$s^0$&$s^0$&$s^0$\tabularnewline
$+, -, +$&$\frac{1}{\sqrt{s}}$&$-$&$\frac{m_W (s+t)}{\sqrt{-t}}$&$-$&$\frac{m_W (s+t)}{\sqrt{-t}}$&$-$&$\sqrt{-t} m_t$&$\sqrt{-t} m_t$&$-$&$-$&$-$\tabularnewline
$+, +, -$&$\frac{1}{s}$&$s^0$&$s^0$&$s^0$&$s^0$&$s^0$&$\sqrt{s (s+t)}$&$\sqrt{s (s+t)}$&$s^0$&$s^0$&$s^0$\tabularnewline
$+, +, +$&$\frac{1}{\sqrt{s}}$&$-$&$\frac{s m_W}{\sqrt{-t}}$&$-$&$\frac{s m_W}{\sqrt{-t}}$&$-$&$ \sqrt{-t} m_t$&$ \sqrt{-t} m_t$&$-$&$-$&$\sqrt{-t} m_W$\tabularnewline
\end{tabular}
\end{center}
}
\caption{\label{tab:tzth}
Helicity amplitudes in the high-energy limit ($s\sim-t\gg v$) for $t Z\to t h$ scattering.  Overall multiplicative factors are dropped and only the schematic energy
dependence is retained for the SM contribution.  Helicity entries marked by `$\ast$' indicate any combination of $\pm,\mp$.
}
\end{table}
\renewcommand{\arraystretch}{1.}

    \renewcommand{\arraystretch}{1.2}
\begin{table}[h!]
{\tiny
\begin{center}
\begin{tabular}{c|cccccc}
$\lambda_{\sss t}$, $\lambda_{\sss \gamma}$, $\lambda_{\sss t}$&SM&$\Op{\phi B}$&$\Op{\phi W}$&$\Op{\phi WB}$&$\Op{tB}$&$\Op{tW}$\tabularnewline
\hline
$-, -, -$&$\frac{1}{\sqrt{s}}$&$\frac{s m_W}{\sqrt{-t}}$&$\frac{s m_W}{\sqrt{-t}}$&$\frac{s m_W}{\sqrt{-t}}$&$\sqrt{-t} m_t$&$\sqrt{-t} m_t$\tabularnewline
$-, -, +$&$\frac{1}{s}$&$s^0$&$-$&$s^0$&$\sqrt{s(s+t)}$&$\sqrt{s (s+t)}$\tabularnewline
$-, +, -$&$\frac{1}{\sqrt{s}}$&$\frac{m_W (s+t)}{\sqrt{-t}}$&$\frac{m_W (s+t)}{\sqrt{-t}}$&$\frac{m_W (s+t)}{\sqrt{-t}}$&$\sqrt{-t} m_t$&$\sqrt{-t} m_t$\tabularnewline
$-, +, +$&$s^0$&$s^0$&$s^0$&$s^0$&$s^0$&$s^0$\tabularnewline
$+, -, -$&$s^0$&$s^0$&$-$&$s^0$&$s^0$&$s^0$\tabularnewline
$+, -, +$&$\frac{1}{\sqrt{s}}$&$\frac{m_W (s+t)}{\sqrt{-t}}$&$-$&$\frac{m_W (s+t)}{\sqrt{-t}}$&$\sqrt{-t} m_t$&$\sqrt{-t} m_t$\tabularnewline
$+, +, -$&$\frac{1}{s}$&$s^0$&$s^0$&$s^0$&$\sqrt{s(s+t)}$&$\sqrt{s (s+t)}$\tabularnewline
$+, +, +$&$\frac{1}{\sqrt{s}}$&$\frac{s m_W}{\sqrt{-t}}$&$-$&$\frac{s m_W}{\sqrt{-t}}$&$\sqrt{-t} m_t$&$\sqrt{-t} m_t$\tabularnewline
\end{tabular}
\end{center}
}
\caption{\label{tab:tath}
Helicity amplitudes in the high-energy limit ($s\sim-t\gg v$) for $t \gamma\to t h$ scattering. Overall multiplicative factors are dropped and only the schematic energy
dependence is retained for the SM contribution. Helicity entries marked by `$\ast$' indicate any combination of $\pm,\mp$.
}
\end{table}
\renewcommand{\arraystretch}{1.}

    \renewcommand{\arraystretch}{1.2}
\begin{table}[h!]
{\tiny
\begin{center}
\begin{tabular}{c|cccc}
$\lambda_{\sss t}$, $\lambda_{\sss t}$,&SM&$\Op{\phi D}$&$\Op{\phi d}$&$\Op{t \phi}$\tabularnewline
\hline
$\pm, \pm$&$s^0$&$s^0$&$s^0$&$s^0$\tabularnewline
$\pm, \mp$&$\frac{1}{\sqrt{s}}$&$\sqrt{-t} m_t$&$\sqrt{-t} m_t$&$\sqrt{-t} v$\tabularnewline
\end{tabular}
\end{center}
}
\caption{\label{tab:thth}
Helicity amplitudes in the high-energy limit ($s\sim-t\gg v$) for $t h\to t h$ scattering. Overall multiplicative factors are dropped and only the schematic energy
dependence is retained for the SM contribution. Helicity entries marked by `$\ast$' indicate any combination of $\pm,\mp$.
}
\end{table}
\renewcommand{\arraystretch}{1.}

\chapter{Operator sensitivities for top-EW processes}
\label{app:topbarplots}

A summary of the findings of Ref.~\cite{Maltoni:2019aot} is presented. Each figure is operator specific and presents a
broad perspective of its individual effects on the set of processes under study. Both LHC and CLIC predictions are reported.

\begin{figure}[h!]
  \centering 
  \includegraphics[width=\linewidth]{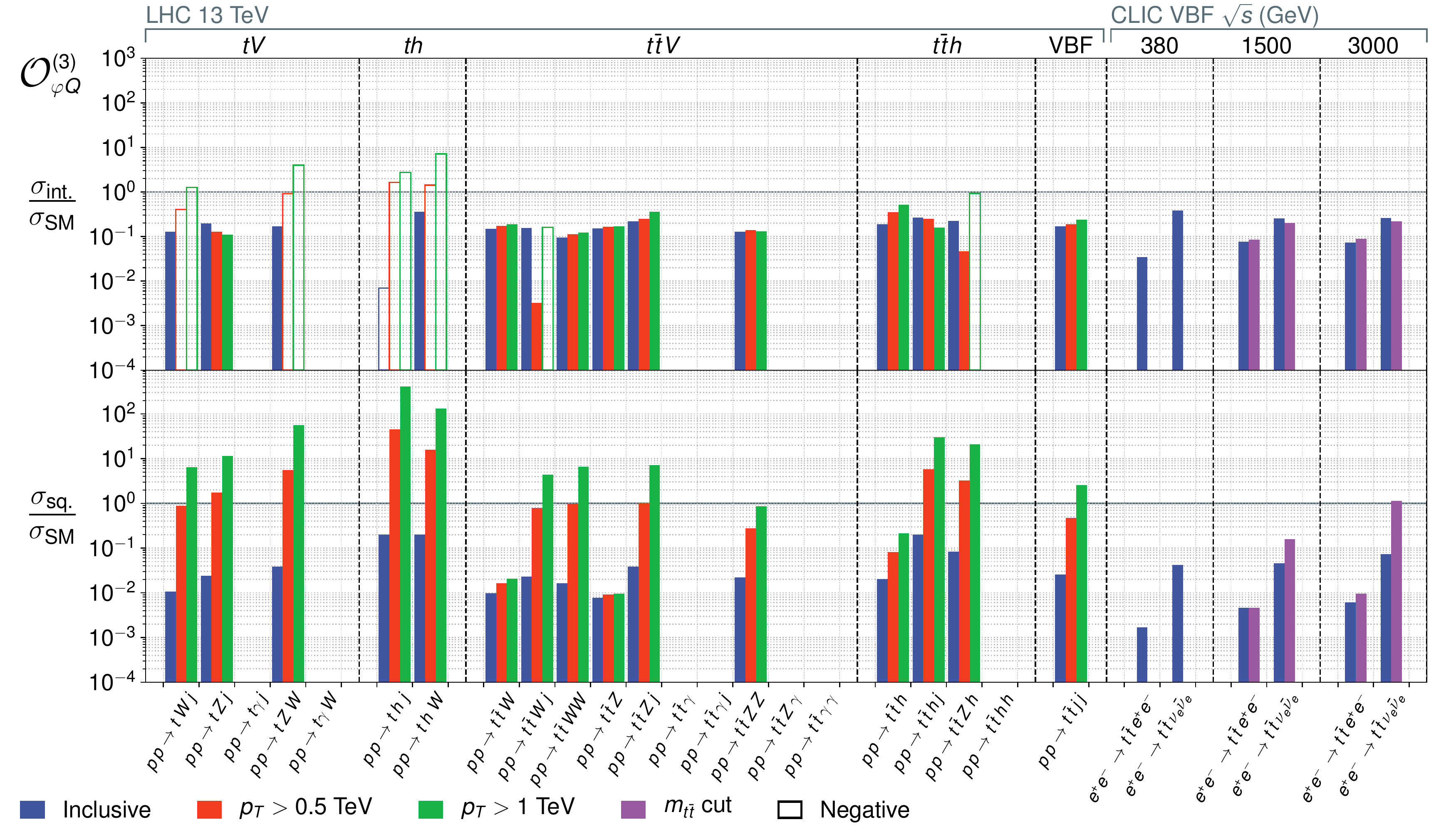}
  \caption{ 
  Summary plot for $\Opp{\phi Q}{(3)}$, displaying the relative sensitivities for the interference $r_i$ (upper row) and squared $r_{i,i}$ (lower row), assuming a Wilson coefficient of 1 TeV$^{-2}$. The sign of the interference term is indicated by filled and unfilled bars. For each process three numbers at increasing high energy cuts are displayed.
  \label{fig:summary_plot_a3phidql}}
\end{figure}
\begin{figure}[h!]
  \centering 
  \includegraphics[width=\linewidth]{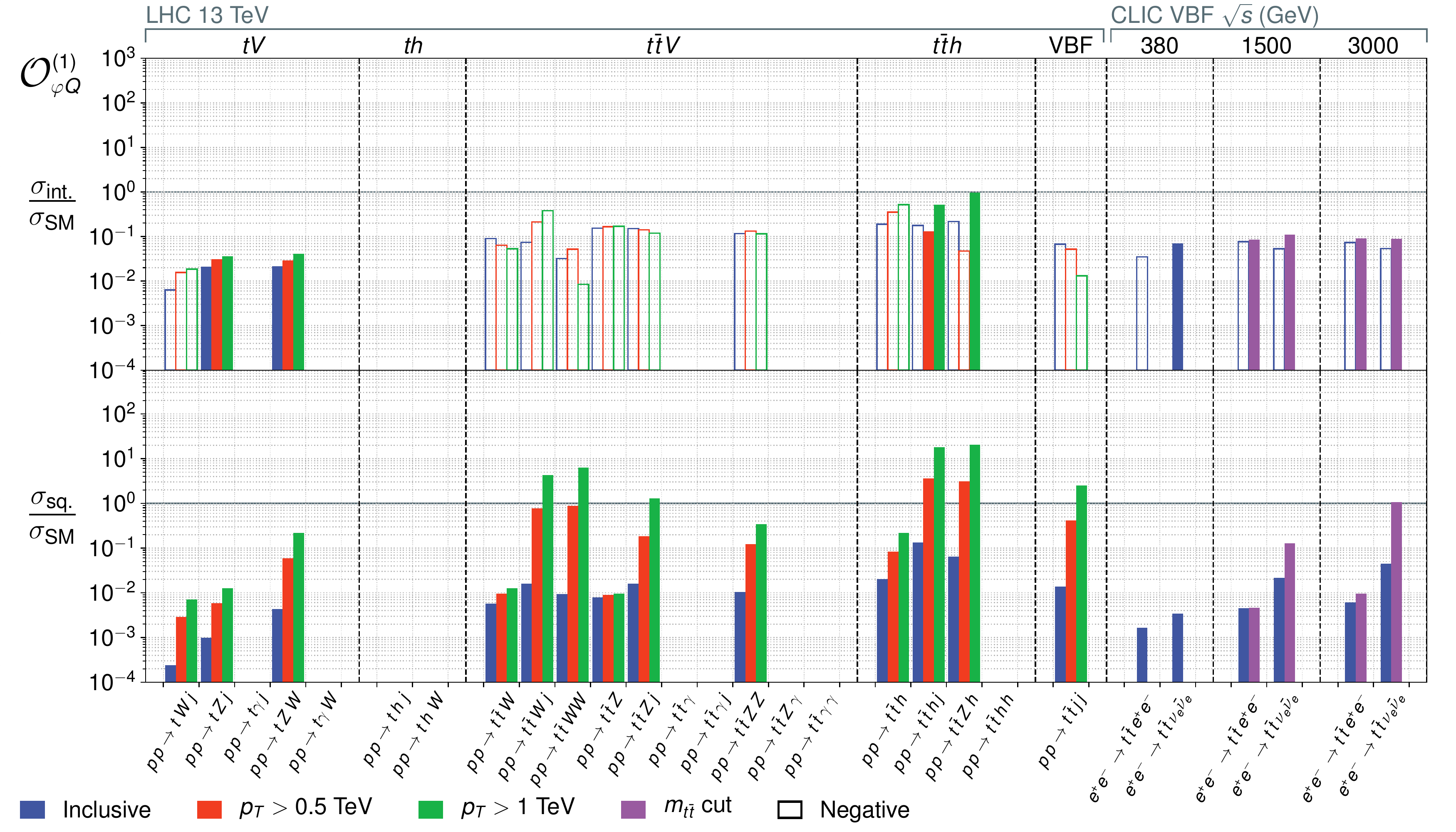}
  \caption{ 
  Same as Figure~\ref{fig:summary_plot_a3phidql} for $\Opp{\phi Q}{(1)}$
  \label{fig:summary_plot_aphidql}}
\end{figure}
\begin{figure}[h!]
  \centering 
  \includegraphics[width=\linewidth]{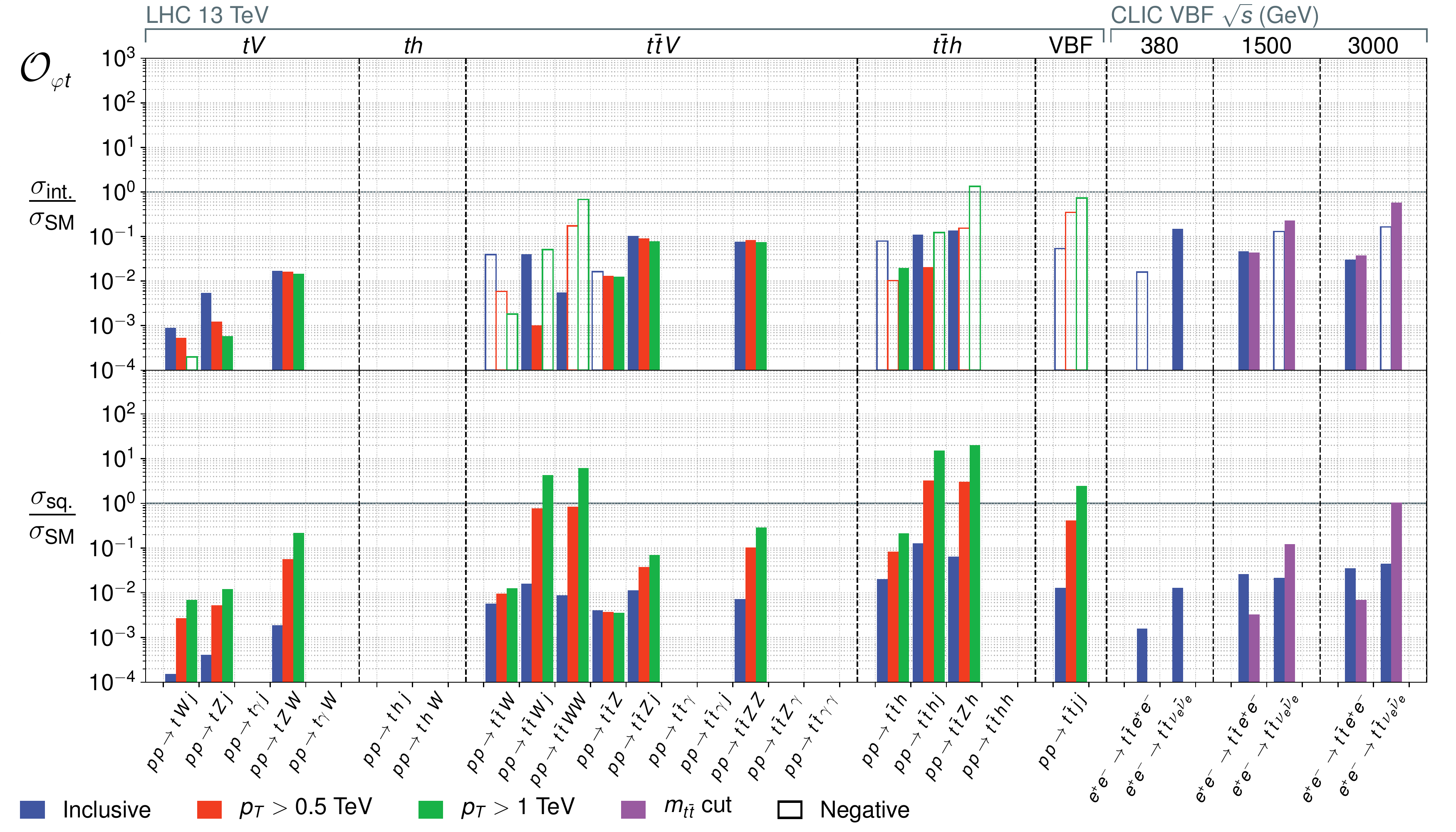}
  \caption{ 
  Same as Figure~\ref{fig:summary_plot_a3phidql} for $\Op{\phi t}$
  \label{fig:summary_plot_aphidtr}}
\end{figure}
\begin{figure}[h!]
  \centering 
  \includegraphics[width=\linewidth]{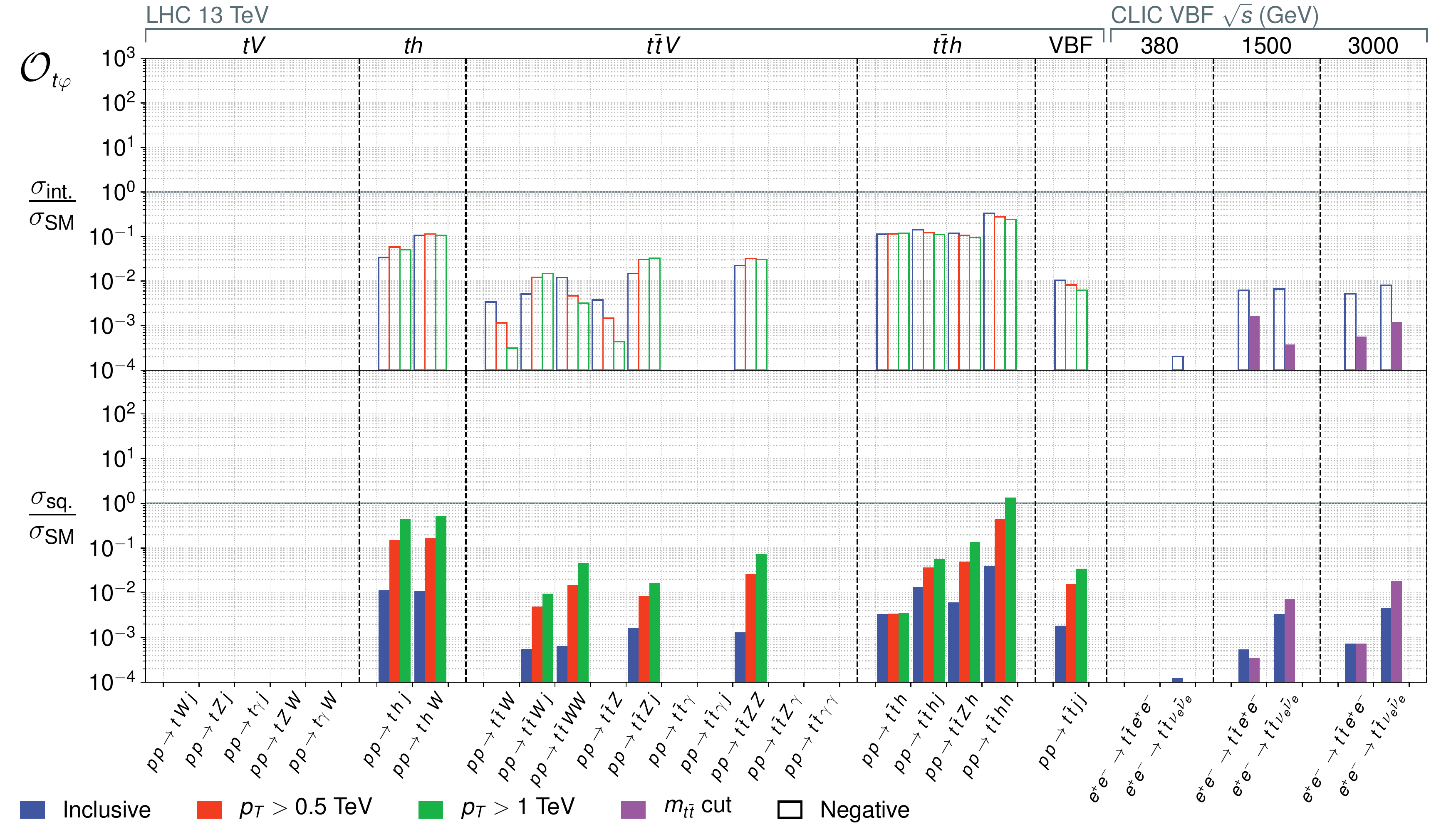}
  \caption{ 
  Same as Figure~\ref{fig:summary_plot_a3phidql} for $\Op{t\phi}$
  \label{fig:summary_plot_atphi}}
\end{figure}
\begin{figure}[h!]
  \centering 
  \includegraphics[width=\linewidth]{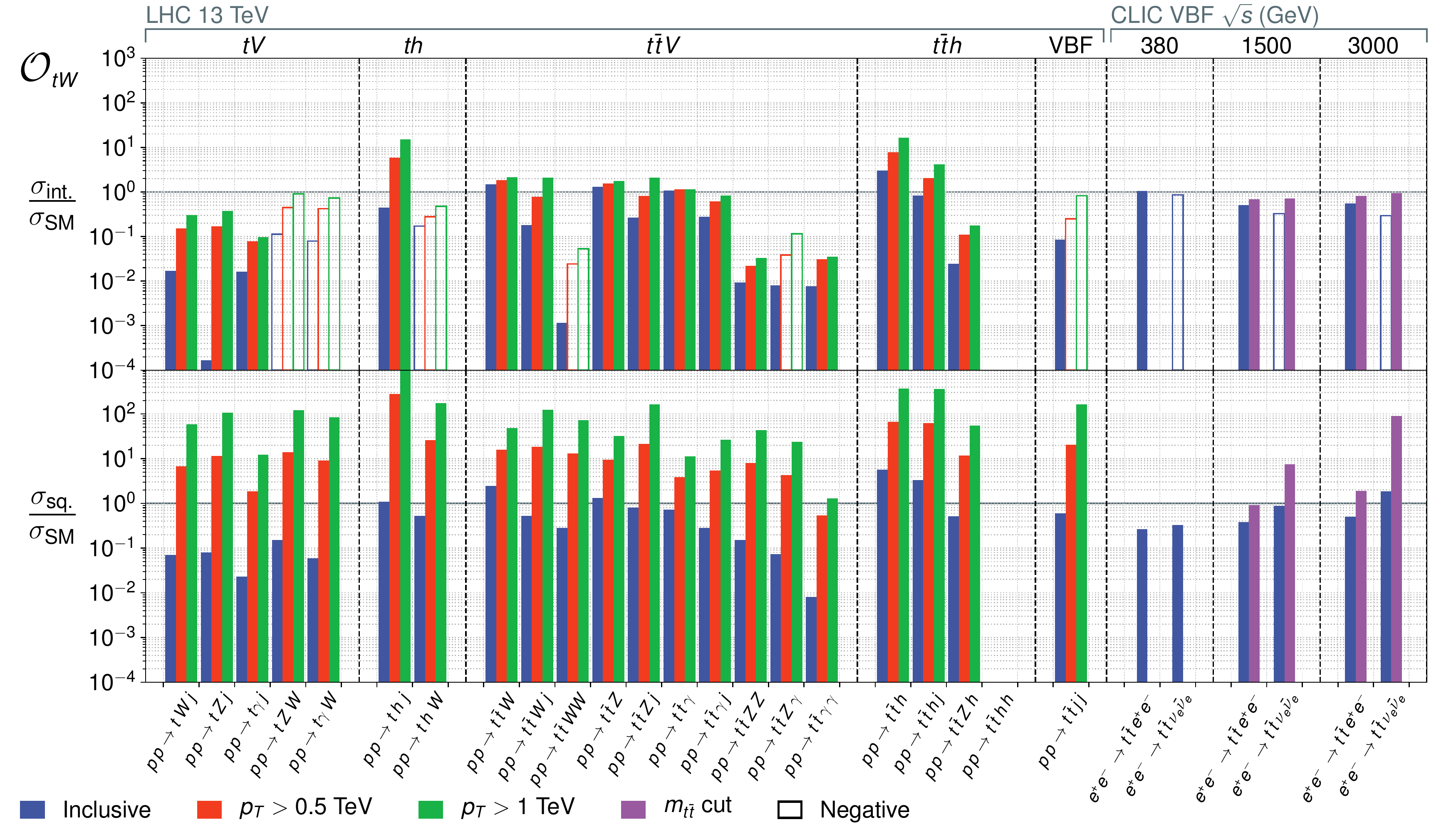}
  \caption{ 
  Same as Figure~\ref{fig:summary_plot_a3phidql} for $\Op{tW}$
  \label{fig:summary_plot_atw}}
\end{figure}
\begin{figure}[h!]
  \centering 
  \includegraphics[width=\linewidth]{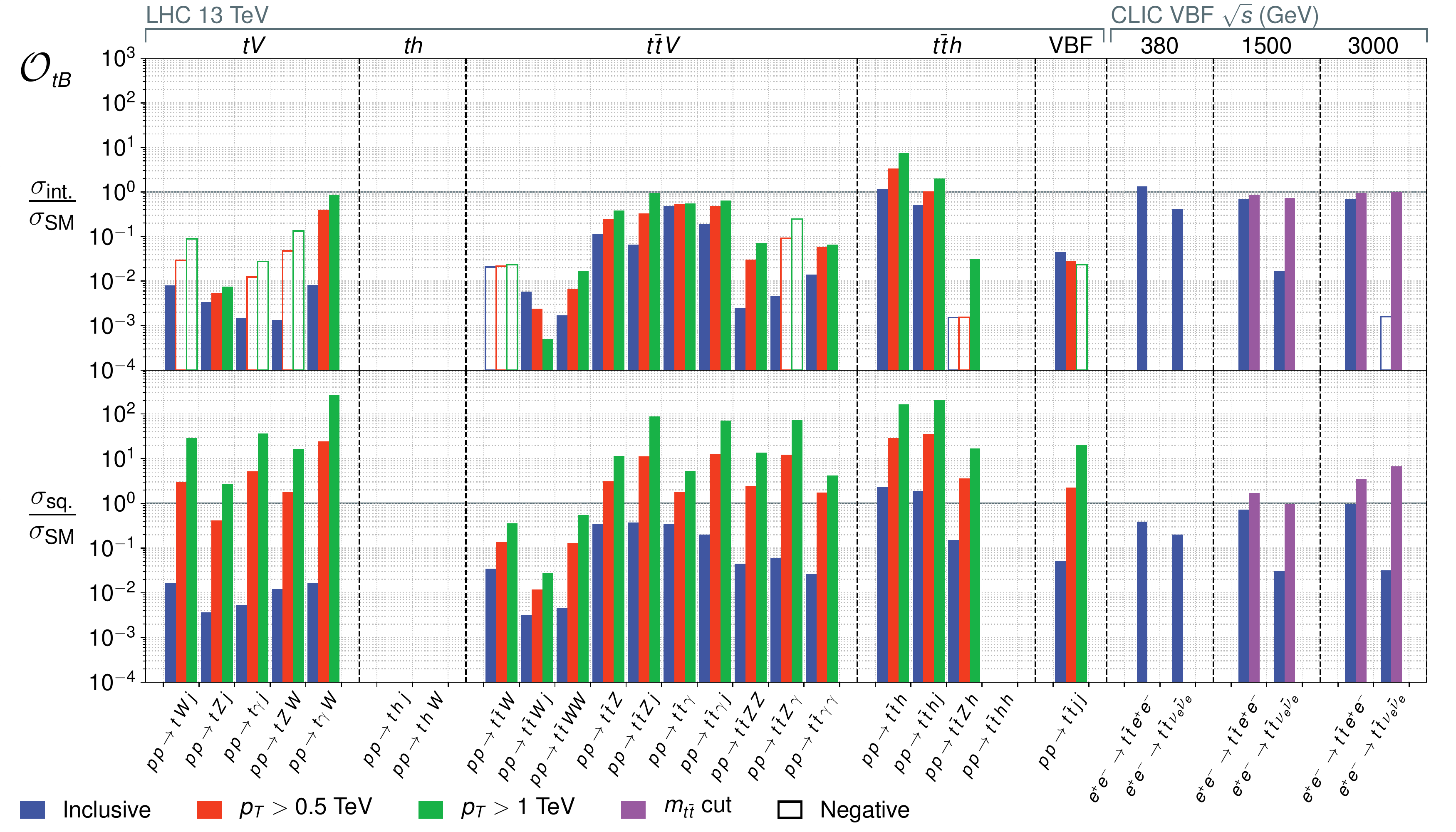}
  \caption{ 
  Same as Figure~\ref{fig:summary_plot_a3phidql} for $\Op{tB}$
  \label{fig:summary_plot_atb}}
\end{figure}
\begin{figure}[h!]
  \centering 
  \includegraphics[width=\linewidth]{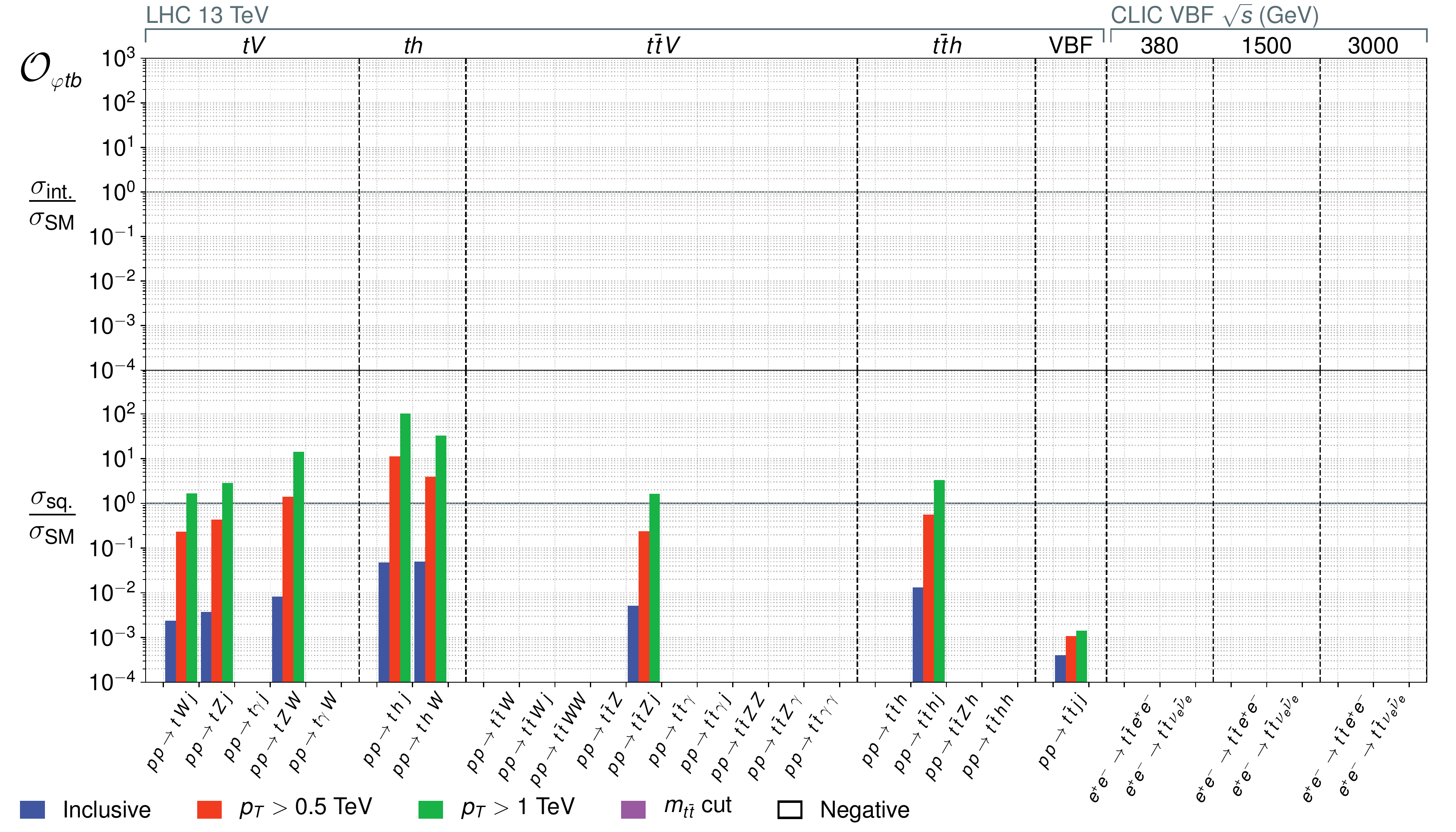}
  \caption{ 
  Same as Figure~\ref{fig:summary_plot_a3phidql} for $\Op{\phi tb}$
  \label{fig:summary_plot_aphitb}}
\end{figure}

\chapter{Global fit data and theory overview}
\label{app:dataoverview}

A summary of the data and theory calculations used for the combined fit of top, Higgs and diboson production data is here
reported. In particular, in Table~\ref{tab:table_dataset_overview} we summarise the dataset considered, dividing the processes in different classes.
In Table~\ref{tab:table-processes-theory} we collect information on the theory predictions, specifying the tools used and the accuracy. Finally, in Table~\ref{tab:operatorprocess},
we report the dependence on the Wilson coefficients of each class of processes.

\begin{table}[htbp]
  \centering
  \scriptsize
   \renewcommand{\arraystretch}{1.0}
   \setlength{\tabcolsep}{1.0pt}
  \begin{tabular}{C{3.9cm}|C{5.6cm}|C{2cm}}
 Category   & Processes    &  $n_{\rm dat}$     \\
    \toprule
    \multirow{6}{*}{Top quark production}   &  $t\bar{t}$ (inclusive)   &  94  \\
    &  $t\bar{t}Z$, $t\bar{t}W$    & 14 \\
    &   single top (inclusive)   & 27 \\
    &  $tZ, tW$   &  9\\
    &  $t\bar{t}t\bar{t}$, $t\bar{t}b\bar{b}$    & 6 \\
    &  {\bf Total}    & {\bf 150 }  \\
    \midrule
    \multirow{3.3}{*}{Higgs production} & Run I signal strengths  &22   \\
    \multirow{3.1}{*}{and decay} & Run II  signal strengths  & 40  \\
    & Run II, differential distributions \& STXS  & 35  \\
    &  {\bf Total}    & {\bf 97}  \\
    \midrule
    \multirow{3}{*}{Diboson production} & LEP-2 &40   \\
     & LHC & 30  \\
    &  {\bf Total}    & {\bf 70}  \\
    \bottomrule
   Baseline dataset     & {\bf Total}      & {\bf 317}  \\
\bottomrule
  \end{tabular}
  \caption{Overview of the data points considered for each class of processes.
 \label{tab:table_dataset_overview}
}
\end{table}


\begin{table}[htbp]
  \centering
  \scriptsize
  \renewcommand{\arraystretch}{1.0}
   \setlength{\tabcolsep}{0.5pt}
    \begin{tabular}{c|c|c|c|c|c}
  Category & Process 
  & SM   
  & Code/Ref  
  & SMEFT 
  & Code 
  \\
  \toprule
 \multirow{10}{*}{Top quark}  &   \multirow{2}{*}{$t\bar{t}$ (incl)}  
  & \multirow{2}{*}{NNLO QCD}   
  & {\tt MG5\_aMC} NLO  
  & \multirow{2}{*}{NLO QCD}  
  & \multirow{2}{*}{{\tt SMEFT@NLO}} 
  \\
\multirow{10}{*}{production}   &    
&  & + NNLO $K$-fact  
  &    
  &  
\\
\cmidrule(lr{0.7em}){2-6}
  &  \multirow{2}{*}{$t\bar{t}+V$} 
  & \multirow{2}{*}{NLO QCD} 
  & \multirow{2}{*}{{\tt MG5\_aMC} NLO}  
  & LO QCD   
  & \multirow{2}{*}{{\tt SMEFT@NLO}} 
  \\
  &  
  &   
  & & + NLO SM $K$-fact  
  & 
  \\
 \cmidrule(lr{0.7em}){2-6}
&    \multirow{2}{*}{single-$t$ (incl)} 
  & \multirow{2}{*}{NNLO QCD} 
  & {\tt MG5\_aMC} NLO  
  & \multirow{2}{*}{NLO QCD}  
  & \multirow{2}{*}{{\tt SMEFT@NLO}} 
  \\
  &   
 & & + NNLO $K$-fact     
  & &
  \\
  \cmidrule(lr{0.7em}){2-6}
  &  \multirow{2}{*}{$t+V$} 
  & \multirow{2}{*}{NLO QCD} 
  & \multirow{2}{*}{{\tt MG5\_aMC} NLO}  
  & LO QCD   
  & \multirow{2}{*}{{\tt SMEFT@NLO}} 
  \\
  &  
  &   
 & & + NLO SM $K$-fact  
  & 
  \\
  \cmidrule(lr{0.7em}){2-6}
 &   \multirow{2}{*}{$t\bar{t}t\bar{t},~t\bar{b}t\bar{b}$} 
  & \multirow{2}{*}{NLO QCD} 
  & \multirow{2}{*}{{\tt MG5\_aMC} NLO}  
  & LO QCD  
  & \multirow{2}{*}{{\tt SMEFT@NLO}} 
  \\
  &  
  &   
 & & + NLO SM $K$-fact  
  &
  \\
   \midrule
 \multirow{10}{*}{Higgs production}  &   \multirow{2}{*}{$gg\to h$}  
  & \multirow{1}{*}{NNLO QCD +}   
  &\multirow{2}{*}{ HXSWG}
  & \multirow{2}{*}{NLO QCD}  
  & \multirow{2}{*}{{\tt SMEFT@NLO}} 
  \\
\multirow{10}{*}{and decay}   &    
&  \multirow{1}{*}{NLO EW}   & 
  &    
  &  
\\
 \cmidrule(lr{0.7em}){2-6}
   &   \multirow{2}{*}{VBF}  
  & \multirow{1}{*}{NNLO QCD +}   
  &\multirow{2}{*}{ HXSWG}
  & \multirow{2}{*}{LO QCD}  
  & \multirow{2}{*}{{\tt SMEFT@NLO}} 
 \\
 &    
&  \multirow{1}{*}{NLO EW}   & 
  &    
  &  
 \\
  \cmidrule(lr{0.7em}){2-6}
   &   \multirow{2}{*}{$h+V$}  
  & \multirow{1}{*}{NNLO QCD +}   
  &\multirow{2}{*}{ HXSWG}
  & \multirow{2}{*}{NLO QCD}  
  & \multirow{2}{*}{{\tt SMEFT@NLO}} 
  \\
   &    
&  \multirow{1}{*}{NLO EW}   & 
  &    
  &  
\\
 \cmidrule(lr{0.7em}){2-6}
  &   \multirow{2}{*}{$ht\bar{t}$}  
  & \multirow{1}{*}{NNLO QCD +}   
  &\multirow{2}{*}{ HXSWG}
  & \multirow{2}{*}{NLO QCD}  
  & \multirow{2}{*}{{\tt SMEFT@NLO}} 
  \\
   &    
&  \multirow{1}{*}{NLO EW}   & 
  &    
  &  
\\
 \cmidrule(lr{0.7em}){2-6}
  &   \multirow{2}{*}{$h\to X$}  
  & \multirow{1}{*}{NNLO QCD +}   
  &\multirow{2}{*}{ HXSWG}
  & \multirow{1}{*}{NLO QCD ($X=b\bar{b}$)}  
  & \multirow{2}{*}{{\tt SMEFT@NLO}} 
 \\
 &    
&  \multirow{1}{*}{NLO EW}   & 
  &    \multirow{1}{*}{LO QCD ($X\ne b \bar{b} $)} 
  &  
\\
 \midrule
 \multirow{4}{*}{Diboson}  &   \multirow{2}{*}{$e^+e^- \to W^+W^-$}  
  & \multirow{1}{*}{NNLO QCD +}   
  &\multirow{2}{*}{ LEP EWWG}
  & \multirow{2}{*}{LO QCD}  
  & \multirow{2}{*}{{\tt SMEFT@NLO}} 
  \\
\multirow{4}{*}{production}   &    
&  \multirow{1}{*}{NLO EW}   & 
  &    
  &  
\\
\cmidrule(lr{0.7em}){2-6}
\multirow{4}{*}{}  &   \multirow{2}{*}{$pp \to VV'$}  
  & \multirow{2}{*}{NNLO QCD}   
  &\multirow{1}{*}{{\tt MG5\_aMC} NLO }
  & \multirow{2}{*}{NLO QCD}  
  & \multirow{2}{*}{{\tt SMEFT@NLO}} 
  \\
\multirow{4}{*}{}   &    
&     & + NNLO $K$-fact  
  &    
  &  
\\
  \bottomrule
 \end{tabular}
 \caption{Overview of the theory predictions for each of the processes considered in this analysis.
 }
  \label{tab:table-processes-theory}
\end{table}


\begin{table}[p]
 \centering
 \scriptsize
 \renewcommand{\arraystretch}{0.8}
   \setlength{\tabcolsep}{0.5pt}
 \begin{tabular}{l|l||C{0.8cm}|C{0.8cm}|C{0.8cm}|C{0.8cm}|C{0.8cm}||C{1.1cm}|C{1.1cm}|C{1.3cm}|C{0.8cm}}
   Class  &
   DoF &
   $\,t\bar{t}\,$
 & $t\bar{t}V$   
 & $\,\, t\,\,$ 
 & $tV$ 
 & $t\bar{t}Q\bar{Q}$ 
 & $h$ ($\mu_i^{f}$, RI)
 & $h$ ($\mu_i^{f}$, RII)
 & $h$ (STXS, RII)
 & $VV$
 \\
 \toprule
\multirow{13.5}{*}{2-heavy-} &  $c_{Qq}^{1,8}$ &  \checkmark  &  \checkmark  &    &    & \checkmark    & \checkmark     &  \checkmark     &    \checkmark  &       \\
\multirow{13.5}{*}{2-light} &  $c_{Qq}^{1,1}$ & \checkmark   & \checkmark  &    &     &  \checkmark   &  \checkmark   & \checkmark     & \checkmark    &       \\
 &  $c_{Qq}^{3,8}$ &   \checkmark  &  \checkmark  &  \checkmark   &  \checkmark    &   \checkmark   &  \checkmark    &   \checkmark    &   \checkmark   &       \\
 &  $c_{Qq}^{3,1}$ &   \checkmark  &    \checkmark   &  \checkmark    &  \checkmark    &  \checkmark   &  \checkmark      &   \checkmark   &   \checkmark   &  \\
 &  $c_{tq}^{8}$ &  \checkmark   &  \checkmark  &    &      & \checkmark    &  \checkmark    &  \checkmark     &  \checkmark    &       \\
 &   $c_{tq}^{1}$&   \checkmark  &  \checkmark  &    &     & \checkmark &  \checkmark    &   \checkmark   &  \checkmark     &           \\
 &  $c_{tu}^{8}$ &   \checkmark  &   \checkmark   &     &     &  \checkmark   &    \checkmark   &  \checkmark    &  \checkmark &     \\
 &  $c_{tu}^{1}$ &    \checkmark &  \checkmark  &    &     &   \checkmark  &   \checkmark   &   \checkmark    &  \checkmark     &       \\
 &  $c_{Qu}^{8}$ &   \checkmark  &  \checkmark  &    &     &   \checkmark  &  \checkmark    &    \checkmark   &  \checkmark    &       \\
 &  $c_{Qu}^{1}$&   \checkmark  &  \checkmark  &    &     &   \checkmark  &  \checkmark    &  \checkmark     &   \checkmark    &       \\
 &  $c_{td}^{8}$ &   \checkmark  &  \checkmark  &    &     &   \checkmark  &   \checkmark   &  \checkmark     &  \checkmark    &       \\
 &  $c_{td}^{1}$ &   \checkmark  &   \checkmark &    &     &   \checkmark  &   \checkmark   &  \checkmark     &  \checkmark    &       \\
 &  $c_{Qd}^{8}$ &   \checkmark  &  \checkmark  &    &     &   \checkmark  &   \checkmark   &    \checkmark   &  \checkmark    &       \\
&  $c_{Qd}^{1}$ &   \checkmark  &  \checkmark  &    &     &   \checkmark  &   \checkmark   &    \checkmark   &   \checkmark    &       \\
\midrule
\multirow{5}{*}{4-heavy} & $c_{QQ}^1$ &    &   &    &     &  \checkmark   &     &      &     &       \\
 &  $c_{QQ}^8$ &    &   &    &     &  \checkmark   &     &      &     &       \\
 &  $c_{Qt}^1$ &    &   &    &     &  \checkmark   &     &      &     &       \\
 &  $c_{Qt}^8$ &    &   &    &     &  \checkmark   &     &      &     &       \\
&  $c_{tt}^1$ &    &   &    &     &  \checkmark   &     &      &     &       \\
\midrule
4-lepton &  $c_{ll}$ &    &   &    &     &    &     &      &     &       \\
\midrule
\multirow{18.5}{*}{2-fermion} &  $c_{t\varphi}$ &    &   &    &     &     &   \checkmark   &   \checkmark    &   \checkmark   &       \\
\multirow{18.5}{*}{+bosonic} &  $c_{tG}$ & \checkmark  & \checkmark   &    &     & \checkmark    &   \checkmark   &   \checkmark    &  \checkmark    &       \\
&  $c_{b\varphi}$ &   &   &    &     &     &  \checkmark   &   \checkmark    &      &       \\
&  $c_{c\varphi}$ &    &   &    &     &     &   \checkmark  &  \checkmark    &     &       \\
&  $c_{\tau\varphi}$ &    &   &    &     &     &   \checkmark  &   \checkmark   &     &       \\
&  $c_{tW}$ & \checkmark   &  \checkmark & \checkmark   &  \checkmark   &     & \checkmark    &  \checkmark    &     &       \\
&  $c_{tZ}$ &    &  \checkmark &    &  \checkmark   &     &   \checkmark  &  \checkmark    &     &       \\[0.1cm]
&  $c_{\varphi Q}^{(3)}$ &  \checkmark  & \checkmark  &  \checkmark  & \checkmark    &     &  \checkmark   &   \checkmark   &  \checkmark   &       \\[0.1cm]
&  $c_{\varphi Q}^{(-)}$ &    & \checkmark  &    &  \checkmark   &     &  \checkmark   &   \checkmark   &   \checkmark  &       \\
&  $c_{\varphi t}$ &    &  \checkmark   &    & \checkmark    &     & \checkmark    & \checkmark     &  \checkmark   &       \\[0.1cm]
&  $c_{\varphi l_i}^{(1)}$ &    &   &    &     &    &  \checkmark   &  \checkmark    &     &       \\[0.1cm]
&  $c_{\varphi l_i}^{(3)}$ &    &   &    &     &    & \checkmark    &  \checkmark    &     &       \\
&  $c_{\varphi e}$ &    &   &    &     &    &   \checkmark  &  \checkmark    &     &       \\
& $c_{\varphi \mu}$  &    &   &    &     &    &  \checkmark    & \checkmark     &    &       \\
& $c_{\varphi \tau}$  &    &   &    &     &    &    \checkmark   &  \checkmark    &   &       \\[0.1cm]
& $c_{\varphi q}^{(3)}$  &    &   &    &     &    &  \checkmark   &  \checkmark    &     &       \\[0.1cm]
&  $c_{\varphi q}^{(-)}$ &    &   &    &     &    &  \checkmark   &  \checkmark    &     &       \\
& $c_{\varphi u}$  &    &   &    &     &    &    \checkmark &  \checkmark    &     &       \\
& $c_{\varphi d}$  &    &   &    &     &    &  \checkmark   &  \checkmark    &     &       \\
\midrule
\multirow{6.5}{*}{purely} &  $c_{\varphi G}$ &    &   &    &     &     &  \checkmark   & \checkmark     &  \checkmark   &       \\
\multirow{6.5}{*}{bosonic} &  $c_{\varphi B}$ &    &   &    &     &     &   \checkmark  &  \checkmark    &  \checkmark   &       \\
&  $c_{\varphi W}$ &    &   &    &     &     &  \checkmark   &  \checkmark    & \checkmark    &       \\
&  $c_{\varphi d}$ &    &   &    &     &     &   \checkmark  &   \checkmark   &  \checkmark   &       \\
&  $c_{\varphi D}$ &    &   &    &     &     &  \checkmark   &  \checkmark    &  \checkmark   &    \checkmark   \\
&  $c_{\varphi W B}$ &    &   &    &     &     &   \checkmark  & \checkmark     & \checkmark    &   \checkmark    \\
&  $c_{WWW}$  &    &   &    &     &     &     &      &     & \checkmark      \\
 \bottomrule
 \end{tabular}
 \caption{Summary of the Wilson coefficient dependence of each class of processes. 
}
\label{tab:operatorprocess}	
\end{table}


\chapter{Global fit coefficient bounds}
\label{app:coeffs_bounds}
\begin{table}[htbp]
  \centering
  \scriptsize
   \renewcommand{\arraystretch}{1.70}
   \resizebox{\textwidth}{!}{
   \begin{tabular}{l|C{0.7cm}|C{2.1cm}|C{2.1cm}|C{3.8cm}|C{3.8cm}}
     \multirow{2}{*}{Class}   &  \multirow{2}{*}{DoF}
     &  \multicolumn{2}{c|}{ 95\% CL bounds, $\mathcal{O}\lp \Lambda^{-2}\rp$} &
     \multicolumn{2}{c}{95\% CL bounds, $\mathcal{O}\lp \Lambda^{-4}\rp$,} \\ 
 &  & Individual & Marginalised &  Individual & Marginalised  \\ \toprule
\multirow{5}{*}{4H}
 &{\tt cQQ1}& [-6.132,23.281] & [-190,189] & [-2.229,2.019] & [-2.995,3.706] \\ \cline{2-6}
 & {\tt cQQ8}  & [-26.471,57.778] & [-190,170] & [-6.812,5.834] & [-11.177,8.170] \\ \cline{2-6}
 & {\tt cQt1}& [-195,159] & [-190,189] & [-1.830,1.862] & [-1.391,1.251] \\ \cline{2-6}
 & {\tt cQt8}& [-5.722,20.105] & [-190,162] & [-4.213,3.346] & [-3.040,2.202] \\ \cline{2-6}
 & {\tt ctt1}& [-2.782,12.114] & [-115,153] & [-1.151,1.025] & [-0.791,0.714] \\ \hline
\multirow{14}{*}{2L2H}
 & {\tt c81qq}& [-0.273,0.509] & [-2.258,4.822] & [-0.373,0.309] & [-0.555,0.236] \\ \cline{2-6}
 & {\tt c11qq}& [-3.603,0.307] & [-8.047,9.400] & [-0.303,0.225] & [-0.354,0.249] \\ \cline{2-6}
 & {\tt c83qq}& [-1.813,0.625] & [-3.014,7.365] & [-0.470,0.439] & [-0.462,0.497] \\ \cline{2-6}
 & {\tt c13qq}& [-0.099,0.155] & [-0.163,0.296] & [-0.088,0.166] & [-0.167,0.197] \\ \cline{2-6}
 & {\tt c8qt}& [-0.396,0.612] & [-4.035,4.394] & [-0.483,0.393] & [-0.687,0.186] \\ \cline{2-6}
 & {\tt c1qt}& [-0.784,2.771] & [-12.382,6.626] & [-0.205,0.271] & [-0.222,0.226] \\ \cline{2-6}
 & {\tt c8ut}& [-0.774,0.607] & [-16.952,0.368] & [-0.911,0.347] & [-1.118,0.260] \\ \cline{2-6}
 & {\tt c1ut}& [-6.046,0.424] & [-15.565,15.379] & [-0.380,0.293] & [-0.383,0.331] \\ \cline{2-6}
 & {\tt c8qu}& [-1.508,1.022] & [-12.745,13.758] & [-1.007,0.521] & [-1.002,0.312] \\ \cline{2-6}
 & {\tt c1qu}& [-0.938,2.462] & [-16.996,1.072] & [-0.281,0.371] & [-0.207,0.339] \\ \cline{2-6}
 & {\tt c8dt}& [-1.458,1.365] & [-5.494,25.358] & [-1.308,0.638] & [-1.329,0.643] \\ \cline{2-6}
 & {\tt c1dt}& [-9.504,-0.086] & [-27.673,11.356] & [-0.449,0.371] & [-0.474,0.347] \\ \cline{2-6}
 & {\tt c8qd}& [-2.393,2.042] & [-24.479,11.233] & [-1.615,0.888] & [-1.256,0.715] \\ \cline{2-6}
 & {\tt c1qd}& [-0.889,6.459] & [-3.239,34.632] & [-0.332,0.436] & [-0.370,0.384] \\ \hline
 \end{tabular}
}
\caption{\small The 95\% CL intervals for the four fermion SMEFT operators considered in this analysis, for both the linear and quadratic fit.
   Individual and marginalised results are presented.
\label{tab:coeff-bounds-baseline}
}
\end{table}

\begin{table}[htbp]
  \centering
  \scriptsize
   \renewcommand{\arraystretch}{1.70}
   \resizebox{\textwidth}{!}{
   \begin{tabular}{l|C{0.7cm}|C{2.1cm}|C{2.1cm}|C{3.8cm}|C{3.8cm}}
     \multirow{2}{*}{Class}   &  \multirow{2}{*}{DoF}
     &  \multicolumn{2}{c|}{ 95\% CL bounds, $\mathcal{O}\lp \Lambda^{-2}\rp$} &
     \multicolumn{2}{c}{95\% CL bounds, $\mathcal{O}\lp \Lambda^{-4}\rp$,} \\ 
 &  & Individual & Marginalised &  Individual & Marginalised  \\ \toprule
\multirow{23}{*}{2FB}
 & {\tt ctp}& [-1.331,0.355] & [-5.739,3.435] & [-1.286,0.348] & [-2.319,2.797] \\ \cline{2-6}
 & {\tt ctG}& [0.007,0.111] & [-0.127,0.403] & [0.006,0.107] & [0.062,0.243] \\ \cline{2-6}
 & {\tt cbp}& [-0.006,0.040] & [-0.033,0.105]& [-0.007,0.035]$\cup$ [-0.403,-0.360] & [-0.035,0.047]$\cup$ [-0.430,-0.338] \\ \cline{2-6}
 & {\tt ccp}& [-0.025,0.117] & [-0.316,0.134] & [-0.004,0.370] & [-0.096,0.484] \\ \cline{2-6}
 & {\tt ctap}& [-0.026,0.035] & [-0.027,0.044] & [-0.027,0.040]$\cup$ [0.395,0.462] & [-0.019,0.037]$\cup$ [0.389,0.480] \\ \cline{2-6}
 & {\tt ctW}& [-0.093,0.026] & [-0.313,0.123] & [-0.084,0.029] & [-0.241,0.086] \\ \cline{2-6}
 & {\tt ctZ}& [-0.039,0.099] & [-15.869,5.636] & [-0.044,0.094] & [-1.129,0.856] \\ \cline{2-6}
 & {\tt cpl1}& [-0.664,1.016] & [-0.244,0.375] & [-0.281,0.343] & [-0.106,0.129] \\ \cline{2-6}
 & {\tt c3pl1}& [-0.472,0.080] & [-0.098,0.120] & [-0.432,0.062] & [-0.209,0.046] \\ \cline{2-6}
 & {\tt cpl2}& [-0.664,1.016] & [-0.244,0.375] & [-0.281,0.343] & [-0.106,0.129] \\ \cline{2-6}
 & {\tt c3pl2}& [-0.472,0.080] & [-0.098,0.120] & [-0.432,0.062] & [-0.209,0.046] \\ \cline{2-6}
 & {\tt cpl3}& [-0.664,1.016] & [-0.244,0.375] & [-0.281,0.343] & [-0.106,0.129] \\ \cline{2-6}
 & {\tt c3pl3}& [-0.472,0.080] & [-0.098,0.120] & [-0.432,0.062] & [-0.209,0.046] \\ \cline{2-6}
 & {\tt cpe}& [-1.329,2.033] & [-0.487,0.749] & [-0.562,0.687] & [-0.213,0.258] \\ \cline{2-6}
 & {\tt cpmu}& [-1.329,2.033] & [-0.487,0.749] & [-0.562,0.687] & [-0.213,0.258] \\ \cline{2-6}
 & {\tt cpta}& [-1.329,2.033] & [-0.487,0.749] & [-0.562,0.687] & [-0.213,0.258] \\ \cline{2-6}
 & {\tt c3pq}& [-0.472,0.080] & [-0.098,0.120] & [-0.432,0.062] & [-0.209,0.046] \\ \cline{2-6}
 & {\tt c3pQ3}& [-0.350,0.353] & [-1.145,0.740] & [-0.375,0.344] & [-0.615,0.481] \\ \cline{2-6}
 & {\tt cpqMi}& [-2.905,0.490] & [-0.171,0.106] & [-2.659,0.381] & [-0.060,0.216] \\ \cline{2-6}
 & {\tt cpQM}& [-0.998,1.441] & [-1.690,11.569] & [-1.147,1.585] & [-2.250,2.855] \\ \cline{2-6}
 & {\tt cpui}& [-1.355,0.886] & [-0.499,0.325] & [-0.458,0.375] & [-0.172,0.142] \\ \cline{2-6}
 & {\tt cpdi}& [-0.443,0.678] & [-0.162,0.250] & [-0.187,0.229] & [-0.071,0.086] \\ \cline{2-6}
 & {\tt cpt}& [-2.087,2.463] & [-3.270,18.267] & [-3.028,2.195] & [-13.260,3.955] \\ \hline
\multirow{7}{*}{B}
 & {\tt cpG}& [-0.002,0.005] & [-0.043,0.012] & [-0.002,0.005] & [-0.019,0.003] \\ \cline{2-6}
 & {\tt cpB}& [-0.005,0.002] & [-0.739,0.289] & [-0.005,0.002]$\cup$ [0.085,0.092] & [-0.114,0.108] \\ \cline{2-6}
 & {\tt cpW}& [-0.018,0.007] & [-0.592,0.677] & [-0.016,0.007]$\cup$ [0.281,0.305] & [-0.145,0.303] \\ \cline{2-6}
 & {\tt cpWB}& [-2.905,0.490] & [-0.462,0.694] & [-2.659,0.381] & [-0.170,0.273] \\ \cline{2-6}
 & {\tt cpd}& [-0.428,1.214] & [-2.002,3.693] & [-0.404,1.199]$\cup$ [-34.04,-32.61] & [-1.523,1.482] \\ \cline{2-6}
 & {\tt cpD}& [-4.066,2.657] & [-1.498,0.974] & [-1.374,1.124] & [-0.516,0.425] \\ \cline{2-6}
 & {\tt cWWW}& [-1.057,1.318] & [-1.049,1.459] & [-0.208,0.236] & [-0.182,0.222] \\ \bottomrule
\end{tabular}
}
\caption{\small Same as \ref{tab:coeff-bounds-baseline} for two fermion and bosonic operators.
}
\end{table}

\end{appendices}

\clearpage
\addcontentsline{toc}{chapter}{Bibliography}
\bibliography{bibliography,refs_topEW,refs_globalfit,refs_muon}

\providecommand{\href}[2]{#2}\begingroup\raggedright\begin{thebibliography}{100}

\bibitem{Maltoni:2019aot}
F.~Maltoni, L.~Mantani, and K.~Mimasu, ``{Top-quark electroweak interactions at
  high energy},'' \href{http://dx.doi.org/10.1007/JHEP10(2019)004}{{\em JHEP}
  {\bfseries 10} (2019) 004}, \href{http://arxiv.org/abs/1904.05637}{{\ttfamily
  arXiv:1904.05637 [hep-ph]}}.

\bibitem{Ethier:2021bye}
J.~J. Ethier, F.~Maltoni, L.~Mantani, E.~R. Nocera, J.~Rojo, E.~Slade,
  E.~Vryonidou, and C.~Zhang, ``{Combined SMEFT interpretation of Higgs,
  diboson, and top quark data from the LHC},''
  \href{http://arxiv.org/abs/2105.00006}{{\ttfamily arXiv:2105.00006
  [hep-ph]}}.

\bibitem{Costantini:2020stv}
A.~Costantini, F.~De~Lillo, F.~Maltoni, L.~Mantani, O.~Mattelaer, R.~Ruiz, and
  X.~Zhao, ``{Vector boson fusion at multi-TeV muon colliders},''
  \href{http://dx.doi.org/10.1007/JHEP09(2020)080}{{\em JHEP} {\bfseries 09}
  (2020) 080}, \href{http://arxiv.org/abs/2005.10289}{{\ttfamily
  arXiv:2005.10289 [hep-ph]}}.

\bibitem{Chiesa:2020awd}
M.~Chiesa, F.~Maltoni, L.~Mantani, B.~Mele, F.~Piccinini, and X.~Zhao,
  ``{Measuring the quartic Higgs self-coupling at a multi-TeV muon collider},''
  \href{http://dx.doi.org/10.1007/JHEP09(2020)098}{{\em JHEP} {\bfseries 09}
  (2020) 098}, \href{http://arxiv.org/abs/2003.13628}{{\ttfamily
  arXiv:2003.13628 [hep-ph]}}.

\bibitem{Ambrogi:2018jqj}
F.~Ambrogi, C.~Arina, M.~Backovic, J.~Heisig, F.~Maltoni, L.~Mantani,
  O.~Mattelaer, and G.~Mohlabeng, ``{MadDM v.3.0: a Comprehensive Tool for Dark
  Matter Studies},'' \href{http://dx.doi.org/10.1016/j.dark.2018.11.009}{{\em
  Phys. Dark Univ.} {\bfseries 24} (2019) 100249},
  \href{http://arxiv.org/abs/1804.00044}{{\ttfamily arXiv:1804.00044
  [hep-ph]}}.

\bibitem{Arina:2020udz}
C.~Arina, B.~Fuks, and L.~Mantani, ``{A universal framework for t-channel dark
  matter models},''
  \href{http://dx.doi.org/10.1140/epjc/s10052-020-7933-7}{{\em Eur. Phys. J. C}
  {\bfseries 80} no.~5, (2020) 409},
  \href{http://arxiv.org/abs/2001.05024}{{\ttfamily arXiv:2001.05024
  [hep-ph]}}.

\bibitem{Arina:2020tuw}
C.~Arina, B.~Fuks, L.~Mantani, H.~Mies, L.~Panizzi, and J.~Salko, ``{Closing in
  on $t$-channel simplified dark matter models},''
  \href{http://dx.doi.org/10.1016/j.physletb.2020.136038}{{\em Phys. Lett. B}
  {\bfseries 813} (2021) 136038},
  \href{http://arxiv.org/abs/2010.07559}{{\ttfamily arXiv:2010.07559
  [hep-ph]}}.

\bibitem{Cermeno:2021rtk}
M.~Cerme\~no, C.~Degrande, and L.~Mantani, ``{Circular polarisation of gamma
  rays as a probe of dark matter interactions with cosmic ray electrons},''
  \href{http://arxiv.org/abs/2103.14658}{{\ttfamily arXiv:2103.14658
  [hep-ph]}}.

\bibitem{Aad_2012}
G.~Aad, T.~Abajyan, B.~Abbott, J.~Abdallah, S.~Abdel~Khalek, A.~Abdelalim,
  O.~Abdinov, R.~Aben, B.~Abi, M.~Abolins, and et~al., ``Observation of a new
  particle in the search for the standard model higgs boson with the atlas
  detector at the lhc,''
  \href{http://dx.doi.org/10.1016/j.physletb.2012.08.020}{{\em Physics Letters
  B} {\bfseries 716} no.~1, (Sep, 2012) 1–29}.
  \url{http://dx.doi.org/10.1016/j.physletb.2012.08.020}.

\bibitem{Chatrchyan_2012}
S.~Chatrchyan, V.~Khachatryan, A.~Sirunyan, A.~Tumasyan, W.~Adam, E.~Aguilo,
  T.~Bergauer, M.~Dragicevic, J.~Erö, C.~Fabjan, and et~al., ``Observation of
  a new boson at a mass of 125 gev with the cms experiment at the lhc,''
  \href{http://dx.doi.org/10.1016/j.physletb.2012.08.021}{{\em Physics Letters
  B} {\bfseries 716} no.~1, (Sep, 2012) 30–61}.
  \url{http://dx.doi.org/10.1016/j.physletb.2012.08.021}.

\bibitem{Glashow:1961tr}
S.~L. Glashow, ``{Partial Symmetries of Weak Interactions},''
  \href{http://dx.doi.org/10.1016/0029-5582(61)90469-2}{{\em Nucl. Phys.}
  {\bfseries 22} (1961) 579--588}.

\bibitem{Weinberg:1967tq}
S.~Weinberg, ``{A Model of Leptons},''
  \href{http://dx.doi.org/10.1103/PhysRevLett.19.1264}{{\em Phys. Rev. Lett.}
  {\bfseries 19} (1967) 1264--1266}.

\bibitem{Salam:1968rm}
A.~Salam, ``{Weak and Electromagnetic Interactions},''
  \href{http://dx.doi.org/10.1142/9789812795915_0034}{{\em Conf. Proc. C}
  {\bfseries 680519} (1968) 367--377}.

\bibitem{PhysRev.109.193}
R.~P. Feynman and M.~Gell-Mann, ``Theory of the fermi interaction,''
  \href{http://dx.doi.org/10.1103/PhysRev.109.193}{{\em Phys. Rev.} {\bfseries
  109} (Jan, 1958) 193--198}.
  \url{https://link.aps.org/doi/10.1103/PhysRev.109.193}.

\bibitem{PhysRev.105.1413}
C.~S. Wu, E.~Ambler, R.~W. Hayward, D.~D. Hoppes, and R.~P. Hudson,
  ``Experimental test of parity conservation in beta decay,''
  \href{http://dx.doi.org/10.1103/PhysRev.105.1413}{{\em Phys. Rev.} {\bfseries
  105} (Feb, 1957) 1413--1415}.
  \url{https://link.aps.org/doi/10.1103/PhysRev.105.1413}.

\bibitem{PhysRevLett.13.321}
F.~Englert and R.~Brout, ``Broken symmetry and the mass of gauge vector
  mesons,'' \href{http://dx.doi.org/10.1103/PhysRevLett.13.321}{{\em Phys. Rev.
  Lett.} {\bfseries 13} (Aug, 1964) 321--323}.
  \url{https://link.aps.org/doi/10.1103/PhysRevLett.13.321}.

\bibitem{PhysRevLett.13.508}
P.~W. Higgs, ``Broken symmetries and the masses of gauge bosons,''
  \href{http://dx.doi.org/10.1103/PhysRevLett.13.508}{{\em Phys. Rev. Lett.}
  {\bfseries 13} (Oct, 1964) 508--509}.
  \url{https://link.aps.org/doi/10.1103/PhysRevLett.13.508}.

\bibitem{Cabibbo:1963yz}
N.~Cabibbo, ``{Unitary Symmetry and Leptonic Decays},''
  \href{http://dx.doi.org/10.1103/PhysRevLett.10.531}{{\em Phys. Rev. Lett.}
  {\bfseries 10} (1963) 531--533}.

\bibitem{osti_4474667}
M.~Kobayashi and T.~Maskawa, ``Cp-violation in the renormalizable theory of
  weak interaction,'' \href{http://dx.doi.org/10.1143/PTP.49.652}{{\em Progr.
  Theor. Phys. (Kyoto), v. 49, no. 2, pp. 652-657} }.
  \url{https://www.osti.gov/biblio/4474667}.

\bibitem{Schwartz:2014sze}
M.~D. Schwartz, {\em {Quantum Field Theory and the Standard Model}}.
\newblock Cambridge University Press, 3, 2014.

\bibitem{Appelquist:1987cf}
T.~Appelquist and M.~S. Chanowitz, ``{Unitarity Bound on the Scale of Fermion
  Mass Generation},'' \href{http://dx.doi.org/10.1103/PhysRevLett.59.2405}{{\em
  Phys. Rev. Lett.} {\bfseries 59} (1987) 2405}. [Erratum: Phys.Rev.Lett. 60,
  1589 (1988)].

\bibitem{Maltoni:2001dc}
F.~Maltoni, J.~M. Niczyporuk, and S.~Willenbrock, ``{The Scale of fermion mass
  generation},'' \href{http://dx.doi.org/10.1103/PhysRevD.65.033004}{{\em Phys.
  Rev. D} {\bfseries 65} (2002) 033004},
  \href{http://arxiv.org/abs/hep-ph/0106281}{{\ttfamily arXiv:hep-ph/0106281}}.

\bibitem{PhysRevD.16.1519}
B.~W. Lee, C.~Quigg, and H.~B. Thacker, ``Weak interactions at very high
  energies: The role of the higgs-boson mass,''
  \href{http://dx.doi.org/10.1103/PhysRevD.16.1519}{{\em Phys. Rev. D}
  {\bfseries 16} (Sep, 1977) 1519--1531}.
  \url{https://link.aps.org/doi/10.1103/PhysRevD.16.1519}.

\bibitem{Popper:34}
K.~R. Popper, {\em The Logic of Scientific Discovery}.
\newblock Hutchinson, London, 1934.

\bibitem{Bahcall:1964gx}
J.~N. Bahcall, ``{Solar neutrinos. I: Theoretical},''
  \href{http://dx.doi.org/10.1103/PhysRevLett.12.300}{{\em Phys. Rev. Lett.}
  {\bfseries 12} (1964) 300--302}.

\bibitem{PhysRevLett.12.303}
R.~Davis, ``Solar neutrinos. ii. experimental,''
  \href{http://dx.doi.org/10.1103/PhysRevLett.12.303}{{\em Phys. Rev. Lett.}
  {\bfseries 12} (Mar, 1964) 303--305}.
  \url{https://link.aps.org/doi/10.1103/PhysRevLett.12.303}.

\bibitem{Ahmad:2001an}
{\bfseries SNO} Collaboration, Q.~R. Ahmad {\em et~al.}, ``{Measurement of the
  rate of $\nu_e+d \to p+p+e^-$ interactions produced by $^8$B solar neutrinos
  at the Sudbury Neutrino Observatory},''
  \href{http://dx.doi.org/10.1103/PhysRevLett.87.071301}{{\em Phys. Rev. Lett.}
  {\bfseries 87} (2001) 071301},
  \href{http://arxiv.org/abs/nucl-ex/0106015}{{\ttfamily
  arXiv:nucl-ex/0106015}}.

\bibitem{Ahmad:2002jz}
{\bfseries SNO} Collaboration, Q.~R. Ahmad {\em et~al.}, ``{Direct evidence for
  neutrino flavor transformation from neutral current interactions in the
  Sudbury Neutrino Observatory},''
  \href{http://dx.doi.org/10.1103/PhysRevLett.89.011301}{{\em Phys. Rev. Lett.}
  {\bfseries 89} (2002) 011301},
  \href{http://arxiv.org/abs/nucl-ex/0204008}{{\ttfamily
  arXiv:nucl-ex/0204008}}.

\bibitem{Fukuda:1998mi}
{\bfseries Super-Kamiokande} Collaboration, Y.~Fukuda {\em et~al.}, ``{Evidence
  for oscillation of atmospheric neutrinos},''
  \href{http://dx.doi.org/10.1103/PhysRevLett.81.1562}{{\em Phys. Rev. Lett.}
  {\bfseries 81} (1998) 1562--1567},
  \href{http://arxiv.org/abs/hep-ex/9807003}{{\ttfamily arXiv:hep-ex/9807003}}.

\bibitem{Zwicky:1933gu}
F.~Zwicky, ``{Die Rotverschiebung von extragalaktischen Nebeln},''
  \href{http://dx.doi.org/10.1007/s10714-008-0707-4}{{\em Helv. Phys. Acta}
  {\bfseries 6} (1933) 110--127}.

\bibitem{1970ApJ...159..379R}
V.~C. {Rubin} and J.~{Ford}, W.~Kent, ``{Rotation of the Andromeda Nebula from
  a Spectroscopic Survey of Emission Regions},''
  \href{http://dx.doi.org/10.1086/150317}{{\em apj} {\bfseries 159} (Feb.,
  1970) 379}.

\bibitem{2016}
P.~A.~R. Ade, N.~Aghanim, M.~Arnaud, M.~Ashdown, J.~Aumont, C.~Baccigalupi,
  A.~J. Banday, R.~B. Barreiro, J.~G. Bartlett, and et~al., ``Planck2015
  results,'' \href{http://dx.doi.org/10.1051/0004-6361/201525830}{{\em
  Astronomy and Astrophysics} {\bfseries 594} (Sep, 2016) A13}.
  \url{http://dx.doi.org/10.1051/0004-6361/201525830}.

\bibitem{Clowe:2003tk}
D.~Clowe, A.~Gonzalez, and M.~Markevitch, ``{Weak lensing mass reconstruction
  of the interacting cluster 1E0657-558: Direct evidence for the existence of
  dark matter},'' \href{http://dx.doi.org/10.1086/381970}{{\em Astrophys. J.}
  {\bfseries 604} (2004) 596--603},
  \href{http://arxiv.org/abs/astro-ph/0312273}{{\ttfamily
  arXiv:astro-ph/0312273}}.

\bibitem{tHooft:1976rip}
G.~'t~Hooft, ``{Symmetry Breaking Through Bell-Jackiw Anomalies},''
  \href{http://dx.doi.org/10.1103/PhysRevLett.37.8}{{\em Phys. Rev. Lett.}
  {\bfseries 37} (1976) 8--11}.

\bibitem{Peccei:1977ur}
R.~D. Peccei and H.~R. Quinn, ``{Constraints Imposed by CP Conservation in the
  Presence of Instantons},''
  \href{http://dx.doi.org/10.1103/PhysRevD.16.1791}{{\em Phys. Rev. D}
  {\bfseries 16} (1977) 1791--1797}.

\bibitem{Callan:1976je}
C.~G. Callan, Jr., R.~F. Dashen, and D.~J. Gross, ``{The Structure of the Gauge
  Theory Vacuum},'' \href{http://dx.doi.org/10.1016/0370-2693(76)90277-X}{{\em
  Phys. Lett. B} {\bfseries 63} (1976) 334--340}.

\bibitem{Belavin:1975fg}
A.~A. Belavin, A.~M. Polyakov, A.~S. Schwartz, and Y.~S. Tyupkin,
  ``{Pseudoparticle Solutions of the Yang-Mills Equations},''
  \href{http://dx.doi.org/10.1016/0370-2693(75)90163-X}{{\em Phys. Lett. B}
  {\bfseries 59} (1975) 85--87}.

\bibitem{Kuzmin:1985mm}
V.~A. Kuzmin, V.~A. Rubakov, and M.~E. Shaposhnikov, ``{On the Anomalous
  Electroweak Baryon Number Nonconservation in the Early Universe},''
  \href{http://dx.doi.org/10.1016/0370-2693(85)91028-7}{{\em Phys. Lett. B}
  {\bfseries 155} (1985) 36}.

\bibitem{Sakharov:1967dj}
A.~D. Sakharov, ``{Violation of CP Invariance, C asymmetry, and baryon
  asymmetry of the universe},''
  \href{http://dx.doi.org/10.1070/PU1991v034n05ABEH002497}{{\em Sov. Phys.
  Usp.} {\bfseries 34} no.~5, (1991) 392--393}.

\bibitem{penco2020introduction}
R.~Penco, ``An introduction to effective field theories,'' 2020.

\bibitem{skiba2010tasi}
W.~Skiba, ``Tasi lectures on effective field theory and precision electroweak
  measurements,'' 2010.

\bibitem{Burgess_2007}
C.~Burgess, ``An introduction to effective field theory,''
  \href{http://dx.doi.org/10.1146/annurev.nucl.56.080805.140508}{{\em Annual
  Review of Nuclear and Particle Science} {\bfseries 57} no.~1, (Nov, 2007)
  329–362}. \url{http://dx.doi.org/10.1146/annurev.nucl.56.080805.140508}.

\bibitem{Brivio_2019}
I.~Brivio and M.~Trott, ``The standard model as an effective field theory,''
  \href{http://dx.doi.org/10.1016/j.physrep.2018.11.002}{{\em Physics Reports}
  {\bfseries 793} (Feb, 2019) 1–98}.
  \url{http://dx.doi.org/10.1016/j.physrep.2018.11.002}.

\bibitem{manohar2018introduction}
A.~V. Manohar, ``Introduction to effective field theories,'' 2018.

\bibitem{Polchinski:1992ed}
J.~Polchinski, ``{Effective field theory and the Fermi surface},'' in {\em
  {Theoretical Advanced Study Institute (TASI 92): From Black Holes and Strings
  to Particles}}, pp.~0235--276.
\newblock 6, 1992.
\newblock \href{http://arxiv.org/abs/hep-th/9210046}{{\ttfamily
  arXiv:hep-th/9210046}}.

\bibitem{McCullough:2018knz}
M.~McCullough, ``{Lectures on Physics Beyond the Standard Model.},'' in {\em
  {6th Tri-Institute Summer School on Elementary Particles}}.
\newblock 2018.

\bibitem{Fermi:1933jpa}
E.~Fermi, ``{Tentativo di una teoria dell'emissione dei raggi beta},'' {\em
  Ric. Sci.} {\bfseries 4} (1933) 491--495.

\bibitem{Feruglio:1992wf}
F.~Feruglio, ``{The Chiral approach to the electroweak interactions},''
  \href{http://dx.doi.org/10.1142/S0217751X93001946}{{\em Int. J. Mod. Phys. A}
  {\bfseries 8} (1993) 4937--4972},
  \href{http://arxiv.org/abs/hep-ph/9301281}{{\ttfamily arXiv:hep-ph/9301281}}.

\bibitem{Grinstein:2007iv}
B.~Grinstein and M.~Trott, ``{A Higgs-Higgs bound state due to new physics at a
  TeV},'' \href{http://dx.doi.org/10.1103/PhysRevD.76.073002}{{\em Phys. Rev.
  D} {\bfseries 76} (2007) 073002},
  \href{http://arxiv.org/abs/0704.1505}{{\ttfamily arXiv:0704.1505 [hep-ph]}}.

\bibitem{Falkowski:2019tft}
A.~Falkowski and R.~Rattazzi, ``{Which EFT},''
  \href{http://dx.doi.org/10.1007/JHEP10(2019)255}{{\em JHEP} {\bfseries 10}
  (2019) 255}, \href{http://arxiv.org/abs/1902.05936}{{\ttfamily
  arXiv:1902.05936 [hep-ph]}}.

\bibitem{Weinberg:1979sa}
S.~Weinberg, ``{Baryon and Lepton Nonconserving Processes},''
  \href{http://dx.doi.org/10.1103/PhysRevLett.43.1566}{{\em Phys. Rev. Lett.}
  {\bfseries 43} (1979) 1566--1570}.

\bibitem{Li:2020gnx}
H.-L. Li, Z.~Ren, J.~Shu, M.-L. Xiao, J.-H. Yu, and Y.-H. Zheng, ``{Complete
  Set of Dimension-8 Operators in the Standard Model Effective Field Theory},''
  \href{http://arxiv.org/abs/2005.00008}{{\ttfamily arXiv:2005.00008
  [hep-ph]}}.

\bibitem{Murphy:2020rsh}
C.~W. Murphy, ``{Dimension-8 operators in the Standard Model Eective Field
  Theory},'' \href{http://dx.doi.org/10.1007/JHEP10(2020)174}{{\em JHEP}
  {\bfseries 10} (2020) 174}, \href{http://arxiv.org/abs/2005.00059}{{\ttfamily
  arXiv:2005.00059 [hep-ph]}}.

\bibitem{Henning:2015daa}
B.~Henning, X.~Lu, T.~Melia, and H.~Murayama, ``{Hilbert series and operator
  bases with derivatives in effective field theories},''
  \href{http://dx.doi.org/10.1007/s00220-015-2518-2}{{\em Commun. Math. Phys.}
  {\bfseries 347} no.~2, (2016) 363--388},
  \href{http://arxiv.org/abs/1507.07240}{{\ttfamily arXiv:1507.07240
  [hep-th]}}.

\bibitem{Henning:2015alf}
B.~Henning, X.~Lu, T.~Melia, and H.~Murayama, ``{2, 84, 30, 993, 560, 15456,
  11962, 261485, ...: Higher dimension operators in the SM EFT},''
  \href{http://dx.doi.org/10.1007/JHEP08(2017)016}{{\em JHEP} {\bfseries 08}
  (2017) 016}, \href{http://arxiv.org/abs/1512.03433}{{\ttfamily
  arXiv:1512.03433 [hep-ph]}}. [Erratum: JHEP 09, 019 (2019)].

\bibitem{Henning:2017fpj}
B.~Henning, X.~Lu, T.~Melia, and H.~Murayama, ``{Operator bases, $S$-matrices,
  and their partition functions},''
  \href{http://dx.doi.org/10.1007/JHEP10(2017)199}{{\em JHEP} {\bfseries 10}
  (2017) 199}, \href{http://arxiv.org/abs/1706.08520}{{\ttfamily
  arXiv:1706.08520 [hep-th]}}.

\bibitem{Buchmuller:1985jz}
W.~Buchmuller and D.~Wyler, ``{Effective Lagrangian Analysis of New
  Interactions and Flavor Conservation},''
  \href{http://dx.doi.org/10.1016/0550-3213(86)90262-2}{{\em Nucl. Phys. B}
  {\bfseries 268} (1986) 621--653}.

\bibitem{Grzadkowski:2010es}
B.~Grzadkowski, M.~Iskrzynski, M.~Misiak, and J.~Rosiek, ``{Dimension-Six Terms
  in the Standard Model Lagrangian},''
  \href{http://dx.doi.org/10.1007/JHEP10(2010)085}{{\em JHEP} {\bfseries 10}
  (2010) 085}, \href{http://arxiv.org/abs/1008.4884}{{\ttfamily arXiv:1008.4884
  [hep-ph]}}.

\bibitem{Falkowski:2001958}
A.~Falkowski and A.~Falkowski, ``{Higgs Basis: Proposal for an EFT basis choice
  for LHC HXSWG},''. \url{https://cds.cern.ch/record/2001958}.

\bibitem{Giudice:2007fh}
G.~F. Giudice, C.~Grojean, A.~Pomarol, and R.~Rattazzi, ``{The
  Strongly-Interacting Light Higgs},''
  \href{http://dx.doi.org/10.1088/1126-6708/2007/06/045}{{\em JHEP} {\bfseries
  06} (2007) 045}, \href{http://arxiv.org/abs/hep-ph/0703164}{{\ttfamily
  arXiv:hep-ph/0703164}}.

\bibitem{Contino:2013kra}
R.~Contino, M.~Ghezzi, C.~Grojean, M.~Muhlleitner, and M.~Spira, ``{Effective
  Lagrangian for a light Higgs-like scalar},''
  \href{http://dx.doi.org/10.1007/JHEP07(2013)035}{{\em JHEP} {\bfseries 07}
  (2013) 035}, \href{http://arxiv.org/abs/1303.3876}{{\ttfamily arXiv:1303.3876
  [hep-ph]}}.

\bibitem{Elias-Miro:2013eta}
J.~Elias-Mir\'o, C.~Grojean, R.~S. Gupta, and D.~Marzocca, ``{Scaling and
  tuning of EW and Higgs observables},''
  \href{http://dx.doi.org/10.1007/JHEP05(2014)019}{{\em JHEP} {\bfseries 05}
  (2014) 019}, \href{http://arxiv.org/abs/1312.2928}{{\ttfamily arXiv:1312.2928
  [hep-ph]}}.

\bibitem{PhysRevD.48.2182}
K.~Hagiwara, S.~Ishihara, R.~Szalapski, and D.~Zeppenfeld, ``Low energy effects
  of new interactions in the electroweak boson sector,''
  \href{http://dx.doi.org/10.1103/PhysRevD.48.2182}{{\em Phys. Rev. D}
  {\bfseries 48} (Sep, 1993) 2182--2203}.
  \url{https://link.aps.org/doi/10.1103/PhysRevD.48.2182}.

\bibitem{Jenkins:2013zja}
E.~E. Jenkins, A.~V. Manohar, and M.~Trott, ``{Renormalization Group Evolution
  of the Standard Model Dimension Six Operators I: Formalism and lambda
  Dependence},'' \href{http://dx.doi.org/10.1007/JHEP10(2013)087}{{\em JHEP}
  {\bfseries 10} (2013) 087}, \href{http://arxiv.org/abs/1308.2627}{{\ttfamily
  arXiv:1308.2627 [hep-ph]}}.

\bibitem{Jenkins:2013wua}
E.~E. Jenkins, A.~V. Manohar, and M.~Trott, ``{Renormalization Group Evolution
  of the Standard Model Dimension Six Operators II: Yukawa Dependence},''
  \href{http://dx.doi.org/10.1007/JHEP01(2014)035}{{\em JHEP} {\bfseries 01}
  (2014) 035}, \href{http://arxiv.org/abs/1310.4838}{{\ttfamily arXiv:1310.4838
  [hep-ph]}}.

\bibitem{Alonso:2013hga}
R.~Alonso, E.~E. Jenkins, A.~V. Manohar, and M.~Trott, ``{Renormalization Group
  Evolution of the Standard Model Dimension Six Operators III: Gauge Coupling
  Dependence and Phenomenology},''
  \href{http://dx.doi.org/10.1007/JHEP04(2014)159}{{\em JHEP} {\bfseries 04}
  (2014) 159}, \href{http://arxiv.org/abs/1312.2014}{{\ttfamily arXiv:1312.2014
  [hep-ph]}}.

\bibitem{Grojean:2013kd}
C.~Grojean, E.~E. Jenkins, A.~V. Manohar, and M.~Trott, ``{Renormalization
  Group Scaling of Higgs Operators and \textbackslash{}Gamma(h -\ensuremath{>}
  \textbackslash{}gamma \textbackslash{}gamma)},''
  \href{http://dx.doi.org/10.1007/JHEP04(2013)016}{{\em JHEP} {\bfseries 04}
  (2013) 016}, \href{http://arxiv.org/abs/1301.2588}{{\ttfamily arXiv:1301.2588
  [hep-ph]}}.

\bibitem{Alonso:2014zka}
R.~Alonso, H.-M. Chang, E.~E. Jenkins, A.~V. Manohar, and B.~Shotwell,
  ``{Renormalization group evolution of dimension-six baryon number violating
  operators},'' \href{http://dx.doi.org/10.1016/j.physletb.2014.05.065}{{\em
  Phys. Lett. B} {\bfseries 734} (2014) 302--307},
  \href{http://arxiv.org/abs/1405.0486}{{\ttfamily arXiv:1405.0486 [hep-ph]}}.

\bibitem{LlewellynSmith:1973yud}
C.~H. Llewellyn~Smith, ``{High-Energy Behavior and Gauge Symmetry},''
\href{http://dx.doi.org/10.1016/0370-2693(73)90692-8}{{\em Phys. Lett.}
  {\bfseries 46B} (1973) 233--236}.

\bibitem{Lee:1977eg}
B.~W. Lee, C.~Quigg, and H.~B. Thacker, ``{Weak Interactions at Very
  High-Energies: The Role of the Higgs Boson Mass},''
\href{http://dx.doi.org/10.1103/PhysRevD.16.1519}{{\em Phys. Rev.} {\bfseries
  D16} (1977) 1519}.

\bibitem{Lee:1977yc}
B.~W. Lee, C.~Quigg, and H.~B. Thacker, ``{The Strength of Weak Interactions at
  Very High-Energies and the Higgs Boson Mass},''
\href{http://dx.doi.org/10.1103/PhysRevLett.38.883}{{\em Phys. Rev. Lett.}
  {\bfseries 38} (1977) 883--885}.

\bibitem{Maltoni:2000iq}
F.~Maltoni, J.~M. Niczyporuk, and S.~Willenbrock, ``{Upper bound on the scale
  of Majorana neutrino mass generation},''
  \href{http://dx.doi.org/10.1103/PhysRevLett.86.212}{{\em Phys. Rev. Lett.}
  {\bfseries 86} (2001) 212--215},
\href{http://arxiv.org/abs/hep-ph/0006358}{{\ttfamily arXiv:hep-ph/0006358
  [hep-ph]}}.

\bibitem{CMS:2018rbc}
{\bfseries CMS} Collaboration, C.~Collaboration,
``{Measurement of the associated production of a Higgs boson and a pair of
  top-antitop quarks with the Higgs boson decaying to two photons in
  proton-proton collisions at $\sqrt{s}=13~\mathrm{TeV}$},''.

\bibitem{Sirunyan:2018mvw}
{\bfseries CMS} Collaboration, A.~M. Sirunyan {\em et~al.}, ``{Search for
  $\mathrm{t\overline{t}}$H production in the $H\to\mathrm{b\overline{b}}$
  decay channel with leptonic $\mathrm{t\overline{t}}$ decays in proton-proton
  collisions at $\sqrt{s}=$ 13 TeV},''
\href{http://arxiv.org/abs/1804.03682}{{\ttfamily arXiv:1804.03682 [hep-ex]}}.

\bibitem{Sirunyan:2018shy}
{\bfseries CMS} Collaboration, A.~M. Sirunyan {\em et~al.}, ``{Evidence for
  associated production of a Higgs boson with a top quark pair in final states
  with electrons, muons, and hadronically decaying $\tau$ leptons at $\sqrt{s}
  =$ 13 TeV},'' \href{http://dx.doi.org/10.1007/JHEP08(2018)066}{{\em JHEP}
  {\bfseries 08} (2018) 066},
\href{http://arxiv.org/abs/1803.05485}{{\ttfamily arXiv:1803.05485 [hep-ex]}}.

\bibitem{Aaboud:2017jvq}
{\bfseries ATLAS} Collaboration, M.~Aaboud {\em et~al.}, ``{Evidence for the
  associated production of the Higgs boson and a top quark pair with the ATLAS
  detector},'' {\em Submitted to: Phys. Rev. D} (2017) ,
\href{http://arxiv.org/abs/1712.08891}{{\ttfamily arXiv:1712.08891 [hep-ex]}}.

\bibitem{Aaboud:2017rss}
{\bfseries ATLAS} Collaboration, M.~Aaboud {\em et~al.}, ``{Search for the
  Standard Model Higgs boson produced in association with top quarks and
  decaying into a $b\bar{b}$ pair in $pp$ collisions at $\sqrt{s}$ = 13 TeV
  with the ATLAS detector},'' {\em Submitted to: Phys. Rev. D} (2017) ,
\href{http://arxiv.org/abs/1712.08895}{{\ttfamily arXiv:1712.08895 [hep-ex]}}.

\bibitem{Degrande:2018fog}
C.~Degrande, F.~Maltoni, K.~Mimasu, E.~Vryonidou, and C.~Zhang, ``{Single-top
  associated production with a $Z$ or $H$ boson at the LHC: the SMEFT
  interpretation},'' {\em Submitted to: JHEP} (2018) ,
\href{http://arxiv.org/abs/1804.07773}{{\ttfamily arXiv:1804.07773 [hep-ph]}}.

\bibitem{Corbett:2014ora}
T.~Corbett, O.~J.~P. {\'E}boli, and M.~C. Gonzalez-Garcia, ``{Unitarity
  Constraints on Dimension-Six Operators},''
  \href{http://dx.doi.org/10.1103/PhysRevD.91.035014}{{\em Phys. Rev.}
  {\bfseries D91} no.~3, (2015) 035014},
\href{http://arxiv.org/abs/1411.5026}{{\ttfamily arXiv:1411.5026 [hep-ph]}}.

\bibitem{Corbett:2017qgl}
T.~Corbett, O.~J.~P. {\'E}boli, and M.~C. Gonzalez-Garcia, ``{Unitarity
  Constraints on Dimension-six Operators II: Including Fermionic Operators},''
  \href{http://dx.doi.org/10.1103/PhysRevD.96.035006}{{\em Phys. Rev.}
  {\bfseries D96} no.~3, (2017) 035006},
\href{http://arxiv.org/abs/1705.09294}{{\ttfamily arXiv:1705.09294 [hep-ph]}}.

\bibitem{Dror:2015nkp}
J.~A. Dror, M.~Farina, E.~Salvioni, and J.~Serra, ``{Strong tW Scattering at
  the LHC},'' \href{http://dx.doi.org/10.1007/JHEP01(2016)071}{{\em JHEP}
  {\bfseries 01} (2016) 071},
\href{http://arxiv.org/abs/1511.03674}{{\ttfamily arXiv:1511.03674 [hep-ph]}}.

\bibitem{AguilarSaavedra:2018nen}
D.~Barducci {\em et~al.}, ``{Interpreting top-quark LHC measurements in the
  standard-model effective field theory},''
\href{http://arxiv.org/abs/1802.07237}{{\ttfamily arXiv:1802.07237 [hep-ph]}}.

\bibitem{Alloul:2013bka}
A.~Alloul, N.~D. Christensen, C.~Degrande, C.~Duhr, and B.~Fuks, ``{FeynRules
  2.0 - A complete toolbox for tree-level phenomenology},''
  \href{http://dx.doi.org/10.1016/j.cpc.2014.04.012}{{\em Comput.Phys.Commun.}
  {\bfseries 185} (2014) 2250--2300},
\href{http://arxiv.org/abs/1310.1921}{{\ttfamily arXiv:1310.1921 [hep-ph]}}.

\bibitem{Degrande:2011ua}
C.~Degrande, C.~Duhr, B.~Fuks, D.~Grellscheid, O.~Mattelaer, {\em et~al.},
  ``{UFO - The Universal FeynRules Output},''
  \href{http://dx.doi.org/10.1016/j.cpc.2012.01.022}{{\em Comput.Phys.Commun.}
  {\bfseries 183} (2012) 1201--1214},
\href{http://arxiv.org/abs/1108.2040}{{\ttfamily arXiv:1108.2040 [hep-ph]}}.

\bibitem{HAHN2001418}
T.~Hahn, ``Generating feynman diagrams and amplitudes with feynarts 3,''
  \href{http://dx.doi.org/https://doi.org/10.1016/S0010-4655(01)00290-9}{{\em
  Computer Physics Communications} {\bfseries 140} no.~3, (2001) 418 -- 431}.
  \url{http://www.sciencedirect.com/science/article/pii/S0010465501002909}.

\bibitem{Shtabovenko:2016sxi}
V.~Shtabovenko, R.~Mertig, and F.~Orellana, ``{New Developments in FeynCalc
  9.0},'' \href{http://dx.doi.org/10.1016/j.cpc.2016.06.008}{{\em Comput. Phys.
  Commun.} {\bfseries 207} (2016) 432--444},
\href{http://arxiv.org/abs/1601.01167}{{\ttfamily arXiv:1601.01167 [hep-ph]}}.

\bibitem{Alwall:2011uj}
J.~Alwall, M.~Herquet, F.~Maltoni, O.~Mattelaer, and T.~Stelzer, ``{MadGraph 5
  : Going Beyond},'' \href{http://dx.doi.org/10.1007/JHEP06(2011)128}{{\em
  JHEP} {\bfseries 06} (2011) 128},
\href{http://arxiv.org/abs/1106.0522}{{\ttfamily arXiv:1106.0522 [hep-ph]}}.

\bibitem{Azatov:2016sqh}
A.~Azatov, R.~Contino, C.~S. Machado, and F.~Riva, ``{Helicity selection rules
  and noninterference for BSM amplitudes},''
  \href{http://dx.doi.org/10.1103/PhysRevD.95.065014}{{\em Phys. Rev.}
  {\bfseries D95} no.~6, (2017) 065014},
\href{http://arxiv.org/abs/1607.05236}{{\ttfamily arXiv:1607.05236 [hep-ph]}}.

\bibitem{Cornwall:1974km}
J.~M. Cornwall, D.~N. Levin, and G.~Tiktopoulos, ``{Derivation of Gauge
  Invariance from High-Energy Unitarity Bounds on the s Matrix},''
  \href{http://dx.doi.org/10.1103/PhysRevD.10.1145,
  10.1103/PhysRevD.11.972}{{\em Phys. Rev.} {\bfseries D10} (1974) 1145}.
[Erratum: Phys. Rev.D11,972(1975)].

\bibitem{Henning:2018kys}
B.~Henning, D.~Lombardo, M.~Riembau, and F.~Riva, ``{Higgs Couplings without
  the Higgs},''
\href{http://arxiv.org/abs/1812.09299}{{\ttfamily arXiv:1812.09299 [hep-ph]}}.

\bibitem{AguilarSaavedra:2008zc}
J.~A. Aguilar-Saavedra, ``{A Minimal set of top anomalous couplings},''
  \href{http://dx.doi.org/10.1016/j.nuclphysb.2008.12.012}{{\em Nucl. Phys.}
  {\bfseries B812} (2009) 181--204},
\href{http://arxiv.org/abs/0811.3842}{{\ttfamily arXiv:0811.3842 [hep-ph]}}.

\bibitem{Dawson:1984gx}
S.~Dawson, ``{The Effective W Approximation},''
\href{http://dx.doi.org/10.1016/0550-3213(85)90038-0}{{\em Nucl. Phys.}
  {\bfseries B249} (1985) 42--60}.

\bibitem{Kunszt:1987tk}
Z.~Kunszt and D.~E. Soper, ``{On the Validity of the Effective $W$
  Approximation},''
\href{http://dx.doi.org/10.1016/0550-3213(88)90673-6}{{\em Nucl. Phys.}
  {\bfseries B296} (1988) 253--289}.

\bibitem{Borel:2012by}
P.~Borel, R.~Franceschini, R.~Rattazzi, and A.~Wulzer, ``{Probing the
  Scattering of Equivalent Electroweak Bosons},''
  \href{http://dx.doi.org/10.1007/JHEP06(2012)122}{{\em JHEP} {\bfseries 06}
  (2012) 122},
\href{http://arxiv.org/abs/1202.1904}{{\ttfamily arXiv:1202.1904 [hep-ph]}}.

\bibitem{Alwall:2014hca}
J.~Alwall, R.~Frederix, S.~Frixione, V.~Hirschi, F.~Maltoni, O.~Mattelaer,
  H.~S. Shao, T.~Stelzer, P.~Torrielli, and M.~Zaro, ``{The automated
  computation of tree-level and next-to-leading order differential cross
  sections, and their matching to parton shower simulations},''
  \href{http://dx.doi.org/10.1007/JHEP07(2014)079}{{\em JHEP} {\bfseries 07}
  (2014) 079},
\href{http://arxiv.org/abs/1405.0301}{{\ttfamily arXiv:1405.0301 [hep-ph]}}.

\bibitem{SMEFTatNLO}
``http://feynrules.irmp.ucl.ac.be/wiki/smeftatnlo.''.

\bibitem{Aaboud:2017ylb}
{\bfseries ATLAS} Collaboration, M.~Aaboud {\em et~al.}, ``{Measurement of the
  production cross-section of a single top quark in association with a $Z$
  boson in proton--proton collisions at 13 TeV with the ATLAS detector},''
\href{http://arxiv.org/abs/1710.03659}{{\ttfamily arXiv:1710.03659 [hep-ex]}}.

\bibitem{Sirunyan:2017nbr}
{\bfseries CMS} Collaboration, A.~M. Sirunyan {\em et~al.}, ``{Measurement of
  the associated production of a single top quark and a Z boson in pp
  collisions at $\sqrt{s} =$ 13 TeV},''
  \href{http://dx.doi.org/10.1016/j.physletb.2018.02.025}{{\em Phys. Lett.}
  {\bfseries B779} (2018) 358--384},
\href{http://arxiv.org/abs/1712.02825}{{\ttfamily arXiv:1712.02825 [hep-ex]}}.

\bibitem{Sirunyan:2018zgs}
{\bfseries CMS} Collaboration, A.~M. Sirunyan {\em et~al.}, ``{Observation of
  single top quark production in association with a Z boson in proton-proton
  collisions at $\sqrt{s} =$ 13 TeV},'' {\em Submitted to: Phys. Rev. Lett.}
  (2018) ,
\href{http://arxiv.org/abs/1812.05900}{{\ttfamily arXiv:1812.05900 [hep-ex]}}.

\bibitem{Sirunyan:2018bsr}
{\bfseries CMS} Collaboration, A.~M. Sirunyan {\em et~al.}, ``{Evidence for the
  associated production of a single top quark and a photon in proton-proton
  collisions at $\sqrt{s}=$ 13 TeV},''
  \href{http://dx.doi.org/10.1103/PhysRevLett.121.221802}{{\em Phys. Rev.
  Lett.} {\bfseries 121} no.~22, (2018) 221802},
\href{http://arxiv.org/abs/1808.02913}{{\ttfamily arXiv:1808.02913 [hep-ex]}}.

\bibitem{Sirunyan:2018lzm}
{\bfseries CMS} Collaboration, A.~M. Sirunyan {\em et~al.}, ``{Search for
  associated production of a Higgs boson and a single top quark in
  proton-proton collisions at $\sqrt{s} =$ 13 TeV},'' {\em Submitted to: Phys.
  Rev.} (2018) ,
\href{http://arxiv.org/abs/1811.09696}{{\ttfamily arXiv:1811.09696 [hep-ex]}}.

\bibitem{Etesami:2017ufk}
S.~M. Etesami, S.~Khatibi, and M.~Mohammadi~Najafabadi, ``{Study of top quark
  dipole interactions in $t\bar{t}$ production associated with two heavy gauge
  bosons at the LHC},''
  \href{http://dx.doi.org/10.1103/PhysRevD.97.075023}{{\em Phys. Rev.}
  {\bfseries D97} no.~7, (2018) 075023},
\href{http://arxiv.org/abs/1712.07184}{{\ttfamily arXiv:1712.07184 [hep-ph]}}.

\bibitem{Bylund:2016phk}
O.~Bessidskaia~Bylund, F.~Maltoni, I.~Tsinikos, E.~Vryonidou, and C.~Zhang,
  ``{Probing top quark neutral couplings in the Standard Model Effective Field
  Theory at NLO in QCD},''
  \href{http://dx.doi.org/10.1007/JHEP05(2016)052}{{\em JHEP} {\bfseries 05}
  (2016) 052},
\href{http://arxiv.org/abs/1601.08193}{{\ttfamily arXiv:1601.08193 [hep-ph]}}.

\bibitem{Abramowicz:2018rjq}
{\bfseries CLICdp} Collaboration, H.~Abramowicz {\em et~al.}, ``{Top-Quark
  Physics at the CLIC Electron-Positron Linear Collider},''
\href{http://arxiv.org/abs/1807.02441}{{\ttfamily arXiv:1807.02441 [hep-ex]}}.

\bibitem{Maltoni:2016yxb}
F.~Maltoni, E.~Vryonidou, and C.~Zhang, ``{Higgs production in association with
  a top-antitop pair in the Standard Model Effective Field Theory at NLO in
  QCD},'' \href{http://dx.doi.org/10.1007/JHEP10(2016)123}{{\em JHEP}
  {\bfseries 10} (2016) 123},
\href{http://arxiv.org/abs/1607.05330}{{\ttfamily arXiv:1607.05330 [hep-ph]}}.

\bibitem{Liu:2015aka}
N.~Liu, Y.~Zhang, J.~Han, and B.~Yang, ``{Enhancing $ t\overline{t}hh $
  production through CP-violating top-Higgs interaction at the LHC and future
  colliders},'' \href{http://dx.doi.org/10.1007/JHEP09(2015)008}{{\em JHEP}
  {\bfseries 09} (2015) 008},
\href{http://arxiv.org/abs/1503.08537}{{\ttfamily arXiv:1503.08537 [hep-ph]}}.

\bibitem{Pich:2019pzg}
A.~Pich, ``{Flavour Anomalies},'' {\em PoS} {\bfseries LHCP2019} (2019) 078,
  \href{http://arxiv.org/abs/1911.06211}{{\ttfamily arXiv:1911.06211
  [hep-ph]}}.

\bibitem{Greljo:2017vvb}
A.~Greljo and D.~Marzocca, ``{High-$p_T$ dilepton tails and flavor physics},''
  \href{http://dx.doi.org/10.1140/epjc/s10052-017-5119-8}{{\em Eur. Phys. J.}
  {\bfseries C77} no.~8, (2017) 548},
\href{http://arxiv.org/abs/1704.09015}{{\ttfamily arXiv:1704.09015 [hep-ph]}}.

\bibitem{Buckley:2016cfg}
A.~Buckley, C.~Englert, J.~Ferrando, D.~J. Miller, L.~Moore, K.~N{\"o}rdstrom,
  M.~Russell, and C.~D. White, ``{Results from TopFitter},'' {\em PoS}
  {\bfseries CKM2016} (2016) 127,
\href{http://arxiv.org/abs/1612.02294}{{\ttfamily arXiv:1612.02294 [hep-ph]}}.

\bibitem{Buckley:2015lku}
A.~Buckley, C.~Englert, J.~Ferrando, D.~J. Miller, L.~Moore, M.~Russell, and
  C.~D. White, ``{Constraining top quark effective theory in the LHC Run II
  era},'' \href{http://dx.doi.org/10.1007/JHEP04(2016)015}{{\em JHEP}
  {\bfseries 04} (2016) 015},
\href{http://arxiv.org/abs/1512.03360}{{\ttfamily arXiv:1512.03360 [hep-ph]}}.

\bibitem{Hartland:2019bjb}
N.~P. Hartland, F.~Maltoni, E.~R. Nocera, J.~Rojo, E.~Slade, E.~Vryonidou, and
  C.~Zhang, ``{A Monte Carlo global analysis of the Standard Model Effective
  Field Theory: the top quark sector},''
\href{http://arxiv.org/abs/1901.05965}{{\ttfamily arXiv:1901.05965 [hep-ph]}}.

\bibitem{Brivio:2019ius}
I.~Brivio, S.~Bruggisser, F.~Maltoni, R.~Moutafis, T.~Plehn, E.~Vryonidou,
  S.~Westhoff, and C.~Zhang, ``{O new physics, where art thou? A global search
  in the top sector},'' \href{http://dx.doi.org/10.1007/JHEP02(2020)131}{{\em
  JHEP} {\bfseries 02} (2020) 131},
  \href{http://arxiv.org/abs/1910.03606}{{\ttfamily arXiv:1910.03606
  [hep-ph]}}.

\bibitem{Biekotter:2018rhp}
A.~Biekötter, T.~Corbett, and T.~Plehn, ``{The Gauge-Higgs Legacy of the LHC
  Run II},'' \href{http://dx.doi.org/10.21468/SciPostPhys.6.6.064}{{\em SciPost
  Phys.} {\bfseries 6} (2019) 064},
\href{http://arxiv.org/abs/1812.07587}{{\ttfamily arXiv:1812.07587 [hep-ph]}}.

\bibitem{Ellis:2018gqa}
J.~Ellis, C.~W. Murphy, V.~Sanz, and T.~You, ``{Updated Global SMEFT Fit to
  Higgs, Diboson and Electroweak Data},''
  \href{http://dx.doi.org/10.1007/JHEP06(2018)146}{{\em JHEP} {\bfseries 06}
  (2018) 146},
\href{http://arxiv.org/abs/1803.03252}{{\ttfamily arXiv:1803.03252 [hep-ph]}}.

\bibitem{Almeida:2018cld}
E.~da~Silva~Almeida, A.~Alves, N.~Rosa~Agostinho, O.~J. \'Eboli, and
  M.~Gonzalez-Garcia, ``{Electroweak Sector Under Scrutiny: A Combined Analysis
  of LHC and Electroweak Precision Data},''
  \href{http://dx.doi.org/10.1103/PhysRevD.99.033001}{{\em Phys. Rev. D}
  {\bfseries 99} no.~3, (2019) 033001},
  \href{http://arxiv.org/abs/1812.01009}{{\ttfamily arXiv:1812.01009
  [hep-ph]}}.

\bibitem{Baglio:2020ibv}
J.~Baglio, S.~Dawson, S.~Homiller, S.~D. Lane, and I.~M. Lewis, ``{Validity of
  standard model EFT studies of VH and VV production at NLO},''
  \href{http://dx.doi.org/10.1103/PhysRevD.101.115004}{{\em Phys. Rev. D}
  {\bfseries 101} no.~11, (2020) 115004}.

\bibitem{Gomez-Ambrosio:2018pnl}
R.~Gomez-Ambrosio, ``{Studies of Dimension-Six EFT effects in Vector Boson
  Scattering},'' \href{http://dx.doi.org/10.1140/epjc/s10052-019-6893-2}{{\em
  Eur. Phys. J. C} {\bfseries 79} no.~5, (2019) 389},
  \href{http://arxiv.org/abs/1809.04189}{{\ttfamily arXiv:1809.04189
  [hep-ph]}}.

\bibitem{Krauss:2016ely}
F.~Krauss, S.~Kuttimalai, and T.~Plehn, ``{LHC multijet events as a probe for
  anomalous dimension-six gluon interactions},''
  \href{http://dx.doi.org/10.1103/PhysRevD.95.035024}{{\em Phys. Rev.}
  {\bfseries D95} no.~3, (2017) 035024},
\href{http://arxiv.org/abs/1611.00767}{{\ttfamily arXiv:1611.00767 [hep-ph]}}.

\bibitem{Alte:2017pme}
S.~Alte, M.~K{\"o}nig, and W.~Shepherd, ``{Consistent Searches for SMEFT
  Effects in Non-Resonant Dijet Events},''
  \href{http://dx.doi.org/10.1007/JHEP01(2018)094}{{\em JHEP} {\bfseries 01}
  (2018) 094},
\href{http://arxiv.org/abs/1711.07484}{{\ttfamily arXiv:1711.07484 [hep-ph]}}.

\bibitem{Aebischer:2018iyb}
J.~Aebischer, J.~Kumar, P.~Stangl, and D.~M. Straub, ``{A Global Likelihood for
  Precision Constraints and Flavour Anomalies},''
  \href{http://dx.doi.org/10.1140/epjc/s10052-019-6977-z}{{\em Eur. Phys. J. C}
  {\bfseries 79} no.~6, (2019) 509},
  \href{http://arxiv.org/abs/1810.07698}{{\ttfamily arXiv:1810.07698
  [hep-ph]}}.

\bibitem{Falkowski:2019xoe}
A.~Falkowski, M.~González-Alonso, and Z.~Tabrizi, ``{Reactor neutrino
  oscillations as constraints on Effective Field Theory},''
  \href{http://dx.doi.org/10.1007/JHEP05(2019)173}{{\em JHEP} {\bfseries 05}
  (2019) 173}, \href{http://arxiv.org/abs/1901.04553}{{\ttfamily
  arXiv:1901.04553 [hep-ph]}}.

\bibitem{Falkowski:2017pss}
A.~Falkowski, M.~González-Alonso, and K.~Mimouni, ``{Compilation of low-energy
  constraints on 4-fermion operators in the SMEFT},''
  \href{http://dx.doi.org/10.1007/JHEP08(2017)123}{{\em JHEP} {\bfseries 08}
  (2017) 123}, \href{http://arxiv.org/abs/1706.03783}{{\ttfamily
  arXiv:1706.03783 [hep-ph]}}.

\bibitem{Ellis:2020unq}
J.~Ellis, M.~Madigan, K.~Mimasu, V.~Sanz, and T.~You, ``{Top, Higgs, Diboson
  and Electroweak Fit to the Standard Model Effective Field Theory},''
  \href{http://dx.doi.org/10.1007/JHEP04(2021)279}{{\em JHEP} {\bfseries 04}
  (2021) 279}, \href{http://arxiv.org/abs/2012.02779}{{\ttfamily
  arXiv:2012.02779 [hep-ph]}}.

\bibitem{ATLAS-CONF-2020-053}
{\bfseries ATLAS Collaboration} Collaboration, ``{Interpretations of the
  combined measurement of Higgs boson production and decay},'' Tech. Rep.
  ATLAS-CONF-2020-053, CERN, Geneva, Oct, 2020.
\newblock \url{http://cds.cern.ch/record/2743067}.

\bibitem{CMS-PAS-TOP-19-001}
{\bfseries CMS Collaboration} Collaboration, ``{Using associated top quark
  production to probe for new physics within the framework of effective field
  theory},'' Tech. Rep. CMS-PAS-TOP-19-001, CERN, Geneva, 2020.
\newblock \url{https://cds.cern.ch/record/2725399}.

\bibitem{Ball:2008by}
{\bfseries The NNPDF} Collaboration, R.~D. Ball {\em et~al.}, ``{A
  determination of parton distributions with faithful uncertainty
  estimation},'' \href{http://dx.doi.org/10.1016/j.nuclphysb.2008.09.037}{{\em
  Nucl. Phys.} {\bfseries B809} (2009) 1--63},
\href{http://arxiv.org/abs/0808.1231}{{\ttfamily arXiv:0808.1231 [hep-ph]}}.

\bibitem{Ball:2010de}
{\bfseries {The NNPDF }} Collaboration, R.~D. Ball {\em et~al.}, ``{A first
  unbiased global NLO determination of parton distributions and their
  uncertainties},''
  \href{http://dx.doi.org/10.1016/j.nuclphysb.2010.05.008}{{\em Nucl. Phys.}
  {\bfseries B838} (2010) 136},
\href{http://arxiv.org/abs/1002.4407}{{\ttfamily arXiv:1002.4407 [hep-ph]}}.

\bibitem{Ball:2014uwa}
{\bfseries NNPDF} Collaboration, R.~D. Ball {\em et~al.}, ``{Parton
  distributions for the LHC Run II},''
  \href{http://dx.doi.org/10.1007/JHEP04(2015)040}{{\em JHEP} {\bfseries 04}
  (2015) 040},
\href{http://arxiv.org/abs/1410.8849}{{\ttfamily arXiv:1410.8849 [hep-ph]}}.

\bibitem{ALEPH:2005ab}
{\bfseries ALEPH, DELPHI, L3, OPAL, SLD, LEP Electroweak Working Group, SLD
  Electroweak Group, SLD Heavy Flavour Group} Collaboration, S.~Schael {\em
  et~al.}, ``{Precision electroweak measurements on the $Z$ resonance},''
  \href{http://dx.doi.org/10.1016/j.physrep.2005.12.006}{{\em Phys. Rept.}
  {\bfseries 427} (2006) 257--454},
\href{http://arxiv.org/abs/hep-ex/0509008}{{\ttfamily arXiv:hep-ex/0509008
  [hep-ex]}}.

\bibitem{Feroz:2013hea}
F.~Feroz, M.~P. Hobson, E.~Cameron, and A.~N. Pettitt, ``{Importance Nested
  Sampling and the MultiNest Algorithm},''
\href{http://arxiv.org/abs/1306.2144}{{\ttfamily arXiv:1306.2144
  [astro-ph.IM]}}.

\bibitem{DAmbrosio:2002vsn}
G.~D'Ambrosio, G.~F. Giudice, G.~Isidori, and A.~Strumia, ``{Minimal flavor
  violation: An Effective field theory approach},''
  \href{http://dx.doi.org/10.1016/S0550-3213(02)00836-2}{{\em Nucl. Phys.}
  {\bfseries B645} (2002) 155--187},
\href{http://arxiv.org/abs/hep-ph/0207036}{{\ttfamily arXiv:hep-ph/0207036
  [hep-ph]}}.

\bibitem{Degrande:2020evl}
C.~Degrande, G.~Durieux, F.~Maltoni, K.~Mimasu, E.~Vryonidou, and C.~Zhang,
  ``{Automated one-loop computations in the SMEFT},''
  \href{http://arxiv.org/abs/2008.11743}{{\ttfamily arXiv:2008.11743
  [hep-ph]}}.

\bibitem{Han:2004az}
Z.~Han and W.~Skiba, ``{Effective theory analysis of precision electroweak
  data},'' \href{http://dx.doi.org/10.1103/PhysRevD.71.075009}{{\em Phys. Rev.}
  {\bfseries D71} (2005) 075009},
\href{http://arxiv.org/abs/hep-ph/0412166}{{\ttfamily arXiv:hep-ph/0412166
  [hep-ph]}}.

\bibitem{Falkowski:2014tna}
A.~Falkowski and F.~Riva, ``{Model-independent precision constraints on
  dimension-6 operators},''
  \href{http://dx.doi.org/10.1007/JHEP02(2015)039}{{\em JHEP} {\bfseries 02}
  (2015) 039},
\href{http://arxiv.org/abs/1411.0669}{{\ttfamily arXiv:1411.0669 [hep-ph]}}.

\bibitem{Grojean:2006nn}
C.~Grojean, W.~Skiba, and J.~Terning, ``{Disguising the oblique parameters},''
  \href{http://dx.doi.org/10.1103/PhysRevD.73.075008}{{\em Phys. Rev.}
  {\bfseries D73} (2006) 075008},
\href{http://arxiv.org/abs/hep-ph/0602154}{{\ttfamily arXiv:hep-ph/0602154
  [hep-ph]}}.

\bibitem{Brivio:2017bnu}
I.~Brivio and M.~Trott, ``{Scheming in the SMEFT... and a reparameterization
  invariance!},'' \href{http://dx.doi.org/10.1007/JHEP07(2017)148}{{\em JHEP}
  {\bfseries 07} (2017) 148},
\href{http://arxiv.org/abs/1701.06424}{{\ttfamily arXiv:1701.06424 [hep-ph]}}.

\bibitem{Zhang:2016zsp}
Z.~Zhang, ``{Time to Go Beyond Triple-Gauge-Boson-Coupling Interpretation of
  $W$ Pair Production},''
  \href{http://dx.doi.org/10.1103/PhysRevLett.118.011803}{{\em Phys. Rev.
  Lett.} {\bfseries 118} no.~1, (2017) 011803},
\href{http://arxiv.org/abs/1610.01618}{{\ttfamily arXiv:1610.01618 [hep-ph]}}.

\bibitem{Grojean:2018dqj}
C.~Grojean, M.~Montull, and M.~Riembau, ``{Diboson at the LHC vs LEP},''
  \href{http://dx.doi.org/10.1007/JHEP03(2019)020}{{\em JHEP} {\bfseries 03}
  (2019) 020}, \href{http://arxiv.org/abs/1810.05149}{{\ttfamily
  arXiv:1810.05149 [hep-ph]}}.

\bibitem{Boughezal:2016wmq}
R.~Boughezal, J.~M. Campbell, R.~K. Ellis, C.~Focke, W.~Giele, X.~Liu,
  F.~Petriello, and C.~Williams, ``{Color singlet production at NNLO in
  MCFM},'' \href{http://dx.doi.org/10.1140/epjc/s10052-016-4558-y}{{\em Eur.
  Phys. J.} {\bfseries C77} no.~1, (2017) 7},
\href{http://arxiv.org/abs/1605.08011}{{\ttfamily arXiv:1605.08011 [hep-ph]}}.

\bibitem{Czakon:2016dgf}
M.~Czakon, D.~Heymes, and A.~Mitov, ``{Dynamical scales for multi-TeV top-pair
  production at the LHC},''
  \href{http://dx.doi.org/10.1007/JHEP04(2017)071}{{\em JHEP} {\bfseries 04}
  (2017) 071},
\href{http://arxiv.org/abs/1606.03350}{{\ttfamily arXiv:1606.03350 [hep-ph]}}.

\bibitem{Czakon:2016olj}
M.~Czakon, N.~P. Hartland, A.~Mitov, E.~R. Nocera, and J.~Rojo, ``{Pinning down
  the large-x gluon with NNLO top-quark pair differential distributions},''
  \href{http://dx.doi.org/10.1007/JHEP04(2017)044}{{\em JHEP} {\bfseries 04}
  (2017) 044},
\href{http://arxiv.org/abs/1611.08609}{{\ttfamily arXiv:1611.08609 [hep-ph]}}.

\bibitem{Ball:2017nwa}
{\bfseries NNPDF} Collaboration, R.~D. Ball {\em et~al.}, ``{Parton
  distributions from high-precision collider data},''
  \href{http://dx.doi.org/10.1140/epjc/s10052-017-5199-5}{{\em Eur. Phys. J.}
  {\bfseries C77} no.~10, (2017) 663},
\href{http://arxiv.org/abs/1706.00428}{{\ttfamily arXiv:1706.00428 [hep-ph]}}.

\bibitem{Aad:2015mbv}
{\bfseries ATLAS} Collaboration, G.~Aad {\em et~al.}, ``{Measurements of
  top-quark pair differential cross-sections in the lepton+jets channel in $pp$
  collisions at $\sqrt{s}=8$ TeV using the ATLAS detector},''
  \href{http://dx.doi.org/10.1140/epjc/s10052-016-4366-4}{{\em Eur. Phys. J.}
  {\bfseries C76} no.~10, (2016) 538},
\href{http://arxiv.org/abs/1511.04716}{{\ttfamily arXiv:1511.04716 [hep-ex]}}.

\bibitem{Khachatryan:2015oqa}
{\bfseries CMS} Collaboration, V.~Khachatryan {\em et~al.}, ``{Measurement of
  the differential cross section for top quark pair production in pp collisions
  at $\sqrt{s} = 8\,\text {TeV} $},''
  \href{http://dx.doi.org/10.1140/epjc/s10052-015-3709-x}{{\em Eur. Phys. J.}
  {\bfseries C75} no.~11, (2015) 542},
\href{http://arxiv.org/abs/1505.04480}{{\ttfamily arXiv:1505.04480 [hep-ex]}}.

\bibitem{Sirunyan:2017azo}
{\bfseries CMS} Collaboration, A.~M. Sirunyan {\em et~al.}, ``{Measurement of
  double-differential cross sections for top quark pair production in pp
  collisions at $\sqrt{s} = 8$ $\,\text {TeV}$ and impact on parton
  distribution functions},''
  \href{http://dx.doi.org/10.1140/epjc/s10052-017-4984-5}{{\em Eur. Phys. J.}
  {\bfseries C77} no.~7, (2017) 459},
\href{http://arxiv.org/abs/1703.01630}{{\ttfamily arXiv:1703.01630 [hep-ex]}}.

\bibitem{Aaboud:2016iot}
{\bfseries ATLAS} Collaboration, M.~Aaboud {\em et~al.}, ``{Measurement of top
  quark pair differential cross-sections in the dilepton channel in $pp$
  collisions at $\sqrt{s}$ = 7 and 8 TeV with ATLAS},''
  \href{http://dx.doi.org/10.1103/PhysRevD.94.092003}{{\em Phys. Rev. D}
  {\bfseries 94} no.~9, (2016) 092003},
  \href{http://arxiv.org/abs/1607.07281}{{\ttfamily arXiv:1607.07281
  [hep-ex]}}. [Addendum: Phys.Rev.D 101, 119901 (2020)].

\bibitem{Khachatryan:2016mnb}
{\bfseries CMS} Collaboration, V.~Khachatryan {\em et~al.}, ``{Measurement of
  differential cross sections for top quark pair production using the
  lepton+jets final state in proton-proton collisions at 13 TeV},''
  \href{http://dx.doi.org/10.1103/PhysRevD.95.092001}{{\em Phys. Rev.}
  {\bfseries D95} no.~9, (2017) 092001},
\href{http://arxiv.org/abs/1610.04191}{{\ttfamily arXiv:1610.04191 [hep-ex]}}.

\bibitem{Sirunyan:2017mzl}
{\bfseries CMS} Collaboration, A.~M. Sirunyan {\em et~al.}, ``{Measurement of
  normalized differential $ \mathrm{t}\overline{\mathrm{t}} $ cross sections in
  the dilepton channel from pp collisions at $ \sqrt{s}=13 $ TeV},''
  \href{http://dx.doi.org/10.1007/JHEP04(2018)060}{{\em JHEP} {\bfseries 04}
  (2018) 060},
\href{http://arxiv.org/abs/1708.07638}{{\ttfamily arXiv:1708.07638 [hep-ex]}}.

\bibitem{Sirunyan:2018wem}
{\bfseries CMS} Collaboration, A.~M. Sirunyan {\em et~al.}, ``{Measurement of
  differential cross sections for the production of top quark pairs and of
  additional jets in lepton+jets events from pp collisions at $\sqrt{s} =$ 13
  TeV},'' \href{http://dx.doi.org/10.1103/PhysRevD.97.112003}{{\em Phys. Rev.}
  {\bfseries D97} no.~11, (2018) 112003},
\href{http://arxiv.org/abs/1803.08856}{{\ttfamily arXiv:1803.08856 [hep-ex]}}.

\bibitem{Sirunyan:2018ucr}
{\bfseries CMS} Collaboration, A.~M. Sirunyan {\em et~al.}, ``{Measurements of
  $\mathrm{t\overline{t}}$ differential cross sections in proton-proton
  collisions at $\sqrt{s}=$ 13 TeV using events containing two leptons},''
  \href{http://dx.doi.org/10.1007/JHEP02(2019)149}{{\em JHEP} {\bfseries 02}
  (2019) 149}, \href{http://arxiv.org/abs/1811.06625}{{\ttfamily
  arXiv:1811.06625 [hep-ex]}}.

\bibitem{Aad:2019ntk}
{\bfseries ATLAS} Collaboration, G.~Aad {\em et~al.}, ``{Measurements of
  top-quark pair differential and double-differential cross-sections in the
  $\ell$+jets channel with $pp$ collisions at $\sqrt{s}=13$ TeV using the ATLAS
  detector},'' \href{http://dx.doi.org/10.1140/epjc/s10052-019-7525-6}{{\em
  Eur. Phys. J. C} {\bfseries 79} no.~12, (2019) 1028},
  \href{http://arxiv.org/abs/1908.07305}{{\ttfamily arXiv:1908.07305
  [hep-ex]}}.

\bibitem{Aaboud:2016hsq}
{\bfseries ATLAS} Collaboration, M.~Aaboud {\em et~al.}, ``{Measurement of the
  W boson polarisation in $t\bar{t}$ events from pp collisions at $\sqrt{s}$ =
  8 TeV in the lepton + jets channel with ATLAS},''
  \href{http://dx.doi.org/10.1140/epjc/s10052-017-4819-4}{{\em Eur. Phys. J.}
  {\bfseries C77} no.~4, (2017) 264},
\href{http://arxiv.org/abs/1612.02577}{{\ttfamily arXiv:1612.02577 [hep-ex]}}.

\bibitem{Khachatryan:2016fky}
{\bfseries CMS} Collaboration, V.~Khachatryan {\em et~al.}, ``{Measurement of
  the W boson helicity fractions in the decays of top quark pairs to lepton $+$
  jets final states produced in pp collisions at $\sqrt s=$ 8TeV},''
  \href{http://dx.doi.org/10.1016/j.physletb.2016.10.007}{{\em Phys. Lett.}
  {\bfseries B762} (2016) 512--534},
\href{http://arxiv.org/abs/1605.09047}{{\ttfamily arXiv:1605.09047 [hep-ex]}}.

\bibitem{Sirunyan:2017lvd}
{\bfseries ATLAS, CMS} Collaboration, M.~Aaboud {\em et~al.}, ``{Combination of
  inclusive and differential $ \mathrm{t}\overline{\mathrm{t}} $ charge
  asymmetry measurements using ATLAS and CMS data at $ \sqrt{s}=7 $ and 8
  TeV},'' \href{http://dx.doi.org/10.1007/JHEP04(2018)033}{{\em JHEP}
  {\bfseries 04} (2018) 033},
\href{http://arxiv.org/abs/1709.05327}{{\ttfamily arXiv:1709.05327 [hep-ex]}}.

\bibitem{ATLAS:2019czt}
{\bfseries ATLAS} Collaboration, ``{Inclusive and differential measurement of
  the charge asymmetry in $t\bar{t}$ events at 13 TeV with the ATLAS
  detector},''.

\bibitem{Aad:2020tmz}
{\bfseries ATLAS} Collaboration, G.~Aad {\em et~al.}, ``{Measurement of the
  $t\bar{t}$ production cross-section in the lepton+jets channel at
  $\sqrt{s}=13\;$TeV with the ATLAS experiment},''
  \href{http://dx.doi.org/10.1016/j.physletb.2020.135797}{{\em Phys. Lett. B}
  {\bfseries 810} (2020) 135797},
  \href{http://arxiv.org/abs/2006.13076}{{\ttfamily arXiv:2006.13076
  [hep-ex]}}.

\bibitem{Sirunyan:2017snr}
{\bfseries CMS} Collaboration, A.~M. Sirunyan {\em et~al.}, ``{Measurements of
  $t\bar{t}$ cross sections in association with $b$ jets and inclusive jets and
  their ratio using dilepton final states in pp collisions at $\sqrt{s}$ = 13
  TeV},'' \href{http://dx.doi.org/10.1016/j.physletb.2017.11.043}{{\em Phys.
  Lett.} {\bfseries B776} (2018) 355--378},
\href{http://arxiv.org/abs/1705.10141}{{\ttfamily arXiv:1705.10141 [hep-ex]}}.

\bibitem{Sirunyan:2019jud}
{\bfseries CMS} Collaboration, A.~M. Sirunyan {\em et~al.}, ``{Measurement of
  the $\mathrm{t\bar{t}}\mathrm{b\bar{b}}$ production cross section in the
  all-jet final state in pp collisions at $\sqrt{s} =$ 13 TeV},''
  \href{http://dx.doi.org/10.1016/j.physletb.2020.135285}{{\em Phys. Lett. B}
  {\bfseries 803} (2020) 135285},
  \href{http://arxiv.org/abs/1909.05306}{{\ttfamily arXiv:1909.05306
  [hep-ex]}}.

\bibitem{Aaboud:2018eki}
{\bfseries ATLAS} Collaboration, M.~Aaboud {\em et~al.}, ``{Measurements of
  inclusive and differential fiducial cross-sections of $ t\overline{t} $
  production with additional heavy-flavour jets in proton-proton collisions at
  $ \sqrt{s} $ = 13 TeV with the ATLAS detector},''
  \href{http://dx.doi.org/10.1007/JHEP04(2019)046}{{\em JHEP} {\bfseries 04}
  (2019) 046}, \href{http://arxiv.org/abs/1811.12113}{{\ttfamily
  arXiv:1811.12113 [hep-ex]}}.

\bibitem{Sirunyan:2017roi}
{\bfseries CMS} Collaboration, A.~M. Sirunyan {\em et~al.}, ``{Search for
  standard model production of four top quarks with same-sign and multilepton
  final states in proton--proton collisions at $\sqrt{s} = 13\,\text {TeV}
  $},'' \href{http://dx.doi.org/10.1140/epjc/s10052-018-5607-5}{{\em Eur. Phys.
  J.} {\bfseries C78} no.~2, (2018) 140},
\href{http://arxiv.org/abs/1710.10614}{{\ttfamily arXiv:1710.10614 [hep-ex]}}.

\bibitem{Sirunyan:2019wxt}
{\bfseries CMS} Collaboration, A.~M. Sirunyan {\em et~al.}, ``{Search for
  production of four top quarks in final states with same-sign or multiple
  leptons in proton-proton collisions at $\sqrt{s}=$ 13 TeV},''
  \href{http://dx.doi.org/10.1140/epjc/s10052-019-7593-7}{{\em Eur. Phys. J. C}
  {\bfseries 80} no.~2, (2020) 75},
  \href{http://arxiv.org/abs/1908.06463}{{\ttfamily arXiv:1908.06463
  [hep-ex]}}.

\bibitem{Aad:2020klt}
{\bfseries ATLAS} Collaboration, G.~Aad {\em et~al.}, ``{Evidence for
  $t\bar{t}t\bar{t}$ production in the multilepton final state in proton-proton
  collisions at $\sqrt{s}$=13 TeV with the ATLAS detector},''
  \href{http://arxiv.org/abs/2007.14858}{{\ttfamily arXiv:2007.14858
  [hep-ex]}}.

\bibitem{Khachatryan:2015sha}
{\bfseries CMS} Collaboration, V.~Khachatryan {\em et~al.}, ``{Observation of
  top quark pairs produced in association with a vector boson in pp collisions
  at sqrt(s) = 8 TeV},''
\href{http://arxiv.org/abs/1510.01131}{{\ttfamily arXiv:1510.01131 [hep-ex]}}.

\bibitem{Sirunyan:2017uzs}
{\bfseries CMS} Collaboration, A.~M. Sirunyan {\em et~al.}, ``{Measurement of
  the cross section for top quark pair production in association with a W or Z
  boson in proton-proton collisions at $\sqrt{s} =$ 13 TeV},''
  \href{http://dx.doi.org/10.1007/JHEP08(2018)011}{{\em JHEP} {\bfseries 08}
  (2018) 011},
\href{http://arxiv.org/abs/1711.02547}{{\ttfamily arXiv:1711.02547 [hep-ex]}}.

\bibitem{CMS:2019too}
{\bfseries CMS} Collaboration, A.~M. Sirunyan {\em et~al.}, ``{Measurement of
  top quark pair production in association with a Z boson in proton-proton
  collisions at $\sqrt{s}=$ 13 TeV},''
  \href{http://dx.doi.org/10.1007/JHEP03(2020)056}{{\em JHEP} {\bfseries 03}
  (2020) 056}, \href{http://arxiv.org/abs/1907.11270}{{\ttfamily
  arXiv:1907.11270 [hep-ex]}}.

\bibitem{Aad:2015eua}
{\bfseries ATLAS} Collaboration, G.~Aad {\em et~al.}, ``{Measurement of the
  $t\bar{t}W$ and $t\bar{t}Z$ production cross sections in $pp$ collisions at
  $\sqrt{s}= 8$TeV with the ATLAS detector},''
\href{http://arxiv.org/abs/1509.05276}{{\ttfamily arXiv:1509.05276 [hep-ex]}}.

\bibitem{Aaboud:2016xve}
{\bfseries ATLAS} Collaboration, M.~Aaboud {\em et~al.}, ``{Measurement of the
  $t\bar{t}Z$ and $t\bar{t}W$ production cross sections in multilepton final
  states using 3.2 fb$^{-1}$ of $pp$ collisions at $\sqrt{s}$ = 13 TeV with the
  ATLAS detector},''
  \href{http://dx.doi.org/10.1140/epjc/s10052-016-4574-y}{{\em Eur. Phys. J.}
  {\bfseries C77} no.~1, (2017) 40},
\href{http://arxiv.org/abs/1609.01599}{{\ttfamily arXiv:1609.01599 [hep-ex]}}.

\bibitem{Aaboud:2019njj}
{\bfseries ATLAS} Collaboration, M.~Aaboud {\em et~al.}, ``{Measurement of the
  $t\bar{t}Z$ and $t\bar{t}W$ cross sections in proton-proton collisions at
  $\sqrt{s}=13$ TeV with the ATLAS detector},''
  \href{http://dx.doi.org/10.1103/PhysRevD.99.072009}{{\em Phys. Rev. D}
  {\bfseries 99} no.~7, (2019) 072009},
  \href{http://arxiv.org/abs/1901.03584}{{\ttfamily arXiv:1901.03584
  [hep-ex]}}.

\bibitem{ATLAS-CONF-2020-028}
{\bfseries ATLAS Collaboration} Collaboration, ``{Measurements of the inclusive
  and differential production cross sections of a top-quark-antiquark pair in
  association with a $Z$ boson at $\sqrt{s} = 13$ TeV with the ATLAS
  detector},'' Tech. Rep. ATLAS-CONF-2020-028, CERN, Geneva, Aug, 2020.
\newblock \url{https://cds.cern.ch/record/2725734}.

\bibitem{Khachatryan:2014iya}
{\bfseries CMS} Collaboration, V.~Khachatryan {\em et~al.}, ``{Measurement of
  the t-channel single-top-quark production cross section and of the $\mid
  V_{tb} \mid$ CKM matrix element in pp collisions at $\sqrt{s}$= 8 TeV},''
  \href{http://dx.doi.org/10.1007/JHEP06(2014)090}{{\em JHEP} {\bfseries 06}
  (2014) 090},
\href{http://arxiv.org/abs/1403.7366}{{\ttfamily arXiv:1403.7366 [hep-ex]}}.

\bibitem{Aaboud:2017pdi}
{\bfseries ATLAS} Collaboration, M.~Aaboud {\em et~al.}, ``{Fiducial, total and
  differential cross-section measurements of $t$-channel single top-quark
  production in $pp$ collisions at 8 TeV using data collected by the ATLAS
  detector},'' \href{http://dx.doi.org/10.1140/epjc/s10052-017-5061-9}{{\em
  Eur. Phys. J.} {\bfseries C77} no.~8, (2017) 531},
\href{http://arxiv.org/abs/1702.02859}{{\ttfamily arXiv:1702.02859 [hep-ex]}}.

\bibitem{CMS-PAS-TOP-14-004}
{\bfseries CMS Collaboration} Collaboration, A.~M. Sirunyan {\em et~al.},
  ``{Single top t-channel differential cross section at 8 TeV},'' {\em
  CMS-PAS-TOP-14-004} (2014) \url{https://cds.cern.ch/record/1956681}.
  \url{https://cds.cern.ch/record/1956681}.

\bibitem{Khachatryan:2016ewo}
{\bfseries CMS} Collaboration, V.~Khachatryan {\em et~al.}, ``{Search for s
  channel single top quark production in pp collisions at $ \sqrt{s}=7 $ and 8
  TeV},'' \href{http://dx.doi.org/10.1007/JHEP09(2016)027}{{\em JHEP}
  {\bfseries 09} (2016) 027},
\href{http://arxiv.org/abs/1603.02555}{{\ttfamily arXiv:1603.02555 [hep-ex]}}.

\bibitem{Aad:2015upn}
{\bfseries ATLAS} Collaboration, G.~Aad {\em et~al.}, ``{Evidence for single
  top-quark production in the $s$-channel in proton-proton collisions at
  $\sqrt{s}=$8 TeV with the ATLAS detector using the Matrix Element Method},''
  \href{http://dx.doi.org/10.1016/j.physletb.2016.03.017}{{\em Phys. Lett.}
  {\bfseries B756} (2016) 228--246},
\href{http://arxiv.org/abs/1511.05980}{{\ttfamily arXiv:1511.05980 [hep-ex]}}.

\bibitem{Aaboud:2016ymp}
{\bfseries ATLAS} Collaboration, M.~Aaboud {\em et~al.}, ``{Measurement of the
  inclusive cross-sections of single top-quark and top-antiquark $t$-channel
  production in $pp$ collisions at $\sqrt{s}$ = 13 TeV with the ATLAS
  detector},'' \href{http://dx.doi.org/10.1007/JHEP04(2017)086}{{\em JHEP}
  {\bfseries 04} (2017) 086},
\href{http://arxiv.org/abs/1609.03920}{{\ttfamily arXiv:1609.03920 [hep-ex]}}.

\bibitem{Sirunyan:2016cdg}
{\bfseries CMS} Collaboration, A.~M. Sirunyan {\em et~al.}, ``{Cross section
  measurement of $t$-channel single top quark production in pp collisions at
  $\sqrt s =$ 13 TeV},''
  \href{http://dx.doi.org/10.1016/j.physletb.2017.07.047}{{\em Phys. Lett.}
  {\bfseries B772} (2017) 752--776},
\href{http://arxiv.org/abs/1610.00678}{{\ttfamily arXiv:1610.00678 [hep-ex]}}.

\bibitem{CMS:2016xnv}
{\bfseries CMS} Collaboration, V.~Khachatryan {\em et~al.}, ``{Measurement of
  the differential cross section for $t$-channel single-top-quark production at
  $\sqrt{s}=13~\mathrm{TeV}$},'' {\em CMS-PAS-TOP-16-004} (2016)
  \url{https://cds.cern.ch/record/2151074}.

\bibitem{Sirunyan:2019hqb}
{\bfseries CMS} Collaboration, A.~M. Sirunyan {\em et~al.}, ``{Measurement of
  differential cross sections and charge ratios for t-channel single top quark
  production in proton\textendash{}proton collisions at $\sqrt{s}=13\,\text
  {Te}\text {V}$},''
  \href{http://dx.doi.org/10.1140/epjc/s10052-020-7858-1}{{\em Eur. Phys. J. C}
  {\bfseries 80} no.~5, (2020) 370},
  \href{http://arxiv.org/abs/1907.08330}{{\ttfamily arXiv:1907.08330
  [hep-ex]}}.

\bibitem{Nocera:2019wyk}
E.~R. Nocera, M.~Ubiali, and C.~Voisey, ``{Single Top Production in PDF
  fits},'' \href{http://dx.doi.org/10.1007/JHEP05(2020)067}{{\em JHEP}
  {\bfseries 05} (2020) 067}, \href{http://arxiv.org/abs/1912.09543}{{\ttfamily
  arXiv:1912.09543 [hep-ph]}}.

\bibitem{Berger:2016oht}
E.~L. Berger, J.~Gao, C.~P. Yuan, and H.~X. Zhu, ``{NNLO QCD Corrections to
  t-channel Single Top-Quark Production and Decay},''
  \href{http://dx.doi.org/10.1103/PhysRevD.94.071501}{{\em Phys. Rev.}
  {\bfseries D94} no.~7, (2016) 071501},
\href{http://arxiv.org/abs/1606.08463}{{\ttfamily arXiv:1606.08463 [hep-ph]}}.

\bibitem{Aad:2015eto}
{\bfseries ATLAS} Collaboration, G.~Aad {\em et~al.}, ``{Measurement of the
  production cross-section of a single top quark in association with a $W$
  boson at 8 TeV with the ATLAS experiment},''
  \href{http://dx.doi.org/10.1007/JHEP01(2016)064}{{\em JHEP} {\bfseries 01}
  (2016) 064},
\href{http://arxiv.org/abs/1510.03752}{{\ttfamily arXiv:1510.03752 [hep-ex]}}.

\bibitem{Aad:2020zhd}
{\bfseries ATLAS} Collaboration, G.~Aad {\em et~al.}, ``{Measurement of single
  top-quark production in association with a $W$ boson in the single-lepton
  channel at $\sqrt{s} = 8$ TeV with the ATLAS detector},''
  \href{http://arxiv.org/abs/2007.01554}{{\ttfamily arXiv:2007.01554
  [hep-ex]}}.

\bibitem{Chatrchyan:2014tua}
{\bfseries CMS} Collaboration, S.~Chatrchyan {\em et~al.}, ``{Observation of
  the associated production of a single top quark and a $W$ boson in $pp$
  collisions at $\sqrt s = $8 TeV},''
  \href{http://dx.doi.org/10.1103/PhysRevLett.112.231802}{{\em Phys. Rev.
  Lett.} {\bfseries 112} no.~23, (2014) 231802},
\href{http://arxiv.org/abs/1401.2942}{{\ttfamily arXiv:1401.2942 [hep-ex]}}.

\bibitem{Aaboud:2016lpj}
{\bfseries ATLAS} Collaboration, M.~Aaboud {\em et~al.}, ``{Measurement of the
  cross-section for producing a W boson in association with a single top quark
  in pp collisions at $ \sqrt{s}=13 $ TeV with ATLAS},''
  \href{http://dx.doi.org/10.1007/JHEP01(2018)063}{{\em JHEP} {\bfseries 01}
  (2018) 063},
\href{http://arxiv.org/abs/1612.07231}{{\ttfamily arXiv:1612.07231 [hep-ex]}}.

\bibitem{Sirunyan:2018lcp}
{\bfseries CMS} Collaboration, A.~M. Sirunyan {\em et~al.}, ``{Measurement of
  the production cross section for single top quarks in association with W
  bosons in proton-proton collisions at $\sqrt{s}=$ 13 TeV},''
  \href{http://dx.doi.org/10.3204/PUBDB-2018-02197}{{\em JHEP} (2018) },
\href{http://arxiv.org/abs/1805.07399}{{\ttfamily arXiv:1805.07399 [hep-ex]}}.

\bibitem{Aad:2020wog}
{\bfseries ATLAS} Collaboration, G.~Aad {\em et~al.}, ``{Observation of the
  associated production of a top quark and a $Z$ boson in $pp$ collisions at
  $\sqrt{s} = 13$ TeV with the ATLAS detector},''
  \href{http://dx.doi.org/10.1007/JHEP07(2020)124}{{\em JHEP} {\bfseries 07}
  (2020) 124}, \href{http://arxiv.org/abs/2002.07546}{{\ttfamily
  arXiv:2002.07546 [hep-ex]}}.

\bibitem{Khachatryan:2016vau}
{\bfseries ATLAS, CMS} Collaboration, G.~Aad {\em et~al.}, ``{Measurements of
  the Higgs boson production and decay rates and constraints on its couplings
  from a combined ATLAS and CMS analysis of the LHC pp collision data at $
  \sqrt{s}=7 $ and 8 TeV},''
  \href{http://dx.doi.org/10.1007/JHEP08(2016)045}{{\em JHEP} {\bfseries 08}
  (2016) 045},
\href{http://arxiv.org/abs/1606.02266}{{\ttfamily arXiv:1606.02266 [hep-ex]}}.

\bibitem{Aad:2015gba}
{\bfseries ATLAS} Collaboration, G.~Aad {\em et~al.}, ``{Measurements of the
  Higgs boson production and decay rates and coupling strengths using pp
  collision data at $\sqrt{s}=7$ and 8 TeV in the ATLAS experiment},''
  \href{http://dx.doi.org/10.1140/epjc/s10052-015-3769-y}{{\em Eur. Phys. J.}
  {\bfseries C76} no.~1, (2016) 6},
\href{http://arxiv.org/abs/1507.04548}{{\ttfamily arXiv:1507.04548 [hep-ex]}}.

\bibitem{Aad:2019mbh}
{\bfseries ATLAS} Collaboration, G.~Aad {\em et~al.}, ``{Combined measurements
  of Higgs boson production and decay using up to $80$ fb$^{-1}$ of
  proton-proton collision data at $\sqrt{s}=$ 13 TeV collected with the ATLAS
  experiment},'' \href{http://dx.doi.org/10.1103/PhysRevD.101.012002}{{\em
  Phys. Rev. D} {\bfseries 101} no.~1, (2020) 012002},
  \href{http://arxiv.org/abs/1909.02845}{{\ttfamily arXiv:1909.02845
  [hep-ex]}}.

\bibitem{Sirunyan:2018koj}
{\bfseries CMS} Collaboration, A.~M. Sirunyan {\em et~al.}, ``{Combined
  measurements of Higgs boson couplings in proton--proton collisions at
  $\sqrt{s}=13\,\text {Te}\text {V} $},''
  \href{http://dx.doi.org/10.1140/epjc/s10052-019-6909-y}{{\em Eur. Phys. J. C}
  {\bfseries 79} no.~5, (2019) 421},
  \href{http://arxiv.org/abs/1809.10733}{{\ttfamily arXiv:1809.10733
  [hep-ex]}}.

\bibitem{deFlorian:2016spz}
{\bfseries LHC Higgs Cross Section Working Group} Collaboration, D.~de~Florian
  {\em et~al.}, ``{Handbook of LHC Higgs Cross Sections: 4. Deciphering the
  Nature of the Higgs Sector},''
\href{http://arxiv.org/abs/1610.07922}{{\ttfamily arXiv:1610.07922 [hep-ph]}}.

\bibitem{Dittmaier:2012vm}
{\bfseries LHC Higgs Cross Section Working Group} Collaboration, S.~Dittmaier
  {\em et~al.}, ``{Handbook of LHC Higgs Cross Sections: 2. Differential
  Distributions},''
\href{http://arxiv.org/abs/1201.3084}{{\ttfamily arXiv:1201.3084 [hep-ph]}}.

\bibitem{Sirunyan:2018sgc}
{\bfseries CMS} Collaboration, A.~M. Sirunyan {\em et~al.}, ``{Measurement and
  interpretation of differential cross sections for Higgs boson production at
  $\sqrt{s} =$ 13 TeV},''
  \href{http://dx.doi.org/10.1016/j.physletb.2019.03.059}{{\em Phys. Lett. B}
  {\bfseries 792} (2019) 369--396},
  \href{http://arxiv.org/abs/1812.06504}{{\ttfamily arXiv:1812.06504
  [hep-ex]}}.

\bibitem{Aaboud:2018ezd}
{\bfseries ATLAS} Collaboration, M.~Aaboud {\em et~al.}, ``{Combined
  measurement of differential and total cross sections in the $H \rightarrow
  \gamma \gamma$ and the $H \rightarrow ZZ^* \rightarrow 4\ell$ decay channels
  at $\sqrt{s} = 13$ TeV with the ATLAS detector},''
  \href{http://dx.doi.org/10.1016/j.physletb.2018.09.019}{{\em Phys. Lett. B}
  {\bfseries 786} (2018) 114--133},
  \href{http://arxiv.org/abs/1805.10197}{{\ttfamily arXiv:1805.10197
  [hep-ex]}}.

\bibitem{Aaboud:2019nan}
{\bfseries ATLAS} Collaboration, M.~Aaboud {\em et~al.}, ``{Measurement of VH,
  $ \mathrm{H}\to \mathrm{b}\overline{\mathrm{b}} $ production as a function of
  the vector-boson transverse momentum in 13 TeV pp collisions with the ATLAS
  detector},'' \href{http://dx.doi.org/10.1007/JHEP05(2019)141}{{\em JHEP}
  {\bfseries 05} (2019) 141},
\href{http://arxiv.org/abs/1903.04618}{{\ttfamily arXiv:1903.04618 [hep-ex]}}.

\bibitem{CMS:1900lgv}
{\bfseries CMS} Collaboration, ``{Measurements of Higgs boson production via
  gluon fusion and vector boson fusion in the diphoton decay channel at
  $\sqrt{s} = 13$ TeV},''.

\bibitem{CMS:2019pyn}
{\bfseries CMS} Collaboration, ``{Measurement of Higgs boson production and
  decay to the $\tau\tau$ final state},''.

\bibitem{ATLAS:2020wny}
{\bfseries ATLAS} Collaboration, G.~Aad {\em et~al.}, ``{Measurements of the
  Higgs boson inclusive and differential fiducial cross sections in the 4$\ell$
  decay channel at $\sqrt{s}$ = 13 TeV},''
  \href{http://arxiv.org/abs/2004.03969}{{\ttfamily arXiv:2004.03969
  [hep-ex]}}.

\bibitem{ATLAS:2019ssu}
{\bfseries ATLAS} Collaboration, ``{Measurements of the Higgs boson inclusive,
  differential and production cross sections in the 4$\ell$ decay channel at
  $\sqrt{s}$ = 13 TeV with the ATLAS detector},''.

\bibitem{Aad:2020jym}
{\bfseries ATLAS} Collaboration, G.~Aad {\em et~al.}, ``{Measurements of $WH$
  and $ZH$ production in the $H \rightarrow b\bar{b}$ decay channel in $pp$
  collisions at 13 TeV with the ATLAS detector},''
  \href{http://arxiv.org/abs/2007.02873}{{\ttfamily arXiv:2007.02873
  [hep-ex]}}.

\bibitem{Sirunyan:2020tzo}
{\bfseries CMS} Collaboration, A.~M. Sirunyan {\em et~al.}, ``{Measurement of
  the inclusive and differential Higgs boson production cross sections in the
  leptonic WW decay mode at $\sqrt{s} =$ 13 TeV},''
  \href{http://arxiv.org/abs/2007.01984}{{\ttfamily arXiv:2007.01984
  [hep-ex]}}.

\bibitem{Schael:2013ita}
{\bfseries ALEPH, DELPHI, L3, OPAL, LEP Electroweak} Collaboration, S.~Schael
  {\em et~al.}, ``{Electroweak Measurements in Electron-Positron Collisions at
  W-Boson-Pair Energies at LEP},''
  \href{http://dx.doi.org/10.1016/j.physrep.2013.07.004}{{\em Phys. Rept.}
  {\bfseries 532} (2013) 119--244},
\href{http://arxiv.org/abs/1302.3415}{{\ttfamily arXiv:1302.3415 [hep-ex]}}.

\bibitem{Aaboud:2019gxl}
{\bfseries ATLAS} Collaboration, M.~Aaboud {\em et~al.}, ``{Measurement of
  $W^{\pm}Z$ production cross sections and gauge boson polarisation in $pp$
  collisions at $\sqrt{s} = 13$ TeV with the ATLAS detector},''
  \href{http://dx.doi.org/10.1140/epjc/s10052-019-7027-6}{{\em Eur. Phys. J. C}
  {\bfseries 79} no.~6, (2019) 535},
  \href{http://arxiv.org/abs/1902.05759}{{\ttfamily arXiv:1902.05759
  [hep-ex]}}.

\bibitem{Aaboud:2019nkz}
{\bfseries ATLAS} Collaboration, M.~Aaboud {\em et~al.}, ``{Measurement of
  fiducial and differential $W^+W^-$ production cross-sections at $\sqrt{s}=13$
  TeV with the ATLAS detector},''
  \href{http://dx.doi.org/10.1140/epjc/s10052-019-7371-6}{{\em Eur. Phys. J. C}
  {\bfseries 79} no.~10, (2019) 884},
  \href{http://arxiv.org/abs/1905.04242}{{\ttfamily arXiv:1905.04242
  [hep-ex]}}.

\bibitem{Sirunyan:2019bez}
{\bfseries CMS} Collaboration, A.~M. Sirunyan {\em et~al.}, ``{Measurements of
  the pp $\to$ WZ inclusive and differential production cross section and
  constraints on charged anomalous triple gauge couplings at $\sqrt{s} =$ 13
  TeV},'' \href{http://dx.doi.org/10.1007/JHEP04(2019)122}{{\em JHEP}
  {\bfseries 04} (2019) 122}, \href{http://arxiv.org/abs/1901.03428}{{\ttfamily
  arXiv:1901.03428 [hep-ex]}}.

\bibitem{ATLAS-CONF-2018-034}
{\bfseries ATLAS Collaboration} Collaboration, ``{Measurement of $W^{\pm}Z$
  production cross sections and gauge boson polarisation in $pp$ collisions at
  $\sqrt{s} = 13$ TeV with the ATLAS detector},'' Tech. Rep.
  ATLAS-CONF-2018-034, CERN, Geneva, Jul, 2018.
\newblock \url{https://cds.cern.ch/record/2630187}.

\bibitem{Grazzini:2017mhc}
M.~Grazzini, S.~Kallweit, and M.~Wiesemann, ``{Fully differential NNLO
  computations with MATRIX},''
  \href{http://dx.doi.org/10.1140/epjc/s10052-018-5771-7}{{\em Eur. Phys. J. C}
  {\bfseries 78} no.~7, (2018) 537},
  \href{http://arxiv.org/abs/1711.06631}{{\ttfamily arXiv:1711.06631
  [hep-ph]}}.

\bibitem{Brehmer:2017lrt}
J.~Brehmer, F.~Kling, T.~Plehn, and T.~M.~P. Tait, ``{Better Higgs-CP Tests
  Through Information Geometry},''
  \href{http://dx.doi.org/10.1103/PhysRevD.97.095017}{{\em Phys. Rev.}
  {\bfseries D97} no.~9, (2018) 095017},
\href{http://arxiv.org/abs/1712.02350}{{\ttfamily arXiv:1712.02350 [hep-ph]}}.

\bibitem{Feroz:2007kg}
F.~Feroz and M.~Hobson, ``{Multimodal nested sampling: an efficient and robust
  alternative to MCMC methods for astronomical data analysis},''
  \href{http://dx.doi.org/10.1111/j.1365-2966.2007.12353.x}{{\em Mon. Not. Roy.
  Astron. Soc.} {\bfseries 384} (2008) 449},
  \href{http://arxiv.org/abs/0704.3704}{{\ttfamily arXiv:0704.3704
  [astro-ph]}}.

\bibitem{AbdulKhalek:2019ihb}
{\bfseries NNPDF} Collaboration, R.~Abdul~Khalek {\em et~al.}, ``{Parton
  Distributions with Theory Uncertainties: General Formalism and First
  Phenomenological Studies},''
  \href{http://dx.doi.org/10.1140/epjc/s10052-019-7401-4}{{\em Eur. Phys. J. C}
  {\bfseries 79} no.~11, (2019) 931},
  \href{http://arxiv.org/abs/1906.10698}{{\ttfamily arXiv:1906.10698
  [hep-ph]}}.

\bibitem{AbdulKhalek:2019bux}
{\bfseries NNPDF} Collaboration, R.~Abdul~Khalek {\em et~al.}, ``{A first
  determination of parton distributions with theoretical uncertainties},''
  \href{http://dx.doi.org/10.1140/epjc/s10052-019-7364-5}{{\em Eur. Phys. J.}
  {\bfseries C} (2019) 79:838},
  \href{http://arxiv.org/abs/1905.04311}{{\ttfamily arXiv:1905.04311
  [hep-ph]}}.

\bibitem{Ethier:2021ydt}
J.~J. Ethier, R.~Gomez-Ambrosio, G.~Magni, and J.~Rojo, ``{SMEFT analysis of
  vector boson scattering and diboson data from the LHC Run II},''
  \href{http://arxiv.org/abs/2101.03180}{{\ttfamily arXiv:2101.03180
  [hep-ph]}}.

\bibitem{Strategy:2019vxc}
R.~K. Ellis and others ({European Strategy for Particle Physics Preparatory
  Group}), ``{Physics Briefing Book}: {Input for the European Strategy for
  Particle Physics Update 2020},''
  \href{http://arxiv.org/abs/1910.11775}{{\ttfamily arXiv:1910.11775
  [hep-ex]}}.

\bibitem{EuropeanStrategyGroup:2020pow}
{\bfseries European Strategy Group} Collaboration,
  \href{http://dx.doi.org/10.17181/ESU2020}{{\em {2020 Update of the European
  Strategy for Particle Physics}}}.
\newblock CERN Council, Geneva, 2020.

\bibitem{Palmer:1996gs}
R.~Palmer {\em et~al.}, ``{Muon collider design},''
  \href{http://dx.doi.org/10.1016/0920-5632(96)00417-3}{{\em Nucl. Phys. Proc.
  Suppl.} {\bfseries 51A} (1996) 61--84},
\href{http://arxiv.org/abs/acc-phys/9604001}{{\ttfamily arXiv:acc-phys/9604001
  [acc-phys]}}.

\bibitem{Ankenbrandt:1999cta}
C.~M. Ankenbrandt {\em et~al.}, ``{Status of muon collider research and
  development and future plans},''
  \href{http://dx.doi.org/10.1103/PhysRevSTAB.2.081001}{{\em Phys. Rev. ST
  Accel. Beams} {\bfseries 2} (1999) 081001},
\href{http://arxiv.org/abs/physics/9901022}{{\ttfamily arXiv:physics/9901022
  [physics]}}.

\bibitem{Palmer:2014nza}
R.~B. Palmer, ``{Muon Colliders},''
\href{http://dx.doi.org/10.1142/S1793626814300072}{{\em Rev. Accel. Sci. Tech.}
  {\bfseries 7} (2014) 137--159}.

\bibitem{Antonelli:2013mmk}
M.~Antonelli and P.~Raimondi, ``{Snowmass Report: Ideas for Muon Production
  from Positron Beam Interaction on a Plasma Target},'' in {\em {Proceedings,
  2013 Community Summer Study on the Future of U.S. Particle Physics: Snowmass
  on the Mississippi (CSS2013): Minneapolis, MN, USA, July 29-August 6, 2013}}.
\newblock 2013.
\newblock
\url{http://www.lnf.infn.it/sis/preprint/detail-new.php?id=5331}.
\newblock

\bibitem{Antonelli:2015nla}
M.~Antonelli, M.~Boscolo, R.~Di~Nardo, and P.~Raimondi, ``{Novel proposal for a
  low emittance muon beam using positron beam on target},''
  \href{http://dx.doi.org/10.1016/j.nima.2015.10.097}{{\em Nucl. Instrum.
  Meth.} {\bfseries A807} (2016) 101--107},
\href{http://arxiv.org/abs/1509.04454}{{\ttfamily arXiv:1509.04454
  [physics.acc-ph]}}.

\bibitem{Barger:1995hr}
V.~D. Barger, M.~S. Berger, J.~F. Gunion, and T.~Han, ``{s channel Higgs boson
  production at a muon muon collider},''
  \href{http://dx.doi.org/10.1103/PhysRevLett.75.1462}{{\em Phys. Rev. Lett.}
  {\bfseries 75} (1995) 1462--1465},
\href{http://arxiv.org/abs/hep-ph/9504330}{{\ttfamily arXiv:hep-ph/9504330
  [hep-ph]}}.

\bibitem{Han:2012rb}
T.~Han and Z.~Liu, ``{Potential precision of a direct measurement of the Higgs
  boson total width at a muon collider},''
  \href{http://dx.doi.org/10.1103/PhysRevD.87.033007}{{\em Phys. Rev. D}
  {\bfseries 87} no.~3, (2013) 033007},
  \href{http://arxiv.org/abs/1210.7803}{{\ttfamily arXiv:1210.7803 [hep-ph]}}.

\bibitem{Chakrabarty:2014pja}
N.~Chakrabarty, T.~Han, Z.~Liu, and B.~Mukhopadhyaya, ``{Radiative Return for
  Heavy Higgs Boson at a Muon Collider},''
  \href{http://dx.doi.org/10.1103/PhysRevD.91.015008}{{\em Phys. Rev. D}
  {\bfseries 91} no.~1, (2015) 015008},
  \href{http://arxiv.org/abs/1408.5912}{{\ttfamily arXiv:1408.5912 [hep-ph]}}.

\bibitem{Greco:2016izi}
M.~Greco, T.~Han, and Z.~Liu, {\em {ISR effects for resonant Higgs production
  at future lepton colliders}}, vol.~763,
  \href{http://dx.doi.org/10.1016/j.physletb.2016.10.078}{pp.~409--415}.
\newblock 12, 2016.
\newblock \href{http://arxiv.org/abs/1607.03210}{{\ttfamily arXiv:1607.03210
  [hep-ph]}}.

\bibitem{Buttazzo:2018qqp}
D.~Buttazzo, D.~Redigolo, F.~Sala, and A.~Tesi, ``{Fusing Vectors into Scalars
  at High Energy Lepton Colliders},''
  \href{http://dx.doi.org/10.1007/JHEP11(2018)144}{{\em JHEP} {\bfseries 11}
  (2018) 144}, \href{http://arxiv.org/abs/1807.04743}{{\ttfamily
  arXiv:1807.04743 [hep-ph]}}.

\bibitem{Delahaye:2019omf}
J.~P. Delahaye, M.~Diemoz, K.~Long, B.~Mansouli{\'e}, N.~Pastrone, L.~Rivkin,
  D.~Schulte, A.~Skrinsky, and A.~Wulzer, ``{Muon Colliders},''
\href{http://arxiv.org/abs/1901.06150}{{\ttfamily arXiv:1901.06150
  [physics.acc-ph]}}.

\bibitem{Ruhdorfer:2019utl}
M.~Ruhdorfer, E.~Salvioni, and A.~Weiler, ``{A Global View of the Off-Shell
  Higgs Portal},'' \href{http://dx.doi.org/10.21468/SciPostPhys.8.2.027}{{\em
  SciPost Phys.} {\bfseries 8} (2020) 027},
  \href{http://arxiv.org/abs/1910.04170}{{\ttfamily arXiv:1910.04170
  [hep-ph]}}.

\bibitem{Dawson:1984ta}
S.~Dawson and J.~L. Rosner, ``{Capabilities of $e^+ e^-$ Collisions for
  Producing Very Heavy Higgs Bosons},''
  \href{http://dx.doi.org/10.1016/0370-2693(84)90746-9}{{\em Phys. Lett. B}
  {\bfseries 148} (1984) 497--501}.

\bibitem{Hikasa:1985ee}
K.-i. Hikasa, ``{Heavy Higgs Production in $e^+ e^-$ and $e^- e^-$
  Collisions},'' \href{http://dx.doi.org/10.1016/0370-2693(85)90346-6}{{\em
  Phys. Lett. B} {\bfseries 164} (1985) 385}. [Erratum: Phys.Lett.B 195, 623
  (1987)].

\bibitem{Altarelli:1987ue}
G.~Altarelli, B.~Mele, and F.~Pitolli, ``{Heavy Higgs Production at Future
  Colliders},''
\href{http://dx.doi.org/10.1016/0550-3213(87)90103-9}{{\em Nucl. Phys.}
  {\bfseries B287} (1987) 205--224}.

\bibitem{Kilian:1995tr}
W.~Kilian, M.~Kramer, and P.~Zerwas, ``{Higgsstrahlung and W W fusion in e+ e-
  collisions},'' \href{http://dx.doi.org/10.1016/0370-2693(96)00100-1}{{\em
  Phys. Lett. B} {\bfseries 373} (1996) 135--140},
  \href{http://arxiv.org/abs/hep-ph/9512355}{{\ttfamily arXiv:hep-ph/9512355}}.

\bibitem{Gunion:1998jc}
J.~Gunion, T.~Han, and R.~Sobey, ``{Measuring the coupling of a Higgs boson to
  Z Z at linear colliders},''
  \href{http://dx.doi.org/10.1016/S0370-2693(98)00450-X}{{\em Phys. Lett. B}
  {\bfseries 429} (1998) 79--86},
  \href{http://arxiv.org/abs/hep-ph/9801317}{{\ttfamily arXiv:hep-ph/9801317}}.

\bibitem{Quigg:2009gg}
C.~Quigg, ``{LHC Physics Potential versus Energy},''
\href{http://arxiv.org/abs/0908.3660}{{\ttfamily arXiv:0908.3660 [hep-ph]}}.

\bibitem{Denner:1999gp}
A.~Denner, S.~Dittmaier, M.~Roth, and D.~Wackeroth, ``{Predictions for all
  processes e+ e- 4 fermions + gamma},''
  \href{http://dx.doi.org/10.1016/S0550-3213(99)00437-X}{{\em Nucl.Phys.}
  {\bfseries B560} (1999) 33--65},
\href{http://arxiv.org/abs/hep-ph/9904472}{{\ttfamily arXiv:hep-ph/9904472
  [hep-ph]}}.

\bibitem{Denner:2005fg}
A.~Denner, S.~Dittmaier, M.~Roth, and L.~Wieders, ``{Electroweak corrections to
  charged-current e+ e- 4 fermion processes: Technical details and further
  results},'' \href{http://dx.doi.org/10.1016/j.nuclphysb.2011.09.001,
  10.1016/j.nuclphysb.2005.06.033}{{\em Nucl.Phys.} {\bfseries B724} (2005)
  247--294},
\href{http://arxiv.org/abs/hep-ph/0505042}{{\ttfamily arXiv:hep-ph/0505042
  [hep-ph]}}.

\bibitem{Chung:2012vg}
D.~J.~H. Chung, A.~J. Long, and L.-T. Wang, ``{125 GeV Higgs boson and
  electroweak phase transition model classes},''
  \href{http://dx.doi.org/10.1103/PhysRevD.87.023509}{{\em Phys. Rev.}
  {\bfseries D87} no.~2, (2013) 023509},
\href{http://arxiv.org/abs/1209.1819}{{\ttfamily arXiv:1209.1819 [hep-ph]}}.

\bibitem{Sirunyan:2017guj}
{\bfseries CMS} Collaboration, A.~M. Sirunyan {\em et~al.}, ``{Search for
  resonant and nonresonant Higgs boson pair production in the $
  \mathrm{b}\overline{\mathrm{b}}\mathit{\ell \nu \ell \nu } $ final state in
  proton-proton collisions at $ \sqrt{s}=13 $ TeV},''
  \href{http://dx.doi.org/10.1007/JHEP01(2018)054}{{\em JHEP} {\bfseries 01}
  (2018) 054},
\href{http://arxiv.org/abs/1708.04188}{{\ttfamily arXiv:1708.04188 [hep-ex]}}.

\bibitem{CMS:2018rig}
{\bfseries CMS} Collaboration, C.~Collaboration,
``{Constraints on the Higgs boson self-coupling from ttH+tH, H to gamma gamma
  differential measurements at the HL-LHC},''.

\bibitem{CMS:2018ccd}
{\bfseries CMS} Collaboration, C.~Collaboration,
``{Prospects for HH measurements at the HL-LHC},''.

\bibitem{ATLAS:2018otd}
{\bfseries ATLAS} Collaboration, T.~A. collaboration,
``{Combination of searches for Higgs boson pairs in $pp$ collisions at 13 TeV
  with the ATLAS experiment.},''.

\bibitem{Aaboud:2018zhh}
{\bfseries ATLAS} Collaboration, M.~Aaboud {\em et~al.}, ``{Search for Higgs
  boson pair production in the $b\bar{b}WW^{*}$ decay mode at $\sqrt{s}=13$ TeV
  with the ATLAS detector},''
  \href{http://dx.doi.org/10.1007/JHEP04(2019)092}{{\em JHEP} {\bfseries 04}
  (2019) 092},
\href{http://arxiv.org/abs/1811.04671}{{\ttfamily arXiv:1811.04671 [hep-ex]}}.

\bibitem{Aad:2019uzh}
{\bfseries ATLAS} Collaboration, G.~Aad {\em et~al.}, ``{Combination of
  searches for Higgs boson pairs in $pp$ collisions at $\sqrt{s} = $13 TeV with
  the ATLAS detector},''
  \href{http://dx.doi.org/10.1016/j.physletb.2019.135103}{{\em Phys. Lett.}
  {\bfseries B800} (2020) 135103},
\href{http://arxiv.org/abs/1906.02025}{{\ttfamily arXiv:1906.02025 [hep-ex]}}.

\bibitem{Aad:2019yxi}
{\bfseries ATLAS} Collaboration, G.~Aad {\em et~al.}, ``{Search for
  non-resonant Higgs boson pair production in the $bb\ell\nu\ell\nu$ final
  state with the ATLAS detector in $pp$ collisions at $\sqrt{s} = 13$ TeV},''
  \href{http://dx.doi.org/10.1016/j.physletb.2019.135145}{{\em Phys. Lett.}
  {\bfseries B801} (2020) 135145},
\href{http://arxiv.org/abs/1908.06765}{{\ttfamily arXiv:1908.06765 [hep-ex]}}.

\bibitem{Sirunyan:2018iwt}
{\bfseries CMS} Collaboration, A.~M. Sirunyan {\em et~al.}, ``{Search for Higgs
  boson pair production in the $\gamma\gamma\mathrm{b\overline{b}}$ final state
  in pp collisions at $\sqrt{s}=$ 13 TeV},''
  \href{http://dx.doi.org/10.1016/j.physletb.2018.10.056}{{\em Phys. Lett.}
  {\bfseries B788} (2019) 7--36},
\href{http://arxiv.org/abs/1806.00408}{{\ttfamily arXiv:1806.00408 [hep-ex]}}.

\bibitem{CMS:2018dvu}
{\bfseries CMS} Collaboration, C.~Collaboration,
``{Search for resonant double Higgs production with $bbZZ$ decays in the
  $b\bar{b} \ell\ell\nu\nu$ final state},''.

\bibitem{ATLAS:2019pbo}
{\bfseries ATLAS} Collaboration, T.~A. collaboration, ``{Constraints on the
  Higgs boson self-coupling from the combination of single-Higgs and
  double-Higgs production analyses performed with the ATLAS experiment},''
  CERN.
\newblock CERN, Geneva,
2019.
\newblock

\bibitem{ATL-PHYS-PUB-2014-019}
T.~A. collaboration,
``{Prospects for measuring Higgs pair production in the channel
  $H(\rightarrow\gamma\gamma)H(\rightarrow b\overline{b}) $ using the ATLAS
  detector at the HL-LHC},''.

\bibitem{ATL-PHYS-PUB-2017-001}
{\bfseries ATLAS} Collaboration, T.~A. collaboration, ``{Study of the double
  Higgs production channel $H(\rightarrow b\bar{b})H(\rightarrow \gamma\gamma)$
  with the ATLAS experiment at the HL-LHC},''
\newblock
2017.
\newblock

\bibitem{Kim:2018cxf}
J.~H. Kim, K.~Kong, K.~T. Matchev, and M.~Park, ``{Probing the Triple Higgs
  Self-Interaction at the Large Hadron Collider},''
  \href{http://dx.doi.org/10.1103/PhysRevLett.122.091801}{{\em Phys. Rev.
  Lett.} {\bfseries 122} no.~9, (2019) 091801},
\href{http://arxiv.org/abs/1807.11498}{{\ttfamily arXiv:1807.11498 [hep-ph]}}.

\bibitem{Roloff:2018dqu}
{\bfseries CLIC, CLICdp} Collaboration, P.~Roloff, R.~Franceschini, U.~Schnoor,
  and A.~Wulzer, ``{The Compact Linear e$^+$e$^-$ Collider (CLIC): Physics
  Potential},''
\href{http://arxiv.org/abs/1812.07986}{{\ttfamily arXiv:1812.07986 [hep-ex]}}.

\bibitem{Vasquez:2019muw}
A.~Vasquez, C.~Degrande, A.~Tonero, and R.~Rosenfeld, ``{New physics in double
  Higgs production at future e$^{+}$e$^{-}$ colliders},''
  \href{http://dx.doi.org/10.1007/JHEP05(2019)020}{{\em JHEP} {\bfseries 05}
  (2019) 020},
\href{http://arxiv.org/abs/1901.05979}{{\ttfamily arXiv:1901.05979 [hep-ph]}}.

\bibitem{Roloff:2019crr}
{\bfseries CLICdp} Collaboration, P.~Roloff, U.~Schnoor, R.~Simoniello, and
  B.~Xu, ``{Double Higgs boson production and Higgs self-coupling extraction at
  CLIC},''
\href{http://arxiv.org/abs/1901.05897}{{\ttfamily arXiv:1901.05897 [hep-ex]}}.

\bibitem{Liu:2018peg}
T.~Liu, K.-F. Lyu, J.~Ren, and H.~X. Zhu, ``{Probing the quartic Higgs boson
  self-interaction},'' \href{http://dx.doi.org/10.1103/PhysRevD.98.093004}{{\em
  Phys. Rev.} {\bfseries D98} no.~9, (2018) 093004},
\href{http://arxiv.org/abs/1803.04359}{{\ttfamily arXiv:1803.04359 [hep-ph]}}.

\bibitem{Maltoni:2018ttu}
F.~Maltoni, D.~Pagani, and X.~Zhao, ``{Constraining the Higgs self-couplings at
  e$^{+}$e$^{-}$ colliders},''
  \href{http://dx.doi.org/10.1007/JHEP07(2018)087}{{\em JHEP} {\bfseries 07}
  (2018) 087},
\href{http://arxiv.org/abs/1802.07616}{{\ttfamily arXiv:1802.07616 [hep-ph]}}.

\bibitem{deBlas:2019rxi}
J.~de~Blas {\em et~al.}, ``{Higgs Boson Studies at Future Particle
  Colliders},'' \href{http://dx.doi.org/10.1007/JHEP01(2020)139}{{\em JHEP}
  {\bfseries 01} (2020) 139},
\href{http://arxiv.org/abs/1905.03764}{{\ttfamily arXiv:1905.03764 [hep-ph]}}.

\bibitem{DiMicco:2019ngk}
J.~Alison {\em et~al.}, ``{Higgs boson pair production at colliders: status and
  perspectives},'' in {\em {Double Higgs Production at Colliders Batavia, IL,
  USA, September 4, 2018-9, 2019}}, B.~Di~Micco, M.~Gouzevitch, J.~Mazzitelli,
  and C.~Vernieri, eds.
\newblock 2019.
\newblock
\href{http://arxiv.org/abs/1910.00012}{{\ttfamily arXiv:1910.00012 [hep-ph]}}.
\newblock

\bibitem{Abada:2019lih}
{\bfseries FCC} Collaboration, A.~Abada {\em et~al.}, ``{FCC Physics
  Opportunities}: {Future Circular Collider Conceptual Design Report Volume
  1},'' \href{http://dx.doi.org/10.1140/epjc/s10052-019-6904-3}{{\em Eur. Phys.
  J. C} {\bfseries 79} no.~6, (2019) 474}.

\bibitem{Abada:2019zxq}
{\bfseries FCC} Collaboration, A.~Abada {\em et~al.}, ``{FCC-ee: The Lepton
  Collider}: {Future Circular Collider Conceptual Design Report Volume 2},''
  \href{http://dx.doi.org/10.1140/epjst/e2019-900045-4}{{\em Eur. Phys. J. ST}
  {\bfseries 228} no.~2, (2019) 261--623}.

\bibitem{Benedikt:2018csr}
{\bfseries FCC} Collaboration, A.~Abada {\em et~al.}, ``{FCC-hh: The Hadron
  Collider}: {Future Circular Collider Conceptual Design Report Volume 3},''
  \href{http://dx.doi.org/10.1140/epjst/e2019-900087-0}{{\em Eur. Phys. J. ST}
  {\bfseries 228} no.~4, (2019) 755--1107}.

\bibitem{Papaefstathiou:2015paa}
A.~Papaefstathiou and K.~Sakurai, ``{Triple Higgs boson production at a 100 TeV
  proton-proton collider},''
  \href{http://dx.doi.org/10.1007/JHEP02(2016)006}{{\em JHEP} {\bfseries 02}
  (2016) 006}, \href{http://arxiv.org/abs/1508.06524}{{\ttfamily
  arXiv:1508.06524 [hep-ph]}}.

\bibitem{Contino:2016spe}
R.~Contino {\em et~al.}, ``{Physics at a 100 TeV pp collider: Higgs and EW
  symmetry breaking studies},''
  \href{http://dx.doi.org/10.23731/CYRM-2017-003.255}{{\em CERN Yellow Report}
  no.~3, (2017) 255--440},
\href{http://arxiv.org/abs/1606.09408}{{\ttfamily arXiv:1606.09408 [hep-ph]}}.

\bibitem{Fuks:2017zkg}
B.~Fuks, J.~H. Kim, and S.~J. Lee, ``{Scrutinizing the Higgs quartic coupling
  at a future 100 TeV proton\textendash{}proton collider with taus and
  b-jets},'' \href{http://dx.doi.org/10.1016/j.physletb.2017.05.075}{{\em Phys.
  Lett. B} {\bfseries 771} (2017) 354--358},
  \href{http://arxiv.org/abs/1704.04298}{{\ttfamily arXiv:1704.04298
  [hep-ph]}}.

\bibitem{Bizon:2018syu}
W.~Bizo\'n, U.~Haisch, and L.~Rottoli, ``{Constraints on the quartic Higgs
  self-coupling from double-Higgs production at future hadron colliders},''
  \href{http://dx.doi.org/10.1007/JHEP10(2019)267}{{\em JHEP} {\bfseries 10}
  (2019) 267}, \href{http://arxiv.org/abs/1810.04665}{{\ttfamily
  arXiv:1810.04665 [hep-ph]}}.

\bibitem{Borowka:2018pxx}
S.~Borowka, C.~Duhr, F.~Maltoni, D.~Pagani, A.~Shivaji, and X.~Zhao, ``{Probing
  the scalar potential via double Higgs boson production at hadron
  colliders},'' \href{http://dx.doi.org/10.1007/JHEP04(2019)016}{{\em JHEP}
  {\bfseries 04} (2019) 016}, \href{http://arxiv.org/abs/1811.12366}{{\ttfamily
  arXiv:1811.12366 [hep-ph]}}.

\bibitem{Moretti:2001zz}
M.~Moretti, T.~Ohl, and J.~Reuter, ``{O'Mega: An optimizing matrix element
  generator},'' \href{http://arxiv.org/abs/hep-ph/0102195}{{\ttfamily
  arXiv:hep-ph/0102195}}.
\url{http://www-spires.dur.ac.uk/spires/find/hep/www?eprint=hep-ph/0102195}.

\bibitem{Kilian:2007gr}
W.~Kilian, T.~Ohl, and J.~Reuter, ``{WHIZARD: Simulating Multi-Particle
  Processes at LHC and ILC},''
  \href{http://dx.doi.org/10.1140/epjc/s10052-011-1742-y}{{\em Eur. Phys. J.}
  {\bfseries C71} (2007) 1742},
  \href{http://arxiv.org/abs/0708.4233}{{\ttfamily arXiv:0708.4233 [hep-ph]}}.
  \url{http://inspirehep.net/record/759495}.

\bibitem{Foster:1995ru}
G.~W. Foster and N.~V. Mokhov, ``{Backgrounds and detector performance at a 2
  $\times$2 TeV $\mu^+ \mu^-$ collider},''
  \href{http://dx.doi.org/10.1063/1.49354}{{\em AIP Conf. Proc.} {\bfseries
  352} (1996) 178--190}.

\bibitem{Johnstone:1996hp}
C.~J. Johnstone and N.~V. Mokhov, ``{Optimization of a muon collider
  interaction region with respect to detector backgrounds and the heat load to
  the cryogenic systems},'' {\em eConf} {\bfseries C960625} (1996) ACC030.

\bibitem{Mokhov:2011zzd}
N.~V. Mokhov and S.~I. Striganov, ``{Detector Background at Muon Colliders},''
  \href{http://dx.doi.org/10.1016/j.phpro.2012.03.761}{{\em Phys. Procedia}
  {\bfseries 37} (2012) 2015--2022},
  \href{http://arxiv.org/abs/1204.6721}{{\ttfamily arXiv:1204.6721
  [physics.ins-det]}}.

\bibitem{Alexahin:2011zz}
Y.~I. Alexahin, E.~Gianfelice-Wendt, V.~V. Kashikhin, N.~V. Mokhov, A.~V.
  Zlobin, and V.~Y. Alexakhin, ``{Muon collider interaction region design},''
  \href{http://dx.doi.org/10.1103/PhysRevSTAB.14.061001}{{\em Phys. Rev. ST
  Accel. Beams} {\bfseries 14} (2011) 061001},
  \href{http://arxiv.org/abs/1204.5739}{{\ttfamily arXiv:1204.5739
  [physics.acc-ph]}}.

\bibitem{Mokhov:2014hza}
N.~V. Mokhov, S.~I. Striganov, and I.~S. Tropin,
  \href{http://dx.doi.org/10.18429/JACoW-IPAC2014-TUPRO029}{``{Reducing
  Backgrounds in the Higgs Factory Muon Collider Detector},''} in {\em {5th
  International Particle Accelerator Conference}}.
\newblock 6, 2014.
\newblock \href{http://arxiv.org/abs/1409.1939}{{\ttfamily arXiv:1409.1939
  [physics.ins-det]}}.

\end{thebibliography}\endgroup
\bibliographystyle{utphys}

\printindex

\end{document}